%% file: ms.tex
\pdfoutput=1
\documentclass[acmsmall,authorversion,nonacm]{acmart}
\usepackage{colortbl}
\usepackage{tabu}
\usepackage{wrapfig}
\usepackage{fancyvrb}
\usepackage{tikz}
\usetikzlibrary{positioning, fit, calc, intersections, shapes.geometric, arrows, decorations.pathreplacing}
\usepackage{adjustbox}
\usepackage[justification=centering]{subfig}
\usepackage{unison}
\appto\UrlBreaks{\do\-}
\usepackage{gnuplot-lua-tikz}
\usepackage{numprint}
\usepackage{mathtools}
\usepackage{multirow}

\acmJournal{TOPLAS}

\setcopyright{acmlicensed}

\acmDOI{?}

\newcommand{\sics}{%
\affiliation{%
  \institution{RISE SICS}
  \streetaddress{Electrum 229}
  \city{Kista}
  \postcode{164 40}
  \country{Sweden}}}
\newcommand{\kth}{%
\affiliation{%
  \institution{KTH Royal Institute of Technology}
  \department{School of Electrical Engineering and Computer
    Science}
  \streetaddress{Electrum 229}
  \city{Kista}
  \postcode{164 40}
  \country{Sweden}}}
\AtBeginDocument{\let\textlabel\label}
\makeatletter
\newcommand{\mylabel}[2]{\raisebox{.7\normalbaselineskip}{\phantomsection}#1%
  \def\@currentlabel{#1}\textlabel{#2}}
\makeatother
\theoremstyle{definition}
\newtheorem{example}{Example}
\newcounter{var}
\newcommand{\varlabel}[1]{\refstepcounter{var}\label{#1}\tag{V\thevar}}
\newcounter{con}
\newcommand{\conlabel}[1]{\refstepcounter{con}\label{#1}\tag{C\thecon}}
\newcommand{\newconlabel}[3]{\label{#1}\tag{\ref*{#2}.#3}}

\input{results/hexagon-speed-summary}

\input{results/arm-speed-summary}

\input{results/mips-speed-summary}

\input{results/hexagon-size-summary}

\input{results/arm-size-summary}

\input{results/mips-size-summary}

\input{results/hexagon-all-summary}

\input{results/arm-all-summary}

\input{results/mips-all-summary}

\input{results/multi-all-improved-summary}

\input{results/multi-all-basic-summary}

\input{results/multi-results}

\input{results/multi-combined-results}

\input{results/combined-results}
\input{results/hexagon-speed-phases-summary}

\input{results/arm-speed-phases-summary}

\input{results/mips-speed-phases-summary}

\input{results/hexagon-size-phases-summary}

\input{results/arm-size-phases-summary}

\input{results/mips-size-phases-summary}

\input{results/hexagon-all-trivialbound-summary}

\input{results/arm-all-trivialbound-summary}

\input{results/mips-all-trivialbound-summary}

\input{results/gap-results}

\input{results/actual-speedup-results}

\input{results/global-stats}

\begin{document}
\title{Combinatorial Register Allocation and Instruction Scheduling}
\author{Roberto Casta\~{n}eda Lozano}
\orcid{0000-0002-2806-7333}
\sics{}
\kth{}
\email{roberto.castaneda@ri.se}
\author{Mats Carlsson}
\orcid{0000-0003-3079-8095}
\sics{}
\email{mats.carlsson@ri.se}
\author{Gabriel Hjort Blindell}
\orcid{0000-0001-6794-6413}
\kth{}
\email{ghb@kth.se}
\author{Christian Schulte}
\orcid{0000-0002-6283-7004}
\kth{}
\sics{}
\email{cschulte@kth.se}

\begin{abstract}
This paper introduces a combinatorial optimization approach to register
allocation and instruction scheduling, two central compiler problems.
Combinatorial optimization has the potential to solve these problems optimally
and to exploit processor-specific features readily.
Our approach is the first to leverage this potential \emph{in practice}: it
captures the \emph{complete} set of program transformations used in
state-of-the-art compilers, \emph{scales} to medium-sized functions of up
to $1000$~instructions, and generates \emph{executable} code.
This level of practicality is reached by using constraint programming, a
particularly suitable combinatorial optimization technique.
Unison, the implementation of our approach, is open source, used in industry,
and integrated with the LLVM toolchain.

An extensive evaluation confirms that Unison generates better code than LLVM
while scaling to medium-sized functions.
The evaluation uses systematically selected benchmarks from
MediaBench and SPEC CPU2006 and different processor architectures
(Hexagon, ARM, MIPS).
Mean estimated speedup ranges from~$\improvedARMSpeedMeanImprovement{}\%$
to~$\improvedHexagonSpeedMeanImprovement{}\%$ and mean code size reduction
ranges from~$\improvedHexagonSizeMeanImprovement{}\%$
to~$\improvedMipsSizeMeanImprovement{}\%$ for the different architectures.
A significant part of this improvement is due to the integrated nature of the
approach.
Executing the generated code on Hexagon confirms that the estimated speedup
results in actual speedup.
Given a fixed time limit, Unison solves optimally functions of up to
$\improvedmultiAllLargestOptimalIns{}$ instructions, nearly an order of
magnitude larger than previous approaches.

The results show that our combinatorial approach can be applied in practice to
trade compilation time for code quality beyond the usual compiler optimization
levels, identify improvement opportunities in heuristic algorithms, and fully
exploit processor-specific features.
\end{abstract}

%
%


%
%


\thanks{%
This paper is partially based on preliminary work presented at the
\emph{Principles and Practice of Constraint Programming}
(2012)~\cite{Castaneda2012}; \emph{Languages, Compilers, and Tools for Embedded
  Systems} (2014)~\cite{Castaneda2014}; and \emph{Compiler Construction}
(2016)~\cite{Castaneda2016} conferences.
Compared to the preliminary work, this paper is completely restructured and
rewritten, completes the combinatorial model with rematerialization, proposes
extensions to capture additional program transformations and processor-specific
features, and contributes a more exhaustive evaluation.
Additions to the evaluation include more benchmarks and processors, evidence of
the fundamental benefit of the integrated approach, an in-depth study of
scalability, and actual execution measurements.}

\newtoggle{applyHighlight}
\toggletrue{applyHighlight}

\input{constraints}

\maketitle

\section{Introduction}\label{sec:introduction}

Register allocation and instruction scheduling are central compiler problems.
Register allocation maps temporaries (program or compiler-generated variables)
to registers or memory.
Instruction scheduling reorders instructions to improve total latency or throughput.
Both problems are key to generating high-quality assembly code, are
computationally complex (NP-hard), and are mutually interdependent: no matter in
which order these problems are approached, aggressively solving one of them
might worsen the outcome of the
other~\cite{Goodman1988,Govindarajan2007,Nandivada2007,Hennessy2011}.

Today's compilers typically solve each problem in isolation
with heuristic algorithms, taking a sequence of greedy decisions
based on local optimization criteria.
This arrangement reduces compilation time but precludes optimal solutions
and complicates the exploitation of processor-specific features.

Combinatorial optimization has been proposed to address these limitations: it
can deliver optimal solutions according to a model, it can accurately capture
problem interdependencies, and it can accommodate specific architectural
features at the expense of increased compilation time~\cite{Castaneda2014b}.
While this approach shows promise for register allocation and
instruction scheduling, its practical significance has not yet
been established.
Typically, combinatorial approaches model only a subset of the program
transformations offered by their heuristic counterparts, do not scale beyond
small problems of up to 100 input instructions, and are not known to generate
executable code.

The goal of this paper is to introduce a combinatorial approach to integrated
register allocation and instruction scheduling that is practical.
By \emph{practical} we mean that the approach is:
\begin{description}
\item[Complete.] It models the same set of program transformations
  as state-of-the-art compilers.
\item[Scalable.] It scales to medium-sized problems of up to 1000 input
  instructions.
\item[Executable.] It generates executable code.
\end{description}
This combination of properties allows us, for the first time, to evaluate the
potential benefit of combinatorial register allocation and instruction
scheduling in practice:
completeness enables a direct comparison with heuristic approaches;
scalability puts most problems from typical benchmark suites within reach; and
executability is a precondition to study the practical significance.
Our evaluation shows that this potential benefit can be achieved, generating
better code than state-of-the-art heuristic approaches for a variety of
benchmarks, processors, and optimization goals.

Our approach has practical applications for both compiler users and developers,
as corroborated by our research partner Ericsson~\cite{ericsson-blog}.
Compiler users can apply it to trade compilation time for code quality beyond
the usual compiler optimization levels.
This is particularly attractive if longer compilation times are tolerable, such
as for compilation of high-quality library releases or embedded systems
software~\cite{Fisher2005}.
If compilation time is more limited, our approach might still be applied to the
most frequently executed parts of a program.
Compiler developers can apply the combinatorial approach as a baseline for
assessment and development of heuristic approaches, exploiting the optimal,
integrated nature of its solutions.
Furthermore, comparing the generated code can reveal improvement opportunities
in heuristic algorithms, even for production-quality compilers such as
LLVM~\cite{UnisonLLVM2017}.
Another application is to generate high-quality code for processor-specific,
irregular features.
Such features can be naturally modeled and fully exploited by combinatorial
approaches, while adapting heuristic approaches tends to involve a major
effort~\cite{Leupers2001}.
The ease of modeling of our approach can also be exploited for rapid compiler
prototyping for new processor revisions.

\begin{figure}%
  \centering%
  \adjustbox{trim=0.75cm 0cm 0cm 0cm,clip=true,scale=0.98}{%
    \scalebox{1}{\input{./figures/factorial-motivation}}}
  \subfloat[\label{fig:factorial-c} Factorial in C.]{\hspace{.2\linewidth}}
  \subfloat[\label{fig:factorial-assignment} Register assignment.]{\hspace{.3\linewidth}}
  \subfloat[\label{fig:factorial-assembly} Simplified assembly code.]{\hspace{.4\linewidth}}
  \subfloat{\hspace{.2\linewidth}}
  \caption{Motivating example: Hexagon assembly code generated by LLVM and
    our approach for factorial.\label{fig:factorial-motivation}}
\end{figure}
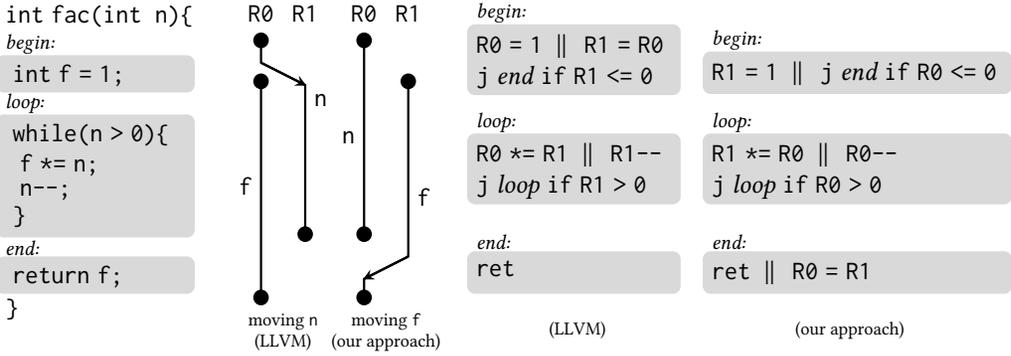

\begin{example}\label{ex:motivation}
Figure~\ref{fig:factorial-motivation} illustrates our combinatorial
approach in practice~\cite{UnisonDemo2016}.
Figure~\ref{fig:factorial-c} shows a C~implementation of the factorial function.
The function takes \code{n} as argument, initializes \code{f} to one,
iterates multiplying and accumulating \code{f} with \code{n} in each
iteration until \code{n} is zero, and returns \code{f}.

Register allocation for this function must assign \code{n} and \code{f} to
different registers within the \emph{loop} basic block, since their values
would be otherwise clobbered (that is, mutually overwritten).
Let us assume that the target processor is Hexagon, a multiple-issue
digital signal processor ubiquitous in modern mobile
platforms~\cite{Codrescu2014}.
Its calling convention forces the assignment of \code{n} and \code{f} to the
same register (\code{R0}) on entry and on return
respectively~\cite{HexagonABI2013}.
These clobbering and calling convention constraints can only be satisfied by
assigning one of the program variables $v \in \set{\code{n},\code{f}}$ to some
other register than \code{R0} within \emph{loop} and moving $v$ to or from
\code{R0} outside of \emph{loop} with a register-to-register move instruction.

Figure~\ref{fig:factorial-assignment} shows how the state-of-the-art heuristic
compiler LLVM~\cite{Lattner2004} and our approach
produce opposite assignments by moving either \code{n} or \code{f}.
From an isolated register allocation perspective, both
assignments incur the same cost: an additional move instruction
at \emph{begin} (LLVM) or \emph{end} (our approach) is
required.
However, our integrated approach properly reflects that
only the move of \code{f} can be parallelized ($\bundled$) with
another instruction (\code{ret}), yielding slightly faster
assembly code (Figure~\ref{fig:factorial-assembly}).
This advanced optimization level requires reasoning about multiple aspects of
global register allocation and instruction scheduling in integration.
\end{example}

\begin{figure}%
  \centering%
  \adjustbox{trim=0.3cm 0cm 0cm 0cm,clip=true,scale=0.98}{%
    \scalebox{1}{\input{./figures/approach}}}
  \caption{Our approach to combinatorial register allocation and instruction scheduling.\label{fig:approach}}
\end{figure}
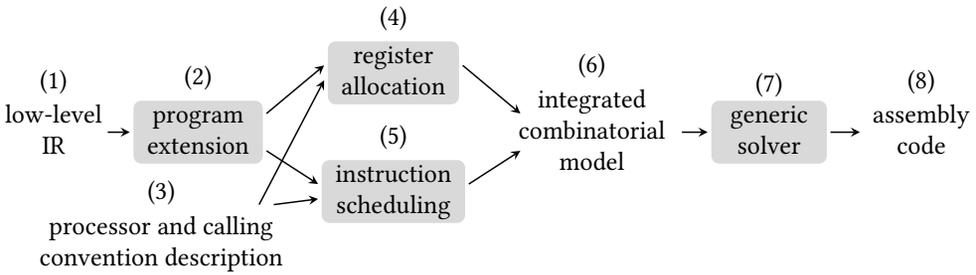

\paragraph{Approach}

Our approach is outlined in Figure~\ref{fig:approach}.
A low-level intermediate representation (IR) of a function with instructions
from a specific processor is taken as input~(\ref{step:input}).
The representation of the function is extended~(\ref{step:construction}) to
expose its structure and the multiple decisions involved in the problem.
From the extended function and a description of the processor and calling
convention~(\ref{step:description}), combinatorial models of register
allocation~(\ref{step:regalloc}) and instruction scheduling~(\ref{step:isched})
are derived.
A combinatorial model consists of variables representing the problem decisions,
program and processor constraints over the variables, and an objective function
to be optimized.
Both models are then integrated into a single model~(\ref{step:problem}) that
precisely captures the interdependencies between the corresponding problems.
Solving the model with a generic solver~(\ref{step:solver}) gives a
register allocation and an instruction schedule used to generate executable
assembly code~(\ref{step:output}).

A combinatorial model is of limited practical value unless complemented with
suitable solving techniques, effective solving methods, and a robust
implementation that gives confidence in the results.
Our approach is implemented in \emph{Unison}~\cite{UnisonTool2018}, a software
tool that uses constraint programming~\cite{Rossi2006} as a modern combinatorial
optimization technique.
Constraint programming is particularly capable of exploiting the structure
underlying the register allocation and instruction scheduling problems.
Unison applies general and problem-specific constraint solving methods with
which medium-sized functions can be solved optimally.
It also deals with the practicalities of generating executable code (such as
calling convention and call stack management), delegating decisions that are
interdependent with register allocation and instruction scheduling to the
combinatorial model.
Unison is a robust tool whose results can be used with
confidence: it is open source, well-documented, systematically
tested, used in industry, and integrated with the LLVM toolchain.

\paragraph{Contributions}

This paper contributes the first combinatorial approach to register allocation
and instruction scheduling that is \emph{complete}, \emph{scales} up to
medium-sized problems, and generates \emph{executable} code.

The combinatorial model is \emph{complete} as it includes spilling (internalized
into the model), register assignment and packing, coalescing, load-store
optimization, live range splitting, rematerialization, and multi-allocation.
This is enabled by a novel combination of abstractions that capture different
aspects of register allocation together with a suitable program representation.
In addition, the paper introduces model extensions for features such as stack
frame elimination, latencies across basic blocks, operand forwarding, and
two-address instructions.

The paper introduces solving methods that are crucial for \emph{scalability}.
Extensive experiments on MediaBench~\cite{Lee1997} and SPEC
CPU2006~\cite{SpecCPU2006} functions for three different processor architectures
(Hexagon, ARM, MIPS) show that, given a time limit of~15 minutes, our approach
solves optimally functions of up to $\improvedmultiAllLargestOptimalIns{}$
instructions.
Under this time limit, the percentage of functions solved optimally ranges
from~$\improvedMipsAllPercentOptimal{}\%$
to~$\improvedHexagonAllPercentOptimal{}\%$ across processors,
and~$\percentOfSolvedFunctionsWithAGapOfTen{}\%$ of the functions on Hexagon are
solved optimally or less than~$10\%$ away from the optimal solution.

The experiments confirm that there is a potential benefit to be gained by solving
register allocation and instruction scheduling in integration.
Our approach exploits this potential, delivering a mean estimated speedup over
LLVM ranging from~$\improvedARMSpeedMeanImprovement{}\%$
to~$\improvedHexagonSpeedMeanImprovement{}\%$ and a mean code size reduction
ranging from~$\improvedHexagonSizeMeanImprovement{}\%$
to~$\improvedMipsSizeMeanImprovement{}\%$, depending on the characteristics of
the processor.
For the first time, the speedup estimation is examined by \emph{executing}
MediaBench applications on Hexagon, the processor with highest estimated speedup
in the experiments.
The results show that the approach achieves significant speedup in practice
($\baseMeanImprovement{}\%$ across functions and $\multiMeanImprovement{}\%$
across whole applications).

\paragraph{Plan of the paper}

Section~\ref{sec:background} covers the background on register
allocation, instruction scheduling, and combinatorial
optimization.
Section~\ref{sec:related-work} reviews related approaches.

Sections~\ref{sec:local-register-allocation}-\ref{sec:model-extensions}
introduce the combinatorial model and its associated program
representation. They incrementally describe local register
allocation (Section~\ref{sec:local-register-allocation}), global
register allocation
(Section~\ref{sec:global-register-allocation}), instruction
scheduling (Section~\ref{sec:instruction-scheduling}), the
integrated model
(Section~\ref{sec:global-register-allocation-and-instruction-scheduling}),
its objective function (Section~\ref{sec:objective-function}),
and additional program transformations
(Section~\ref{sec:model-extensions}).
Appendix~\ref{app:combinatorial-model} complements this incremental description
with a summary of the parameters, variables, constraints, and objective function
in the combinatorial model.

Section~\ref{sec:solving-in-unison} outlines the solving methods
employed by Unison.
Section~\ref{sec:experimental-evaluation} presents the experimental evaluation,
where Appendix~\ref{app:accuracy} studies the accuracy of the speedup estimation
and Appendix~\ref{app:functions} describes the functions used in the evaluation.
Section~\ref{sec:conclusion-and-future-work} concludes and proposes future work.

\section{Background}\label{sec:background}

This section reviews the input program representation assumed in the paper, the
register allocation and instruction scheduling problems, and constraint
programming as the key technique for modeling and solving register allocation
and instruction scheduling.

\subsection{Input Program Representation}\label{sec:background-input-program-representation}

This paper assumes as input a function represented by its control-flow graph
(CFG), with instructions for a specific processor already selected.
Instructions \emph{define} (assign) and \emph{use} (access) program and
compiler-generated variables.
These variables are referred to as \emph{temporaries} and are associated with a
bit-width derived from their source data type.

\emph{Program points} are locations between consecutive instructions.
A temporary~$t$ defined by an instruction~$i$ is \emph{live} at a
program point if $t$~holds a value that might be used later by
another instruction~$j$.
In this situation, instruction~$j$ is said to be \emph{dependent} on~$i$.
A temporary~$t$ is \emph{live-in} (\emph{live-out}) in a basic
block~$b$ if $t$ is live at the entry (exit) of $b$.
The \emph{live range} of a temporary~$t$ is the set of program
points where~$t$ is live.
Two temporaries \emph{interfere} if their live ranges are not disjoint.

The paper assumes that all temporaries that are live-in into the function (such
as function arguments) are defined by a \emph{virtual entry instruction} at the
entry basic block.
Similarly, \emph{virtual exit instructions} use temporaries that are live-out at
the exit basic blocks of the function.
\emph{Virtual} (as opposed to \emph{real}) instructions are pseudo-instructions
introduced to support compilation and do not appear in the generated assembly
code.
Temporaries being live-in into the function and live-out of the
function are pre-assigned according to the calling convention.
A pre-assignment of a temporary~$t$ to a register~$r$ is denoted as
$\preAssigned{t}{r}$.

\begin{figure}%
  \centering%
  \adjustbox{trim=0cm 0cm 0cm 0cm,clip=true,scale=0.98}{%
    \scalebox{1}{\input{./figures/factorial-input}}}
  \caption{CFG of the factorial function with Hexagon instructions.\label{fig:factorial-input}}
\end{figure}
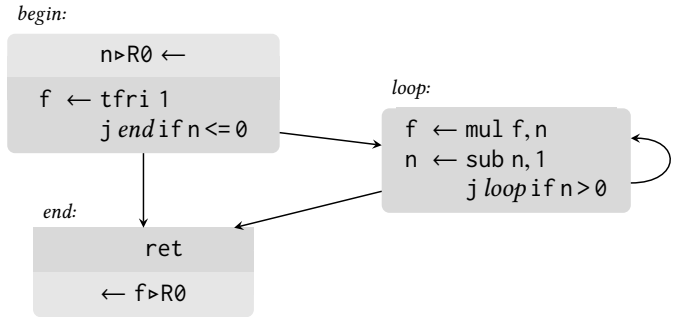

\begin{example}\label{ex:register-allocation-input}
Figure~\ref{fig:factorial-input} shows the CFG from Example~\ref{ex:motivation}
with Hexagon instructions.
For readability, Hexagon immediate transfer
(\code{\transferImmediateOpcode{}}), conditional jump
(\code{\jumpIfOpcode{}}), multiply (\code{\multiplyOpcode{}}), subtract
(\code{\subtractOpcode{}}), and jump to return address
(\code{\indirectJumpOpcode{}}) instructions are shown in simplified
syntax.
The body of a basic block is shown in dark gray while the boundaries containing
entry and exit instructions are shown in light gray.
For simplicity, \emph{critical edges} (arcs from basic blocks with multiple
successors to basic blocks with multiple predecessors) are preserved in the
example CFG.
In practice, such edges are often split by earlier compiler transformations.

The 32-bit temporaries $\TempN{}$ and $\TempF{}$ correspond to the
$\code{int}$ variables in
Example~\ref{ex:motivation}.
$\TempN{}$ is live-in and hence defined by the
entry instruction in the \emph{begin} basic block.
$\TempF{}$ is live-out and hence used by the exit instruction in the \emph{end}
basic block.
Hexagon's calling convention requires the pre-assignment
$\preAssigned{\TempN{}}{\register{R0}}$ on the function's entry, and
$\preAssigned{\TempF{}}{\register{R0}}$ on the function's exit.
The live range of~$\TempN{}$ starts with its definition by the entry instruction
in the \emph{begin} basic block and spans the entire \emph{loop} basic block.
The live range of~$\TempF{}$ starts with its definition by
\code{\transferImmediateOpcode{}} in the \emph{begin} basic block and
extends until the exit instruction in the \emph{end} basic block.
$\TempN{}$~and~$\TempF{}$ interfere as their live ranges are not disjoint.
\end{example}

\subsection{Register Allocation}\label{sec:background-register-allocation}

Register allocation maps temporaries to either processor registers or memory.
While processor registers have shorter access times, they are limited in number.
This shortage of registers leads to a high \emph{register pressure} for
functions where many temporaries interfere.
In the worst case, this may force the allocation (\emph{spilling}) of some
of the temporaries to memory.
The spill of a temporary~$t$ is implemented by inserting code (often costly
store and load instructions) to move~$t$'s value to and from memory.
In its most basic form (called \emph{spill-everywhere}), \emph{spill code} is
generated by inserting store instructions after each definition of~$t$ and load
instructions before each use of~$t$.

Register allocation applies a number of program transformations to reduce
register pressure, spilling, or --- if the latter is unavoidable --- the
overhead of the spill code:
\begin{description}
\item[Register assignment] maps non-spilled temporaries to
  individual registers, reducing register pressure by reusing
  registers among non-interfering temporaries.
\item[Live range splitting] allocates temporaries to different
  locations in different parts of their live ranges.
\item[Coalescing] allocates a temporary's split live range to the
  same location to eliminate the corresponding copy instructions
  (dual of live range splitting).
\item[Load-store optimization] avoids reloading spilled values by
  reusing values loaded in previous parts of the spill code.
\item[Multi-allocation] allocates temporaries simultaneously
  to registers as well as memory, reducing the overhead of
  spilling in certain scenarios~\cite{Colombet2015}.
\item[Register packing] assigns temporaries of small bit-widths
  to different parts of larger-width registers (for processors
  supporting such assignments) to improve register utilization.
\item[Rematerialization] recomputes values at their use points rather than
  spilling them when the recomputation instructions are deemed less costly than
  the spill code.
\end{description}

\paragraph{Scope}
Register allocation can be solved locally or globally.
\emph{Local} register allocation handles single basic blocks, spilling all
temporaries that are live across basic block boundaries.
\emph{Global} register allocation handles entire functions, potentially
improving code quality by keeping temporaries in registers across basic block
boundaries.

\paragraph{Calling convention management}
Register allocation is responsible to enforce the pre-assignment of arguments
and return values to certain registers and the preservation of
\emph{callee-saved} registers across function calls as dictated by the calling
convention.
A common approach to handle callee-saved registers is to hold their values in
temporaries that are live through the entire function and that can be spilled at
the function's entry and exit.
This approach can be refined by moving the spill code closer to the program
points where the callee-saved registers are needed
(\emph{shrink-wrapping}~\cite{Chow1988}).

\paragraph{SSA-based register allocation}
Static single assignment (SSA)~\cite{Cytron1991} is a program form in which
temporaries are defined exactly once.
This form is useful for register allocation as it allows register assignment to
be solved optimally in polynomial time~\cite{Hack2006}.
SSA places $\phi$-functions of the form $\phiOperation{t_n}{t_1, t_{2},
  \dots, t_{n-1}}$ at the entry of basic blocks with multiple incoming arcs
to distinguish which definition among $\set{t_1, t_2, \dots, t_{n-1}}$
reaches $t_n$ depending on the program execution.
Temporaries related transitively by $\phi$-functions are called
\emph{$\phi$-congruent}~\cite{Sreedhar1999}.
This paper uses a generalized form of SSA for global register allocation, as
Section~\ref{sec:global-register-allocation-program-representation} explains.

\subsection{Instruction Scheduling}\label{sec:background-instruction-scheduling}

Instruction scheduling assigns issue cycles to program instructions.
Valid instruction schedules must satisfy instruction dependencies and
constraints imposed by limited processor resources.

\paragraph{Latency}
Dependencies between instructions are often associated with a
latency indicating the minimum number of cycles that must elapse
between the issue of the depending instructions.
Variable latencies (such as those arising from cache memory accesses) are
typically handled by assuming the best case and relying on the processor to
stall the execution otherwise~\cite{Govindarajan2007}.

\paragraph{Resources}
The organization of hardware resources varies heavily among
different processors and has a profound impact on the complexity
of instruction scheduling.
This paper assumes a resource model where each resource $s$ has a capacity
$\capacity{s}$ and each instruction $i$ consumes $\units{i}{s}$ units of
each resource $s$ during $\duration{i}{s}$ cycles.
This model is sufficiently general to capture the resources of the processors
studied in this paper, including Very Long Instruction Word (VLIW) processors
such as Hexagon that can issue multiple instructions in each cycle.
These processors can be modeled by an additional resource with capacity equal to
the processor's issue width.

\paragraph{Order}
Non-integrated compilers perform instruction scheduling before or after
register allocation.
Pre-register-allocation scheduling typically seeks a schedule that reduces
register pressure during register allocation, while post-register-allocation
scheduling often aims at minimizing the duration of the schedule.
The latter is particularly relevant for \emph{in-order} processors which
issue instructions according to the schedule produced by the compiler,
although \emph{out-of-order} processors can also benefit from compile-time
instruction scheduling~\cite{Govindarajan2007}.

\paragraph{Scope}
Instruction scheduling can be solved at different program scopes.
This paper is concerned with \emph{local} instruction scheduling, where the
instructions from each basic block are scheduled independently.
Larger scopes to which instruction scheduling can be applied include
\emph{superblocks} (consecutive basic blocks with multiple exit points but a
single entry point) and \emph{loops} (where the instructions of multiple
iterations are scheduled simultaneously in a pipelined fashion).

\subsection{Constraint Programming}\label{sec:combinatorial-optimization}

Combinatorial optimization is a collection of techniques to solve
computationally hard combinatorial optimization problems such as register
allocation and instruction scheduling.
Examples of these techniques are constraint programming (CP)~\cite{Rossi2006},
integer programming (IP)~\cite{Nemhauser1999}, Boolean satisfiability
(SAT)~\cite{Biere2009}, partitioned Boolean quadratic programming
(PBQP)~\cite{Scholz2002}, and dynamic programming (DP)~\cite{Cormen2009}.
Their key property is \emph{completeness}: they automatically explore the full
solution space and, given enough time, guarantee to find the optimal solution if
there is one.

Combinatorial optimization techniques capture the problems to be solved as
combinatorial models.
A combinatorial model consists of \emph{variables} capturing problem decisions,
\emph{constraints} expressing relations over the variables that must be
satisfied by a solution, and an \emph{objective function} to be minimized (or
maximized) expressed in terms of the variables.
A \emph{solution} to the model is an assignment of values to the variables that
satisfies all the constraints, and an \emph{optimal solution} minimizes the
value of the objective function.
In this paper variables are written in bold (e.g.\ $\modelSingleVariable{x}$),
and indexable variables are written as $\modelVariable{x}{i}$.

Combinatorial optimization techniques differ significantly in the level of
abstraction of their models, underlying solving methods, and problem classes for
which they are particularly well-suited.
This paper uses CP as a technique that is particularly suitable for handling
register allocation and instruction scheduling problems.
In CP, variables usually range over finite subsets of the integers or Booleans,
and constraints and objective function are expressed by general relations.
The purpose of this section is to provide enough information so that the models
developed in
Sections~\mbox{\ref{sec:local-register-allocation}-\ref{sec:model-extensions}}
are understandable.
Some additional modeling and solving methods are presented in
Section~\ref{sec:solving-in-unison}.

Constraint solvers proceed by interleaving \emph{constraint propagation} and
\emph{branch-and-bound search}.
Constraint propagation discards value assignments that cannot be part of a
solution to reduce the search space.
Constraint propagation is applied iteratively until no more
value assignments can be discarded~\cite{Bessiere2006}.
If several values can still be assigned to a variable, search is
used to decompose the problem into alternative subproblems on
which propagation and search are repeated.
Solutions found during solving are exploited in a
branch-and-bound fashion to further reduce the search
space~\cite{VanBeek2006}: after a solution is found, constraints
are added such that the next solution must be better according to the
objective function.

As is common in combinatorial optimization, constraint solvers are usually run
with a time limit as it can be prohibitive to find the optimal solution to
large problems.
Constraint solvers exhibit \emph{anytime behavior} in that they can often
deliver suboptimal solutions if they time out.
Even when no solution is delivered, the solvers always provide the
\emph{optimality gap} (hereafter just \emph{gap}), an upper bound on the
distance to the optimal solution.
The gap can be used as an estimation of the computational effort
to obtain the optimal solution, as a certificate of the quality
of suboptimal solutions, or as a criterion for solver
termination.

\newcommand{\cumulativeStart}[1]{\modelVariable{s}{#1}}
\newcommand{\cumulativeCon}[1]{\modelVariable{c}{#1}}
\newcommand{\cumulativeDur}[1]{\modelVariable{d}{#1}}

\newcommand{\nooverlapLeft}[1]{\modelVariable{l}{#1}}
\newcommand{\nooverlapRight}[1]{\modelVariable{r}{#1}}
\newcommand{\nooverlapTop}[1]{\modelVariable{t}{#1}}
\newcommand{\nooverlapBottom}[1]{\modelVariable{b}{#1}}

\newcommand{\elementIndex}[1]{\modelSingleVariable{x}}
\newcommand{\elementValue}[1]{\modelSingleVariable{y}}

\paragraph{Global constraints}
A key feature of CP is the use of \emph{global constraints} that capture
recurring modeling substructures involving multiple variables.
Besides being convenient for modeling, global constraints are essential for
efficient solving since they are implemented by dedicated propagation
algorithms that further reduce the search space~\cite{VanHoeve2006}.
The model introduced in this paper
(Sections~\mbox{\ref{sec:local-register-allocation}-\ref{sec:model-extensions}})
uses the following three global constraints:
\begin{itemize}
\item $\cumulative{\setBuilder{\tuple{\cumulativeStart{i},\cumulativeDur{i},\cumulativeCon{i}}}{i \squeeze{\in} \sequence{1, n}}\!, b}$
ensures that a set of $n$ tasks does not exceed a given resource capacity $b$, where
each task $i$ is defined by its start time $\cumulativeStart{i}$, duration
$\cumulativeDur{i}$, and resource units consumed $\cumulativeCon{i}$~\cite{Aggoun1993}:
\begin{equation}
  \sum_{\mathclap{i \in \sequence{1, n} \suchThat
    \cumulativeStart{i} \le k \land
    \cumulativeStart{i} + \cumulativeDur{i} > k}}
      {\cumulativeCon{i}} \le b \quad
      \forall k \squeeze{\in} \integerNumbers{}\,.
\end{equation}
\item
$\disjoint{\setBuilder{\tuple{\nooverlapLeft{i},\nooverlapRight{i},\nooverlapTop{i},\nooverlapBottom{i}}}{i \squeeze{\in} \sequence{1, n}}}$
(also known as \emph{diffn}) ensures that a set of $n$ rectangles does
not overlap, where each rectangle $i$ is defined by its left
$\nooverlapLeft{i}$, right $\nooverlapRight{i}$, top $\nooverlapTop{i}$,
and bottom $\nooverlapBottom{i}$ coordinates~\cite{Beldiceanu1994}:
\begin{equation}
  \nooverlapRight{i} \squeeze{\le} \nooverlapLeft{j} \lor
  \nooverlapLeft{i} \squeeze{\ge} \nooverlapRight{j} \lor
  \nooverlapBottom{i} \squeeze{\le} \nooverlapTop{j} \lor
  \nooverlapTop{i} \squeeze{\ge} \nooverlapBottom{j} \quad
  \forallIn{i,j}{\sequence{1, n}} \suchThat i \ne j.
\end{equation}
\item
$\element{\elementIndex{}, a, \elementValue{}}$ generalizes array
  access to variables by ensuring that the variable $\elementValue{}$ is
  equal to the $\elementIndex{}$\textsuperscript{th} element of an array
  $a$ of integers or integer variables~\cite{VanHentenryck1988}:
\begin{equation}
  a\elementMark{\elementIndex{}} = \elementValue{}.
\end{equation}
As a compromise between readability and precision, the rest of the paper
expresses this constraint using the array notation.
This can be extended to multi-dimensional arrays provided that only a single
integer variable is used for indexing.
Constraints using lookup functions, such as $f(\elementIndex{}) =
\elementValue{}$, can be expressed as $f\elementMark{\elementIndex{}} =
\elementValue{}$ by converting $f$ into an array.
Using this constraint, we can also implement the constraint
\begin{equation}
  \elementValue{} \in a\elementMark{\elementIndex{}}
\end{equation}
by introducing, for each such constraint, a new array~$b$ of integer variables,
where each integer variable~$b[i]$ has a domain equal to $a[i]$, and enforcing
$b\elementMark{\elementIndex{}} = \elementValue{}$.

\end{itemize}

\section{Related Approaches}\label{sec:related-work}

This section provides an overview of combinatorial approaches to register
allocation and instruction scheduling.
The overview is complemented by more specific discussions in the rest of the
paper.
A more comprehensive review is provided by \citet{Castaneda2014b}.

\paragraph{Combinatorial register allocation}
Combinatorial approaches to register allocation in isolation
(Table~\ref{tab:combinatorial-register-allocation}) have been proposed that
satisfy all properties required to be \emph{practical}: they model most or all
of the standard program transformations (\emph{completeness}, columns
\textbf{SP}-\textbf{MA}), scale to medium-sized problems
(\emph{scalability}, column \textbf{SZ}), and generate executable code
(\emph{executability}, column \textbf{EX}).
Furthermore, their ability to accommodate specific architectural features and
alternative optimization objectives has been demonstrated in numerous
studies~\cite{Naik2002,Nandivada2006,Barik2007,Falk2011}.
A particular focus has been to study the trade-off between solution quality and
scalability.
Numerous approaches~\cite{Appel2001,Ebner2009,Colombet2015} advocate solving
spilling first (including relevant aspects of coalescing in the case of Colombet
\etal{}~\cite{Colombet2015}), followed by register assignment and coalescing.
This arrangement can improve scalability with virtually no performance loss for
single-issue and out-of-order processors, but is less suitable when register
assignment and coalescing have high impact on code quality, such as in code size
optimization~\cite{Koes2009} or in speed optimization for VLIW
processors~\cite{Colombet2015}.

\begin{table}
  \newcommand{\raApproach}[9]{#7 & #6 & #2 & #1 & #3 & #8 & #4 & #9}
  \newcommand{\newTableSection}{\hline\\[-0.3cm]}
  \setlength{\tabcolsep}{2pt}
  \setlength{\tabulinesep}{0.03cm}
  \newcommand{\columnWidth}{0.5cm}
  \caption{Combinatorial register allocation approaches ordered
    chronologically: technique, scope, spilling (\textbf{SP}), register
    assignment (\textbf{RA}), coalescing (\textbf{CO}), load-store
    optimization (\textbf{LO}), register packing (\textbf{RP}), live range
    splitting (\textbf{LS}), rematerialization (\textbf{RM}), multi-allocation
    (\textbf{MA}), size of largest problem solved optimally
    (\textbf{SZ}) in input instructions, and whether it is demonstrated to
    generate executable code
    (\textbf{EX}).\label{tab:combinatorial-register-allocation}}
  \begin{tabu}{%
        l%
        c%
        c%
        >{\centering\arraybackslash}p{\columnWidth}%
        >{\centering\arraybackslash}p{\columnWidth}%
        >{\centering\arraybackslash}p{\columnWidth}%
        >{\centering\arraybackslash}p{\columnWidth}%
        >{\centering\arraybackslash}p{\columnWidth}%
        >{\centering\arraybackslash}p{\columnWidth}%
        >{\centering\arraybackslash}p{\columnWidth}%
        >{\centering\arraybackslash}p{\columnWidth}%
        c%
        >{\centering\arraybackslash}p{\columnWidth}}\hline
    \rowfont{\bfseries}
    approach & technique & scope & \raApproach{LO}{CO}{RP}{RM}{MB}{RA}{SP}{LS}{MA} & SZ & EX \\
    \hline
    \rowcolor{tblrow} Goodwin~\etal{}~\cite{Goodwin1996} & IP & global & \raApproach{\yes}{\yes}{\yes}{\yes}{\yes}{\yes}{\yes}{\yes}{\yes} & $\sim{}$2000 & \yes\\
    Appel~\etal{}~\cite{Appel2001} & IP & global & \raApproach{\yes}{\no}{\no}{\no}{\no}{\no}{\yes}{\yes}{\no} & $\sim{}$2000 & \yes \\
    \rowcolor{tblrow} Scholz~\etal{}~\cite{Scholz2002} & PBQP & global & \raApproach{\no}{\yes}{\yes}{\yes}{\no}{\yes}{\yes}{\no}{\no} & $\sim{}$180 & \yes \\
    Nandivada~\etal{}~\cite{Nandivada2006} & IP & global & \raApproach{\yes}{\no}{\no}{\no}{\no}{\yes}{\yes}{\yes}{\no} & ? & \yes \\
    \rowcolor{tblrow} Koes~\etal{}~\cite{Koes2006} & IP & global & \raApproach{\yes}{\no}{\no}{\yes}{\yes}{\yes}{\yes}{\yes}{\no} & ? & \yes \\
    Barik~\etal{}~\cite{Barik2007} & IP & global & \raApproach{\yes}{\no}{\yes}{\yes}{\yes}{\yes}{\yes}{\yes}{\no} & 302 & \no \\
    \rowcolor{tblrow} Ebner~\etal{}~\cite{Ebner2009} & IP & global & \raApproach{\yes}{\no}{\no}{\no}{\no}{\no}{\yes}{\yes}{\no} & ? & \yes \\
    Falk~\etal{}~\cite{Falk2011} & IP & global & \raApproach{\yes}{\no}{\no}{\no}{\yes}{\yes}{\yes}{\yes}{\yes} & $\sim{}$1000 & \yes \\
    \rowcolor{tblrow} Colombet~\etal{}~\cite{Colombet2015} & IP & global & \raApproach{\yes}{\no}{\no}{\yes}{\no}{\no}{\yes}{\yes}{\yes} & ? & \yes \\
    \end{tabu}%
\end{table}

\paragraph{Combinatorial instruction scheduling}
A large number of combinatorial instruction scheduling approaches have been
proposed, where the underlying resource-constrained project scheduling problem
has already been solved with IP in the early
1960s~\cite{Bowman1959,Wagner1959,Manne1960}.
\emph{Practical} approaches have been proposed for both
local~\cite{Shobaki2013} and global~\cite{Winkel2007,Barany2013}
instruction scheduling.
Other approaches that scale to medium-sized problems but do not demonstrate
executable code have been proposed for local~\cite{Wilken2000,Malik2008} and
superblock~\cite{Malik2008b} instruction scheduling.
A detailed review of combinatorial instruction scheduling is provided by
\citet{Castaneda2014b}.

\paragraph{Combinatorial register allocation and instruction scheduling}
Integrated combinatorial approaches
(Table~\ref{tab:combinatorial-integrated}) capture the
interdependencies between register allocation and instruction
scheduling in a single combinatorial model. They typically are
less scalable but can deliver better solutions than isolated
approaches.
A particular focus has been on VLIW processors, for which the interdependencies
between both problems are strong~\cite{Kessler2010}.
As Table~\ref{tab:combinatorial-integrated} shows, our integrated
approach is the first that matches the practicality of isolated register
allocation approaches.  This concerns the program transformations modeled
(\emph{completeness}, columns \textbf{SP}-\textbf{MA}), the \emph{scalability}
(column \textbf{SZ}), and the \emph{executability} (column \textbf{EX}).
While earlier integrated approaches might be able to generate executable code,
this fact is not demonstrated in the respective publications, which precludes an
evaluation of their practical significance.
Moreover, only the approach by Wilson \etal{}~\cite{Wilson1994} models register
allocation at the global scope.

A particularly challenging problem not addressed by this paper is
to incorporate instruction selection (column \textbf{SE}).
The complexity of instruction selection and its interdependencies
with register allocation and instruction scheduling are
well-understood~\cite{HjortBlindell2016}. However, the potential
benefits and scalability challenges of incorporating instruction
selection into our integrated approach have not yet been
explored.

\begin{table}
  \newcommand{\inApproach}[9]{#7 & #6 & #2 & #1 & #3 & #8 & #4 & #9}
  \newcommand{\newTableSection}{\hline\\[-0.3cm]}
  \setlength{\tabcolsep}{1.85pt}
  \setlength{\tabulinesep}{0.03cm}
  \newcommand{\columnWidth}{0.5cm}
  \caption{Combinatorial register allocation and instruction scheduling
    approaches ordered chronologically: technique, scope, spilling
    (\textbf{SP}), register assignment (\textbf{RA}), coalescing
    (\textbf{CO}), load-store optimization (\textbf{LO}), register packing
    (\textbf{RP}), live range splitting (\textbf{LS}), rematerialization
    (\textbf{RM}), multi-allocation (\textbf{MA}), instruction selection
    (\textbf{SE}), size of largest problem solved optimally (\textbf{SZ})
    in input instructions, and whether it is demonstrated to generate
    executable code (\textbf{EX}).\label{tab:combinatorial-integrated}}
  \begin{tabu}{%
        l%
        c%
        c%
        >{\centering\arraybackslash}p{\columnWidth}%
        >{\centering\arraybackslash}p{\columnWidth}%
        >{\centering\arraybackslash}p{\columnWidth}%
        >{\centering\arraybackslash}p{\columnWidth}%
        >{\centering\arraybackslash}p{\columnWidth}%
        >{\centering\arraybackslash}p{\columnWidth}%
        >{\centering\arraybackslash}p{\columnWidth}%
        >{\centering\arraybackslash}p{\columnWidth}%
        >{\centering\arraybackslash}p{\columnWidth}%
        c%
        >{\centering\arraybackslash}p{\columnWidth}}\hline
    \rowfont{\bfseries}
    approach & technique & scope & \inApproach{LO}{CO}{RP}{RM}{MB}{RA}{SP}{LS}{MA} & SE & SZ & EX \\
    \hline
    \rowcolor{tblrow} Wilson~\etal{}~\cite{Wilson1994} & IP & global & \inApproach{\no}{\yes}{\no}{\no}{\no}{\yes}{\yes}{\yes}{\no} & \yes & 30 & \no \\
    Gebotys~\cite{Gebotys1997} & IP & local & \inApproach{\yes}{\no}{\no}{\no}{\yes}{\yes}{\yes}{\yes}{\no} & \yes & 108 & \no \\
    \rowcolor{tblrow} Chang~\etal{}~\cite{Chang1997} & IP & local & \inApproach{\yes}{\no}{\no}{\no}{\no}{\no}{\yes}{\no}{\no} & \no & $\sim{}$10 & \no \\
    Bashford~\etal{}~\cite{Bashford1999} & CP & local & \inApproach{\yes}{\yes}{\no}{\no}{\yes}{\yes}{\yes}{\yes}{\no} & \yes & 23 & \no \\
    \rowcolor{tblrow} K\"{a}stner~\cite{Kastner2001} & IP & superblock & \inApproach{\no}{\no}{\no}{\no}{\yes}{\yes}{\no}{\no}{\no} & \no & 39 & \no \\
    Kessler~\etal{}~\cite{Kessler2006} & DP & local & \inApproach{\no}{\no}{\no}{\no}{\yes}{\no}{\no}{\yes}{\yes} & \yes & 42 & \no \\
    \rowcolor{tblrow} Nagarakatte~\etal{}~\cite{Nagarakatte2007} & IP & loop & \inApproach{\yes}{\no}{\no}{\no}{\no}{\yes}{\yes}{\yes}{\yes} & \no & ? & \no \\
    Eriksson~\etal{}~\cite{Eriksson2012} & IP & loop & \inApproach{\yes}{\no}{\no}{\no}{\yes}{\no}{\yes}{\yes}{\yes} & \yes & 100 & \no \\
    \rowfont{\bfseries} \rowcolor{tblrow} (this paper) & CP & global & \inApproach{\yes}{\yes}{\yes}{\yes}{\yes}{\yes}{\yes}{\yes}{\yes} & \no & \improvedmultiAllLargestOptimalIns{} & \yes \\
  \end{tabu}%
\end{table}

\paragraph{Optimization techniques}
IP is the most widely used optimization technique with the exception of Scholz
and Eckstein~\cite{Scholz2002} using PBQP, Bashford and
Leupers~\cite{Bashford1999} using CP, and Kessler and
Bednarski~\cite{Kessler2006} using DP.
For combinatorial instruction scheduling in isolation a broader set of
techniques has been applied, including the use of special-purpose
branch-and-bound techniques~\cite{Castaneda2014b}.

\section{Local Register Allocation}\label{sec:local-register-allocation}

This section introduces the combinatorial model for local register
allocation and its underlying program transformations.
The model is introduced step-by-step: starting from a definition
of the input programs
(Section~\ref{sec:local-program-representation}) and register
assignment (Section~\ref{sec:register-assignment}), the model is
extended with alternative instructions
(Section~\ref{sec:alternative-instructions}); spilling, live
range splitting, and coalescing (Section~\ref{sec:spilling}); and
rematerialization (Section~\ref{sec:rematerialization}).
Section~\ref{sec:local-register-allocation-summary} briefly
summarizes the main contributions in the model and program
representation.

In the remainder of the paper, each variable added to the model is numbered
by~(V$n$) where~$n$ is a number. Likewise, constraints are numbered by~(C$n$),
where~$n$ is either a number or a period-separated number pair in case the
constraint has been refined or extended.

\subsection{Program Representation}\label{sec:local-program-representation}
The model for local register allocation is parameterized with
respect to the input program and processor.
This section defines the basic program parameters that describe
an input program.
Processor parameters and the remaining program parameters are
introduced as needed.

The input program consists of a sequence of operations in a single basic block.
A key idea in the program representation is that it distinguishes between
\emph{operations} and \emph{instructions} as well as between \emph{operands} and
\emph{temporaries} to make the model more powerful.
Operations are seen as abstract instructions (such as addition) that are
implemented by specific processor instructions (such as
$\instruction{\addOpcode{}}$).
Distinguishing between operations and instructions enables supporting
alternative instructions, see Section~\ref{sec:alternative-instructions}.
Operations contain \emph{operands} that denote particular use and definition
points of temporaries.
Distinguishing between operands and temporaries enables register allocation
transformations beyond register assignment and packing, as
Sections~\ref{sec:spilling} and~\ref{sec:rematerialization} show.

Live-in (live-out) temporaries in a basic block are defined
(used) by an entry (exit) operation, corresponding to the virtual
instructions discussed in
Section~\ref{sec:background-input-program-representation}.
This ensures that the live range of a temporary can be derived simply from the
operations defining and using the temporary.
An operation using or defining a temporary $t$ is called a \emph{user}
respectively the \emph{definer} of~$t$.

Programs in this representation are described by a set of operations
$\allOperations{}$, operands $\allOperands{}$, and temporaries
$\allTemporaries{}$\!.
An operation implemented by instruction $i$ that uses temporaries $t_1$ to $t_n$
and defines temporaries $t_{n+1}$ to $t_{m}$ through its corresponding operands
is represented as
\begin{equation*}
  \naturalOperation
      {\Operand{\Oper{n + 1}}{\Temp{n + 1}},\ldots,\Operand{\Oper{m}}{\Temp{m}}}
      {i}
      {\Operand{\Oper{1}}{\Temp{1}},\ldots,\Operand{\Oper{n}}{\Temp{n}}}.
\end{equation*}
The temporary used or defined by each operand $p$ is denoted as
$\temporary{p}$.
For simplicity, operand identifiers $\Oper{1}$, $\Oper{2}$,
\dots, $\Oper{m}$ are omitted if possible.

As the input program is assumed to be in SSA, each temporary is defined exactly
once.
The program point immediately after the single definition of a
temporary~$t$ is denoted as $\liveStart{t}$, whereas the program
point immediately before the last use of~$t$ is denoted as
$\liveEnd{t}$. In our setup, the live range of a temporary~$t$ is
indeed a range (or interval) $\sequence{\liveStart{t},
  \liveEnd{t}}$, and can be enforced straightforwardly in single
basic blocks by local value numbering~\cite{Cooper2012}.
Live ranges being simple intervals is essential for modeling
register assignment as Section~\ref{sec:register-assignment}
shows.

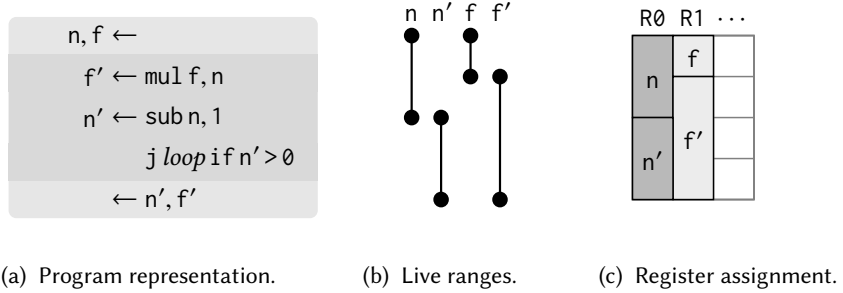
\begin{figure}%
  \adjustbox{trim=1.3cm 0cm 0.15cm 0.1cm,clip=true,scale=0.98,center}{%
    \scalebox{1}{\input{./figures/local-register-assignment}}}
  \subfloat{\hspace{0.07\linewidth}}
  \setcounter{subfigure}{0}
  \subfloat[\label{fig:local-register-assignment-input} Program representation.]{\hspace{.3\linewidth}}
  \subfloat{\hspace{0.017\linewidth}}
  \setcounter{subfigure}{1}
  \subfloat[\label{fig:local-register-assignment-live-ranges} Live ranges.]{\hspace{.23\linewidth}}
  \subfloat{\hspace{0.01\linewidth}}
  \setcounter{subfigure}{2}
  \subfloat[\label{fig:local-register-assignment-solution} Register assignment.]{\hspace{.27\linewidth}}
  \caption{Local register assignment for the \emph{loop} basic block from Example~\ref{ex:register-allocation-input}.\label{fig:local-register-assignment}}
\end{figure}

\begin{example}\label{ex:local-register-allocation-input}
Figure~\ref{fig:local-register-assignment-input} illustrates the input
representation of the \emph{loop} basic block from
Example~\ref{ex:register-allocation-input} for local register allocation.
Operand identifiers are omitted for simplicity.
The live-in temporaries $\TempN{}$ and $\TempF{}$ are defined by an
entry operation.
The redefinitions of $\TempF{}$ (by $\instruction{\multiplyOpcode{}}$) and
$\TempN{}$ (by $\instruction{\subtractOpcode{}}$) are renamed as $\TempFAfter{}$
and $\TempNAfter{}$ to enforce SSA.
Single definitions result in interval live ranges as shown in
Figure~\ref{fig:local-register-assignment-live-ranges}.
The live-out temporaries $\TempNAfter{}$ and $\TempFAfter{}$
are used by an exit operation.

\end{example}

\subsection{Register Assignment}\label{sec:register-assignment}

This section introduces a simple model for register assignment, where
temporaries are mapped to individual registers.
The mapping is modeled by a variable~$\temporaryRegister{t}$ for each temporary
$t$ giving its assigned register from a register set $R$.
This paper uses symbolic domains for clarity; the actual model maps symbolic
values to integer values:
\begin{equation*}\varlabel{var:reg}
  \temporaryRegister{t} \in R
  \quad
  \forallIn{t}{\allTemporaries{}}.
\end{equation*}

Register assignment is constrained by interference among temporaries.
For example, $\TempN{}$ and $\TempF{}$ in
Figure~\ref{fig:local-register-assignment-input} interfere and therefore
cannot be assigned to the same register.
Register assignment and interference are captured by a simple geometric
interpretation that can be modeled by global constraints.
This interpretation, based on Pereira and Palsberg's \emph{puzzle solving}
approach~\cite{Pereira2008}, projects the register assignment of a basic block
onto a rectangular area, where the horizontal dimension corresponds to an
ordering of~$R$ (called \emph{register array}) and the vertical dimension
corresponds to program points.
Each temporary $t$ yields a rectangle of $\width{t} = 1$, where the top and
bottom borders correspond to the $\liveStart{t}$ and $\liveEnd{t}$ program
points of $t$'s live range and the left border corresponds to the register
$\temporaryRegister{t}$ to which $t$ is assigned.
The direct mapping from temporaries to rectangles is possible due
to the interval
shape of the temporary live ranges.
In this interpretation, the rectangles of interfering temporaries overlap
vertically.
A single \emph{\disjointName{}} constraint in the model ensures
that interfering temporaries are assigned to different registers:
\begin{equation*}\conlabel{con:local-disjoint-live-ranges}
  \localDisjointLiveRangesConstraint{}\,.
\end{equation*}

\begin{example}\label{ex:local-register-assignment}
Figure~\ref{fig:local-register-assignment-solution} shows the geometric
interpretation of a register assignment for the \emph{loop} basic block where
$\TempN{}$ and $\TempNAfter{}$ are
assigned to $\code{R0}$ and $\TempF{}$ and $\TempFAfter{}$ are assigned to
$\code{R1}$.
Assuming the program points are numbered from one to five, the corresponding
instantiation of constraint~\ref{con:local-disjoint-live-ranges} becomes
$
  \disjoint{
    \set*{
      \tuple*{
        \temporaryRegister{\TempF},
        \temporaryRegister{\TempF} + 1,
        1,
        2
      },
      \tuple*{
        \temporaryRegister{\TempFAfter},
        \temporaryRegister{\TempFAfter} + 1,
        2,
        5
      },
      \tuple*{
        \temporaryRegister{\TempN},
        \temporaryRegister{\TempN} + 1,
        1,
        3
      },\linebreak
      \tuple*{
        \temporaryRegister{\TempNAfter},
        \temporaryRegister{\TempNAfter} + 1,
        3,
        5
      }
    }
  }
$.
\end{example}

\paragraph{Register packing}

\begin{wrapfigure}{r}{0.2\textwidth}
  \vspace{-0.4cm}
  \centering
  \adjustbox{trim=1.82cm 0cm 0.15cm 0.1cm,clip=true,scale=0.98}{%
    \scalebox{1.0}{\input{./figures/local-register-packing}}}
  \vspace{-0.2cm}%
  \caption{Reg.~packing.}
  \label{fig:local-register-packing}
  \vspace{-0.5cm}%
\end{wrapfigure}
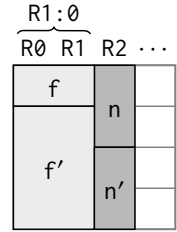
Register packing assigns temporaries of small bit-widths to different parts
of larger-width registers to improve register utilization.
This program transformation can be easily accommodated within the geometric
representation of local register assignment.
The registers in $R$ are decomposed into their underlying register \emph{atoms}
(minimum register parts addressable by instructions), the function $\width{t}$
is extended to give the number of adjacent atoms that temporary $t$ occupies,
and the variable $\temporaryRegister{t}$ is reinterpreted as the leftmost atom
to which $t$ is assigned.
This extension generalizes the geometric representation as a rectangle
packing problem with rectangles of different widths and heights, which is
precisely captured by constraint~\ref{con:local-disjoint-live-ranges}
as is.

\begin{example}\label{ex:local-register-packing}
Figure~\ref{fig:local-register-packing} illustrates register packing for a
version of the \emph{loop} basic block in
Example~\ref{ex:local-register-allocation-input} where $\TempF{}$ and
$\TempFAfter{}$ have a width of 64 rather than 32 bits.
Hexagon's register atoms ($\register{R0}$, $\register{R1}$, \dots) have a
width of 32 bits and can be combined into 64-bit registers
($\register{R1:0}$, $\register{R3:2}$, \dots), hence $\width{\TempF{}} =
\width{\TempFAfter{}} = 2$.
In the example register packing, $\TempF{}$ and $\TempFAfter{}$ are assigned to
the combined register $\register{R1:0}$.
This is modeled by the assignment of the variables
$\temporaryRegister{\TempF{}}$ and $\temporaryRegister{\TempFAfter{}}$ to
$\register{R0}$, the leftmost atom of $\register{R1:0}$.
$\TempNAfter{}$ and $\TempNAfter{}$ are assigned to $\register{R2}$.
\end{example}

\paragraph{Pre-assignments}

As discussed in
Section~\ref{sec:background-input-program-representation}, an
operand~$p$ might be pre-assigned to a certain register~$r$
(given by $\preAssigned{p\hspace{0.025cm}}{r}$).
Temporaries used or defined by such operands are pre-assigned to the
corresponding registers:
\begin{equation*}\conlabel{con:fixed-pre-assignment}
  \fixedPreAssignmentConstraint{}.
\end{equation*}

\paragraph{Register classes}

Register files tend to be irregular in that different instructions are
allowed to access different subsets of the available registers.
A subset of the registers available to a particular instruction or group of
instructions is called a \emph{register class}.
The processor parameter $\fixedRegisterClass{p}$ gives the register class of
operand $p$.
The registers to which a temporary can be assigned are determined by the operands
that define and use the temporary:
\begin{equation*}\conlabel{con:fixed-instruction-selection}
  \fixedInstructionSelectionConstraint{}.
\end{equation*}
Section~\ref{sec:spilling} below exploits register classes for spilling,
where memory is treated as an additional register class, albeit with an
unlimited number of registers.

\paragraph{Contributions}

This section contributes a novel use of Pereira and Palsberg's register
assignment and packing approach~\cite{Pereira2008}, by incorporating it into a
combinatorial model.
This is possible as constraint programming readily provides
efficient solving methods for rectangle packing in the form of
dedicated propagation algorithms for the \emph{\disjointName{}}
constraint.
Integer programming, an alternative technique that is popular for register
allocation (see Table~\ref{tab:combinatorial-register-allocation}), has
difficulties in dealing with the disjunctive nature of rectangle
packing~\cite{Lodi2002}.

\subsection{Alternative Instructions}\label{sec:alternative-instructions}

The model is extended by \emph{alternative instructions} that can be selected to
implement operations relevant for register allocation.
For example, ARM's Thumb-2 instruction set extension allows 16-~and 32-bit
instructions to be freely mixed, where the 16-bit instructions reduce code size
but can only access a subset of the registers.
The selection of 16- or 32-bit instructions is interdependent with register
allocation as it has a significant effect on register
pressure~\cite{EdlervonKoch2010}.
Alternative instructions are also central to model spilling,
live range splitting, and coalescing as Section~\ref{sec:spilling}
explains.

The model supports alternative instructions as
follows.
The set of instructions that can implement an operation $o$ is given by the
parameter $\instructions{o}$, and the function $\registerClass{p}{i}$ is
redefined to give the register class of operand $p$ if implemented by
instruction $i$.
An operation that can be implemented by alternative instructions
$i_1,\ldots,i_n$ is represented as
$\naturalOperation{\cdots}{\set{i_1,\ldots,i_n}}{\cdots}\,$.

A variable $\operationInstruction{o}$ for each operation $o$ gives the
instruction that implements $o$:
\begin{equation*}\varlabel{var:ins}
  \operationInstruction{o} \in \instructions{o}
  \quad
  \forallIn{o}{\allOperations{}}.
\end{equation*}
The register class of each operand $p$ is linked to the instruction that
implements $p$'s operation in constraint~\ref{con:fixed-instruction-selection}
(changes to constraint~\ref{con:fixed-instruction-selection} are highlighted
in gray):
\begin{equation*}\newconlabel{con:active-instruction-selection}{con:fixed-instruction-selection}{1}
  \activeInstructionSelectionConstraint{}
\end{equation*}
where the parameter $\operandOperation{p}$ gives the operation of operand $p$.
Multi-dimensional array access and set membership can be expressed with
\emph{\elementName{}} constraints as described in
Section~\ref{sec:combinatorial-optimization}.

\subsection{Spilling, Live Range Splitting, and Coalescing}\label{sec:spilling}

Spilling (allocate temporaries into memory, including load-store optimization
and multi-allocation), live range splitting (allocate temporaries to different
locations along their live ranges), and coalescing (assign copied temporaries to
the same register) are supported by a single extension of the program
representation and the model.
The extension is based on the notion of optional copy operations.

A \cop{} $\naturalOperation{\Temp{d}}{\set{i_1,\ldots,i_n}}{\Temp{s}}$ defines a
destination temporary $\Temp{d}$ with the value of a source temporary $\Temp{s}$
using a copy instruction among a set of alternatives $\set{i_1,\ldots,i_n}$.
Another operation using $\Temp{s}$ in the original program might after the
introduction of the copy use $\Temp{d}$ as an alternative, because~$\Temp{s}$
and~$\Temp{d}$ hold the same value as they are \emph{copy-related}.
If some operation uses $\Temp{d}$, the temporary is considered \emph{live} and
its definer \cop{} \emph{active}.
Otherwise, $\Temp{d}$ is considered \emph{dead} and its definer \cop{}
\emph{inactive}, effectively coalescing $\Temp{d}$ into $\Temp{s}$.
Inactive operations and dead temporaries do not affect the constraints in which
they occur.
\Cops{} can be: implemented by store and load instructions to support spilling;
implemented by register-to-register move instructions to support live range
splitting; or made inactive to support coalescing.

To handle store and load instructions and register-to-register move instructions
uniformly, the register array is extended with a \emph{memory} register class
containing an unbound number of registers $\set{\code{M0},\code{M1},\dots}$
corresponding to memory locations in the function's stack frame.
Unifying registers and memory yields a simple and expressive model that
internalizes spilling by reducing register allocation to register assignment,
where the register assigned to a temporary $t$ implies $t$'s allocation.

\paragraph{Copy extension}

The number, type, and relation among the \cops{} with which a program is
extended to support spilling and live range splitting depends on the
targeted processor architecture.
This paper assumes load-store processors where values can be copied between
registers and memory by single instructions.
However, the model supports
more complex architectures such as clustered VLIW processors by
introducing additional \cops{}.

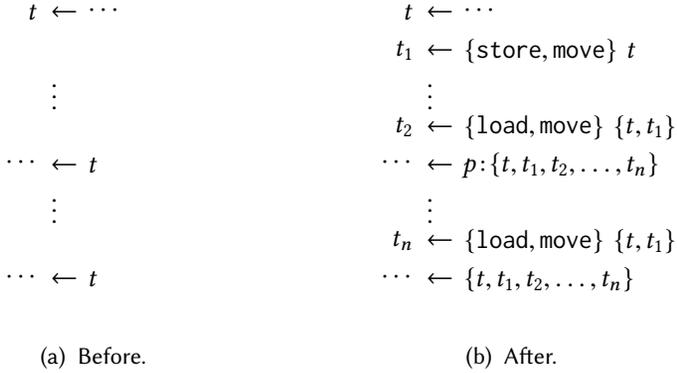
\begin{figure}%
  \centering%
  \adjustbox{trim=1cm 0cm 0cm 0cm,clip=true,scale=1}{%
    \scalebox{1}{\input{./figures/copy-extension}}}
  \subfloat{\hspace{.1\linewidth}}
  \setcounter{subfigure}{0}
  \subfloat[\label{fig:copy-extension-before} Before.]{\hspace{.4\linewidth}}
  \subfloat[\label{fig:copy-extension-after} After.]{\hspace{.4\linewidth}}
  \subfloat{\hspace{.1\linewidth}}
  \caption{Copy extension.\label{fig:copy-extension}}
\end{figure}

For load-store processor architectures, a program is extended by visiting each
temporary $t$ once and introducing a \emph{store-move} \cop{} with instructions
$\instruction{\genericStoreOpcode{}}$ and $\instruction{\genericMoveOpcode{}}$
at the program point immediately after the the definition of $t$ and a
\emph{load-move} \cop{} with instructions $\instruction{\genericLoadOpcode{}}$
and $\instruction{\genericMoveOpcode{}}$ at the program point immediately before
each use of~$t$.
The store-move can only use~$t$, each load-move might use~$t$ or the destination
temporary of the store-move, and each original user of~$t$ might use~$t$ or any
of the introduced temporaries.
Figure~\ref{fig:copy-extension} illustrates the transformation.

Note that copy extension suggests the use of some alternative temporaries at
points where they are not yet defined, for example $\Temp{n}$ used by operand
$p$ in Figure~\ref{fig:copy-extension}.
Such uses are invalid in isolated register allocation where operations are
totally ordered.
However, they become relevant in the integration with instruction scheduling
(see Section~\ref{sec:global-register-allocation-and-instruction-scheduling}),
where operations (including \cops{}) can be rearranged.

\paragraph{Model extension}

The new decisions that follow from the introduction of \cops{} are captured by
three classes of variables.
A variable $\operandTemporary{p}$ for each operand $p$ gives the temporary that
is used or defined by operand $p$ among a set of copy-related alternatives
$\temps{p}$:
\begin{equation*}\varlabel{var:temp}
  \operandTemporary{p} \in \temps{p}
  \quad
  \forallIn{p}{\allOperands{}};
\end{equation*}
a Boolean variable $\liveTemporary{t}$ indicates whether temporary $t$ is live:
\begin{equation*}\varlabel{var:live}
  \liveTemporary{t} \in \booleans{}
  \quad
  \forallIn{t}{\allTemporaries{}};
\end{equation*}
and a Boolean variable $\activeOperation{o}$ indicates whether operation $o$ is
active:
\begin{equation*}\varlabel{var:active}
  \activeOperation{o} \in \booleans{}
  \quad
  \forallIn{o}{\allOperations{}}.
\end{equation*}
As is common in constraint programming, we define the set
$\booleans$ as $\{0,1\}$ and abbreviate, for example,
$\activeOperation{o}=1$ as $\activeOperation{o}$ and
$\activeOperation{o}=0$ as $\neg\activeOperation{o}$.

For uniformity, $\operandTemporary{p}$ is defined for both use and define
operands even though no alternative is available for the latter.
Similarly, $\activeOperation{o}$ is defined for both copy and non-copy
operations even though the latter must always be active:
\begin{equation*}\conlabel{con:active-operation}
  \activeOperationConstraint{}\,.
\end{equation*}
For a live temporary $t$, the definer of $t$ and at least one user of $t$ must
be active:
\begin{equation*}\conlabel{con:live-temporary}
  \liveTemporaryConstraint{}
\end{equation*}
where $\definer{t}$ and $\users{t}$ are the operand(s) that might define
and use a temporary $t$, and $\operandOperation{p}$ is the operation of
operand $p$.

With the introduction of alternative temporaries, the temporary used or
defined by operands involved in pre-assignments becomes variable:
\begin{equation*}\newconlabel{con:pre-assignment}{con:fixed-pre-assignment}{1}
  \preAssignmentConstraint{}
\end{equation*}
Additionally, the register class constraint only needs to consider
active operations:
\begin{equation*}\newconlabel{con:instruction-selection}{con:fixed-instruction-selection}{2}
  \instructionSelectionConstraint{}.
\end{equation*}

\newcommand{\copyMargin}{0.07cm}

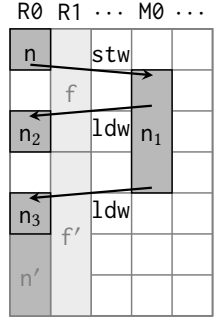
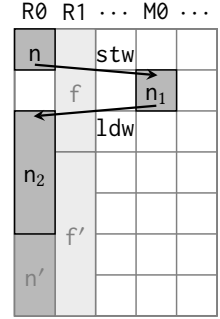
\begin{figure}%
  \centering%
  \adjustbox{trim=1cm 0cm 0cm 0cm,clip=true,scale=0.98}{%
    \scalebox{1}{\input{./figures/local-register-allocation}}}
  \subfloat[\label{fig:local-register-allocation-extension} Copy extension of $\TempN{}$.]{\hspace{.4\linewidth}}
  \subfloat{\hspace{.025\linewidth}}
  \setcounter{subfigure}{1}
  \subfloat[\label{fig:local-register-allocation-spill-everywhere} Spill-everywhere.]{\hspace{.25\linewidth}}
  \subfloat{\hspace{.025\linewidth}}
  \setcounter{subfigure}{2}
  \subfloat[\label{fig:local-register-allocation-load-store-optimization} Load-store optimization.]{\hspace{.27\linewidth}}
  \caption{Copy extension and spilling for $\TempN{}$ in the \emph{loop} basic block.\label{fig:local-register-allocation}}
\end{figure}

\begin{example}
Figure~\ref{fig:local-register-allocation} illustrates the application of copy
extension to support different implementations of spilling for the running
example.
Figure~\ref{fig:local-register-allocation-extension} shows the result of
applying copy extension to~$\TempN{}$ using Hexagon's store
($\instruction{\storeOpcode{}}$), load ($\instruction{\loadOpcode{}}$), and
register-to-register move ($\instruction{\moveOpcode{}}$) instructions.

Figure~\ref{fig:local-register-allocation-spill-everywhere} illustrates the
spill of $\TempN{}$ on the register array extended with memory registers.
$\TempN{}$ is stored directly after its definition (copied to $\TempNStore{}$
which is assigned to memory register $\code{M0}$) and loaded before each use as
$\TempNLoadFst{}$ and $\TempNLoadSnd{}$ in a spill-everywhere fashion.
This illustrates how the approach supports multi-allocation:
immediately before $\instruction{\multiplyOpcode{}}$, $\TempNStore{}$ and
$\TempNLoadFst{}$ are live simultaneously in memory ($\code{M0}$) and in a
processor register ($\code{R0}$), simulating multiple
allocations of their original temporary $\TempN{}$.

Figure~\ref{fig:local-register-allocation-load-store-optimization} illustrates
the spill of $\TempN{}$ where load-store optimization is applied, rendering the
second load-move inactive.
Load-store optimization is only possible due to the availability of
$\TempNLoadFst{}$ as an alternative use for
$\instruction{\subtractOpcode{}}$.

Live range splitting is supported similarly to spilling.
For example, the live range of $\TempN{}$ can be split by implementing the
\cops{} that are active in
Figure~\ref{fig:local-register-allocation-load-store-optimization} with
$\instruction{\moveOpcode}$ instructions, and letting
constraint~\ref{con:instruction-selection} force the assignment of
$\TempNStore{}$ and $\TempNLoadFst{}$ to processor registers.
\end{example}

\paragraph{Discussion}

The local register allocation model introduced in this section is the first
to exploit \cops{}, copy extension, and memory registers as a uniform
abstraction to capture spilling, live range splitting, coalescing, and
rematerialization (to be explained in Section~\ref{sec:rematerialization}).
Only the related approaches of Wilson \etal{}~\cite{Wilson1994,
  Wilson2002} and Chang \etal{}~\cite{Chang1997} extend the program
representation with explicit \cops{} and treat them exactly as
original program operations.
However, they do not model memory and processor registers uniformly, which
requires special treatment of spilling and hence results in a more complicated
model.

Alternative temporaries are related to Colombet \etal{}'s \emph{local
  equivalence classes}~\cite{Colombet2015} in their definition and targeted
program transformations, and to Wilson \etal{}'s \emph{alternative
  implementations}~\cite{Wilson1994, Wilson2002} in their way of extending the
original program representation with implementation choices to be considered by
a solver.

Copy extension for load-store processors (Figure~\ref{fig:copy-extension})
allows a temporary's live range to be split at each definition and use point.
Introducing even more splitting choices at other program points might enable
better code.
The number of these choices and their distribution among program points differs
considerably among combinatorial register allocation
approaches. This ranges from no live
range splitting (as in Scholz and Eckstein~\cite{Scholz2002}) to
\emph{split-everywhere} at every program point (as introduced by Appel and
George~\cite{Appel2001}).
The potential loss in code quality by bounding live range splitting to definition and use points
is not yet established.
In any case, the potential loss is mitigated as the model is expanded to global
register allocation in Section~\ref{sec:global-register-allocation} (where
temporaries can be split at the boundaries of each basic block) and integrated
with instruction scheduling in
Section~\ref{sec:global-register-allocation-and-instruction-scheduling} (where
the \cops{} performing the splits can be rearranged).

\subsection{Rematerialization}\label{sec:rematerialization}

Rematerialization recomputes values at their use points rather
than spilling them if the recomputation instructions are
sufficiently cheap.
A common case is rematerialization of \emph{never-killed}
values~\cite{Chaitin1981} that can be recomputed by a
single instruction from operands that are always available.
Typical examples are numeric constants and
variable addresses in the stack frame.
This paper captures rematerialization of never-killed values by a simple
adjustment of copy extension and the register array and does not require any
changes to the actual model.
The level of rematerialization is similar to that of related approaches in
Table~\ref{tab:combinatorial-register-allocation}.

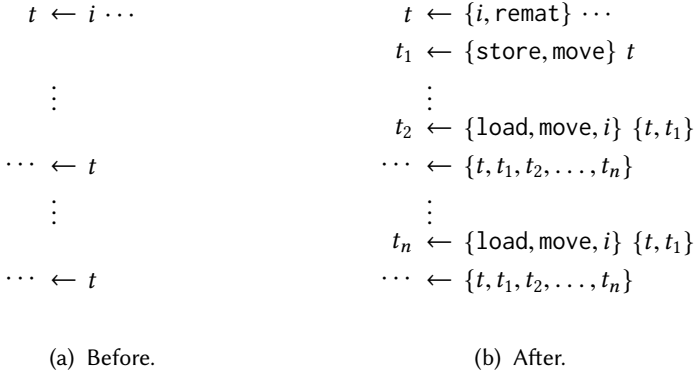
\begin{figure}%
  \centering%
  \adjustbox{trim=1cm 0cm 0cm 0cm,clip=true,scale=1}{%
    \scalebox{1}{\input{./figures/remat-extension}}}
  \subfloat{\hspace{.1\linewidth}}
  \setcounter{subfigure}{0}
  \subfloat[\label{fig:remat-extension-before} Before.]{\hspace{.4\linewidth}}
  \subfloat[\label{fig:remat-extension-after} After.]{\hspace{.4\linewidth}}
  \subfloat{\hspace{.1\linewidth}}
  \caption{Adjusted copy extension for a rematerializable temporary.\label{fig:remat-extension}}
\end{figure}

Rematerializable temporaries are identified using the data-flow analysis
of Briggs \etal{}~\cite{Briggs1992}.
The register array is extended with a \emph{rematerialization} register class
containing an unbounded number of registers
$(\code{RM0},\code{RM1},\dots)$. Rematerialization of a temporary $t$ is modeled
as if $t$ were defined in a rematerialization register by a virtual instruction
$\instruction{\genericDematOpcode{}}$ and loaded into a processor register by
$t$'s actual definer.
This is achieved by adjusting the copy extension of each rematerializable
temporary~$t$ to include the virtual instruction $\instruction{\genericDematOpcode{}}$ as an
  alternative to~$t$'s original definer instruction~$i$, and include~$i$ as an alternative to the instructions of each
  load-move.
Figure~\ref{fig:remat-extension} illustrates this adjustment.

Modeling rematerialization as an alternative load-move instruction is inspired
by Colombet \etal{}~\cite{Colombet2015}.
Unlike Colombet \etal{}, we use \cops{} and extend the
register array to avoid additional variables and constraints.

\subsection{Summary}\label{sec:local-register-allocation-summary}

This section has introduced a program representation and model
that enable all transformations offered by local, state-of-the-art register
allocation using few variables and constraints.
The compact, yet expressive model is based on two significant
contributions to the state of the art in combinatorial register
allocation. First, the model incorporates Pereira and Palsberg's
register assignment and packing approach~\cite{Pereira2008} into
a combinatorial setup.
Second, it introduces \cops{}, copy extension, and memory
registers as a uniform abstraction to model spilling, live range
splitting, coalescing, and rematerialization.

\section{Global Register Allocation}\label{sec:global-register-allocation}

This section extends the model from local to global register allocation for
entire functions.

\subsection{Program Representation}\label{sec:global-register-allocation-program-representation}

Global register allocation represents an input function as a set of basic blocks
$\allBlocks{}$ where each basic block is defined as in
Section~\ref{sec:local-program-representation}.
For a basic block $b$, $\operations{b}$ is the set of operations
in~$b$ and $\temporaries{b}$ is the set of temporaries
in~$b$. The disjoint unions of these sets are denoted as
$\allOperations{}$ and $\allTemporaries{}$.

The input function is assumed to be in \emph{linear static single
  assignment} (\emph{LSSA}) form.
LSSA generalizes SSA by enforcing that each
temporary is defined once and only used within the basic block
where it is defined~\cite{Appel1998,Aycock2000}.
LSSA construction decomposes each temporary $t$ with a live range spanning
multiple blocks $\B{1}, \B{2}, \ldots, \B{n}$ into a set of \emph{congruent}
temporaries $\Temp{\B{1}} \congruence \Temp{\B{2}} \congruence \cdots
\congruence \Temp{\B{n}}$, where each temporary is local to its respective basic
block.
Hence the congruences denote that two or more temporaries originate from the
same temporary and must therefore be assigned the same register.
Live-in (live-out) temporaries resulting from LSSA construction are defined
(used) by entry (exit) operations.
The conventional form~\cite{Sreedhar1999} of LSSA is assumed, where replacing
all congruent temporaries by a single representative preserves the program
semantics.
LSSA is also referred to as \emph{the really-crude approach} to
SSA~\cite{Appel1998,Aycock2000} and \emph{maximal SSA}~\cite{Buchwald2016}.
LSSA's single definitions and local temporaries are crucial for modeling global
register allocation: these properties preserve live ranges as simple intervals,
which is required for modeling register assignment and packing geometrically.

\newcommand{\blockPredRaw}{pred}
\newcommand{\blockSuccRaw}{succ}
\newcommand{\blockPred}{\text{\emph{\blockPredRaw{}}}}
\newcommand{\blockSucc}{\text{\emph{\blockSuccRaw{}}}}
\newcommand{\tempPred}{t_{\blockPred{}}}
\newcommand{\tempSucc}{t_{\blockSucc{}}}
\newcommand{\operPred}{p_{\blockPred{}}}
\newcommand{\operSucc}{p_{\blockSucc{}}}
\newcommand{\operPredIdx}[1]{p_{#1.\blockPred{}}}
\newcommand{\operSuccIdx}[1]{p_{#1.\blockSucc{}}}
\newcommand{\tempSuccIdx}[1]{t_{#1.\blockSucc{}}}
\newcommand{\tempPredStore}{t_{\blockPred{}.1}}
\newcommand{\tempSuccLoad}{t_{\blockSucc{}.1}}

After LSSA construction, each basic block in the input function is copy-extended
and adjusted for rematerialization as in
Section~\ref{sec:local-register-allocation}.
Copy extension induces alternative temporaries that can be used by each
operation, making it necessary to lift the LSSA congruence from temporaries to
operands as illustrated in Figure~\ref{fig:congruence-lifting}.
Assume that $\blockPred{}$ and $\blockSucc{}$ are adjacent basic blocks.
After copy extension, each congruence $\congruent{\tempPred{}}{\tempSucc{}}$ is
lifted to the congruence $\congruent{\operPred{}}{\operSucc{}}$ where
$\operPred{}$ is the exit operand using $\tempPred{}$ (or any of its
alternatives $t_{\blockPred{}.1}$, \dots, $t_{\blockPred{}.n}$) and
$\operSucc{}$ is the entry operand defining $\tempSucc{}$.

\begin{figure}%
  \centering%
  \adjustbox{trim=0.4cm 0cm 0cm 0cm,clip=true,scale=1}{%
    \scalebox{1}{\input{./figures/congruence-lifting}}}
  \subfloat[\label{fig:congruence-lifting-before} Before.]{\hspace{.4\linewidth}}
  \subfloat[\label{fig:congruence-lifting-after} After.]{\hspace{.5\linewidth}}
  \subfloat{\hspace{.1\linewidth}}
  \caption{Lifting of the LSSA congruence from temporaries to
operands.\label{fig:congruence-lifting}}
\end{figure}
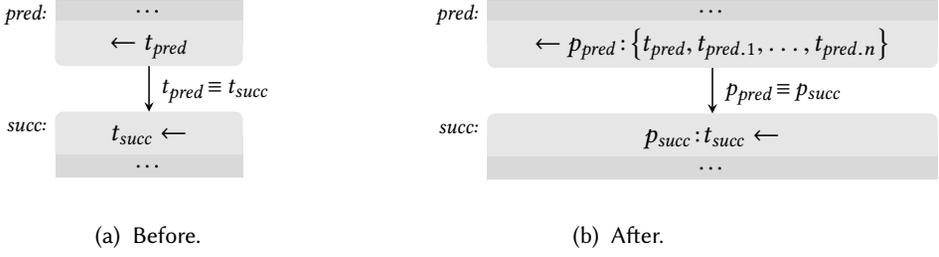

\begin{figure}%
  \centering%
  \adjustbox{trim=0cm 0cm 0cm 0cm,clip=true,scale=0.98}{%
    \scalebox{1}{\input{./figures/factorial-lssa}}}
  \caption{CFG of the factorial function from
    Example~\ref{ex:register-allocation-input} in LSSA where $\TempFEnd{}$ is
    copy-extended.\label{fig:global-register-allocation-input}}
\end{figure}
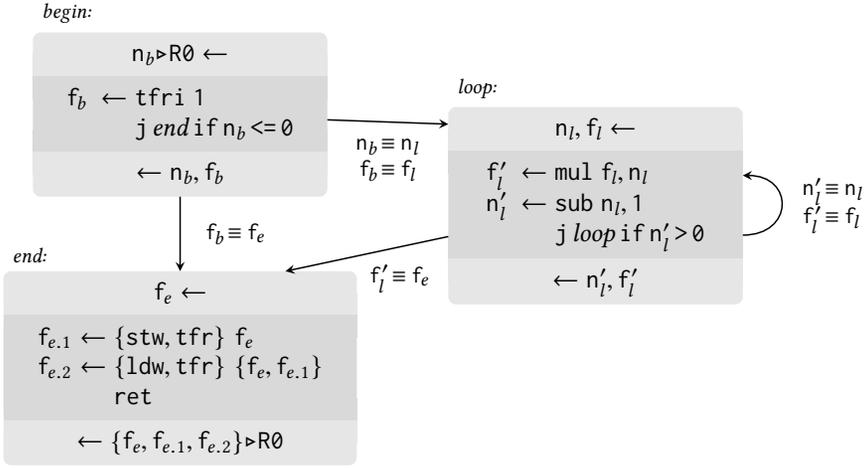

\begin{example}\label{ex:global-register-allocation-input}
Figure~\ref{fig:global-register-allocation-input} shows the input
representation of the factorial function from
Example~\ref{ex:register-allocation-input} for global register allocation.
The function is in LSSA: each original temporary $t$ in
Example~\ref{ex:register-allocation-input} is decomposed into a temporary $t_b$
for each block $b$ in which the temporary is live.
For example, the temporary $\TempN{}$ is decomposed into temporaries
$\TempNBegin{}$ for the \emph{begin} basic block and $\TempNLoop{}$ for the
\emph{loop} basic block (for conciseness, only the first letter
of a block
name is used as index).
To enforce single definitions, the redefinition of $\TempNLoop{}$ is further
renamed as $\TempNLoopAfter{}$ similarly to
Example~\ref{ex:local-register-allocation-input}.
The resulting temporaries are local and defined only once.
For example, $\TempNLoop{}$ is defined once by the entry operation of
\emph{loop} and only used locally by $\instruction{\multiplyOpcode{}}$ and
$\instruction{\subtractOpcode{}}$.

The input representation to global register allocation assumes that all
temporaries are copy-extended, but the example limits copy extension to
$\TempFEnd{}$ for simplicity, which yields the alternative temporaries
$\TempFEndStore{}$ and $\TempFEndLoad{}$.
Also for simplicity, the LSSA congruence (displayed on the CFG arcs) is defined
on temporaries and not on operands as shown in
Figure~\ref{fig:congruence-lifting}.
\end{example}

\subsection{Model}\label{sec:global-register-allocation-combinatorial-model}

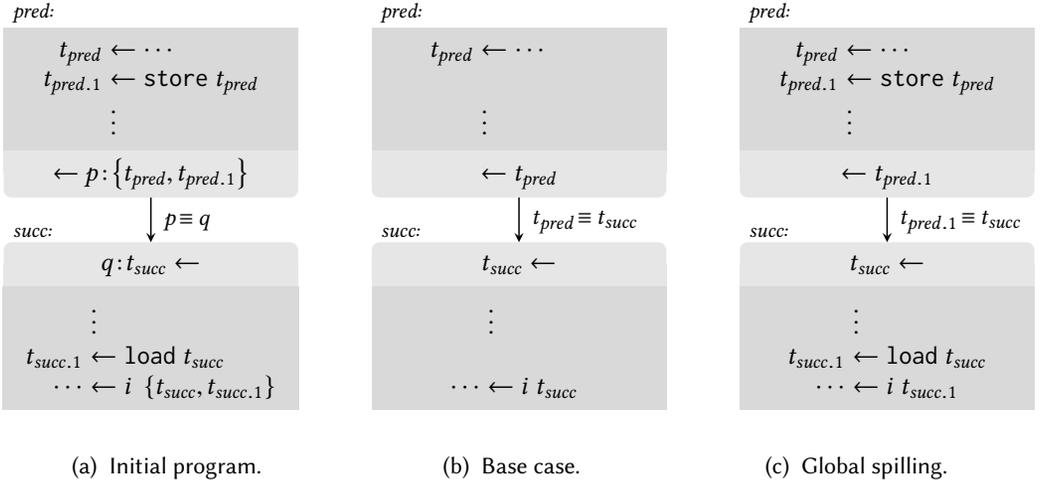
\begin{figure}%
  \centering%
  \adjustbox{trim=0.7cm 0cm 0cm 0cm,clip=true,scale=0.98}{%
    \scalebox{1}{\input{./figures/congruence-spilling}}}
  \subfloat[\label{fig:congruence-initial} Initial program.]{\hspace{.33\linewidth}}
  \subfloat[\label{fig:congruence-registers} Base case.]{\hspace{.33\linewidth}}
  \subfloat[\label{fig:congruence-memory} Global spilling.]{\hspace{.33\linewidth}}
  \caption{Interplay of live temporary, register class, and congruence constraints.\label{fig:congruence-constraints-example}}
\end{figure}

The program representation for global register allocation preserves the
structure of the local model of each basic block described in
Section~\ref{sec:local-register-allocation} since operations, operands, and
temporaries are local and basic blocks are only related by operand
congruences.
Therefore, a global model of register allocation is simply composed of the
variables and constraints of each local model and linked with constraints that
assign congruent operands to the same register:
\begin{equation*}\conlabel{con:congruence}
  \congruenceConstraint{}\,.
\end{equation*}
The congruence constraints extend the scope of all register allocation
transformations to whole functions by propagating the impact of local decisions
across basic blocks.
Figure~\ref{fig:congruence-constraints-example} illustrates the interplay of
live temporary (\ref{con:live-temporary}), register class
(\ref{con:instruction-selection}), and congruence (\ref{con:congruence})
constraints across two basic blocks \emph{pred} and \emph{succ}.

Figure~\ref{fig:congruence-initial} displays a program fragment where the
original temporary $t$ is live across the boundaries of \emph{pred} and
\emph{succ}.
Temporary $t$ is decomposed into $\tempPred{}$ and $\tempSucc{}$ and
copy-extended with a store-move \cop{} ($\instruction{\genericStoreOpcode{}}$)
at \emph{pred} and a load-move \cop{} ($\instruction{\genericLoadOpcode{}}$) at
\emph{succ} (the remaining \cops{} and register-to-register move instructions
are omitted for simplicity).

Figure~\ref{fig:congruence-registers} shows the base case where $\tempPred{}$ is
not spilled (hence the store-move is deactivated).
Since $\tempPredStore{}$ is dead, the exit operation of \emph{pred} uses
$\tempPred{}$ and the congruence constraints propagate the processor register of
$\tempPred{}$ to $\tempSucc{}$.
Now the processor register of $\tempSucc{}$ is incompatible with the register
class of $\instruction{\genericLoadOpcode{}}$, which deactivates the load-move
and forces $i$ to use~$\tempSucc{}$.

Figure~\ref{fig:congruence-memory} shows the global spill of $t$ where the
store-move is active and the exit operation of \emph{pred} uses
$\tempPredStore{}$ instead of $\tempPred{}$.
In this case the congruence constraints propagate the memory register of
$\tempPredStore{}$ to $\tempSucc{}$, which forces $i$ to use the temporary
$\tempSuccLoad{}$ copied from $\tempSucc{}$ by the load-move.

In general, entry and exit operations define and use multiple temporaries.
All temporaries used or defined by the same boundary operation interfere and are
thus necessarily assigned to different registers according to
constraint~\ref{con:global-disjoint-live-ranges}.
The use of LSSA in its conventional form~\cite{Sreedhar1999} guarantees the
absence of conflicts between this and the congruence constraints
(\ref{con:congruence}).

The same rectangle assignment and packing constraint defined for a single basic
block in Section~\ref{sec:register-assignment} is used for each basic block:
\begin{equation*}\newconlabel{con:global-disjoint-live-ranges}{con:local-disjoint-live-ranges}{1}
  \globalDisjointLiveRangesConstraint{}.
\end{equation*}

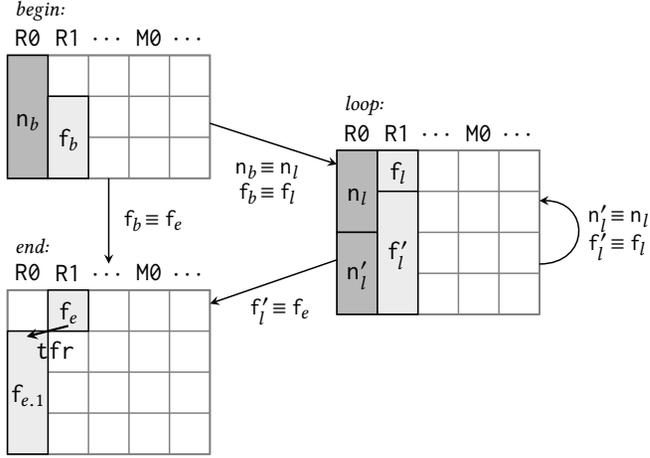
\begin{figure}%
  \centering%
  \adjustbox{trim=1cm 0cm 0cm 0cm,clip=true,scale=0.98}{%
    \scalebox{1}{\input{./figures/global-register-allocation}}}
  \caption{Global register allocation for the factorial function from Example~\ref{ex:global-register-allocation-input}.\label{fig:global-register-allocation}}
\end{figure}

\begin{example}\label{ex:global-register-allocation-solution}
Figure~\ref{fig:global-register-allocation} illustrates global register allocation
for the input function from Example~\ref{ex:global-register-allocation-input}.
It corresponds to the better solution in the introductory
Example~\ref{ex:motivation}: the temporaries derived from \code{n} are assigned
to \code{R0} and the temporaries derived from \code{f} are assigned to \code{R1}
with the exception of $\TempFEndStore{}$. $\TempFEndStore{}$ is copied to
\code{R0} in the \emph{end} basic block by activating its defining store-move
and implementing it with the $\instruction{\moveOpcode{}}$ instruction.
The load-move of $\TempFEnd{}$ is inactive, which renders $\TempFEndLoad{}$ dead
(\ref{con:live-temporary}).

The congruence constraints (\ref{con:congruence}) guarantee the compatibility of
the local register assignments across boundaries.
For example, the live-out temporaries of \emph{begin} ($\tuple{\TempNBegin{},
  \TempFBegin{}}$) and the live-in temporaries of \emph{loop}
($\tuple{\TempNLoop{}, \TempFLoop{}}$) are pairwise congruent and
thus assigned to the same registers ($\tuple{\register{R0}, \register{R1}}$).
\end{example}

\paragraph{Discussion}

The introduced model is the first to exploit the properties of
LSSA for register allocation.
LSSA and similar representations have been used to establish
theoretical connections between programming paradigms~\cite{Appel1998}, as an
intermediate form in SSA construction~\cite{Aycock2000}, and as a basis for
intermediate-level program analysis and optimization~\cite{Gange2015}.

The model has a limitation in that it does not support multi-allocation across
basic blocks.
This is due to the fact that only one temporary among possibly many copied
temporaries can be used by exit operands as seen in
Figure~\ref{fig:congruence-lifting}.
This limitation is shared with most combinatorial register allocation approaches
(see Table~\ref{tab:combinatorial-register-allocation}), and its impact on
scalability and code quality is unclear.
Among integrated approaches (Table~\ref{tab:combinatorial-integrated}), only
Wilson \etal{}'s model~\cite{Wilson1994} is global.
Wilson \etal{} propose a similar concept to congruence constraints but their
model does not support multi-allocation.

Multi-allocation in a combinatorial setup is discussed in detail by
Colombet \etal{}~\cite{Colombet2015}.
Among the three scenarios discussed by Colombet \etal{}, our model
supports the optimizations illustrated in Figures~1 and~2 from Colombet
\etal{}'s paper~\cite{Colombet2015} but not in Figure~3.

\section{Instruction Scheduling}\label{sec:instruction-scheduling}

This section describes the model for local instruction scheduling.
It is mostly based on previous work on constraint-based
scheduling by Malik \etal{}~\cite{Malik2008} and serves as an intermediate step
towards the integrated register allocation and instruction scheduling model.
It captures pre-register allocation scheduling and assumes the same basic block
representation as in Section~\ref{sec:local-register-allocation}, with the only
difference that operations are not totally ordered.
Post-register allocation scheduling can be easily captured by introducing
additional dependencies caused by register assignment.

\paragraph{Variables}

Instruction scheduling assigns issue cycles to operations in a basic block such
that dependencies between operations and processor resource constraints are
satisfied.
In the model, a variable
$\operationIssueCycle{o}$ defines the cycle in which $o$ is
issued relative to the beginning of the basic block:
\begin{equation*}\varlabel{var:cycle}
  \operationIssueCycle{o} \in \naturalNumbersZero{}
  \quad
  \forallIn{o}{\allOperations{}}.
\end{equation*}

\paragraph{Dependency constraints}

Valid schedules must preserve the input order among dependent operations.
An operation $u$ depends on another operation $d$ if $u$ uses a temporary $t$
defined by $d$.
If operation $d$ is issued at cycle $\operationIssueCycle{d}$, its result
becomes available at cycle $\operationIssueCycle{d} +
\latency{\fixedOperationInstruction{d}}$, where $\fixedOperationInstruction{o}$
is the instruction that implements operation $o$ and $\latency{i}$ is the
latency of instruction $i$.
To satisfy the dependency, $u$ must be issued after the result of $d$ is
available, that is: $\operationIssueCycle{u} \ge \operationIssueCycle{d} +
\latency{\fixedOperationInstruction{d}}$.
The model includes one such inequality for each use of a
temporary $t$, where $u$ uses $t$ through operand $q$ and $d$
defines $t$ through operand $p$:
\begin{equation*}\conlabel{con:fixed-data-precedences}
  \fixedDataPrecedencesConstraint{}
\end{equation*}
The dependency constraints deviate slightly from early constraint
models such as that of Malik \etal{}~\cite{Malik2008} in that
they make the underlying uses and definitions of temporaries
explicit.
This is essential for the integration with register allocation as
explained in
Section~\ref{sec:global-register-allocation-and-instruction-scheduling}.
The integration also requires that the virtual entry instruction is given a
latency of one.
This requirement ensures that live-in temporaries in a basic block have
non-empty live ranges and are thus assigned to different registers according to
constraint~\ref{con:global-disjoint-live-ranges}.

\paragraph{Processor resource constraints}

The capacity of processor resources such as functional units and buses cannot be
exceeded.
As discussed in Section~\ref{sec:background-instruction-scheduling}, this paper
assumes a set of processor resources $S$ where each resource $s \in S$ has a
capacity $\capacity{s}$ and each instruction $i$ consumes $\units{i}{s}$ units of
each resource $s$ during $\duration{i}{s}$ cycles.
The model includes a \emph{\cumulativeName{}} constraint for each resource $s$
to ensure that the capacity of $s$ is never exceeded by the scheduled
operations:
\begin{equation*}\conlabel{con:fixed-processor-resources}
  \fixedProcessorResourcesConstraint{}.
\end{equation*}
This paper assumes that resource consumption always starts at the issue of the
consumer.
The model can be easily extended to capture more complex scenarios (for example,
modeling each stage in a processor pipeline) by adding an offset $\offset{i}{s}$
to the issue cycle of the consumer ($\operationIssueCycle{o}$) in the resource
constraint (\ref{con:fixed-processor-resources})~\cite{Unison2017}.

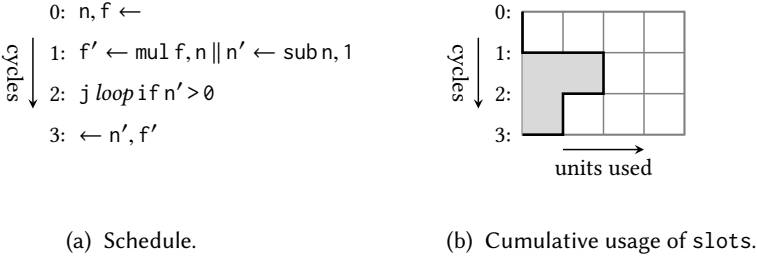
\begin{figure}%
  \centering%
  \adjustbox{trim=1cm 0cm 0cm 0cm,clip=true,scale=0.98}{%
    \scalebox{1}{\input{./figures/instruction-scheduling}}}
  \subfloat{\hspace{0.1\linewidth}}
  \setcounter{subfigure}{0}
  \subfloat[\label{fig:instruction-schedule} Schedule.]{\hspace{.4\linewidth}}
  \subfloat[\label{fig:cumulative-usage} Cumulative usage of $\code{slots}$.]{\hspace{.5\linewidth}}
  \caption{Instruction schedule for the \emph{loop} basic block from Example~\ref{ex:local-register-allocation-input}.\label{fig:instruction-scheduling}}
\end{figure}

\begin{example}\label{ex:instruction-scheduling}
Figure~\ref{fig:instruction-schedule} shows a schedule of the \emph{loop}
basic block from Example~\ref{ex:local-register-allocation-input} where
all instructions are assumed to have a latency of one cycle.
The vertical dimension ranges over cycles as opposed to program points in
register allocation.

Hexagon is a VLIW processor that allows up to four instructions to be issued
in parallel.
In the example, only $\instruction{\multiplyOpcode{}}$ and
$\instruction{\subtractOpcode{}}$ can be issued in parallel ($\bundled$) as
they do not depend on each other.
The multiple-issue capacity of Hexagon is modeled as a resource \code{slots}
with $\capacity{\code{slots}} = 4$, $\units{i}{\code{slots}} = 1$, and
$\duration{i}{\code{slots}} = 1$ for each instruction $i \in
\set{\instruction{\multiplyOpcode{}},\instruction{\subtractOpcode{}},\instruction{\jumpIfOpcode{}}}$.
With this resource, constraint~\ref{con:fixed-processor-resources} is
instantiated as
$
  \cumulative{
    \set*{
      \tuple*{
        \operationIssueCycle{\instruction{\multiplyOpcode{}}},
        1,
        1
      },
      \tuple*{
        \operationIssueCycle{\instruction{\subtractOpcode{}}},
        1,
        1
      },\linebreak
      \tuple*{
        \operationIssueCycle{\instruction{\jumpIfOpcode{}}},
        1,
        1
      }
    },
    4
  }
$.
Figure~\ref{fig:cumulative-usage} shows the cumulative usage of the
\code{slots} resource over time.
The figure shows that parallelism in the example is not limited by
the \code{slots} resource constraint (\ref{con:fixed-processor-resources}) but
by the dependency constraint
$\operationIssueCycle{\instruction{\jumpIfOpcode{}}} \ge
\operationIssueCycle{\instruction{\subtractOpcode{}}} + 1$ over temporary
$\TempNAfter{}$ (\ref{con:fixed-data-precedences}).
\end{example}

\paragraph{Discussion}

Constraint programming is a popular technique for resource-constrained
scheduling problems~\cite{Baptiste2006}, as it provides efficient and dedicated
propagation algorithms for scheduling constraints.
As shown here, it enables a simple model using
a single variable per operation and a single constraint for
each dependency and resource.
For comparison, integer programming typically requires a
decomposition into~$n$ variables with $\set{0,1}$ domain for each
operation and~$n$ linear constraints for each resource, where $n$
is an upper bound on the number of issue cycles of the basic
block.
This decomposition limits the scalability of approaches based on integer
programming~\cite{Castaneda2014b}.

\section{Global Register Allocation and Instruction Scheduling}\label{sec:global-register-allocation-and-instruction-scheduling}

This section combines the models for global register allocation
and instruction scheduling into an integrated model that
precisely captures their interdependencies.
The integrated model is a significant contribution to
combinatorial code generation as, for the first time, it captures
the same program transformations as state-of-the-art heuristic
approaches (see Table~\ref{tab:combinatorial-integrated}).

The model assumes the program representation described in
Section~\ref{sec:global-register-allocation-program-representation},
except that operations within basic blocks are not totally
ordered.
The model uses the variables and constraints from the global register
allocation model and the
instruction scheduling model
generalized to handle copy-extended programs.
This generalization allows \cops{} (introduced to support spilling, live range
splitting, and rematerialization) to be scheduled in the same manner as the
original program operations, by means of issue cycle variables.
Compared to the register allocation model without instruction scheduling, the
ability to schedule and thus rearrange copy operations increases the freedom of
their supported program transformations.

\paragraph{Live ranges and issue cycles}

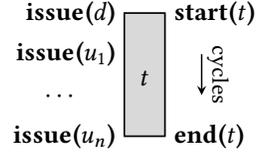
\begin{wrapfigure}{r}{0.26\textwidth}
  \vspace{-0.5cm}
  \centering
  \adjustbox{trim=0.7cm 0.1cm 0.2cm 0.1cm,clip=true,scale=0.98}{%
    \scalebox{1.0}{\input{./figures/live-range}}}
  \vspace{-0.3cm}%
  \caption{Live range of $t$.}
  \label{fig:live-range}
  \vspace{-0.4cm}%
\end{wrapfigure}
In the integrated model, live ranges link register allocation and instruction
scheduling since they relate registers assigned to temporaries with the issue
cycles of their definers and users.
The live start of a temporary $t$ is linked to the issue of the operation~$d$
defining~$t$, and the live end is linked to the issue of the last
operation~$u_n$ using~$t$ as shown in Figure~\ref{fig:live-range}.
Variables $\temporaryLiveStart{t}$ and $\temporaryLiveEnd{t}$ are introduced for
each temporary $t$ giving its live start and end cycles:
\begin{equation*}\varlabel{var:start-end}
  \temporaryLiveStart{t}, \temporaryLiveEnd{t} \in \naturalNumbersZero{}
  \quad
  \forallIn{t}{\allTemporaries{}}.
\end{equation*}
The live start of a (live) temporary $t$ corresponds to the issue of its
defining operation:
\begin{equation*}\conlabel{con:live-start}
  \liveStartConstraint{},
\end{equation*}
while the live end of $t$ corresponds to the issue of the last user operation:
\begin{equation*}\conlabel{con:live-end}
  \liveEndConstraint{}.
\end{equation*}

\paragraph{Generalized instruction scheduling}

The generalized model captures alternative instructions, optional (active or
inactive) operations, and alternative temporaries.
Alternative instructions are handled by using the variable
$\operationInstruction{o}$ instead of the parameter
$\fixedOperationInstruction{o}$ for each operation~$o$.
The generalized dependency constraints handle inactive operations and
alternative temporaries by making each dependency involving an operation that
potentially uses temporary~$t$ through operand~$q$ conditional on whether the
operation is active and actually
uses~$t$:
\begin{equation*}\newconlabel{con:data-precedences}{con:fixed-data-precedences}{1}
  \dataPrecedencesConstraint{}
\end{equation*}

The generalized processor resource constraints handle optional operations by
making their resource usage conditional on whether they are
active. Additionally, they are extended to global scope by including a
\emph{\cumulativeName{}} constraint for each resource $s$ and basic block $b$:
\begin{equation*}\newconlabel{con:global-processor-resources}{con:fixed-processor-resources}{1}
  \globalProcessorResourcesConstraint{}
\end{equation*}

\begin{figure}%
  \centering%
  \adjustbox{trim=1.15cm 0cm 0.2cm 0cm,clip=true,scale=0.98}{%
    \scalebox{1}{\input{./figures/integrated-solution}}}
  \caption{Integrated solution for the factorial function from Example~\ref{ex:global-register-allocation-input}.\label{fig:integrated-solution}}
\end{figure}
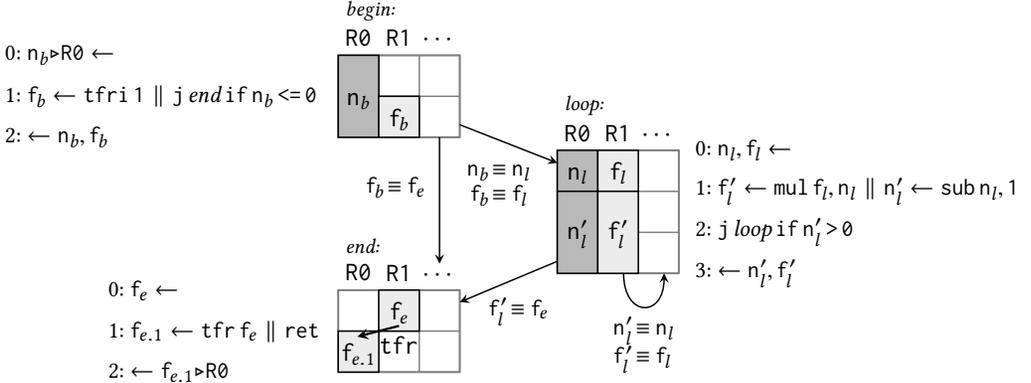

\begin{example}\label{ex:integrated-solution}
  Figure~\ref{fig:integrated-solution} shows a solution to the integrated
  register allocation and instruction scheduling problem for the function in
  Example~\ref{ex:global-register-allocation-input}.
  The scheduled operations are shown besides the register assignment of each
  basic block. The memory registers $\code{M0},\code{M1},\dots$ are omitted for
  conciseness.
  The solution corresponds to the faster assembly code given in
  Example~\ref{ex:motivation}.
  The final assembly code shown in Example~\ref{ex:motivation} can be generated
  by simply removing the entry and exit virtual operations and replacing each
  temporary with its assigned register.
\end{example}

\paragraph{Discussion}

While the scope of register allocation in the integrated model is global,
instruction scheduling is local.
This limitation is shared with the only available integrated approach that is
global (Wilson \etal{}~\cite{Wilson1994}, see
Table~\ref{tab:combinatorial-integrated}).
Extending instruction scheduling beyond basic blocks is known to improve code
quality, particularly for VLIW processors~\cite{Fisher1983}.
A number of combinatorial approaches have explored such extensions
successfully~\cite{Castaneda2014b}.
The model presented in this section might be readily expanded to superblocks
(consecutive basic blocks with multiple exit points but a single entry point)
using a corresponding number of exit operations similarly to Malik
\etal{}~\cite{Malik2008b}.

\section{Objective Function}\label{sec:objective-function}

This section completes the integrated model with a generic
objective function. The entire model, including the objective
function, is presented in Appendix~\ref{app:combinatorial-model}.

The objective function is orthogonal to the variables and constraints in the
model and can be chosen to optimize for different goals.
In this paper, the objective is to minimize the weighted sum of the cost of each
basic block:
\begin{equation}\label{eq:generic-objective-function}
  \genericObjectiveFunction{}
\end{equation}
where $\blockWeight{b}$ and $\blockCost{b}$ can be defined to optimize for
different goals.

\paragraph{Optimizing speed}
To optimize for speed, the weight of a basic block $b$ is set as its execution
frequency $\blockFrequency{b}$ and its cost is defined as the issue cycle of the
exit operation of~$b$ (indicated by $\exitOperation{b}$), which by construction
is the last scheduled operation in~$b$:
\begin{equation}\label{eq:speed-optimization}
  \blockWeight{b} = \blockFrequency{b}; \qquad
  \blockCost{b} = \operationIssueCycle{\exitOperation{b}} - 1 \quad
  \forall b \in \allBlocks{}.
\end{equation}
Note that subtracting one to $\operationIssueCycle{\exitOperation{b}}$ accounts
for the cycle occupied by the virtual entry operation, as discussed in
Section~\ref{sec:instruction-scheduling}.

\paragraph{Optimizing code size}
To minimize code size, it is sufficient to choose one as weight
for all blocks, as they contribute proportionally to the total
code size. The cost is defined as the sum of the size of
the instruction $i$ (given by $\instructionSize{i}$) implementing each
active operation:
\begin{equation}\label{eq:size-optimization}
  \blockWeight{b} = 1; \qquad
  \blockCost{b} =
  \sum_{\mathclap{o \in \operations{b} \,\suchThat\, \activeOperation{o}}}
      {\instructionSizeNoParen{\elementMark{\operationInstruction{o}}}} \quad
  \forall b \in \allBlocks{}.
\end{equation}

Other optimization criteria such as spill code or energy
consumption minimization can be expressed by defining
$\blockWeight{b}$ and $\blockCost{b}$ accordingly.
Constraint programming also supports optimizing for multiple criteria (for
example, first for speed and then for code size) by using a tuple as the
objective function.

\begin{example}\label{ex:integrated-solution-cost}
  Assuming the estimated execution frequencies
  $\blockFrequency{\text{\emph{begin}}} = 1$,
  $\blockFrequency{\text{\emph{loop}}} = 10$, and
  $\blockFrequency{\text{\emph{end}}} = 1$ and a uniform size of 32 bits for all
  real instructions, the value of the speed objective function
  (equation~\eqref{eq:speed-optimization}) for the solution given in
  Example~\ref{ex:integrated-solution} is:
\begin{equation}\label{eq:speed-example}
  1 \times \p{2 - 1} + 10 \times \p{3 - 1} + 1 \times \p{2 - 1} = 1 + 20 +
  1 = 22\;\text{execution cycles,}
\end{equation}
while the value of the code size objective function
(equation~\eqref{eq:size-optimization}) is:
\begin{equation}\label{eq:size-example}
  1 \times 64 + 1 \times 96 + 1 \times 64 = 64 + 96 + 64 = 224\;\text{bits}.
\end{equation}
\end{example}

\paragraph{Discussion}

The cost model that underlies the objective function for speed optimization
(equation~\eqref{eq:speed-optimization}) assumes a processor with constant
instruction latencies.
This assumption, common in combinatorial approaches, may underestimate the
contribution of variable latencies to the actual execution time.
Generally, the less predictable the targeted processor is, the
lower the accuracy of the speed cost model.
In extreme cases (out-of-order, superscalar, speculative
processors with long pipelines and deep memory hierarchies),
other optimization criteria such as spill code minimization might
be chosen as a better approximation to speed optimization.

\section{Additional Program Transformations}\label{sec:model-extensions}

A key advantage of combinatorial approaches is the ease with
which additional program transformations and processor-specific
features can be captured in a compositional manner.
This section contributes model extensions to handle program transformations that
depend on register allocation and instruction scheduling but are usually
approached in isolation by today's state-of-the-art heuristic compilers.

\paragraph{Stack frame elimination}
One of the main responsibilities of code generation is to manage
the call stack by
placing special operations at entry, return, and function call points.
\emph{Stack frame elimination} is a common optimization that avoids generating
a stack frame for spill-free functions that meet some additional conditions (for
example not containing calls or other stack-allocated temporaries).
Heuristic compilers typically apply this optimization \emph{opportunistically}
after deciding whether to spill in register allocation.
In contrast, our approach captures the optimization in integration with register
allocation and instruction scheduling, hence taking into account the overhead in
generating a frame by spilling in functions where the frame could be otherwise
avoided.

The model is easily extended to capture stack frame elimination by introducing
a single variable $\functionFrame{}$ that indicates whether the function under
compilation requires a frame:
\begin{equation*}\varlabel{var:frame}
  \functionFrame{} \in \booleans{}.
\end{equation*}
This variable is true if there exists an active operation~$o$ implemented by
an instruction~$i$ requiring a frame (indicated by $\requiresFrame{o}{i}$):
\begin{equation*}\conlabel{con:requires-frame}
  \requiresFrameConstraint{}.
\end{equation*}
Typical examples of operations requiring a frame are calls and \cops{}
implemented by spill instructions.
If the function requires a frame, each optional operation~$o$ implemented
by instruction~$i$ managing the stack (indicated by $\managesFrame{o}{i}$)
must be active:
\begin{equation*}\conlabel{con:manages-frame}
  \managesFrameConstraint{}.
\end{equation*}
Operations that manage the stack typically include pushes, pops,
and adjustments of the pointers that delimit the stack frame.

\paragraph{Scheduling with latencies across basic blocks}

Local instruction scheduling as described in
Section~\ref{sec:instruction-scheduling} conservatively assumes that a temporary
defined within a certain basic block might be used at the beginning of a
successor basic block.
This assumption can force long-latency instructions such as integer divisions to
be scheduled unnecessarily early, which limits the solution space.
While this paper is confined to local instruction scheduling, latencies across
basic blocks (including loops) are captured by a simple model extension.

\begin{figure}%
  \centering%
  \adjustbox{trim=0.3cm 0cm 0cm 0cm,clip=true,scale=0.98}{%
    \scalebox{1}{\input{./figures/slack-balancing-constraints}}}
  \caption{Distribution of latency slack among the boundaries of two adjacent basic blocks.\label{fig:slack-balancing-constraints}}
\end{figure}
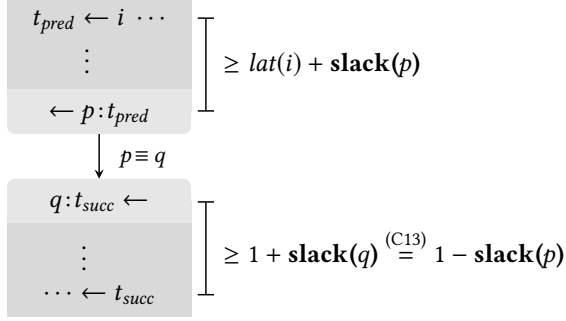

Assume a temporary $t$ live across basic blocks before LSSA construction.
The key idea is to distribute the latency of $t$ between the use and the
definition of the corresponding LSSA temporaries at the basic block boundaries.
The operands of virtual boundary operations that are not placed at the function
boundaries are called \emph{boundary operands}.
A boundary operand $p$ can be classified as either entry or exit (indicated by
$\entryOperand{p}$ or $\exitOperand{p}$).
A variable $\operandSlack{p}$ for each boundary operand $p$ gives the amount of
latency \emph{slack} assigned to~$p$.
The slack can be either zero (no latency across boundaries) or negative for exit
operands and positive for entry operands.
The latter case corresponds to a temporary that is defined \emph{late} in the
predecessor basic block:
\begin{equation*}\varlabel{var:slack}
  \operandSlack{p} \in
  \begin{cases}
    \set{0} \union \negativeIntegerNumbers{}\!, & \exitOperand{p} \\
    \set{0} \union \positiveIntegerNumbers{}\!, & \entryOperand{p} \\
    \set{0}\!, & \text{otherwise}
  \end{cases}
  \quad
  \forallIn{p}{\allOperands{}},
\end{equation*}
To avoid that pre-LSSA temporaries defined late in a basic block are used too
early in successor basic blocks, the slack of the boundary operands relating the
derived LSSA temporaries across basic blocks boundaries must be balanced:
\begin{equation*}\conlabel{con:slack-balancing}
  \slackBalancingConstraint{}.
\end{equation*}
Finally, the latency added or subtracted by the slack is reflected in the
dependency constraints:
\begin{equation*}\newconlabel{con:slack-data-precedences}{con:fixed-data-precedences}{2}
  \slackDataPrecedencesConstraint{}
\end{equation*}
Figure~\ref{fig:slack-balancing-constraints} illustrates how the slack is
distributed among the boundary operands relating the congruent temporaries
$\tempPred{}$ and $\tempSucc{}$.
Because of the slack balancing constraints (\ref{con:slack-balancing}), the
slack of operand $p$ is the opposite of the slack of $q$.
Note that the latency of the entry operation in the successor basic block is one
as discussed in Section~\ref{sec:instruction-scheduling}.

\paragraph{Scheduling with operand forwarding}

Operand forwarding is a processor optimization that makes a value available
before the end of the execution of its definer instruction.
Some processors such as Hexagon exploit this optimization by providing
operand-forwarding instructions that access values from registers in the same
cycle as they are defined.
These instructions reduce the latency of their dependencies but
may impose additional constraints on instruction scheduling,
hence it is desirable to capture their effect in the model.

An operation implemented by an instruction $i$ forwarding its operand $p$ (which
is indicated by $\forwarded{i}{p}$) is scheduled in parallel with the definer of
the temporary used by $p$:
\begin{equation*}\conlabel{con:forwarding}
  \forwardingConstraint{}
\end{equation*}
If an operand is forwarded, its corresponding dependency constraint that
includes the latency of the definer instruction has no effect:
\begin{equation*}\newconlabel{con:fwd-data-precedences}{con:fixed-data-precedences}{3}
  \fwdDataPrecedencesConstraint{}
\end{equation*}
For clarity, this refinement of the dependency constraints is presented
independently of that required to capture latencies across basic blocks
(\ref{con:slack-data-precedences}), but both are compatible.

\paragraph{Selection of two- and three-address instructions}

Two-address instructions are instructions of the form
$\naturalOperation{\register{Rd}}{i}{\register{Rd},\register{Rs}}$ where the
register $\register{Rd}$ gets overwritten.
These instructions can often be encoded more compactly but impose additional
constraints on register allocation.
In particular, the definition and use operands $p, q$ that correspond to the
register overwrite of a two-address instruction $i$ (indicated by
$\twoAddress{i}{p}{q}$) are assigned to the same register:
\begin{equation*}\conlabel{con:two-address}
  \twoAddressConstraint{}
\end{equation*}
Some instruction sets like ARM's Thumb-2 allow two- and three-address
instructions to be freely intermixed.
The selection of two- and three-address instructions is captured by
constraint~\ref{con:two-address} together with alternative instructions as
introduced in Section~\ref{sec:alternative-instructions}.

\paragraph{Selection of double-load and double-store instructions}

ARM v5TE and later versions provide load ($\instruction{\armDoubleLoad{}}$) and
store ($\instruction{\armDoubleStore{}}$) instructions that access two 32-bits
values at consecutive addresses in memory.
Combining pairs of 32-bits load ($\instruction{\armSingleLoad{}}$) or store
instructions ($\instruction{\armSingleStore{}}$) into their double counterparts
can improve memory bandwidth but might increase register pressure by making the
two 32-bits values interfere.
Single/double-load instruction selection can be integrated into the model using
alternative temporaries (the process is analogous for double-store
instructions).

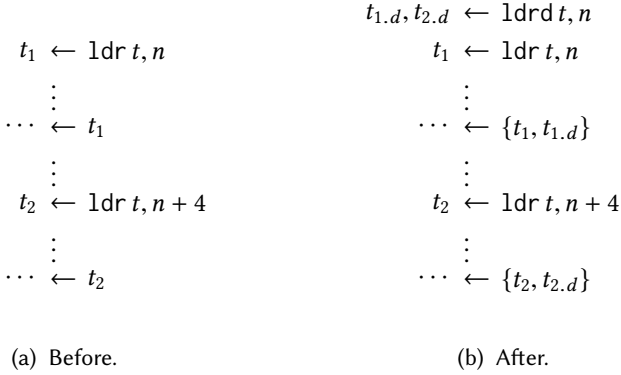
\begin{figure}%
  \centering%
  \adjustbox{trim=1cm 0cm 0cm 0cm,clip=true,scale=1}{%
    \scalebox{1}{\input{./figures/double-load-extension}}}
  \subfloat{\hspace{.1\linewidth}}
  \setcounter{subfigure}{0}
  \subfloat[\label{fig:double-load-extension-before} Before.]{\hspace{.40\linewidth}}
  \subfloat[\label{fig:double-load-extension-after} After.]{\hspace{.44\linewidth}}
  \subfloat{\hspace{.06\linewidth}}
  \caption{Extension for double-load instruction selection.\label{fig:double-load-extension}}
\end{figure}

The input program is extended with an optional $\instruction{\armDoubleLoad{}}$
operation for each pair of $\instruction{\armSingleLoad{}}$ operations accessing
consecutive addresses $\sequence{t + n, t + n + 4}$, and the temporaries
loaded by $\instruction{\armDoubleLoad{}}$
are presented as alternatives to the temporaries loaded by each
individual $\instruction{\armSingleLoad{}}$.
Figure~\ref{fig:double-load-extension} illustrates the extension.

Nandivada and Palsberg~\cite{Nandivada2006} propose exploiting double-store and
double-load instructions for spilling.
Incorporating this feature into our model is in principle possible since memory
registers capture the assignment of individual stack frame locations.

\section{Solving in Unison}\label{sec:solving-in-unison}

Our integrated, combinatorial approach to register allocation and instruction
scheduling is implemented in the \emph{Unison} software
tool~\cite{UnisonTool2018}.
The core feature of Unison is its use of constraint programming
(CP)~\cite{Rossi2006} to exploit the structure of these compiler problems.
This section outlines the main methods used to scale to medium-sized problems,
as well as Unison's implementation.
The practical impact of these methods is studied in
Section~\ref{sec:scalability}.

\subsection{External Solver Portfolio}\label{sec:solver-portfolio}

Constraint solvers guarantee to find the optimal solution to a problem if
there is one.
However, they typically exhibit a large variability in solving time, even for
problems of similar size~\cite{Gomes2000}.
Furthermore, different solvers tend to perform best for
problems of different sizes and structure, particularly when the
solving methods differ.
This variability can be exploited to reduce solving time by
running a portfolio of multiple solvers in parallel on the same
problem~\cite{Gomes2001}.
Unison employs two different portfolios, an \emph{internal} and
an \emph{external} portfolio. The internal portfolio is discussed
in Section~\ref{sec:solving-improvements}.
The external portfolio runs two solvers in parallel each using a
single operating system process: a
\emph{decomposition-based} solver that exploits the problem
structure for scalability and an off-the-shelf hybrid CP-Boolean
satisfiability solver.
As Section~\ref{sec:scalability} discusses, the two solvers complement each
other and thus yield a portfolio that is superior to each of them in
isolation.

\paragraph{Decomposition-based solver}

The decomposition-based solver is based on a novel use of the LSSA form (see
Section~\ref{sec:global-register-allocation-program-representation}) for
increased scalability and anytime behavior.
The key insight is that, in the combinatorial model of a LSSA function, the only
connection between basic blocks is established by the congruence constraints
(\ref{con:congruence}) on boundary operands.
If registers are assigned to all boundary operands, the rest of the register
allocation and instruction scheduling problem can be decomposed and solved
independently for each basic block.

The decomposition-based solver exploits this structure to solve the
combinatorial problems iteratively at two levels: first, a \emph{master problem}
is solved where $\temporaryRegister{\operandTemporary{p}}$ is assigned for each
boundary operand~$p$ such that no constraint violation is detected by
propagation.
If the model is extended as in Section~\ref{sec:model-extensions},
$\functionFrame{}$ and $\operandSlack{p}$ for each boundary operand $p$ must
also be assigned as part of the master problem.
Then, a \emph{subproblem} is solved for each basic block $b$ by assigning its
remaining variables such that all but the congruence constraints
(\ref{con:congruence}) are satisfied and $\blockCost{b}$ is minimized.
Finally, the solution $m$ to the master problem and the solutions to all
subproblems are combined into a full solution $s$, the value of $s$ is evaluated
according to equation~\eqref{eq:generic-objective-function}, and a new iteration
is run.
In the new iteration, the objective function is constrained to be less than that
of $s$ and $\neg m$ is added to the set of constraints to avoid finding the same
solution to the master problem again.
When the solver proves optimality or times out, the last full solution is
returned.

\subsection{Solving Improvements}\label{sec:solving-improvements}

The model introduced through
Sections~\ref{sec:local-register-allocation}-\ref{sec:model-extensions} can be
directly implemented and solved with CP.
However, as is common in combinatorial optimization a wide array of additional
modeling and solving improvements are required in practice to scale beyond small
problems.
This section outlines the most successful methods used in Unison.

\paragraph{Implied constraints}

An effective improving method in CP is to add \emph{implied constraints}, which
are constraints that are logically redundant but yield additional propagation
hence reducing the amount of search required~\cite{Smith2006}.
Implied constraints in Unison include:
\begin{itemize}
\item Different temporaries used by an operation must be assigned to different
  registers.
\item The live range of the temporary used by the first operand of an operation
  $o$ implemented by a two-address instruction (see
  Section~\ref{sec:model-extensions}) ends at the issue cycle of $o$.
\item If definition operands $p, q$ are pre-assigned to the same register
  and it is known statically that $\operationIssueCycle{\operandOperation{p}} <
  \operationIssueCycle{\operandOperation{q}}$, then
  $\temporaryLiveEndNoParen{\varElementMark{\operandTemporary{p}}} \leq
  \temporaryLiveStartNoParen{\varElementMark{\operandTemporary{q}}}$ holds.
\item \emph{\CumulativeName{}} \emph{constraints} derived from the interference
  constraints (\ref{con:global-disjoint-live-ranges}) where the live range
  of each temporary $t$ yields a task that consumes $\width{t}$ units of a
  register resource~\cite{Simonis2008}.
\item \emph{Distance constraints} of the form $\operationIssueCycle{o_2} \geq
  \operationIssueCycle{o_1} + \modelParameter{d}{o_1,o_2}$ derived from a static
  \emph{region analysis}~\cite{Wilken2000,Malik2008}, where operation $o_1$ is
  statically known to precede operation $o_2$ and their derived issue cycle
  distance $\modelParameter{d}{o_1,o_2}$ is greater than the distance that would
  be otherwise inferred by propagation of the dependency constraints.
\item Constraints enumerating all allowed combinations of temporaries used or
  defined by sets of copy-related operands.
  For each such set, the allowed combinations of used and defined temporaries
  are derived from a relaxation of the combinatorial model that only includes
  the constraints over $\activeOperation{o}$, $\liveTemporary{t}$, and
  $\operandTemporary{p}$.
\item \emph{Cost lower bound constraints} of the form $\blockCost{b} \geq c$
  where $c$ is the optimal cost to a relaxation of the subproblem for basic
  block $b$ (see Section~\ref{sec:solver-portfolio}) where the registers of the
  boundary operands are unassigned.
\end{itemize}

\paragraph{Symmetry breaking constraints}

Combinatorial models often admit multiple solutions that are \emph{symmetric} in
that they are all considered equivalent~\cite{Smith2006}.
Example sources of symmetries in our model are \emph{interchangeable registers}
(where the registers assigned to two temporaries can be swapped) and
\emph{interchangeable \cops{}} (where the \cops{} that support a spill or a live
range split can be swapped).
A common improving method is to add \emph{symmetry breaking constraints} that
reduce the search effort by avoiding searching for symmetric solutions.
Symmetries caused by both interchangeable registers and \cops{} are broken
in Unison with global \emph{value precedence constraints}~\cite{Law2004},
which enforce a precedence among values in a sequence of variables.

\paragraph{Dominance breaking constraints}

Combinatorial models often admit solutions that are \emph{dominated} in that
they can be mapped to solutions that are always of at most the same cost.
Similarly to symmetries, dominated solutions can be discarded by
\emph{dominance breaking constraints} to reduce the search
effort~\cite{Smith2006}.
An example of dominance in our model is in the registers of the source
$\Temp{s}$ and destination $\Temp{d}$ temporaries of an active \cop{} $o$: for
$o$ to be useful, constraints are added to enforce the assignment of $\Temp{s}$
and $\Temp{d}$ to different registers.
Another example involves the variables related to inactive operations
(issue cycles, instructions, and temporaries) and dead temporaries
(registers and live range cycles): such variables are assigned to arbitrary
values as they do not affect the constraints in which they occur.

\paragraph{Probing}

Probing is a common method in combinatorial optimization that tests variable
assignments and discards the values for which propagation proves the absence of
solutions~\cite{Savelsbergh1994}.
The method is performed only before search starts, as it has
a high computational cost and its benefit is highest at this phase.
Probing is applied both to the instructions of $\operationInstruction{o}$ for
each operation $o$ and to the values of the objective function
(equation~\eqref{eq:generic-objective-function}) and $\blockCost{b}$ for each
basic block $b$.

\paragraph{Internal portfolio}\label{sec:internal-portfolio}

The decomposition-based solver discussed above solves each subproblem with a
portfolio of \emph{search strategies}.
A search strategy defines how a problem is decomposed into alternative
subproblems and the order in which search explores the subproblems.
The strategies included in the internal portfolio complement each other (hence
achieving better robustness) and differ in two main aspects.
First, they differ in the order in which they select variables to try values on.
For example, some strategies try to assign instructions and temporaries for one
operation at a time while others try to assign instructions for all operations
first.
Second, the strategies differ in the order in which they select values to be
tried for a variable.
For example, different strategies try to schedule operations either as
early as possible or as close as possible to their users.
Each search strategy runs as a portfolio asset, but in contrast to the external
portfolio the assets communicate by adding constraints to their search based on
the cost of solutions found by the other assets.

\subsection{Implementation}\label{sec:implementation}

Unison is implemented as an open-source tool~\cite{UnisonTool2018} that can be
used as a complement to the LLVM compiler~\cite{Lattner2004}: the LLVM
interface, the transformations to the input program (such as LSSA construction
and copy extension), and the processor and calling convention descriptions are
implemented in Haskell; the decomposition-based solver and the solving
improvements are implemented in C\texttt{++} using the constraint programming
system Gecode~\cite{Gecode2018}; and the off-the-shelf solver
Chuffed~\cite{Chu2011} is interfaced through the MiniZinc modeling
language~\cite{Nethercote2007}.
The tool includes in total~\numprint{16569}~and~\numprint{19456}~lines of
hand-written Haskell and C\texttt{++} code.
Each of the supported processors is described in around one thousand lines of
hand-written Haskell code and complemented by tens of thousands of lines of
Haskell code automatically generated from LLVM's processor descriptions.
Further detail about Unison's architecture and interface can be found in its
manual~\cite{Unison2017}.

\newcommand{\plotScale}{0.57}

\section{Experimental Evaluation}\label{sec:experimental-evaluation}

This section presents experimental results for Unison, with a focus on
scalability and code quality.
Section~\ref{sec:setup} defines the setup of the experimental evaluation,
Sections~\ref{sec:estimated-code-quality} and~\ref{sec:scalability} analyze the
estimated code quality and scalability of Unison for different goals and
processors, and Section~\ref{sec:actual-speedup} presents a study of the actual
speedup achieved for Hexagon V4 on MediaBench applications.

\subsection{Setup}\label{sec:setup}

\paragraph{Evaluation processors and benchmarks}

The experimental evaluation includes three processors with different
architectures and application domains: Hexagon V4~\cite{Codrescu2014} (a
four-way VLIW digital signal processor), ARM1156T2F-S~\cite{ARM2007} (a
single-issue general-purpose processor), and a generic MIPS32
processor~\cite{MIPS2016} (a single-issue general-purpose processor used
predominantly in embedded applications).
The specifics of the MIPS32 processor are defined as in the LLVM~3.8
compiler~\cite{Lattner2004}.
The rest of the section refers to these processor as \emph{Hexagon}, \emph{ARM},
and \emph{MIPS}.
The three processors are in-order, although they include unpredictable features
such as cache memories and branch predictors.
Evaluating Unison for out-of-order processors such as x86~\cite{x862017}
is part of future work.

The evaluation uses functions from the C applications in
MediaBench~\cite{Lee1997} and SPEC CPU2006~\cite{SpecCPU2006}.
These benchmark suites match the application domain of the selected processors
and are widely employed in embedded and general-purpose compiler research.
The \code{perlbench} application in SPEC CPU2006 is excluded due to
cross-compilation problems~\cite{SpecInstructions2018}.

\paragraph{Baseline}

The state-of-the-art heuristic compiler LLVM 3.8 is used both to generate the
input low-level IR and as a baseline for comparison.
LLVM performs register allocation and instruction scheduling heuristically by
priority-based coloring~\cite{Chow1984} and list scheduling~\cite{Rau1993}, and
applies the additional program transformations described in
Section~\ref{sec:model-extensions} opportunistically.
The cost of LLVM's solution minus one is used as an upper bound on Unison's
objective function (equation~\eqref{eq:generic-objective-function}), which
guarantees that any solution found by Unison improves that of LLVM.
The execution frequency of each basic block is estimated using LLVM's standard
analysis.

Different flags are used to compile for speed and code size optimization.
Table~\ref{tab:flags} shows the flags used for each goal and LLVM component
(\code{clang}, \code{opt}, and \code{llc} are LLVM's front-, middle-, and
back-end).
The neutral optimization flag \code{O2} is chosen for \code{clang} to avoid
favoring one optimization goal at the expense of the other.
\code{O2} is also chosen for code size optimization with \code{llc} since this
LLVM component does not provide a more specific \code{Oz} flag for this goal.
Certain CFG transformations that are orthogonal to the problems solved by Unison
are disabled from \code{llc} to produce a more accurate comparison (see last row
in Table~\ref{tab:flags}).
The \emph{clustering} column is discussed in
Section~\ref{sec:estimated-code-quality}.

\begin{table}
  \caption{Flags for each goal and LLVM component.}
  \label{tab:flags}
  \begin{minipage}{\columnwidth}
  \begin{center}
    \begin{tabu}{lccc}
      \hline
      \rowfont{\bfseries}
      & speed optimization & code size optimization & clustering\\\hline
      \code{clang} & \code{O2} & \code{O2} & \code{O2}\\\hline
      \code{opt} & \code{O3} & \code{Oz} & \code{O2}\\\hline
      \code{llc} & \code{O3} & \code{O2} & \code{O2}\\
      & \multicolumn{3}{p{10cm}}{prefixed with \path{disable-}: \path{post-ra},
        \path{tail-duplicate}, \path{branch-fold}, \path{block-placement},
        \path{phi-elim-edge-splitting}, \path{if-conversion},
        \path{hexagon-cfgopt} (Hexagon),
        \path{arm-optimize-thumb2-in-cp-islands} (ARM),
        \path{skip-mips-long-branch} (MIPS)}\\\hline
    \end{tabu}
  \end{center}
  \end{minipage}
\end{table}%

\paragraph{Unison configuration}

\newcommand{\numberOfRepetitions}{5}

Unison is run on a Linux machine equipped with an Intel Xeon E5-2680 processor
with 12 hyperthreaded cores and 64 GB of main memory.
Each of the solvers in the external portfolio runs in its own operating
system process with a time limit of 15 minutes.
The time limit has been chosen to make it feasible to run a large number
of benchmarks. Time has been chosen as a limit as other measures
(such as number of search nodes) vary widely across the
benchmark set.
This limit does not only determine the solving time but also the quality of
the generated code: as the solvers exhibit anytime behavior, a larger time
limit translates into better code.
The potential dispersion in code quality measurements due to timeouts is
controlled by using the median of~$\numberOfRepetitions{}$~repetitions.
Across these repetitions, code quality varies for
$\percentDifferentFunctions{}\%$ of the functions.
For the most variable function, the quality between the worst and the best
solutions differs
by~$\improvedmultiAllMaxUnisonCostDistanceFromMedianInReps{}\%$.

To make benchmarking feasible, the internal portfolio in the decomposition-based
solver is run sequentially in a round-robin fashion.
The same holds true when solving the subproblems.
We have verified that running the internal portfolio and solving the subproblems
in parallel is only slightly advantageous in that it improves the estimated
speedup by a factor of less than $1.1$.
Instead, the experiments devote the large number of cores available to solving
multiple functions in parallel.

The experiments use the revision \code{33bdc9b} of Unison~\cite{UnisonTool2018},
Gecode 6.0.0~\cite{Gecode2018} for the decomposition-based solver, and Chuffed as distributed with
MiniZinc 2.1.2~\cite{MiniZinc2018}.

\subsection{Estimated Code Quality}\label{sec:estimated-code-quality}

This section compares the quality of the code generated by Unison to that of
LLVM as a state-of-the-art representative of heuristic approaches.
The approaches are compared for speed and code size optimization on the three
evaluation processors.
The quality of Unison and LLVM solutions is estimated according to
equation~\eqref{eq:generic-objective-function}.
As discussed in Section~\ref{sec:objective-function}, the speed estimation is
subject to inaccuracies due to its dynamic nature, while code size reduction is
a static goal which leads to a naturally accurate estimation~\cite{Koes2009}.
Actual speedup measurements are given in Section~\ref{sec:actual-speedup}, while
the accuracy of the speed estimation is studied in detail in
Appendix~\ref{app:accuracy}.

The quality of a Unison solution is presented as its improvement (speedup and
code size reduction) over LLVM's corresponding solution.
The improvement of a Unison solution where
equation~\eqref{eq:generic-objective-function} has value $u$ over an LLVM solution
where equation~\eqref{eq:generic-objective-function} has value $l$ is computed as:
\begin{equation}\label{eq:improvement}
  \improvement{u}{l} =
  \begin{cases}
    \frac{l}{u} - 1, & \text{if } l \ge u \\
    1 - \frac{u}{l}, & \text{otherwise}.
  \end{cases}
\end{equation}
Hence the magnitude of the improvement is independent of its sign (negative
improvements occur in the improvement estimation of isolated approaches and in
the actual speedup measurements).

The gap of a Unison solution where the solver provides a lower bound $u^{*}$ on
equation~\eqref{eq:generic-objective-function} and $u$ and $l$ are defined as
above is computed as:
\begin{equation}\label{eq:gap}
  \gap{u}{l}{u^{*}} = \improvement{u^{*}}{l} - \improvement{u}{l}
\end{equation}
Improvement and gap results are summarized using the geometric mean.

\begin{example}\label{ex:speedup-computation}
  According to equation~\eqref{eq:improvement}, the speedup of Unison over LLVM in
  Example~\ref{ex:motivation} (assuming the execution frequencies from
  Example~\ref{ex:integrated-solution-cost}) is:
\begin{equation}\label{eq:improvement-example}
  \improvement{u}{l} = \improvement{22}{23} = 23 \div 22 - 1 = 4.5 \%.
\end{equation}
Let us assume that the solver times out without proving that the solution in
Example~\ref{ex:motivation} is optimal and provides a lower bound of $u^{*} =
20$.
Then the gap is:
\begin{equation}\label{eq:gap-example}
  \gap{u}{l}{u^{*}} = \improvement{20}{23} - \improvement{22}{23} = 15 \% - 4.5 \% = 10.5 \%.
\end{equation}
\end{example}

\paragraph{Input functions}

As input for the estimated code quality and scalability experiments,
100~functions are sampled out of a pool of~\numprint{10874} functions from~22
MediaBench and SPEC CPU2006 applications.
The purpose of sampling is to keep the experimental evaluation feasible
while ensuring that the selected functions are sufficiently diverse to
exercise different aspects of Unison.
The size of the sampled functions ranges from small (up to 100~input
instructions) to medium (between~100 and~1000 input instructions).

The sampling procedure splits all benchmark functions into 100 clusters with a
$k$-means clustering algorithm~\cite{MacQueen1967} and selects the most central
function from each cluster~\cite{Phansalkar2005}.
Functions are clustered by size (input instructions, input instructions per
basic block), key register allocation and instruction scheduling features
(register pressure, instruction-level parallelism, and call instruction ratio),
and parent application (to improve the spread across the benchmark suites).
Register pressure is approximated by the spill ratio of LLVM's register
allocator, and instruction-level parallelism is approximated by the number of
instructions scheduled per cycle by LLVM's instruction scheduler.
The neutral optimization flag \code{O2} is used to extract the clustering
features as Table~\ref{tab:flags} shows.
All features are averaged over the evaluation processors.
Appendix~\ref{app:functions} details the
features and cluster size of each selected function.

\paragraph{Speedup over LLVM}

The speedup gained by Unison over LLVM is significant for Hexagon
($\improvedHexagonSpeedMeanImprovement{}\%$~mean speedup), moderate for MIPS
($\improvedMipsSpeedMeanImprovement{}\%$), and only slight for ARM
($\improvedARMSpeedMeanImprovement{}\%$).
Figure~\ref{fig:estimated-speedup} shows the improvement for each function
(black) and the gap for the functions where Unison times out (gray).
Each function is identified by a number, the details can be found in
Appendix~\ref{app:functions}.

The significant improvement for Hexagon is due to a better exploitation of its
VLIW architecture, where Unison generally schedules more instructions in
parallel than LLVM's list scheduling algorithm.
In doing so, Unison spills on average
$\improvedHexagonSpeedMeanSpillNumberWorsening{}\%$~more temporaries, but
manages to minimize the overhead by scheduling the spill code in parallel with
other instructions.
Similarly, the overhead of live range splitting is minimized as its precise cost
in terms of instruction scheduling is taken into account.
Furthermore, Unison tends to spill less callee-saved registers in loopless
functions than LLVM.
This type of spilling has a high overhead, because in loopless functions the
callee-saved spill code is placed in the most frequent basic blocks.
The extreme case is function~$72$, for which Unison almost doubles the execution
speed of LLVM by reducing the number of live range splits by two thirds and the
number of callee-saved spills from five to two.

The moderate improvement for MIPS is almost entirely achieved on loopless
functions, by spilling short live ranges in low-frequency basic blocks rather
than callee-saved registers.
For example, Unison speeds up function~$79$ by~$82\%$ by spilling~$17$
temporaries in low-frequency basic blocks instead of eight callee-saved
registers in the high-frequency entry and exit basic blocks.
On average, this strategy
spills~$\improvedMipsSpeedMeanSpillNumberWorsening{}\%$ more temporaries but
reduces the total spill code overhead
by~$\improvedMipsSpeedMeanSpillOverheadImprovement{}\%$.

For MIPS on functions with loops and for ARM in general, LLVM's heuristic
approach suffices to generate code of near-optimal speed.
The occasional speedups achieved for ARM are due to reducing the amount of
spilling by $\improvedARMSpeedMeanSpillNumberImprovement{}\%$~on average (for
example in functions~$77$, $79$ and~$100$), adjusting the aggressiveness of
coalescing to the frequency of the corresponding basic block (for example in
functions~$77$, $79$ and~$55$), and rematerializing more aggressively (for
example in function~$88$).

\newtoggle{showPlotNames}
\toggletrue{showPlotNames}
\newcommand{\plotVertDist}{-2.5cm}

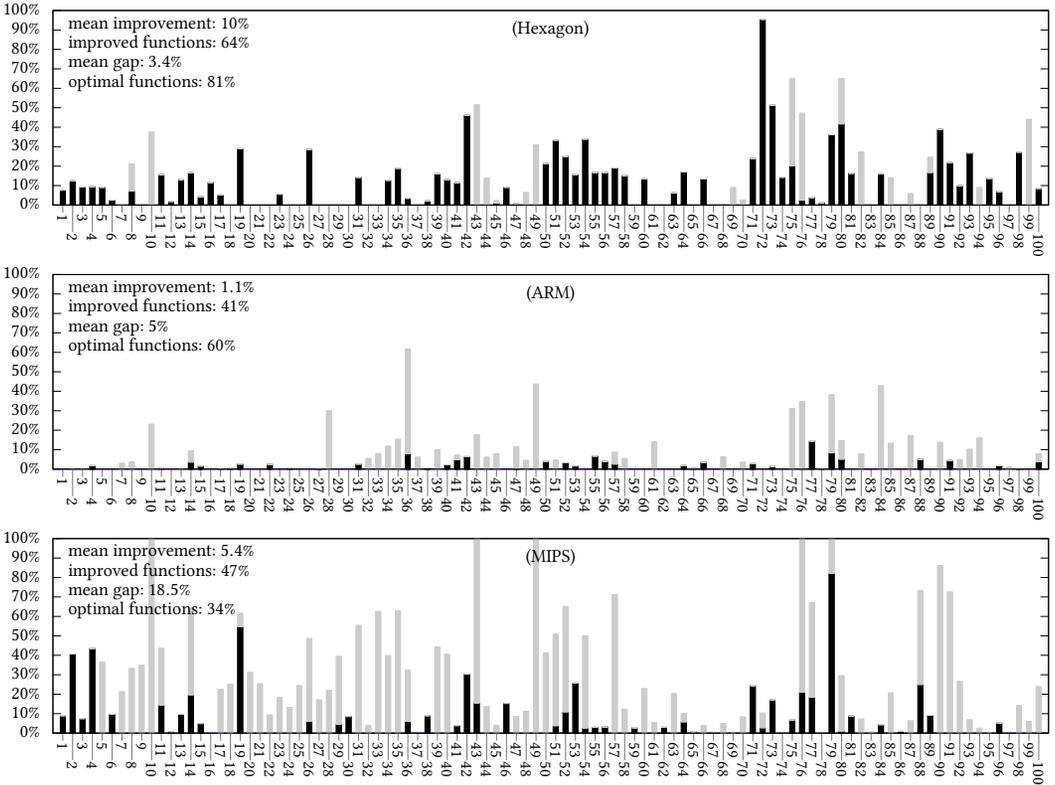
\begin{figure}
  \togglefalse{showPlotNames}%
  \adjustbox{trim=0.15cm 0cm 0cm 0cm,clip=true,scale=\plotScale}{\input{./results/hexagon-speed-improvement}}\\
  \vspace{\plotVertDist}
  \adjustbox{trim=0.15cm 0cm 0cm 0cm,clip=true,scale=\plotScale}{\input{./results/arm-speed-improvement}}\\
  \vspace{\plotVertDist}
  \adjustbox{trim=0.15cm 0cm 0cm 0cm,clip=true,scale=\plotScale}{\input{./results/mips-speed-improvement}}\\
  \vspace{\plotVertDist}
  \caption{Estimated speedup over LLVM (black) and gap (gray) for each function.}
  \label{fig:estimated-speedup}
\end{figure}

\paragraph{Code size reduction over LLVM}

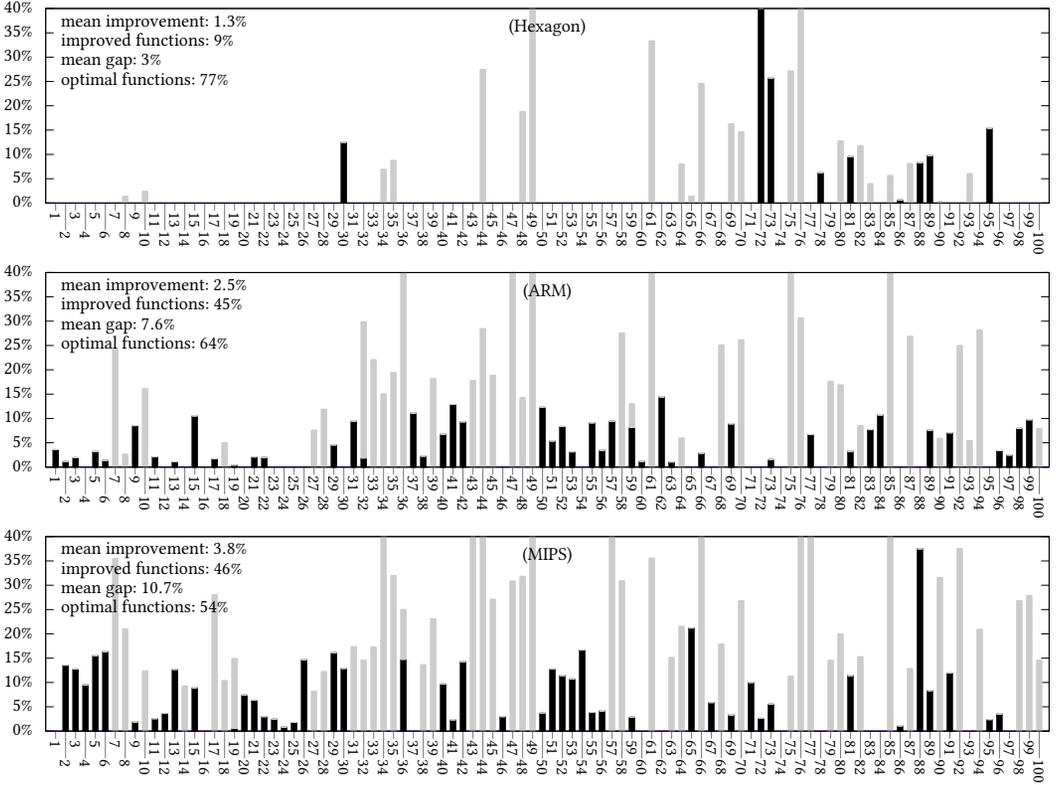
\begin{figure}
  \togglefalse{showPlotNames}%
  \adjustbox{trim=0.1cm 0cm 0cm 0cm,clip=true,scale=\plotScale}{\input{./results/hexagon-size-improvement}}\\
  \vspace{\plotVertDist}
  \adjustbox{trim=0.1cm 0cm 0cm 0cm,clip=true,scale=\plotScale}{\input{./results/arm-size-improvement}}\\
  \vspace{\plotVertDist}
  \adjustbox{trim=0.1cm 0cm 0cm 0cm,clip=true,scale=\plotScale}{\input{./results/mips-size-improvement}}\\
  \vspace{\plotVertDist}
  \caption{Code size reduction over LLVM (black) and gap (gray) for each function.}
  \label{fig:estimated-size}
\end{figure}

The code size reduction achieved by Unison over LLVM is moderate for MIPS
($\improvedMipsSizeMeanImprovement{}\%$ mean code size reduction) and ARM
($\improvedARMSizeMeanImprovement{}\%$), and only slight for Hexagon
($\improvedHexagonSizeMeanImprovement{}\%$).
Although the estimated code size reduction is in general modest, the results
directly translate into actual improvement unlike the case of speed
optimization.
Figure~\ref{fig:estimated-size} shows the improvement for each function (black)
and the gap for the functions where Unison times out (gray).

The moderate improvement for MIPS is mostly due to aggressive coalescing that is
boosted by the ability to reorder instructions simultaneously.
In the extreme case (function~$88$), Unison eliminates all~$10$
register-to-register move instructions placed by LLVM.
Unison recognizes the high impact of coalescing on code size and spills
$\improvedMipsSizeMeanSpillNumberWorsening{}\%$ more often than LLVM just to
facilitate additional coalescing.
For example, in function~$65$ Unison places two load instructions more than LLVM
but in return it eliminates nine register-to-register move instructions.

ARM also benefits from Unison's aggressive coalescing, albeit to a lesser
extent.
Additional code size reduction for this processor is achieved by exploiting the
Thumb-2 instruction set extension which allows 16- and 32-bit instructions to be
freely mixed (see Section~\ref{sec:alternative-instructions}).
For example, Unison reduces significantly the size of function~$41$ by
selecting~$14$ more 16-bit instructions than LLVM, at the expense of an
additional move instruction.

For Hexagon, LLVM's heuristic approach generates code of near-optimal size.
Unison is only able to improve nine functions by coalescing aggressively (for
example function~$30$), by reducing the amount of spilling (for example
function~$89$), or by a combination of both (for example function~$73$).

\paragraph{Impact of integration}

Unison's improvement over LLVM can be partially attributed to its integrated
approach to register allocation and instruction scheduling.
To evaluate the fundamental benefit of solving these problems in integration, we
measure the improvement of the optimal solutions generated by Unison over those
generated by solving register allocation and instruction scheduling optimally
but in isolation.
We call this optimal but isolated variant \emph{Disunion}.

Disunion proceeds as follows.
First, global register allocation is solved optimally according to the model in
Section~\ref{sec:global-register-allocation}, assuming LLVM's given operation
order.
The objective function is similar to that of earlier combinatorial register
allocation approaches (see Section~\ref{sec:related-work}).
For speed optimization, the objective function minimizes total instruction
latency weighted by execution frequency:
\begin{equation}\label{eq:speed-optimization-for-register-allocation}
  \blockWeight{b} = \blockFrequency{b}; \qquad
  \blockCost{b} =
  \sum_{\mathclap{o \in \operations{b} \,\suchThat\, \activeOperation{o}}}
      {\latencyNoParen{\elementMark{\operationInstruction{o}}}} \quad
  \forall b \in \allBlocks{}.
\end{equation}
For code size optimization, Unison's usual objective function
(equation~\eqref{eq:size-optimization}) is employed.
After global register allocation, Disunion solves instruction scheduling
optimally according to the model in Section~\ref{sec:instruction-scheduling}.
The model is completed with additional dependencies caused by register
assignment.
The objective function for isolated instruction scheduling is the same as for
Unison for both goals.

The mean improvement of solving register allocation and instruction scheduling
in integration over the isolated approach ranges from a
slight~$\bothphasesARMSpeedMeanIntegrationImprovementOverTwophase{}\%$ to a
significant~$\bothphasesHexagonSpeedMeanIntegrationImprovementOverTwophase{}\%$ for the
different goals and processors, as summarized in Table~\ref{tab:integration}.
These results confirm the benefit of integrating register allocation and
instruction scheduling, although the extent depends on the goal and processor.
In general, the integrated approach obtains better code through more effective
spilling and coalescing, two program transformations that particularly benefit
from simultaneous instruction scheduling.
Additionally, the integrated approach generates shorter schedules for speed
optimization, either by exploiting Hexagon's VLIW architecture or by hiding long
latencies in the case of ARM and MIPS.

\begin{table}
  \caption{Mean improvement of Unison's optimal solutions over combinatorial
    approach in isolation.}
  \label{tab:integration}
  \begin{minipage}{\columnwidth}
  \begin{center}
    \begin{tabu}{lccc}
      \hline
      \rowfont{\bfseries}
      goal & Hexagon & ARM & MIPS \\\hline
      speed &
      $\bothphasesHexagonSpeedMeanIntegrationImprovementOverTwophase{}\%$ &
      $\bothphasesARMSpeedMeanIntegrationImprovementOverTwophase{}\%$ &
      $\bothphasesMipsSpeedMeanIntegrationImprovementOverTwophase{}\%$ \\\hline
      code size optimization &
      $\bothphasesHexagonSizeMeanIntegrationImprovementOverTwophase{}\%$ &
      $\bothphasesARMSizeMeanIntegrationImprovementOverTwophase{}\%$ &
      $\bothphasesMipsSizeMeanIntegrationImprovementOverTwophase{}\%$ \\\hline
    \end{tabu}
  \end{center}
  \end{minipage}
\end{table}

Comparing the improvement of Unison and Disunion over LLVM for the same subset
of functions solved optimally reveals that the integration is a significant
factor in the overall Unison improvement.
For example, the mean speedup of Unison over LLVM for Hexagon on this function
subset is~$\bothphasesHexagonSpeedMeanIntegrationImprovementOverLLVM{}\%$, while
that of Disunion over LLVM is
only~$\bothphasesHexagonSpeedMeanTwophaseImprovementOverLLVM{}\%$.
In two cases, Disunion generates on average worse code than LLVM
($\bothphasesHexagonSizeMeanLLVMImprovementOverTwophase\%$ mean code size
increase for Hexagon and~$\bothphasesARMSpeedMeanLLVMImprovementOverTwophase\%$
mean slowdown for ARM).
In these cases, the objectives of register allocation and instruction scheduling
conflict to the extent that solving each problem optimally only decreases the
quality of the generated code.
Unison avoids this issue by solving register allocation and instruction
scheduling simultaneously according to a single objective function.

The code quality benefits of Unison's integrated approach come at a price in
terms of scalability compared to Disunion's decomposed approach.
Given the same time limit, Disunion is able to solve optimally
between~$\bothphasesHexagonSpeedTwophaseTimesOptimalOverIntegrated{}$
and~$\bothphasesMipsSpeedTwophaseTimesOptimalOverIntegrated{}$ more functions
than Unison, depending on the processor and goal.

\subsection{Scalability}\label{sec:scalability}

This section studies the scalability of Unison for the three evaluation
processors and the same functions as in
Section~\ref{sec:estimated-code-quality}.
The section aggregates speed and code size optimization results (unless
otherwise stated), as the scalability trends are similar for both goals.

\paragraph{Overall scalability}

\newcommand{\percentOfSmallAndMediumSizedFunctions}{96}

\begin{figure}%
  \centering%
  \subfloat[\label{fig:scalability} Solving time to optimality by function size.]
           {\adjustbox{trim=0.25cm 0cm 0.2cm 0cm,clip=true,scale=\plotScale}{\input{./results/scalability}}}
   \subfloat[\label{fig:optimality} Accumulated \% of optimal solutions over time.]
           {\adjustbox{trim=0.25cm 0cm 0.2cm 0cm,clip=true,scale=\plotScale}{\input{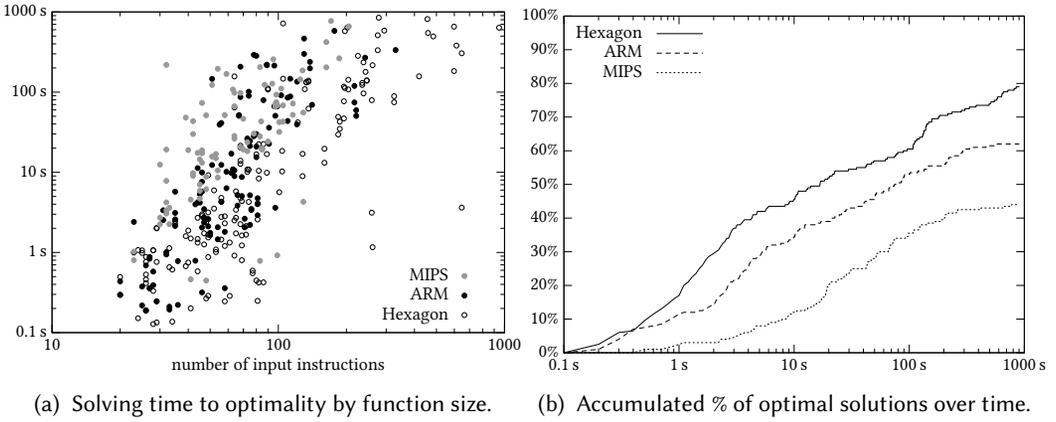}}}
  \caption{Solving complexity for different processors.\label{fig:solving-complexity}}
\end{figure}

The results indicate that Unison scales up to medium-sized functions, which puts
$\percentOfSmallAndMediumSizedFunctions{}\%$ of all MediaBench and SPEC CPU2006
functions within reach.
Figure~\ref{fig:solving-complexity} summarizes the solving complexity of
the experiment functions.
Figure~\ref{fig:scalability} shows the solving time of each function solved
optimally.
Depending on the processor, Unison solves functions of up to
$\improvedmultiAllLargestOptimalIns{}$ input instructions optimally in tens to
hundreds of seconds.
At the same time, Unison can time out for functions with as few as
$\improvedmultiAllSmallestNonZeroGapIns{}$ input instructions.
In such cases, high-quality solutions with moderate gaps are often obtained
independently of the function size (mean
gap~$\improvedmultiAllMeanNonZeroGap{}\%$).

\begin{figure}%
  \adjustbox{trim=0.0cm 0.0cm 0.0cm 0.0cm,clip=true,scale=\plotScale}{\input{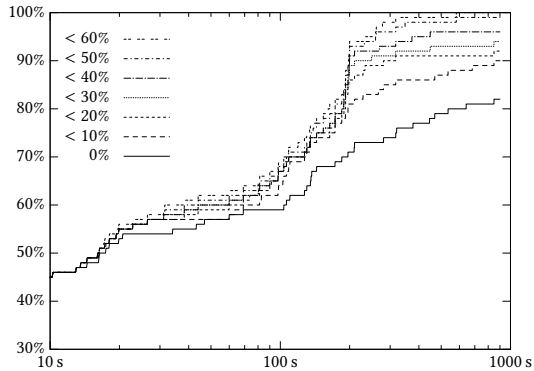}}
  \caption{Accumulated \% of solutions over time for different gap requirements (speed optimization on Hexagon).\label{fig:gap-requirements}}
\end{figure}

The gap information does not only provide insight into the quality of a
solution, but can be used to push the scalability of our approach
further while
preserving its code quality guarantees.
Figure~\ref{fig:gap-requirements} shows the accumulated percentage of acceptable
solutions obtained over time for different gap requirements, when optimizing for
speed on Hexagon.
The solutions to $\improvedHexagonSpeedPercentOptimal{}\%$ of the functions are
proven optimal ($0\%$ gap) within the time limit.
If the optimality requirement is relaxed to a certain gap limit, the scalability
of Unison improves as more functions can be solved acceptably.
For example, $\percentOfSolvedFunctionsWithAGapOfTen{}\%$ of the
functions are solved with less than $10\%$ gap within the time limit.
The percentage of acceptable solutions increases with diminishing returns for
larger gap limits, as Figure~\ref{fig:gap-requirements} shows.

\paragraph{Impact of different processors}

The scalability of Unison does not only depend on the function under
compilation, but is also significantly affected by the targeted processor.
The results show that the best scalability is achieved for Hexagon: Unison can
solve optimally~$\improvedmultiAllHexagonTimesMoreOptimalOverHundredThanArm{}$
and~$\improvedmultiAllHexagonTimesMoreOptimalOverHundredThanMips{}$ times more
medium-sized functions (between 100 and 1000~input instructions) for this
processor than for ARM and MIPS.
Furthermore, the largest function solved optimally for Hexagon has
$\improvedHexagonAllLargestOptimalIns{}$ input instructions, compared to
$\improvedARMAllLargestOptimalIns{}$ input instructions for ARM and
$\improvedMipsAllLargestOptimalIns{}$ input instructions for MIPS (see
Figure~\ref{fig:scalability}).
The same trend can be seen for small functions (up to 100~input instructions),
where Unison times out in $\improvedmultiAllMipsPercentNonZeroGapUnderHundred\%$
and $\improvedmultiAllArmPercentNonZeroGapUnderHundred\%$ of the cases for MIPS
and ARM but not a single time for Hexagon.
The median size of the functions for which Unison times out
is~$\improvedHexagonAllMedianNonZeroGapIns{}$,
$\improvedARMAllMedianNonZeroGapIns{}$,
and~$\improvedMipsAllMedianNonZeroGapIns{}$ input instructions for Hexagon, ARM,
and MIPS.
Given a fixed size, more functions are solved optimally and in less time for
Hexagon while MIPS yields the fewest optimal solutions and the longest solving
times (see Figure~\ref{fig:scalability}).
As a consequence, given a time limit, Unison solves more functions optimally for
Hexagon, followed by ARM and MIPS as Figure~\ref{fig:optimality} corroborates.
Furthermore, for functions where Unison times out, the smallest and largest gaps
are achieved for Hexagon and MIPS respectively, independently of function size.

The fact that Unison scales better for Hexagon (a VLIW processor) than for ARM
and MIPS (single-issue processors) is surprising as the latter are considered
easier to handle by heuristic approaches~\cite{Kessler2010}.
The fact that Unison scales better for ARM than for MIPS is also unexpected, as
ARM includes more features for the solver to deal with such as selection of 16-
and 32-bit instructions (see Section~\ref{sec:alternative-instructions}) and
double-load and double-store instructions (see
Section~\ref{sec:model-extensions}).

A factor that has been found to affect Unison's scalability differently
depending on the targeted processor is the use of LLVM's solution cost as an
upper bound on the objective function.
Table~\ref{tab:initial-final-gaps} compares Unison's scalability (in percentage
of optimal solutions and mean gap) using LLVM's and a trivial upper bound.
With the LLVM upper bound (the default configuration), Unison clearly scales
best for Hexagon, followed by ARM, and followed by MIPS.
With the trivial upper bound, the results for Hexagon and ARM tend to even out,
indicating that differences in LLVM's upper bound are the main cause of
scalability divergence among these processors.
In this alternative configuration, the relative difference between MIPS and
these two processors persists, indicating that Unison is simply less effective
in solving functions for MIPS.

\begin{table}
  \caption{Percentage of optimal solutions and mean gap with LLVM and trivial upper bounds.}
  \label{tab:initial-final-gaps}
  \begin{minipage}{\columnwidth}
  \begin{center}
    \begin{tabu}{lcccc}
      \hline
      \rowfont{\bfseries}
      \multirow{2}{*}{processor} & \multicolumn{2}{c}{LLVM upper bound} & \multicolumn{2}{c}{trivial upper bound} \\
      \rowfont{\bfseries}
      & optimal solutions & mean gap & optimal solutions & mean gap \\\hline
      Hexagon &
      $\improvedHexagonAllPercentOptimal{}\%$ &
      $\improvedHexagonAllMeanGap{}\%$ &
      $\trivialboundHexagonAllPercentOptimal{}\%$ &
      $\trivialboundHexagonAllMeanGap{}\%$\\\hline
      ARM &
      $\improvedARMAllPercentOptimal{}\%$ &
      $\improvedARMAllMeanGap{}\%$ &
      $\trivialboundARMAllPercentOptimal{}\%$ &
      $\trivialboundARMAllMeanGap{}\%$\\\hline
      MIPS &
      $\improvedMipsAllPercentOptimal{}\%$ &
      $\improvedMipsAllMeanGap{}\%$ &
      $\trivialboundMipsAllPercentOptimal{}\%$ &
      $\trivialboundMipsAllMeanGap{}\%$\\\hline
    \end{tabu}
  \end{center}
  \end{minipage}
\end{table}

A major factor that limits Unison's scalability for MIPS is the high overhead of
modeling the preservation of its callee-saved registers.
If these registers are not preserved, the scalability for MIPS improves
significantly ($\mipsNoCCTimesMoreOptimal{}$~times more functions are solved
optimally), surpassing that for ARM and approaching that for Hexagon, which are
only slightly affected.
The reason for this difference is that MIPS requires 14~temporaries to hold the
value of all callee-saved registers, whereas Hexagon and ARM only require six
and three temporaries (by using double-store and double-load instructions for
Hexagon and \code{push} and \code{pop} instructions for ARM).
The number of such temporaries has a high impact on Unison's scalability as,
after LSSA construction, it is multiplied by the number of basic blocks (as
callee-saved temporaries are live through the entire function).

Another factor that also limits Unison's scalability for MIPS is the
processor's wide range of instruction latencies (from one cycle for common
arithmetic-logic instructions to 38 cycles for integer division).
If a uniform, one-cycle latency model is assumed instead for all processors, the
scalability for MIPS improves significantly
($\mipsUnitLatencyTimesMoreOptimal{}$~times more functions are solved
optimally), approaching that of ARM which is less affected.
The scalability for Hexagon remains unaltered, as this processor does not
provide floating-point or costly integer arithmetic instructions.

A factor that has been evaluated and ruled out as a potential cause of
divergence in Unison's scalability between Hexagon and the other two processors
is the issue width.
The scalability differences do not decrease if a single-issue pipeline is
assumed for Hexagon.
On the contrary, $\hexagonSingleIssueTimesMoreOptimal{}$~times more functions
are solved optimally for a single-issue version of Hexagon compared to the
original VLIW architecture.

A more comprehensive evaluation of the effect of processor features and their
interactions on Unison's scalability is part of future work.

\paragraph{Impact of solving methods}

As is common in combinatorial optimization, the solver portfolio and
solving improvements described in Section~\ref{sec:solving-in-unison} are
crucial for Unison's scalability.

The two solvers included in the portfolio
(Section~\ref{sec:solver-portfolio}) are highly complementary:
$\improvedmultiAllPercentGecodeSolverTotal{}\%$ of the solutions are
delivered by the decomposition-based solver and the remaining ones by the
off-the-shelf solver.
The amount of solutions delivered by each solver is distributed rather evenly as
function size increases, but the largest function solved by the
decomposition-based solver ($\improvedmultiAllLargestGecodeWin{}$~input
instructions) is larger than the corresponding one by the off-the-shelf solver
($\improvedmultiAllLargestMinizincWin{}$~input instructions).
The better scalability of the decomposition-based solver is achieved by
exploiting the structure of the LSSA form as explained in
Section~\ref{sec:solver-portfolio}.

\begin{figure}%
  \centering%
  \subfloat[\label{fig:combined-scalability} Solving time to optimality by function size.]
           {\adjustbox{trim=0.25cm 0cm 0.2cm 0cm,clip=true,scale=\plotScale}{\input{./results/combined-scalability}}}
  \subfloat[\label{fig:combined-optimality} Accumulated \% of optimal solutions over time.]
           {\adjustbox{trim=0.25cm 0cm 0.2cm 0cm,clip=true,scale=\plotScale}{\input{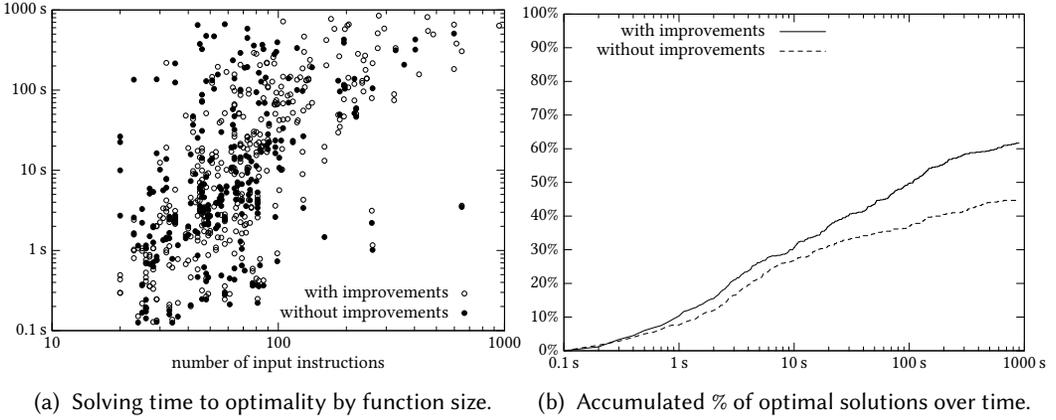}}}
           \caption{Solving complexity with and without solving improvements.\label{fig:combined-solving-complexity}}
\end{figure}

The solving improvements described in Section~\ref{sec:solving-improvements}
allow Unison to find significantly more solutions, in particular for larger
functions.
Figure~\ref{fig:combined-solving-complexity} summarizes the solving
complexity of the experiment functions for all processors with and without
solving improvements.
The percentage of functions that are solved optimally grows from
$\basicmultiAllPercentOptimal{}\%$ to $\improvedmultiAllPercentOptimal\%$ when
the solving improvements are applied.
Many of the functions that are additionally solved are medium-sized: the solver
with improvements can solve optimally
$\combinedmultiAllImprovedTimesMoreOptimalOverHundredThanBasic{}$ times more
functions of this size than without improvements.
The largest functions obtained with and without improvements have
$\improvedmultiAllLargestOptimalIns{}$ and $\basicmultiAllLargestOptimalIns{}$
input instructions respectively (see Figure~\ref{fig:combined-scalability}).
The same trend can be observed for small functions, where Unison times out
$\combinedmultiAllBasicTimesMoreNonZeroGapUnderHundredThanImproved{}$ times more
often when the improvements are not applied.
The median size of the functions for which Unison times out is
of~$\improvedmultiAllMedianNonZeroGapIns{}$ input instructions with improvements
and~$\basicmultiAllMedianNonZeroGapIns{}$ input instructions without.

In general, when the improvements are enabled more functions are solved
optimally and in less time for any size (see
Figure~\ref{fig:combined-scalability}).
Furthermore, given a time limit, significantly more functions are solved
optimally (see Figure~\ref{fig:combined-optimality}).
An exception is for the most trivial functions that are solved optimally within
$\combinedmultiAllImprovedCrossesBasicTime{}$ seconds, for which the overhead
introduced by the solving improvements is not amortized.
Furthermore, for functions that cannot be solved optimally the improvements
reduce the gap significantly: from $\basicmultiAllMeanNonZeroGap{}\%$ to
$\improvedmultiAllMeanNonZeroSmallGap{}\%$ mean gap.

\subsection{Actual Speedup}\label{sec:actual-speedup}

This section compares the \emph{actual} (as opposed to \emph{estimated})
execution speed of the code generated by Unison to that of LLVM.
The section contributes the first actual speedup results for a combinatorial
register allocation and instruction scheduling approach.

\paragraph{Processor, input functions, and execution platform}

For the actual speedup experiments, we focus on Hexagon as the processor for
which Unison improves most functions and estimates the highest speedup (see
Figure~\ref{fig:estimated-speedup}).
We select functions from MediaBench as this benchmark suite characterizes best
the multimedia and communications applications targeted by Hexagon.
MediaBench consists of~$11$ applications where most applications can be run in
two modes (typically encoding and decoding, represented with \texttt{+} and
\texttt{-} respectively).
The applications \code{ghostscript} and \code{mesa} are excluded due to
compilation errors in LLVM.

Each application and mode is compiled and profiled with LLVM using the reference
input loads provided by MediaBench, and the top-five hottest functions (those
that account for most execution time) are selected for compilation with Unison.
This methodology is applied to ensure that the selected functions are
sufficiently exercised during the execution and to enable a whole-application
speedup comparison.
The same methodology could be used in a practical setup to strike a balance
between compilation time and code quality, exploiting the Pareto Principle which
is commonly observed in software (``20\% of the code accounts for 80\% of the
execution time'').
In the case of MediaBench, the selected functions account for more than
$\minExecPercentAcrossApps{}\%$ of the execution time in all applications
and modes except \code{epic+}, \code{epic-}, \code{mpeg2+}, and
\code{rasta} where most execution time is spent in floating-point emulation
library calls.

Appendix~\ref{app:functions} details the features and percentage of the
execution time of each selected function within its application and mode.
The applications are executed on Qualcomm's Hexagon
Simulator~6.4~\cite{HexagonSim2013}.
Using a simulator eases reproducibility, provides detailed execution statistics,
and permits experimenting with architectural variations (as in
Appendix~\ref{app:accuracy}) at the cost of a slight accuracy loss (cache
simulation induces an error of up to 2\% from perfect
cycle-accuracy~\cite{HexagonSim2013}).
This paper assumes that this error is propagated similarly for LLVM and Unison's
generated code.

\paragraph{Function speedup over LLVM}

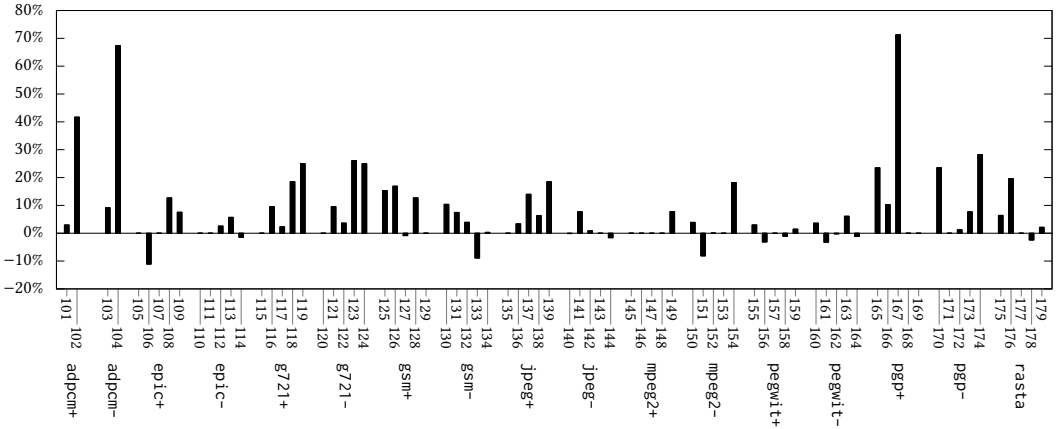
\begin{figure}
  \togglefalse{showPlotNames}%
  \toggletrue{showAppNames}%
  \adjustbox{trim=0.0cm 2.3cm 0.0cm 0.0cm,clip=true,scale=\plotScale}{\input{./results/base-speedup}}
  \caption{Actual function speedup over LLVM grouped by application and mode.}
  \label{fig:actual-speedup-single}
\end{figure}

The results show that the speedup gained by Unison over LLVM for the hottest
MediaBench functions is significant ($\baseMeanImprovement{}\%$ mean speedup),
despite a few individual slowdowns.
Figure~\ref{fig:actual-speedup-single} shows the actual speedup over LLVM for
each hot function in MediaBench, grouped by application and mode.
\pgfmathparse{(-1)*\baseWorst}
For $\basePercentOfBetterInstances{}\%$ of the functions, Unison speeds up the
code generated by LLVM (up to $\baseBest{}\%$), while
$\basePercentOfWorseInstances{}\%$ of the functions are actually slowed down
(down by $\pgfmathprintnumber{\pgfmathresult}\%$), contradicting Unison's
estimation.
Appendix~\ref{app:accuracy} studies the factors behind this contradiction.

\newtoggle{showAppNames}
\toggletrue{showAppNames}

\paragraph{Application speedup over LLVM}

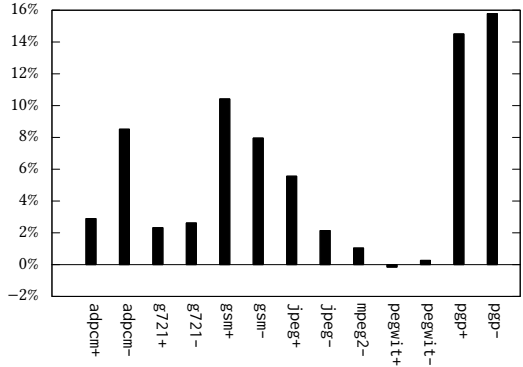
\begin{figure}
  \toggletrue{showPlotNames}%
  \adjustbox{trim=0.0cm 0.0cm 0.0cm 0.0cm,clip=true,scale=\plotScale}{\input{./results/multi-speedup}}
  \caption{Actual application speedup over LLVM.}
  \label{fig:actual-speedup-multi}
\end{figure}

The results show that the actual speedup demonstrated by Unison for individual
functions propagates proportionally to whole applications
($\multiMeanImprovement{}\%$ mean speedup).
Figure~\ref{fig:actual-speedup-multi} shows the actual speedup when Unison is
applied to the five hottest functions of each application.
Unison speeds up all applications (up to $\multiBest{}\%$), except
\code{pegwit+} which is slightly slowed down ($\multiWorst{}\%$).
\code{epic+}, \code{epic-}, \code{mpeg2+}, and \code{rasta} are excluded
from the comparison as less than $5\%$ of their execution time is spent on
code generated by Unison.
In general, Amdahl's law applies to the speedup results, as only a fraction of
each application is compiled by Unison.
Additional speedup can be expected if this fraction is increased.

\section{Conclusion and Future Work}\label{sec:conclusion-and-future-work}

This paper has introduced a combinatorial approach to integrated register
allocation and instruction scheduling.
It is the first approach of its kind to be \emph{practical}, as it is
\emph{complete} (modeling all program transformations used in state-of-the art
compilers), \emph{scalable} (scaling to medium-sized functions of up to
$1000$~instructions), and \emph{executable} (generating executable code).
Key to satisfying these properties is the use of constraint programming to
capture and exploit the structure underlying register allocation and instruction
scheduling.

The approach is implemented in the open-source tool Unison.
A thorough experimental evaluation on Hexagon, ARM, and MIPS confirms that
Unison generates better code than LLVM (in terms of estimated speedup and code
size reduction) while scaling to medium-sized functions.
A significant part of this improvement is found to stem from the integrated
nature of Unison.
For the first time, the evaluation confirms that the speedup estimate for
MediaBench benchmarks on Hexagon results in actual execution speedup.

Our approach can be used in practice to trade compilation time for code quality
beyond the usual compiler optimization levels, fully exploit processor-specific
features, and identify improvement opportunities in existing heuristic
algorithms.

\paragraph{Future work}

Future work includes addressing the model limitations discussed in
Section~\ref{sec:global-register-allocation} (lack of global support for
multi-allocation) and Section~\ref{sec:instruction-scheduling} (local
instruction scheduling).
A first step towards global instruction scheduling could be to follow the
superblock approach proposed by Malik \etal{}~\cite{Malik2008b}.
Additional improvements can also be expected from extending the scope of
register allocation to multiple functions~\cite{Wall1986}.

The actual speedup delivered by our approach can be improved by addressing the
model inaccuracies that in the worst case can lead to slowdowns, as seen in
Section~\ref{sec:actual-speedup} and discussed in Appendix~\ref{app:accuracy}.
This line of work includes characterizing those processor
features responsible for the inaccuracy and capturing their
effect in the cost model that underlies the objective function.
A greater challenge is to devise accurate cost models for out-of-order
processors such as x86~\cite{x862017}.
%
%
Supporting x86 in Unison is an ongoing effort~\cite{UnisonX862018}.

Another potential direction is to improve scalability by
exploring a broader array of solving
methods and combinatorial optimization techniques.
For example, hybrid integer and constraint programming techniques tend to
perform better than each of the techniques in isolation for a wide range of
resource allocation and scheduling problems~\cite{Lombardi2012}.

Finally, a longer-term goal is to integrate register allocation and instruction
scheduling with instruction selection.
Such an approach would not only deliver additional improvements, but also
provide a framework for studying the trade-offs and interactions between
different configurations of the three problems.
A constraint-based approach to instruction selection is available together with
a discussion of how the models could be integrated~\cite{GHB:PHD}.
The main challenge lies in devising modeling and solving improvements to handle
the considerable size of the resulting solution space.

\begin{acks}
The authors are grateful for helpful comments from David Broman, Mattias
Eriksson, Aly Metwaly, Martin Nilsson, Martin Persson, and the anonymous
reviewers.
This work has been partially funded by
\grantsponsor{Ericsson}{Ericsson AB}{https://www.ericsson.com/}
and the \grantsponsor{VR}{Swedish Research Council
  (VR)}{https://www.vr.se/} under
grant~\grantnum{VR}{621-2011-6229}.
\end{acks}

\bibliographystyle{ACM-Reference-Format}
\bibliography{paper}

\newpage

\appendix

\section{Complete Model}\label{app:combinatorial-model}

\begin{table}
  \caption{Parameters of global register allocation and instruction scheduling
    (without extensions).}
  \label{tab:model-parameters}
  \raggedright
  \newcommand{\newSectionSep}[1]{& \\[-0.27cm]}
  \renewcommand{\arraystretch}{1.1}
  \begin{tabular}{p{0.3cm}@{}p{2cm} p{12cm}}
    \multicolumn{3}{l}{\textbf{Program parameters:}} \\
    &\parameter{B, O, P, T}{sets of basic blocks, operations, operands and temporaries}
    &\parameter{\operations{b}, \temporaries{b}}{sets of operations and temporaries in basic block $b$}
    &\parameter{\operandOperation{p}}{operation to which operand $p$ belongs}
    &\parameter{\definer{t}}{operand that potentially defines temporary $t$}
    &\parameter{\users{t}}{operands that potentially use temporary $t$}
    &\parameter{\copySemantics{o}}{whether operation $o$ is a \cop{}}
    &\parameter{\width{t}}{number of register atoms that temporary $t$ occupies}
    &\parameter{\preAssigned{p\hspace{0.025cm}}{r}}
              {whether operand $p$ is pre-assigned to register $r$}
    &\parameter{\congruent{p}{q}}{whether operands $p$ and $q$ are congruent}
    \newSectionSep{}
    \multicolumn{3}{l}{\textbf{Processor parameters:}} \\
    &\parameter{R, S}{sets of registers and resources}
    &\parameter{\instructions{o}}{set of instructions that can implement operation
      $o$}
    &\parameter{\registerClass{p}{i}}{register class of operand $p$ when implemented by instruction $i$}
    &\parameter{\latency{i}}{latency of instruction $i$}
    &\parameter{\units{i}{s}}{consumption of processor resource $s$ by instruction $i$}
    &\parameter{\duration{i}{s}}{duration of usage of processor resource $s$ by instruction $i$}
    &\parameter{\capacity{s}}{capacity of processor resource $s$}
    \newSectionSep{}
    \multicolumn{3}{l}{\textbf{Objective function parameters:}} \\
    &\parameter{\blockWeight{b}}{weight of basic block $b$}
    &\parameter{\blockCost{b}}{cost of basic block $b$}
    \end{tabular}
\end{table}

\newcommand{\tableVariable}[4]{& $#1 \in #2$ & #3 & (\ref{var:#4})}
\newcommand{\tableConstraint}[2]{%
  & \multicolumn{2}{c}{$\displaystyle#1$}
  & (\ref{con:#2})}

\begin{table}
  \caption{Combinatorial model of global register allocation and instruction
    scheduling (without extensions).}
  \label{tab:model-variables-constraints}
  \newcommand{\newSectionSep}[1]{& & & \\[-0.3cm]}
  \newcommand{\afterSectionSep}[1]{& & & \\[-0.4cm]}
  \newcommand{\littleSep}[1]{& & & \\[-0.4cm]}
  \togglefalse{applyHighlight}
  \renewcommand{\arraystretch}{1.3}
  \begin{tabular}{p{0.3cm}@{}p{3.4cm}@{}p{9cm}@{}r@{}}
    \multicolumn{4}{l}{\textbf{Variables:}}\\
    \afterSectionSep{}
    \tableVariable{\temporaryRegister{t}}{R}{register to which temporary $t$ is assigned}{reg}\\
    \tableVariable{\operationInstruction{o}}{\instructions{o}}{instruction that implements operation $o$}{ins}\\
    \tableVariable{\operandTemporary{p}}{\temps{p}}{temporary used or defined by operand $p$}{temp}\\
    \tableVariable{\liveTemporary{t}}{\booleans{}}{whether temporary $t$ is live}{live}\\
    \tableVariable{\activeOperation{o}}{\booleans{}}{whether operation $o$ is active}{active}\\
    \tableVariable{\operationIssueCycle{o}}{\naturalNumbersZero}{issue cycle of operation $o$ from the beginning of its basic block}{cycle}\\
    \tableVariable{\temporaryLiveStart{t}, \temporaryLiveEnd{t}}{\naturalNumbersZero}{live start and end cycles of temporary $t$}{start-end}\\
    \newSectionSep{}
    \multicolumn{4}{l}{\textbf{Register allocation constraints:}}\\
    \afterSectionSep{}
    \tableConstraint{\globalDisjointLiveRangesConstraint{}}{global-disjoint-live-ranges}\\
    \tableConstraint{\preAssignmentConstraint{}}{pre-assignment}\\
    \tableConstraint{\instructionSelectionConstraint{}}{instruction-selection}\\
    \tableConstraint{\activeOperationConstraint{}}{active-operation}\\
    \littleSep{}
    \tableConstraint{\liveTemporaryConstraint{}}{live-temporary}\\
    \littleSep{}
    \tableConstraint{\congruenceConstraint{}}{congruence}\\
    \newSectionSep{}
    \multicolumn{4}{l}{\textbf{Instruction scheduling constraints:}}\\
    \afterSectionSep{}
    \littleSep{}
    \tableConstraint{\dataPrecedencesConstraint{}}{data-precedences}\\
    \littleSep{}
    \littleSep{}
    \tableConstraint{\globalProcessorResourcesConstraint{}}{global-processor-resources}\\
    \newSectionSep{}
    \multicolumn{4}{l}{\textbf{Integration constraints:}}\\
    \afterSectionSep{}
    \tableConstraint{\liveStartConstraint{}}{live-start}\\
    \tableConstraint{\liveEndConstraint{}}{live-end}\\
    \newSectionSep{}
    \multicolumn{4}{l}{\textbf{Objective function:}}\\
    \afterSectionSep{}
    &\multicolumn{2}{c}{$\displaystyle\text{\emph{minimize}} \; \genericObjectiveFunction{}$}
    &(\eqref{eq:generic-objective-function})\\
  \end{tabular}
\end{table}

This appendix summarizes the complete model as introduced in
Sections~\mbox{\ref{sec:local-register-allocation}-\ref{sec:objective-function}},
excluding the additional program transformations from
Section~\ref{sec:model-extensions}.  Table~\ref{tab:model-parameters} lists all
parameters for the input program, the processor, and the objective function,
whereas Table~\ref{tab:model-variables-constraints} lists all variables and
constraints.

\section{Accuracy of the Speedup Estimation}\label{app:accuracy}

This appendix studies the accuracy of Unison's speedup estimation which is based
on equations~\eqref{eq:generic-objective-function} and~\eqref{eq:speed-optimization}
by comparing to the actual speed measurements for MediaBench functions on
Hexagon.
The appendix extends the methodology and results from a prestudy by
Persson~\cite{Persson2017}.

\paragraph{Accuracy}

\begin{figure}%
  \adjustbox{trim=0.0cm 0.0cm 0.4cm 0.0cm,clip=true,scale=\plotScale}{\input{./results/speedup-multi-correlation}}
  \caption{Relation between estimated and actual speedup over LLVM.}
  \label{fig:speedup-relation}
\end{figure}
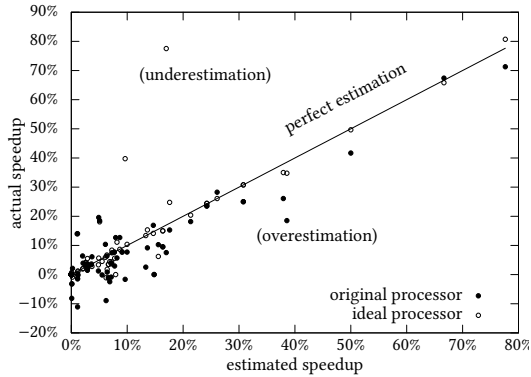

Figure~\ref{fig:speedup-relation} (\emph{original processor} points) depicts the
relation between estimated and actual speedup for each of the MediaBench
functions studied in Section~\ref{sec:actual-speedup}.
As the figure shows, Unison's speedup estimation suffers from inaccuracies.
This is confirmed by a Wilcoxon signed-rank test~\cite{Wilcoxon1945} which is
used as an alternative to the more common \emph{t}-test since the speedup
difference (estimated minus actual speedup) does not follow a normal
distribution.
A significance level of $1\%$ is required for all statistical tests.
The estimation error (the absolute value of the speedup difference) follows an
exponential distribution where the error is less than $1\%$ for
$\baseSpeedupErrorSmallPercent{}\%$ of the functions, between $1\%$ and $10\%$
for $\baseSpeedupErrorMediumPercent{}\%$ of the functions, and greater than
$10\%$ (up to $\baseLargestSpeedupError{}\%$) for
$\baseSpeedupErrorLargePercent{}\%$ of the functions.
Unison's speedup estimation is biased in that it overestimates the actual
speedup for $\baseSpeedupOverestimationPercent{}\%$ of the functions where there
is an estimation error.

Despite its inaccuracy, the estimation is monotonically related to the actual
speedup, in that higher estimated speedup often leads to higher actual speedup.
A Spearman correlation coefficient~\cite{Spearman1904}
of~$\baseSpeedupDiffSpearmanCoefficient{}$ confirms that the monotonic
relationship is strong ($0$ indicates no relationship and~$1$ indicates a
perfect monotonically increasing relationship).
Similarly to the Wilcoxon test, this measure is used as an alternative to the
more common Pearson correlation coefficient because the speedup difference is
not normally distributed.

\paragraph{Accuracy factors}

The potential sources of inaccuracy in a static speedup estimation are:
\begin{description}
\item[Late program transformations.] Late compiler, assembler, and link-time
  optimizations can modify the program under compilation after Unison and
  invalidate the speedup estimation.
\item[Dynamic processor behavior.] Unpredictable processor features such as
  cache memories and branch prediction are not captured by Unison's speedup
  estimation and can lead to processor stalls that are unaccounted for.
\item[Dynamic program behavior.] The estimated execution frequency of each basic
  block (equation~\eqref{eq:speed-optimization}) might deviate from the actual
  execution frequency.
\end{description}
The first source (late program transformations) is controlled in the
experiments by simply disabling all optimizations that run after Unison
(see Table~\ref{tab:flags}).
The influence of the other sources is determined by measuring the actual speedup
on an \emph{ideal} Hexagon processor without pipeline stalls.
Estimation errors for such a processor are solely due to the execution frequency
estimation provided by LLVM.
The study of the effect of enabling late program transformations in the speedup
estimation's accuracy is part of future work.

The results show that the dynamic processor behavior is the main source of
inaccuracy in Unison's speedup estimation and the sole responsible for
overestimation.
The \emph{ideal processor} points in Figure~\ref{fig:speedup-relation} depict
the relation between estimated and actual speedup on a Hexagon processor without
stalls, where the dynamic program behavior is the sole source of inaccuracy in
the estimated speedup.
For this configuration, the percentage of the functions that are slowed down
drops to $\nostallsPercentOfWorseInstances{}\%$, the percentage of functions for
which the estimation error is less than $1\%$ increases to
$\nostallsSpeedupErrorSmallPercent{}\%$, and the Spearman correlation
coefficient increases to a very strong
$\nostallsSpeedupDiffSpearmanCoefficient{}$.
Furthermore, the overestimation bias in the speedup estimation is dampened as
seen in Figure~\ref{fig:speedup-relation} (\emph{ideal processor} points).

\paragraph{Implications}

Unison's estimated speedup is generally greater than the actual speedup, but
both are strongly monotonically related.
This fact is key to the combinatorial approach as it motivates investing
additional computational effort in finding solutions of increasing quality.
In practical terms, for MediaBench functions on Hexagon the actual speedup
can be expected to be lower than estimated, but solutions of increasing
estimated speedup can be expected to indeed yield higher speedup.
The accuracy of the speedup estimation can be expected to improve for more
predictable processors than Hexagon (for example, with scratchpad rather
than cache memories) and vice versa.
For applications exhibiting more irregular control behavior than those
present in MediaBench, the speedup estimation can be improved by
profile-guided optimization.

\section{Functions}\label{app:functions}

\begin{table}
  \caption{MediaBench and SPEC CPU2006 functions for the evaluation of estimated
    code quality and scalability.}
  \label{tab:functions}
  \tiny
  \setlength{\tabcolsep}{3pt}
  \begin{tabular}{l l}
    \input{results/functions.tex}
  \end{tabular}
\end{table}

Table~\ref{tab:functions} lists the MediaBench and SPEC CPU2006 functions for
the evaluation of estimated code quality
(Section~\ref{sec:estimated-code-quality}) and scalability
(Section~\ref{sec:scalability}). Their features are: input instructions
(\textbf{I}), input instructions per basic block (\textbf{I/B}), register
pressure (\textbf{RP}), instruction-level parallelism (\textbf{ILP}), call
instruction ratio (\textbf{CR}), and cluster size (\textbf{CS}) in the sampling
algorithm.
All features are averaged over the three studied processors.

\begin{table}
  \caption{MediaBench functions for the evaluation of actual speedup
     and accuracy of the speedup estimation on Hexagon V4.}
  \label{tab:mediabench-functions}
  \tiny
  \setlength{\tabcolsep}{3pt}
  \begin{tabular}{l l}
    \input{results/mediabench-functions.tex}
  \end{tabular}
\end{table}

Table~\ref{tab:mediabench-functions} lists the MediaBench functions for the
evaluation of actual speedup (Section~\ref{sec:actual-speedup}) and accuracy of
the speedup estimation (Appendix~\ref{app:accuracy}) on Hexagon V4. Their
features are: input instructions (\textbf{I}), input instructions per basic
block (\textbf{I/B}), register pressure (\textbf{RP}), instruction-level
parallelism (\textbf{ILP}), call instruction ratio (\textbf{CR}), and percentage
of the execution time (\textbf{EX}).

\end{document}

%% file: results/hexagon-speed-summary.tex
\newcommand{\improvedHexagonSpeedMeanImprovement}{10}

\newcommand{\improvedHexagonSpeedPercentOptimal}{81}

\newcommand{\improvedHexagonSpeedMeanSpillNumberWorsening}{28.2}

%% file: results/arm-speed-summary.tex
\newcommand{\improvedARMSpeedMeanImprovement}{1.1}

\newcommand{\improvedARMSpeedMeanSpillNumberImprovement}{7.5}

%% file: results/mips-speed-summary.tex
\newcommand{\improvedMipsSpeedMeanImprovement}{5.4}

\newcommand{\improvedMipsSpeedMeanSpillNumberWorsening}{15.8}
\newcommand{\improvedMipsSpeedMeanSpillOverheadImprovement}{49.1}

%% file: results/hexagon-size-summary.tex
\newcommand{\improvedHexagonSizeMeanImprovement}{1.3}

%% file: results/arm-size-summary.tex
\newcommand{\improvedARMSizeMeanImprovement}{2.5}

%% file: results/mips-size-summary.tex
\newcommand{\improvedMipsSizeMeanImprovement}{3.8}

\newcommand{\improvedMipsSizeMeanSpillNumberWorsening}{5.5}

%% file: results/hexagon-all-summary.tex
\newcommand{\improvedHexagonAllPercentOptimal}{79}
\newcommand{\improvedHexagonAllMeanGap}{3.2}

\newcommand{\improvedHexagonAllLargestOptimalIns}{946}

\newcommand{\improvedHexagonAllMedianNonZeroGapIns}{277}

%% file: results/arm-all-summary.tex
\newcommand{\improvedARMAllPercentOptimal}{62}
\newcommand{\improvedARMAllMeanGap}{6.3}

\newcommand{\improvedARMAllLargestOptimalIns}{330}

\newcommand{\improvedARMAllMedianNonZeroGapIns}{220}

%% file: results/mips-all-summary.tex
\newcommand{\improvedMipsAllPercentOptimal}{44}
\newcommand{\improvedMipsAllMeanGap}{14.5}

\newcommand{\improvedMipsAllLargestOptimalIns}{205}

\newcommand{\improvedMipsAllMedianNonZeroGapIns}{184}

%% file: results/multi-all-improved-summary.tex
\newcommand{\improvedmultiAllPercentOptimal}{61.7}

\newcommand{\improvedmultiAllMeanNonZeroGap}{22.2}
\newcommand{\improvedmultiAllMeanNonZeroSmallGap}{22.6}

\newcommand{\improvedmultiAllLargestOptimalIns}{946}
\newcommand{\improvedmultiAllSmallestNonZeroGapIns}{30}
\newcommand{\improvedmultiAllPercentGecodeSolverTotal}{49.7}
\newcommand{\improvedmultiAllLargestGecodeWin}{946}
\newcommand{\improvedmultiAllLargestMinizincWin}{874}

\newcommand{\improvedmultiAllMedianNonZeroGapIns}{218}

%% file: results/multi-all-basic-summary.tex
\newcommand{\basicmultiAllPercentOptimal}{44.7}

\newcommand{\basicmultiAllMeanNonZeroGap}{71.3}

\newcommand{\basicmultiAllLargestOptimalIns}{647}

\newcommand{\basicmultiAllMedianNonZeroGapIns}{184}

%% file: results/multi-results.tex
\newcommand{\improvedmultiAllMipsPercentNonZeroGapUnderHundred}{73.2}

\newcommand{\improvedmultiAllArmPercentNonZeroGapUnderHundred}{26.8}

\newcommand{\improvedmultiAllHexagonTimesMoreOptimalOverHundredThanMips}{3.5}
\newcommand{\improvedmultiAllHexagonTimesMoreOptimalOverHundredThanArm}{2.9}

\newcommand{\improvedmultiAllMaxUnisonCostDistanceFromMedianInReps}{21.1}

%% file: results/multi-combined-results.tex
\newcommand{\combinedmultiAllImprovedTimesMoreOptimalOverHundredThanBasic}{2.5}
\newcommand{\combinedmultiAllBasicTimesMoreNonZeroGapUnderHundredThanImproved}{2.1}

%% file: results/combined-results.tex
\newcommand{\combinedmultiAllImprovedCrossesBasicTime}{0.3}

%% file: results/hexagon-speed-phases-summary.tex
\newcommand{\bothphasesHexagonSpeedMeanIntegrationImprovementOverTwophase}{7.2}
\newcommand{\bothphasesHexagonSpeedMeanIntegrationImprovementOverLLVM}{8.8}
\newcommand{\bothphasesHexagonSpeedMeanTwophaseImprovementOverLLVM}{1}

\newcommand{\bothphasesHexagonSpeedTwophaseTimesOptimalOverIntegrated}{1.4}

%% file: results/arm-speed-phases-summary.tex
\newcommand{\bothphasesARMSpeedMeanIntegrationImprovementOverTwophase}{0.7}

\newcommand{\bothphasesARMSpeedMeanLLVMImprovementOverTwophase}{0.3}

%% file: results/mips-speed-phases-summary.tex
\newcommand{\bothphasesMipsSpeedMeanIntegrationImprovementOverTwophase}{1.8}

\newcommand{\bothphasesMipsSpeedTwophaseTimesOptimalOverIntegrated}{2.8}

%% file: results/hexagon-size-phases-summary.tex
\newcommand{\bothphasesHexagonSizeMeanIntegrationImprovementOverTwophase}{2.4}

\newcommand{\bothphasesHexagonSizeMeanLLVMImprovementOverTwophase}{2.2}

%% file: results/arm-size-phases-summary.tex
\newcommand{\bothphasesARMSizeMeanIntegrationImprovementOverTwophase}{1.4}

%% file: results/mips-size-phases-summary.tex
\newcommand{\bothphasesMipsSizeMeanIntegrationImprovementOverTwophase}{2.2}

%% file: results/hexagon-all-trivialbound-summary.tex
\newcommand{\trivialboundHexagonAllPercentOptimal}{61.5}
\newcommand{\trivialboundHexagonAllMeanGap}{8.3}

%% file: results/arm-all-trivialbound-summary.tex
\newcommand{\trivialboundARMAllPercentOptimal}{58}
\newcommand{\trivialboundARMAllMeanGap}{9.1}

%% file: results/mips-all-trivialbound-summary.tex
\newcommand{\trivialboundMipsAllPercentOptimal}{42}
\newcommand{\trivialboundMipsAllMeanGap}{20.1}

%% file: results/gap-results.tex
\newcommand{\percentOfSolvedFunctionsWithAGapOfTen}{90}

%% file: results/actual-speedup-results.tex
\newcommand{\minExecPercentAcrossApps}{66.7}

\newcommand{\basePercentOfBetterInstances}{60.8}
\newcommand{\baseBest}{71.3}
\newcommand{\basePercentOfWorseInstances}{17.7}
\newcommand{\baseWorst}{-11.1}
\newcommand{\baseMeanImprovement}{6.6}

\newcommand{\multiBest}{15.8}

\newcommand{\multiWorst}{-0.2}
\newcommand{\multiMeanImprovement}{5.6}

\newcommand{\nostallsPercentOfWorseInstances}{3.8}

\newcommand{\baseSpeedupErrorSmallPercent}{36.7}
\newcommand{\baseSpeedupErrorMediumPercent}{50.6}
\newcommand{\baseSpeedupErrorLargePercent}{12.7}
\newcommand{\baseLargestSpeedupError}{20}

\newcommand{\baseSpeedupOverestimationPercent}{71}

\newcommand{\baseSpeedupDiffSpearmanCoefficient}{0.7}

\newcommand{\nostallsSpeedupErrorSmallPercent}{60.8}

\newcommand{\nostallsSpeedupDiffSpearmanCoefficient}{0.89}

%% file: results/global-stats.tex
\newcommand{\percentDifferentFunctions}{5}
\newcommand{\hexagonSingleIssueTimesMoreOptimal}{1.1}
\newcommand{\mipsUnitLatencyTimesMoreOptimal}{1.3}
\newcommand{\mipsNoCCTimesMoreOptimal}{1.6}

%% file: constraints.tex
\newcommand{\localDisjointLiveRangesConstraint}{%
  \disjoint{
    \set{\tuple{
        \temporaryRegister{t},
        \temporaryRegister{t} + \width{t},
        \liveStart{t},
        \liveEnd{t}}
      \! \suchThat \!
      t \in \allTemporaries{}}}}

\newcommand{\fixedPreAssignmentConstraint}{%
  \temporaryRegister{\temporary{p}} = r
  \quad
  \forallIn{p}{\allOperands{}} \suchThat \preAssigned{p\hspace{0.025cm}}{r}}

\newcommand{\fixedInstructionSelectionConstraint}{%
  \temporaryRegister{\temporary{p}}
  \in
  \fixedRegisterClass{p}
  \quad
  \forallIn{p}{\allOperands{}}}

\newcommand{\activeInstructionSelectionConstraint}{%
  \temporaryRegister{\temporary{p}}
  \in
  \registerClassNoParen{\highlight{\elementMark{p, \operationInstruction{\operandOperation{p}}}}}
  \quad
  \forallIn{p}{\allOperands{}}}

\newcommand{\activeOperationConstraint}{%
  \activeOperation{o}
  \quad
  \forallIn{o}{\allOperations{}} \suchThat \neg \copySemantics{o}}

\newcommand{\liveTemporaryConstraint}{%
  \begin{aligned}
    \liveTemporary{t}
    &\iff
    \activeOperation{\operandOperation{\definer{t}}}\\
    &\iff
    \existsIn{p}{\users{t}} \suchThat \activeOperation{\operandOperation{p}} \land \operandTemporary{p} = t
    \quad
    \forallIn{t}{\allTemporaries{}}\iftoggle{applyHighlight}{,}{}
  \end{aligned}}

\newcommand{\preAssignmentConstraint}{%
  \temporaryRegisterNoParen{\highlight{\varElementMark{\operandTemporary{p}}}} = r
  \quad
  \forallIn{p}{\allOperands{}} \suchThat \preAssigned{p\hspace{0.025cm}}{r}}

\newcommand{\instructionSelectionConstraint}{%
  \temporaryRegisterNoParen{\highlight{\varElementMark{\operandTemporary{p}}}}
  \in
  \registerClassNoParen{\elementMark{p, \operationInstruction{\operandOperation{p}}}}
  \quad
  \forallIn{p}{\allOperands{}} \highlight{\suchThat \activeOperation{\operandOperation{p}}}}

\newcommand{\congruenceConstraint}{%
  \temporaryRegisterNoParen{\varElementMark{\operandTemporary{p}}} =
  \temporaryRegisterNoParen{\varElementMark{\operandTemporary{q}}}
  \quad
  \forallIn{p, q}{\allOperands{}} \suchThat \congruent{p}{q}}

\newcommand{\globalDisjointLiveRangesConstraint}{%
  \disjoint{
    \flatSet{\tuple{
        \temporaryRegister{t},
        \temporaryRegister{t} + \width{t},
        \temporaryLiveStart{t},
        \temporaryLiveEnd{t}}
      \! \suchThat \!
      t \in \highlight{\temporaries{b}} \land \liveTemporary{t}}}
  \;\;
  \highlight{\forallIn{b}{\allBlocks{}}}}

\newcommand{\fixedDataPrecedencesConstraint}{%
  \begin{aligned}
    \operationIssueCycle{\operandOperation{q}}
    &\ge \operationIssueCycle{\operandOperation{p}} +
    \latency{\fixedOperationInstruction{\operandOperation{p}}}\\
    &\forall t \in \allTemporaries{}\!,
    \forall p \in \set{\definer{t}}\!,
    \forall q \in \users{t}
    \suchThat \temporary{q} = t.
  \end{aligned}}

\newcommand{\fixedProcessorResourcesConstraint}{%
  \cumulative{
    \setBuilder
        {\tuple{
            \operationIssueCycle{o},
            \duration{\fixedOperationInstruction{o}}{s},
            \units{\fixedOperationInstruction{o}}{s}
            \eat}}
        {\eat o \in \allOperations{}}\!,
        \capacity{s}}
  \quad
  \forallIn{s}{S}}

\newcommand{\dataPrecedencesConstraint}{%
  \begin{aligned}
    &\operationIssueCycle{\operandOperation{q}}
    \ge \operationIssueCycle{\operandOperation{p}} +
    \latencyNoParen{\highlight{\elementMark{\operationInstruction{\operandOperation{p}}}}}\\
    &\hspace{1cm}\forall t \in \allTemporaries{},
    \forall p \in \set{\definer{t}}\!,
    \forall q \in \users{t}
    \suchThat \highlight{\activeOperation{\operandOperation{q}} \land \operandTemporary{q} = t}\iftoggle{applyHighlight}{.}{}
  \end{aligned}}

\newcommand{\globalProcessorResourcesConstraint}{%
  \begin{aligned}
    \cumulative{
      \setBuilder
          {\tuple{
              \operationIssueCycle{o},
              \durationNoParen{\highlight{\elementMark{\operationInstruction{o}, s}}},
              \unitsNoParen{\highlight{\elementMark{\operationInstruction{o}, s}}}
              \eat}}
          {\eat o \in \highlight{\operations{b} \land \activeOperation{o}}}\!,
          \capacity{s}}&\\
    \highlight{\forallIn{b}{\allBlocks{}}},
    \forallIn{s}{S}\iftoggle{applyHighlight}{.}{}&
  \end{aligned}}

\newcommand{\liveStartConstraint}{%
  \temporaryLiveStart{t} =
  \operationIssueCycle{\operandOperation{\definer{t}}}
  \quad
  \forallIn{t}{\allTemporaries{}} \suchThat \liveTemporary{t}}

\newcommand{\liveEndConstraint}{%
  \temporaryLiveEnd{t} =
  \maximum{p \in \users{t} \,\suchThat\, \operandTemporary{p} \,=\, t}{\operationIssueCycle{\operandOperation{p}}}
  \quad
  \forallIn{t}{\allTemporaries{}} \suchThat \liveTemporary{t}}

\newcommand{\genericObjectiveFunction}{
  \sum_{b \in B}{\blockWeight{b} \times \blockCost{b}}
}

\newcommand{\requiresFrameConstraint}{%
  \exists o \in O \suchThat \activeOperation{o} \land \requiresFrameNoParen{\elementMark{o, \operationInstruction{o}}} \implies \functionFrame{}}

\newcommand{\managesFrameConstraint}{%
  \functionFrame{} \implies \activeOperation{o}
  \quad
  \forallIn{o}{\allOperations{}}
  \suchThat \managesFrameNoParen{\elementMark{o, \operationInstruction{o}}}
}

\newcommand{\slackBalancingConstraint}{%
  \operandSlack{p} + \operandSlack{q} = 0
  \quad
  \forallIn{p, q}{\allOperands{}} \suchThat \congruent{p}{q}}

\newcommand{\slackDataPrecedencesConstraint}{%
  \begin{aligned}
    &\operationIssueCycle{\operandOperation{q}}
    \ge \operationIssueCycle{\operandOperation{p}} +
    \latencyNoParen{\elementMark{\operationInstruction{\operandOperation{p}}}}
    \highlight{+\,\operandSlack{p} + \operandSlack{q}}\\
    &\hspace{1cm}\forall t \in \allTemporaries{},
    \forall p \in \set{\definer{t}}\!,
    \forall q \in \users{t}
    \suchThat \activeOperation{\operandOperation{q}} \land \operandTemporary{q} = t\iftoggle{applyHighlight}{.}{}
  \end{aligned}}

\newcommand{\forwardingConstraint}{%
  \begin{aligned}
    \operationIssueCycle{o} &=
    \operationIssueCycle{\operandOperation{\definerNoParen{\elementMark{\operandTemporary{p}}}}}\\
    &\forallIn{p}{\allOperands{}}\!, \forallIn{o}{\set{\operandOperation{p}}} \suchThat \activeOperation{o} \land \forwardedNoParen{\elementMark{\operationInstruction{o}, p}}\iftoggle{applyHighlight}{.}{}
\end{aligned}}

\newcommand{\twoAddressConstraint}{%
  \begin{aligned}
    &\temporaryRegisterNoParen{\varElementMark{\operandTemporary{p}}} =
     \temporaryRegisterNoParen{\varElementMark{\operandTemporary{q}}}\\
    &\hspace{1cm}\forallIn{p, q}{\allOperands{}}\!, \forallIn{o}{\set{\operandOperation{p}}} \suchThat \activeOperation{o} \land \twoAddressNoParen{\elementMark{\operationInstruction{o}, p, q}}\iftoggle{applyHighlight}{.}{}
  \end{aligned}}

\newcommand{\offsetProcessorResourcesConstraint}{%
  \begin{aligned}
    \cumulative{
      \setBuilder
          {\tuple{
              \operationIssueCycle{o} \highlight{+ \offset{\operationInstruction{o}}{s}},
              \duration{\operationInstruction{o}}{s},
              \units{\operationInstruction{o}}{s}
              \eat}}
          {\eat o \in \operations{b} \land \activeOperation{o}},
          \capacity{s}}&\\
    \forallIn{b}{\allBlocks{}},
    \forallIn{s}{S}\iftoggle{applyHighlight}{\,.}{}&
  \end{aligned}}

\newcommand{\fwdDataPrecedencesConstraint}{%
  \begin{aligned}
    &\operationIssueCycle{\operandOperation{q}}
    \ge \operationIssueCycle{\operandOperation{p}} +
    \latencyNoParen{\elementMark{\operationInstruction{\operandOperation{p}}}}\\
    &\hspace{1cm}\forall t \in \allTemporaries{}\!,
    \forall p \in \set{\definer{t}}\!,
    \forall q \in \users{t} \suchThat\\
    &\hspace{2cm}\activeOperation{\operandOperation{q}} \land \operandTemporary{q} = t \highlight{\land \neg \forwardedNoParen{\elementMark{\operationInstruction{\operandOperation{q}}, q}}}\iftoggle{applyHighlight}{.}{}
  \end{aligned}}

%% file: figures/factorial-motivation.tex
\input{./figures/styles}%
\newcommand{\rzero}{\texttt{R0}}
\newcommand{\rone}{\texttt{R1}}
\newcommand{\cs}{\hspace{0.1cm}}
\begin{SaveVerbatim}[commandchars=\\\[\]]{fac-in}
int\cs[]fac(int n){
\end{SaveVerbatim}
\begin{SaveVerbatim}[commandchars=\\\[\]]{fac-b1}
\cs[]int\cs[]f\cs[]=\cs[]1;
\end{SaveVerbatim}
\begin{SaveVerbatim}[commandchars=\\\[\]]{fac-b2s1}
\cs[]while(n\cs[]>\cs[]0){
\end{SaveVerbatim}
\begin{SaveVerbatim}[commandchars=\\\[\]]{fac-b2s2}
\cs[]\cs[]f\cs[]*=\cs[]n;
\end{SaveVerbatim}
\begin{SaveVerbatim}[commandchars=\\\[\]]{fac-b2s3}
\cs[]\cs[]n--;
\end{SaveVerbatim}
\begin{SaveVerbatim}[commandchars=\\\[\]]{fac-b2s4}
\cs[]}
\end{SaveVerbatim}
\begin{SaveVerbatim}[commandchars=\\\[\]]{fac-b3}
\cs[]return\cs[]f;
\end{SaveVerbatim}
\begin{SaveVerbatim}{fac-out}
}
\end{SaveVerbatim}

\newcommand{\ws}{\hspace{0.1cm}}
\newcommand{\asmlabel}[1]{{\it{#1}}}
\newcommand{\asmbundled}{$\hspace{0.17cm}\bundled{}\hspace{0.12cm}$}

\begin{SaveVerbatim}[commandchars=\\\{\}]{unison-b1}
R1\ws{}=\ws{}1\asmbundled{}\ws{}j\ws{}\asmlabel{end}\ws{}if\ws{}R0\ws{}<=\ws{}0
\end{SaveVerbatim}
\begin{SaveVerbatim}[commandchars=\\\{\}]{unison-b2}
R1\ws{}*=\ws{}R0\asmbundled{}\ws{}R0--
j\ws{}\asmlabel{loop}\ws{}if\ws{}R0\ws{}>\ws{}0
\end{SaveVerbatim}
\begin{SaveVerbatim}[commandchars=\\\{\}]{unison-b3}
ret\asmbundled{}\ws{}R0\ws{}=\ws{}R1
\end{SaveVerbatim}

\begin{SaveVerbatim}[commandchars=\\\{\}]{llvm-b1}
R0\ws{}=\ws{}1\asmbundled{}\ws{}R1\ws{}=\ws{}R0
j\ws{}\asmlabel{end}\ws{}if\ws{}R1\ws{}<=\ws{}0
\end{SaveVerbatim}
\begin{SaveVerbatim}[commandchars=\\\{\}]{llvm-b2}
R0\ws{}*=\ws{}R1\asmbundled{}\ws{}R1--
j\ws{}\asmlabel{loop}\ws{}if\ws{}R1\ws{}>\ws{}0
\end{SaveVerbatim}
\begin{SaveVerbatim}[commandchars=\\\{\}]{llvm-b3}
ret
\end{SaveVerbatim}

\begin{tikzpicture}
  \coordinate (bin);
  \coordinate [below=0.8cm of bin]  (b1);
  \coordinate [below=0.8cm of b1]   (b2s1);
  \coordinate [below=0.4cm of b2s1] (b2s2);
  \coordinate [below=0.4cm of b2s2] (b2s3);
  \coordinate [below=0.3cm of b2s3] (b2s4);
  \coordinate [below=0.8cm of b2s4] (b3);
  \coordinate [below=0.45cm of b3]   (bout);
  \node [cfg block, right=of bin, minimum height=0.5cm, invisible] (blockin) {};
  \node [below right] at (blockin.north west) {\BUseVerbatim{fac-in}};
  \node [cfg block, right=of b1, minimum width=2.7cm, minimum height=0.5cm] (block1) {};
  \node [below right] at (block1.north west) {\BUseVerbatim{fac-b1}};
  \node [cfg block, right=of b2s2, minimum width=2.7cm, minimum height=1.65cm, yshift=-0.17cm] (block2) {};
  \node [cfg block, right=of b2s1, minimum height=0.5cm, invisible] (block2s1) {};
  \node [below right] at (block2s1.north west) {\BUseVerbatim{fac-b2s1}};
  \node [cfg block, right=of b2s2, minimum height=0.5cm, invisible] (block2s2) {};
  \node [below right] at (block2s2.north west) {\BUseVerbatim{fac-b2s2}};
  \node [cfg block, right=of b2s3, minimum height=0.5cm, invisible] (block2s3) {};
  \node [below right] at (block2s3.north west) {\BUseVerbatim{fac-b2s3}};
  \node [cfg block, right=of b2s4, minimum height=0.5cm, invisible] (block2s4) {};
  \node [below right] at (block2s4.north west) {\BUseVerbatim{fac-b2s4}};
  \node [cfg block, right=of b3, minimum width=2.7cm, minimum height=0.5cm] (block3) {};
  \node [below right] at (block3.north west) {\BUseVerbatim{fac-b3}};
  \node [cfg block, right=of bout, minimum height=0.5cm, invisible] (blockout) {};
  \node [below right] at (blockout.north west) {\BUseVerbatim{fac-out}};
  \node [block label, above right, yshift=-0.1cm] at (block1.north west) {begin:};
  \node [block label, above right, yshift=-0.1cm] at (block2.north west) {loop:};
  \node [block label, above right, yshift=-0.1cm] at (block3.north west) {end:};
  \coordinate (entry)  at (blockin.north west);
  \coordinate (bin-p1) at (blockin.south west);
  \coordinate (b1-p1)  at (block1.south west);
  \coordinate (b2-p1)  at (block2s1.south west);
  \coordinate (b2-p2)  at (block2s2.south west);
  \coordinate (b2-p3)  at (block2s3.south west);
  \coordinate (b2-p4)  at (block2s4.south west);
  \coordinate (b3-p1)  at (block3.south west);
  \coordinate [below=0.5cm of b3-p1] (exit);
  \coordinate [below=0.1cm of bin-p1] (bin-entry);
  \coordinate [above=0.25cm of b1-p1] (b1-mid);
  \coordinate [below=0.1cm of b1-mid] (b1-mid-low);
  \coordinate [above=0.35cm of b3-p1] (b3-mid);
  \coordinate [below=0.05cm of b3-p1] (bout-exit);
  \newcommand{\copyDistance}{0.15cm}
  \newcommand{\captionDistance}{0.25cm}
  \newcommand{\doubleCaptionDistance}{0.4cm}
  \newcommand{\llvmRZeroDistance}{3.6cm}
  \coordinate [right=\llvmRZeroDistance of entry] (llvmRZero);
  \node [below=0cm of llvmRZero] {\rzero{}};
  \newcommand{\llvmROneDistance}{4.2cm}
  \coordinate [right=\llvmROneDistance of entry] (llvmROne);
  \node [below=0cm of llvmROne] {\texttt{\rone{}}};
  \newcommand{\llvmRBetweenDistance}{3.9cm}
  \coordinate [right=\llvmRZeroDistance of bin-entry] (beginLlvmNZero);
  \coordinate [right=\llvmRZeroDistance of b1-mid, yshift=\copyDistance] (endLlvmNZero);
  \coordinate [right=\llvmROneDistance of b1-mid, yshift=-\copyDistance] (beginLlvmNOne);
  \coordinate [right=\llvmROneDistance of b2-p4]   (endLlvmNOne);
  \begin{scope}[live range]
    \draw [<->] (beginLlvmNZero) -- (endLlvmNZero) -- (beginLlvmNOne) -- (endLlvmNOne);
  \end{scope}
  \begin{scope}[copy]
    \draw (endLlvmNZero) -- (beginLlvmNOne);
  \end{scope}
  \node [right=0cm of endLlvmNOne, yshift=1.8cm] {\texttt{n}};
  \coordinate [right=\llvmRZeroDistance of b1-mid-low] (beginLlvmFZero);
  \coordinate [right=\llvmRZeroDistance of bout-exit] (endLlvmFZero);
  \begin{scope}[live range]
    \draw [<->] (beginLlvmFZero) -- (endLlvmFZero);
  \end{scope}
  \node [left=0cm of endLlvmFZero, yshift=1.5cm] {\texttt{f}};
  \coordinate [right=\llvmRBetweenDistance of exit] (llvmRA);
  \node [caption, below=0cm of llvmRA, yshift=\doubleCaptionDistance] {%
    \begin{tabular}{c}moving \code{n}\\(LLVM)\end{tabular}};
  \newcommand{\unisonRZeroDistance}{5cm}
  \coordinate [right=\unisonRZeroDistance of entry] (unisonRZero);
  \node [below=0cm of unisonRZero] {\rzero{}};
  \newcommand{\unisonROneDistance}{5.6cm}
  \coordinate [right=\unisonROneDistance of entry] (unisonROne);
  \node [below=0cm of unisonROne] {\rone{}};
  \newcommand{\unisonRBetweenDistance}{5.3cm}
  \coordinate [right=\unisonRZeroDistance of bin-entry] (beginUnisonNZero);
  \coordinate [right=\unisonRZeroDistance of b2-p4] (endUnisonNZero);
  \begin{scope}[live range]
    \draw [<->] (beginUnisonNZero) -- (endUnisonNZero);
  \end{scope}
  \node [left=0cm of endUnisonNZero, yshift=1.3cm] {\texttt{n}};
  \coordinate [right=\unisonROneDistance of b1-mid-low] (beginUnisonFOne);
  \coordinate [right=\unisonROneDistance of b3-mid, yshift=\copyDistance] (endUnisonFOne);
  \coordinate [right=\unisonRZeroDistance of b3-mid, yshift=-\copyDistance] (beginUnisonFZero);
  \coordinate [right=\unisonRZeroDistance of bout-exit] (endUnisonFZero);
  \begin{scope}[live range]
    \draw [<->] (beginUnisonFOne) -- (endUnisonFOne) -- (beginUnisonFZero) -- (endUnisonFZero);
  \end{scope}
  \begin{scope}[copy]
    \draw (endUnisonFOne) -- (beginUnisonFZero);
  \end{scope}
  \node [right=0cm of endUnisonFOne, yshift=0.8cm] {\texttt{f}};
  \coordinate [right=\unisonRBetweenDistance of exit] (unisonRA);
  \node [caption, below=0cm of unisonRA, yshift=\doubleCaptionDistance] {%
    \begin{tabular}{c}moving \code{f}\\(our approach)\end{tabular}};
  \newcommand{\llvmAsmDistance}{7.4cm}
  \coordinate [right=\llvmAsmDistance of b1, yshift=0.665cm] (llvmAsmB1);
  \node [anchor=north west, cfg block, minimum width=2.9cm, minimum height=0.95cm] at (llvmAsmB1) (llvmAsmBlock1) {};
  \node [below right] at (llvmAsmBlock1.north west) {\BUseVerbatim{llvm-b1}};
  \coordinate [right=\llvmAsmDistance of b2s1] (llvmAsmB2);
  \node [anchor=north west, cfg block, minimum width=2.9cm, minimum height=0.95cm] at (llvmAsmB2) (llvmAsmBlock2) {};
  \node [below right] at (llvmAsmBlock2.north west) {\BUseVerbatim{llvm-b2}};
  \coordinate [right=\llvmAsmDistance of b3, yshift=0.32cm] (llvmAsmB3);
  \node [anchor=north west, cfg block, minimum width=2.9cm, minimum height=0.57cm] at (llvmAsmB3) (llvmAsmBlock3) {};
  \node [below right] at (llvmAsmBlock3.north west) {\BUseVerbatim{llvm-b3}};
  \coordinate [right=\llvmAsmDistance of exit] (llvmAsm);
  \node [block label, above right, yshift=-0.1cm] at (llvmAsmBlock1.north west) {begin:};
  \node [block label, above right, yshift=-0.1cm] at (llvmAsmBlock2.north west) {loop:};
  \node [block label, above right, yshift=-0.1cm] at (llvmAsmBlock3.north west) {end:};
  \node [caption, below=0cm of llvmAsm, yshift=\captionDistance, xshift=0.5cm] {(LLVM)};
  \newcommand{\unisonAsmDistance}{10.6cm}
  \coordinate [right=\unisonAsmDistance of b1, yshift=0.3cm] (unisonAsmB1);
  \node [anchor=north west, cfg block, minimum height=0.57cm, minimum width=4.2cm] at (unisonAsmB1) (unisonAsmBlock1) {};
  \node [below right] at (unisonAsmBlock1.north west) {\BUseVerbatim{unison-b1}};
  \coordinate [right=\unisonAsmDistance of b2s1] (unisonAsmB2);
  \node [anchor=north west, cfg block, minimum height=0.95cm, minimum width=4.2cm] at (unisonAsmB2) (unisonAsmBlock2) {};
  \node [below right] at (unisonAsmBlock2.north west) {\BUseVerbatim{unison-b2}};
  \coordinate [right=\unisonAsmDistance of b3, yshift=0.32cm] (unisonAsmB3);
  \node [anchor=north west, cfg block, minimum height=0.57cm, minimum width=4.2cm] at (unisonAsmB3) (unisonAsmBlock3) {};
  \node [below right] at (unisonAsmBlock3.north west) {\BUseVerbatim{unison-b3}};
  \coordinate [right=\unisonAsmDistance of exit] (unisonAsm);
  \node [block label, above right, yshift=-0.1cm] at (unisonAsmBlock1.north west) {begin:};
  \node [block label, above right, yshift=-0.1cm] at (unisonAsmBlock2.north west) {loop:};
  \node [block label, above right, yshift=-0.1cm] at (unisonAsmBlock3.north west) {end:};
  \node [caption, below=0cm of unisonAsm, yshift=\captionDistance, xshift=1cm] {(our approach)};
\end{tikzpicture}

%% file: figures/approach.tex
\input{./figures/styles}%
\pgfdeclarelayer{layer0}%
\pgfdeclarelayer{layer1}%
\pgfdeclarelayer{layer2}%
\pgfdeclarelayer{layer3}%
\pgfdeclarelayer{layer4}%
\pgfdeclarelayer{layer5}%
\pgfsetlayers{layer0,layer1,layer2,layer3,layer4,layer5}%
\begin{tikzpicture}%
  \begin{pgfonlayer}{layer2}
    \node [input, inner xsep=-0.1cm] (input) {%
    \begin{tabular}{c}
        low-level\\
        IR
      \end{tabular}
    };
    \node [unison stage, right=0.3cm of input] (construction) {%
      \begin{tabular}{c}
        program\\
        extension
      \end{tabular}
    };
    \node [invisible, right=1.6cm of construction] (center) {};
    \node [unison stage, above=of center, yshift=-0.8cm] (regalloc) {%
      \begin{tabular}{c}
        register\\
        allocation
      \end{tabular}
    };
    \node [unison stage, below=of center, yshift=0.8cm] (isched) {%
      \begin{tabular}{c}
        instruction\\
        scheduling
      \end{tabular}
    };
    \node [file, right=1.2cm of center, inner xsep=0.2cm]
      (problem) {%
      \begin{tabular}{c}
        integrated\\
        combinatorial\\
        model
      \end{tabular}
    };
    \node [unison stage, right=0.2cm of problem, minimum width=0cm, inner xsep=0.08cm] (solver) {%
    \begin{tabular}{c}
        generic\\
        solver
      \end{tabular}
      };
    \node [output, inner xsep=0cm, right=0.4cm of solver] (output) {%
      \begin{tabular}{c}
        assembly\\
        code
      \end{tabular}
    };
    \node [input, below=0.5cm of construction, xshift=-0.5cm] (description) {%
      \begin{tabular}{c}
        processor and calling\\
        convention description
      \end{tabular}
    };
  \end{pgfonlayer}

  \begin{pgfonlayer}{layer2}
    \begin{scope}[unison step]
      \node [above= of input] {(\mylabel{1}{step:input})};
      \node [above= of construction, yshift=0.1cm] {(\mylabel{2}{step:construction})};
      \node [above= of description] {(\mylabel{3}{step:description})};
      \node [above= of regalloc, yshift=0.1cm] {(\mylabel{4}{step:regalloc})};
      \node [above= of isched, yshift=0.1cm] {(\mylabel{5}{step:isched})};
      \node [above= of problem] {(\mylabel{6}{step:problem})};
      \node [above= of solver] {(\mylabel{7}{step:solver})};
      \node [above= of output] {(\mylabel{8}{step:output})};
    \end{scope}
  \end{pgfonlayer}

  \begin{pgfonlayer}{layer2}
    \begin{scope}[flow]
      \draw (input) -- (construction);

      \draw ([yshift=2.5mm] construction.east) -- ([yshift=1mm]regalloc.west);
      \draw ([yshift=-2.5mm] construction.east) -- ([yshift=1mm]isched.west);

      \draw ([yshift=1mm]regalloc.east) -- ([yshift=2.5mm,xshift=0.4cm] problem.west);
      \draw ([yshift=1mm]isched.east) -- ([yshift=-2.5mm,xshift=0.4cm] problem.west);
      \draw ([xshift=-0.15cm] problem.east) -- (solver);
      \draw (solver) -- (output);
    \end{scope}
  \end{pgfonlayer}

  \begin{pgfonlayer}{layer1}
    \begin{scope}[flow]
      \draw ([xshift=-6mm]description.north east) -- ([yshift=-1mm]regalloc.west);
      \draw ([xshift=-4mm]description.north east) -- ([yshift=-1mm]isched.west);
    \end{scope}
  \end{pgfonlayer}

\end{tikzpicture}

%% file: figures/factorial-input.tex
\input{./figures/styles}%
\begin{tikzpicture}
\pgfsetlayers{main,layer0}%

\begin{pgfonlayer}{layer0}

    \node [block code] (blockBegin-pre-code) {%
      \begin{basicBlock}
        \alignCentered{\inOperation{\preAssigned{\TempN{}}{\register{R0}}}}
      \end{basicBlock}
    };
    \node [block code, below=of blockBegin-pre-code] (blockBegin-code) {%
      \begin{basicBlock}
        \operation{\TempF{}}{\transferImmediateOpcode{}}{\code{1}}
        \freeBranchOperation{\code{j} \, \emph{end} \, \code{if} \, \TempN{} \, \code{<=} \, \code{0}}
      \end{basicBlock}
    };

    \node [block code, right=1.5cm of blockBegin-code, yshift=-0.6cm] (blockLoop-code) {%
      \begin{basicBlock}
        \operation{\TempF{}}{\multiplyOpcode{}}{\TempF{}, \TempN{}}
        \operation{\TempN{}}{\subtractOpcode{}}{\TempN{}, \code{1}}
        \freeBranchOperation{\code{j} \, \emph{loop} \, \code{if} \, \TempN{} \, \code{>} \, \code{0}}
      \end{basicBlock}
    };

    \node [block code, below=1cm of blockBegin-code] (blockEnd-code) {%
      \begin{basicBlock}
        \branchOperation{\indirectJumpOpcode{}}{}
      \end{basicBlock}
    };
    \node [block code, below=of blockEnd-code] (blockEnd-post-code) {%
      \begin{basicBlock}
        \alignCentered{\outOperation{\preAssigned{\TempF{}}{\register{R0}}}}
      \end{basicBlock}
    };

\end{pgfonlayer}

\newcommand{\blockBeginWidth}{3.7cm}
\newcommand{\blockLoopWidth}{3cm}
\newcommand{\blockEndWidth}{3cm}
\begin{pgfonlayer}{main}
    \node [cfg block boundary, fit=(blockBegin-pre-code), minimum width=\blockBeginWidth] (blockBegin-pre) {};
    \node [cfg boundary filler, above=0cm of blockBegin-pre.south, minimum width=\blockBeginWidth] {};

    \node [cfg block, fit=(blockBegin-code), minimum width=\blockBeginWidth] (blockBegin) {};
    \node [cfg block filler, below=0cm of blockBegin.north, yshift=0.01cm, minimum width=\blockBeginWidth] {};

    \node [cfg block, fit=(blockLoop-code), minimum width=\blockLoopWidth] (blockLoop) {};
    \node [cfg block filler, below=0cm of blockLoop.north, yshift=0.01cm, minimum width=\blockLoopWidth] {};

    \node [cfg block, fit=(blockEnd-code), minimum width=\blockEndWidth] (blockEnd) {};
    \node [cfg block filler, below=0cm of blockEnd.north, yshift=0.01cm, minimum width=\blockEndWidth] {};
    \node [cfg block filler, above=0cm of blockEnd.south, yshift=-0.01cm, minimum width=\blockEndWidth] {};

    \node [cfg block boundary, fit=(blockEnd-post-code), minimum width=\blockEndWidth] (blockEnd-post) {};
    \node [cfg boundary filler, below=0cm of blockEnd-post.north, minimum width=\blockEndWidth] {};

    \begin{scope}[flow]
      \draw (blockBegin) -- (blockLoop);
      \draw (blockBegin) -- (blockEnd);
      \draw ([yshift=-0.3cm]blockLoop.east) to[bend left=-90, distance=0.7cm] ([yshift=0.3cm]blockLoop.east);
      \draw (blockLoop) -- (blockEnd);
    \end{scope}

    \begin{scope}[block label, above right, yshift=-0.1cm]
    \node at (blockBegin-pre.north west) {begin:};
    \node at (blockLoop.north west) {loop:};
    \node at (blockEnd.north west) {end:};
    \end{scope}

\end{pgfonlayer}

\end{tikzpicture}

%% file: figures/local-register-assignment.tex
\input{./figures/styles}%
\input{./figures/register-array}%
\pgfsetlayers{main,layer0,layer1,layer2,layer3,layer4,layer5}%

\newcommand{\drawRegisterArrayLabels}[1]{%
  \node [register, above=0.0cm of #1-c1-r1] (#1-reg1) {\texttt{R0}};
  \node [register, above=0.0cm of #1-c1-r2] (#1-reg2) {\texttt{R1}};
  \node [register, above=0.0cm of #1-c1-r3] (#1-reg3) {$\cdots$};
}

\newcommand{\drawLiveRange}[6]{%
  \coordinate [left=#1 of blockLoop-c#4-r1.north west] (begin-#2);
  \coordinate [left=#1 of blockLoop-c#5-r1.south west] (end-#2);
  \draw [live range, <->] (begin-#2) -- (end-#2);
  \node [left=#1 of blockLoop-reg1, xshift=#6, font=\strut, yshift=0.05cm] {$#3$};
}

\newcommand{\defsDistance}{8.4cm}
\newcommand{\arrowDistance}{7.2cm}
\newcommand{\usesDistance}{6.75cm}
\newcommand{\drawOperation}[5]{%
  \coordinate [left=\defsDistance of blockLoop-c#2-r1.north west] (defs#1);
  \coordinate [left=\arrowDistance of blockLoop-c#2-r1.north west] (arrow#1);
  \coordinate [left=\usesDistance of blockLoop-c#2-r1.north west] (uses#1);
  \node [right=0cm of defs#1, align=right, text width=1.1cm] {#3};
  \node [right=0cm of arrow#1] {#4};
  \node [right=0cm of uses#1] {#5};
}

\begin{tikzpicture}

  \begin{pgfonlayer}{layer0}
  \drawEmptyRegisterArray{blockLoop}{4}{3}
  \drawRegisterArrayLabels{blockLoop}
  \drawTemporary[tempNColor]{\TempN{}}{blockLoop-c1-r1}{blockLoop-c2-r1}
  \drawTemporary[tempFColor]{\TempF{}}{blockLoop-c1-r2}{blockLoop-c1-r2}
  \drawTemporary[tempFColor]{\TempFAfter{}}{blockLoop-c2-r2}{blockLoop-c4-r2}
  \drawTemporary[tempNColor]{\TempNAfter{}}{blockLoop-c3-r1}{blockLoop-c4-r1}

  \drawLiveRange{3cm}{Four}{\TempN{}}{1}{2}{0.23cm}
  \drawLiveRange{2.6cm}{Seven}{\TempNAfter{}}{3}{4}{0.31cm}
  \drawLiveRange{2.2cm}{Five}{\TempF{}}{1}{1}{0.24cm}
  \drawLiveRange{1.8cm}{Six}{\TempFAfter{}}{2}{4}{0.31cm}

  \drawOperation{One}{1}{$\TempN{}, \TempF{}$}{$\leftarrow$}{}
  \drawOperation{Two}{2}{$\TempFAfter{}$}{$\leftarrow$}{$\instruction{\multiplyOpcode{}} \,\TempF{}, \TempN{}$}
  \drawOperation{Three}{3}{$\TempNAfter{}$}{$\leftarrow$}{$\instruction{\subtractOpcode{}} \,\TempN{}, \code{1}$}
  \drawOperation{Four}{4}{}{}{$\code{j} \, \emph{loop} \, \code{if} \, \TempNAfter{} \, \code{>} \, \code{0}$}
  \coordinate [left=\defsDistance of blockLoop-c4-r1.south west] (defsFive);
  \coordinate [left=\arrowDistance of blockLoop-c4-r1.south west] (arrowFive);
  \coordinate [left=\usesDistance of blockLoop-c4-r1.south west] (usesFive);
  \node [right=0cm of arrowFive] {$\leftarrow$};
  \node [right=0cm of usesFive] {$\TempNAfter{}, \TempFAfter{}$};

  \end{pgfonlayer}

  \newcommand{\blockShift}{1.2cm}
  \newcommand{\blockWidth}{4.2cm}
  \begin{pgfonlayer}{main}
    \node [cfg block boundary, fit=(defsOne) (usesOne), minimum height=0.9cm, yshift=-0.2cm, minimum width=\blockWidth, xshift=\blockShift] (blockLoop-pre) {};
    \node [cfg block boundary, fit=(defsFive) (usesFive), minimum height=0.9cm, yshift=0.2cm, minimum width=\blockWidth, xshift=\blockShift] (blockLoop-post) {};
    \node [cfg block, fit=(defsTwo) (usesFour), inner ysep=0.3cm, minimum width=\blockWidth, xshift=\blockShift] (blockLoop-body) {};
    \node [cfg block filler, below=0cm of blockLoop-body.north, minimum height=0.5cm, minimum width=\blockWidth] {};
    \node [cfg block filler, above=0cm of blockLoop-body.south, minimum height=0.5cm, minimum width=\blockWidth] {};
  \end{pgfonlayer}

\end{tikzpicture}

%% file: figures/local-register-packing.tex
\input{./figures/styles}%
\input{./figures/register-array}%
\pgfsetlayers{main,layer0,layer1,layer2,layer3,layer4,layer5}%

\newcommand{\drawRegisterArrayLabels}[1]{%
  \node [register, above=0.0cm of #1-c1-r1] (#1-reg1) {\texttt{R0}};
  \node [register, above=0.0cm of #1-c1-r2] (#1-reg2) {\texttt{R1}};
  \node [register, above=0.0cm of #1-c1-r3] (#1-reg3) {\texttt{R2}};
  \node [register, above=0.0cm of #1-c1-r4] (#1-reg4) {$\cdots$};
}

\begin{tikzpicture}

  \drawEmptyRegisterArray{blockLoop}{4}{4}
  \drawRegisterArrayLabels{blockLoop}
  \drawTemporary[tempNColor]{\TempN{}}{blockLoop-c1-r3}{blockLoop-c2-r3}
  \drawTemporary[tempFColor]{\TempF{}}{blockLoop-c1-r1}{blockLoop-c1-r2}
  \drawTemporary[tempFColor]{\TempFAfter{}}{blockLoop-c2-r1}{blockLoop-c4-r2}
  \drawTemporary[tempNColor]{\TempNAfter{}}{blockLoop-c3-r3}{blockLoop-c4-r3}
  \draw [decorate,decoration=brace,line width=0.2mm]
        ([xshift=0.08cm, yshift=-0.05cm]blockLoop-reg1.north west)
        -- node[above,yshift=0.04cm] {$\register{R1:0}$}
        ([xshift=-0.08cm, yshift=-0.05cm]blockLoop-reg2.north east);
\end{tikzpicture}

%% file: figures/copy-extension.tex
\input{./figures/styles}%
\input{./figures/operations}%
\pgfsetlayers{main,layer0,layer1,layer2,layer3,layer4,layer5}%

\begin{tikzpicture}

\begin{pgfonlayer}{layer0}

\coordinate (p1);
\foreach \p in {2, ..., 8} {%
  \pgfmathtruncatemacro{\prevP}{\p - 1}
  \coordinate [below=0.5cm of p\prevP] (p\p);
}

\drawSimpleDefinerBeforeAfter{p1}
\drawVDots{p3}
\drawUserBefore{p5}
\drawVDots{p6}
\drawUserBefore{p8}

\foreach \a in {1, ..., 8} {%
  \coordinate [right=5cm of p\a] (a\a);
}

\drawSimpleDefinerBeforeAfter{a1}
\drawStoreAfter{a2}
\drawVDots{a3}
\drawLoadAfter{a4}{2}
\drawFirstUserAfter{a5}
\drawVDots{a6}
\drawLoadAfter{a7}{n}
\drawUserAfter{a8}

\end{pgfonlayer}

\end{tikzpicture}

%% file: figures/local-register-allocation.tex
\input{./figures/styles}%
\input{./figures/register-array}%
\pgfsetlayers{main,layer0,layer1,layer2,layer3,layer4,layer5}%

\newcommand{\drawRegisterArrayLabels}[1]{%
  \node [register, above=0.0cm of #1-c1-r1] (#1-reg1) {\texttt{R0}};
  \node [register, above=0.0cm of #1-c1-r2] (#1-reg2) {\texttt{R1}};
  \node [register, above=0.0cm of #1-c1-r3] (#1-reg3) {$\cdots$};
  \node [register, above=0.0cm of #1-c1-r4] (#1-reg4) {\texttt{M0}};
  \node [register, above=0.0cm of #1-c1-r5] (#1-reg5) {$\cdots$};
}

\newcommand{\defsDistance}{6.5cm}
\newcommand{\arrowDistance}{5.3cm}
\newcommand{\usesDistance}{4.85cm}
\newcommand{\drawOperation}[5]{%
  \coordinate [left=\defsDistance of blockLoop-c#2-r1.north west] (defs#1);
  \coordinate [left=\arrowDistance of blockLoop-c#2-r1.north west] (arrow#1);
  \coordinate [left=\usesDistance of blockLoop-c#2-r1.north west] (uses#1);
  \node [right=0cm of defs#1, align=right, text width=1.1cm] {#3};
  \node [right=0cm of arrow#1] {#4};
  \node [right=0cm of uses#1] {#5};
}

\begin{tikzpicture}

  \begin{pgfonlayer}{layer0}

  \drawEmptyRegisterArray{blockLoop}{7}{5}
  \drawRegisterArrayLabels{blockLoop}
  \drawBgTemporary[tempFColor]{\TempF{}}{blockLoop-c1-r2}{blockLoop-c3-r2}
  \drawBgTemporary[tempFColor]{\TempFAfter{}}{blockLoop-c4-r2}{blockLoop-c7-r2}
  \drawBgTemporary[tempNColor]{\TempNAfter{}}{blockLoop-c6-r1}{blockLoop-c7-r1}
  \drawTemporary[tempNColor]{\TempN{}}{blockLoop-c1-r1}{blockLoop-c1-r1}
  \drawTemporary[tempNColor]{\TempNStore{}}{blockLoop-c2-r4}{blockLoop-c4-r4}
  \drawTemporary[tempNColor]{\TempNLoadFst{}}{blockLoop-c3-r1}{blockLoop-c3-r1}
  \drawTemporary[tempNColor]{\TempNLoadSnd{}}{blockLoop-c5-r1}{blockLoop-c5-r1}

  \begin{scope}[copy]
    \draw ([yshift=\copyMargin]blockLoop-c1-r1.south) -- ([yshift=-\copyMargin]blockLoop-c2-r4.north) node [pos=0.66, above] {$\instruction{\storeOpcode{}}$};
    \draw ([yshift=\copyMargin]blockLoop-c2-r4.south) -- ([yshift=-\copyMargin]blockLoop-c3-r1.north) node [pos=0.34, below] {$\instruction{\loadOpcode{}}$};
    \draw ([yshift=\copyMargin]blockLoop-c4-r4.south) -- ([yshift=-\copyMargin]blockLoop-c5-r1.north) node [pos=0.34, below] {$\instruction{\loadOpcode{}}$};
  \end{scope}

  \drawEmptyRegisterArray[right=4cm of blockLoop-c0-r1]{blockLoopTwo}{7}{5}
  \drawRegisterArrayLabels{blockLoopTwo}
  \drawBgTemporary[tempFColor]{\TempF{}}{blockLoopTwo-c1-r2}{blockLoopTwo-c3-r2}
  \drawBgTemporary[tempFColor]{\TempFAfter{}}{blockLoopTwo-c4-r2}{blockLoopTwo-c7-r2}
  \drawBgTemporary[tempNColor]{\TempNAfter{}}{blockLoopTwo-c6-r1}{blockLoopTwo-c7-r1}
  \drawTemporary[tempNColor]{\TempN{}}{blockLoopTwo-c1-r1}{blockLoopTwo-c1-r1}
  \drawTemporary[tempNColor]{\TempNStore{}}{blockLoopTwo-c2-r4}{blockLoopTwo-c2-r4}
  \drawTemporary[tempNColor]{\TempNLoadFst{}}{blockLoopTwo-c3-r1}{blockLoopTwo-c5-r1}

  \begin{scope}[copy]
    \draw ([yshift=\copyMargin]blockLoopTwo-c1-r1.south) -- ([yshift=-\copyMargin]blockLoopTwo-c2-r4.north) node [pos=0.66, above] {$\instruction{\storeOpcode{}}$};
    \draw ([yshift=\copyMargin]blockLoopTwo-c2-r4.south) -- ([yshift=-\copyMargin]blockLoopTwo-c3-r1.north) node [pos=0.34, below] {$\instruction{\loadOpcode{}}$};
  \end{scope}

  \drawOperation{One}{1}{$\TempN{}, \TempF{}$}{$\leftarrow$}{}
  \drawOperation{Two}{2}{$\TempNStore{}$}{$\leftarrow$}{$\set{\instruction{\storeOpcode{}},\instruction{\moveOpcode{}}} \,\TempN{}$}
  \drawOperation{Three}{3}{$\TempNLoadFst{}$}{$\leftarrow$}{$\set{\instruction{\loadOpcode{}},\instruction{\moveOpcode{}}} \,\set{\TempN{},\TempNStore{}}$}
  \drawOperation{Four}{4}{$\TempFAfter{}$}{$\leftarrow$}{$\instruction{\multiplyOpcode{}} \,\TempF{}, \set{\TempN{},\TempNStore{},\TempNLoadFst{},\TempNLoadSnd{}}$}
  \drawOperation{Five}{5}{$\TempNLoadSnd$}{$\leftarrow$}{$\set{\instruction{\loadOpcode{}},\instruction{\moveOpcode{}}} \,\set{\TempN{},\TempNStore{}}$}
  \drawOperation{Six}{6}{$\TempNAfter{}$}{$\leftarrow$}{$\instruction{\subtractOpcode{}} \,\set{\TempN{},\TempNStore{},\TempNLoadFst{},\TempNLoadSnd{}}, \code{1}$}
  \drawOperation{Seven}{7}{}{}{$\code{j} \, \emph{loop} \, \code{if} \, \TempNAfter{} \, \code{>} \, \code{0}$}
  \coordinate [left=\defsDistance of blockLoop-c7-r1.south west] (defsEight);
  \coordinate [left=\arrowDistance of blockLoop-c7-r1.south west] (arrowEight);
  \coordinate [left=\usesDistance of blockLoop-c7-r1.south west] (usesEight);
  \node [right=0cm of arrowEight] {$\leftarrow$};
  \node [right=0cm of usesEight] {$\TempNAfter{}, \TempFAfter{}$};
  \end{pgfonlayer}

  \newcommand{\blockShift}{1.85cm}
  \newcommand{\blockWidth}{5.3cm}
  \begin{pgfonlayer}{main}
    \node [cfg block boundary, fit=(defsOne) (usesOne), minimum height=0.9cm, yshift=-0.2cm, minimum width=\blockWidth, xshift=\blockShift] (blockLoop-pre) {};
    \node [cfg block boundary, fit=(defsEight) (usesEight), minimum height=0.9cm, yshift=0.2cm, minimum width=\blockWidth, xshift=\blockShift] (blockLoop-post) {};
    \node [cfg block, fit=(defsTwo) (usesSeven), inner ysep=0.3cm, minimum width=\blockWidth, xshift=\blockShift] (blockLoop-body) {};
    \node [cfg block filler, below=0cm of blockLoop-body.north, minimum height=0.5cm, minimum width=\blockWidth] {};
    \node [cfg block filler, above=0cm of blockLoop-body.south, minimum height=0.5cm, minimum width=\blockWidth] {};
  \end{pgfonlayer}

\end{tikzpicture}

%% file: figures/remat-extension.tex
\input{./figures/styles}%
\input{./figures/operations}%
\pgfsetlayers{main,layer0,layer1,layer2,layer3,layer4,layer5}%

\begin{tikzpicture}

\begin{pgfonlayer}{layer0}

\coordinate (p1);
\foreach \p in {2, ..., 8} {%
  \pgfmathtruncatemacro{\prevP}{\p - 1}
  \coordinate [below=0.5cm of p\prevP] (p\p);
}

\drawDefinerBeforeAfter{p1}
\drawVDots{p3}
\drawUserBefore{p5}
\drawVDots{p6}
\drawUserBefore{p8}

\foreach \r in {1, ..., 8} {%
  \coordinate [right=5cm of p\r] (r\r);
}

\drawDefinerDemat{r1}
\drawStoreAfter{r2}
\drawVDots{r3}
\drawLoadRemat{r4}{2}
\drawUserAfter{r5}
\drawVDots{r6}
\drawLoadRemat{r7}{n}
\drawUserAfter{r8}

\end{pgfonlayer}

\end{tikzpicture}

%% file: figures/congruence-lifting.tex
\input{./figures/styles}%
\input{./figures/boundaries}%

\begin{tikzpicture}

\pgfsetlayers{main,layer0}%

\coordinate (p1);
\coordinate [right=7.5cm of p1](p2);

\drawCongruence{p1}{\tempPred{}}{\tempSucc{}}{\congruent{\tempPred{}}{\tempSucc}}{2.5cm}{0.88cm}{true}{0.6cm}

\drawCongruence{p2}{\Operand{\operPred{}}{\set{t_{\blockPred{}}, t_{\blockPred{}.1}, \dots, t_{\blockPred{}.n}}}}
                   {\Operand{\operSucc{}}{\tempSucc{}}}
                   {\congruent{\operPred{}}{\operSucc{}}}{6cm}{0.95cm}{true}{0.6cm}

\end{tikzpicture}

%% file: figures/factorial-lssa.tex
\input{./figures/styles}%
\begin{tikzpicture}
\pgfsetlayers{main,layer0}%

\begin{pgfonlayer}{layer0}

    \node [block code] (blockBegin-pre-code) {%
      \begin{basicBlock}
        \alignCentered{\inOperation{\preAssigned{\TempNBegin{}}{\register{R0}}}}
      \end{basicBlock}
    };
    \node [block code, below=of blockBegin-pre-code] (blockBegin-code) {%
      \begin{basicBlock}
        \operation{\TempFBegin{}}{\transferImmediateOpcode{}}{\code{1}}
        \freeBranchOperation{\code{j} \, \emph{end} \, \code{if} \, \TempNBegin{} \, \code{<=} \, \code{0}}
      \end{basicBlock}
    };
    \node [block code, below=of blockBegin-code] (blockBegin-post-code) {%
      \begin{basicBlock}
        \alignCentered{\outOperation{\TempNBegin{}, \TempFBegin{}}}
      \end{basicBlock}
    };

    \node [block code, right=3.8cm of blockBegin-pre-code, yshift=-1cm] (blockLoop-pre-code) {%
      \begin{basicBlock}
        \alignCentered{\inOperation{\TempNLoop{}, \TempFLoop{}}}
      \end{basicBlock}
    };
    \node [block code, below=of blockLoop-pre-code] (blockLoop-code) {%
      \begin{basicBlock}
        \operation{\TempFLoopAfter{}}{\multiplyOpcode{}}{\TempFLoop{}, \TempNLoop{}}
        \operation{\TempNLoopAfter{}}{\subtractOpcode{}}{\TempNLoop{}, \code{1}}
        \freeBranchOperation{\code{j} \, \emph{loop} \, \code{if} \, \TempNLoopAfter{} \, \code{>} \, \code{0}}
      \end{basicBlock}
    };
    \node [block code, below=of blockLoop-code] (blockLoop-post-code) {%
      \begin{basicBlock}
        \alignCentered{\outOperation{\TempNLoopAfter{}, \TempFLoopAfter{}}}
      \end{basicBlock}
    };

    \node [block code, below=1cm of blockBegin-post-code] (blockEnd-pre-code) {%
      \begin{basicBlock}
        \inOperation{\TempFEnd{}}
      \end{basicBlock}
    };
    \node [block code, below=of blockEnd-pre-code] (blockEnd-code) {%
      \begin{basicBlock}
        \copyOperation{\TempFEndStore{}}{\instruction{\storeOpcode{}},\instruction{\moveOpcode{}}}{\TempFEnd{}}
        \copyOperation{\TempFEndLoad{}}{\instruction{\loadOpcode{}},\instruction{\moveOpcode{}}}{\set{\TempFEnd{}, \TempFEndStore{}}}
        \branchOperation{\indirectJumpOpcode{}}{}
      \end{basicBlock}
    };
    \node [block code, below=of blockEnd-code] (blockEnd-post-code) {%
      \begin{basicBlock}
        \alignCentered{\outOperation{\preAssigned{\set{\TempFEnd{}, \TempFEndStore{}, \TempFEndLoad{}}}{\register{R0}}}}
      \end{basicBlock}
    };

\end{pgfonlayer}

\newcommand{\blockBeginWidth}{4cm}
\newcommand{\blockLoopWidth}{4cm}
\newcommand{\blockEndWidth}{4.8cm}
\begin{pgfonlayer}{main}
    \node [cfg block boundary, fit=(blockBegin-pre-code), minimum width=\blockBeginWidth] (blockBegin-pre) {};
    \node [cfg boundary filler, above=0cm of blockBegin-pre.south, minimum width=\blockBeginWidth] {};

    \node [cfg block, fit=(blockBegin-code), minimum width=\blockBeginWidth] (blockBegin) {};
    \node [cfg block filler, below=0cm of blockBegin.north, yshift=0.01cm, minimum width=\blockBeginWidth] {};
    \node [cfg block filler, above=0cm of blockBegin.south, yshift=-0.01cm, minimum width=\blockBeginWidth] {};

    \node [cfg block boundary, fit=(blockBegin-post-code), minimum width=\blockBeginWidth] (blockBegin-post) {};
    \node [cfg boundary filler, below=0cm of blockBegin-post.north, minimum width=\blockBeginWidth] {};

    \node [cfg block boundary, fit=(blockLoop-pre-code), minimum width=\blockLoopWidth] (blockLoop-pre) {};
    \node [cfg boundary filler, above=0cm of blockLoop-pre.south, minimum width=\blockLoopWidth] {};

    \node [cfg block, fit=(blockLoop-code), minimum width=\blockLoopWidth] (blockLoop) {};
    \node [cfg block filler, below=0cm of blockLoop.north, yshift=0.01cm, minimum width=\blockLoopWidth] {};
    \node [cfg block filler, above=0cm of blockLoop.south, yshift=-0.01cm, minimum width=\blockLoopWidth] {};

    \node [cfg block boundary, fit=(blockLoop-post-code), minimum width=\blockLoopWidth] (blockLoop-post) {};
    \node [cfg boundary filler, below=0cm of blockLoop-post.north, minimum width=\blockLoopWidth] {};

    \node [cfg block boundary, fit=(blockEnd-pre-code), minimum width=\blockEndWidth] (blockEnd-pre) {};
    \node [cfg boundary filler, above=0cm of blockEnd-pre.south, minimum width=\blockEndWidth] {};

    \node [cfg block, fit=(blockEnd-code), minimum width=\blockEndWidth] (blockEnd) {};
    \node [cfg block filler, below=0cm of blockEnd.north, yshift=0.01cm, minimum width=\blockEndWidth] {};
    \node [cfg block filler, above=0cm of blockEnd.south, yshift=-0.01cm, minimum width=\blockEndWidth] {};

    \node [cfg block boundary, fit=(blockEnd-post-code), minimum width=\blockEndWidth] (blockEnd-post) {};
    \node [cfg boundary filler, below=0cm of blockEnd-post.north, minimum width=\blockEndWidth] {};

    \begin{scope}[flow]
      \draw (blockBegin) -- node[congruences, below]{%
        \begin{congruences}
          \congruent{\TempNBegin{}}{\TempNLoop{}}
          \congruent{\TempFBegin{}}{\TempFLoop{}}
        \end{congruences}
      } (blockLoop-pre);
      \draw (blockBegin-post) -- node[congruences, xshift=0.75cm]{%
        \begin{congruences}
          \congruent{\TempFBegin{}}{\TempFEnd{}}
        \end{congruences}
      } (blockEnd-pre);
      \draw ([yshift=-0.4cm]blockLoop.east) to[bend left=-90, distance=0.7cm] node[congruences, xshift=0.7cm]{%
        \begin{congruences}
          \congruent{\TempNLoopAfter{}}{\TempNLoop{}}
          \congruent{\TempFLoopAfter{}}{\TempFLoop{}}
        \end{congruences}
      } ([yshift=0.4cm]blockLoop.east);
      \draw (blockLoop) -- node[congruences, below, pos=0.3, yshift=-0.1cm]{%
        \begin{congruences}
          \congruent{\TempFLoopAfter{}}{\TempFEnd{}}
        \end{congruences}
      } (blockEnd-pre);
    \end{scope}

    \begin{scope}[block label, above right, yshift=-0.1cm]
    \node at (blockBegin-pre.north west) {begin:};
    \node at (blockLoop-pre.north west) {loop:};
    \node at (blockEnd-pre.north west) {end:};
    \end{scope}

\end{pgfonlayer}

\end{tikzpicture}

%% file: figures/congruence-spilling.tex
\input{./figures/styles}%

\newcommand{\blockWidth}{4cm}
\newcommand{\bracesShift}{0.2cm}
\newcommand{\mathShift}{0.1cm}

\newcommand{\drawBoundaries}[8]{%
\begin{pgfonlayer}{layer0}
    \node [below=0cm of #1, block code] (pred) {%
      \begin{basicBlock}
      \simpleOperation{\tempPred{}}{\cdots}
      #2
      \vdotsOperation
      \end{basicBlock}
    };
    \node [below=0cm of pred, block code] (post-pred) {%
      \begin{basicBlock}
        \alignCentered{\outOperation{#3}}
      \end{basicBlock}
    };
    \node [below=0.6cm of post-pred, block code] (pre-succ) {%
      \begin{basicBlock}
        \alignCentered{\inOperation{#5}}
      \end{basicBlock}
    };
    \node [below=0cm of pre-succ, block code] (succ) {%
      \begin{basicBlock}
      \vdotsOperation
      #6
      #7
      \end{basicBlock}
    };
\end{pgfonlayer}
\begin{pgfonlayer}{main}
    \node [cfg block filler, fit=(pred), minimum width=\blockWidth] (predDots) {};
    \node [cfg block boundary, fit=(post-pred), minimum width=\blockWidth] (post) {};
    \node [cfg boundary filler, below=0cm of post.north, minimum width=\blockWidth] {};

    \node [cfg block boundary, fit=(pre-succ), minimum width=\blockWidth] (pre) {};
    \node [cfg boundary filler, above=0cm of pre.south, minimum width=\blockWidth] {};
    \node [cfg block filler, fit=(succ), minimum width=\blockWidth] (succDots) {};

    \begin{scope}[flow]
      \draw (post) -- node[congruences, xshift=#8] (congruences){%
        \begin{congruences}
          #4
        \end{congruences}
      } (pre);
    \end{scope}

    \begin{scope}[block label, above right]
      \node at ([yshift=-0.05cm]predDots.north west) {\blockPredRaw{}:};
      \node at ([yshift=-0.05cm]pre.north west) {\blockSuccRaw{}:};
    \end{scope}
\end{pgfonlayer}
}

\begin{tikzpicture}

\pgfsetlayers{main,layer0}%

\coordinate (p1);
\coordinate [right=5cm of p1](p2);
\coordinate [right=5cm of p2](p3);

\drawBoundaries
  {p1}
  {\naturalOperation{\tempPredStore{}}{\instruction{\genericStoreOpcode{}}}{\tempPred{}}}
  {\Operand{p}{\set{\tempPred{}, \tempPredStore{}}}}
  {\congruent{p}{q}}
  {\Operand{q}{\tempSucc{}}}
  {\naturalOperation{\tempSuccLoad{}}{\instruction{\genericLoadOpcode{}}}{\tempSucc{}}}
  {\naturalOperation{\cdots}{i}{\set{\tempSucc{},\tempSuccLoad{}}}}
  {0.5cm}

\drawBoundaries
  {p2}
  {\phantomOperation{\tempPredStore{}}{\instruction{\genericStoreOpcode{}}}{\tempPred{}}}
  {\tempPred{}\vphantom{\set{\tempPred{}, \tempPredStore{}}}}
  {\congruent{\tempPred{}}{\tempSucc{}}}
  {\tempSucc{}}
  {\phantomOperation{\tempSuccLoad{}}{\instruction{\genericLoadOpcode{}}}{\tempSucc{}}}
  {\naturalOperation{\cdots}{i}{\tempSucc{}}}
  {0.9cm}

\drawBoundaries
  {p3}
  {\naturalOperation{\tempPredStore{}}{\instruction{\genericStoreOpcode{}}}{\tempPred{}}}
  {\tempPredStore{}\vphantom{\set{\tempPred{}, \tempPredStore{}}}}
  {\congruent{\tempPredStore{}}{\tempSucc{}}}
  {\tempSucc{}}
  {\naturalOperation{\tempSuccLoad{}}{\instruction{\genericLoadOpcode{}}}{\tempSucc{}}}
  {\naturalOperation{\cdots}{i}{\tempSuccLoad{}}}
  {1cm}

\end{tikzpicture}

%% file: figures/global-register-allocation.tex
\input{./figures/styles}%
\input{./figures/register-array}%
\pgfsetlayers{main,layer0,layer1,layer2,layer3,layer4,layer5}%

\newcommand{\drawRegisterArrayLabels}[1]{%
  \node [register, above=0.0cm of #1-c1-r1] (#1-reg1) {\texttt{R0}};
  \node [register, above=0.0cm of #1-c1-r2] (#1-reg2) {\texttt{R1}};
  \node [register, above=0.0cm of #1-c1-r3] (#1-reg3) {$\cdots$};
  \node [register, above=0.0cm of #1-c1-r4] (#1-reg4) {\texttt{M0}};
  \node [register, above=0.0cm of #1-c1-r5] (#1-reg5) {$\cdots$};
}

\begin{tikzpicture}
  \drawEmptyRegisterArray{blockBegin}{3}{5}
  \drawRegisterArrayLabels{blockBegin}
  \drawTemporary[tempNColor]{\TempNBegin{}}{blockBegin-c1-r1}{blockBegin-c3-r1}
  \drawTemporary[tempFColor]{\TempFBegin{}}{blockBegin-c2-r2}{blockBegin-c3-r2}

  \drawEmptyRegisterArray[right=4.2cm of blockBegin-c1-r1, yshift=-1cm]{blockLoop}{4}{5}
  \drawRegisterArrayLabels{blockLoop}
  \drawTemporary[tempNColor]{\TempNLoop{}}{blockLoop-c1-r1}{blockLoop-c2-r1}
  \drawTemporary[tempNColor]{\TempNLoopAfter{}}{blockLoop-c3-r1}{blockLoop-c4-r1}
  \drawTemporary[tempFColor]{\TempFLoop{}}{blockLoop-c1-r2}{blockLoop-c1-r2}
  \drawTemporary[tempFColor]{\TempFLoopAfter{}}{blockLoop-c2-r2}{blockLoop-c4-r2}

  \drawEmptyRegisterArray[below=1.5cm of blockBegin-c3-r1]{blockEnd}{4}{5}
  \drawRegisterArrayLabels{blockEnd}
  \drawTemporary[tempFColor]{\TempFEnd{}}{blockEnd-c1-r2}{blockEnd-c1-r2}
  \drawTemporary[tempFColor]{\TempFEndStore{}}{blockEnd-c2-r1}{blockEnd-c4-r1}

  \begin{scope}[flow]
      \draw (blockBegin-c2-r5) -- node[congruences, below, pos=0.45, yshift=-0.05cm]{%
        \begin{congruences}
          \congruent{\TempNBegin{}}{\TempNLoop{}}
          \congruent{\TempFBegin{}}{\TempFLoop{}}
        \end{congruences}
      } (blockLoop-c1-r1);
      \draw (blockBegin-c3-r3) -- node[congruences, xshift=0.6cm]{%
        \begin{congruences}
          \congruent{\TempFBegin{}}{\TempFEnd{}}
        \end{congruences}
      } (blockEnd-reg3);
      \draw ([yshift=-0.15cm]blockLoop-c3-r5.east) to[bend left=-90, distance=0.7cm] node[congruences, xshift=0.55cm]{%
        \begin{congruences}
          \congruent{\TempNLoopAfter{}}{\TempNLoop{}}
          \congruent{\TempFLoopAfter{}}{\TempFLoop{}}
        \end{congruences}
      } ([yshift=0.15cm]blockLoop-c2-r5.east);
      \draw (blockLoop-c3-r1) -- node[congruences, below, pos=0.45, yshift=-0.1cm]{%
        \begin{congruences}
          \congruent{\TempFLoopAfter{}}{\TempFEnd{}}
        \end{congruences}
      } (blockEnd-c1-r5);
  \end{scope}

  \begin{scope}[block label, above right]
  \node at ([yshift=-0.13cm]blockBegin-reg1.north west) {begin:};
  \node at ([yshift=-0.13cm]blockLoop-reg1.north west) {loop:};
  \node at ([yshift=-0.1cm]blockEnd-reg1.north west) {end:};
  \end{scope}

  \begin{scope}[copy]
    \draw ([yshift=\copyMargin]blockEnd-c1-r2.south) -- ([yshift=-\copyMargin]blockEnd-c2-r1.north) node [midway, below, xshift=0.1cm] {$\instruction{\moveOpcode{}}$};
  \end{scope}

\end{tikzpicture}

%% file: figures/instruction-scheduling.tex
\input{./figures/styles}%
\input{./figures/register-array}%
\pgfsetlayers{main,layer0,layer1,layer2,layer3,layer4,layer5}%

\newcommand{\drawOperationLeft}[3]{%
  \coordinate [left=6.15cm of #1-c#2-r1.north west] (o-#1-#2);
  \node [font=\small, right=0cm of o-#1-#2, align=flush left, text width=4.9cm] {#3};
}

\newcommand{\drawCycleLeft}[2]{%
  \coordinate [left=0.4cm of #1] (c-#1);
  \node [font=\small, right=0cm of c-#1, align=flush left] {#2:};
}

\newcommand{\cycleLabelShift}{0.1cm}
\newcommand{\unitsLabelShift}{0.2cm}
\begin{tikzpicture}

  \drawEmptyRegisterArray{blockLoop}{3}{4}
  \coordinate [below=0cm of blockLoop-c3-r1.south east, xshift=-0.55cm] (blockLoop-c4-r1);
  \begin{scope}[draw, line width=0.35mm]
  \draw[fill=blockcolor]
        (blockLoop-c1-r1.north west) --
        (blockLoop-c2-r1.north west) --
        (blockLoop-c2-r2.north east) --
        (blockLoop-c2-r2.south east) --
        (blockLoop-c2-r2.south west) --
        (blockLoop-c3-r2.south west) --
        (blockLoop-c3-r1.south west);
  \end{scope}

  \drawOperationLeft{blockLoop}{1}{$\inOperation{\TempN{}, \TempF{}}$}
  \drawOperationLeft{blockLoop}{2}{$\TempFAfter{}\leftarrow\instruction{\multiplyOpcode{}}\,\TempF{}, \TempN{}\parbundled{}\TempNAfter{}\leftarrow\instruction{\subtractOpcode{}}\,\TempN{}, \code{1}$}
  \drawOperationLeft{blockLoop}{3}{$\code{j} \, \emph{loop} \, \code{if} \, \TempNAfter{} \, \code{>} \, \code{0}$}
  \drawOperationLeft{blockLoop}{4}{$\outOperation{\TempNAfter{}, \TempFAfter{}}$}

  \foreach \c in {0, ..., 3} {%
    \pgfmathtruncatemacro{\gridCycle}{\c + 1}
    \coordinate [left=0.1cm of blockLoop-c\gridCycle-r1.north west] (blockLoop-l\c);
    \drawCycleLeft{blockLoop-l\c}{\c}
    \drawCycleLeft{o-blockLoop-\gridCycle}{\c}
  }

  \coordinate [left=\cycleLabelShift of c-o-blockLoop-2, yshift=0.2cm] (cycleStart);
  \coordinate [left=\cycleLabelShift of c-o-blockLoop-3, yshift=-0.2cm] (cycleEnd);
  \draw [flow] (cycleStart) -- node[rotate=-90, font=\small, below] {cycles} (cycleEnd);

  \coordinate [left=\cycleLabelShift of c-blockLoop-l1, yshift=0.2cm] (cycleCStart);
  \coordinate [left=\cycleLabelShift of c-blockLoop-l2, yshift=-0.2cm] (cycleCEnd);
  \draw [flow] (cycleCStart) -- node[rotate=-90, font=\small, below] {cycles} (cycleCEnd);

  \coordinate [below=\unitsLabelShift of blockLoop-c3-r2.south west] (cycleUStart);
  \coordinate [below=\unitsLabelShift of blockLoop-c3-r3.south east] (cycleUEnd);
  \draw [flow] (cycleUStart) -- node[font=\small, below] {units used} (cycleUEnd);

\end{tikzpicture}

%% file: figures/live-range.tex
\input{./figures/styles}%
\input{./figures/register-array}%
\pgfsetlayers{main,layer0,layer1,layer2,layer3,layer4,layer5}%

\newcommand{\cycleShift}{0.5cm}
\begin{tikzpicture}
  \drawEmptyRegisterArray{liveRange}{3}{1}
  \drawTemporary{t}{liveRange-c1-r1}{liveRange-c3-r1}
  \node [left=0cm of liveRange-c1-r1.north west, align=right] {$\operationIssueCycle{d}$};
  \node [left=0cm of liveRange-c2-r1.north west, align=right] {$\operationIssueCycle{u_1}$};
  \node [left=0.5cm of liveRange-c3-r1.north west, align=right, yshift=-0.05cm] {$\cdots$};
  \node [left=0cm of liveRange-c3-r1.south west, align=right] {$\operationIssueCycle{u_n}$};
  \node [right=0cm of liveRange-c1-r1.north east] {$\temporaryLiveStart{t}$};
  \node [right=0cm of liveRange-c3-r1.south east] {$\temporaryLiveEnd{t}$};
  \coordinate [right=\cycleShift of liveRange-c2-r1.north east] (cycleStart);
  \coordinate [right=\cycleShift of liveRange-c3-r1.north east] (cycleEnd);
  \draw [flow] (cycleStart) -- node[rotate=-90, font=\small, above, yshift=-0.05cm] {cycles} (cycleEnd);
\end{tikzpicture}

%% file: figures/integrated-solution.tex
\input{./figures/styles}%
\input{./figures/register-array}%
\pgfsetlayers{main,layer0,layer1,layer2,layer3,layer4,layer5}%

\newcommand{\drawOperationLeft}[4]{%
  \pgfmathtruncatemacro{\cycleLabel}{#2 - 1}
  \coordinate [left=#4 of #1-c#2-r1.north west] (o-#1-#2);
  \node [font=\small, right=0cm of o-#1-#2] {\cycleLabel: #3};
}

\newcommand{\drawOperationRight}[3]{%
  \pgfmathtruncatemacro{\cycleLabel}{#2 - 1}
  \coordinate [right=0.1cm of #1-c#2-r3.north east] (o-#1-#2);
  \node [font=\small, right=0cm of o-#1-#2] {\cycleLabel: #3};
}

\newcommand{\drawRegisterArrayLabels}[1]{%
  \node [register, above=0.0cm of #1-c1-r1] (#1-reg1) {\texttt{R0}};
  \node [register, above=0.0cm of #1-c1-r2] (#1-reg2) {\texttt{R1}};
  \node [register, above=0.0cm of #1-c1-r3] (#1-reg3) {$\cdots$};
}

\newcommand{\blockBeginXShift}{4.65cm}
\newcommand{\blockEndXShift}{3.25cm}

\begin{tikzpicture}
  \drawEmptyRegisterArray{blockBegin}{2}{3}
  \drawRegisterArrayLabels{blockBegin}
  \drawTemporary[tempNColor]{\TempNBegin{}}{blockBegin-c1-r1}{blockBegin-c2-r1}
  \drawTemporary[tempFColor]{\TempFBegin{}}{blockBegin-c2-r2}{blockBegin-c2-r2}

  \drawEmptyRegisterArray[right=2.7cm of blockBegin-c1-r1, yshift=-1cm]{blockLoop}{3}{3}
  \drawRegisterArrayLabels{blockLoop}
  \drawTemporary[tempNColor]{\TempNLoop{}}{blockLoop-c1-r1}{blockLoop-c1-r1}
  \drawTemporary[tempNColor]{\TempNLoopAfter{}}{blockLoop-c2-r1}{blockLoop-c3-r1}
  \drawTemporary[tempFColor]{\TempFLoop{}}{blockLoop-c1-r2}{blockLoop-c1-r2}
  \drawTemporary[tempFColor]{\TempFLoopAfter{}}{blockLoop-c2-r2}{blockLoop-c3-r2}

  \drawEmptyRegisterArray[below=1.5cm of blockBegin-c3-r1]{blockEnd}{2}{3}
  \drawRegisterArrayLabels{blockEnd}
  \drawTemporary[tempFColor]{\TempFEnd{}}{blockEnd-c1-r2}{blockEnd-c1-r2}
  \drawTemporary[tempFColor]{\TempFEndStore{}}{blockEnd-c2-r1}{blockEnd-c2-r1}

  \begin{scope}[flow]
      \draw (blockBegin-c2-r3) -- node[congruences, below, pos=0.4, yshift=-0.1cm]{%
        \begin{congruences}
          \congruent{\TempNBegin{}}{\TempNLoop{}}
          \congruent{\TempFBegin{}}{\TempFLoop{}}
        \end{congruences}
      } (blockLoop-c1-r1);
      \draw (blockBegin-c2-r3) -- node[congruences, pos=0.4, xshift=-0.6cm]{%
        \begin{congruences}
          \congruent{\TempFBegin{}}{\TempFEnd{}}
        \end{congruences}
      } (blockEnd-reg3);
      \draw ([xshift=-0.2cm]blockLoop-c3-r3.south west) to[bend left=-90, distance=0.6cm] node[congruences, below]{%
        \begin{congruences}
          \congruent{\TempNLoopAfter{}}{\TempNLoop{}}
          \congruent{\TempFLoopAfter{}}{\TempFLoop{}}
        \end{congruences}
      } ([xshift=-0.2cm]blockLoop-c3-r3.south east);
      \draw (blockLoop-c3-r1) -- node[congruences, below, pos=0.4, yshift=-0.13cm]{%
        \begin{congruences}
          \congruent{\TempFLoopAfter{}}{\TempFEnd{}}
        \end{congruences}
      } (blockEnd-c1-r3);
  \end{scope}

  \begin{scope}[block label, above right]
  \node at ([yshift=-0.13cm]blockBegin-reg1.north west) {begin:};
  \node at ([yshift=-0.13cm]blockLoop-reg1.north west) {loop:};
  \node at ([yshift=-0.1cm]blockEnd-reg1.north west) {end:};
  \end{scope}

  \begin{scope}[copy]
    \draw ([yshift=\copyMargin]blockEnd-c1-r2.south) -- ([yshift=-\copyMargin]blockEnd-c2-r1.north) node [midway, below, xshift=0.27cm, yshift=0.07cm] {$\instruction{\moveOpcode{}}$};
  \end{scope}

  \drawOperationLeft{blockBegin}{1}{$\inOperation{\preAssigned{\TempNBegin{}}{\register{R0}}}$}{\blockBeginXShift}
  \drawOperationLeft{blockBegin}{2}{$\TempFBegin{}\leftarrow\instruction{\transferImmediateOpcode{}}\,\code{1}\,\parbundled{}\,\code{j} \, \emph{end} \, \code{if} \, \TempNBegin{} \, \code{<=} \, \code{0}$}{\blockBeginXShift}
  \drawOperationLeft{blockBegin}{3}{$\outOperation{\TempNBegin{}, \TempFBegin{}}$}{\blockBeginXShift}

  \drawOperationRight{blockLoop}{1}{$\inOperation{\TempNLoop{}, \TempFLoop{}}$}
  \drawOperationRight{blockLoop}{2}{$\TempFLoopAfter{}\leftarrow\instruction{\multiplyOpcode{}}\,\TempFLoop{}, \TempNLoop{}\,\parbundled{}\,\TempNLoopAfter{}\leftarrow\instruction{\subtractOpcode{}}\,\TempNLoop{}, \code{1}$}
  \drawOperationRight{blockLoop}{3}{$\code{j} \, \emph{loop} \, \code{if} \, \TempNLoopAfter{} \, \code{>} \, \code{0}$}
  \coordinate [right=0.55cm of blockLoop-c3-r3.south west] (blockLoop-c4-r3);
  \drawOperationRight{blockLoop}{4}{$\outOperation{\TempNLoopAfter{}, \TempFLoopAfter{}}$}

  \drawOperationLeft{blockEnd}{1}{$\inOperation{\TempFEnd{}}$}{\blockEndXShift}
  \drawOperationLeft{blockEnd}{2}{$\TempFEndStore{}\leftarrow\instruction{\moveOpcode{}}\,\TempFEnd{}\,\parbundled{}\,\instruction{\indirectJumpOpcode{}}$}{\blockEndXShift}
  \drawOperationLeft{blockEnd}{3}{$\outOperation{\preAssigned{\TempFEndStore{}}{\register{R0}}}$}{\blockEndXShift}

\end{tikzpicture}

%% file: figures/slack-balancing-constraints.tex
\input{./figures/styles}%
\input{./figures/boundaries}%

\newcommand{\blockWidth}{2.5cm}
\newcommand{\bracesShift}{0.2cm}
\newcommand{\mathShift}{0.1cm}
\begin{tikzpicture}

\pgfsetlayers{main,layer0}%

\begin{pgfonlayer}{layer0}
    \node [block code] (pred) {%
      \begin{basicBlock}
      \naturalOperation{\tempPred{}}{i}{\cdots}
      \vdotsOperation
      \end{basicBlock}
    };
    \node [below=0cm of pred, block code] (post-pred) {%
      \begin{basicBlock}
        \alignCentered{\outOperation{\Operand{p}{\tempPred{}}}}
      \end{basicBlock}
    };
    \node [below=0.6cm of post-pred, block code] (pre-succ) {%
      \begin{basicBlock}
        \alignCentered{\inOperation{\Operand{q}{\tempSucc{}}}}
      \end{basicBlock}
    };
    \node [below=0cm of pre-succ, block code] (succ) {%
      \begin{basicBlock}
      \vdotsOperation
      \simpleOperation{\cdots}{\tempSucc{}}
      \end{basicBlock}
    };
\end{pgfonlayer}
\begin{pgfonlayer}{main}
    \node [cfg block filler, fit=(pred), minimum width=\blockWidth] (predDots) {};
    \node [cfg block boundary, fit=(post-pred), minimum width=\blockWidth] (post) {};
    \node [cfg boundary filler, below=0cm of post.north, minimum width=\blockWidth] {};

    \node [cfg block boundary, fit=(pre-succ), minimum width=\blockWidth] (pre) {};
    \node [cfg boundary filler, above=0cm of pre.south, minimum width=\blockWidth] {};
    \node [cfg block filler, fit=(succ), minimum width=\blockWidth] (succDots) {};

    \begin{scope}[flow]
      \draw (post) -- node[congruences, xshift=0.6cm] (congruences){%
        \begin{congruences}
          \congruent{p}{q}
        \end{congruences}
      } (pre);
    \end{scope}

    \begin{scope}[flow,|-|]
    \draw ([xshift=\bracesShift, yshift=0.34cm]predDots.east)
          -- node[right,xshift=\mathShift] {$\ge \latency{i} + \operandSlack{p}$}
          ([xshift=\bracesShift]post.east);

    \draw ([xshift=\bracesShift]pre.east)
          -- node[right,xshift=\mathShift] {$\ge 1 + \operandSlack{q} \stackrel{\text{(\ref{con:slack-balancing})}}{=} 1 - \operandSlack{p}$}
          ([xshift=\bracesShift, yshift=-0.34cm]succDots.east);
    \end{scope}

\end{pgfonlayer}

\end{tikzpicture}

%% file: figures/double-load-extension.tex
\input{./figures/styles}%
\input{./figures/operations}%
\pgfsetlayers{main,layer0,layer1,layer2,layer3,layer4,layer5}%

\begin{tikzpicture}

\begin{pgfonlayer}{layer0}

\coordinate (p1);
\foreach \p in {2, ..., 8} {%
  \pgfmathtruncatemacro{\prevP}{\p - 1}
  \coordinate [below=0.5cm of p\prevP] (p\p);
}

\drawFirstLoad{p2}
\drawVDots{p3}
\drawFirstLoadUser{p4}
\drawVDots{p5}
\drawSecondLoad{p6}
\drawVDots{p7}
\drawSecondLoadUser{p8}

\foreach \a in {1, ..., 8} {%
  \coordinate [right=5.5cm of p\a] (a\a);
}

\drawDoubleLoad{a1}
\drawFirstLoad{a2}
\drawVDots{a3}
\drawFirstLoadUserAfter{a4}
\drawVDots{a5}
\drawSecondLoad{a6}
\drawVDots{a7}
\drawSecondLoadUserAfter{a8}

\end{pgfonlayer}

\end{tikzpicture}

%% file: results/hexagon-speed-improvement.tex
\begin{tikzpicture}[gnuplot]
\path (0.000,0.000) rectangle (12.500,8.750);
\gpcolor{color=gp lt color border}
\gpsetlinetype{gp lt border}
\gpsetdashtype{gp dt solid}
\gpsetlinewidth{1.00}
\draw[gp path] (1.196,5.644)--(1.376,5.644);
\draw[gp path] (24.446,5.644)--(24.266,5.644);
\node[gp node right,font={\fontsize{8.0pt}{9.6pt}\selectfont}] at (1.012,5.644) {\plotPercentage{0}};
\draw[gp path] (1.196,6.093)--(1.376,6.093);
\draw[gp path] (24.446,6.093)--(24.266,6.093);
\node[gp node right,font={\fontsize{8.0pt}{9.6pt}\selectfont}] at (1.012,6.093) {\plotPercentage{10}};
\draw[gp path] (1.196,6.541)--(1.376,6.541);
\draw[gp path] (24.446,6.541)--(24.266,6.541);
\node[gp node right,font={\fontsize{8.0pt}{9.6pt}\selectfont}] at (1.012,6.541) {\plotPercentage{20}};
\draw[gp path] (1.196,6.990)--(1.376,6.990);
\draw[gp path] (24.446,6.990)--(24.266,6.990);
\node[gp node right,font={\fontsize{8.0pt}{9.6pt}\selectfont}] at (1.012,6.990) {\plotPercentage{30}};
\draw[gp path] (1.196,7.439)--(1.376,7.439);
\draw[gp path] (24.446,7.439)--(24.266,7.439);
\node[gp node right,font={\fontsize{8.0pt}{9.6pt}\selectfont}] at (1.012,7.439) {\plotPercentage{40}};
\draw[gp path] (1.196,7.888)--(1.376,7.888);
\draw[gp path] (24.446,7.888)--(24.266,7.888);
\node[gp node right,font={\fontsize{8.0pt}{9.6pt}\selectfont}] at (1.012,7.888) {\plotPercentage{50}};
\draw[gp path] (1.196,8.336)--(1.376,8.336);
\draw[gp path] (24.446,8.336)--(24.266,8.336);
\node[gp node right,font={\fontsize{8.0pt}{9.6pt}\selectfont}] at (1.012,8.336) {\plotPercentage{60}};
\draw[gp path] (1.196,8.785)--(1.376,8.785);
\draw[gp path] (24.446,8.785)--(24.266,8.785);
\node[gp node right,font={\fontsize{8.0pt}{9.6pt}\selectfont}] at (1.012,8.785) {\plotPercentage{70}};
\draw[gp path] (1.196,9.234)--(1.376,9.234);
\draw[gp path] (24.446,9.234)--(24.266,9.234);
\node[gp node right,font={\fontsize{8.0pt}{9.6pt}\selectfont}] at (1.012,9.234) {\plotPercentage{80}};
\draw[gp path] (1.196,9.682)--(1.376,9.682);
\draw[gp path] (24.446,9.682)--(24.266,9.682);
\node[gp node right,font={\fontsize{8.0pt}{9.6pt}\selectfont}] at (1.012,9.682) {\plotPercentage{90}};
\draw[gp path] (1.196,10.131)--(1.376,10.131);
\draw[gp path] (24.446,10.131)--(24.266,10.131);
\node[gp node right,font={\fontsize{8.0pt}{9.6pt}\selectfont}] at (1.012,10.131) {\plotPercentage{100}};
\node[gp node left,rotate=-90,font={\fontsize{7.0pt}{8.4pt}\selectfont}] at (1.426,5.721) {\functionId{1}\functionName{handle_noinline_attribute}};
\node[gp node left,rotate=-90,font={\fontsize{7.0pt}{8.4pt}\selectfont}] at (1.656,5.721) {\functionId{2}\functionName{control_flow_insn_p}};
\node[gp node left,rotate=-90,font={\fontsize{7.0pt}{8.4pt}\selectfont}] at (1.887,5.721) {\functionId{3}\functionName{insert_insn_on_edge}};
\node[gp node left,rotate=-90,font={\fontsize{7.0pt}{8.4pt}\selectfont}] at (2.117,5.721) {\functionId{4}\functionName{update_br_prob_note}};
\node[gp node left,rotate=-90,font={\fontsize{7.0pt}{8.4pt}\selectfont}] at (2.347,5.721) {\functionId{5}\functionName{_cpp_init_internal_pragma.}};
\node[gp node left,rotate=-90,font={\fontsize{7.0pt}{8.4pt}\selectfont}] at (2.577,5.721) {\functionId{6}\functionName{lex_macro_node}};
\node[gp node left,rotate=-90,font={\fontsize{7.0pt}{8.4pt}\selectfont}] at (2.807,5.721) {\functionId{7}\functionName{cse_basic_block}};
\node[gp node left,rotate=-90,font={\fontsize{7.0pt}{8.4pt}\selectfont}] at (3.038,5.721) {\functionId{8}\functionName{rtx_equal_for_cselib_p}};
\node[gp node left,rotate=-90,font={\fontsize{7.0pt}{8.4pt}\selectfont}] at (3.268,5.721) {\functionId{9}\functionName{debug_df_chain}};
\node[gp node left,rotate=-90,font={\fontsize{7.0pt}{8.4pt}\selectfont}] at (3.498,5.721) {\functionId{10}\functionName{modified_type_die}};
\node[gp node left,rotate=-90,font={\fontsize{7.0pt}{8.4pt}\selectfont}] at (3.728,5.721) {\functionId{11}\functionName{emit_note}};
\node[gp node left,rotate=-90,font={\fontsize{7.0pt}{8.4pt}\selectfont}] at (3.958,5.721) {\functionId{12}\functionName{gen_sequence}};
\node[gp node left,rotate=-90,font={\fontsize{7.0pt}{8.4pt}\selectfont}] at (4.189,5.721) {\functionId{13}\functionName{subreg_hard_regno}};
\node[gp node left,rotate=-90,font={\fontsize{7.0pt}{8.4pt}\selectfont}] at (4.419,5.721) {\functionId{14}\functionName{split_double}};
\node[gp node left,rotate=-90,font={\fontsize{7.0pt}{8.4pt}\selectfont}] at (4.649,5.721) {\functionId{15}\functionName{add_to_mem_set_list}};
\node[gp node left,rotate=-90,font={\fontsize{7.0pt}{8.4pt}\selectfont}] at (4.879,5.721) {\functionId{16}\functionName{find_regno_partial}};
\node[gp node left,rotate=-90,font={\fontsize{7.0pt}{8.4pt}\selectfont}] at (5.109,5.721) {\functionId{17}\functionName{use_return_register}};
\node[gp node left,rotate=-90,font={\fontsize{7.0pt}{8.4pt}\selectfont}] at (5.340,5.721) {\functionId{18}\functionName{ix86_expand_move}};
\node[gp node left,rotate=-90,font={\fontsize{7.0pt}{8.4pt}\selectfont}] at (5.570,5.721) {\functionId{19}\functionName{legitimate_pic_address_di.}};
\node[gp node left,rotate=-90,font={\fontsize{7.0pt}{8.4pt}\selectfont}] at (5.800,5.721) {\functionId{20}\functionName{gen_extendsfdf2}};
\node[gp node left,rotate=-90,font={\fontsize{7.0pt}{8.4pt}\selectfont}] at (6.030,5.721) {\functionId{21}\functionName{gen_mulsidi3}};
\node[gp node left,rotate=-90,font={\fontsize{7.0pt}{8.4pt}\selectfont}] at (6.260,5.721) {\functionId{22}\functionName{gen_peephole2_1255}};
\node[gp node left,rotate=-90,font={\fontsize{7.0pt}{8.4pt}\selectfont}] at (6.491,5.721) {\functionId{23}\functionName{gen_peephole2_1271}};
\node[gp node left,rotate=-90,font={\fontsize{7.0pt}{8.4pt}\selectfont}] at (6.721,5.721) {\functionId{24}\functionName{gen_peephole2_1277}};
\node[gp node left,rotate=-90,font={\fontsize{7.0pt}{8.4pt}\selectfont}] at (6.951,5.721) {\functionId{25}\functionName{gen_pfnacc}};
\node[gp node left,rotate=-90,font={\fontsize{7.0pt}{8.4pt}\selectfont}] at (7.181,5.721) {\functionId{26}\functionName{gen_rotlsi3}};
\node[gp node left,rotate=-90,font={\fontsize{7.0pt}{8.4pt}\selectfont}] at (7.411,5.721) {\functionId{27}\functionName{gen_split_1001}};
\node[gp node left,rotate=-90,font={\fontsize{7.0pt}{8.4pt}\selectfont}] at (7.642,5.721) {\functionId{28}\functionName{gen_split_1028}};
\node[gp node left,rotate=-90,font={\fontsize{7.0pt}{8.4pt}\selectfont}] at (7.872,5.721) {\functionId{29}\functionName{gen_sse_nandti3}};
\node[gp node left,rotate=-90,font={\fontsize{7.0pt}{8.4pt}\selectfont}] at (8.102,5.721) {\functionId{30}\functionName{gen_sunge}};
\node[gp node left,rotate=-90,font={\fontsize{7.0pt}{8.4pt}\selectfont}] at (8.332,5.721) {\functionId{31}\functionName{insert_loop_mem}};
\node[gp node left,rotate=-90,font={\fontsize{7.0pt}{8.4pt}\selectfont}] at (8.562,5.721) {\functionId{32}\functionName{eiremain}};
\node[gp node left,rotate=-90,font={\fontsize{7.0pt}{8.4pt}\selectfont}] at (8.793,5.721) {\functionId{33}\functionName{elimination_effects}};
\node[gp node left,rotate=-90,font={\fontsize{7.0pt}{8.4pt}\selectfont}] at (9.023,5.721) {\functionId{34}\functionName{gen_reload}};
\node[gp node left,rotate=-90,font={\fontsize{7.0pt}{8.4pt}\selectfont}] at (9.253,5.721) {\functionId{35}\functionName{reload_cse_simplify_set}};
\node[gp node left,rotate=-90,font={\fontsize{7.0pt}{8.4pt}\selectfont}] at (9.483,5.721) {\functionId{36}\functionName{simplify_binary_is2orm1}};
\node[gp node left,rotate=-90,font={\fontsize{7.0pt}{8.4pt}\selectfont}] at (9.713,5.721) {\functionId{37}\functionName{remove_phi_alternative}};
\node[gp node left,rotate=-90,font={\fontsize{7.0pt}{8.4pt}\selectfont}] at (9.944,5.721) {\functionId{38}\functionName{contains_placeholder_p}};
\node[gp node left,rotate=-90,font={\fontsize{7.0pt}{8.4pt}\selectfont}] at (10.174,5.721) {\functionId{39}\functionName{assemble_end_function}};
\node[gp node left,rotate=-90,font={\fontsize{7.0pt}{8.4pt}\selectfont}] at (10.404,5.721) {\functionId{40}\functionName{default_named_section_asm.}};
\node[gp node left,rotate=-90,font={\fontsize{7.0pt}{8.4pt}\selectfont}] at (10.634,5.721) {\functionId{41}\functionName{sample_unpack_12}};
\node[gp node left,rotate=-90,font={\fontsize{7.0pt}{8.4pt}\selectfont}] at (10.864,5.721) {\functionId{42}\functionName{autohelperattpat10}};
\node[gp node left,rotate=-90,font={\fontsize{7.0pt}{8.4pt}\selectfont}] at (11.095,5.721) {\functionId{43}\functionName{autohelperbarrierspat126}};
\node[gp node left,rotate=-90,font={\fontsize{7.0pt}{8.4pt}\selectfont}] at (11.325,5.721) {\functionId{44}\functionName{atari_atari_attack_callba.}};
\node[gp node left,rotate=-90,font={\fontsize{7.0pt}{8.4pt}\selectfont}] at (11.555,5.721) {\functionId{45}\functionName{compute_aa_status}};
\node[gp node left,rotate=-90,font={\fontsize{7.0pt}{8.4pt}\selectfont}] at (11.785,5.721) {\functionId{46}\functionName{dragon_weak}};
\node[gp node left,rotate=-90,font={\fontsize{7.0pt}{8.4pt}\selectfont}] at (12.015,5.721) {\functionId{47}\functionName{get_saved_worms}};
\node[gp node left,rotate=-90,font={\fontsize{7.0pt}{8.4pt}\selectfont}] at (12.246,5.721) {\functionId{48}\functionName{read_eye}};
\node[gp node left,rotate=-90,font={\fontsize{7.0pt}{8.4pt}\selectfont}] at (12.476,5.721) {\functionId{49}\functionName{topological_eye}};
\node[gp node left,rotate=-90,font={\fontsize{7.0pt}{8.4pt}\selectfont}] at (12.706,5.721) {\functionId{50}\functionName{autohelperowl_attackpat19.}};
\node[gp node left,rotate=-90,font={\fontsize{7.0pt}{8.4pt}\selectfont}] at (12.936,5.721) {\functionId{51}\functionName{autohelperowl_attackpat29.}};
\node[gp node left,rotate=-90,font={\fontsize{7.0pt}{8.4pt}\selectfont}] at (13.166,5.721) {\functionId{52}\functionName{autohelperowl_defendpat28.}};
\node[gp node left,rotate=-90,font={\fontsize{7.0pt}{8.4pt}\selectfont}] at (13.396,5.721) {\functionId{53}\functionName{autohelperowl_defendpat38.}};
\node[gp node left,rotate=-90,font={\fontsize{7.0pt}{8.4pt}\selectfont}] at (13.627,5.721) {\functionId{54}\functionName{autohelperpat1114}};
\node[gp node left,rotate=-90,font={\fontsize{7.0pt}{8.4pt}\selectfont}] at (13.857,5.721) {\functionId{55}\functionName{autohelperpat335}};
\node[gp node left,rotate=-90,font={\fontsize{7.0pt}{8.4pt}\selectfont}] at (14.087,5.721) {\functionId{56}\functionName{autohelperpat508}};
\node[gp node left,rotate=-90,font={\fontsize{7.0pt}{8.4pt}\selectfont}] at (14.317,5.721) {\functionId{57}\functionName{autohelperpat83}};
\node[gp node left,rotate=-90,font={\fontsize{7.0pt}{8.4pt}\selectfont}] at (14.547,5.721) {\functionId{58}\functionName{simple_showboard}};
\node[gp node left,rotate=-90,font={\fontsize{7.0pt}{8.4pt}\selectfont}] at (14.778,5.721) {\functionId{59}\functionName{skip_intrabk_SAD}};
\node[gp node left,rotate=-90,font={\fontsize{7.0pt}{8.4pt}\selectfont}] at (15.008,5.721) {\functionId{60}\functionName{free_orig_planes}};
\node[gp node left,rotate=-90,font={\fontsize{7.0pt}{8.4pt}\selectfont}] at (15.238,5.721) {\functionId{61}\functionName{GetSkipCostMB}};
\node[gp node left,rotate=-90,font={\fontsize{7.0pt}{8.4pt}\selectfont}] at (15.468,5.721) {\functionId{62}\functionName{writeSyntaxElement_Level_.}};
\node[gp node left,rotate=-90,font={\fontsize{7.0pt}{8.4pt}\selectfont}] at (15.698,5.721) {\functionId{63}\functionName{GSIAddKeyToIndex}};
\node[gp node left,rotate=-90,font={\fontsize{7.0pt}{8.4pt}\selectfont}] at (15.929,5.721) {\functionId{64}\functionName{EVDBasicFit}};
\node[gp node left,rotate=-90,font={\fontsize{7.0pt}{8.4pt}\selectfont}] at (16.159,5.721) {\functionId{65}\functionName{SampleDirichlet}};
\node[gp node left,rotate=-90,font={\fontsize{7.0pt}{8.4pt}\selectfont}] at (16.389,5.721) {\functionId{66}\functionName{DegenerateSymbolScore}};
\node[gp node left,rotate=-90,font={\fontsize{7.0pt}{8.4pt}\selectfont}] at (16.619,5.721) {\functionId{67}\functionName{Plan7SetCtime}};
\node[gp node left,rotate=-90,font={\fontsize{7.0pt}{8.4pt}\selectfont}] at (16.849,5.721) {\functionId{68}\functionName{MSAToSqinfo}};
\node[gp node left,rotate=-90,font={\fontsize{7.0pt}{8.4pt}\selectfont}] at (17.080,5.721) {\functionId{69}\functionName{null_convert}};
\node[gp node left,rotate=-90,font={\fontsize{7.0pt}{8.4pt}\selectfont}] at (17.310,5.721) {\functionId{70}\functionName{jinit_c_prep_controller}};
\node[gp node left,rotate=-90,font={\fontsize{7.0pt}{8.4pt}\selectfont}] at (17.540,5.721) {\functionId{71}\functionName{glFogf}};
\node[gp node left,rotate=-90,font={\fontsize{7.0pt}{8.4pt}\selectfont}] at (17.770,5.721) {\functionId{72}\functionName{glNormal3d}};
\node[gp node left,rotate=-90,font={\fontsize{7.0pt}{8.4pt}\selectfont}] at (18.000,5.721) {\functionId{73}\functionName{glRasterPos3d}};
\node[gp node left,rotate=-90,font={\fontsize{7.0pt}{8.4pt}\selectfont}] at (18.231,5.721) {\functionId{74}\functionName{glTexCoord2d}};
\node[gp node left,rotate=-90,font={\fontsize{7.0pt}{8.4pt}\selectfont}] at (18.461,5.721) {\functionId{75}\functionName{gl_stippled_bresenham}};
\node[gp node left,rotate=-90,font={\fontsize{7.0pt}{8.4pt}\selectfont}] at (18.691,5.721) {\functionId{76}\functionName{gl_save_Frustum}};
\node[gp node left,rotate=-90,font={\fontsize{7.0pt}{8.4pt}\selectfont}] at (18.921,5.721) {\functionId{77}\functionName{gl_save_LineWidth}};
\node[gp node left,rotate=-90,font={\fontsize{7.0pt}{8.4pt}\selectfont}] at (19.151,5.721) {\functionId{78}\functionName{translate_id}};
\node[gp node left,rotate=-90,font={\fontsize{7.0pt}{8.4pt}\selectfont}] at (19.382,5.721) {\functionId{79}\functionName{gl_Map1f}};
\node[gp node left,rotate=-90,font={\fontsize{7.0pt}{8.4pt}\selectfont}] at (19.612,5.721) {\functionId{80}\functionName{smooth_ci_line}};
\node[gp node left,rotate=-90,font={\fontsize{7.0pt}{8.4pt}\selectfont}] at (19.842,5.721) {\functionId{81}\functionName{free_unified_knots}};
\node[gp node left,rotate=-90,font={\fontsize{7.0pt}{8.4pt}\selectfont}] at (20.072,5.721) {\functionId{82}\functionName{tess_test_polygon}};
\node[gp node left,rotate=-90,font={\fontsize{7.0pt}{8.4pt}\selectfont}] at (20.302,5.721) {\functionId{83}\functionName{auxWireBox}};
\node[gp node left,rotate=-90,font={\fontsize{7.0pt}{8.4pt}\selectfont}] at (20.533,5.721) {\functionId{84}\functionName{gl_ColorPointer}};
\node[gp node left,rotate=-90,font={\fontsize{7.0pt}{8.4pt}\selectfont}] at (20.763,5.721) {\functionId{85}\functionName{r_serial}};
\node[gp node left,rotate=-90,font={\fontsize{7.0pt}{8.4pt}\selectfont}] at (20.993,5.721) {\functionId{86}\functionName{scalar_mult_sub_su3_matri.}};
\node[gp node left,rotate=-90,font={\fontsize{7.0pt}{8.4pt}\selectfont}] at (21.223,5.721) {\functionId{87}\functionName{Decode_MPEG1_Non_Intra_Bl.}};
\node[gp node left,rotate=-90,font={\fontsize{7.0pt}{8.4pt}\selectfont}] at (21.453,5.721) {\functionId{88}\functionName{cpDecodeSecret}};
\node[gp node left,rotate=-90,font={\fontsize{7.0pt}{8.4pt}\selectfont}] at (21.684,5.721) {\functionId{89}\functionName{vlShortLshift}};
\node[gp node left,rotate=-90,font={\fontsize{7.0pt}{8.4pt}\selectfont}] at (21.914,5.721) {\functionId{90}\functionName{encryptfile}};
\node[gp node left,rotate=-90,font={\fontsize{7.0pt}{8.4pt}\selectfont}] at (22.144,5.721) {\functionId{91}\functionName{make_canonical}};
\node[gp node left,rotate=-90,font={\fontsize{7.0pt}{8.4pt}\selectfont}] at (22.374,5.721) {\functionId{92}\functionName{LANG}};
\node[gp node left,rotate=-90,font={\fontsize{7.0pt}{8.4pt}\selectfont}] at (22.604,5.721) {\functionId{93}\functionName{MD5Transform}};
\node[gp node left,rotate=-90,font={\fontsize{7.0pt}{8.4pt}\selectfont}] at (22.835,5.721) {\functionId{94}\functionName{mp_display}};
\node[gp node left,rotate=-90,font={\fontsize{7.0pt}{8.4pt}\selectfont}] at (23.065,5.721) {\functionId{95}\functionName{comp_Jboundaries}};
\node[gp node left,rotate=-90,font={\fontsize{7.0pt}{8.4pt}\selectfont}] at (23.295,5.721) {\functionId{96}\functionName{is_draw}};
\node[gp node left,rotate=-90,font={\fontsize{7.0pt}{8.4pt}\selectfont}] at (23.525,5.721) {\functionId{97}\functionName{push_king}};
\node[gp node left,rotate=-90,font={\fontsize{7.0pt}{8.4pt}\selectfont}] at (23.755,5.721) {\functionId{98}\functionName{stat_retry}};
\node[gp node left,rotate=-90,font={\fontsize{7.0pt}{8.4pt}\selectfont}] at (23.986,5.721) {\functionId{99}\functionName{lextree_subtree_print}};
\node[gp node left,rotate=-90,font={\fontsize{7.0pt}{8.4pt}\selectfont}] at (24.216,5.721) {\functionId{100}\functionName{lm_tg_score}};
\draw[gp path] (1.196,10.131)--(1.196,5.644)--(24.446,5.644)--(24.446,10.131)--cycle;
\gpcolor{rgb color={0.580,0.000,0.827}}
\draw[gp path] (1.196,5.644)--(1.431,5.644)--(1.666,5.644)--(1.901,5.644)--(2.135,5.644)%
  --(2.370,5.644)--(2.605,5.644)--(2.840,5.644)--(3.075,5.644)--(3.310,5.644)--(3.544,5.644)%
  --(3.779,5.644)--(4.014,5.644)--(4.249,5.644)--(4.484,5.644)--(4.719,5.644)--(4.954,5.644)%
  --(5.188,5.644)--(5.423,5.644)--(5.658,5.644)--(5.893,5.644)--(6.128,5.644)--(6.363,5.644)%
  --(6.598,5.644)--(6.832,5.644)--(7.067,5.644)--(7.302,5.644)--(7.537,5.644)--(7.772,5.644)%
  --(8.007,5.644)--(8.241,5.644)--(8.476,5.644)--(8.711,5.644)--(8.946,5.644)--(9.181,5.644)%
  --(9.416,5.644)--(9.651,5.644)--(9.885,5.644)--(10.120,5.644)--(10.355,5.644)--(10.590,5.644)%
  --(10.825,5.644)--(11.060,5.644)--(11.294,5.644)--(11.529,5.644)--(11.764,5.644)--(11.999,5.644)%
  --(12.234,5.644)--(12.469,5.644)--(12.704,5.644)--(12.938,5.644)--(13.173,5.644)--(13.408,5.644)%
  --(13.643,5.644)--(13.878,5.644)--(14.113,5.644)--(14.348,5.644)--(14.582,5.644)--(14.817,5.644)%
  --(15.052,5.644)--(15.287,5.644)--(15.522,5.644)--(15.757,5.644)--(15.991,5.644)--(16.226,5.644)%
  --(16.461,5.644)--(16.696,5.644)--(16.931,5.644)--(17.166,5.644)--(17.401,5.644)--(17.635,5.644)%
  --(17.870,5.644)--(18.105,5.644)--(18.340,5.644)--(18.575,5.644)--(18.810,5.644)--(19.044,5.644)%
  --(19.279,5.644)--(19.514,5.644)--(19.749,5.644)--(19.984,5.644)--(20.219,5.644)--(20.454,5.644)%
  --(20.688,5.644)--(20.923,5.644)--(21.158,5.644)--(21.393,5.644)--(21.628,5.644)--(21.863,5.644)%
  --(22.098,5.644)--(22.332,5.644)--(22.567,5.644)--(22.802,5.644)--(23.037,5.644)--(23.272,5.644)%
  --(23.507,5.644)--(23.741,5.644)--(23.976,5.644)--(24.211,5.644)--(24.446,5.644);
\gpfill{rgb color={0.000,0.000,0.000}} (1.369,5.644)--(1.485,5.644)--(1.485,5.990)--(1.369,5.990)--cycle;
\gpcolor{rgb color={0.000,0.000,0.000}}
\draw[gp path] (1.369,5.644)--(1.369,5.989)--(1.484,5.989)--(1.484,5.644)--cycle;
\gpfill{rgb color={0.000,0.000,0.000}} (1.599,5.644)--(1.715,5.644)--(1.715,6.202)--(1.599,6.202)--cycle;
\draw[gp path] (1.599,5.644)--(1.599,6.201)--(1.714,6.201)--(1.714,5.644)--cycle;
\gpfill{rgb color={0.000,0.000,0.000}} (1.829,5.644)--(1.945,5.644)--(1.945,6.060)--(1.829,6.060)--cycle;
\draw[gp path] (1.829,5.644)--(1.829,6.059)--(1.944,6.059)--(1.944,5.644)--cycle;
\gpfill{rgb color={0.000,0.000,0.000}} (2.059,5.644)--(2.175,5.644)--(2.175,6.068)--(2.059,6.068)--cycle;
\draw[gp path] (2.059,5.644)--(2.059,6.067)--(2.174,6.067)--(2.174,5.644)--cycle;
\gpfill{rgb color={0.000,0.000,0.000}} (2.289,5.644)--(2.406,5.644)--(2.406,6.053)--(2.289,6.053)--cycle;
\draw[gp path] (2.289,5.644)--(2.289,6.052)--(2.405,6.052)--(2.405,5.644)--cycle;
\gpfill{rgb color={0.000,0.000,0.000}} (2.520,5.644)--(2.636,5.644)--(2.636,5.756)--(2.520,5.756)--cycle;
\draw[gp path] (2.520,5.644)--(2.520,5.755)--(2.635,5.755)--(2.635,5.644)--cycle;
\gpfill{rgb color={0.000,0.000,0.000}} (2.980,5.644)--(3.096,5.644)--(3.096,5.967)--(2.980,5.967)--cycle;
\draw[gp path] (2.980,5.644)--(2.980,5.966)--(3.095,5.966)--(3.095,5.644)--cycle;
\gpfill{rgb color={0.000,0.000,0.000}} (3.671,5.644)--(3.787,5.644)--(3.787,6.343)--(3.671,6.343)--cycle;
\draw[gp path] (3.671,5.644)--(3.671,6.342)--(3.786,6.342)--(3.786,5.644)--cycle;
\gpfill{rgb color={0.000,0.000,0.000}} (3.901,5.644)--(4.017,5.644)--(4.017,5.728)--(3.901,5.728)--cycle;
\draw[gp path] (3.901,5.644)--(3.901,5.727)--(4.016,5.727)--(4.016,5.644)--cycle;
\gpfill{rgb color={0.000,0.000,0.000}} (4.131,5.644)--(4.247,5.644)--(4.247,6.233)--(4.131,6.233)--cycle;
\draw[gp path] (4.131,5.644)--(4.131,6.232)--(4.246,6.232)--(4.246,5.644)--cycle;
\gpfill{rgb color={0.000,0.000,0.000}} (4.361,5.644)--(4.477,5.644)--(4.477,6.387)--(4.361,6.387)--cycle;
\draw[gp path] (4.361,5.644)--(4.361,6.386)--(4.476,6.386)--(4.476,5.644)--cycle;
\gpfill{rgb color={0.000,0.000,0.000}} (4.591,5.644)--(4.708,5.644)--(4.708,5.834)--(4.591,5.834)--cycle;
\draw[gp path] (4.591,5.644)--(4.591,5.833)--(4.707,5.833)--(4.707,5.644)--cycle;
\gpfill{rgb color={0.000,0.000,0.000}} (4.822,5.644)--(4.938,5.644)--(4.938,6.165)--(4.822,6.165)--cycle;
\draw[gp path] (4.822,5.644)--(4.822,6.164)--(4.937,6.164)--(4.937,5.644)--cycle;
\gpfill{rgb color={0.000,0.000,0.000}} (5.052,5.644)--(5.168,5.644)--(5.168,5.875)--(5.052,5.875)--cycle;
\draw[gp path] (5.052,5.644)--(5.052,5.874)--(5.167,5.874)--(5.167,5.644)--cycle;
\gpfill{rgb color={0.000,0.000,0.000}} (5.512,5.644)--(5.628,5.644)--(5.628,6.942)--(5.512,6.942)--cycle;
\draw[gp path] (5.512,5.644)--(5.512,6.941)--(5.627,6.941)--(5.627,5.644)--cycle;
\gpfill{rgb color={0.000,0.000,0.000}} (6.433,5.644)--(6.549,5.644)--(6.549,5.894)--(6.433,5.894)--cycle;
\draw[gp path] (6.433,5.644)--(6.433,5.893)--(6.548,5.893)--(6.548,5.644)--cycle;
\gpfill{rgb color={0.000,0.000,0.000}} (7.124,5.644)--(7.240,5.644)--(7.240,6.927)--(7.124,6.927)--cycle;
\draw[gp path] (7.124,5.644)--(7.124,6.926)--(7.239,6.926)--(7.239,5.644)--cycle;
\gpfill{rgb color={0.000,0.000,0.000}} (8.275,5.644)--(8.391,5.644)--(8.391,6.283)--(8.275,6.283)--cycle;
\draw[gp path] (8.275,5.644)--(8.275,6.282)--(8.390,6.282)--(8.390,5.644)--cycle;
\gpfill{rgb color={0.000,0.000,0.000}} (8.965,5.644)--(9.081,5.644)--(9.081,6.215)--(8.965,6.215)--cycle;
\draw[gp path] (8.965,5.644)--(8.965,6.214)--(9.080,6.214)--(9.080,5.644)--cycle;
\gpfill{rgb color={0.000,0.000,0.000}} (9.195,5.644)--(9.311,5.644)--(9.311,6.493)--(9.195,6.493)--cycle;
\draw[gp path] (9.195,5.644)--(9.195,6.492)--(9.310,6.492)--(9.310,5.644)--cycle;
\gpfill{rgb color={0.000,0.000,0.000}} (9.426,5.644)--(9.542,5.644)--(9.542,5.800)--(9.426,5.800)--cycle;
\draw[gp path] (9.426,5.644)--(9.426,5.799)--(9.541,5.799)--(9.541,5.644)--cycle;
\gpfill{rgb color={0.000,0.000,0.000}} (9.886,5.644)--(10.002,5.644)--(10.002,5.738)--(9.886,5.738)--cycle;
\draw[gp path] (9.886,5.644)--(9.886,5.737)--(10.001,5.737)--(10.001,5.644)--cycle;
\gpfill{rgb color={0.000,0.000,0.000}} (10.116,5.644)--(10.232,5.644)--(10.232,6.364)--(10.116,6.364)--cycle;
\draw[gp path] (10.116,5.644)--(10.116,6.363)--(10.231,6.363)--(10.231,5.644)--cycle;
\gpfill{rgb color={0.000,0.000,0.000}} (10.346,5.644)--(10.462,5.644)--(10.462,6.227)--(10.346,6.227)--cycle;
\draw[gp path] (10.346,5.644)--(10.346,6.226)--(10.461,6.226)--(10.461,5.644)--cycle;
\gpfill{rgb color={0.000,0.000,0.000}} (10.577,5.644)--(10.693,5.644)--(10.693,6.158)--(10.577,6.158)--cycle;
\draw[gp path] (10.577,5.644)--(10.577,6.157)--(10.692,6.157)--(10.692,5.644)--cycle;
\gpfill{rgb color={0.000,0.000,0.000}} (10.807,5.644)--(10.923,5.644)--(10.923,7.716)--(10.807,7.716)--cycle;
\draw[gp path] (10.807,5.644)--(10.807,7.715)--(10.922,7.715)--(10.922,5.644)--cycle;
\gpfill{rgb color={0.000,0.000,0.000}} (11.497,5.644)--(11.613,5.644)--(11.613,5.689)--(11.497,5.689)--cycle;
\draw[gp path] (11.497,5.644)--(11.497,5.688)--(11.612,5.688)--(11.612,5.644)--cycle;
\gpfill{rgb color={0.000,0.000,0.000}} (11.728,5.644)--(11.844,5.644)--(11.844,6.053)--(11.728,6.053)--cycle;
\draw[gp path] (11.728,5.644)--(11.728,6.052)--(11.843,6.052)--(11.843,5.644)--cycle;
\gpfill{rgb color={0.000,0.000,0.000}} (11.958,5.644)--(12.074,5.644)--(12.074,5.671)--(11.958,5.671)--cycle;
\draw[gp path] (11.958,5.644)--(11.958,5.670)--(12.073,5.670)--(12.073,5.644)--cycle;
\gpfill{rgb color={0.000,0.000,0.000}} (12.648,5.644)--(12.764,5.644)--(12.764,6.599)--(12.648,6.599)--cycle;
\draw[gp path] (12.648,5.644)--(12.648,6.598)--(12.763,6.598)--(12.763,5.644)--cycle;
\gpfill{rgb color={0.000,0.000,0.000}} (12.879,5.644)--(12.995,5.644)--(12.995,7.141)--(12.879,7.141)--cycle;
\draw[gp path] (12.879,5.644)--(12.879,7.140)--(12.994,7.140)--(12.994,5.644)--cycle;
\gpfill{rgb color={0.000,0.000,0.000}} (13.109,5.644)--(13.225,5.644)--(13.225,6.767)--(13.109,6.767)--cycle;
\draw[gp path] (13.109,5.644)--(13.109,6.766)--(13.224,6.766)--(13.224,5.644)--cycle;
\gpfill{rgb color={0.000,0.000,0.000}} (13.339,5.644)--(13.455,5.644)--(13.455,6.348)--(13.339,6.348)--cycle;
\draw[gp path] (13.339,5.644)--(13.339,6.347)--(13.454,6.347)--(13.454,5.644)--cycle;
\gpfill{rgb color={0.000,0.000,0.000}} (13.569,5.644)--(13.685,5.644)--(13.685,7.165)--(13.569,7.165)--cycle;
\draw[gp path] (13.569,5.644)--(13.569,7.164)--(13.684,7.164)--(13.684,5.644)--cycle;
\gpfill{rgb color={0.000,0.000,0.000}} (13.799,5.644)--(13.915,5.644)--(13.915,6.393)--(13.799,6.393)--cycle;
\draw[gp path] (13.799,5.644)--(13.799,6.392)--(13.914,6.392)--(13.914,5.644)--cycle;
\gpfill{rgb color={0.000,0.000,0.000}} (14.030,5.644)--(14.146,5.644)--(14.146,6.393)--(14.030,6.393)--cycle;
\draw[gp path] (14.030,5.644)--(14.030,6.392)--(14.145,6.392)--(14.145,5.644)--cycle;
\gpfill{rgb color={0.000,0.000,0.000}} (14.260,5.644)--(14.376,5.644)--(14.376,6.498)--(14.260,6.498)--cycle;
\draw[gp path] (14.260,5.644)--(14.260,6.497)--(14.375,6.497)--(14.375,5.644)--cycle;
\gpfill{rgb color={0.000,0.000,0.000}} (14.490,5.644)--(14.606,5.644)--(14.606,6.324)--(14.490,6.324)--cycle;
\draw[gp path] (14.490,5.644)--(14.490,6.323)--(14.605,6.323)--(14.605,5.644)--cycle;
\gpfill{rgb color={0.000,0.000,0.000}} (14.950,5.644)--(15.066,5.644)--(15.066,6.247)--(14.950,6.247)--cycle;
\draw[gp path] (14.950,5.644)--(14.950,6.246)--(15.065,6.246)--(15.065,5.644)--cycle;
\gpfill{rgb color={0.000,0.000,0.000}} (15.641,5.644)--(15.757,5.644)--(15.757,5.923)--(15.641,5.923)--cycle;
\draw[gp path] (15.641,5.644)--(15.641,5.922)--(15.756,5.922)--(15.756,5.644)--cycle;
\gpfill{rgb color={0.000,0.000,0.000}} (15.871,5.644)--(15.987,5.644)--(15.987,6.406)--(15.871,6.406)--cycle;
\draw[gp path] (15.871,5.644)--(15.871,6.405)--(15.986,6.405)--(15.986,5.644)--cycle;
\gpfill{rgb color={0.000,0.000,0.000}} (16.332,5.644)--(16.448,5.644)--(16.448,6.245)--(16.332,6.245)--cycle;
\draw[gp path] (16.332,5.644)--(16.332,6.244)--(16.447,6.244)--(16.447,5.644)--cycle;
\gpfill{rgb color={0.000,0.000,0.000}} (16.792,5.644)--(16.908,5.644)--(16.908,5.648)--(16.792,5.648)--cycle;
\draw[gp path] (16.792,5.644)--(16.792,5.647)--(16.907,5.647)--(16.907,5.644)--cycle;
\gpfill{rgb color={0.000,0.000,0.000}} (17.483,5.644)--(17.599,5.644)--(17.599,6.716)--(17.483,6.716)--cycle;
\draw[gp path] (17.483,5.644)--(17.483,6.715)--(17.598,6.715)--(17.598,5.644)--cycle;
\gpfill{rgb color={0.000,0.000,0.000}} (17.713,5.644)--(17.829,5.644)--(17.829,9.928)--(17.713,9.928)--cycle;
\draw[gp path] (17.713,5.644)--(17.713,9.927)--(17.828,9.927)--(17.828,5.644)--cycle;
\gpfill{rgb color={0.000,0.000,0.000}} (17.943,5.644)--(18.059,5.644)--(18.059,7.949)--(17.943,7.949)--cycle;
\draw[gp path] (17.943,5.644)--(17.943,7.948)--(18.058,7.948)--(18.058,5.644)--cycle;
\gpfill{rgb color={0.000,0.000,0.000}} (18.173,5.644)--(18.289,5.644)--(18.289,6.286)--(18.173,6.286)--cycle;
\draw[gp path] (18.173,5.644)--(18.173,6.285)--(18.288,6.285)--(18.288,5.644)--cycle;
\gpfill{rgb color={0.000,0.000,0.000}} (18.403,5.644)--(18.519,5.644)--(18.519,6.552)--(18.403,6.552)--cycle;
\draw[gp path] (18.403,5.644)--(18.403,6.551)--(18.518,6.551)--(18.518,5.644)--cycle;
\gpfill{rgb color={0.000,0.000,0.000}} (18.634,5.644)--(18.750,5.644)--(18.750,5.757)--(18.634,5.757)--cycle;
\draw[gp path] (18.634,5.644)--(18.634,5.756)--(18.749,5.756)--(18.749,5.644)--cycle;
\gpfill{rgb color={0.000,0.000,0.000}} (18.864,5.644)--(18.980,5.644)--(18.980,5.818)--(18.864,5.818)--cycle;
\draw[gp path] (18.864,5.644)--(18.864,5.817)--(18.979,5.817)--(18.979,5.644)--cycle;
\gpfill{rgb color={0.000,0.000,0.000}} (19.094,5.644)--(19.210,5.644)--(19.210,5.698)--(19.094,5.698)--cycle;
\draw[gp path] (19.094,5.644)--(19.094,5.697)--(19.209,5.697)--(19.209,5.644)--cycle;
\gpfill{rgb color={0.000,0.000,0.000}} (19.324,5.644)--(19.440,5.644)--(19.440,7.266)--(19.324,7.266)--cycle;
\draw[gp path] (19.324,5.644)--(19.324,7.265)--(19.439,7.265)--(19.439,5.644)--cycle;
\gpfill{rgb color={0.000,0.000,0.000}} (19.554,5.644)--(19.670,5.644)--(19.670,7.516)--(19.554,7.516)--cycle;
\draw[gp path] (19.554,5.644)--(19.554,7.515)--(19.669,7.515)--(19.669,5.644)--cycle;
\gpfill{rgb color={0.000,0.000,0.000}} (19.784,5.644)--(19.901,5.644)--(19.901,6.376)--(19.784,6.376)--cycle;
\draw[gp path] (19.784,5.644)--(19.784,6.375)--(19.900,6.375)--(19.900,5.644)--cycle;
\gpfill{rgb color={0.000,0.000,0.000}} (20.475,5.644)--(20.591,5.644)--(20.591,6.363)--(20.475,6.363)--cycle;
\draw[gp path] (20.475,5.644)--(20.475,6.362)--(20.590,6.362)--(20.590,5.644)--cycle;
\gpfill{rgb color={0.000,0.000,0.000}} (21.626,5.644)--(21.742,5.644)--(21.742,6.394)--(21.626,6.394)--cycle;
\draw[gp path] (21.626,5.644)--(21.626,6.393)--(21.741,6.393)--(21.741,5.644)--cycle;
\gpfill{rgb color={0.000,0.000,0.000}} (21.856,5.644)--(21.972,5.644)--(21.972,7.388)--(21.856,7.388)--cycle;
\draw[gp path] (21.856,5.644)--(21.856,7.387)--(21.971,7.387)--(21.971,5.644)--cycle;
\gpfill{rgb color={0.000,0.000,0.000}} (22.086,5.644)--(22.203,5.644)--(22.203,6.620)--(22.086,6.620)--cycle;
\draw[gp path] (22.086,5.644)--(22.086,6.619)--(22.202,6.619)--(22.202,5.644)--cycle;
\gpfill{rgb color={0.000,0.000,0.000}} (22.317,5.644)--(22.433,5.644)--(22.433,6.089)--(22.317,6.089)--cycle;
\draw[gp path] (22.317,5.644)--(22.317,6.088)--(22.432,6.088)--(22.432,5.644)--cycle;
\gpfill{rgb color={0.000,0.000,0.000}} (22.547,5.644)--(22.663,5.644)--(22.663,6.845)--(22.547,6.845)--cycle;
\draw[gp path] (22.547,5.644)--(22.547,6.844)--(22.662,6.844)--(22.662,5.644)--cycle;
\gpfill{rgb color={0.000,0.000,0.000}} (23.007,5.644)--(23.123,5.644)--(23.123,6.259)--(23.007,6.259)--cycle;
\draw[gp path] (23.007,5.644)--(23.007,6.258)--(23.122,6.258)--(23.122,5.644)--cycle;
\gpfill{rgb color={0.000,0.000,0.000}} (23.237,5.644)--(23.354,5.644)--(23.354,5.956)--(23.237,5.956)--cycle;
\draw[gp path] (23.237,5.644)--(23.237,5.955)--(23.353,5.955)--(23.353,5.644)--cycle;
\gpfill{rgb color={0.000,0.000,0.000}} (23.698,5.644)--(23.814,5.644)--(23.814,6.864)--(23.698,6.864)--cycle;
\draw[gp path] (23.698,5.644)--(23.698,6.863)--(23.813,6.863)--(23.813,5.644)--cycle;
\gpfill{rgb color={0.000,0.000,0.000}} (24.158,5.644)--(24.274,5.644)--(24.274,6.028)--(24.158,6.028)--cycle;
\draw[gp path] (24.158,5.644)--(24.158,6.027)--(24.273,6.027)--(24.273,5.644)--cycle;
\gpfill{rgb color={0.800,0.800,0.800}} (1.369,5.989)--(1.485,5.989)--(1.485,5.990)--(1.369,5.990)--cycle;
\gpcolor{rgb color={0.800,0.800,0.800}}
\draw[gp path] (1.369,5.989)--(1.484,5.989)--cycle;
\gpfill{rgb color={0.800,0.800,0.800}} (1.599,6.201)--(1.715,6.201)--(1.715,6.202)--(1.599,6.202)--cycle;
\draw[gp path] (1.599,6.201)--(1.714,6.201)--cycle;
\gpfill{rgb color={0.800,0.800,0.800}} (1.829,6.059)--(1.945,6.059)--(1.945,6.060)--(1.829,6.060)--cycle;
\draw[gp path] (1.829,6.059)--(1.944,6.059)--cycle;
\gpfill{rgb color={0.800,0.800,0.800}} (2.059,6.067)--(2.175,6.067)--(2.175,6.068)--(2.059,6.068)--cycle;
\draw[gp path] (2.059,6.067)--(2.174,6.067)--cycle;
\gpfill{rgb color={0.800,0.800,0.800}} (2.289,6.052)--(2.406,6.052)--(2.406,6.053)--(2.289,6.053)--cycle;
\draw[gp path] (2.289,6.052)--(2.405,6.052)--cycle;
\gpfill{rgb color={0.800,0.800,0.800}} (2.520,5.755)--(2.636,5.755)--(2.636,5.756)--(2.520,5.756)--cycle;
\draw[gp path] (2.520,5.755)--(2.635,5.755)--cycle;
\gpfill{rgb color={0.800,0.800,0.800}} (2.750,5.644)--(2.866,5.644)--(2.866,5.646)--(2.750,5.646)--cycle;
\draw[gp path] (2.750,5.644)--(2.750,5.645)--(2.865,5.645)--(2.865,5.644)--cycle;
\gpfill{rgb color={0.800,0.800,0.800}} (2.980,5.966)--(3.096,5.966)--(3.096,6.591)--(2.980,6.591)--cycle;
\draw[gp path] (2.980,5.966)--(2.980,6.590)--(3.095,6.590)--(3.095,5.966)--cycle;
\gpfill{rgb color={0.800,0.800,0.800}} (3.440,5.644)--(3.557,5.644)--(3.557,7.323)--(3.440,7.323)--cycle;
\draw[gp path] (3.440,5.644)--(3.440,7.322)--(3.556,7.322)--(3.556,5.644)--cycle;
\gpfill{rgb color={0.800,0.800,0.800}} (3.671,6.342)--(3.787,6.342)--(3.787,6.343)--(3.671,6.343)--cycle;
\draw[gp path] (3.671,6.342)--(3.786,6.342)--cycle;
\gpfill{rgb color={0.800,0.800,0.800}} (3.901,5.727)--(4.017,5.727)--(4.017,5.728)--(3.901,5.728)--cycle;
\draw[gp path] (3.901,5.727)--(4.016,5.727)--cycle;
\gpfill{rgb color={0.800,0.800,0.800}} (4.131,6.232)--(4.247,6.232)--(4.247,6.233)--(4.131,6.233)--cycle;
\draw[gp path] (4.131,6.232)--(4.246,6.232)--cycle;
\gpfill{rgb color={0.800,0.800,0.800}} (4.361,6.386)--(4.477,6.386)--(4.477,6.387)--(4.361,6.387)--cycle;
\draw[gp path] (4.361,6.386)--(4.476,6.386)--cycle;
\gpfill{rgb color={0.800,0.800,0.800}} (4.591,5.833)--(4.708,5.833)--(4.708,5.834)--(4.591,5.834)--cycle;
\draw[gp path] (4.591,5.833)--(4.707,5.833)--cycle;
\gpfill{rgb color={0.800,0.800,0.800}} (4.822,6.164)--(4.938,6.164)--(4.938,6.165)--(4.822,6.165)--cycle;
\draw[gp path] (4.822,6.164)--(4.937,6.164)--cycle;
\gpfill{rgb color={0.800,0.800,0.800}} (5.052,5.874)--(5.168,5.874)--(5.168,5.875)--(5.052,5.875)--cycle;
\draw[gp path] (5.052,5.874)--(5.167,5.874)--cycle;
\gpfill{rgb color={0.800,0.800,0.800}} (5.512,6.941)--(5.628,6.941)--(5.628,6.942)--(5.512,6.942)--cycle;
\draw[gp path] (5.512,6.941)--(5.627,6.941)--cycle;
\gpfill{rgb color={0.800,0.800,0.800}} (6.433,5.893)--(6.549,5.893)--(6.549,5.894)--(6.433,5.894)--cycle;
\draw[gp path] (6.433,5.893)--(6.548,5.893)--cycle;
\gpfill{rgb color={0.800,0.800,0.800}} (7.124,6.926)--(7.240,6.926)--(7.240,6.927)--(7.124,6.927)--cycle;
\draw[gp path] (7.124,6.926)--(7.239,6.926)--cycle;
\gpfill{rgb color={0.800,0.800,0.800}} (8.275,6.282)--(8.391,6.282)--(8.391,6.283)--(8.275,6.283)--cycle;
\draw[gp path] (8.275,6.282)--(8.390,6.282)--cycle;
\gpfill{rgb color={0.800,0.800,0.800}} (8.965,6.214)--(9.081,6.214)--(9.081,6.215)--(8.965,6.215)--cycle;
\draw[gp path] (8.965,6.214)--(9.080,6.214)--cycle;
\gpfill{rgb color={0.800,0.800,0.800}} (9.195,6.492)--(9.311,6.492)--(9.311,6.493)--(9.195,6.493)--cycle;
\draw[gp path] (9.195,6.492)--(9.310,6.492)--cycle;
\gpfill{rgb color={0.800,0.800,0.800}} (9.426,5.799)--(9.542,5.799)--(9.542,5.800)--(9.426,5.800)--cycle;
\draw[gp path] (9.426,5.799)--(9.541,5.799)--cycle;
\gpfill{rgb color={0.800,0.800,0.800}} (9.886,5.737)--(10.002,5.737)--(10.002,5.738)--(9.886,5.738)--cycle;
\draw[gp path] (9.886,5.737)--(10.001,5.737)--cycle;
\gpfill{rgb color={0.800,0.800,0.800}} (10.116,6.363)--(10.232,6.363)--(10.232,6.364)--(10.116,6.364)--cycle;
\draw[gp path] (10.116,6.363)--(10.231,6.363)--cycle;
\gpfill{rgb color={0.800,0.800,0.800}} (10.346,6.226)--(10.462,6.226)--(10.462,6.227)--(10.346,6.227)--cycle;
\draw[gp path] (10.346,6.226)--(10.461,6.226)--cycle;
\gpfill{rgb color={0.800,0.800,0.800}} (10.577,6.157)--(10.693,6.157)--(10.693,6.158)--(10.577,6.158)--cycle;
\draw[gp path] (10.577,6.157)--(10.692,6.157)--cycle;
\gpfill{rgb color={0.800,0.800,0.800}} (10.807,7.715)--(10.923,7.715)--(10.923,7.716)--(10.807,7.716)--cycle;
\draw[gp path] (10.807,7.715)--(10.922,7.715)--cycle;
\gpfill{rgb color={0.800,0.800,0.800}} (11.037,5.644)--(11.153,5.644)--(11.153,7.959)--(11.037,7.959)--cycle;
\draw[gp path] (11.037,5.644)--(11.037,7.958)--(11.152,7.958)--(11.152,5.644)--cycle;
\gpfill{rgb color={0.800,0.800,0.800}} (11.267,5.644)--(11.383,5.644)--(11.383,6.264)--(11.267,6.264)--cycle;
\draw[gp path] (11.267,5.644)--(11.267,6.263)--(11.382,6.263)--(11.382,5.644)--cycle;
\gpfill{rgb color={0.800,0.800,0.800}} (11.497,5.688)--(11.613,5.688)--(11.613,5.747)--(11.497,5.747)--cycle;
\draw[gp path] (11.497,5.688)--(11.497,5.746)--(11.612,5.746)--(11.612,5.688)--cycle;
\gpfill{rgb color={0.800,0.800,0.800}} (11.728,6.052)--(11.844,6.052)--(11.844,6.053)--(11.728,6.053)--cycle;
\draw[gp path] (11.728,6.052)--(11.843,6.052)--cycle;
\gpfill{rgb color={0.800,0.800,0.800}} (11.958,5.670)--(12.074,5.670)--(12.074,5.671)--(11.958,5.671)--cycle;
\draw[gp path] (11.958,5.670)--(12.073,5.670)--cycle;
\gpfill{rgb color={0.800,0.800,0.800}} (12.188,5.644)--(12.304,5.644)--(12.304,5.931)--(12.188,5.931)--cycle;
\draw[gp path] (12.188,5.644)--(12.188,5.930)--(12.303,5.930)--(12.303,5.644)--cycle;
\gpfill{rgb color={0.800,0.800,0.800}} (12.418,5.644)--(12.534,5.644)--(12.534,7.033)--(12.418,7.033)--cycle;
\draw[gp path] (12.418,5.644)--(12.418,7.032)--(12.533,7.032)--(12.533,5.644)--cycle;
\gpfill{rgb color={0.800,0.800,0.800}} (12.648,6.598)--(12.764,6.598)--(12.764,6.599)--(12.648,6.599)--cycle;
\draw[gp path] (12.648,6.598)--(12.763,6.598)--cycle;
\gpfill{rgb color={0.800,0.800,0.800}} (12.879,7.140)--(12.995,7.140)--(12.995,7.141)--(12.879,7.141)--cycle;
\draw[gp path] (12.879,7.140)--(12.994,7.140)--cycle;
\gpfill{rgb color={0.800,0.800,0.800}} (13.109,6.766)--(13.225,6.766)--(13.225,6.767)--(13.109,6.767)--cycle;
\draw[gp path] (13.109,6.766)--(13.224,6.766)--cycle;
\gpfill{rgb color={0.800,0.800,0.800}} (13.339,6.347)--(13.455,6.347)--(13.455,6.348)--(13.339,6.348)--cycle;
\draw[gp path] (13.339,6.347)--(13.454,6.347)--cycle;
\gpfill{rgb color={0.800,0.800,0.800}} (13.569,7.164)--(13.685,7.164)--(13.685,7.165)--(13.569,7.165)--cycle;
\draw[gp path] (13.569,7.164)--(13.684,7.164)--cycle;
\gpfill{rgb color={0.800,0.800,0.800}} (13.799,6.392)--(13.915,6.392)--(13.915,6.393)--(13.799,6.393)--cycle;
\draw[gp path] (13.799,6.392)--(13.914,6.392)--cycle;
\gpfill{rgb color={0.800,0.800,0.800}} (14.030,6.392)--(14.146,6.392)--(14.146,6.393)--(14.030,6.393)--cycle;
\draw[gp path] (14.030,6.392)--(14.145,6.392)--cycle;
\gpfill{rgb color={0.800,0.800,0.800}} (14.260,6.497)--(14.376,6.497)--(14.376,6.498)--(14.260,6.498)--cycle;
\draw[gp path] (14.260,6.497)--(14.375,6.497)--cycle;
\gpfill{rgb color={0.800,0.800,0.800}} (14.490,6.323)--(14.606,6.323)--(14.606,6.324)--(14.490,6.324)--cycle;
\draw[gp path] (14.490,6.323)--(14.605,6.323)--cycle;
\gpfill{rgb color={0.800,0.800,0.800}} (14.950,6.246)--(15.066,6.246)--(15.066,6.247)--(14.950,6.247)--cycle;
\draw[gp path] (14.950,6.246)--(15.065,6.246)--cycle;
\gpfill{rgb color={0.800,0.800,0.800}} (15.641,5.922)--(15.757,5.922)--(15.757,5.923)--(15.641,5.923)--cycle;
\draw[gp path] (15.641,5.922)--(15.756,5.922)--cycle;
\gpfill{rgb color={0.800,0.800,0.800}} (15.871,6.405)--(15.987,6.405)--(15.987,6.406)--(15.871,6.406)--cycle;
\draw[gp path] (15.871,6.405)--(15.986,6.405)--cycle;
\gpfill{rgb color={0.800,0.800,0.800}} (16.332,6.244)--(16.448,6.244)--(16.448,6.245)--(16.332,6.245)--cycle;
\draw[gp path] (16.332,6.244)--(16.447,6.244)--cycle;
\gpfill{rgb color={0.800,0.800,0.800}} (16.792,5.647)--(16.908,5.647)--(16.908,5.648)--(16.792,5.648)--cycle;
\draw[gp path] (16.792,5.647)--(16.907,5.647)--cycle;
\gpfill{rgb color={0.800,0.800,0.800}} (17.022,5.644)--(17.138,5.644)--(17.138,6.046)--(17.022,6.046)--cycle;
\draw[gp path] (17.022,5.644)--(17.022,6.045)--(17.137,6.045)--(17.137,5.644)--cycle;
\gpfill{rgb color={0.800,0.800,0.800}} (17.252,5.644)--(17.368,5.644)--(17.368,5.761)--(17.252,5.761)--cycle;
\draw[gp path] (17.252,5.644)--(17.252,5.760)--(17.367,5.760)--(17.367,5.644)--cycle;
\gpfill{rgb color={0.800,0.800,0.800}} (17.483,6.715)--(17.599,6.715)--(17.599,6.716)--(17.483,6.716)--cycle;
\draw[gp path] (17.483,6.715)--(17.598,6.715)--cycle;
\gpfill{rgb color={0.800,0.800,0.800}} (17.713,9.927)--(17.829,9.927)--(17.829,9.928)--(17.713,9.928)--cycle;
\draw[gp path] (17.713,9.927)--(17.828,9.927)--cycle;
\gpfill{rgb color={0.800,0.800,0.800}} (17.943,7.948)--(18.059,7.948)--(18.059,7.949)--(17.943,7.949)--cycle;
\draw[gp path] (17.943,7.948)--(18.058,7.948)--cycle;
\gpfill{rgb color={0.800,0.800,0.800}} (18.173,6.285)--(18.289,6.285)--(18.289,6.286)--(18.173,6.286)--cycle;
\draw[gp path] (18.173,6.285)--(18.288,6.285)--cycle;
\gpfill{rgb color={0.800,0.800,0.800}} (18.403,6.551)--(18.519,6.551)--(18.519,8.556)--(18.403,8.556)--cycle;
\draw[gp path] (18.403,6.551)--(18.403,8.555)--(18.518,8.555)--(18.518,6.551)--cycle;
\gpfill{rgb color={0.800,0.800,0.800}} (18.634,5.756)--(18.750,5.756)--(18.750,7.759)--(18.634,7.759)--cycle;
\draw[gp path] (18.634,5.756)--(18.634,7.758)--(18.749,7.758)--(18.749,5.756)--cycle;
\gpfill{rgb color={0.800,0.800,0.800}} (18.864,5.817)--(18.980,5.817)--(18.980,5.818)--(18.864,5.818)--cycle;
\draw[gp path] (18.864,5.817)--(18.979,5.817)--cycle;
\gpfill{rgb color={0.800,0.800,0.800}} (19.094,5.697)--(19.210,5.697)--(19.210,5.698)--(19.094,5.698)--cycle;
\draw[gp path] (19.094,5.697)--(19.209,5.697)--cycle;
\gpfill{rgb color={0.800,0.800,0.800}} (19.324,7.265)--(19.440,7.265)--(19.440,7.266)--(19.324,7.266)--cycle;
\draw[gp path] (19.324,7.265)--(19.439,7.265)--cycle;
\gpfill{rgb color={0.800,0.800,0.800}} (19.554,7.515)--(19.670,7.515)--(19.670,8.560)--(19.554,8.560)--cycle;
\draw[gp path] (19.554,7.515)--(19.554,8.559)--(19.669,8.559)--(19.669,7.515)--cycle;
\gpfill{rgb color={0.800,0.800,0.800}} (19.784,6.375)--(19.901,6.375)--(19.901,6.376)--(19.784,6.376)--cycle;
\draw[gp path] (19.784,6.375)--(19.900,6.375)--cycle;
\gpfill{rgb color={0.800,0.800,0.800}} (20.015,5.644)--(20.131,5.644)--(20.131,6.869)--(20.015,6.869)--cycle;
\draw[gp path] (20.015,5.644)--(20.015,6.868)--(20.130,6.868)--(20.130,5.644)--cycle;
\gpfill{rgb color={0.800,0.800,0.800}} (20.475,6.362)--(20.591,6.362)--(20.591,6.363)--(20.475,6.363)--cycle;
\draw[gp path] (20.475,6.362)--(20.590,6.362)--cycle;
\gpfill{rgb color={0.800,0.800,0.800}} (20.705,5.644)--(20.821,5.644)--(20.821,6.270)--(20.705,6.270)--cycle;
\draw[gp path] (20.705,5.644)--(20.705,6.269)--(20.820,6.269)--(20.820,5.644)--cycle;
\gpfill{rgb color={0.800,0.800,0.800}} (21.166,5.644)--(21.282,5.644)--(21.282,5.905)--(21.166,5.905)--cycle;
\draw[gp path] (21.166,5.644)--(21.166,5.904)--(21.281,5.904)--(21.281,5.644)--cycle;
\gpfill{rgb color={0.800,0.800,0.800}} (21.626,6.393)--(21.742,6.393)--(21.742,6.746)--(21.626,6.746)--cycle;
\draw[gp path] (21.626,6.393)--(21.626,6.745)--(21.741,6.745)--(21.741,6.393)--cycle;
\gpfill{rgb color={0.800,0.800,0.800}} (21.856,7.387)--(21.972,7.387)--(21.972,7.388)--(21.856,7.388)--cycle;
\draw[gp path] (21.856,7.387)--(21.971,7.387)--cycle;
\gpfill{rgb color={0.800,0.800,0.800}} (22.086,6.619)--(22.203,6.619)--(22.203,6.620)--(22.086,6.620)--cycle;
\draw[gp path] (22.086,6.619)--(22.202,6.619)--cycle;
\gpfill{rgb color={0.800,0.800,0.800}} (22.317,6.088)--(22.433,6.088)--(22.433,6.089)--(22.317,6.089)--cycle;
\draw[gp path] (22.317,6.088)--(22.432,6.088)--cycle;
\gpfill{rgb color={0.800,0.800,0.800}} (22.547,6.844)--(22.663,6.844)--(22.663,6.845)--(22.547,6.845)--cycle;
\draw[gp path] (22.547,6.844)--(22.662,6.844)--cycle;
\gpfill{rgb color={0.800,0.800,0.800}} (22.777,5.644)--(22.893,5.644)--(22.893,6.046)--(22.777,6.046)--cycle;
\draw[gp path] (22.777,5.644)--(22.777,6.045)--(22.892,6.045)--(22.892,5.644)--cycle;
\gpfill{rgb color={0.800,0.800,0.800}} (23.007,6.258)--(23.123,6.258)--(23.123,6.259)--(23.007,6.259)--cycle;
\draw[gp path] (23.007,6.258)--(23.122,6.258)--cycle;
\gpfill{rgb color={0.800,0.800,0.800}} (23.237,5.955)--(23.354,5.955)--(23.354,5.956)--(23.237,5.956)--cycle;
\draw[gp path] (23.237,5.955)--(23.353,5.955)--cycle;
\gpfill{rgb color={0.800,0.800,0.800}} (23.698,6.863)--(23.814,6.863)--(23.814,6.864)--(23.698,6.864)--cycle;
\draw[gp path] (23.698,6.863)--(23.813,6.863)--cycle;
\gpfill{rgb color={0.800,0.800,0.800}} (23.928,5.644)--(24.044,5.644)--(24.044,7.619)--(23.928,7.619)--cycle;
\draw[gp path] (23.928,5.644)--(23.928,7.618)--(24.043,7.618)--(24.043,5.644)--cycle;
\gpfill{rgb color={0.800,0.800,0.800}} (24.158,6.027)--(24.274,6.027)--(24.274,6.028)--(24.158,6.028)--cycle;
\draw[gp path] (24.158,6.027)--(24.273,6.027)--cycle;
\gpcolor{color=gp lt color border}
\draw[gp path] (1.196,10.131)--(1.196,5.644)--(24.446,5.644)--(24.446,10.131)--cycle;
\node[gp node center] at (12.821,9.682) {\plotLegend{(Hexagon)}};
\node[gp node left,font={\fontsize{11.0pt}{13.2pt}\selectfont}] at (1.429,9.817) {\plotLegend{mean improvement: $10\%$}};
\node[gp node left,font={\fontsize{11.0pt}{13.2pt}\selectfont}] at (1.429,9.368) {\plotLegend{improved functions: $64\%$}};
\node[gp node left,font={\fontsize{11.0pt}{13.2pt}\selectfont}] at (1.429,8.920) {\plotLegend{mean gap: $3.4\%$}};
\node[gp node left,font={\fontsize{11.0pt}{13.2pt}\selectfont}] at (1.429,8.471) {\plotLegend{optimal functions: $81\%$}};
\gpdefrectangularnode{gp plot 1}{\pgfpoint{1.196cm}{5.644cm}}{\pgfpoint{24.446cm}{10.131cm}}
\end{tikzpicture}

%% file: results/arm-speed-improvement.tex
\begin{tikzpicture}[gnuplot]
\path (0.000,0.000) rectangle (12.500,8.750);
\gpcolor{color=gp lt color border}
\gpsetlinetype{gp lt border}
\gpsetdashtype{gp dt solid}
\gpsetlinewidth{1.00}
\draw[gp path] (1.196,5.644)--(1.376,5.644);
\draw[gp path] (24.446,5.644)--(24.266,5.644);
\node[gp node right,font={\fontsize{8.0pt}{9.6pt}\selectfont}] at (1.012,5.644) {\plotPercentage{0}};
\draw[gp path] (1.196,6.093)--(1.376,6.093);
\draw[gp path] (24.446,6.093)--(24.266,6.093);
\node[gp node right,font={\fontsize{8.0pt}{9.6pt}\selectfont}] at (1.012,6.093) {\plotPercentage{10}};
\draw[gp path] (1.196,6.541)--(1.376,6.541);
\draw[gp path] (24.446,6.541)--(24.266,6.541);
\node[gp node right,font={\fontsize{8.0pt}{9.6pt}\selectfont}] at (1.012,6.541) {\plotPercentage{20}};
\draw[gp path] (1.196,6.990)--(1.376,6.990);
\draw[gp path] (24.446,6.990)--(24.266,6.990);
\node[gp node right,font={\fontsize{8.0pt}{9.6pt}\selectfont}] at (1.012,6.990) {\plotPercentage{30}};
\draw[gp path] (1.196,7.439)--(1.376,7.439);
\draw[gp path] (24.446,7.439)--(24.266,7.439);
\node[gp node right,font={\fontsize{8.0pt}{9.6pt}\selectfont}] at (1.012,7.439) {\plotPercentage{40}};
\draw[gp path] (1.196,7.888)--(1.376,7.888);
\draw[gp path] (24.446,7.888)--(24.266,7.888);
\node[gp node right,font={\fontsize{8.0pt}{9.6pt}\selectfont}] at (1.012,7.888) {\plotPercentage{50}};
\draw[gp path] (1.196,8.336)--(1.376,8.336);
\draw[gp path] (24.446,8.336)--(24.266,8.336);
\node[gp node right,font={\fontsize{8.0pt}{9.6pt}\selectfont}] at (1.012,8.336) {\plotPercentage{60}};
\draw[gp path] (1.196,8.785)--(1.376,8.785);
\draw[gp path] (24.446,8.785)--(24.266,8.785);
\node[gp node right,font={\fontsize{8.0pt}{9.6pt}\selectfont}] at (1.012,8.785) {\plotPercentage{70}};
\draw[gp path] (1.196,9.234)--(1.376,9.234);
\draw[gp path] (24.446,9.234)--(24.266,9.234);
\node[gp node right,font={\fontsize{8.0pt}{9.6pt}\selectfont}] at (1.012,9.234) {\plotPercentage{80}};
\draw[gp path] (1.196,9.682)--(1.376,9.682);
\draw[gp path] (24.446,9.682)--(24.266,9.682);
\node[gp node right,font={\fontsize{8.0pt}{9.6pt}\selectfont}] at (1.012,9.682) {\plotPercentage{90}};
\draw[gp path] (1.196,10.131)--(1.376,10.131);
\draw[gp path] (24.446,10.131)--(24.266,10.131);
\node[gp node right,font={\fontsize{8.0pt}{9.6pt}\selectfont}] at (1.012,10.131) {\plotPercentage{100}};
\node[gp node left,rotate=-90,font={\fontsize{7.0pt}{8.4pt}\selectfont}] at (1.426,5.721) {\functionId{1}\functionName{handle_noinline_attribute}};
\node[gp node left,rotate=-90,font={\fontsize{7.0pt}{8.4pt}\selectfont}] at (1.656,5.721) {\functionId{2}\functionName{control_flow_insn_p}};
\node[gp node left,rotate=-90,font={\fontsize{7.0pt}{8.4pt}\selectfont}] at (1.887,5.721) {\functionId{3}\functionName{insert_insn_on_edge}};
\node[gp node left,rotate=-90,font={\fontsize{7.0pt}{8.4pt}\selectfont}] at (2.117,5.721) {\functionId{4}\functionName{update_br_prob_note}};
\node[gp node left,rotate=-90,font={\fontsize{7.0pt}{8.4pt}\selectfont}] at (2.347,5.721) {\functionId{5}\functionName{_cpp_init_internal_pragma.}};
\node[gp node left,rotate=-90,font={\fontsize{7.0pt}{8.4pt}\selectfont}] at (2.577,5.721) {\functionId{6}\functionName{lex_macro_node}};
\node[gp node left,rotate=-90,font={\fontsize{7.0pt}{8.4pt}\selectfont}] at (2.807,5.721) {\functionId{7}\functionName{cse_basic_block}};
\node[gp node left,rotate=-90,font={\fontsize{7.0pt}{8.4pt}\selectfont}] at (3.038,5.721) {\functionId{8}\functionName{rtx_equal_for_cselib_p}};
\node[gp node left,rotate=-90,font={\fontsize{7.0pt}{8.4pt}\selectfont}] at (3.268,5.721) {\functionId{9}\functionName{debug_df_chain}};
\node[gp node left,rotate=-90,font={\fontsize{7.0pt}{8.4pt}\selectfont}] at (3.498,5.721) {\functionId{10}\functionName{modified_type_die}};
\node[gp node left,rotate=-90,font={\fontsize{7.0pt}{8.4pt}\selectfont}] at (3.728,5.721) {\functionId{11}\functionName{emit_note}};
\node[gp node left,rotate=-90,font={\fontsize{7.0pt}{8.4pt}\selectfont}] at (3.958,5.721) {\functionId{12}\functionName{gen_sequence}};
\node[gp node left,rotate=-90,font={\fontsize{7.0pt}{8.4pt}\selectfont}] at (4.189,5.721) {\functionId{13}\functionName{subreg_hard_regno}};
\node[gp node left,rotate=-90,font={\fontsize{7.0pt}{8.4pt}\selectfont}] at (4.419,5.721) {\functionId{14}\functionName{split_double}};
\node[gp node left,rotate=-90,font={\fontsize{7.0pt}{8.4pt}\selectfont}] at (4.649,5.721) {\functionId{15}\functionName{add_to_mem_set_list}};
\node[gp node left,rotate=-90,font={\fontsize{7.0pt}{8.4pt}\selectfont}] at (4.879,5.721) {\functionId{16}\functionName{find_regno_partial}};
\node[gp node left,rotate=-90,font={\fontsize{7.0pt}{8.4pt}\selectfont}] at (5.109,5.721) {\functionId{17}\functionName{use_return_register}};
\node[gp node left,rotate=-90,font={\fontsize{7.0pt}{8.4pt}\selectfont}] at (5.340,5.721) {\functionId{18}\functionName{ix86_expand_move}};
\node[gp node left,rotate=-90,font={\fontsize{7.0pt}{8.4pt}\selectfont}] at (5.570,5.721) {\functionId{19}\functionName{legitimate_pic_address_di.}};
\node[gp node left,rotate=-90,font={\fontsize{7.0pt}{8.4pt}\selectfont}] at (5.800,5.721) {\functionId{20}\functionName{gen_extendsfdf2}};
\node[gp node left,rotate=-90,font={\fontsize{7.0pt}{8.4pt}\selectfont}] at (6.030,5.721) {\functionId{21}\functionName{gen_mulsidi3}};
\node[gp node left,rotate=-90,font={\fontsize{7.0pt}{8.4pt}\selectfont}] at (6.260,5.721) {\functionId{22}\functionName{gen_peephole2_1255}};
\node[gp node left,rotate=-90,font={\fontsize{7.0pt}{8.4pt}\selectfont}] at (6.491,5.721) {\functionId{23}\functionName{gen_peephole2_1271}};
\node[gp node left,rotate=-90,font={\fontsize{7.0pt}{8.4pt}\selectfont}] at (6.721,5.721) {\functionId{24}\functionName{gen_peephole2_1277}};
\node[gp node left,rotate=-90,font={\fontsize{7.0pt}{8.4pt}\selectfont}] at (6.951,5.721) {\functionId{25}\functionName{gen_pfnacc}};
\node[gp node left,rotate=-90,font={\fontsize{7.0pt}{8.4pt}\selectfont}] at (7.181,5.721) {\functionId{26}\functionName{gen_rotlsi3}};
\node[gp node left,rotate=-90,font={\fontsize{7.0pt}{8.4pt}\selectfont}] at (7.411,5.721) {\functionId{27}\functionName{gen_split_1001}};
\node[gp node left,rotate=-90,font={\fontsize{7.0pt}{8.4pt}\selectfont}] at (7.642,5.721) {\functionId{28}\functionName{gen_split_1028}};
\node[gp node left,rotate=-90,font={\fontsize{7.0pt}{8.4pt}\selectfont}] at (7.872,5.721) {\functionId{29}\functionName{gen_sse_nandti3}};
\node[gp node left,rotate=-90,font={\fontsize{7.0pt}{8.4pt}\selectfont}] at (8.102,5.721) {\functionId{30}\functionName{gen_sunge}};
\node[gp node left,rotate=-90,font={\fontsize{7.0pt}{8.4pt}\selectfont}] at (8.332,5.721) {\functionId{31}\functionName{insert_loop_mem}};
\node[gp node left,rotate=-90,font={\fontsize{7.0pt}{8.4pt}\selectfont}] at (8.562,5.721) {\functionId{32}\functionName{eiremain}};
\node[gp node left,rotate=-90,font={\fontsize{7.0pt}{8.4pt}\selectfont}] at (8.793,5.721) {\functionId{33}\functionName{elimination_effects}};
\node[gp node left,rotate=-90,font={\fontsize{7.0pt}{8.4pt}\selectfont}] at (9.023,5.721) {\functionId{34}\functionName{gen_reload}};
\node[gp node left,rotate=-90,font={\fontsize{7.0pt}{8.4pt}\selectfont}] at (9.253,5.721) {\functionId{35}\functionName{reload_cse_simplify_set}};
\node[gp node left,rotate=-90,font={\fontsize{7.0pt}{8.4pt}\selectfont}] at (9.483,5.721) {\functionId{36}\functionName{simplify_binary_is2orm1}};
\node[gp node left,rotate=-90,font={\fontsize{7.0pt}{8.4pt}\selectfont}] at (9.713,5.721) {\functionId{37}\functionName{remove_phi_alternative}};
\node[gp node left,rotate=-90,font={\fontsize{7.0pt}{8.4pt}\selectfont}] at (9.944,5.721) {\functionId{38}\functionName{contains_placeholder_p}};
\node[gp node left,rotate=-90,font={\fontsize{7.0pt}{8.4pt}\selectfont}] at (10.174,5.721) {\functionId{39}\functionName{assemble_end_function}};
\node[gp node left,rotate=-90,font={\fontsize{7.0pt}{8.4pt}\selectfont}] at (10.404,5.721) {\functionId{40}\functionName{default_named_section_asm.}};
\node[gp node left,rotate=-90,font={\fontsize{7.0pt}{8.4pt}\selectfont}] at (10.634,5.721) {\functionId{41}\functionName{sample_unpack_12}};
\node[gp node left,rotate=-90,font={\fontsize{7.0pt}{8.4pt}\selectfont}] at (10.864,5.721) {\functionId{42}\functionName{autohelperattpat10}};
\node[gp node left,rotate=-90,font={\fontsize{7.0pt}{8.4pt}\selectfont}] at (11.095,5.721) {\functionId{43}\functionName{autohelperbarrierspat126}};
\node[gp node left,rotate=-90,font={\fontsize{7.0pt}{8.4pt}\selectfont}] at (11.325,5.721) {\functionId{44}\functionName{atari_atari_attack_callba.}};
\node[gp node left,rotate=-90,font={\fontsize{7.0pt}{8.4pt}\selectfont}] at (11.555,5.721) {\functionId{45}\functionName{compute_aa_status}};
\node[gp node left,rotate=-90,font={\fontsize{7.0pt}{8.4pt}\selectfont}] at (11.785,5.721) {\functionId{46}\functionName{dragon_weak}};
\node[gp node left,rotate=-90,font={\fontsize{7.0pt}{8.4pt}\selectfont}] at (12.015,5.721) {\functionId{47}\functionName{get_saved_worms}};
\node[gp node left,rotate=-90,font={\fontsize{7.0pt}{8.4pt}\selectfont}] at (12.246,5.721) {\functionId{48}\functionName{read_eye}};
\node[gp node left,rotate=-90,font={\fontsize{7.0pt}{8.4pt}\selectfont}] at (12.476,5.721) {\functionId{49}\functionName{topological_eye}};
\node[gp node left,rotate=-90,font={\fontsize{7.0pt}{8.4pt}\selectfont}] at (12.706,5.721) {\functionId{50}\functionName{autohelperowl_attackpat19.}};
\node[gp node left,rotate=-90,font={\fontsize{7.0pt}{8.4pt}\selectfont}] at (12.936,5.721) {\functionId{51}\functionName{autohelperowl_attackpat29.}};
\node[gp node left,rotate=-90,font={\fontsize{7.0pt}{8.4pt}\selectfont}] at (13.166,5.721) {\functionId{52}\functionName{autohelperowl_defendpat28.}};
\node[gp node left,rotate=-90,font={\fontsize{7.0pt}{8.4pt}\selectfont}] at (13.396,5.721) {\functionId{53}\functionName{autohelperowl_defendpat38.}};
\node[gp node left,rotate=-90,font={\fontsize{7.0pt}{8.4pt}\selectfont}] at (13.627,5.721) {\functionId{54}\functionName{autohelperpat1114}};
\node[gp node left,rotate=-90,font={\fontsize{7.0pt}{8.4pt}\selectfont}] at (13.857,5.721) {\functionId{55}\functionName{autohelperpat335}};
\node[gp node left,rotate=-90,font={\fontsize{7.0pt}{8.4pt}\selectfont}] at (14.087,5.721) {\functionId{56}\functionName{autohelperpat508}};
\node[gp node left,rotate=-90,font={\fontsize{7.0pt}{8.4pt}\selectfont}] at (14.317,5.721) {\functionId{57}\functionName{autohelperpat83}};
\node[gp node left,rotate=-90,font={\fontsize{7.0pt}{8.4pt}\selectfont}] at (14.547,5.721) {\functionId{58}\functionName{simple_showboard}};
\node[gp node left,rotate=-90,font={\fontsize{7.0pt}{8.4pt}\selectfont}] at (14.778,5.721) {\functionId{59}\functionName{skip_intrabk_SAD}};
\node[gp node left,rotate=-90,font={\fontsize{7.0pt}{8.4pt}\selectfont}] at (15.008,5.721) {\functionId{60}\functionName{free_orig_planes}};
\node[gp node left,rotate=-90,font={\fontsize{7.0pt}{8.4pt}\selectfont}] at (15.238,5.721) {\functionId{61}\functionName{GetSkipCostMB}};
\node[gp node left,rotate=-90,font={\fontsize{7.0pt}{8.4pt}\selectfont}] at (15.468,5.721) {\functionId{62}\functionName{writeSyntaxElement_Level_.}};
\node[gp node left,rotate=-90,font={\fontsize{7.0pt}{8.4pt}\selectfont}] at (15.698,5.721) {\functionId{63}\functionName{GSIAddKeyToIndex}};
\node[gp node left,rotate=-90,font={\fontsize{7.0pt}{8.4pt}\selectfont}] at (15.929,5.721) {\functionId{64}\functionName{EVDBasicFit}};
\node[gp node left,rotate=-90,font={\fontsize{7.0pt}{8.4pt}\selectfont}] at (16.159,5.721) {\functionId{65}\functionName{SampleDirichlet}};
\node[gp node left,rotate=-90,font={\fontsize{7.0pt}{8.4pt}\selectfont}] at (16.389,5.721) {\functionId{66}\functionName{DegenerateSymbolScore}};
\node[gp node left,rotate=-90,font={\fontsize{7.0pt}{8.4pt}\selectfont}] at (16.619,5.721) {\functionId{67}\functionName{Plan7SetCtime}};
\node[gp node left,rotate=-90,font={\fontsize{7.0pt}{8.4pt}\selectfont}] at (16.849,5.721) {\functionId{68}\functionName{MSAToSqinfo}};
\node[gp node left,rotate=-90,font={\fontsize{7.0pt}{8.4pt}\selectfont}] at (17.080,5.721) {\functionId{69}\functionName{null_convert}};
\node[gp node left,rotate=-90,font={\fontsize{7.0pt}{8.4pt}\selectfont}] at (17.310,5.721) {\functionId{70}\functionName{jinit_c_prep_controller}};
\node[gp node left,rotate=-90,font={\fontsize{7.0pt}{8.4pt}\selectfont}] at (17.540,5.721) {\functionId{71}\functionName{glFogf}};
\node[gp node left,rotate=-90,font={\fontsize{7.0pt}{8.4pt}\selectfont}] at (17.770,5.721) {\functionId{72}\functionName{glNormal3d}};
\node[gp node left,rotate=-90,font={\fontsize{7.0pt}{8.4pt}\selectfont}] at (18.000,5.721) {\functionId{73}\functionName{glRasterPos3d}};
\node[gp node left,rotate=-90,font={\fontsize{7.0pt}{8.4pt}\selectfont}] at (18.231,5.721) {\functionId{74}\functionName{glTexCoord2d}};
\node[gp node left,rotate=-90,font={\fontsize{7.0pt}{8.4pt}\selectfont}] at (18.461,5.721) {\functionId{75}\functionName{gl_stippled_bresenham}};
\node[gp node left,rotate=-90,font={\fontsize{7.0pt}{8.4pt}\selectfont}] at (18.691,5.721) {\functionId{76}\functionName{gl_save_Frustum}};
\node[gp node left,rotate=-90,font={\fontsize{7.0pt}{8.4pt}\selectfont}] at (18.921,5.721) {\functionId{77}\functionName{gl_save_LineWidth}};
\node[gp node left,rotate=-90,font={\fontsize{7.0pt}{8.4pt}\selectfont}] at (19.151,5.721) {\functionId{78}\functionName{translate_id}};
\node[gp node left,rotate=-90,font={\fontsize{7.0pt}{8.4pt}\selectfont}] at (19.382,5.721) {\functionId{79}\functionName{gl_Map1f}};
\node[gp node left,rotate=-90,font={\fontsize{7.0pt}{8.4pt}\selectfont}] at (19.612,5.721) {\functionId{80}\functionName{smooth_ci_line}};
\node[gp node left,rotate=-90,font={\fontsize{7.0pt}{8.4pt}\selectfont}] at (19.842,5.721) {\functionId{81}\functionName{free_unified_knots}};
\node[gp node left,rotate=-90,font={\fontsize{7.0pt}{8.4pt}\selectfont}] at (20.072,5.721) {\functionId{82}\functionName{tess_test_polygon}};
\node[gp node left,rotate=-90,font={\fontsize{7.0pt}{8.4pt}\selectfont}] at (20.302,5.721) {\functionId{83}\functionName{auxWireBox}};
\node[gp node left,rotate=-90,font={\fontsize{7.0pt}{8.4pt}\selectfont}] at (20.533,5.721) {\functionId{84}\functionName{gl_ColorPointer}};
\node[gp node left,rotate=-90,font={\fontsize{7.0pt}{8.4pt}\selectfont}] at (20.763,5.721) {\functionId{85}\functionName{r_serial}};
\node[gp node left,rotate=-90,font={\fontsize{7.0pt}{8.4pt}\selectfont}] at (20.993,5.721) {\functionId{86}\functionName{scalar_mult_sub_su3_matri.}};
\node[gp node left,rotate=-90,font={\fontsize{7.0pt}{8.4pt}\selectfont}] at (21.223,5.721) {\functionId{87}\functionName{Decode_MPEG1_Non_Intra_Bl.}};
\node[gp node left,rotate=-90,font={\fontsize{7.0pt}{8.4pt}\selectfont}] at (21.453,5.721) {\functionId{88}\functionName{cpDecodeSecret}};
\node[gp node left,rotate=-90,font={\fontsize{7.0pt}{8.4pt}\selectfont}] at (21.684,5.721) {\functionId{89}\functionName{vlShortLshift}};
\node[gp node left,rotate=-90,font={\fontsize{7.0pt}{8.4pt}\selectfont}] at (21.914,5.721) {\functionId{90}\functionName{encryptfile}};
\node[gp node left,rotate=-90,font={\fontsize{7.0pt}{8.4pt}\selectfont}] at (22.144,5.721) {\functionId{91}\functionName{make_canonical}};
\node[gp node left,rotate=-90,font={\fontsize{7.0pt}{8.4pt}\selectfont}] at (22.374,5.721) {\functionId{92}\functionName{LANG}};
\node[gp node left,rotate=-90,font={\fontsize{7.0pt}{8.4pt}\selectfont}] at (22.604,5.721) {\functionId{93}\functionName{MD5Transform}};
\node[gp node left,rotate=-90,font={\fontsize{7.0pt}{8.4pt}\selectfont}] at (22.835,5.721) {\functionId{94}\functionName{mp_display}};
\node[gp node left,rotate=-90,font={\fontsize{7.0pt}{8.4pt}\selectfont}] at (23.065,5.721) {\functionId{95}\functionName{comp_Jboundaries}};
\node[gp node left,rotate=-90,font={\fontsize{7.0pt}{8.4pt}\selectfont}] at (23.295,5.721) {\functionId{96}\functionName{is_draw}};
\node[gp node left,rotate=-90,font={\fontsize{7.0pt}{8.4pt}\selectfont}] at (23.525,5.721) {\functionId{97}\functionName{push_king}};
\node[gp node left,rotate=-90,font={\fontsize{7.0pt}{8.4pt}\selectfont}] at (23.755,5.721) {\functionId{98}\functionName{stat_retry}};
\node[gp node left,rotate=-90,font={\fontsize{7.0pt}{8.4pt}\selectfont}] at (23.986,5.721) {\functionId{99}\functionName{lextree_subtree_print}};
\node[gp node left,rotate=-90,font={\fontsize{7.0pt}{8.4pt}\selectfont}] at (24.216,5.721) {\functionId{100}\functionName{lm_tg_score}};
\draw[gp path] (1.196,10.131)--(1.196,5.644)--(24.446,5.644)--(24.446,10.131)--cycle;
\gpcolor{rgb color={0.580,0.000,0.827}}
\draw[gp path] (1.196,5.644)--(1.431,5.644)--(1.666,5.644)--(1.901,5.644)--(2.135,5.644)%
  --(2.370,5.644)--(2.605,5.644)--(2.840,5.644)--(3.075,5.644)--(3.310,5.644)--(3.544,5.644)%
  --(3.779,5.644)--(4.014,5.644)--(4.249,5.644)--(4.484,5.644)--(4.719,5.644)--(4.954,5.644)%
  --(5.188,5.644)--(5.423,5.644)--(5.658,5.644)--(5.893,5.644)--(6.128,5.644)--(6.363,5.644)%
  --(6.598,5.644)--(6.832,5.644)--(7.067,5.644)--(7.302,5.644)--(7.537,5.644)--(7.772,5.644)%
  --(8.007,5.644)--(8.241,5.644)--(8.476,5.644)--(8.711,5.644)--(8.946,5.644)--(9.181,5.644)%
  --(9.416,5.644)--(9.651,5.644)--(9.885,5.644)--(10.120,5.644)--(10.355,5.644)--(10.590,5.644)%
  --(10.825,5.644)--(11.060,5.644)--(11.294,5.644)--(11.529,5.644)--(11.764,5.644)--(11.999,5.644)%
  --(12.234,5.644)--(12.469,5.644)--(12.704,5.644)--(12.938,5.644)--(13.173,5.644)--(13.408,5.644)%
  --(13.643,5.644)--(13.878,5.644)--(14.113,5.644)--(14.348,5.644)--(14.582,5.644)--(14.817,5.644)%
  --(15.052,5.644)--(15.287,5.644)--(15.522,5.644)--(15.757,5.644)--(15.991,5.644)--(16.226,5.644)%
  --(16.461,5.644)--(16.696,5.644)--(16.931,5.644)--(17.166,5.644)--(17.401,5.644)--(17.635,5.644)%
  --(17.870,5.644)--(18.105,5.644)--(18.340,5.644)--(18.575,5.644)--(18.810,5.644)--(19.044,5.644)%
  --(19.279,5.644)--(19.514,5.644)--(19.749,5.644)--(19.984,5.644)--(20.219,5.644)--(20.454,5.644)%
  --(20.688,5.644)--(20.923,5.644)--(21.158,5.644)--(21.393,5.644)--(21.628,5.644)--(21.863,5.644)%
  --(22.098,5.644)--(22.332,5.644)--(22.567,5.644)--(22.802,5.644)--(23.037,5.644)--(23.272,5.644)%
  --(23.507,5.644)--(23.741,5.644)--(23.976,5.644)--(24.211,5.644)--(24.446,5.644);
\gpfill{rgb color={0.000,0.000,0.000}} (1.599,5.644)--(1.715,5.644)--(1.715,5.645)--(1.599,5.645)--cycle;
\gpcolor{rgb color={0.000,0.000,0.000}}
\draw[gp path] (1.599,5.644)--(1.714,5.644)--cycle;
\gpfill{rgb color={0.000,0.000,0.000}} (1.829,5.644)--(1.945,5.644)--(1.945,5.645)--(1.829,5.645)--cycle;
\draw[gp path] (1.829,5.644)--(1.944,5.644)--cycle;
\gpfill{rgb color={0.000,0.000,0.000}} (2.059,5.644)--(2.175,5.644)--(2.175,5.722)--(2.059,5.722)--cycle;
\draw[gp path] (2.059,5.644)--(2.059,5.721)--(2.174,5.721)--(2.174,5.644)--cycle;
\gpfill{rgb color={0.000,0.000,0.000}} (3.671,5.644)--(3.787,5.644)--(3.787,5.663)--(3.671,5.663)--cycle;
\draw[gp path] (3.671,5.644)--(3.671,5.662)--(3.786,5.662)--(3.786,5.644)--cycle;
\gpfill{rgb color={0.000,0.000,0.000}} (4.131,5.644)--(4.247,5.644)--(4.247,5.645)--(4.131,5.645)--cycle;
\draw[gp path] (4.131,5.644)--(4.246,5.644)--cycle;
\gpfill{rgb color={0.000,0.000,0.000}} (4.361,5.644)--(4.477,5.644)--(4.477,5.807)--(4.361,5.807)--cycle;
\draw[gp path] (4.361,5.644)--(4.361,5.806)--(4.476,5.806)--(4.476,5.644)--cycle;
\gpfill{rgb color={0.000,0.000,0.000}} (4.591,5.644)--(4.708,5.644)--(4.708,5.715)--(4.591,5.715)--cycle;
\draw[gp path] (4.591,5.644)--(4.591,5.714)--(4.707,5.714)--(4.707,5.644)--cycle;
\gpfill{rgb color={0.000,0.000,0.000}} (5.052,5.644)--(5.168,5.644)--(5.168,5.649)--(5.052,5.649)--cycle;
\draw[gp path] (5.052,5.644)--(5.052,5.648)--(5.167,5.648)--(5.167,5.644)--cycle;
\gpfill{rgb color={0.000,0.000,0.000}} (5.282,5.644)--(5.398,5.644)--(5.398,5.663)--(5.282,5.663)--cycle;
\draw[gp path] (5.282,5.644)--(5.282,5.662)--(5.397,5.662)--(5.397,5.644)--cycle;
\gpfill{rgb color={0.000,0.000,0.000}} (5.512,5.644)--(5.628,5.644)--(5.628,5.760)--(5.512,5.760)--cycle;
\draw[gp path] (5.512,5.644)--(5.512,5.759)--(5.627,5.759)--(5.627,5.644)--cycle;
\gpfill{rgb color={0.000,0.000,0.000}} (6.203,5.644)--(6.319,5.644)--(6.319,5.752)--(6.203,5.752)--cycle;
\draw[gp path] (6.203,5.644)--(6.203,5.751)--(6.318,5.751)--(6.318,5.644)--cycle;
\gpfill{rgb color={0.000,0.000,0.000}} (6.663,5.644)--(6.779,5.644)--(6.779,5.668)--(6.663,5.668)--cycle;
\draw[gp path] (6.663,5.644)--(6.663,5.667)--(6.778,5.667)--(6.778,5.644)--cycle;
\gpfill{rgb color={0.000,0.000,0.000}} (7.354,5.644)--(7.470,5.644)--(7.470,5.650)--(7.354,5.650)--cycle;
\draw[gp path] (7.354,5.644)--(7.354,5.649)--(7.469,5.649)--(7.469,5.644)--cycle;
\gpfill{rgb color={0.000,0.000,0.000}} (8.275,5.644)--(8.391,5.644)--(8.391,5.760)--(8.275,5.760)--cycle;
\draw[gp path] (8.275,5.644)--(8.275,5.759)--(8.390,5.759)--(8.390,5.644)--cycle;
\gpfill{rgb color={0.000,0.000,0.000}} (9.426,5.644)--(9.542,5.644)--(9.542,5.997)--(9.426,5.997)--cycle;
\draw[gp path] (9.426,5.644)--(9.426,5.996)--(9.541,5.996)--(9.541,5.644)--cycle;
\gpfill{rgb color={0.000,0.000,0.000}} (9.886,5.644)--(10.002,5.644)--(10.002,5.671)--(9.886,5.671)--cycle;
\draw[gp path] (9.886,5.644)--(9.886,5.670)--(10.001,5.670)--(10.001,5.644)--cycle;
\gpfill{rgb color={0.000,0.000,0.000}} (10.346,5.644)--(10.462,5.644)--(10.462,5.743)--(10.346,5.743)--cycle;
\draw[gp path] (10.346,5.644)--(10.346,5.742)--(10.461,5.742)--(10.461,5.644)--cycle;
\gpfill{rgb color={0.000,0.000,0.000}} (10.577,5.644)--(10.693,5.644)--(10.693,5.859)--(10.577,5.859)--cycle;
\draw[gp path] (10.577,5.644)--(10.577,5.858)--(10.692,5.858)--(10.692,5.644)--cycle;
\gpfill{rgb color={0.000,0.000,0.000}} (10.807,5.644)--(10.923,5.644)--(10.923,5.930)--(10.807,5.930)--cycle;
\draw[gp path] (10.807,5.644)--(10.807,5.929)--(10.922,5.929)--(10.922,5.644)--cycle;
\gpfill{rgb color={0.000,0.000,0.000}} (11.958,5.644)--(12.074,5.644)--(12.074,5.653)--(11.958,5.653)--cycle;
\draw[gp path] (11.958,5.644)--(11.958,5.652)--(12.073,5.652)--(12.073,5.644)--cycle;
\gpfill{rgb color={0.000,0.000,0.000}} (12.648,5.644)--(12.764,5.644)--(12.764,5.826)--(12.648,5.826)--cycle;
\draw[gp path] (12.648,5.644)--(12.648,5.825)--(12.763,5.825)--(12.763,5.644)--cycle;
\gpfill{rgb color={0.000,0.000,0.000}} (13.109,5.644)--(13.225,5.644)--(13.225,5.790)--(13.109,5.790)--cycle;
\draw[gp path] (13.109,5.644)--(13.109,5.789)--(13.224,5.789)--(13.224,5.644)--cycle;
\gpfill{rgb color={0.000,0.000,0.000}} (13.339,5.644)--(13.455,5.644)--(13.455,5.718)--(13.339,5.718)--cycle;
\draw[gp path] (13.339,5.644)--(13.339,5.717)--(13.454,5.717)--(13.454,5.644)--cycle;
\gpfill{rgb color={0.000,0.000,0.000}} (13.799,5.644)--(13.915,5.644)--(13.915,5.944)--(13.799,5.944)--cycle;
\draw[gp path] (13.799,5.644)--(13.799,5.943)--(13.914,5.943)--(13.914,5.644)--cycle;
\gpfill{rgb color={0.000,0.000,0.000}} (14.030,5.644)--(14.146,5.644)--(14.146,5.818)--(14.030,5.818)--cycle;
\draw[gp path] (14.030,5.644)--(14.030,5.817)--(14.145,5.817)--(14.145,5.644)--cycle;
\gpfill{rgb color={0.000,0.000,0.000}} (14.260,5.644)--(14.376,5.644)--(14.376,5.761)--(14.260,5.761)--cycle;
\draw[gp path] (14.260,5.644)--(14.260,5.760)--(14.375,5.760)--(14.375,5.644)--cycle;
\gpfill{rgb color={0.000,0.000,0.000}} (15.871,5.644)--(15.987,5.644)--(15.987,5.724)--(15.871,5.724)--cycle;
\draw[gp path] (15.871,5.644)--(15.871,5.723)--(15.986,5.723)--(15.986,5.644)--cycle;
\gpfill{rgb color={0.000,0.000,0.000}} (16.101,5.644)--(16.217,5.644)--(16.217,5.657)--(16.101,5.657)--cycle;
\draw[gp path] (16.101,5.644)--(16.101,5.656)--(16.216,5.656)--(16.216,5.644)--cycle;
\gpfill{rgb color={0.000,0.000,0.000}} (16.332,5.644)--(16.448,5.644)--(16.448,5.799)--(16.332,5.799)--cycle;
\draw[gp path] (16.332,5.644)--(16.332,5.798)--(16.447,5.798)--(16.447,5.644)--cycle;
\gpfill{rgb color={0.000,0.000,0.000}} (17.483,5.644)--(17.599,5.644)--(17.599,5.769)--(17.483,5.769)--cycle;
\draw[gp path] (17.483,5.644)--(17.483,5.768)--(17.598,5.768)--(17.598,5.644)--cycle;
\gpfill{rgb color={0.000,0.000,0.000}} (17.943,5.644)--(18.059,5.644)--(18.059,5.707)--(17.943,5.707)--cycle;
\draw[gp path] (17.943,5.644)--(17.943,5.706)--(18.058,5.706)--(18.058,5.644)--cycle;
\gpfill{rgb color={0.000,0.000,0.000}} (18.634,5.644)--(18.750,5.644)--(18.750,5.645)--(18.634,5.645)--cycle;
\draw[gp path] (18.634,5.644)--(18.749,5.644)--cycle;
\gpfill{rgb color={0.000,0.000,0.000}} (18.864,5.644)--(18.980,5.644)--(18.980,6.290)--(18.864,6.290)--cycle;
\draw[gp path] (18.864,5.644)--(18.864,6.289)--(18.979,6.289)--(18.979,5.644)--cycle;
\gpfill{rgb color={0.000,0.000,0.000}} (19.094,5.644)--(19.210,5.644)--(19.210,5.650)--(19.094,5.650)--cycle;
\draw[gp path] (19.094,5.644)--(19.094,5.649)--(19.209,5.649)--(19.209,5.644)--cycle;
\gpfill{rgb color={0.000,0.000,0.000}} (19.324,5.644)--(19.440,5.644)--(19.440,6.020)--(19.324,6.020)--cycle;
\draw[gp path] (19.324,5.644)--(19.324,6.019)--(19.439,6.019)--(19.439,5.644)--cycle;
\gpfill{rgb color={0.000,0.000,0.000}} (19.554,5.644)--(19.670,5.644)--(19.670,5.874)--(19.554,5.874)--cycle;
\draw[gp path] (19.554,5.644)--(19.554,5.873)--(19.669,5.873)--(19.669,5.644)--cycle;
\gpfill{rgb color={0.000,0.000,0.000}} (21.396,5.644)--(21.512,5.644)--(21.512,5.869)--(21.396,5.869)--cycle;
\draw[gp path] (21.396,5.644)--(21.396,5.868)--(21.511,5.868)--(21.511,5.644)--cycle;
\gpfill{rgb color={0.000,0.000,0.000}} (22.086,5.644)--(22.203,5.644)--(22.203,5.840)--(22.086,5.840)--cycle;
\draw[gp path] (22.086,5.644)--(22.086,5.839)--(22.202,5.839)--(22.202,5.644)--cycle;
\gpfill{rgb color={0.000,0.000,0.000}} (23.237,5.644)--(23.354,5.644)--(23.354,5.721)--(23.237,5.721)--cycle;
\draw[gp path] (23.237,5.644)--(23.237,5.720)--(23.353,5.720)--(23.353,5.644)--cycle;
\gpfill{rgb color={0.000,0.000,0.000}} (23.698,5.644)--(23.814,5.644)--(23.814,5.653)--(23.698,5.653)--cycle;
\draw[gp path] (23.698,5.644)--(23.698,5.652)--(23.813,5.652)--(23.813,5.644)--cycle;
\gpfill{rgb color={0.000,0.000,0.000}} (24.158,5.644)--(24.274,5.644)--(24.274,5.815)--(24.158,5.815)--cycle;
\draw[gp path] (24.158,5.644)--(24.158,5.814)--(24.273,5.814)--(24.273,5.644)--cycle;
\gpfill{rgb color={0.800,0.800,0.800}} (1.599,5.644)--(1.715,5.644)--(1.715,5.645)--(1.599,5.645)--cycle;
\gpcolor{rgb color={0.800,0.800,0.800}}
\draw[gp path] (1.599,5.644)--(1.714,5.644)--cycle;
\gpfill{rgb color={0.800,0.800,0.800}} (1.829,5.644)--(1.945,5.644)--(1.945,5.645)--(1.829,5.645)--cycle;
\draw[gp path] (1.829,5.644)--(1.944,5.644)--cycle;
\gpfill{rgb color={0.800,0.800,0.800}} (2.059,5.721)--(2.175,5.721)--(2.175,5.722)--(2.059,5.722)--cycle;
\draw[gp path] (2.059,5.721)--(2.174,5.721)--cycle;
\gpfill{rgb color={0.800,0.800,0.800}} (2.750,5.644)--(2.866,5.644)--(2.866,5.776)--(2.750,5.776)--cycle;
\draw[gp path] (2.750,5.644)--(2.750,5.775)--(2.865,5.775)--(2.865,5.644)--cycle;
\gpfill{rgb color={0.800,0.800,0.800}} (2.980,5.644)--(3.096,5.644)--(3.096,5.809)--(2.980,5.809)--cycle;
\draw[gp path] (2.980,5.644)--(2.980,5.808)--(3.095,5.808)--(3.095,5.644)--cycle;
\gpfill{rgb color={0.800,0.800,0.800}} (3.440,5.644)--(3.557,5.644)--(3.557,6.679)--(3.440,6.679)--cycle;
\draw[gp path] (3.440,5.644)--(3.440,6.678)--(3.556,6.678)--(3.556,5.644)--cycle;
\gpfill{rgb color={0.800,0.800,0.800}} (3.671,5.662)--(3.787,5.662)--(3.787,5.663)--(3.671,5.663)--cycle;
\draw[gp path] (3.671,5.662)--(3.786,5.662)--cycle;
\gpfill{rgb color={0.800,0.800,0.800}} (4.131,5.644)--(4.247,5.644)--(4.247,5.645)--(4.131,5.645)--cycle;
\draw[gp path] (4.131,5.644)--(4.246,5.644)--cycle;
\gpfill{rgb color={0.800,0.800,0.800}} (4.361,5.806)--(4.477,5.806)--(4.477,6.065)--(4.361,6.065)--cycle;
\draw[gp path] (4.361,5.806)--(4.361,6.064)--(4.476,6.064)--(4.476,5.806)--cycle;
\gpfill{rgb color={0.800,0.800,0.800}} (4.591,5.714)--(4.708,5.714)--(4.708,5.715)--(4.591,5.715)--cycle;
\draw[gp path] (4.591,5.714)--(4.707,5.714)--cycle;
\gpfill{rgb color={0.800,0.800,0.800}} (5.052,5.648)--(5.168,5.648)--(5.168,5.649)--(5.052,5.649)--cycle;
\draw[gp path] (5.052,5.648)--(5.167,5.648)--cycle;
\gpfill{rgb color={0.800,0.800,0.800}} (5.282,5.662)--(5.398,5.662)--(5.398,5.663)--(5.282,5.663)--cycle;
\draw[gp path] (5.282,5.662)--(5.397,5.662)--cycle;
\gpfill{rgb color={0.800,0.800,0.800}} (5.512,5.759)--(5.628,5.759)--(5.628,5.760)--(5.512,5.760)--cycle;
\draw[gp path] (5.512,5.759)--(5.627,5.759)--cycle;
\gpfill{rgb color={0.800,0.800,0.800}} (6.203,5.751)--(6.319,5.751)--(6.319,5.752)--(6.203,5.752)--cycle;
\draw[gp path] (6.203,5.751)--(6.318,5.751)--cycle;
\gpfill{rgb color={0.800,0.800,0.800}} (6.663,5.667)--(6.779,5.667)--(6.779,5.668)--(6.663,5.668)--cycle;
\draw[gp path] (6.663,5.667)--(6.778,5.667)--cycle;
\gpfill{rgb color={0.800,0.800,0.800}} (7.354,5.649)--(7.470,5.649)--(7.470,5.650)--(7.354,5.650)--cycle;
\draw[gp path] (7.354,5.649)--(7.469,5.649)--cycle;
\gpfill{rgb color={0.800,0.800,0.800}} (7.584,5.644)--(7.700,5.644)--(7.700,6.987)--(7.584,6.987)--cycle;
\draw[gp path] (7.584,5.644)--(7.584,6.986)--(7.699,6.986)--(7.699,5.644)--cycle;
\gpfill{rgb color={0.800,0.800,0.800}} (8.275,5.759)--(8.391,5.759)--(8.391,5.760)--(8.275,5.760)--cycle;
\draw[gp path] (8.275,5.759)--(8.390,5.759)--cycle;
\gpfill{rgb color={0.800,0.800,0.800}} (8.505,5.644)--(8.621,5.644)--(8.621,5.887)--(8.505,5.887)--cycle;
\draw[gp path] (8.505,5.644)--(8.505,5.886)--(8.620,5.886)--(8.620,5.644)--cycle;
\gpfill{rgb color={0.800,0.800,0.800}} (8.735,5.644)--(8.851,5.644)--(8.851,6.001)--(8.735,6.001)--cycle;
\draw[gp path] (8.735,5.644)--(8.735,6.000)--(8.850,6.000)--(8.850,5.644)--cycle;
\gpfill{rgb color={0.800,0.800,0.800}} (8.965,5.644)--(9.081,5.644)--(9.081,6.170)--(8.965,6.170)--cycle;
\draw[gp path] (8.965,5.644)--(8.965,6.169)--(9.080,6.169)--(9.080,5.644)--cycle;
\gpfill{rgb color={0.800,0.800,0.800}} (9.195,5.644)--(9.311,5.644)--(9.311,6.333)--(9.195,6.333)--cycle;
\draw[gp path] (9.195,5.644)--(9.195,6.332)--(9.310,6.332)--(9.310,5.644)--cycle;
\gpfill{rgb color={0.800,0.800,0.800}} (9.426,5.996)--(9.542,5.996)--(9.542,8.416)--(9.426,8.416)--cycle;
\draw[gp path] (9.426,5.996)--(9.426,8.415)--(9.541,8.415)--(9.541,5.996)--cycle;
\gpfill{rgb color={0.800,0.800,0.800}} (9.656,5.644)--(9.772,5.644)--(9.772,5.918)--(9.656,5.918)--cycle;
\draw[gp path] (9.656,5.644)--(9.656,5.917)--(9.771,5.917)--(9.771,5.644)--cycle;
\gpfill{rgb color={0.800,0.800,0.800}} (9.886,5.670)--(10.002,5.670)--(10.002,5.671)--(9.886,5.671)--cycle;
\draw[gp path] (9.886,5.670)--(10.001,5.670)--cycle;
\gpfill{rgb color={0.800,0.800,0.800}} (10.116,5.644)--(10.232,5.644)--(10.232,6.086)--(10.116,6.086)--cycle;
\draw[gp path] (10.116,5.644)--(10.116,6.085)--(10.231,6.085)--(10.231,5.644)--cycle;
\gpfill{rgb color={0.800,0.800,0.800}} (10.346,5.742)--(10.462,5.742)--(10.462,5.743)--(10.346,5.743)--cycle;
\draw[gp path] (10.346,5.742)--(10.461,5.742)--cycle;
\gpfill{rgb color={0.800,0.800,0.800}} (10.577,5.858)--(10.693,5.858)--(10.693,5.964)--(10.577,5.964)--cycle;
\draw[gp path] (10.577,5.858)--(10.577,5.963)--(10.692,5.963)--(10.692,5.858)--cycle;
\gpfill{rgb color={0.800,0.800,0.800}} (10.807,5.929)--(10.923,5.929)--(10.923,5.930)--(10.807,5.930)--cycle;
\draw[gp path] (10.807,5.929)--(10.922,5.929)--cycle;
\gpfill{rgb color={0.800,0.800,0.800}} (11.037,5.644)--(11.153,5.644)--(11.153,6.437)--(11.037,6.437)--cycle;
\draw[gp path] (11.037,5.644)--(11.037,6.436)--(11.152,6.436)--(11.152,5.644)--cycle;
\gpfill{rgb color={0.800,0.800,0.800}} (11.267,5.644)--(11.383,5.644)--(11.383,5.917)--(11.267,5.917)--cycle;
\draw[gp path] (11.267,5.644)--(11.267,5.916)--(11.382,5.916)--(11.382,5.644)--cycle;
\gpfill{rgb color={0.800,0.800,0.800}} (11.497,5.644)--(11.613,5.644)--(11.613,5.997)--(11.497,5.997)--cycle;
\draw[gp path] (11.497,5.644)--(11.497,5.996)--(11.612,5.996)--(11.612,5.644)--cycle;
\gpfill{rgb color={0.800,0.800,0.800}} (11.958,5.652)--(12.074,5.652)--(12.074,6.160)--(11.958,6.160)--cycle;
\draw[gp path] (11.958,5.652)--(11.958,6.159)--(12.073,6.159)--(12.073,5.652)--cycle;
\gpfill{rgb color={0.800,0.800,0.800}} (12.188,5.644)--(12.304,5.644)--(12.304,5.845)--(12.188,5.845)--cycle;
\draw[gp path] (12.188,5.644)--(12.188,5.844)--(12.303,5.844)--(12.303,5.644)--cycle;
\gpfill{rgb color={0.800,0.800,0.800}} (12.418,5.644)--(12.534,5.644)--(12.534,7.604)--(12.418,7.604)--cycle;
\draw[gp path] (12.418,5.644)--(12.418,7.603)--(12.533,7.603)--(12.533,5.644)--cycle;
\gpfill{rgb color={0.800,0.800,0.800}} (12.648,5.825)--(12.764,5.825)--(12.764,5.826)--(12.648,5.826)--cycle;
\draw[gp path] (12.648,5.825)--(12.763,5.825)--cycle;
\gpfill{rgb color={0.800,0.800,0.800}} (12.879,5.644)--(12.995,5.644)--(12.995,5.849)--(12.879,5.849)--cycle;
\draw[gp path] (12.879,5.644)--(12.879,5.848)--(12.994,5.848)--(12.994,5.644)--cycle;
\gpfill{rgb color={0.800,0.800,0.800}} (13.109,5.789)--(13.225,5.789)--(13.225,5.790)--(13.109,5.790)--cycle;
\draw[gp path] (13.109,5.789)--(13.224,5.789)--cycle;
\gpfill{rgb color={0.800,0.800,0.800}} (13.339,5.717)--(13.455,5.717)--(13.455,5.718)--(13.339,5.718)--cycle;
\draw[gp path] (13.339,5.717)--(13.454,5.717)--cycle;
\gpfill{rgb color={0.800,0.800,0.800}} (13.799,5.943)--(13.915,5.943)--(13.915,5.944)--(13.799,5.944)--cycle;
\draw[gp path] (13.799,5.943)--(13.914,5.943)--cycle;
\gpfill{rgb color={0.800,0.800,0.800}} (14.030,5.817)--(14.146,5.817)--(14.146,5.818)--(14.030,5.818)--cycle;
\draw[gp path] (14.030,5.817)--(14.145,5.817)--cycle;
\gpfill{rgb color={0.800,0.800,0.800}} (14.260,5.760)--(14.376,5.760)--(14.376,6.032)--(14.260,6.032)--cycle;
\draw[gp path] (14.260,5.760)--(14.260,6.031)--(14.375,6.031)--(14.375,5.760)--cycle;
\gpfill{rgb color={0.800,0.800,0.800}} (14.490,5.644)--(14.606,5.644)--(14.606,5.887)--(14.490,5.887)--cycle;
\draw[gp path] (14.490,5.644)--(14.490,5.886)--(14.605,5.886)--(14.605,5.644)--cycle;
\gpfill{rgb color={0.800,0.800,0.800}} (15.181,5.644)--(15.297,5.644)--(15.297,6.275)--(15.181,6.275)--cycle;
\draw[gp path] (15.181,5.644)--(15.181,6.274)--(15.296,6.274)--(15.296,5.644)--cycle;
\gpfill{rgb color={0.800,0.800,0.800}} (15.871,5.723)--(15.987,5.723)--(15.987,5.730)--(15.871,5.730)--cycle;
\draw[gp path] (15.871,5.723)--(15.871,5.729)--(15.986,5.729)--(15.986,5.723)--cycle;
\gpfill{rgb color={0.800,0.800,0.800}} (16.101,5.656)--(16.217,5.656)--(16.217,5.657)--(16.101,5.657)--cycle;
\draw[gp path] (16.101,5.656)--(16.216,5.656)--cycle;
\gpfill{rgb color={0.800,0.800,0.800}} (16.332,5.798)--(16.448,5.798)--(16.448,5.799)--(16.332,5.799)--cycle;
\draw[gp path] (16.332,5.798)--(16.447,5.798)--cycle;
\gpfill{rgb color={0.800,0.800,0.800}} (16.792,5.644)--(16.908,5.644)--(16.908,5.925)--(16.792,5.925)--cycle;
\draw[gp path] (16.792,5.644)--(16.792,5.924)--(16.907,5.924)--(16.907,5.644)--cycle;
\gpfill{rgb color={0.800,0.800,0.800}} (17.252,5.644)--(17.368,5.644)--(17.368,5.800)--(17.252,5.800)--cycle;
\draw[gp path] (17.252,5.644)--(17.252,5.799)--(17.367,5.799)--(17.367,5.644)--cycle;
\gpfill{rgb color={0.800,0.800,0.800}} (17.483,5.768)--(17.599,5.768)--(17.599,5.769)--(17.483,5.769)--cycle;
\draw[gp path] (17.483,5.768)--(17.598,5.768)--cycle;
\gpfill{rgb color={0.800,0.800,0.800}} (17.943,5.706)--(18.059,5.706)--(18.059,5.707)--(17.943,5.707)--cycle;
\draw[gp path] (17.943,5.706)--(18.058,5.706)--cycle;
\gpfill{rgb color={0.800,0.800,0.800}} (18.403,5.644)--(18.519,5.644)--(18.519,7.043)--(18.403,7.043)--cycle;
\draw[gp path] (18.403,5.644)--(18.403,7.042)--(18.518,7.042)--(18.518,5.644)--cycle;
\gpfill{rgb color={0.800,0.800,0.800}} (18.634,5.644)--(18.750,5.644)--(18.750,7.198)--(18.634,7.198)--cycle;
\draw[gp path] (18.634,5.644)--(18.634,7.197)--(18.749,7.197)--(18.749,5.644)--cycle;
\gpfill{rgb color={0.800,0.800,0.800}} (18.864,6.289)--(18.980,6.289)--(18.980,6.290)--(18.864,6.290)--cycle;
\draw[gp path] (18.864,6.289)--(18.979,6.289)--cycle;
\gpfill{rgb color={0.800,0.800,0.800}} (19.094,5.649)--(19.210,5.649)--(19.210,5.650)--(19.094,5.650)--cycle;
\draw[gp path] (19.094,5.649)--(19.209,5.649)--cycle;
\gpfill{rgb color={0.800,0.800,0.800}} (19.324,6.019)--(19.440,6.019)--(19.440,7.357)--(19.324,7.357)--cycle;
\draw[gp path] (19.324,6.019)--(19.324,7.356)--(19.439,7.356)--(19.439,6.019)--cycle;
\gpfill{rgb color={0.800,0.800,0.800}} (19.554,5.873)--(19.670,5.873)--(19.670,6.300)--(19.554,6.300)--cycle;
\draw[gp path] (19.554,5.873)--(19.554,6.299)--(19.669,6.299)--(19.669,5.873)--cycle;
\gpfill{rgb color={0.800,0.800,0.800}} (20.015,5.644)--(20.131,5.644)--(20.131,5.993)--(20.015,5.993)--cycle;
\draw[gp path] (20.015,5.644)--(20.015,5.992)--(20.130,5.992)--(20.130,5.644)--cycle;
\gpfill{rgb color={0.800,0.800,0.800}} (20.475,5.644)--(20.591,5.644)--(20.591,7.569)--(20.475,7.569)--cycle;
\draw[gp path] (20.475,5.644)--(20.475,7.568)--(20.590,7.568)--(20.590,5.644)--cycle;
\gpfill{rgb color={0.800,0.800,0.800}} (20.705,5.644)--(20.821,5.644)--(20.821,6.241)--(20.705,6.241)--cycle;
\draw[gp path] (20.705,5.644)--(20.705,6.240)--(20.820,6.240)--(20.820,5.644)--cycle;
\gpfill{rgb color={0.800,0.800,0.800}} (21.166,5.644)--(21.282,5.644)--(21.282,6.418)--(21.166,6.418)--cycle;
\draw[gp path] (21.166,5.644)--(21.166,6.417)--(21.281,6.417)--(21.281,5.644)--cycle;
\gpfill{rgb color={0.800,0.800,0.800}} (21.396,5.868)--(21.512,5.868)--(21.512,5.869)--(21.396,5.869)--cycle;
\draw[gp path] (21.396,5.868)--(21.511,5.868)--cycle;
\gpfill{rgb color={0.800,0.800,0.800}} (21.856,5.644)--(21.972,5.644)--(21.972,6.260)--(21.856,6.260)--cycle;
\draw[gp path] (21.856,5.644)--(21.856,6.259)--(21.971,6.259)--(21.971,5.644)--cycle;
\gpfill{rgb color={0.800,0.800,0.800}} (22.086,5.839)--(22.203,5.839)--(22.203,5.840)--(22.086,5.840)--cycle;
\draw[gp path] (22.086,5.839)--(22.202,5.839)--cycle;
\gpfill{rgb color={0.800,0.800,0.800}} (22.317,5.644)--(22.433,5.644)--(22.433,5.863)--(22.317,5.863)--cycle;
\draw[gp path] (22.317,5.644)--(22.317,5.862)--(22.432,5.862)--(22.432,5.644)--cycle;
\gpfill{rgb color={0.800,0.800,0.800}} (22.547,5.644)--(22.663,5.644)--(22.663,6.100)--(22.547,6.100)--cycle;
\draw[gp path] (22.547,5.644)--(22.547,6.099)--(22.662,6.099)--(22.662,5.644)--cycle;
\gpfill{rgb color={0.800,0.800,0.800}} (22.777,5.644)--(22.893,5.644)--(22.893,6.361)--(22.777,6.361)--cycle;
\draw[gp path] (22.777,5.644)--(22.777,6.360)--(22.892,6.360)--(22.892,5.644)--cycle;
\gpfill{rgb color={0.800,0.800,0.800}} (23.237,5.720)--(23.354,5.720)--(23.354,5.721)--(23.237,5.721)--cycle;
\draw[gp path] (23.237,5.720)--(23.353,5.720)--cycle;
\gpfill{rgb color={0.800,0.800,0.800}} (23.468,5.644)--(23.584,5.644)--(23.584,5.696)--(23.468,5.696)--cycle;
\draw[gp path] (23.468,5.644)--(23.468,5.695)--(23.583,5.695)--(23.583,5.644)--cycle;
\gpfill{rgb color={0.800,0.800,0.800}} (23.698,5.652)--(23.814,5.652)--(23.814,5.653)--(23.698,5.653)--cycle;
\draw[gp path] (23.698,5.652)--(23.813,5.652)--cycle;
\gpfill{rgb color={0.800,0.800,0.800}} (24.158,5.814)--(24.274,5.814)--(24.274,5.997)--(24.158,5.997)--cycle;
\draw[gp path] (24.158,5.814)--(24.158,5.996)--(24.273,5.996)--(24.273,5.814)--cycle;
\gpcolor{color=gp lt color border}
\draw[gp path] (1.196,10.131)--(1.196,5.644)--(24.446,5.644)--(24.446,10.131)--cycle;
\node[gp node center] at (12.821,9.682) {\plotLegend{(ARM)}};
\node[gp node left,font={\fontsize{11.0pt}{13.2pt}\selectfont}] at (1.429,9.817) {\plotLegend{mean improvement: $1.1\%$}};
\node[gp node left,font={\fontsize{11.0pt}{13.2pt}\selectfont}] at (1.429,9.368) {\plotLegend{improved functions: $41\%$}};
\node[gp node left,font={\fontsize{11.0pt}{13.2pt}\selectfont}] at (1.429,8.920) {\plotLegend{mean gap: $5\%$}};
\node[gp node left,font={\fontsize{11.0pt}{13.2pt}\selectfont}] at (1.429,8.471) {\plotLegend{optimal functions: $60\%$}};
\gpdefrectangularnode{gp plot 1}{\pgfpoint{1.196cm}{5.644cm}}{\pgfpoint{24.446cm}{10.131cm}}
\end{tikzpicture}

%% file: results/mips-speed-improvement.tex
\begin{tikzpicture}[gnuplot]
\path (0.000,0.000) rectangle (12.500,8.750);
\gpcolor{color=gp lt color border}
\gpsetlinetype{gp lt border}
\gpsetdashtype{gp dt solid}
\gpsetlinewidth{1.00}
\draw[gp path] (1.196,5.644)--(1.376,5.644);
\draw[gp path] (24.446,5.644)--(24.266,5.644);
\node[gp node right,font={\fontsize{8.0pt}{9.6pt}\selectfont}] at (1.012,5.644) {\plotPercentage{0}};
\draw[gp path] (1.196,6.093)--(1.376,6.093);
\draw[gp path] (24.446,6.093)--(24.266,6.093);
\node[gp node right,font={\fontsize{8.0pt}{9.6pt}\selectfont}] at (1.012,6.093) {\plotPercentage{10}};
\draw[gp path] (1.196,6.541)--(1.376,6.541);
\draw[gp path] (24.446,6.541)--(24.266,6.541);
\node[gp node right,font={\fontsize{8.0pt}{9.6pt}\selectfont}] at (1.012,6.541) {\plotPercentage{20}};
\draw[gp path] (1.196,6.990)--(1.376,6.990);
\draw[gp path] (24.446,6.990)--(24.266,6.990);
\node[gp node right,font={\fontsize{8.0pt}{9.6pt}\selectfont}] at (1.012,6.990) {\plotPercentage{30}};
\draw[gp path] (1.196,7.439)--(1.376,7.439);
\draw[gp path] (24.446,7.439)--(24.266,7.439);
\node[gp node right,font={\fontsize{8.0pt}{9.6pt}\selectfont}] at (1.012,7.439) {\plotPercentage{40}};
\draw[gp path] (1.196,7.888)--(1.376,7.888);
\draw[gp path] (24.446,7.888)--(24.266,7.888);
\node[gp node right,font={\fontsize{8.0pt}{9.6pt}\selectfont}] at (1.012,7.888) {\plotPercentage{50}};
\draw[gp path] (1.196,8.336)--(1.376,8.336);
\draw[gp path] (24.446,8.336)--(24.266,8.336);
\node[gp node right,font={\fontsize{8.0pt}{9.6pt}\selectfont}] at (1.012,8.336) {\plotPercentage{60}};
\draw[gp path] (1.196,8.785)--(1.376,8.785);
\draw[gp path] (24.446,8.785)--(24.266,8.785);
\node[gp node right,font={\fontsize{8.0pt}{9.6pt}\selectfont}] at (1.012,8.785) {\plotPercentage{70}};
\draw[gp path] (1.196,9.234)--(1.376,9.234);
\draw[gp path] (24.446,9.234)--(24.266,9.234);
\node[gp node right,font={\fontsize{8.0pt}{9.6pt}\selectfont}] at (1.012,9.234) {\plotPercentage{80}};
\draw[gp path] (1.196,9.682)--(1.376,9.682);
\draw[gp path] (24.446,9.682)--(24.266,9.682);
\node[gp node right,font={\fontsize{8.0pt}{9.6pt}\selectfont}] at (1.012,9.682) {\plotPercentage{90}};
\draw[gp path] (1.196,10.131)--(1.376,10.131);
\draw[gp path] (24.446,10.131)--(24.266,10.131);
\node[gp node right,font={\fontsize{8.0pt}{9.6pt}\selectfont}] at (1.012,10.131) {\plotPercentage{100}};
\node[gp node left,rotate=-90,font={\fontsize{7.0pt}{8.4pt}\selectfont}] at (1.426,5.721) {\functionId{1}\functionName{handle_noinline_attribute}};
\node[gp node left,rotate=-90,font={\fontsize{7.0pt}{8.4pt}\selectfont}] at (1.656,5.721) {\functionId{2}\functionName{control_flow_insn_p}};
\node[gp node left,rotate=-90,font={\fontsize{7.0pt}{8.4pt}\selectfont}] at (1.887,5.721) {\functionId{3}\functionName{insert_insn_on_edge}};
\node[gp node left,rotate=-90,font={\fontsize{7.0pt}{8.4pt}\selectfont}] at (2.117,5.721) {\functionId{4}\functionName{update_br_prob_note}};
\node[gp node left,rotate=-90,font={\fontsize{7.0pt}{8.4pt}\selectfont}] at (2.347,5.721) {\functionId{5}\functionName{_cpp_init_internal_pragma.}};
\node[gp node left,rotate=-90,font={\fontsize{7.0pt}{8.4pt}\selectfont}] at (2.577,5.721) {\functionId{6}\functionName{lex_macro_node}};
\node[gp node left,rotate=-90,font={\fontsize{7.0pt}{8.4pt}\selectfont}] at (2.807,5.721) {\functionId{7}\functionName{cse_basic_block}};
\node[gp node left,rotate=-90,font={\fontsize{7.0pt}{8.4pt}\selectfont}] at (3.038,5.721) {\functionId{8}\functionName{rtx_equal_for_cselib_p}};
\node[gp node left,rotate=-90,font={\fontsize{7.0pt}{8.4pt}\selectfont}] at (3.268,5.721) {\functionId{9}\functionName{debug_df_chain}};
\node[gp node left,rotate=-90,font={\fontsize{7.0pt}{8.4pt}\selectfont}] at (3.498,5.721) {\functionId{10}\functionName{modified_type_die}};
\node[gp node left,rotate=-90,font={\fontsize{7.0pt}{8.4pt}\selectfont}] at (3.728,5.721) {\functionId{11}\functionName{emit_note}};
\node[gp node left,rotate=-90,font={\fontsize{7.0pt}{8.4pt}\selectfont}] at (3.958,5.721) {\functionId{12}\functionName{gen_sequence}};
\node[gp node left,rotate=-90,font={\fontsize{7.0pt}{8.4pt}\selectfont}] at (4.189,5.721) {\functionId{13}\functionName{subreg_hard_regno}};
\node[gp node left,rotate=-90,font={\fontsize{7.0pt}{8.4pt}\selectfont}] at (4.419,5.721) {\functionId{14}\functionName{split_double}};
\node[gp node left,rotate=-90,font={\fontsize{7.0pt}{8.4pt}\selectfont}] at (4.649,5.721) {\functionId{15}\functionName{add_to_mem_set_list}};
\node[gp node left,rotate=-90,font={\fontsize{7.0pt}{8.4pt}\selectfont}] at (4.879,5.721) {\functionId{16}\functionName{find_regno_partial}};
\node[gp node left,rotate=-90,font={\fontsize{7.0pt}{8.4pt}\selectfont}] at (5.109,5.721) {\functionId{17}\functionName{use_return_register}};
\node[gp node left,rotate=-90,font={\fontsize{7.0pt}{8.4pt}\selectfont}] at (5.340,5.721) {\functionId{18}\functionName{ix86_expand_move}};
\node[gp node left,rotate=-90,font={\fontsize{7.0pt}{8.4pt}\selectfont}] at (5.570,5.721) {\functionId{19}\functionName{legitimate_pic_address_di.}};
\node[gp node left,rotate=-90,font={\fontsize{7.0pt}{8.4pt}\selectfont}] at (5.800,5.721) {\functionId{20}\functionName{gen_extendsfdf2}};
\node[gp node left,rotate=-90,font={\fontsize{7.0pt}{8.4pt}\selectfont}] at (6.030,5.721) {\functionId{21}\functionName{gen_mulsidi3}};
\node[gp node left,rotate=-90,font={\fontsize{7.0pt}{8.4pt}\selectfont}] at (6.260,5.721) {\functionId{22}\functionName{gen_peephole2_1255}};
\node[gp node left,rotate=-90,font={\fontsize{7.0pt}{8.4pt}\selectfont}] at (6.491,5.721) {\functionId{23}\functionName{gen_peephole2_1271}};
\node[gp node left,rotate=-90,font={\fontsize{7.0pt}{8.4pt}\selectfont}] at (6.721,5.721) {\functionId{24}\functionName{gen_peephole2_1277}};
\node[gp node left,rotate=-90,font={\fontsize{7.0pt}{8.4pt}\selectfont}] at (6.951,5.721) {\functionId{25}\functionName{gen_pfnacc}};
\node[gp node left,rotate=-90,font={\fontsize{7.0pt}{8.4pt}\selectfont}] at (7.181,5.721) {\functionId{26}\functionName{gen_rotlsi3}};
\node[gp node left,rotate=-90,font={\fontsize{7.0pt}{8.4pt}\selectfont}] at (7.411,5.721) {\functionId{27}\functionName{gen_split_1001}};
\node[gp node left,rotate=-90,font={\fontsize{7.0pt}{8.4pt}\selectfont}] at (7.642,5.721) {\functionId{28}\functionName{gen_split_1028}};
\node[gp node left,rotate=-90,font={\fontsize{7.0pt}{8.4pt}\selectfont}] at (7.872,5.721) {\functionId{29}\functionName{gen_sse_nandti3}};
\node[gp node left,rotate=-90,font={\fontsize{7.0pt}{8.4pt}\selectfont}] at (8.102,5.721) {\functionId{30}\functionName{gen_sunge}};
\node[gp node left,rotate=-90,font={\fontsize{7.0pt}{8.4pt}\selectfont}] at (8.332,5.721) {\functionId{31}\functionName{insert_loop_mem}};
\node[gp node left,rotate=-90,font={\fontsize{7.0pt}{8.4pt}\selectfont}] at (8.562,5.721) {\functionId{32}\functionName{eiremain}};
\node[gp node left,rotate=-90,font={\fontsize{7.0pt}{8.4pt}\selectfont}] at (8.793,5.721) {\functionId{33}\functionName{elimination_effects}};
\node[gp node left,rotate=-90,font={\fontsize{7.0pt}{8.4pt}\selectfont}] at (9.023,5.721) {\functionId{34}\functionName{gen_reload}};
\node[gp node left,rotate=-90,font={\fontsize{7.0pt}{8.4pt}\selectfont}] at (9.253,5.721) {\functionId{35}\functionName{reload_cse_simplify_set}};
\node[gp node left,rotate=-90,font={\fontsize{7.0pt}{8.4pt}\selectfont}] at (9.483,5.721) {\functionId{36}\functionName{simplify_binary_is2orm1}};
\node[gp node left,rotate=-90,font={\fontsize{7.0pt}{8.4pt}\selectfont}] at (9.713,5.721) {\functionId{37}\functionName{remove_phi_alternative}};
\node[gp node left,rotate=-90,font={\fontsize{7.0pt}{8.4pt}\selectfont}] at (9.944,5.721) {\functionId{38}\functionName{contains_placeholder_p}};
\node[gp node left,rotate=-90,font={\fontsize{7.0pt}{8.4pt}\selectfont}] at (10.174,5.721) {\functionId{39}\functionName{assemble_end_function}};
\node[gp node left,rotate=-90,font={\fontsize{7.0pt}{8.4pt}\selectfont}] at (10.404,5.721) {\functionId{40}\functionName{default_named_section_asm.}};
\node[gp node left,rotate=-90,font={\fontsize{7.0pt}{8.4pt}\selectfont}] at (10.634,5.721) {\functionId{41}\functionName{sample_unpack_12}};
\node[gp node left,rotate=-90,font={\fontsize{7.0pt}{8.4pt}\selectfont}] at (10.864,5.721) {\functionId{42}\functionName{autohelperattpat10}};
\node[gp node left,rotate=-90,font={\fontsize{7.0pt}{8.4pt}\selectfont}] at (11.095,5.721) {\functionId{43}\functionName{autohelperbarrierspat126}};
\node[gp node left,rotate=-90,font={\fontsize{7.0pt}{8.4pt}\selectfont}] at (11.325,5.721) {\functionId{44}\functionName{atari_atari_attack_callba.}};
\node[gp node left,rotate=-90,font={\fontsize{7.0pt}{8.4pt}\selectfont}] at (11.555,5.721) {\functionId{45}\functionName{compute_aa_status}};
\node[gp node left,rotate=-90,font={\fontsize{7.0pt}{8.4pt}\selectfont}] at (11.785,5.721) {\functionId{46}\functionName{dragon_weak}};
\node[gp node left,rotate=-90,font={\fontsize{7.0pt}{8.4pt}\selectfont}] at (12.015,5.721) {\functionId{47}\functionName{get_saved_worms}};
\node[gp node left,rotate=-90,font={\fontsize{7.0pt}{8.4pt}\selectfont}] at (12.246,5.721) {\functionId{48}\functionName{read_eye}};
\node[gp node left,rotate=-90,font={\fontsize{7.0pt}{8.4pt}\selectfont}] at (12.476,5.721) {\functionId{49}\functionName{topological_eye}};
\node[gp node left,rotate=-90,font={\fontsize{7.0pt}{8.4pt}\selectfont}] at (12.706,5.721) {\functionId{50}\functionName{autohelperowl_attackpat19.}};
\node[gp node left,rotate=-90,font={\fontsize{7.0pt}{8.4pt}\selectfont}] at (12.936,5.721) {\functionId{51}\functionName{autohelperowl_attackpat29.}};
\node[gp node left,rotate=-90,font={\fontsize{7.0pt}{8.4pt}\selectfont}] at (13.166,5.721) {\functionId{52}\functionName{autohelperowl_defendpat28.}};
\node[gp node left,rotate=-90,font={\fontsize{7.0pt}{8.4pt}\selectfont}] at (13.396,5.721) {\functionId{53}\functionName{autohelperowl_defendpat38.}};
\node[gp node left,rotate=-90,font={\fontsize{7.0pt}{8.4pt}\selectfont}] at (13.627,5.721) {\functionId{54}\functionName{autohelperpat1114}};
\node[gp node left,rotate=-90,font={\fontsize{7.0pt}{8.4pt}\selectfont}] at (13.857,5.721) {\functionId{55}\functionName{autohelperpat335}};
\node[gp node left,rotate=-90,font={\fontsize{7.0pt}{8.4pt}\selectfont}] at (14.087,5.721) {\functionId{56}\functionName{autohelperpat508}};
\node[gp node left,rotate=-90,font={\fontsize{7.0pt}{8.4pt}\selectfont}] at (14.317,5.721) {\functionId{57}\functionName{autohelperpat83}};
\node[gp node left,rotate=-90,font={\fontsize{7.0pt}{8.4pt}\selectfont}] at (14.547,5.721) {\functionId{58}\functionName{simple_showboard}};
\node[gp node left,rotate=-90,font={\fontsize{7.0pt}{8.4pt}\selectfont}] at (14.778,5.721) {\functionId{59}\functionName{skip_intrabk_SAD}};
\node[gp node left,rotate=-90,font={\fontsize{7.0pt}{8.4pt}\selectfont}] at (15.008,5.721) {\functionId{60}\functionName{free_orig_planes}};
\node[gp node left,rotate=-90,font={\fontsize{7.0pt}{8.4pt}\selectfont}] at (15.238,5.721) {\functionId{61}\functionName{GetSkipCostMB}};
\node[gp node left,rotate=-90,font={\fontsize{7.0pt}{8.4pt}\selectfont}] at (15.468,5.721) {\functionId{62}\functionName{writeSyntaxElement_Level_.}};
\node[gp node left,rotate=-90,font={\fontsize{7.0pt}{8.4pt}\selectfont}] at (15.698,5.721) {\functionId{63}\functionName{GSIAddKeyToIndex}};
\node[gp node left,rotate=-90,font={\fontsize{7.0pt}{8.4pt}\selectfont}] at (15.929,5.721) {\functionId{64}\functionName{EVDBasicFit}};
\node[gp node left,rotate=-90,font={\fontsize{7.0pt}{8.4pt}\selectfont}] at (16.159,5.721) {\functionId{65}\functionName{SampleDirichlet}};
\node[gp node left,rotate=-90,font={\fontsize{7.0pt}{8.4pt}\selectfont}] at (16.389,5.721) {\functionId{66}\functionName{DegenerateSymbolScore}};
\node[gp node left,rotate=-90,font={\fontsize{7.0pt}{8.4pt}\selectfont}] at (16.619,5.721) {\functionId{67}\functionName{Plan7SetCtime}};
\node[gp node left,rotate=-90,font={\fontsize{7.0pt}{8.4pt}\selectfont}] at (16.849,5.721) {\functionId{68}\functionName{MSAToSqinfo}};
\node[gp node left,rotate=-90,font={\fontsize{7.0pt}{8.4pt}\selectfont}] at (17.080,5.721) {\functionId{69}\functionName{null_convert}};
\node[gp node left,rotate=-90,font={\fontsize{7.0pt}{8.4pt}\selectfont}] at (17.310,5.721) {\functionId{70}\functionName{jinit_c_prep_controller}};
\node[gp node left,rotate=-90,font={\fontsize{7.0pt}{8.4pt}\selectfont}] at (17.540,5.721) {\functionId{71}\functionName{glFogf}};
\node[gp node left,rotate=-90,font={\fontsize{7.0pt}{8.4pt}\selectfont}] at (17.770,5.721) {\functionId{72}\functionName{glNormal3d}};
\node[gp node left,rotate=-90,font={\fontsize{7.0pt}{8.4pt}\selectfont}] at (18.000,5.721) {\functionId{73}\functionName{glRasterPos3d}};
\node[gp node left,rotate=-90,font={\fontsize{7.0pt}{8.4pt}\selectfont}] at (18.231,5.721) {\functionId{74}\functionName{glTexCoord2d}};
\node[gp node left,rotate=-90,font={\fontsize{7.0pt}{8.4pt}\selectfont}] at (18.461,5.721) {\functionId{75}\functionName{gl_stippled_bresenham}};
\node[gp node left,rotate=-90,font={\fontsize{7.0pt}{8.4pt}\selectfont}] at (18.691,5.721) {\functionId{76}\functionName{gl_save_Frustum}};
\node[gp node left,rotate=-90,font={\fontsize{7.0pt}{8.4pt}\selectfont}] at (18.921,5.721) {\functionId{77}\functionName{gl_save_LineWidth}};
\node[gp node left,rotate=-90,font={\fontsize{7.0pt}{8.4pt}\selectfont}] at (19.151,5.721) {\functionId{78}\functionName{translate_id}};
\node[gp node left,rotate=-90,font={\fontsize{7.0pt}{8.4pt}\selectfont}] at (19.382,5.721) {\functionId{79}\functionName{gl_Map1f}};
\node[gp node left,rotate=-90,font={\fontsize{7.0pt}{8.4pt}\selectfont}] at (19.612,5.721) {\functionId{80}\functionName{smooth_ci_line}};
\node[gp node left,rotate=-90,font={\fontsize{7.0pt}{8.4pt}\selectfont}] at (19.842,5.721) {\functionId{81}\functionName{free_unified_knots}};
\node[gp node left,rotate=-90,font={\fontsize{7.0pt}{8.4pt}\selectfont}] at (20.072,5.721) {\functionId{82}\functionName{tess_test_polygon}};
\node[gp node left,rotate=-90,font={\fontsize{7.0pt}{8.4pt}\selectfont}] at (20.302,5.721) {\functionId{83}\functionName{auxWireBox}};
\node[gp node left,rotate=-90,font={\fontsize{7.0pt}{8.4pt}\selectfont}] at (20.533,5.721) {\functionId{84}\functionName{gl_ColorPointer}};
\node[gp node left,rotate=-90,font={\fontsize{7.0pt}{8.4pt}\selectfont}] at (20.763,5.721) {\functionId{85}\functionName{r_serial}};
\node[gp node left,rotate=-90,font={\fontsize{7.0pt}{8.4pt}\selectfont}] at (20.993,5.721) {\functionId{86}\functionName{scalar_mult_sub_su3_matri.}};
\node[gp node left,rotate=-90,font={\fontsize{7.0pt}{8.4pt}\selectfont}] at (21.223,5.721) {\functionId{87}\functionName{Decode_MPEG1_Non_Intra_Bl.}};
\node[gp node left,rotate=-90,font={\fontsize{7.0pt}{8.4pt}\selectfont}] at (21.453,5.721) {\functionId{88}\functionName{cpDecodeSecret}};
\node[gp node left,rotate=-90,font={\fontsize{7.0pt}{8.4pt}\selectfont}] at (21.684,5.721) {\functionId{89}\functionName{vlShortLshift}};
\node[gp node left,rotate=-90,font={\fontsize{7.0pt}{8.4pt}\selectfont}] at (21.914,5.721) {\functionId{90}\functionName{encryptfile}};
\node[gp node left,rotate=-90,font={\fontsize{7.0pt}{8.4pt}\selectfont}] at (22.144,5.721) {\functionId{91}\functionName{make_canonical}};
\node[gp node left,rotate=-90,font={\fontsize{7.0pt}{8.4pt}\selectfont}] at (22.374,5.721) {\functionId{92}\functionName{LANG}};
\node[gp node left,rotate=-90,font={\fontsize{7.0pt}{8.4pt}\selectfont}] at (22.604,5.721) {\functionId{93}\functionName{MD5Transform}};
\node[gp node left,rotate=-90,font={\fontsize{7.0pt}{8.4pt}\selectfont}] at (22.835,5.721) {\functionId{94}\functionName{mp_display}};
\node[gp node left,rotate=-90,font={\fontsize{7.0pt}{8.4pt}\selectfont}] at (23.065,5.721) {\functionId{95}\functionName{comp_Jboundaries}};
\node[gp node left,rotate=-90,font={\fontsize{7.0pt}{8.4pt}\selectfont}] at (23.295,5.721) {\functionId{96}\functionName{is_draw}};
\node[gp node left,rotate=-90,font={\fontsize{7.0pt}{8.4pt}\selectfont}] at (23.525,5.721) {\functionId{97}\functionName{push_king}};
\node[gp node left,rotate=-90,font={\fontsize{7.0pt}{8.4pt}\selectfont}] at (23.755,5.721) {\functionId{98}\functionName{stat_retry}};
\node[gp node left,rotate=-90,font={\fontsize{7.0pt}{8.4pt}\selectfont}] at (23.986,5.721) {\functionId{99}\functionName{lextree_subtree_print}};
\node[gp node left,rotate=-90,font={\fontsize{7.0pt}{8.4pt}\selectfont}] at (24.216,5.721) {\functionId{100}\functionName{lm_tg_score}};
\draw[gp path] (1.196,10.131)--(1.196,5.644)--(24.446,5.644)--(24.446,10.131)--cycle;
\gpcolor{rgb color={0.580,0.000,0.827}}
\draw[gp path] (1.196,5.644)--(1.431,5.644)--(1.666,5.644)--(1.901,5.644)--(2.135,5.644)%
  --(2.370,5.644)--(2.605,5.644)--(2.840,5.644)--(3.075,5.644)--(3.310,5.644)--(3.544,5.644)%
  --(3.779,5.644)--(4.014,5.644)--(4.249,5.644)--(4.484,5.644)--(4.719,5.644)--(4.954,5.644)%
  --(5.188,5.644)--(5.423,5.644)--(5.658,5.644)--(5.893,5.644)--(6.128,5.644)--(6.363,5.644)%
  --(6.598,5.644)--(6.832,5.644)--(7.067,5.644)--(7.302,5.644)--(7.537,5.644)--(7.772,5.644)%
  --(8.007,5.644)--(8.241,5.644)--(8.476,5.644)--(8.711,5.644)--(8.946,5.644)--(9.181,5.644)%
  --(9.416,5.644)--(9.651,5.644)--(9.885,5.644)--(10.120,5.644)--(10.355,5.644)--(10.590,5.644)%
  --(10.825,5.644)--(11.060,5.644)--(11.294,5.644)--(11.529,5.644)--(11.764,5.644)--(11.999,5.644)%
  --(12.234,5.644)--(12.469,5.644)--(12.704,5.644)--(12.938,5.644)--(13.173,5.644)--(13.408,5.644)%
  --(13.643,5.644)--(13.878,5.644)--(14.113,5.644)--(14.348,5.644)--(14.582,5.644)--(14.817,5.644)%
  --(15.052,5.644)--(15.287,5.644)--(15.522,5.644)--(15.757,5.644)--(15.991,5.644)--(16.226,5.644)%
  --(16.461,5.644)--(16.696,5.644)--(16.931,5.644)--(17.166,5.644)--(17.401,5.644)--(17.635,5.644)%
  --(17.870,5.644)--(18.105,5.644)--(18.340,5.644)--(18.575,5.644)--(18.810,5.644)--(19.044,5.644)%
  --(19.279,5.644)--(19.514,5.644)--(19.749,5.644)--(19.984,5.644)--(20.219,5.644)--(20.454,5.644)%
  --(20.688,5.644)--(20.923,5.644)--(21.158,5.644)--(21.393,5.644)--(21.628,5.644)--(21.863,5.644)%
  --(22.098,5.644)--(22.332,5.644)--(22.567,5.644)--(22.802,5.644)--(23.037,5.644)--(23.272,5.644)%
  --(23.507,5.644)--(23.741,5.644)--(23.976,5.644)--(24.211,5.644)--(24.446,5.644);
\gpfill{rgb color={0.000,0.000,0.000}} (1.369,5.644)--(1.485,5.644)--(1.485,6.047)--(1.369,6.047)--cycle;
\gpcolor{rgb color={0.000,0.000,0.000}}
\draw[gp path] (1.369,5.644)--(1.369,6.046)--(1.484,6.046)--(1.484,5.644)--cycle;
\gpfill{rgb color={0.000,0.000,0.000}} (1.599,5.644)--(1.715,5.644)--(1.715,7.469)--(1.599,7.469)--cycle;
\draw[gp path] (1.599,5.644)--(1.599,7.468)--(1.714,7.468)--(1.714,5.644)--cycle;
\gpfill{rgb color={0.000,0.000,0.000}} (1.829,5.644)--(1.945,5.644)--(1.945,5.980)--(1.829,5.980)--cycle;
\draw[gp path] (1.829,5.644)--(1.829,5.979)--(1.944,5.979)--(1.944,5.644)--cycle;
\gpfill{rgb color={0.000,0.000,0.000}} (2.059,5.644)--(2.175,5.644)--(2.175,7.595)--(2.059,7.595)--cycle;
\draw[gp path] (2.059,5.644)--(2.059,7.594)--(2.174,7.594)--(2.174,5.644)--cycle;
\gpfill{rgb color={0.000,0.000,0.000}} (2.520,5.644)--(2.636,5.644)--(2.636,6.077)--(2.520,6.077)--cycle;
\draw[gp path] (2.520,5.644)--(2.520,6.076)--(2.635,6.076)--(2.635,5.644)--cycle;
\gpfill{rgb color={0.000,0.000,0.000}} (3.671,5.644)--(3.787,5.644)--(3.787,6.294)--(3.671,6.294)--cycle;
\draw[gp path] (3.671,5.644)--(3.671,6.293)--(3.786,6.293)--(3.786,5.644)--cycle;
\gpfill{rgb color={0.000,0.000,0.000}} (3.901,5.644)--(4.017,5.644)--(4.017,5.679)--(3.901,5.679)--cycle;
\draw[gp path] (3.901,5.644)--(3.901,5.678)--(4.016,5.678)--(4.016,5.644)--cycle;
\gpfill{rgb color={0.000,0.000,0.000}} (4.131,5.644)--(4.247,5.644)--(4.247,6.078)--(4.131,6.078)--cycle;
\draw[gp path] (4.131,5.644)--(4.131,6.077)--(4.246,6.077)--(4.246,5.644)--cycle;
\gpfill{rgb color={0.000,0.000,0.000}} (4.361,5.644)--(4.477,5.644)--(4.477,6.528)--(4.361,6.528)--cycle;
\draw[gp path] (4.361,5.644)--(4.361,6.527)--(4.476,6.527)--(4.476,5.644)--cycle;
\gpfill{rgb color={0.000,0.000,0.000}} (4.591,5.644)--(4.708,5.644)--(4.708,5.867)--(4.591,5.867)--cycle;
\draw[gp path] (4.591,5.644)--(4.591,5.866)--(4.707,5.866)--(4.707,5.644)--cycle;
\gpfill{rgb color={0.000,0.000,0.000}} (5.512,5.644)--(5.628,5.644)--(5.628,8.099)--(5.512,8.099)--cycle;
\draw[gp path] (5.512,5.644)--(5.512,8.098)--(5.627,8.098)--(5.627,5.644)--cycle;
\gpfill{rgb color={0.000,0.000,0.000}} (7.124,5.644)--(7.240,5.644)--(7.240,5.920)--(7.124,5.920)--cycle;
\draw[gp path] (7.124,5.644)--(7.124,5.919)--(7.239,5.919)--(7.239,5.644)--cycle;
\gpfill{rgb color={0.000,0.000,0.000}} (7.814,5.644)--(7.930,5.644)--(7.930,5.849)--(7.814,5.849)--cycle;
\draw[gp path] (7.814,5.644)--(7.814,5.848)--(7.929,5.848)--(7.929,5.644)--cycle;
\gpfill{rgb color={0.000,0.000,0.000}} (8.044,5.644)--(8.160,5.644)--(8.160,6.029)--(8.044,6.029)--cycle;
\draw[gp path] (8.044,5.644)--(8.044,6.028)--(8.159,6.028)--(8.159,5.644)--cycle;
\gpfill{rgb color={0.000,0.000,0.000}} (9.426,5.644)--(9.542,5.644)--(9.542,5.916)--(9.426,5.916)--cycle;
\draw[gp path] (9.426,5.644)--(9.426,5.915)--(9.541,5.915)--(9.541,5.644)--cycle;
\gpfill{rgb color={0.000,0.000,0.000}} (9.886,5.644)--(10.002,5.644)--(10.002,6.048)--(9.886,6.048)--cycle;
\draw[gp path] (9.886,5.644)--(9.886,6.047)--(10.001,6.047)--(10.001,5.644)--cycle;
\gpfill{rgb color={0.000,0.000,0.000}} (10.577,5.644)--(10.693,5.644)--(10.693,5.819)--(10.577,5.819)--cycle;
\draw[gp path] (10.577,5.644)--(10.577,5.818)--(10.692,5.818)--(10.692,5.644)--cycle;
\gpfill{rgb color={0.000,0.000,0.000}} (10.807,5.644)--(10.923,5.644)--(10.923,7.004)--(10.807,7.004)--cycle;
\draw[gp path] (10.807,5.644)--(10.807,7.003)--(10.922,7.003)--(10.922,5.644)--cycle;
\gpfill{rgb color={0.000,0.000,0.000}} (11.037,5.644)--(11.153,5.644)--(11.153,6.337)--(11.037,6.337)--cycle;
\draw[gp path] (11.037,5.644)--(11.037,6.336)--(11.152,6.336)--(11.152,5.644)--cycle;
\gpfill{rgb color={0.000,0.000,0.000}} (11.728,5.644)--(11.844,5.644)--(11.844,6.331)--(11.728,6.331)--cycle;
\draw[gp path] (11.728,5.644)--(11.728,6.330)--(11.843,6.330)--(11.843,5.644)--cycle;
\gpfill{rgb color={0.000,0.000,0.000}} (12.879,5.644)--(12.995,5.644)--(12.995,5.816)--(12.879,5.816)--cycle;
\draw[gp path] (12.879,5.644)--(12.879,5.815)--(12.994,5.815)--(12.994,5.644)--cycle;
\gpfill{rgb color={0.000,0.000,0.000}} (13.109,5.644)--(13.225,5.644)--(13.225,6.134)--(13.109,6.134)--cycle;
\draw[gp path] (13.109,5.644)--(13.109,6.133)--(13.224,6.133)--(13.224,5.644)--cycle;
\gpfill{rgb color={0.000,0.000,0.000}} (13.339,5.644)--(13.455,5.644)--(13.455,6.799)--(13.339,6.799)--cycle;
\draw[gp path] (13.339,5.644)--(13.339,6.798)--(13.454,6.798)--(13.454,5.644)--cycle;
\gpfill{rgb color={0.000,0.000,0.000}} (13.569,5.644)--(13.685,5.644)--(13.685,5.761)--(13.569,5.761)--cycle;
\draw[gp path] (13.569,5.644)--(13.569,5.760)--(13.684,5.760)--(13.684,5.644)--cycle;
\gpfill{rgb color={0.000,0.000,0.000}} (13.799,5.644)--(13.915,5.644)--(13.915,5.777)--(13.799,5.777)--cycle;
\draw[gp path] (13.799,5.644)--(13.799,5.776)--(13.914,5.776)--(13.914,5.644)--cycle;
\gpfill{rgb color={0.000,0.000,0.000}} (14.030,5.644)--(14.146,5.644)--(14.146,5.785)--(14.030,5.785)--cycle;
\draw[gp path] (14.030,5.644)--(14.030,5.784)--(14.145,5.784)--(14.145,5.644)--cycle;
\gpfill{rgb color={0.000,0.000,0.000}} (14.720,5.644)--(14.836,5.644)--(14.836,5.766)--(14.720,5.766)--cycle;
\draw[gp path] (14.720,5.644)--(14.720,5.765)--(14.835,5.765)--(14.835,5.644)--cycle;
\gpfill{rgb color={0.000,0.000,0.000}} (15.411,5.644)--(15.527,5.644)--(15.527,5.778)--(15.411,5.778)--cycle;
\draw[gp path] (15.411,5.644)--(15.411,5.777)--(15.526,5.777)--(15.526,5.644)--cycle;
\gpfill{rgb color={0.000,0.000,0.000}} (15.871,5.644)--(15.987,5.644)--(15.987,5.905)--(15.871,5.905)--cycle;
\draw[gp path] (15.871,5.644)--(15.871,5.904)--(15.986,5.904)--(15.986,5.644)--cycle;
\gpfill{rgb color={0.000,0.000,0.000}} (16.101,5.644)--(16.217,5.644)--(16.217,5.672)--(16.101,5.672)--cycle;
\draw[gp path] (16.101,5.644)--(16.101,5.671)--(16.216,5.671)--(16.216,5.644)--cycle;
\gpfill{rgb color={0.000,0.000,0.000}} (17.022,5.644)--(17.138,5.644)--(17.138,5.646)--(17.022,5.646)--cycle;
\draw[gp path] (17.022,5.644)--(17.022,5.645)--(17.137,5.645)--(17.137,5.644)--cycle;
\gpfill{rgb color={0.000,0.000,0.000}} (17.483,5.644)--(17.599,5.644)--(17.599,6.739)--(17.483,6.739)--cycle;
\draw[gp path] (17.483,5.644)--(17.483,6.738)--(17.598,6.738)--(17.598,5.644)--cycle;
\gpfill{rgb color={0.000,0.000,0.000}} (17.713,5.644)--(17.829,5.644)--(17.829,5.768)--(17.713,5.768)--cycle;
\draw[gp path] (17.713,5.644)--(17.713,5.767)--(17.828,5.767)--(17.828,5.644)--cycle;
\gpfill{rgb color={0.000,0.000,0.000}} (17.943,5.644)--(18.059,5.644)--(18.059,6.412)--(17.943,6.412)--cycle;
\draw[gp path] (17.943,5.644)--(17.943,6.411)--(18.058,6.411)--(18.058,5.644)--cycle;
\gpfill{rgb color={0.000,0.000,0.000}} (18.403,5.644)--(18.519,5.644)--(18.519,5.948)--(18.403,5.948)--cycle;
\draw[gp path] (18.403,5.644)--(18.403,5.947)--(18.518,5.947)--(18.518,5.644)--cycle;
\gpfill{rgb color={0.000,0.000,0.000}} (18.634,5.644)--(18.750,5.644)--(18.750,6.589)--(18.634,6.589)--cycle;
\draw[gp path] (18.634,5.644)--(18.634,6.588)--(18.749,6.588)--(18.749,5.644)--cycle;
\gpfill{rgb color={0.000,0.000,0.000}} (18.864,5.644)--(18.980,5.644)--(18.980,6.471)--(18.864,6.471)--cycle;
\draw[gp path] (18.864,5.644)--(18.864,6.470)--(18.979,6.470)--(18.979,5.644)--cycle;
\gpfill{rgb color={0.000,0.000,0.000}} (19.324,5.644)--(19.440,5.644)--(19.440,9.328)--(19.324,9.328)--cycle;
\draw[gp path] (19.324,5.644)--(19.324,9.327)--(19.439,9.327)--(19.439,5.644)--cycle;
\gpfill{rgb color={0.000,0.000,0.000}} (19.554,5.644)--(19.670,5.644)--(19.670,5.692)--(19.554,5.692)--cycle;
\draw[gp path] (19.554,5.644)--(19.554,5.691)--(19.669,5.691)--(19.669,5.644)--cycle;
\gpfill{rgb color={0.000,0.000,0.000}} (19.784,5.644)--(19.901,5.644)--(19.901,6.039)--(19.784,6.039)--cycle;
\draw[gp path] (19.784,5.644)--(19.784,6.038)--(19.900,6.038)--(19.900,5.644)--cycle;
\gpfill{rgb color={0.000,0.000,0.000}} (20.475,5.644)--(20.591,5.644)--(20.591,5.840)--(20.475,5.840)--cycle;
\draw[gp path] (20.475,5.644)--(20.475,5.839)--(20.590,5.839)--(20.590,5.644)--cycle;
\gpfill{rgb color={0.000,0.000,0.000}} (20.935,5.644)--(21.052,5.644)--(21.052,5.684)--(20.935,5.684)--cycle;
\draw[gp path] (20.935,5.644)--(20.935,5.683)--(21.051,5.683)--(21.051,5.644)--cycle;
\gpfill{rgb color={0.000,0.000,0.000}} (21.396,5.644)--(21.512,5.644)--(21.512,6.767)--(21.396,6.767)--cycle;
\draw[gp path] (21.396,5.644)--(21.396,6.766)--(21.511,6.766)--(21.511,5.644)--cycle;
\gpfill{rgb color={0.000,0.000,0.000}} (21.626,5.644)--(21.742,5.644)--(21.742,6.054)--(21.626,6.054)--cycle;
\draw[gp path] (21.626,5.644)--(21.626,6.053)--(21.741,6.053)--(21.741,5.644)--cycle;
\gpfill{rgb color={0.000,0.000,0.000}} (23.007,5.644)--(23.123,5.644)--(23.123,5.651)--(23.007,5.651)--cycle;
\draw[gp path] (23.007,5.644)--(23.007,5.650)--(23.122,5.650)--(23.122,5.644)--cycle;
\gpfill{rgb color={0.000,0.000,0.000}} (23.237,5.644)--(23.354,5.644)--(23.354,5.875)--(23.237,5.875)--cycle;
\draw[gp path] (23.237,5.644)--(23.237,5.874)--(23.353,5.874)--(23.353,5.644)--cycle;
\gpfill{rgb color={0.000,0.000,0.000}} (24.158,5.644)--(24.274,5.644)--(24.274,5.655)--(24.158,5.655)--cycle;
\draw[gp path] (24.158,5.644)--(24.158,5.654)--(24.273,5.654)--(24.273,5.644)--cycle;
\gpfill{rgb color={0.800,0.800,0.800}} (1.369,6.046)--(1.485,6.046)--(1.485,6.047)--(1.369,6.047)--cycle;
\gpcolor{rgb color={0.800,0.800,0.800}}
\draw[gp path] (1.369,6.046)--(1.484,6.046)--cycle;
\gpfill{rgb color={0.800,0.800,0.800}} (1.599,7.468)--(1.715,7.468)--(1.715,7.469)--(1.599,7.469)--cycle;
\draw[gp path] (1.599,7.468)--(1.714,7.468)--cycle;
\gpfill{rgb color={0.800,0.800,0.800}} (1.829,5.979)--(1.945,5.979)--(1.945,5.980)--(1.829,5.980)--cycle;
\draw[gp path] (1.829,5.979)--(1.944,5.979)--cycle;
\gpfill{rgb color={0.800,0.800,0.800}} (2.059,7.594)--(2.175,7.594)--(2.175,7.595)--(2.059,7.595)--cycle;
\draw[gp path] (2.059,7.594)--(2.174,7.594)--cycle;
\gpfill{rgb color={0.800,0.800,0.800}} (2.289,5.644)--(2.406,5.644)--(2.406,7.283)--(2.289,7.283)--cycle;
\draw[gp path] (2.289,5.644)--(2.289,7.282)--(2.405,7.282)--(2.405,5.644)--cycle;
\gpfill{rgb color={0.800,0.800,0.800}} (2.520,6.076)--(2.636,6.076)--(2.636,6.077)--(2.520,6.077)--cycle;
\draw[gp path] (2.520,6.076)--(2.635,6.076)--cycle;
\gpfill{rgb color={0.800,0.800,0.800}} (2.750,5.644)--(2.866,5.644)--(2.866,6.599)--(2.750,6.599)--cycle;
\draw[gp path] (2.750,5.644)--(2.750,6.598)--(2.865,6.598)--(2.865,5.644)--cycle;
\gpfill{rgb color={0.800,0.800,0.800}} (2.980,5.644)--(3.096,5.644)--(3.096,7.133)--(2.980,7.133)--cycle;
\draw[gp path] (2.980,5.644)--(2.980,7.132)--(3.095,7.132)--(3.095,5.644)--cycle;
\gpfill{rgb color={0.800,0.800,0.800}} (3.210,5.644)--(3.326,5.644)--(3.326,7.216)--(3.210,7.216)--cycle;
\draw[gp path] (3.210,5.644)--(3.210,7.215)--(3.325,7.215)--(3.325,5.644)--cycle;
\gpfill{rgb color={0.800,0.800,0.800}} (3.440,5.644)--(3.557,5.644)--(3.557,10.132)--(3.440,10.132)--cycle;
\draw[gp path] (3.440,5.644)--(3.440,10.131)--(3.556,10.131)--(3.556,5.644)--cycle;
\gpfill{rgb color={0.800,0.800,0.800}} (3.671,6.293)--(3.787,6.293)--(3.787,7.598)--(3.671,7.598)--cycle;
\draw[gp path] (3.671,6.293)--(3.671,7.597)--(3.786,7.597)--(3.786,6.293)--cycle;
\gpfill{rgb color={0.800,0.800,0.800}} (3.901,5.678)--(4.017,5.678)--(4.017,5.679)--(3.901,5.679)--cycle;
\draw[gp path] (3.901,5.678)--(4.016,5.678)--cycle;
\gpfill{rgb color={0.800,0.800,0.800}} (4.131,6.077)--(4.247,6.077)--(4.247,6.078)--(4.131,6.078)--cycle;
\draw[gp path] (4.131,6.077)--(4.246,6.077)--cycle;
\gpfill{rgb color={0.800,0.800,0.800}} (4.361,6.527)--(4.477,6.527)--(4.477,8.535)--(4.361,8.535)--cycle;
\draw[gp path] (4.361,6.527)--(4.361,8.534)--(4.476,8.534)--(4.476,6.527)--cycle;
\gpfill{rgb color={0.800,0.800,0.800}} (4.591,5.866)--(4.708,5.866)--(4.708,5.867)--(4.591,5.867)--cycle;
\draw[gp path] (4.591,5.866)--(4.707,5.866)--cycle;
\gpfill{rgb color={0.800,0.800,0.800}} (5.052,5.644)--(5.168,5.644)--(5.168,6.654)--(5.052,6.654)--cycle;
\draw[gp path] (5.052,5.644)--(5.052,6.653)--(5.167,6.653)--(5.167,5.644)--cycle;
\gpfill{rgb color={0.800,0.800,0.800}} (5.282,5.644)--(5.398,5.644)--(5.398,6.769)--(5.282,6.769)--cycle;
\draw[gp path] (5.282,5.644)--(5.282,6.768)--(5.397,6.768)--(5.397,5.644)--cycle;
\gpfill{rgb color={0.800,0.800,0.800}} (5.512,8.098)--(5.628,8.098)--(5.628,8.407)--(5.512,8.407)--cycle;
\draw[gp path] (5.512,8.098)--(5.512,8.406)--(5.627,8.406)--(5.627,8.098)--cycle;
\gpfill{rgb color={0.800,0.800,0.800}} (5.742,5.644)--(5.859,5.644)--(5.859,7.047)--(5.742,7.047)--cycle;
\draw[gp path] (5.742,5.644)--(5.742,7.046)--(5.858,7.046)--(5.858,5.644)--cycle;
\gpfill{rgb color={0.800,0.800,0.800}} (5.973,5.644)--(6.089,5.644)--(6.089,6.783)--(5.973,6.783)--cycle;
\draw[gp path] (5.973,5.644)--(5.973,6.782)--(6.088,6.782)--(6.088,5.644)--cycle;
\gpfill{rgb color={0.800,0.800,0.800}} (6.203,5.644)--(6.319,5.644)--(6.319,6.062)--(6.203,6.062)--cycle;
\draw[gp path] (6.203,5.644)--(6.203,6.061)--(6.318,6.061)--(6.318,5.644)--cycle;
\gpfill{rgb color={0.800,0.800,0.800}} (6.433,5.644)--(6.549,5.644)--(6.549,6.465)--(6.433,6.465)--cycle;
\draw[gp path] (6.433,5.644)--(6.433,6.464)--(6.548,6.464)--(6.548,5.644)--cycle;
\gpfill{rgb color={0.800,0.800,0.800}} (6.663,5.644)--(6.779,5.644)--(6.779,6.239)--(6.663,6.239)--cycle;
\draw[gp path] (6.663,5.644)--(6.663,6.238)--(6.778,6.238)--(6.778,5.644)--cycle;
\gpfill{rgb color={0.800,0.800,0.800}} (6.893,5.644)--(7.010,5.644)--(7.010,6.739)--(6.893,6.739)--cycle;
\draw[gp path] (6.893,5.644)--(6.893,6.738)--(7.009,6.738)--(7.009,5.644)--cycle;
\gpfill{rgb color={0.800,0.800,0.800}} (7.124,5.919)--(7.240,5.919)--(7.240,7.824)--(7.124,7.824)--cycle;
\draw[gp path] (7.124,5.919)--(7.124,7.823)--(7.239,7.823)--(7.239,5.919)--cycle;
\gpfill{rgb color={0.800,0.800,0.800}} (7.354,5.644)--(7.470,5.644)--(7.470,6.410)--(7.354,6.410)--cycle;
\draw[gp path] (7.354,5.644)--(7.354,6.409)--(7.469,6.409)--(7.469,5.644)--cycle;
\gpfill{rgb color={0.800,0.800,0.800}} (7.584,5.644)--(7.700,5.644)--(7.700,6.628)--(7.584,6.628)--cycle;
\draw[gp path] (7.584,5.644)--(7.584,6.627)--(7.699,6.627)--(7.699,5.644)--cycle;
\gpfill{rgb color={0.800,0.800,0.800}} (7.814,5.848)--(7.930,5.848)--(7.930,7.413)--(7.814,7.413)--cycle;
\draw[gp path] (7.814,5.848)--(7.814,7.412)--(7.929,7.412)--(7.929,5.848)--cycle;
\gpfill{rgb color={0.800,0.800,0.800}} (8.044,6.028)--(8.160,6.028)--(8.160,6.029)--(8.044,6.029)--cycle;
\draw[gp path] (8.044,6.028)--(8.159,6.028)--cycle;
\gpfill{rgb color={0.800,0.800,0.800}} (8.275,5.644)--(8.391,5.644)--(8.391,8.127)--(8.275,8.127)--cycle;
\draw[gp path] (8.275,5.644)--(8.275,8.126)--(8.390,8.126)--(8.390,5.644)--cycle;
\gpfill{rgb color={0.800,0.800,0.800}} (8.505,5.644)--(8.621,5.644)--(8.621,5.821)--(8.505,5.821)--cycle;
\draw[gp path] (8.505,5.644)--(8.505,5.820)--(8.620,5.820)--(8.620,5.644)--cycle;
\gpfill{rgb color={0.800,0.800,0.800}} (8.735,5.644)--(8.851,5.644)--(8.851,8.451)--(8.735,8.451)--cycle;
\draw[gp path] (8.735,5.644)--(8.735,8.450)--(8.850,8.450)--(8.850,5.644)--cycle;
\gpfill{rgb color={0.800,0.800,0.800}} (8.965,5.644)--(9.081,5.644)--(9.081,7.429)--(8.965,7.429)--cycle;
\draw[gp path] (8.965,5.644)--(8.965,7.428)--(9.080,7.428)--(9.080,5.644)--cycle;
\gpfill{rgb color={0.800,0.800,0.800}} (9.195,5.644)--(9.311,5.644)--(9.311,8.462)--(9.195,8.462)--cycle;
\draw[gp path] (9.195,5.644)--(9.195,8.461)--(9.310,8.461)--(9.310,5.644)--cycle;
\gpfill{rgb color={0.800,0.800,0.800}} (9.426,5.915)--(9.542,5.915)--(9.542,7.092)--(9.426,7.092)--cycle;
\draw[gp path] (9.426,5.915)--(9.426,7.091)--(9.541,7.091)--(9.541,5.915)--cycle;
\gpfill{rgb color={0.800,0.800,0.800}} (9.886,6.047)--(10.002,6.047)--(10.002,6.048)--(9.886,6.048)--cycle;
\draw[gp path] (9.886,6.047)--(10.001,6.047)--cycle;
\gpfill{rgb color={0.800,0.800,0.800}} (10.116,5.644)--(10.232,5.644)--(10.232,7.627)--(10.116,7.627)--cycle;
\draw[gp path] (10.116,5.644)--(10.116,7.626)--(10.231,7.626)--(10.231,5.644)--cycle;
\gpfill{rgb color={0.800,0.800,0.800}} (10.346,5.644)--(10.462,5.644)--(10.462,7.460)--(10.346,7.460)--cycle;
\draw[gp path] (10.346,5.644)--(10.346,7.459)--(10.461,7.459)--(10.461,5.644)--cycle;
\gpfill{rgb color={0.800,0.800,0.800}} (10.577,5.818)--(10.693,5.818)--(10.693,5.819)--(10.577,5.819)--cycle;
\draw[gp path] (10.577,5.818)--(10.692,5.818)--cycle;
\gpfill{rgb color={0.800,0.800,0.800}} (10.807,7.003)--(10.923,7.003)--(10.923,7.004)--(10.807,7.004)--cycle;
\draw[gp path] (10.807,7.003)--(10.922,7.003)--cycle;
\gpfill{rgb color={0.800,0.800,0.800}} (11.037,6.336)--(11.153,6.336)--(11.153,10.132)--(11.037,10.132)--cycle;
\draw[gp path] (11.037,6.336)--(11.037,10.131)--(11.152,10.131)--(11.152,6.336)--cycle;
\gpfill{rgb color={0.800,0.800,0.800}} (11.267,5.644)--(11.383,5.644)--(11.383,6.255)--(11.267,6.255)--cycle;
\draw[gp path] (11.267,5.644)--(11.267,6.254)--(11.382,6.254)--(11.382,5.644)--cycle;
\gpfill{rgb color={0.800,0.800,0.800}} (11.497,5.644)--(11.613,5.644)--(11.613,5.818)--(11.497,5.818)--cycle;
\draw[gp path] (11.497,5.644)--(11.497,5.817)--(11.612,5.817)--(11.612,5.644)--cycle;
\gpfill{rgb color={0.800,0.800,0.800}} (11.728,6.330)--(11.844,6.330)--(11.844,6.331)--(11.728,6.331)--cycle;
\draw[gp path] (11.728,6.330)--(11.843,6.330)--cycle;
\gpfill{rgb color={0.800,0.800,0.800}} (11.958,5.644)--(12.074,5.644)--(12.074,6.030)--(11.958,6.030)--cycle;
\draw[gp path] (11.958,5.644)--(11.958,6.029)--(12.073,6.029)--(12.073,5.644)--cycle;
\gpfill{rgb color={0.800,0.800,0.800}} (12.188,5.644)--(12.304,5.644)--(12.304,6.148)--(12.188,6.148)--cycle;
\draw[gp path] (12.188,5.644)--(12.188,6.147)--(12.303,6.147)--(12.303,5.644)--cycle;
\gpfill{rgb color={0.800,0.800,0.800}} (12.418,5.644)--(12.534,5.644)--(12.534,10.132)--(12.418,10.132)--cycle;
\draw[gp path] (12.418,5.644)--(12.418,10.131)--(12.533,10.131)--(12.533,5.644)--cycle;
\gpfill{rgb color={0.800,0.800,0.800}} (12.648,5.644)--(12.764,5.644)--(12.764,7.496)--(12.648,7.496)--cycle;
\draw[gp path] (12.648,5.644)--(12.648,7.495)--(12.763,7.495)--(12.763,5.644)--cycle;
\gpfill{rgb color={0.800,0.800,0.800}} (12.879,5.815)--(12.995,5.815)--(12.995,7.924)--(12.879,7.924)--cycle;
\draw[gp path] (12.879,5.815)--(12.879,7.923)--(12.994,7.923)--(12.994,5.815)--cycle;
\gpfill{rgb color={0.800,0.800,0.800}} (13.109,6.133)--(13.225,6.133)--(13.225,8.555)--(13.109,8.555)--cycle;
\draw[gp path] (13.109,6.133)--(13.109,8.554)--(13.224,8.554)--(13.224,6.133)--cycle;
\gpfill{rgb color={0.800,0.800,0.800}} (13.339,6.798)--(13.455,6.798)--(13.455,6.799)--(13.339,6.799)--cycle;
\draw[gp path] (13.339,6.798)--(13.454,6.798)--cycle;
\gpfill{rgb color={0.800,0.800,0.800}} (13.569,5.760)--(13.685,5.760)--(13.685,7.889)--(13.569,7.889)--cycle;
\draw[gp path] (13.569,5.760)--(13.569,7.888)--(13.684,7.888)--(13.684,5.760)--cycle;
\gpfill{rgb color={0.800,0.800,0.800}} (13.799,5.776)--(13.915,5.776)--(13.915,5.777)--(13.799,5.777)--cycle;
\draw[gp path] (13.799,5.776)--(13.914,5.776)--cycle;
\gpfill{rgb color={0.800,0.800,0.800}} (14.030,5.784)--(14.146,5.784)--(14.146,5.785)--(14.030,5.785)--cycle;
\draw[gp path] (14.030,5.784)--(14.145,5.784)--cycle;
\gpfill{rgb color={0.800,0.800,0.800}} (14.260,5.644)--(14.376,5.644)--(14.376,8.841)--(14.260,8.841)--cycle;
\draw[gp path] (14.260,5.644)--(14.260,8.840)--(14.375,8.840)--(14.375,5.644)--cycle;
\gpfill{rgb color={0.800,0.800,0.800}} (14.490,5.644)--(14.606,5.644)--(14.606,6.194)--(14.490,6.194)--cycle;
\draw[gp path] (14.490,5.644)--(14.490,6.193)--(14.605,6.193)--(14.605,5.644)--cycle;
\gpfill{rgb color={0.800,0.800,0.800}} (14.720,5.765)--(14.836,5.765)--(14.836,5.766)--(14.720,5.766)--cycle;
\draw[gp path] (14.720,5.765)--(14.835,5.765)--cycle;
\gpfill{rgb color={0.800,0.800,0.800}} (14.950,5.644)--(15.066,5.644)--(15.066,6.668)--(14.950,6.668)--cycle;
\draw[gp path] (14.950,5.644)--(14.950,6.667)--(15.065,6.667)--(15.065,5.644)--cycle;
\gpfill{rgb color={0.800,0.800,0.800}} (15.181,5.644)--(15.297,5.644)--(15.297,5.886)--(15.181,5.886)--cycle;
\draw[gp path] (15.181,5.644)--(15.181,5.885)--(15.296,5.885)--(15.296,5.644)--cycle;
\gpfill{rgb color={0.800,0.800,0.800}} (15.411,5.777)--(15.527,5.777)--(15.527,5.778)--(15.411,5.778)--cycle;
\draw[gp path] (15.411,5.777)--(15.526,5.777)--cycle;
\gpfill{rgb color={0.800,0.800,0.800}} (15.641,5.644)--(15.757,5.644)--(15.757,6.554)--(15.641,6.554)--cycle;
\draw[gp path] (15.641,5.644)--(15.641,6.553)--(15.756,6.553)--(15.756,5.644)--cycle;
\gpfill{rgb color={0.800,0.800,0.800}} (15.871,5.904)--(15.987,5.904)--(15.987,6.096)--(15.871,6.096)--cycle;
\draw[gp path] (15.871,5.904)--(15.871,6.095)--(15.986,6.095)--(15.986,5.904)--cycle;
\gpfill{rgb color={0.800,0.800,0.800}} (16.101,5.671)--(16.217,5.671)--(16.217,5.672)--(16.101,5.672)--cycle;
\draw[gp path] (16.101,5.671)--(16.216,5.671)--cycle;
\gpfill{rgb color={0.800,0.800,0.800}} (16.332,5.644)--(16.448,5.644)--(16.448,5.817)--(16.332,5.817)--cycle;
\draw[gp path] (16.332,5.644)--(16.332,5.816)--(16.447,5.816)--(16.447,5.644)--cycle;
\gpfill{rgb color={0.800,0.800,0.800}} (16.792,5.644)--(16.908,5.644)--(16.908,5.861)--(16.792,5.861)--cycle;
\draw[gp path] (16.792,5.644)--(16.792,5.860)--(16.907,5.860)--(16.907,5.644)--cycle;
\gpfill{rgb color={0.800,0.800,0.800}} (17.022,5.645)--(17.138,5.645)--(17.138,5.646)--(17.022,5.646)--cycle;
\draw[gp path] (17.022,5.645)--(17.137,5.645)--cycle;
\gpfill{rgb color={0.800,0.800,0.800}} (17.252,5.644)--(17.368,5.644)--(17.368,6.017)--(17.252,6.017)--cycle;
\draw[gp path] (17.252,5.644)--(17.252,6.016)--(17.367,6.016)--(17.367,5.644)--cycle;
\gpfill{rgb color={0.800,0.800,0.800}} (17.483,6.738)--(17.599,6.738)--(17.599,6.739)--(17.483,6.739)--cycle;
\draw[gp path] (17.483,6.738)--(17.598,6.738)--cycle;
\gpfill{rgb color={0.800,0.800,0.800}} (17.713,5.767)--(17.829,5.767)--(17.829,6.107)--(17.713,6.107)--cycle;
\draw[gp path] (17.713,5.767)--(17.713,6.106)--(17.828,6.106)--(17.828,5.767)--cycle;
\gpfill{rgb color={0.800,0.800,0.800}} (17.943,6.411)--(18.059,6.411)--(18.059,6.412)--(17.943,6.412)--cycle;
\draw[gp path] (17.943,6.411)--(18.058,6.411)--cycle;
\gpfill{rgb color={0.800,0.800,0.800}} (18.403,5.947)--(18.519,5.947)--(18.519,5.948)--(18.403,5.948)--cycle;
\draw[gp path] (18.403,5.947)--(18.518,5.947)--cycle;
\gpfill{rgb color={0.800,0.800,0.800}} (18.634,6.588)--(18.750,6.588)--(18.750,10.132)--(18.634,10.132)--cycle;
\draw[gp path] (18.634,6.588)--(18.634,10.131)--(18.749,10.131)--(18.749,6.588)--cycle;
\gpfill{rgb color={0.800,0.800,0.800}} (18.864,6.470)--(18.980,6.470)--(18.980,8.657)--(18.864,8.657)--cycle;
\draw[gp path] (18.864,6.470)--(18.864,8.656)--(18.979,8.656)--(18.979,6.470)--cycle;
\gpfill{rgb color={0.800,0.800,0.800}} (19.324,9.327)--(19.440,9.327)--(19.440,10.132)--(19.324,10.132)--cycle;
\draw[gp path] (19.324,9.327)--(19.324,10.131)--(19.439,10.131)--(19.439,9.327)--cycle;
\gpfill{rgb color={0.800,0.800,0.800}} (19.554,5.691)--(19.670,5.691)--(19.670,6.967)--(19.554,6.967)--cycle;
\draw[gp path] (19.554,5.691)--(19.554,6.966)--(19.669,6.966)--(19.669,5.691)--cycle;
\gpfill{rgb color={0.800,0.800,0.800}} (19.784,6.038)--(19.901,6.038)--(19.901,6.039)--(19.784,6.039)--cycle;
\draw[gp path] (19.784,6.038)--(19.900,6.038)--cycle;
\gpfill{rgb color={0.800,0.800,0.800}} (20.015,5.644)--(20.131,5.644)--(20.131,5.967)--(20.015,5.967)--cycle;
\draw[gp path] (20.015,5.644)--(20.015,5.966)--(20.130,5.966)--(20.130,5.644)--cycle;
\gpfill{rgb color={0.800,0.800,0.800}} (20.245,5.644)--(20.361,5.644)--(20.361,5.668)--(20.245,5.668)--cycle;
\draw[gp path] (20.245,5.644)--(20.245,5.667)--(20.360,5.667)--(20.360,5.644)--cycle;
\gpfill{rgb color={0.800,0.800,0.800}} (20.475,5.839)--(20.591,5.839)--(20.591,5.840)--(20.475,5.840)--cycle;
\draw[gp path] (20.475,5.839)--(20.590,5.839)--cycle;
\gpfill{rgb color={0.800,0.800,0.800}} (20.705,5.644)--(20.821,5.644)--(20.821,6.572)--(20.705,6.572)--cycle;
\draw[gp path] (20.705,5.644)--(20.705,6.571)--(20.820,6.571)--(20.820,5.644)--cycle;
\gpfill{rgb color={0.800,0.800,0.800}} (20.935,5.683)--(21.052,5.683)--(21.052,5.684)--(20.935,5.684)--cycle;
\draw[gp path] (20.935,5.683)--(21.051,5.683)--cycle;
\gpfill{rgb color={0.800,0.800,0.800}} (21.166,5.644)--(21.282,5.644)--(21.282,5.927)--(21.166,5.927)--cycle;
\draw[gp path] (21.166,5.644)--(21.166,5.926)--(21.281,5.926)--(21.281,5.644)--cycle;
\gpfill{rgb color={0.800,0.800,0.800}} (21.396,6.766)--(21.512,6.766)--(21.512,8.924)--(21.396,8.924)--cycle;
\draw[gp path] (21.396,6.766)--(21.396,8.923)--(21.511,8.923)--(21.511,6.766)--cycle;
\gpfill{rgb color={0.800,0.800,0.800}} (21.626,6.053)--(21.742,6.053)--(21.742,6.054)--(21.626,6.054)--cycle;
\draw[gp path] (21.626,6.053)--(21.741,6.053)--cycle;
\gpfill{rgb color={0.800,0.800,0.800}} (21.856,5.644)--(21.972,5.644)--(21.972,9.505)--(21.856,9.505)--cycle;
\draw[gp path] (21.856,5.644)--(21.856,9.504)--(21.971,9.504)--(21.971,5.644)--cycle;
\gpfill{rgb color={0.800,0.800,0.800}} (22.086,5.644)--(22.203,5.644)--(22.203,8.898)--(22.086,8.898)--cycle;
\draw[gp path] (22.086,5.644)--(22.086,8.897)--(22.202,8.897)--(22.202,5.644)--cycle;
\gpfill{rgb color={0.800,0.800,0.800}} (22.317,5.644)--(22.433,5.644)--(22.433,6.842)--(22.317,6.842)--cycle;
\draw[gp path] (22.317,5.644)--(22.317,6.841)--(22.432,6.841)--(22.432,5.644)--cycle;
\gpfill{rgb color={0.800,0.800,0.800}} (22.547,5.644)--(22.663,5.644)--(22.663,5.953)--(22.547,5.953)--cycle;
\draw[gp path] (22.547,5.644)--(22.547,5.952)--(22.662,5.952)--(22.662,5.644)--cycle;
\gpfill{rgb color={0.800,0.800,0.800}} (22.777,5.644)--(22.893,5.644)--(22.893,5.757)--(22.777,5.757)--cycle;
\draw[gp path] (22.777,5.644)--(22.777,5.756)--(22.892,5.756)--(22.892,5.644)--cycle;
\gpfill{rgb color={0.800,0.800,0.800}} (23.007,5.650)--(23.123,5.650)--(23.123,5.651)--(23.007,5.651)--cycle;
\draw[gp path] (23.007,5.650)--(23.122,5.650)--cycle;
\gpfill{rgb color={0.800,0.800,0.800}} (23.237,5.874)--(23.354,5.874)--(23.354,5.875)--(23.237,5.875)--cycle;
\draw[gp path] (23.237,5.874)--(23.353,5.874)--cycle;
\gpfill{rgb color={0.800,0.800,0.800}} (23.698,5.644)--(23.814,5.644)--(23.814,6.279)--(23.698,6.279)--cycle;
\draw[gp path] (23.698,5.644)--(23.698,6.278)--(23.813,6.278)--(23.813,5.644)--cycle;
\gpfill{rgb color={0.800,0.800,0.800}} (23.928,5.644)--(24.044,5.644)--(24.044,5.912)--(23.928,5.912)--cycle;
\draw[gp path] (23.928,5.644)--(23.928,5.911)--(24.043,5.911)--(24.043,5.644)--cycle;
\gpfill{rgb color={0.800,0.800,0.800}} (24.158,5.654)--(24.274,5.654)--(24.274,6.714)--(24.158,6.714)--cycle;
\draw[gp path] (24.158,5.654)--(24.158,6.713)--(24.273,6.713)--(24.273,5.654)--cycle;
\gpcolor{color=gp lt color border}
\draw[gp path] (1.196,10.131)--(1.196,5.644)--(24.446,5.644)--(24.446,10.131)--cycle;
\node[gp node center] at (12.821,9.682) {\plotLegend{(MIPS)}};
\node[gp node left,font={\fontsize{11.0pt}{13.2pt}\selectfont}] at (1.429,9.817) {\plotLegend{mean improvement: $5.4\%$}};
\node[gp node left,font={\fontsize{11.0pt}{13.2pt}\selectfont}] at (1.429,9.368) {\plotLegend{improved functions: $47\%$}};
\node[gp node left,font={\fontsize{11.0pt}{13.2pt}\selectfont}] at (1.429,8.920) {\plotLegend{mean gap: $18.5\%$}};
\node[gp node left,font={\fontsize{11.0pt}{13.2pt}\selectfont}] at (1.429,8.471) {\plotLegend{optimal functions: $34\%$}};
\gpdefrectangularnode{gp plot 1}{\pgfpoint{1.196cm}{5.644cm}}{\pgfpoint{24.446cm}{10.131cm}}
\end{tikzpicture}

%% file: results/hexagon-size-improvement.tex
\begin{tikzpicture}[gnuplot]
\path (0.000,0.000) rectangle (12.500,8.750);
\gpcolor{color=gp lt color border}
\gpsetlinetype{gp lt border}
\gpsetdashtype{gp dt solid}
\gpsetlinewidth{1.00}
\draw[gp path] (1.012,5.644)--(1.192,5.644);
\draw[gp path] (24.446,5.644)--(24.266,5.644);
\node[gp node right,font={\fontsize{8.0pt}{9.6pt}\selectfont}] at (0.828,5.644) {\plotPercentage{0}};
\draw[gp path] (1.012,6.205)--(1.192,6.205);
\draw[gp path] (24.446,6.205)--(24.266,6.205);
\node[gp node right,font={\fontsize{8.0pt}{9.6pt}\selectfont}] at (0.828,6.205) {\plotPercentage{5}};
\draw[gp path] (1.012,6.766)--(1.192,6.766);
\draw[gp path] (24.446,6.766)--(24.266,6.766);
\node[gp node right,font={\fontsize{8.0pt}{9.6pt}\selectfont}] at (0.828,6.766) {\plotPercentage{10}};
\draw[gp path] (1.012,7.327)--(1.192,7.327);
\draw[gp path] (24.446,7.327)--(24.266,7.327);
\node[gp node right,font={\fontsize{8.0pt}{9.6pt}\selectfont}] at (0.828,7.327) {\plotPercentage{15}};
\draw[gp path] (1.012,7.888)--(1.192,7.888);
\draw[gp path] (24.446,7.888)--(24.266,7.888);
\node[gp node right,font={\fontsize{8.0pt}{9.6pt}\selectfont}] at (0.828,7.888) {\plotPercentage{20}};
\draw[gp path] (1.012,8.448)--(1.192,8.448);
\draw[gp path] (24.446,8.448)--(24.266,8.448);
\node[gp node right,font={\fontsize{8.0pt}{9.6pt}\selectfont}] at (0.828,8.448) {\plotPercentage{25}};
\draw[gp path] (1.012,9.009)--(1.192,9.009);
\draw[gp path] (24.446,9.009)--(24.266,9.009);
\node[gp node right,font={\fontsize{8.0pt}{9.6pt}\selectfont}] at (0.828,9.009) {\plotPercentage{30}};
\draw[gp path] (1.012,9.570)--(1.192,9.570);
\draw[gp path] (24.446,9.570)--(24.266,9.570);
\node[gp node right,font={\fontsize{8.0pt}{9.6pt}\selectfont}] at (0.828,9.570) {\plotPercentage{35}};
\draw[gp path] (1.012,10.131)--(1.192,10.131);
\draw[gp path] (24.446,10.131)--(24.266,10.131);
\node[gp node right,font={\fontsize{8.0pt}{9.6pt}\selectfont}] at (0.828,10.131) {\plotPercentage{40}};
\node[gp node left,rotate=-90,font={\fontsize{7.0pt}{8.4pt}\selectfont}] at (1.244,5.721) {\functionId{1}\functionName{handle_noinline_attribute}};
\node[gp node left,rotate=-90,font={\fontsize{7.0pt}{8.4pt}\selectfont}] at (1.476,5.721) {\functionId{2}\functionName{control_flow_insn_p}};
\node[gp node left,rotate=-90,font={\fontsize{7.0pt}{8.4pt}\selectfont}] at (1.708,5.721) {\functionId{3}\functionName{insert_insn_on_edge}};
\node[gp node left,rotate=-90,font={\fontsize{7.0pt}{8.4pt}\selectfont}] at (1.940,5.721) {\functionId{4}\functionName{update_br_prob_note}};
\node[gp node left,rotate=-90,font={\fontsize{7.0pt}{8.4pt}\selectfont}] at (2.172,5.721) {\functionId{5}\functionName{_cpp_init_internal_pragma.}};
\node[gp node left,rotate=-90,font={\fontsize{7.0pt}{8.4pt}\selectfont}] at (2.404,5.721) {\functionId{6}\functionName{lex_macro_node}};
\node[gp node left,rotate=-90,font={\fontsize{7.0pt}{8.4pt}\selectfont}] at (2.636,5.721) {\functionId{7}\functionName{cse_basic_block}};
\node[gp node left,rotate=-90,font={\fontsize{7.0pt}{8.4pt}\selectfont}] at (2.868,5.721) {\functionId{8}\functionName{rtx_equal_for_cselib_p}};
\node[gp node left,rotate=-90,font={\fontsize{7.0pt}{8.4pt}\selectfont}] at (3.100,5.721) {\functionId{9}\functionName{debug_df_chain}};
\node[gp node left,rotate=-90,font={\fontsize{7.0pt}{8.4pt}\selectfont}] at (3.332,5.721) {\functionId{10}\functionName{modified_type_die}};
\node[gp node left,rotate=-90,font={\fontsize{7.0pt}{8.4pt}\selectfont}] at (3.564,5.721) {\functionId{11}\functionName{emit_note}};
\node[gp node left,rotate=-90,font={\fontsize{7.0pt}{8.4pt}\selectfont}] at (3.796,5.721) {\functionId{12}\functionName{gen_sequence}};
\node[gp node left,rotate=-90,font={\fontsize{7.0pt}{8.4pt}\selectfont}] at (4.028,5.721) {\functionId{13}\functionName{subreg_hard_regno}};
\node[gp node left,rotate=-90,font={\fontsize{7.0pt}{8.4pt}\selectfont}] at (4.260,5.721) {\functionId{14}\functionName{split_double}};
\node[gp node left,rotate=-90,font={\fontsize{7.0pt}{8.4pt}\selectfont}] at (4.492,5.721) {\functionId{15}\functionName{add_to_mem_set_list}};
\node[gp node left,rotate=-90,font={\fontsize{7.0pt}{8.4pt}\selectfont}] at (4.724,5.721) {\functionId{16}\functionName{find_regno_partial}};
\node[gp node left,rotate=-90,font={\fontsize{7.0pt}{8.4pt}\selectfont}] at (4.956,5.721) {\functionId{17}\functionName{use_return_register}};
\node[gp node left,rotate=-90,font={\fontsize{7.0pt}{8.4pt}\selectfont}] at (5.188,5.721) {\functionId{18}\functionName{ix86_expand_move}};
\node[gp node left,rotate=-90,font={\fontsize{7.0pt}{8.4pt}\selectfont}] at (5.420,5.721) {\functionId{19}\functionName{legitimate_pic_address_di.}};
\node[gp node left,rotate=-90,font={\fontsize{7.0pt}{8.4pt}\selectfont}] at (5.652,5.721) {\functionId{20}\functionName{gen_extendsfdf2}};
\node[gp node left,rotate=-90,font={\fontsize{7.0pt}{8.4pt}\selectfont}] at (5.884,5.721) {\functionId{21}\functionName{gen_mulsidi3}};
\node[gp node left,rotate=-90,font={\fontsize{7.0pt}{8.4pt}\selectfont}] at (6.116,5.721) {\functionId{22}\functionName{gen_peephole2_1255}};
\node[gp node left,rotate=-90,font={\fontsize{7.0pt}{8.4pt}\selectfont}] at (6.348,5.721) {\functionId{23}\functionName{gen_peephole2_1271}};
\node[gp node left,rotate=-90,font={\fontsize{7.0pt}{8.4pt}\selectfont}] at (6.580,5.721) {\functionId{24}\functionName{gen_peephole2_1277}};
\node[gp node left,rotate=-90,font={\fontsize{7.0pt}{8.4pt}\selectfont}] at (6.812,5.721) {\functionId{25}\functionName{gen_pfnacc}};
\node[gp node left,rotate=-90,font={\fontsize{7.0pt}{8.4pt}\selectfont}] at (7.045,5.721) {\functionId{26}\functionName{gen_rotlsi3}};
\node[gp node left,rotate=-90,font={\fontsize{7.0pt}{8.4pt}\selectfont}] at (7.277,5.721) {\functionId{27}\functionName{gen_split_1001}};
\node[gp node left,rotate=-90,font={\fontsize{7.0pt}{8.4pt}\selectfont}] at (7.509,5.721) {\functionId{28}\functionName{gen_split_1028}};
\node[gp node left,rotate=-90,font={\fontsize{7.0pt}{8.4pt}\selectfont}] at (7.741,5.721) {\functionId{29}\functionName{gen_sse_nandti3}};
\node[gp node left,rotate=-90,font={\fontsize{7.0pt}{8.4pt}\selectfont}] at (7.973,5.721) {\functionId{30}\functionName{gen_sunge}};
\node[gp node left,rotate=-90,font={\fontsize{7.0pt}{8.4pt}\selectfont}] at (8.205,5.721) {\functionId{31}\functionName{insert_loop_mem}};
\node[gp node left,rotate=-90,font={\fontsize{7.0pt}{8.4pt}\selectfont}] at (8.437,5.721) {\functionId{32}\functionName{eiremain}};
\node[gp node left,rotate=-90,font={\fontsize{7.0pt}{8.4pt}\selectfont}] at (8.669,5.721) {\functionId{33}\functionName{elimination_effects}};
\node[gp node left,rotate=-90,font={\fontsize{7.0pt}{8.4pt}\selectfont}] at (8.901,5.721) {\functionId{34}\functionName{gen_reload}};
\node[gp node left,rotate=-90,font={\fontsize{7.0pt}{8.4pt}\selectfont}] at (9.133,5.721) {\functionId{35}\functionName{reload_cse_simplify_set}};
\node[gp node left,rotate=-90,font={\fontsize{7.0pt}{8.4pt}\selectfont}] at (9.365,5.721) {\functionId{36}\functionName{simplify_binary_is2orm1}};
\node[gp node left,rotate=-90,font={\fontsize{7.0pt}{8.4pt}\selectfont}] at (9.597,5.721) {\functionId{37}\functionName{remove_phi_alternative}};
\node[gp node left,rotate=-90,font={\fontsize{7.0pt}{8.4pt}\selectfont}] at (9.829,5.721) {\functionId{38}\functionName{contains_placeholder_p}};
\node[gp node left,rotate=-90,font={\fontsize{7.0pt}{8.4pt}\selectfont}] at (10.061,5.721) {\functionId{39}\functionName{assemble_end_function}};
\node[gp node left,rotate=-90,font={\fontsize{7.0pt}{8.4pt}\selectfont}] at (10.293,5.721) {\functionId{40}\functionName{default_named_section_asm.}};
\node[gp node left,rotate=-90,font={\fontsize{7.0pt}{8.4pt}\selectfont}] at (10.525,5.721) {\functionId{41}\functionName{sample_unpack_12}};
\node[gp node left,rotate=-90,font={\fontsize{7.0pt}{8.4pt}\selectfont}] at (10.757,5.721) {\functionId{42}\functionName{autohelperattpat10}};
\node[gp node left,rotate=-90,font={\fontsize{7.0pt}{8.4pt}\selectfont}] at (10.989,5.721) {\functionId{43}\functionName{autohelperbarrierspat126}};
\node[gp node left,rotate=-90,font={\fontsize{7.0pt}{8.4pt}\selectfont}] at (11.221,5.721) {\functionId{44}\functionName{atari_atari_attack_callba.}};
\node[gp node left,rotate=-90,font={\fontsize{7.0pt}{8.4pt}\selectfont}] at (11.453,5.721) {\functionId{45}\functionName{compute_aa_status}};
\node[gp node left,rotate=-90,font={\fontsize{7.0pt}{8.4pt}\selectfont}] at (11.685,5.721) {\functionId{46}\functionName{dragon_weak}};
\node[gp node left,rotate=-90,font={\fontsize{7.0pt}{8.4pt}\selectfont}] at (11.917,5.721) {\functionId{47}\functionName{get_saved_worms}};
\node[gp node left,rotate=-90,font={\fontsize{7.0pt}{8.4pt}\selectfont}] at (12.149,5.721) {\functionId{48}\functionName{read_eye}};
\node[gp node left,rotate=-90,font={\fontsize{7.0pt}{8.4pt}\selectfont}] at (12.381,5.721) {\functionId{49}\functionName{topological_eye}};
\node[gp node left,rotate=-90,font={\fontsize{7.0pt}{8.4pt}\selectfont}] at (12.613,5.721) {\functionId{50}\functionName{autohelperowl_attackpat19.}};
\node[gp node left,rotate=-90,font={\fontsize{7.0pt}{8.4pt}\selectfont}] at (12.845,5.721) {\functionId{51}\functionName{autohelperowl_attackpat29.}};
\node[gp node left,rotate=-90,font={\fontsize{7.0pt}{8.4pt}\selectfont}] at (13.077,5.721) {\functionId{52}\functionName{autohelperowl_defendpat28.}};
\node[gp node left,rotate=-90,font={\fontsize{7.0pt}{8.4pt}\selectfont}] at (13.309,5.721) {\functionId{53}\functionName{autohelperowl_defendpat38.}};
\node[gp node left,rotate=-90,font={\fontsize{7.0pt}{8.4pt}\selectfont}] at (13.541,5.721) {\functionId{54}\functionName{autohelperpat1114}};
\node[gp node left,rotate=-90,font={\fontsize{7.0pt}{8.4pt}\selectfont}] at (13.773,5.721) {\functionId{55}\functionName{autohelperpat335}};
\node[gp node left,rotate=-90,font={\fontsize{7.0pt}{8.4pt}\selectfont}] at (14.005,5.721) {\functionId{56}\functionName{autohelperpat508}};
\node[gp node left,rotate=-90,font={\fontsize{7.0pt}{8.4pt}\selectfont}] at (14.237,5.721) {\functionId{57}\functionName{autohelperpat83}};
\node[gp node left,rotate=-90,font={\fontsize{7.0pt}{8.4pt}\selectfont}] at (14.469,5.721) {\functionId{58}\functionName{simple_showboard}};
\node[gp node left,rotate=-90,font={\fontsize{7.0pt}{8.4pt}\selectfont}] at (14.701,5.721) {\functionId{59}\functionName{skip_intrabk_SAD}};
\node[gp node left,rotate=-90,font={\fontsize{7.0pt}{8.4pt}\selectfont}] at (14.933,5.721) {\functionId{60}\functionName{free_orig_planes}};
\node[gp node left,rotate=-90,font={\fontsize{7.0pt}{8.4pt}\selectfont}] at (15.165,5.721) {\functionId{61}\functionName{GetSkipCostMB}};
\node[gp node left,rotate=-90,font={\fontsize{7.0pt}{8.4pt}\selectfont}] at (15.397,5.721) {\functionId{62}\functionName{writeSyntaxElement_Level_.}};
\node[gp node left,rotate=-90,font={\fontsize{7.0pt}{8.4pt}\selectfont}] at (15.629,5.721) {\functionId{63}\functionName{GSIAddKeyToIndex}};
\node[gp node left,rotate=-90,font={\fontsize{7.0pt}{8.4pt}\selectfont}] at (15.861,5.721) {\functionId{64}\functionName{EVDBasicFit}};
\node[gp node left,rotate=-90,font={\fontsize{7.0pt}{8.4pt}\selectfont}] at (16.093,5.721) {\functionId{65}\functionName{SampleDirichlet}};
\node[gp node left,rotate=-90,font={\fontsize{7.0pt}{8.4pt}\selectfont}] at (16.325,5.721) {\functionId{66}\functionName{DegenerateSymbolScore}};
\node[gp node left,rotate=-90,font={\fontsize{7.0pt}{8.4pt}\selectfont}] at (16.557,5.721) {\functionId{67}\functionName{Plan7SetCtime}};
\node[gp node left,rotate=-90,font={\fontsize{7.0pt}{8.4pt}\selectfont}] at (16.789,5.721) {\functionId{68}\functionName{MSAToSqinfo}};
\node[gp node left,rotate=-90,font={\fontsize{7.0pt}{8.4pt}\selectfont}] at (17.021,5.721) {\functionId{69}\functionName{null_convert}};
\node[gp node left,rotate=-90,font={\fontsize{7.0pt}{8.4pt}\selectfont}] at (17.253,5.721) {\functionId{70}\functionName{jinit_c_prep_controller}};
\node[gp node left,rotate=-90,font={\fontsize{7.0pt}{8.4pt}\selectfont}] at (17.485,5.721) {\functionId{71}\functionName{glFogf}};
\node[gp node left,rotate=-90,font={\fontsize{7.0pt}{8.4pt}\selectfont}] at (17.717,5.721) {\functionId{72}\functionName{glNormal3d}};
\node[gp node left,rotate=-90,font={\fontsize{7.0pt}{8.4pt}\selectfont}] at (17.949,5.721) {\functionId{73}\functionName{glRasterPos3d}};
\node[gp node left,rotate=-90,font={\fontsize{7.0pt}{8.4pt}\selectfont}] at (18.181,5.721) {\functionId{74}\functionName{glTexCoord2d}};
\node[gp node left,rotate=-90,font={\fontsize{7.0pt}{8.4pt}\selectfont}] at (18.413,5.721) {\functionId{75}\functionName{gl_stippled_bresenham}};
\node[gp node left,rotate=-90,font={\fontsize{7.0pt}{8.4pt}\selectfont}] at (18.646,5.721) {\functionId{76}\functionName{gl_save_Frustum}};
\node[gp node left,rotate=-90,font={\fontsize{7.0pt}{8.4pt}\selectfont}] at (18.878,5.721) {\functionId{77}\functionName{gl_save_LineWidth}};
\node[gp node left,rotate=-90,font={\fontsize{7.0pt}{8.4pt}\selectfont}] at (19.110,5.721) {\functionId{78}\functionName{translate_id}};
\node[gp node left,rotate=-90,font={\fontsize{7.0pt}{8.4pt}\selectfont}] at (19.342,5.721) {\functionId{79}\functionName{gl_Map1f}};
\node[gp node left,rotate=-90,font={\fontsize{7.0pt}{8.4pt}\selectfont}] at (19.574,5.721) {\functionId{80}\functionName{smooth_ci_line}};
\node[gp node left,rotate=-90,font={\fontsize{7.0pt}{8.4pt}\selectfont}] at (19.806,5.721) {\functionId{81}\functionName{free_unified_knots}};
\node[gp node left,rotate=-90,font={\fontsize{7.0pt}{8.4pt}\selectfont}] at (20.038,5.721) {\functionId{82}\functionName{tess_test_polygon}};
\node[gp node left,rotate=-90,font={\fontsize{7.0pt}{8.4pt}\selectfont}] at (20.270,5.721) {\functionId{83}\functionName{auxWireBox}};
\node[gp node left,rotate=-90,font={\fontsize{7.0pt}{8.4pt}\selectfont}] at (20.502,5.721) {\functionId{84}\functionName{gl_ColorPointer}};
\node[gp node left,rotate=-90,font={\fontsize{7.0pt}{8.4pt}\selectfont}] at (20.734,5.721) {\functionId{85}\functionName{r_serial}};
\node[gp node left,rotate=-90,font={\fontsize{7.0pt}{8.4pt}\selectfont}] at (20.966,5.721) {\functionId{86}\functionName{scalar_mult_sub_su3_matri.}};
\node[gp node left,rotate=-90,font={\fontsize{7.0pt}{8.4pt}\selectfont}] at (21.198,5.721) {\functionId{87}\functionName{Decode_MPEG1_Non_Intra_Bl.}};
\node[gp node left,rotate=-90,font={\fontsize{7.0pt}{8.4pt}\selectfont}] at (21.430,5.721) {\functionId{88}\functionName{cpDecodeSecret}};
\node[gp node left,rotate=-90,font={\fontsize{7.0pt}{8.4pt}\selectfont}] at (21.662,5.721) {\functionId{89}\functionName{vlShortLshift}};
\node[gp node left,rotate=-90,font={\fontsize{7.0pt}{8.4pt}\selectfont}] at (21.894,5.721) {\functionId{90}\functionName{encryptfile}};
\node[gp node left,rotate=-90,font={\fontsize{7.0pt}{8.4pt}\selectfont}] at (22.126,5.721) {\functionId{91}\functionName{make_canonical}};
\node[gp node left,rotate=-90,font={\fontsize{7.0pt}{8.4pt}\selectfont}] at (22.358,5.721) {\functionId{92}\functionName{LANG}};
\node[gp node left,rotate=-90,font={\fontsize{7.0pt}{8.4pt}\selectfont}] at (22.590,5.721) {\functionId{93}\functionName{MD5Transform}};
\node[gp node left,rotate=-90,font={\fontsize{7.0pt}{8.4pt}\selectfont}] at (22.822,5.721) {\functionId{94}\functionName{mp_display}};
\node[gp node left,rotate=-90,font={\fontsize{7.0pt}{8.4pt}\selectfont}] at (23.054,5.721) {\functionId{95}\functionName{comp_Jboundaries}};
\node[gp node left,rotate=-90,font={\fontsize{7.0pt}{8.4pt}\selectfont}] at (23.286,5.721) {\functionId{96}\functionName{is_draw}};
\node[gp node left,rotate=-90,font={\fontsize{7.0pt}{8.4pt}\selectfont}] at (23.518,5.721) {\functionId{97}\functionName{push_king}};
\node[gp node left,rotate=-90,font={\fontsize{7.0pt}{8.4pt}\selectfont}] at (23.750,5.721) {\functionId{98}\functionName{stat_retry}};
\node[gp node left,rotate=-90,font={\fontsize{7.0pt}{8.4pt}\selectfont}] at (23.982,5.721) {\functionId{99}\functionName{lextree_subtree_print}};
\node[gp node left,rotate=-90,font={\fontsize{7.0pt}{8.4pt}\selectfont}] at (24.214,5.721) {\functionId{100}\functionName{lm_tg_score}};
\draw[gp path] (1.012,10.131)--(1.012,5.644)--(24.446,5.644)--(24.446,10.131)--cycle;
\gpcolor{rgb color={0.580,0.000,0.827}}
\draw[gp path] (1.012,5.644)--(1.249,5.644)--(1.485,5.644)--(1.722,5.644)--(1.959,5.644)%
  --(2.196,5.644)--(2.432,5.644)--(2.669,5.644)--(2.906,5.644)--(3.142,5.644)--(3.379,5.644)%
  --(3.616,5.644)--(3.852,5.644)--(4.089,5.644)--(4.326,5.644)--(4.563,5.644)--(4.799,5.644)%
  --(5.036,5.644)--(5.273,5.644)--(5.509,5.644)--(5.746,5.644)--(5.983,5.644)--(6.220,5.644)%
  --(6.456,5.644)--(6.693,5.644)--(6.930,5.644)--(7.166,5.644)--(7.403,5.644)--(7.640,5.644)%
  --(7.877,5.644)--(8.113,5.644)--(8.350,5.644)--(8.587,5.644)--(8.823,5.644)--(9.060,5.644)%
  --(9.297,5.644)--(9.533,5.644)--(9.770,5.644)--(10.007,5.644)--(10.244,5.644)--(10.480,5.644)%
  --(10.717,5.644)--(10.954,5.644)--(11.190,5.644)--(11.427,5.644)--(11.664,5.644)--(11.901,5.644)%
  --(12.137,5.644)--(12.374,5.644)--(12.611,5.644)--(12.847,5.644)--(13.084,5.644)--(13.321,5.644)%
  --(13.557,5.644)--(13.794,5.644)--(14.031,5.644)--(14.268,5.644)--(14.504,5.644)--(14.741,5.644)%
  --(14.978,5.644)--(15.214,5.644)--(15.451,5.644)--(15.688,5.644)--(15.925,5.644)--(16.161,5.644)%
  --(16.398,5.644)--(16.635,5.644)--(16.871,5.644)--(17.108,5.644)--(17.345,5.644)--(17.581,5.644)%
  --(17.818,5.644)--(18.055,5.644)--(18.292,5.644)--(18.528,5.644)--(18.765,5.644)--(19.002,5.644)%
  --(19.238,5.644)--(19.475,5.644)--(19.712,5.644)--(19.949,5.644)--(20.185,5.644)--(20.422,5.644)%
  --(20.659,5.644)--(20.895,5.644)--(21.132,5.644)--(21.369,5.644)--(21.606,5.644)--(21.842,5.644)%
  --(22.079,5.644)--(22.316,5.644)--(22.552,5.644)--(22.789,5.644)--(23.026,5.644)--(23.262,5.644)%
  --(23.499,5.644)--(23.736,5.644)--(23.973,5.644)--(24.209,5.644)--(24.446,5.644);
\gpfill{rgb color={0.000,0.000,0.000}} (7.915,5.644)--(8.032,5.644)--(8.032,7.047)--(7.915,7.047)--cycle;
\gpcolor{rgb color={0.000,0.000,0.000}}
\draw[gp path] (7.915,5.644)--(7.915,7.046)--(8.031,7.046)--(8.031,5.644)--cycle;
\gpfill{rgb color={0.000,0.000,0.000}} (17.659,5.644)--(17.776,5.644)--(17.776,10.132)--(17.659,10.132)--cycle;
\draw[gp path] (17.659,5.644)--(17.659,10.131)--(17.775,10.131)--(17.775,5.644)--cycle;
\gpfill{rgb color={0.000,0.000,0.000}} (17.891,5.644)--(18.008,5.644)--(18.008,8.530)--(17.891,8.530)--cycle;
\draw[gp path] (17.891,5.644)--(17.891,8.529)--(18.007,8.529)--(18.007,5.644)--cycle;
\gpfill{rgb color={0.000,0.000,0.000}} (19.052,5.644)--(19.169,5.644)--(19.169,6.346)--(19.052,6.346)--cycle;
\draw[gp path] (19.052,5.644)--(19.052,6.345)--(19.168,6.345)--(19.168,5.644)--cycle;
\gpfill{rgb color={0.000,0.000,0.000}} (19.748,5.644)--(19.865,5.644)--(19.865,6.713)--(19.748,6.713)--cycle;
\draw[gp path] (19.748,5.644)--(19.748,6.712)--(19.864,6.712)--(19.864,5.644)--cycle;
\gpfill{rgb color={0.000,0.000,0.000}} (20.908,5.644)--(21.025,5.644)--(21.025,5.717)--(20.908,5.717)--cycle;
\draw[gp path] (20.908,5.644)--(20.908,5.716)--(21.024,5.716)--(21.024,5.644)--cycle;
\gpfill{rgb color={0.000,0.000,0.000}} (21.372,5.644)--(21.489,5.644)--(21.489,6.580)--(21.372,6.580)--cycle;
\draw[gp path] (21.372,5.644)--(21.372,6.579)--(21.488,6.579)--(21.488,5.644)--cycle;
\gpfill{rgb color={0.000,0.000,0.000}} (21.604,5.644)--(21.721,5.644)--(21.721,6.742)--(21.604,6.742)--cycle;
\draw[gp path] (21.604,5.644)--(21.604,6.741)--(21.720,6.741)--(21.720,5.644)--cycle;
\gpfill{rgb color={0.000,0.000,0.000}} (22.996,5.644)--(23.113,5.644)--(23.113,7.371)--(22.996,7.371)--cycle;
\draw[gp path] (22.996,5.644)--(22.996,7.370)--(23.112,7.370)--(23.112,5.644)--cycle;
\gpfill{rgb color={0.800,0.800,0.800}} (2.810,5.644)--(2.927,5.644)--(2.927,5.800)--(2.810,5.800)--cycle;
\gpcolor{rgb color={0.800,0.800,0.800}}
\draw[gp path] (2.810,5.644)--(2.810,5.799)--(2.926,5.799)--(2.926,5.644)--cycle;
\gpfill{rgb color={0.800,0.800,0.800}} (3.274,5.644)--(3.391,5.644)--(3.391,5.914)--(3.274,5.914)--cycle;
\draw[gp path] (3.274,5.644)--(3.274,5.913)--(3.390,5.913)--(3.390,5.644)--cycle;
\gpfill{rgb color={0.800,0.800,0.800}} (7.915,7.046)--(8.032,7.046)--(8.032,7.047)--(7.915,7.047)--cycle;
\draw[gp path] (7.915,7.046)--(8.031,7.046)--cycle;
\gpfill{rgb color={0.800,0.800,0.800}} (8.843,5.644)--(8.960,5.644)--(8.960,6.420)--(8.843,6.420)--cycle;
\draw[gp path] (8.843,5.644)--(8.843,6.419)--(8.959,6.419)--(8.959,5.644)--cycle;
\gpfill{rgb color={0.800,0.800,0.800}} (9.075,5.644)--(9.192,5.644)--(9.192,6.624)--(9.075,6.624)--cycle;
\draw[gp path] (9.075,5.644)--(9.075,6.623)--(9.191,6.623)--(9.191,5.644)--cycle;
\gpfill{rgb color={0.800,0.800,0.800}} (11.163,5.644)--(11.280,5.644)--(11.280,8.720)--(11.163,8.720)--cycle;
\draw[gp path] (11.163,5.644)--(11.163,8.719)--(11.279,8.719)--(11.279,5.644)--cycle;
\gpfill{rgb color={0.800,0.800,0.800}} (12.091,5.644)--(12.208,5.644)--(12.208,7.755)--(12.091,7.755)--cycle;
\draw[gp path] (12.091,5.644)--(12.091,7.754)--(12.207,7.754)--(12.207,5.644)--cycle;
\gpfill{rgb color={0.800,0.800,0.800}} (12.323,5.644)--(12.440,5.644)--(12.440,10.132)--(12.323,10.132)--cycle;
\draw[gp path] (12.323,5.644)--(12.323,10.131)--(12.439,10.131)--(12.439,5.644)--cycle;
\gpfill{rgb color={0.800,0.800,0.800}} (15.107,5.644)--(15.224,5.644)--(15.224,9.384)--(15.107,9.384)--cycle;
\draw[gp path] (15.107,5.644)--(15.107,9.383)--(15.223,9.383)--(15.223,5.644)--cycle;
\gpfill{rgb color={0.800,0.800,0.800}} (15.803,5.644)--(15.920,5.644)--(15.920,6.542)--(15.803,6.542)--cycle;
\draw[gp path] (15.803,5.644)--(15.803,6.541)--(15.919,6.541)--(15.919,5.644)--cycle;
\gpfill{rgb color={0.800,0.800,0.800}} (16.035,5.644)--(16.152,5.644)--(16.152,5.803)--(16.035,5.803)--cycle;
\draw[gp path] (16.035,5.644)--(16.035,5.802)--(16.151,5.802)--(16.151,5.644)--cycle;
\gpfill{rgb color={0.800,0.800,0.800}} (16.267,5.644)--(16.384,5.644)--(16.384,8.403)--(16.267,8.403)--cycle;
\draw[gp path] (16.267,5.644)--(16.267,8.402)--(16.383,8.402)--(16.383,5.644)--cycle;
\gpfill{rgb color={0.800,0.800,0.800}} (16.963,5.644)--(17.080,5.644)--(17.080,7.475)--(16.963,7.475)--cycle;
\draw[gp path] (16.963,5.644)--(16.963,7.474)--(17.079,7.474)--(17.079,5.644)--cycle;
\gpfill{rgb color={0.800,0.800,0.800}} (17.195,5.644)--(17.312,5.644)--(17.312,7.287)--(17.195,7.287)--cycle;
\draw[gp path] (17.195,5.644)--(17.195,7.286)--(17.311,7.286)--(17.311,5.644)--cycle;
\gpfill{rgb color={0.800,0.800,0.800}} (17.659,10.131)--(17.776,10.131)--(17.776,10.132)--(17.659,10.132)--cycle;
\draw[gp path] (17.659,10.131)--(17.775,10.131)--cycle;
\gpfill{rgb color={0.800,0.800,0.800}} (17.891,8.529)--(18.008,8.529)--(18.008,8.530)--(17.891,8.530)--cycle;
\draw[gp path] (17.891,8.529)--(18.007,8.529)--cycle;
\gpfill{rgb color={0.800,0.800,0.800}} (18.355,5.644)--(18.472,5.644)--(18.472,8.689)--(18.355,8.689)--cycle;
\draw[gp path] (18.355,5.644)--(18.355,8.688)--(18.471,8.688)--(18.471,5.644)--cycle;
\gpfill{rgb color={0.800,0.800,0.800}} (18.588,5.644)--(18.705,5.644)--(18.705,10.132)--(18.588,10.132)--cycle;
\draw[gp path] (18.588,5.644)--(18.588,10.131)--(18.704,10.131)--(18.704,5.644)--cycle;
\gpfill{rgb color={0.800,0.800,0.800}} (19.052,6.345)--(19.169,6.345)--(19.169,6.346)--(19.052,6.346)--cycle;
\draw[gp path] (19.052,6.345)--(19.168,6.345)--cycle;
\gpfill{rgb color={0.800,0.800,0.800}} (19.516,5.644)--(19.633,5.644)--(19.633,7.081)--(19.516,7.081)--cycle;
\draw[gp path] (19.516,5.644)--(19.516,7.080)--(19.632,7.080)--(19.632,5.644)--cycle;
\gpfill{rgb color={0.800,0.800,0.800}} (19.748,6.712)--(19.865,6.712)--(19.865,6.713)--(19.748,6.713)--cycle;
\draw[gp path] (19.748,6.712)--(19.864,6.712)--cycle;
\gpfill{rgb color={0.800,0.800,0.800}} (19.980,5.644)--(20.097,5.644)--(20.097,6.967)--(19.980,6.967)--cycle;
\draw[gp path] (19.980,5.644)--(19.980,6.966)--(20.096,6.966)--(20.096,5.644)--cycle;
\gpfill{rgb color={0.800,0.800,0.800}} (20.212,5.644)--(20.329,5.644)--(20.329,6.089)--(20.212,6.089)--cycle;
\draw[gp path] (20.212,5.644)--(20.212,6.088)--(20.328,6.088)--(20.328,5.644)--cycle;
\gpfill{rgb color={0.800,0.800,0.800}} (20.676,5.644)--(20.793,5.644)--(20.793,6.279)--(20.676,6.279)--cycle;
\draw[gp path] (20.676,5.644)--(20.676,6.278)--(20.792,6.278)--(20.792,5.644)--cycle;
\gpfill{rgb color={0.800,0.800,0.800}} (20.908,5.716)--(21.025,5.716)--(21.025,5.717)--(20.908,5.717)--cycle;
\draw[gp path] (20.908,5.716)--(21.024,5.716)--cycle;
\gpfill{rgb color={0.800,0.800,0.800}} (21.140,5.644)--(21.257,5.644)--(21.257,6.555)--(21.140,6.555)--cycle;
\draw[gp path] (21.140,5.644)--(21.140,6.554)--(21.256,6.554)--(21.256,5.644)--cycle;
\gpfill{rgb color={0.800,0.800,0.800}} (21.372,6.579)--(21.489,6.579)--(21.489,6.580)--(21.372,6.580)--cycle;
\draw[gp path] (21.372,6.579)--(21.488,6.579)--cycle;
\gpfill{rgb color={0.800,0.800,0.800}} (21.604,6.741)--(21.721,6.741)--(21.721,6.742)--(21.604,6.742)--cycle;
\draw[gp path] (21.604,6.741)--(21.720,6.741)--cycle;
\gpfill{rgb color={0.800,0.800,0.800}} (21.836,5.644)--(21.953,5.644)--(21.953,5.680)--(21.836,5.680)--cycle;
\draw[gp path] (21.836,5.644)--(21.836,5.679)--(21.952,5.679)--(21.952,5.644)--cycle;
\gpfill{rgb color={0.800,0.800,0.800}} (22.532,5.644)--(22.649,5.644)--(22.649,6.320)--(22.532,6.320)--cycle;
\draw[gp path] (22.532,5.644)--(22.532,6.319)--(22.648,6.319)--(22.648,5.644)--cycle;
\gpfill{rgb color={0.800,0.800,0.800}} (22.996,7.370)--(23.113,7.370)--(23.113,7.371)--(22.996,7.371)--cycle;
\draw[gp path] (22.996,7.370)--(23.112,7.370)--cycle;
\gpcolor{color=gp lt color border}
\draw[gp path] (1.012,10.131)--(1.012,5.644)--(24.446,5.644)--(24.446,10.131)--cycle;
\node[gp node center] at (12.729,9.682) {\plotLegend{(Hexagon)}};
\node[gp node left,font={\fontsize{11.0pt}{13.2pt}\selectfont}] at (1.246,9.817) {\plotLegend{mean improvement: $1.3\%$}};
\node[gp node left,font={\fontsize{11.0pt}{13.2pt}\selectfont}] at (1.246,9.368) {\plotLegend{improved functions: $9\%$}};
\node[gp node left,font={\fontsize{11.0pt}{13.2pt}\selectfont}] at (1.246,8.920) {\plotLegend{mean gap: $3\%$}};
\node[gp node left,font={\fontsize{11.0pt}{13.2pt}\selectfont}] at (1.246,8.471) {\plotLegend{optimal functions: $77\%$}};
\gpdefrectangularnode{gp plot 1}{\pgfpoint{1.012cm}{5.644cm}}{\pgfpoint{24.446cm}{10.131cm}}
\end{tikzpicture}

%% file: results/arm-size-improvement.tex
\begin{tikzpicture}[gnuplot]
\path (0.000,0.000) rectangle (12.500,8.750);
\gpcolor{color=gp lt color border}
\gpsetlinetype{gp lt border}
\gpsetdashtype{gp dt solid}
\gpsetlinewidth{1.00}
\draw[gp path] (1.012,5.644)--(1.192,5.644);
\draw[gp path] (24.446,5.644)--(24.266,5.644);
\node[gp node right,font={\fontsize{8.0pt}{9.6pt}\selectfont}] at (0.828,5.644) {\plotPercentage{0}};
\draw[gp path] (1.012,6.205)--(1.192,6.205);
\draw[gp path] (24.446,6.205)--(24.266,6.205);
\node[gp node right,font={\fontsize{8.0pt}{9.6pt}\selectfont}] at (0.828,6.205) {\plotPercentage{5}};
\draw[gp path] (1.012,6.766)--(1.192,6.766);
\draw[gp path] (24.446,6.766)--(24.266,6.766);
\node[gp node right,font={\fontsize{8.0pt}{9.6pt}\selectfont}] at (0.828,6.766) {\plotPercentage{10}};
\draw[gp path] (1.012,7.327)--(1.192,7.327);
\draw[gp path] (24.446,7.327)--(24.266,7.327);
\node[gp node right,font={\fontsize{8.0pt}{9.6pt}\selectfont}] at (0.828,7.327) {\plotPercentage{15}};
\draw[gp path] (1.012,7.888)--(1.192,7.888);
\draw[gp path] (24.446,7.888)--(24.266,7.888);
\node[gp node right,font={\fontsize{8.0pt}{9.6pt}\selectfont}] at (0.828,7.888) {\plotPercentage{20}};
\draw[gp path] (1.012,8.448)--(1.192,8.448);
\draw[gp path] (24.446,8.448)--(24.266,8.448);
\node[gp node right,font={\fontsize{8.0pt}{9.6pt}\selectfont}] at (0.828,8.448) {\plotPercentage{25}};
\draw[gp path] (1.012,9.009)--(1.192,9.009);
\draw[gp path] (24.446,9.009)--(24.266,9.009);
\node[gp node right,font={\fontsize{8.0pt}{9.6pt}\selectfont}] at (0.828,9.009) {\plotPercentage{30}};
\draw[gp path] (1.012,9.570)--(1.192,9.570);
\draw[gp path] (24.446,9.570)--(24.266,9.570);
\node[gp node right,font={\fontsize{8.0pt}{9.6pt}\selectfont}] at (0.828,9.570) {\plotPercentage{35}};
\draw[gp path] (1.012,10.131)--(1.192,10.131);
\draw[gp path] (24.446,10.131)--(24.266,10.131);
\node[gp node right,font={\fontsize{8.0pt}{9.6pt}\selectfont}] at (0.828,10.131) {\plotPercentage{40}};
\node[gp node left,rotate=-90,font={\fontsize{7.0pt}{8.4pt}\selectfont}] at (1.244,5.721) {\functionId{1}\functionName{handle_noinline_attribute}};
\node[gp node left,rotate=-90,font={\fontsize{7.0pt}{8.4pt}\selectfont}] at (1.476,5.721) {\functionId{2}\functionName{control_flow_insn_p}};
\node[gp node left,rotate=-90,font={\fontsize{7.0pt}{8.4pt}\selectfont}] at (1.708,5.721) {\functionId{3}\functionName{insert_insn_on_edge}};
\node[gp node left,rotate=-90,font={\fontsize{7.0pt}{8.4pt}\selectfont}] at (1.940,5.721) {\functionId{4}\functionName{update_br_prob_note}};
\node[gp node left,rotate=-90,font={\fontsize{7.0pt}{8.4pt}\selectfont}] at (2.172,5.721) {\functionId{5}\functionName{_cpp_init_internal_pragma.}};
\node[gp node left,rotate=-90,font={\fontsize{7.0pt}{8.4pt}\selectfont}] at (2.404,5.721) {\functionId{6}\functionName{lex_macro_node}};
\node[gp node left,rotate=-90,font={\fontsize{7.0pt}{8.4pt}\selectfont}] at (2.636,5.721) {\functionId{7}\functionName{cse_basic_block}};
\node[gp node left,rotate=-90,font={\fontsize{7.0pt}{8.4pt}\selectfont}] at (2.868,5.721) {\functionId{8}\functionName{rtx_equal_for_cselib_p}};
\node[gp node left,rotate=-90,font={\fontsize{7.0pt}{8.4pt}\selectfont}] at (3.100,5.721) {\functionId{9}\functionName{debug_df_chain}};
\node[gp node left,rotate=-90,font={\fontsize{7.0pt}{8.4pt}\selectfont}] at (3.332,5.721) {\functionId{10}\functionName{modified_type_die}};
\node[gp node left,rotate=-90,font={\fontsize{7.0pt}{8.4pt}\selectfont}] at (3.564,5.721) {\functionId{11}\functionName{emit_note}};
\node[gp node left,rotate=-90,font={\fontsize{7.0pt}{8.4pt}\selectfont}] at (3.796,5.721) {\functionId{12}\functionName{gen_sequence}};
\node[gp node left,rotate=-90,font={\fontsize{7.0pt}{8.4pt}\selectfont}] at (4.028,5.721) {\functionId{13}\functionName{subreg_hard_regno}};
\node[gp node left,rotate=-90,font={\fontsize{7.0pt}{8.4pt}\selectfont}] at (4.260,5.721) {\functionId{14}\functionName{split_double}};
\node[gp node left,rotate=-90,font={\fontsize{7.0pt}{8.4pt}\selectfont}] at (4.492,5.721) {\functionId{15}\functionName{add_to_mem_set_list}};
\node[gp node left,rotate=-90,font={\fontsize{7.0pt}{8.4pt}\selectfont}] at (4.724,5.721) {\functionId{16}\functionName{find_regno_partial}};
\node[gp node left,rotate=-90,font={\fontsize{7.0pt}{8.4pt}\selectfont}] at (4.956,5.721) {\functionId{17}\functionName{use_return_register}};
\node[gp node left,rotate=-90,font={\fontsize{7.0pt}{8.4pt}\selectfont}] at (5.188,5.721) {\functionId{18}\functionName{ix86_expand_move}};
\node[gp node left,rotate=-90,font={\fontsize{7.0pt}{8.4pt}\selectfont}] at (5.420,5.721) {\functionId{19}\functionName{legitimate_pic_address_di.}};
\node[gp node left,rotate=-90,font={\fontsize{7.0pt}{8.4pt}\selectfont}] at (5.652,5.721) {\functionId{20}\functionName{gen_extendsfdf2}};
\node[gp node left,rotate=-90,font={\fontsize{7.0pt}{8.4pt}\selectfont}] at (5.884,5.721) {\functionId{21}\functionName{gen_mulsidi3}};
\node[gp node left,rotate=-90,font={\fontsize{7.0pt}{8.4pt}\selectfont}] at (6.116,5.721) {\functionId{22}\functionName{gen_peephole2_1255}};
\node[gp node left,rotate=-90,font={\fontsize{7.0pt}{8.4pt}\selectfont}] at (6.348,5.721) {\functionId{23}\functionName{gen_peephole2_1271}};
\node[gp node left,rotate=-90,font={\fontsize{7.0pt}{8.4pt}\selectfont}] at (6.580,5.721) {\functionId{24}\functionName{gen_peephole2_1277}};
\node[gp node left,rotate=-90,font={\fontsize{7.0pt}{8.4pt}\selectfont}] at (6.812,5.721) {\functionId{25}\functionName{gen_pfnacc}};
\node[gp node left,rotate=-90,font={\fontsize{7.0pt}{8.4pt}\selectfont}] at (7.045,5.721) {\functionId{26}\functionName{gen_rotlsi3}};
\node[gp node left,rotate=-90,font={\fontsize{7.0pt}{8.4pt}\selectfont}] at (7.277,5.721) {\functionId{27}\functionName{gen_split_1001}};
\node[gp node left,rotate=-90,font={\fontsize{7.0pt}{8.4pt}\selectfont}] at (7.509,5.721) {\functionId{28}\functionName{gen_split_1028}};
\node[gp node left,rotate=-90,font={\fontsize{7.0pt}{8.4pt}\selectfont}] at (7.741,5.721) {\functionId{29}\functionName{gen_sse_nandti3}};
\node[gp node left,rotate=-90,font={\fontsize{7.0pt}{8.4pt}\selectfont}] at (7.973,5.721) {\functionId{30}\functionName{gen_sunge}};
\node[gp node left,rotate=-90,font={\fontsize{7.0pt}{8.4pt}\selectfont}] at (8.205,5.721) {\functionId{31}\functionName{insert_loop_mem}};
\node[gp node left,rotate=-90,font={\fontsize{7.0pt}{8.4pt}\selectfont}] at (8.437,5.721) {\functionId{32}\functionName{eiremain}};
\node[gp node left,rotate=-90,font={\fontsize{7.0pt}{8.4pt}\selectfont}] at (8.669,5.721) {\functionId{33}\functionName{elimination_effects}};
\node[gp node left,rotate=-90,font={\fontsize{7.0pt}{8.4pt}\selectfont}] at (8.901,5.721) {\functionId{34}\functionName{gen_reload}};
\node[gp node left,rotate=-90,font={\fontsize{7.0pt}{8.4pt}\selectfont}] at (9.133,5.721) {\functionId{35}\functionName{reload_cse_simplify_set}};
\node[gp node left,rotate=-90,font={\fontsize{7.0pt}{8.4pt}\selectfont}] at (9.365,5.721) {\functionId{36}\functionName{simplify_binary_is2orm1}};
\node[gp node left,rotate=-90,font={\fontsize{7.0pt}{8.4pt}\selectfont}] at (9.597,5.721) {\functionId{37}\functionName{remove_phi_alternative}};
\node[gp node left,rotate=-90,font={\fontsize{7.0pt}{8.4pt}\selectfont}] at (9.829,5.721) {\functionId{38}\functionName{contains_placeholder_p}};
\node[gp node left,rotate=-90,font={\fontsize{7.0pt}{8.4pt}\selectfont}] at (10.061,5.721) {\functionId{39}\functionName{assemble_end_function}};
\node[gp node left,rotate=-90,font={\fontsize{7.0pt}{8.4pt}\selectfont}] at (10.293,5.721) {\functionId{40}\functionName{default_named_section_asm.}};
\node[gp node left,rotate=-90,font={\fontsize{7.0pt}{8.4pt}\selectfont}] at (10.525,5.721) {\functionId{41}\functionName{sample_unpack_12}};
\node[gp node left,rotate=-90,font={\fontsize{7.0pt}{8.4pt}\selectfont}] at (10.757,5.721) {\functionId{42}\functionName{autohelperattpat10}};
\node[gp node left,rotate=-90,font={\fontsize{7.0pt}{8.4pt}\selectfont}] at (10.989,5.721) {\functionId{43}\functionName{autohelperbarrierspat126}};
\node[gp node left,rotate=-90,font={\fontsize{7.0pt}{8.4pt}\selectfont}] at (11.221,5.721) {\functionId{44}\functionName{atari_atari_attack_callba.}};
\node[gp node left,rotate=-90,font={\fontsize{7.0pt}{8.4pt}\selectfont}] at (11.453,5.721) {\functionId{45}\functionName{compute_aa_status}};
\node[gp node left,rotate=-90,font={\fontsize{7.0pt}{8.4pt}\selectfont}] at (11.685,5.721) {\functionId{46}\functionName{dragon_weak}};
\node[gp node left,rotate=-90,font={\fontsize{7.0pt}{8.4pt}\selectfont}] at (11.917,5.721) {\functionId{47}\functionName{get_saved_worms}};
\node[gp node left,rotate=-90,font={\fontsize{7.0pt}{8.4pt}\selectfont}] at (12.149,5.721) {\functionId{48}\functionName{read_eye}};
\node[gp node left,rotate=-90,font={\fontsize{7.0pt}{8.4pt}\selectfont}] at (12.381,5.721) {\functionId{49}\functionName{topological_eye}};
\node[gp node left,rotate=-90,font={\fontsize{7.0pt}{8.4pt}\selectfont}] at (12.613,5.721) {\functionId{50}\functionName{autohelperowl_attackpat19.}};
\node[gp node left,rotate=-90,font={\fontsize{7.0pt}{8.4pt}\selectfont}] at (12.845,5.721) {\functionId{51}\functionName{autohelperowl_attackpat29.}};
\node[gp node left,rotate=-90,font={\fontsize{7.0pt}{8.4pt}\selectfont}] at (13.077,5.721) {\functionId{52}\functionName{autohelperowl_defendpat28.}};
\node[gp node left,rotate=-90,font={\fontsize{7.0pt}{8.4pt}\selectfont}] at (13.309,5.721) {\functionId{53}\functionName{autohelperowl_defendpat38.}};
\node[gp node left,rotate=-90,font={\fontsize{7.0pt}{8.4pt}\selectfont}] at (13.541,5.721) {\functionId{54}\functionName{autohelperpat1114}};
\node[gp node left,rotate=-90,font={\fontsize{7.0pt}{8.4pt}\selectfont}] at (13.773,5.721) {\functionId{55}\functionName{autohelperpat335}};
\node[gp node left,rotate=-90,font={\fontsize{7.0pt}{8.4pt}\selectfont}] at (14.005,5.721) {\functionId{56}\functionName{autohelperpat508}};
\node[gp node left,rotate=-90,font={\fontsize{7.0pt}{8.4pt}\selectfont}] at (14.237,5.721) {\functionId{57}\functionName{autohelperpat83}};
\node[gp node left,rotate=-90,font={\fontsize{7.0pt}{8.4pt}\selectfont}] at (14.469,5.721) {\functionId{58}\functionName{simple_showboard}};
\node[gp node left,rotate=-90,font={\fontsize{7.0pt}{8.4pt}\selectfont}] at (14.701,5.721) {\functionId{59}\functionName{skip_intrabk_SAD}};
\node[gp node left,rotate=-90,font={\fontsize{7.0pt}{8.4pt}\selectfont}] at (14.933,5.721) {\functionId{60}\functionName{free_orig_planes}};
\node[gp node left,rotate=-90,font={\fontsize{7.0pt}{8.4pt}\selectfont}] at (15.165,5.721) {\functionId{61}\functionName{GetSkipCostMB}};
\node[gp node left,rotate=-90,font={\fontsize{7.0pt}{8.4pt}\selectfont}] at (15.397,5.721) {\functionId{62}\functionName{writeSyntaxElement_Level_.}};
\node[gp node left,rotate=-90,font={\fontsize{7.0pt}{8.4pt}\selectfont}] at (15.629,5.721) {\functionId{63}\functionName{GSIAddKeyToIndex}};
\node[gp node left,rotate=-90,font={\fontsize{7.0pt}{8.4pt}\selectfont}] at (15.861,5.721) {\functionId{64}\functionName{EVDBasicFit}};
\node[gp node left,rotate=-90,font={\fontsize{7.0pt}{8.4pt}\selectfont}] at (16.093,5.721) {\functionId{65}\functionName{SampleDirichlet}};
\node[gp node left,rotate=-90,font={\fontsize{7.0pt}{8.4pt}\selectfont}] at (16.325,5.721) {\functionId{66}\functionName{DegenerateSymbolScore}};
\node[gp node left,rotate=-90,font={\fontsize{7.0pt}{8.4pt}\selectfont}] at (16.557,5.721) {\functionId{67}\functionName{Plan7SetCtime}};
\node[gp node left,rotate=-90,font={\fontsize{7.0pt}{8.4pt}\selectfont}] at (16.789,5.721) {\functionId{68}\functionName{MSAToSqinfo}};
\node[gp node left,rotate=-90,font={\fontsize{7.0pt}{8.4pt}\selectfont}] at (17.021,5.721) {\functionId{69}\functionName{null_convert}};
\node[gp node left,rotate=-90,font={\fontsize{7.0pt}{8.4pt}\selectfont}] at (17.253,5.721) {\functionId{70}\functionName{jinit_c_prep_controller}};
\node[gp node left,rotate=-90,font={\fontsize{7.0pt}{8.4pt}\selectfont}] at (17.485,5.721) {\functionId{71}\functionName{glFogf}};
\node[gp node left,rotate=-90,font={\fontsize{7.0pt}{8.4pt}\selectfont}] at (17.717,5.721) {\functionId{72}\functionName{glNormal3d}};
\node[gp node left,rotate=-90,font={\fontsize{7.0pt}{8.4pt}\selectfont}] at (17.949,5.721) {\functionId{73}\functionName{glRasterPos3d}};
\node[gp node left,rotate=-90,font={\fontsize{7.0pt}{8.4pt}\selectfont}] at (18.181,5.721) {\functionId{74}\functionName{glTexCoord2d}};
\node[gp node left,rotate=-90,font={\fontsize{7.0pt}{8.4pt}\selectfont}] at (18.413,5.721) {\functionId{75}\functionName{gl_stippled_bresenham}};
\node[gp node left,rotate=-90,font={\fontsize{7.0pt}{8.4pt}\selectfont}] at (18.646,5.721) {\functionId{76}\functionName{gl_save_Frustum}};
\node[gp node left,rotate=-90,font={\fontsize{7.0pt}{8.4pt}\selectfont}] at (18.878,5.721) {\functionId{77}\functionName{gl_save_LineWidth}};
\node[gp node left,rotate=-90,font={\fontsize{7.0pt}{8.4pt}\selectfont}] at (19.110,5.721) {\functionId{78}\functionName{translate_id}};
\node[gp node left,rotate=-90,font={\fontsize{7.0pt}{8.4pt}\selectfont}] at (19.342,5.721) {\functionId{79}\functionName{gl_Map1f}};
\node[gp node left,rotate=-90,font={\fontsize{7.0pt}{8.4pt}\selectfont}] at (19.574,5.721) {\functionId{80}\functionName{smooth_ci_line}};
\node[gp node left,rotate=-90,font={\fontsize{7.0pt}{8.4pt}\selectfont}] at (19.806,5.721) {\functionId{81}\functionName{free_unified_knots}};
\node[gp node left,rotate=-90,font={\fontsize{7.0pt}{8.4pt}\selectfont}] at (20.038,5.721) {\functionId{82}\functionName{tess_test_polygon}};
\node[gp node left,rotate=-90,font={\fontsize{7.0pt}{8.4pt}\selectfont}] at (20.270,5.721) {\functionId{83}\functionName{auxWireBox}};
\node[gp node left,rotate=-90,font={\fontsize{7.0pt}{8.4pt}\selectfont}] at (20.502,5.721) {\functionId{84}\functionName{gl_ColorPointer}};
\node[gp node left,rotate=-90,font={\fontsize{7.0pt}{8.4pt}\selectfont}] at (20.734,5.721) {\functionId{85}\functionName{r_serial}};
\node[gp node left,rotate=-90,font={\fontsize{7.0pt}{8.4pt}\selectfont}] at (20.966,5.721) {\functionId{86}\functionName{scalar_mult_sub_su3_matri.}};
\node[gp node left,rotate=-90,font={\fontsize{7.0pt}{8.4pt}\selectfont}] at (21.198,5.721) {\functionId{87}\functionName{Decode_MPEG1_Non_Intra_Bl.}};
\node[gp node left,rotate=-90,font={\fontsize{7.0pt}{8.4pt}\selectfont}] at (21.430,5.721) {\functionId{88}\functionName{cpDecodeSecret}};
\node[gp node left,rotate=-90,font={\fontsize{7.0pt}{8.4pt}\selectfont}] at (21.662,5.721) {\functionId{89}\functionName{vlShortLshift}};
\node[gp node left,rotate=-90,font={\fontsize{7.0pt}{8.4pt}\selectfont}] at (21.894,5.721) {\functionId{90}\functionName{encryptfile}};
\node[gp node left,rotate=-90,font={\fontsize{7.0pt}{8.4pt}\selectfont}] at (22.126,5.721) {\functionId{91}\functionName{make_canonical}};
\node[gp node left,rotate=-90,font={\fontsize{7.0pt}{8.4pt}\selectfont}] at (22.358,5.721) {\functionId{92}\functionName{LANG}};
\node[gp node left,rotate=-90,font={\fontsize{7.0pt}{8.4pt}\selectfont}] at (22.590,5.721) {\functionId{93}\functionName{MD5Transform}};
\node[gp node left,rotate=-90,font={\fontsize{7.0pt}{8.4pt}\selectfont}] at (22.822,5.721) {\functionId{94}\functionName{mp_display}};
\node[gp node left,rotate=-90,font={\fontsize{7.0pt}{8.4pt}\selectfont}] at (23.054,5.721) {\functionId{95}\functionName{comp_Jboundaries}};
\node[gp node left,rotate=-90,font={\fontsize{7.0pt}{8.4pt}\selectfont}] at (23.286,5.721) {\functionId{96}\functionName{is_draw}};
\node[gp node left,rotate=-90,font={\fontsize{7.0pt}{8.4pt}\selectfont}] at (23.518,5.721) {\functionId{97}\functionName{push_king}};
\node[gp node left,rotate=-90,font={\fontsize{7.0pt}{8.4pt}\selectfont}] at (23.750,5.721) {\functionId{98}\functionName{stat_retry}};
\node[gp node left,rotate=-90,font={\fontsize{7.0pt}{8.4pt}\selectfont}] at (23.982,5.721) {\functionId{99}\functionName{lextree_subtree_print}};
\node[gp node left,rotate=-90,font={\fontsize{7.0pt}{8.4pt}\selectfont}] at (24.214,5.721) {\functionId{100}\functionName{lm_tg_score}};
\draw[gp path] (1.012,10.131)--(1.012,5.644)--(24.446,5.644)--(24.446,10.131)--cycle;
\gpcolor{rgb color={0.580,0.000,0.827}}
\draw[gp path] (1.012,5.644)--(1.249,5.644)--(1.485,5.644)--(1.722,5.644)--(1.959,5.644)%
  --(2.196,5.644)--(2.432,5.644)--(2.669,5.644)--(2.906,5.644)--(3.142,5.644)--(3.379,5.644)%
  --(3.616,5.644)--(3.852,5.644)--(4.089,5.644)--(4.326,5.644)--(4.563,5.644)--(4.799,5.644)%
  --(5.036,5.644)--(5.273,5.644)--(5.509,5.644)--(5.746,5.644)--(5.983,5.644)--(6.220,5.644)%
  --(6.456,5.644)--(6.693,5.644)--(6.930,5.644)--(7.166,5.644)--(7.403,5.644)--(7.640,5.644)%
  --(7.877,5.644)--(8.113,5.644)--(8.350,5.644)--(8.587,5.644)--(8.823,5.644)--(9.060,5.644)%
  --(9.297,5.644)--(9.533,5.644)--(9.770,5.644)--(10.007,5.644)--(10.244,5.644)--(10.480,5.644)%
  --(10.717,5.644)--(10.954,5.644)--(11.190,5.644)--(11.427,5.644)--(11.664,5.644)--(11.901,5.644)%
  --(12.137,5.644)--(12.374,5.644)--(12.611,5.644)--(12.847,5.644)--(13.084,5.644)--(13.321,5.644)%
  --(13.557,5.644)--(13.794,5.644)--(14.031,5.644)--(14.268,5.644)--(14.504,5.644)--(14.741,5.644)%
  --(14.978,5.644)--(15.214,5.644)--(15.451,5.644)--(15.688,5.644)--(15.925,5.644)--(16.161,5.644)%
  --(16.398,5.644)--(16.635,5.644)--(16.871,5.644)--(17.108,5.644)--(17.345,5.644)--(17.581,5.644)%
  --(17.818,5.644)--(18.055,5.644)--(18.292,5.644)--(18.528,5.644)--(18.765,5.644)--(19.002,5.644)%
  --(19.238,5.644)--(19.475,5.644)--(19.712,5.644)--(19.949,5.644)--(20.185,5.644)--(20.422,5.644)%
  --(20.659,5.644)--(20.895,5.644)--(21.132,5.644)--(21.369,5.644)--(21.606,5.644)--(21.842,5.644)%
  --(22.079,5.644)--(22.316,5.644)--(22.552,5.644)--(22.789,5.644)--(23.026,5.644)--(23.262,5.644)%
  --(23.499,5.644)--(23.736,5.644)--(23.973,5.644)--(24.209,5.644)--(24.446,5.644);
\gpfill{rgb color={0.000,0.000,0.000}} (1.186,5.644)--(1.303,5.644)--(1.303,6.046)--(1.186,6.046)--cycle;
\gpcolor{rgb color={0.000,0.000,0.000}}
\draw[gp path] (1.186,5.644)--(1.186,6.045)--(1.302,6.045)--(1.302,5.644)--cycle;
\gpfill{rgb color={0.000,0.000,0.000}} (1.418,5.644)--(1.535,5.644)--(1.535,5.772)--(1.418,5.772)--cycle;
\draw[gp path] (1.418,5.644)--(1.418,5.771)--(1.534,5.771)--(1.534,5.644)--cycle;
\gpfill{rgb color={0.000,0.000,0.000}} (1.650,5.644)--(1.767,5.644)--(1.767,5.861)--(1.650,5.861)--cycle;
\draw[gp path] (1.650,5.644)--(1.650,5.860)--(1.766,5.860)--(1.766,5.644)--cycle;
\gpfill{rgb color={0.000,0.000,0.000}} (2.114,5.644)--(2.231,5.644)--(2.231,6.001)--(2.114,6.001)--cycle;
\draw[gp path] (2.114,5.644)--(2.114,6.000)--(2.230,6.000)--(2.230,5.644)--cycle;
\gpfill{rgb color={0.000,0.000,0.000}} (2.346,5.644)--(2.463,5.644)--(2.463,5.797)--(2.346,5.797)--cycle;
\draw[gp path] (2.346,5.644)--(2.346,5.796)--(2.462,5.796)--(2.462,5.644)--cycle;
\gpfill{rgb color={0.000,0.000,0.000}} (3.042,5.644)--(3.159,5.644)--(3.159,6.600)--(3.042,6.600)--cycle;
\draw[gp path] (3.042,5.644)--(3.042,6.599)--(3.158,6.599)--(3.158,5.644)--cycle;
\gpfill{rgb color={0.000,0.000,0.000}} (3.506,5.644)--(3.623,5.644)--(3.623,5.881)--(3.506,5.881)--cycle;
\draw[gp path] (3.506,5.644)--(3.506,5.880)--(3.622,5.880)--(3.622,5.644)--cycle;
\gpfill{rgb color={0.000,0.000,0.000}} (3.970,5.644)--(4.087,5.644)--(4.087,5.766)--(3.970,5.766)--cycle;
\draw[gp path] (3.970,5.644)--(3.970,5.765)--(4.086,5.765)--(4.086,5.644)--cycle;
\gpfill{rgb color={0.000,0.000,0.000}} (4.434,5.644)--(4.551,5.644)--(4.551,6.826)--(4.434,6.826)--cycle;
\draw[gp path] (4.434,5.644)--(4.434,6.825)--(4.550,6.825)--(4.550,5.644)--cycle;
\gpfill{rgb color={0.000,0.000,0.000}} (4.898,5.644)--(5.015,5.644)--(5.015,5.835)--(4.898,5.835)--cycle;
\draw[gp path] (4.898,5.644)--(4.898,5.834)--(5.014,5.834)--(5.014,5.644)--cycle;
\gpfill{rgb color={0.000,0.000,0.000}} (5.362,5.644)--(5.479,5.644)--(5.479,5.692)--(5.362,5.692)--cycle;
\draw[gp path] (5.362,5.644)--(5.362,5.691)--(5.478,5.691)--(5.478,5.644)--cycle;
\gpfill{rgb color={0.000,0.000,0.000}} (5.826,5.644)--(5.943,5.644)--(5.943,5.879)--(5.826,5.879)--cycle;
\draw[gp path] (5.826,5.644)--(5.826,5.878)--(5.942,5.878)--(5.942,5.644)--cycle;
\gpfill{rgb color={0.000,0.000,0.000}} (6.058,5.644)--(6.175,5.644)--(6.175,5.865)--(6.058,5.865)--cycle;
\draw[gp path] (6.058,5.644)--(6.058,5.864)--(6.174,5.864)--(6.174,5.644)--cycle;
\gpfill{rgb color={0.000,0.000,0.000}} (7.683,5.644)--(7.800,5.644)--(7.800,6.155)--(7.683,6.155)--cycle;
\draw[gp path] (7.683,5.644)--(7.683,6.154)--(7.799,6.154)--(7.799,5.644)--cycle;
\gpfill{rgb color={0.000,0.000,0.000}} (8.147,5.644)--(8.264,5.644)--(8.264,6.705)--(8.147,6.705)--cycle;
\draw[gp path] (8.147,5.644)--(8.147,6.704)--(8.263,6.704)--(8.263,5.644)--cycle;
\gpfill{rgb color={0.000,0.000,0.000}} (8.379,5.644)--(8.496,5.644)--(8.496,5.854)--(8.379,5.854)--cycle;
\draw[gp path] (8.379,5.644)--(8.379,5.853)--(8.495,5.853)--(8.495,5.644)--cycle;
\gpfill{rgb color={0.000,0.000,0.000}} (9.539,5.644)--(9.656,5.644)--(9.656,6.891)--(9.539,6.891)--cycle;
\draw[gp path] (9.539,5.644)--(9.539,6.890)--(9.655,6.890)--(9.655,5.644)--cycle;
\gpfill{rgb color={0.000,0.000,0.000}} (9.771,5.644)--(9.888,5.644)--(9.888,5.898)--(9.771,5.898)--cycle;
\draw[gp path] (9.771,5.644)--(9.771,5.897)--(9.887,5.897)--(9.887,5.644)--cycle;
\gpfill{rgb color={0.000,0.000,0.000}} (10.235,5.644)--(10.352,5.644)--(10.352,6.403)--(10.235,6.403)--cycle;
\draw[gp path] (10.235,5.644)--(10.235,6.402)--(10.351,6.402)--(10.351,5.644)--cycle;
\gpfill{rgb color={0.000,0.000,0.000}} (10.467,5.644)--(10.584,5.644)--(10.584,7.089)--(10.467,7.089)--cycle;
\draw[gp path] (10.467,5.644)--(10.467,7.088)--(10.583,7.088)--(10.583,5.644)--cycle;
\gpfill{rgb color={0.000,0.000,0.000}} (10.699,5.644)--(10.816,5.644)--(10.816,6.688)--(10.699,6.688)--cycle;
\draw[gp path] (10.699,5.644)--(10.699,6.687)--(10.815,6.687)--(10.815,5.644)--cycle;
\gpfill{rgb color={0.000,0.000,0.000}} (12.555,5.644)--(12.672,5.644)--(12.672,7.023)--(12.555,7.023)--cycle;
\draw[gp path] (12.555,5.644)--(12.555,7.022)--(12.671,7.022)--(12.671,5.644)--cycle;
\gpfill{rgb color={0.000,0.000,0.000}} (12.787,5.644)--(12.904,5.644)--(12.904,6.246)--(12.787,6.246)--cycle;
\draw[gp path] (12.787,5.644)--(12.787,6.245)--(12.903,6.245)--(12.903,5.644)--cycle;
\gpfill{rgb color={0.000,0.000,0.000}} (13.019,5.644)--(13.136,5.644)--(13.136,6.580)--(13.019,6.580)--cycle;
\draw[gp path] (13.019,5.644)--(13.019,6.579)--(13.135,6.579)--(13.135,5.644)--cycle;
\gpfill{rgb color={0.000,0.000,0.000}} (13.251,5.644)--(13.368,5.644)--(13.368,5.996)--(13.251,5.996)--cycle;
\draw[gp path] (13.251,5.644)--(13.251,5.995)--(13.367,5.995)--(13.367,5.644)--cycle;
\gpfill{rgb color={0.000,0.000,0.000}} (13.715,5.644)--(13.832,5.644)--(13.832,6.665)--(13.715,6.665)--cycle;
\draw[gp path] (13.715,5.644)--(13.715,6.664)--(13.831,6.664)--(13.831,5.644)--cycle;
\gpfill{rgb color={0.000,0.000,0.000}} (13.947,5.644)--(14.064,5.644)--(14.064,6.032)--(13.947,6.032)--cycle;
\draw[gp path] (13.947,5.644)--(13.947,6.031)--(14.063,6.031)--(14.063,5.644)--cycle;
\gpfill{rgb color={0.000,0.000,0.000}} (14.179,5.644)--(14.296,5.644)--(14.296,6.703)--(14.179,6.703)--cycle;
\draw[gp path] (14.179,5.644)--(14.179,6.702)--(14.295,6.702)--(14.295,5.644)--cycle;
\gpfill{rgb color={0.000,0.000,0.000}} (14.643,5.644)--(14.760,5.644)--(14.760,6.557)--(14.643,6.557)--cycle;
\draw[gp path] (14.643,5.644)--(14.643,6.556)--(14.759,6.556)--(14.759,5.644)--cycle;
\gpfill{rgb color={0.000,0.000,0.000}} (14.875,5.644)--(14.992,5.644)--(14.992,5.777)--(14.875,5.777)--cycle;
\draw[gp path] (14.875,5.644)--(14.875,5.776)--(14.991,5.776)--(14.991,5.644)--cycle;
\gpfill{rgb color={0.000,0.000,0.000}} (15.339,5.644)--(15.456,5.644)--(15.456,7.264)--(15.339,7.264)--cycle;
\draw[gp path] (15.339,5.644)--(15.339,7.263)--(15.455,7.263)--(15.455,5.644)--cycle;
\gpfill{rgb color={0.000,0.000,0.000}} (15.571,5.644)--(15.688,5.644)--(15.688,5.762)--(15.571,5.762)--cycle;
\draw[gp path] (15.571,5.644)--(15.571,5.761)--(15.687,5.761)--(15.687,5.644)--cycle;
\gpfill{rgb color={0.000,0.000,0.000}} (16.267,5.644)--(16.384,5.644)--(16.384,5.965)--(16.267,5.965)--cycle;
\draw[gp path] (16.267,5.644)--(16.267,5.964)--(16.383,5.964)--(16.383,5.644)--cycle;
\gpfill{rgb color={0.000,0.000,0.000}} (16.963,5.644)--(17.080,5.644)--(17.080,6.642)--(16.963,6.642)--cycle;
\draw[gp path] (16.963,5.644)--(16.963,6.641)--(17.079,6.641)--(17.079,5.644)--cycle;
\gpfill{rgb color={0.000,0.000,0.000}} (17.891,5.644)--(18.008,5.644)--(18.008,5.818)--(17.891,5.818)--cycle;
\draw[gp path] (17.891,5.644)--(17.891,5.817)--(18.007,5.817)--(18.007,5.644)--cycle;
\gpfill{rgb color={0.000,0.000,0.000}} (18.820,5.644)--(18.937,5.644)--(18.937,6.393)--(18.820,6.393)--cycle;
\draw[gp path] (18.820,5.644)--(18.820,6.392)--(18.936,6.392)--(18.936,5.644)--cycle;
\gpfill{rgb color={0.000,0.000,0.000}} (19.748,5.644)--(19.865,5.644)--(19.865,6.007)--(19.748,6.007)--cycle;
\draw[gp path] (19.748,5.644)--(19.748,6.006)--(19.864,6.006)--(19.864,5.644)--cycle;
\gpfill{rgb color={0.000,0.000,0.000}} (20.212,5.644)--(20.329,5.644)--(20.329,6.508)--(20.212,6.508)--cycle;
\draw[gp path] (20.212,5.644)--(20.212,6.507)--(20.328,6.507)--(20.328,5.644)--cycle;
\gpfill{rgb color={0.000,0.000,0.000}} (20.444,5.644)--(20.561,5.644)--(20.561,6.847)--(20.444,6.847)--cycle;
\draw[gp path] (20.444,5.644)--(20.444,6.846)--(20.560,6.846)--(20.560,5.644)--cycle;
\gpfill{rgb color={0.000,0.000,0.000}} (21.604,5.644)--(21.721,5.644)--(21.721,6.495)--(21.604,6.495)--cycle;
\draw[gp path] (21.604,5.644)--(21.604,6.494)--(21.720,6.494)--(21.720,5.644)--cycle;
\gpfill{rgb color={0.000,0.000,0.000}} (22.068,5.644)--(22.185,5.644)--(22.185,6.435)--(22.068,6.435)--cycle;
\draw[gp path] (22.068,5.644)--(22.068,6.434)--(22.184,6.434)--(22.184,5.644)--cycle;
\gpfill{rgb color={0.000,0.000,0.000}} (23.228,5.644)--(23.345,5.644)--(23.345,6.022)--(23.228,6.022)--cycle;
\draw[gp path] (23.228,5.644)--(23.228,6.021)--(23.344,6.021)--(23.344,5.644)--cycle;
\gpfill{rgb color={0.000,0.000,0.000}} (23.460,5.644)--(23.577,5.644)--(23.577,5.922)--(23.460,5.922)--cycle;
\draw[gp path] (23.460,5.644)--(23.460,5.921)--(23.576,5.921)--(23.576,5.644)--cycle;
\gpfill{rgb color={0.000,0.000,0.000}} (23.692,5.644)--(23.809,5.644)--(23.809,6.535)--(23.692,6.535)--cycle;
\draw[gp path] (23.692,5.644)--(23.692,6.534)--(23.808,6.534)--(23.808,5.644)--cycle;
\gpfill{rgb color={0.000,0.000,0.000}} (23.924,5.644)--(24.041,5.644)--(24.041,6.736)--(23.924,6.736)--cycle;
\draw[gp path] (23.924,5.644)--(23.924,6.735)--(24.040,6.735)--(24.040,5.644)--cycle;
\gpfill{rgb color={0.800,0.800,0.800}} (1.186,6.045)--(1.303,6.045)--(1.303,6.046)--(1.186,6.046)--cycle;
\gpcolor{rgb color={0.800,0.800,0.800}}
\draw[gp path] (1.186,6.045)--(1.302,6.045)--cycle;
\gpfill{rgb color={0.800,0.800,0.800}} (1.418,5.771)--(1.535,5.771)--(1.535,5.772)--(1.418,5.772)--cycle;
\draw[gp path] (1.418,5.771)--(1.534,5.771)--cycle;
\gpfill{rgb color={0.800,0.800,0.800}} (1.650,5.860)--(1.767,5.860)--(1.767,5.861)--(1.650,5.861)--cycle;
\draw[gp path] (1.650,5.860)--(1.766,5.860)--cycle;
\gpfill{rgb color={0.800,0.800,0.800}} (2.114,6.000)--(2.231,6.000)--(2.231,6.001)--(2.114,6.001)--cycle;
\draw[gp path] (2.114,6.000)--(2.230,6.000)--cycle;
\gpfill{rgb color={0.800,0.800,0.800}} (2.346,5.796)--(2.463,5.796)--(2.463,5.797)--(2.346,5.797)--cycle;
\draw[gp path] (2.346,5.796)--(2.462,5.796)--cycle;
\gpfill{rgb color={0.800,0.800,0.800}} (2.578,5.644)--(2.695,5.644)--(2.695,8.369)--(2.578,8.369)--cycle;
\draw[gp path] (2.578,5.644)--(2.578,8.368)--(2.694,8.368)--(2.694,5.644)--cycle;
\gpfill{rgb color={0.800,0.800,0.800}} (2.810,5.644)--(2.927,5.644)--(2.927,5.940)--(2.810,5.940)--cycle;
\draw[gp path] (2.810,5.644)--(2.810,5.939)--(2.926,5.939)--(2.926,5.644)--cycle;
\gpfill{rgb color={0.800,0.800,0.800}} (3.042,6.599)--(3.159,6.599)--(3.159,6.600)--(3.042,6.600)--cycle;
\draw[gp path] (3.042,6.599)--(3.158,6.599)--cycle;
\gpfill{rgb color={0.800,0.800,0.800}} (3.274,5.644)--(3.391,5.644)--(3.391,7.450)--(3.274,7.450)--cycle;
\draw[gp path] (3.274,5.644)--(3.274,7.449)--(3.390,7.449)--(3.390,5.644)--cycle;
\gpfill{rgb color={0.800,0.800,0.800}} (3.506,5.880)--(3.623,5.880)--(3.623,5.881)--(3.506,5.881)--cycle;
\draw[gp path] (3.506,5.880)--(3.622,5.880)--cycle;
\gpfill{rgb color={0.800,0.800,0.800}} (3.970,5.765)--(4.087,5.765)--(4.087,5.766)--(3.970,5.766)--cycle;
\draw[gp path] (3.970,5.765)--(4.086,5.765)--cycle;
\gpfill{rgb color={0.800,0.800,0.800}} (4.434,6.825)--(4.551,6.825)--(4.551,6.826)--(4.434,6.826)--cycle;
\draw[gp path] (4.434,6.825)--(4.550,6.825)--cycle;
\gpfill{rgb color={0.800,0.800,0.800}} (4.898,5.834)--(5.015,5.834)--(5.015,5.835)--(4.898,5.835)--cycle;
\draw[gp path] (4.898,5.834)--(5.014,5.834)--cycle;
\gpfill{rgb color={0.800,0.800,0.800}} (5.130,5.644)--(5.247,5.644)--(5.247,6.206)--(5.130,6.206)--cycle;
\draw[gp path] (5.130,5.644)--(5.130,6.205)--(5.246,6.205)--(5.246,5.644)--cycle;
\gpfill{rgb color={0.800,0.800,0.800}} (5.362,5.691)--(5.479,5.691)--(5.479,5.692)--(5.362,5.692)--cycle;
\draw[gp path] (5.362,5.691)--(5.478,5.691)--cycle;
\gpfill{rgb color={0.800,0.800,0.800}} (5.826,5.878)--(5.943,5.878)--(5.943,5.879)--(5.826,5.879)--cycle;
\draw[gp path] (5.826,5.878)--(5.942,5.878)--cycle;
\gpfill{rgb color={0.800,0.800,0.800}} (6.058,5.864)--(6.175,5.864)--(6.175,5.865)--(6.058,5.865)--cycle;
\draw[gp path] (6.058,5.864)--(6.174,5.864)--cycle;
\gpfill{rgb color={0.800,0.800,0.800}} (7.219,5.644)--(7.336,5.644)--(7.336,6.494)--(7.219,6.494)--cycle;
\draw[gp path] (7.219,5.644)--(7.219,6.493)--(7.335,6.493)--(7.335,5.644)--cycle;
\gpfill{rgb color={0.800,0.800,0.800}} (7.451,5.644)--(7.568,5.644)--(7.568,6.977)--(7.451,6.977)--cycle;
\draw[gp path] (7.451,5.644)--(7.451,6.976)--(7.567,6.976)--(7.567,5.644)--cycle;
\gpfill{rgb color={0.800,0.800,0.800}} (7.683,6.154)--(7.800,6.154)--(7.800,6.155)--(7.683,6.155)--cycle;
\draw[gp path] (7.683,6.154)--(7.799,6.154)--cycle;
\gpfill{rgb color={0.800,0.800,0.800}} (8.147,6.704)--(8.264,6.704)--(8.264,6.705)--(8.147,6.705)--cycle;
\draw[gp path] (8.147,6.704)--(8.263,6.704)--cycle;
\gpfill{rgb color={0.800,0.800,0.800}} (8.379,5.853)--(8.496,5.853)--(8.496,8.994)--(8.379,8.994)--cycle;
\draw[gp path] (8.379,5.853)--(8.379,8.993)--(8.495,8.993)--(8.495,5.853)--cycle;
\gpfill{rgb color={0.800,0.800,0.800}} (8.611,5.644)--(8.728,5.644)--(8.728,8.120)--(8.611,8.120)--cycle;
\draw[gp path] (8.611,5.644)--(8.611,8.119)--(8.727,8.119)--(8.727,5.644)--cycle;
\gpfill{rgb color={0.800,0.800,0.800}} (8.843,5.644)--(8.960,5.644)--(8.960,7.330)--(8.843,7.330)--cycle;
\draw[gp path] (8.843,5.644)--(8.843,7.329)--(8.959,7.329)--(8.959,5.644)--cycle;
\gpfill{rgb color={0.800,0.800,0.800}} (9.075,5.644)--(9.192,5.644)--(9.192,7.820)--(9.075,7.820)--cycle;
\draw[gp path] (9.075,5.644)--(9.075,7.819)--(9.191,7.819)--(9.191,5.644)--cycle;
\gpfill{rgb color={0.800,0.800,0.800}} (9.307,5.644)--(9.424,5.644)--(9.424,10.132)--(9.307,10.132)--cycle;
\draw[gp path] (9.307,5.644)--(9.307,10.131)--(9.423,10.131)--(9.423,5.644)--cycle;
\gpfill{rgb color={0.800,0.800,0.800}} (9.539,6.890)--(9.656,6.890)--(9.656,6.891)--(9.539,6.891)--cycle;
\draw[gp path] (9.539,6.890)--(9.655,6.890)--cycle;
\gpfill{rgb color={0.800,0.800,0.800}} (9.771,5.897)--(9.888,5.897)--(9.888,5.898)--(9.771,5.898)--cycle;
\draw[gp path] (9.771,5.897)--(9.887,5.897)--cycle;
\gpfill{rgb color={0.800,0.800,0.800}} (10.003,5.644)--(10.120,5.644)--(10.120,7.685)--(10.003,7.685)--cycle;
\draw[gp path] (10.003,5.644)--(10.003,7.684)--(10.119,7.684)--(10.119,5.644)--cycle;
\gpfill{rgb color={0.800,0.800,0.800}} (10.235,6.402)--(10.352,6.402)--(10.352,6.403)--(10.235,6.403)--cycle;
\draw[gp path] (10.235,6.402)--(10.351,6.402)--cycle;
\gpfill{rgb color={0.800,0.800,0.800}} (10.467,7.088)--(10.584,7.088)--(10.584,7.089)--(10.467,7.089)--cycle;
\draw[gp path] (10.467,7.088)--(10.583,7.088)--cycle;
\gpfill{rgb color={0.800,0.800,0.800}} (10.699,6.687)--(10.816,6.687)--(10.816,6.688)--(10.699,6.688)--cycle;
\draw[gp path] (10.699,6.687)--(10.815,6.687)--cycle;
\gpfill{rgb color={0.800,0.800,0.800}} (10.931,5.644)--(11.048,5.644)--(11.048,7.637)--(10.931,7.637)--cycle;
\draw[gp path] (10.931,5.644)--(10.931,7.636)--(11.047,7.636)--(11.047,5.644)--cycle;
\gpfill{rgb color={0.800,0.800,0.800}} (11.163,5.644)--(11.280,5.644)--(11.280,8.836)--(11.163,8.836)--cycle;
\draw[gp path] (11.163,5.644)--(11.163,8.835)--(11.279,8.835)--(11.279,5.644)--cycle;
\gpfill{rgb color={0.800,0.800,0.800}} (11.395,5.644)--(11.512,5.644)--(11.512,7.760)--(11.395,7.760)--cycle;
\draw[gp path] (11.395,5.644)--(11.395,7.759)--(11.511,7.759)--(11.511,5.644)--cycle;
\gpfill{rgb color={0.800,0.800,0.800}} (11.859,5.644)--(11.976,5.644)--(11.976,10.132)--(11.859,10.132)--cycle;
\draw[gp path] (11.859,5.644)--(11.859,10.131)--(11.975,10.131)--(11.975,5.644)--cycle;
\gpfill{rgb color={0.800,0.800,0.800}} (12.091,5.644)--(12.208,5.644)--(12.208,7.247)--(12.091,7.247)--cycle;
\draw[gp path] (12.091,5.644)--(12.091,7.246)--(12.207,7.246)--(12.207,5.644)--cycle;
\gpfill{rgb color={0.800,0.800,0.800}} (12.323,5.644)--(12.440,5.644)--(12.440,10.132)--(12.323,10.132)--cycle;
\draw[gp path] (12.323,5.644)--(12.323,10.131)--(12.439,10.131)--(12.439,5.644)--cycle;
\gpfill{rgb color={0.800,0.800,0.800}} (12.555,7.022)--(12.672,7.022)--(12.672,7.023)--(12.555,7.023)--cycle;
\draw[gp path] (12.555,7.022)--(12.671,7.022)--cycle;
\gpfill{rgb color={0.800,0.800,0.800}} (12.787,6.245)--(12.904,6.245)--(12.904,6.246)--(12.787,6.246)--cycle;
\draw[gp path] (12.787,6.245)--(12.903,6.245)--cycle;
\gpfill{rgb color={0.800,0.800,0.800}} (13.019,6.579)--(13.136,6.579)--(13.136,6.580)--(13.019,6.580)--cycle;
\draw[gp path] (13.019,6.579)--(13.135,6.579)--cycle;
\gpfill{rgb color={0.800,0.800,0.800}} (13.251,5.995)--(13.368,5.995)--(13.368,5.996)--(13.251,5.996)--cycle;
\draw[gp path] (13.251,5.995)--(13.367,5.995)--cycle;
\gpfill{rgb color={0.800,0.800,0.800}} (13.715,6.664)--(13.832,6.664)--(13.832,6.665)--(13.715,6.665)--cycle;
\draw[gp path] (13.715,6.664)--(13.831,6.664)--cycle;
\gpfill{rgb color={0.800,0.800,0.800}} (13.947,6.031)--(14.064,6.031)--(14.064,6.032)--(13.947,6.032)--cycle;
\draw[gp path] (13.947,6.031)--(14.063,6.031)--cycle;
\gpfill{rgb color={0.800,0.800,0.800}} (14.179,6.702)--(14.296,6.702)--(14.296,6.703)--(14.179,6.703)--cycle;
\draw[gp path] (14.179,6.702)--(14.295,6.702)--cycle;
\gpfill{rgb color={0.800,0.800,0.800}} (14.411,5.644)--(14.528,5.644)--(14.528,8.739)--(14.411,8.739)--cycle;
\draw[gp path] (14.411,5.644)--(14.411,8.738)--(14.527,8.738)--(14.527,5.644)--cycle;
\gpfill{rgb color={0.800,0.800,0.800}} (14.643,6.556)--(14.760,6.556)--(14.760,7.103)--(14.643,7.103)--cycle;
\draw[gp path] (14.643,6.556)--(14.643,7.102)--(14.759,7.102)--(14.759,6.556)--cycle;
\gpfill{rgb color={0.800,0.800,0.800}} (14.875,5.776)--(14.992,5.776)--(14.992,5.777)--(14.875,5.777)--cycle;
\draw[gp path] (14.875,5.776)--(14.991,5.776)--cycle;
\gpfill{rgb color={0.800,0.800,0.800}} (15.107,5.644)--(15.224,5.644)--(15.224,10.132)--(15.107,10.132)--cycle;
\draw[gp path] (15.107,5.644)--(15.107,10.131)--(15.223,10.131)--(15.223,5.644)--cycle;
\gpfill{rgb color={0.800,0.800,0.800}} (15.339,7.263)--(15.456,7.263)--(15.456,7.264)--(15.339,7.264)--cycle;
\draw[gp path] (15.339,7.263)--(15.455,7.263)--cycle;
\gpfill{rgb color={0.800,0.800,0.800}} (15.571,5.761)--(15.688,5.761)--(15.688,5.762)--(15.571,5.762)--cycle;
\draw[gp path] (15.571,5.761)--(15.687,5.761)--cycle;
\gpfill{rgb color={0.800,0.800,0.800}} (15.803,5.644)--(15.920,5.644)--(15.920,6.311)--(15.803,6.311)--cycle;
\draw[gp path] (15.803,5.644)--(15.803,6.310)--(15.919,6.310)--(15.919,5.644)--cycle;
\gpfill{rgb color={0.800,0.800,0.800}} (16.267,5.964)--(16.384,5.964)--(16.384,5.965)--(16.267,5.965)--cycle;
\draw[gp path] (16.267,5.964)--(16.383,5.964)--cycle;
\gpfill{rgb color={0.800,0.800,0.800}} (16.731,5.644)--(16.848,5.644)--(16.848,8.463)--(16.731,8.463)--cycle;
\draw[gp path] (16.731,5.644)--(16.731,8.462)--(16.847,8.462)--(16.847,5.644)--cycle;
\gpfill{rgb color={0.800,0.800,0.800}} (16.963,6.641)--(17.080,6.641)--(17.080,6.642)--(16.963,6.642)--cycle;
\draw[gp path] (16.963,6.641)--(17.079,6.641)--cycle;
\gpfill{rgb color={0.800,0.800,0.800}} (17.195,5.644)--(17.312,5.644)--(17.312,8.578)--(17.195,8.578)--cycle;
\draw[gp path] (17.195,5.644)--(17.195,8.577)--(17.311,8.577)--(17.311,5.644)--cycle;
\gpfill{rgb color={0.800,0.800,0.800}} (17.891,5.817)--(18.008,5.817)--(18.008,5.818)--(17.891,5.818)--cycle;
\draw[gp path] (17.891,5.817)--(18.007,5.817)--cycle;
\gpfill{rgb color={0.800,0.800,0.800}} (18.355,5.644)--(18.472,5.644)--(18.472,10.132)--(18.355,10.132)--cycle;
\draw[gp path] (18.355,5.644)--(18.355,10.131)--(18.471,10.131)--(18.471,5.644)--cycle;
\gpfill{rgb color={0.800,0.800,0.800}} (18.588,5.644)--(18.705,5.644)--(18.705,9.081)--(18.588,9.081)--cycle;
\draw[gp path] (18.588,5.644)--(18.588,9.080)--(18.704,9.080)--(18.704,5.644)--cycle;
\gpfill{rgb color={0.800,0.800,0.800}} (18.820,6.392)--(18.937,6.392)--(18.937,6.393)--(18.820,6.393)--cycle;
\draw[gp path] (18.820,6.392)--(18.936,6.392)--cycle;
\gpfill{rgb color={0.800,0.800,0.800}} (19.284,5.644)--(19.401,5.644)--(19.401,7.617)--(19.284,7.617)--cycle;
\draw[gp path] (19.284,5.644)--(19.284,7.616)--(19.400,7.616)--(19.400,5.644)--cycle;
\gpfill{rgb color={0.800,0.800,0.800}} (19.516,5.644)--(19.633,5.644)--(19.633,7.538)--(19.516,7.538)--cycle;
\draw[gp path] (19.516,5.644)--(19.516,7.537)--(19.632,7.537)--(19.632,5.644)--cycle;
\gpfill{rgb color={0.800,0.800,0.800}} (19.748,6.006)--(19.865,6.006)--(19.865,6.007)--(19.748,6.007)--cycle;
\draw[gp path] (19.748,6.006)--(19.864,6.006)--cycle;
\gpfill{rgb color={0.800,0.800,0.800}} (19.980,5.644)--(20.097,5.644)--(20.097,6.596)--(19.980,6.596)--cycle;
\draw[gp path] (19.980,5.644)--(19.980,6.595)--(20.096,6.595)--(20.096,5.644)--cycle;
\gpfill{rgb color={0.800,0.800,0.800}} (20.212,6.507)--(20.329,6.507)--(20.329,6.508)--(20.212,6.508)--cycle;
\draw[gp path] (20.212,6.507)--(20.328,6.507)--cycle;
\gpfill{rgb color={0.800,0.800,0.800}} (20.444,6.846)--(20.561,6.846)--(20.561,6.847)--(20.444,6.847)--cycle;
\draw[gp path] (20.444,6.846)--(20.560,6.846)--cycle;
\gpfill{rgb color={0.800,0.800,0.800}} (20.676,5.644)--(20.793,5.644)--(20.793,10.132)--(20.676,10.132)--cycle;
\draw[gp path] (20.676,5.644)--(20.676,10.131)--(20.792,10.131)--(20.792,5.644)--cycle;
\gpfill{rgb color={0.800,0.800,0.800}} (21.140,5.644)--(21.257,5.644)--(21.257,8.660)--(21.140,8.660)--cycle;
\draw[gp path] (21.140,5.644)--(21.140,8.659)--(21.256,8.659)--(21.256,5.644)--cycle;
\gpfill{rgb color={0.800,0.800,0.800}} (21.604,6.494)--(21.721,6.494)--(21.721,6.495)--(21.604,6.495)--cycle;
\draw[gp path] (21.604,6.494)--(21.720,6.494)--cycle;
\gpfill{rgb color={0.800,0.800,0.800}} (21.836,5.644)--(21.953,5.644)--(21.953,6.305)--(21.836,6.305)--cycle;
\draw[gp path] (21.836,5.644)--(21.836,6.304)--(21.952,6.304)--(21.952,5.644)--cycle;
\gpfill{rgb color={0.800,0.800,0.800}} (22.068,6.434)--(22.185,6.434)--(22.185,6.435)--(22.068,6.435)--cycle;
\draw[gp path] (22.068,6.434)--(22.184,6.434)--cycle;
\gpfill{rgb color={0.800,0.800,0.800}} (22.300,5.644)--(22.417,5.644)--(22.417,8.445)--(22.300,8.445)--cycle;
\draw[gp path] (22.300,5.644)--(22.300,8.444)--(22.416,8.444)--(22.416,5.644)--cycle;
\gpfill{rgb color={0.800,0.800,0.800}} (22.532,5.644)--(22.649,5.644)--(22.649,6.255)--(22.532,6.255)--cycle;
\draw[gp path] (22.532,5.644)--(22.532,6.254)--(22.648,6.254)--(22.648,5.644)--cycle;
\gpfill{rgb color={0.800,0.800,0.800}} (22.764,5.644)--(22.881,5.644)--(22.881,8.800)--(22.764,8.800)--cycle;
\draw[gp path] (22.764,5.644)--(22.764,8.799)--(22.880,8.799)--(22.880,5.644)--cycle;
\gpfill{rgb color={0.800,0.800,0.800}} (23.228,6.021)--(23.345,6.021)--(23.345,6.022)--(23.228,6.022)--cycle;
\draw[gp path] (23.228,6.021)--(23.344,6.021)--cycle;
\gpfill{rgb color={0.800,0.800,0.800}} (23.460,5.921)--(23.577,5.921)--(23.577,5.922)--(23.460,5.922)--cycle;
\draw[gp path] (23.460,5.921)--(23.576,5.921)--cycle;
\gpfill{rgb color={0.800,0.800,0.800}} (23.692,6.534)--(23.809,6.534)--(23.809,6.535)--(23.692,6.535)--cycle;
\draw[gp path] (23.692,6.534)--(23.808,6.534)--cycle;
\gpfill{rgb color={0.800,0.800,0.800}} (23.924,6.735)--(24.041,6.735)--(24.041,6.736)--(23.924,6.736)--cycle;
\draw[gp path] (23.924,6.735)--(24.040,6.735)--cycle;
\gpfill{rgb color={0.800,0.800,0.800}} (24.156,5.644)--(24.273,5.644)--(24.273,6.529)--(24.156,6.529)--cycle;
\draw[gp path] (24.156,5.644)--(24.156,6.528)--(24.272,6.528)--(24.272,5.644)--cycle;
\gpcolor{color=gp lt color border}
\draw[gp path] (1.012,10.131)--(1.012,5.644)--(24.446,5.644)--(24.446,10.131)--cycle;
\node[gp node center] at (12.729,9.682) {\plotLegend{(ARM)}};
\node[gp node left,font={\fontsize{11.0pt}{13.2pt}\selectfont}] at (1.246,9.817) {\plotLegend{mean improvement: $2.5\%$}};
\node[gp node left,font={\fontsize{11.0pt}{13.2pt}\selectfont}] at (1.246,9.368) {\plotLegend{improved functions: $45\%$}};
\node[gp node left,font={\fontsize{11.0pt}{13.2pt}\selectfont}] at (1.246,8.920) {\plotLegend{mean gap: $7.6\%$}};
\node[gp node left,font={\fontsize{11.0pt}{13.2pt}\selectfont}] at (1.246,8.471) {\plotLegend{optimal functions: $64\%$}};
\gpdefrectangularnode{gp plot 1}{\pgfpoint{1.012cm}{5.644cm}}{\pgfpoint{24.446cm}{10.131cm}}
\end{tikzpicture}

%% file: results/mips-size-improvement.tex
\begin{tikzpicture}[gnuplot]
\path (0.000,0.000) rectangle (12.500,8.750);
\gpcolor{color=gp lt color border}
\gpsetlinetype{gp lt border}
\gpsetdashtype{gp dt solid}
\gpsetlinewidth{1.00}
\draw[gp path] (1.012,5.644)--(1.192,5.644);
\draw[gp path] (24.446,5.644)--(24.266,5.644);
\node[gp node right,font={\fontsize{8.0pt}{9.6pt}\selectfont}] at (0.828,5.644) {\plotPercentage{0}};
\draw[gp path] (1.012,6.205)--(1.192,6.205);
\draw[gp path] (24.446,6.205)--(24.266,6.205);
\node[gp node right,font={\fontsize{8.0pt}{9.6pt}\selectfont}] at (0.828,6.205) {\plotPercentage{5}};
\draw[gp path] (1.012,6.766)--(1.192,6.766);
\draw[gp path] (24.446,6.766)--(24.266,6.766);
\node[gp node right,font={\fontsize{8.0pt}{9.6pt}\selectfont}] at (0.828,6.766) {\plotPercentage{10}};
\draw[gp path] (1.012,7.327)--(1.192,7.327);
\draw[gp path] (24.446,7.327)--(24.266,7.327);
\node[gp node right,font={\fontsize{8.0pt}{9.6pt}\selectfont}] at (0.828,7.327) {\plotPercentage{15}};
\draw[gp path] (1.012,7.888)--(1.192,7.888);
\draw[gp path] (24.446,7.888)--(24.266,7.888);
\node[gp node right,font={\fontsize{8.0pt}{9.6pt}\selectfont}] at (0.828,7.888) {\plotPercentage{20}};
\draw[gp path] (1.012,8.448)--(1.192,8.448);
\draw[gp path] (24.446,8.448)--(24.266,8.448);
\node[gp node right,font={\fontsize{8.0pt}{9.6pt}\selectfont}] at (0.828,8.448) {\plotPercentage{25}};
\draw[gp path] (1.012,9.009)--(1.192,9.009);
\draw[gp path] (24.446,9.009)--(24.266,9.009);
\node[gp node right,font={\fontsize{8.0pt}{9.6pt}\selectfont}] at (0.828,9.009) {\plotPercentage{30}};
\draw[gp path] (1.012,9.570)--(1.192,9.570);
\draw[gp path] (24.446,9.570)--(24.266,9.570);
\node[gp node right,font={\fontsize{8.0pt}{9.6pt}\selectfont}] at (0.828,9.570) {\plotPercentage{35}};
\draw[gp path] (1.012,10.131)--(1.192,10.131);
\draw[gp path] (24.446,10.131)--(24.266,10.131);
\node[gp node right,font={\fontsize{8.0pt}{9.6pt}\selectfont}] at (0.828,10.131) {\plotPercentage{40}};
\node[gp node left,rotate=-90,font={\fontsize{7.0pt}{8.4pt}\selectfont}] at (1.244,5.721) {\functionId{1}\functionName{handle_noinline_attribute}};
\node[gp node left,rotate=-90,font={\fontsize{7.0pt}{8.4pt}\selectfont}] at (1.476,5.721) {\functionId{2}\functionName{control_flow_insn_p}};
\node[gp node left,rotate=-90,font={\fontsize{7.0pt}{8.4pt}\selectfont}] at (1.708,5.721) {\functionId{3}\functionName{insert_insn_on_edge}};
\node[gp node left,rotate=-90,font={\fontsize{7.0pt}{8.4pt}\selectfont}] at (1.940,5.721) {\functionId{4}\functionName{update_br_prob_note}};
\node[gp node left,rotate=-90,font={\fontsize{7.0pt}{8.4pt}\selectfont}] at (2.172,5.721) {\functionId{5}\functionName{_cpp_init_internal_pragma.}};
\node[gp node left,rotate=-90,font={\fontsize{7.0pt}{8.4pt}\selectfont}] at (2.404,5.721) {\functionId{6}\functionName{lex_macro_node}};
\node[gp node left,rotate=-90,font={\fontsize{7.0pt}{8.4pt}\selectfont}] at (2.636,5.721) {\functionId{7}\functionName{cse_basic_block}};
\node[gp node left,rotate=-90,font={\fontsize{7.0pt}{8.4pt}\selectfont}] at (2.868,5.721) {\functionId{8}\functionName{rtx_equal_for_cselib_p}};
\node[gp node left,rotate=-90,font={\fontsize{7.0pt}{8.4pt}\selectfont}] at (3.100,5.721) {\functionId{9}\functionName{debug_df_chain}};
\node[gp node left,rotate=-90,font={\fontsize{7.0pt}{8.4pt}\selectfont}] at (3.332,5.721) {\functionId{10}\functionName{modified_type_die}};
\node[gp node left,rotate=-90,font={\fontsize{7.0pt}{8.4pt}\selectfont}] at (3.564,5.721) {\functionId{11}\functionName{emit_note}};
\node[gp node left,rotate=-90,font={\fontsize{7.0pt}{8.4pt}\selectfont}] at (3.796,5.721) {\functionId{12}\functionName{gen_sequence}};
\node[gp node left,rotate=-90,font={\fontsize{7.0pt}{8.4pt}\selectfont}] at (4.028,5.721) {\functionId{13}\functionName{subreg_hard_regno}};
\node[gp node left,rotate=-90,font={\fontsize{7.0pt}{8.4pt}\selectfont}] at (4.260,5.721) {\functionId{14}\functionName{split_double}};
\node[gp node left,rotate=-90,font={\fontsize{7.0pt}{8.4pt}\selectfont}] at (4.492,5.721) {\functionId{15}\functionName{add_to_mem_set_list}};
\node[gp node left,rotate=-90,font={\fontsize{7.0pt}{8.4pt}\selectfont}] at (4.724,5.721) {\functionId{16}\functionName{find_regno_partial}};
\node[gp node left,rotate=-90,font={\fontsize{7.0pt}{8.4pt}\selectfont}] at (4.956,5.721) {\functionId{17}\functionName{use_return_register}};
\node[gp node left,rotate=-90,font={\fontsize{7.0pt}{8.4pt}\selectfont}] at (5.188,5.721) {\functionId{18}\functionName{ix86_expand_move}};
\node[gp node left,rotate=-90,font={\fontsize{7.0pt}{8.4pt}\selectfont}] at (5.420,5.721) {\functionId{19}\functionName{legitimate_pic_address_di.}};
\node[gp node left,rotate=-90,font={\fontsize{7.0pt}{8.4pt}\selectfont}] at (5.652,5.721) {\functionId{20}\functionName{gen_extendsfdf2}};
\node[gp node left,rotate=-90,font={\fontsize{7.0pt}{8.4pt}\selectfont}] at (5.884,5.721) {\functionId{21}\functionName{gen_mulsidi3}};
\node[gp node left,rotate=-90,font={\fontsize{7.0pt}{8.4pt}\selectfont}] at (6.116,5.721) {\functionId{22}\functionName{gen_peephole2_1255}};
\node[gp node left,rotate=-90,font={\fontsize{7.0pt}{8.4pt}\selectfont}] at (6.348,5.721) {\functionId{23}\functionName{gen_peephole2_1271}};
\node[gp node left,rotate=-90,font={\fontsize{7.0pt}{8.4pt}\selectfont}] at (6.580,5.721) {\functionId{24}\functionName{gen_peephole2_1277}};
\node[gp node left,rotate=-90,font={\fontsize{7.0pt}{8.4pt}\selectfont}] at (6.812,5.721) {\functionId{25}\functionName{gen_pfnacc}};
\node[gp node left,rotate=-90,font={\fontsize{7.0pt}{8.4pt}\selectfont}] at (7.045,5.721) {\functionId{26}\functionName{gen_rotlsi3}};
\node[gp node left,rotate=-90,font={\fontsize{7.0pt}{8.4pt}\selectfont}] at (7.277,5.721) {\functionId{27}\functionName{gen_split_1001}};
\node[gp node left,rotate=-90,font={\fontsize{7.0pt}{8.4pt}\selectfont}] at (7.509,5.721) {\functionId{28}\functionName{gen_split_1028}};
\node[gp node left,rotate=-90,font={\fontsize{7.0pt}{8.4pt}\selectfont}] at (7.741,5.721) {\functionId{29}\functionName{gen_sse_nandti3}};
\node[gp node left,rotate=-90,font={\fontsize{7.0pt}{8.4pt}\selectfont}] at (7.973,5.721) {\functionId{30}\functionName{gen_sunge}};
\node[gp node left,rotate=-90,font={\fontsize{7.0pt}{8.4pt}\selectfont}] at (8.205,5.721) {\functionId{31}\functionName{insert_loop_mem}};
\node[gp node left,rotate=-90,font={\fontsize{7.0pt}{8.4pt}\selectfont}] at (8.437,5.721) {\functionId{32}\functionName{eiremain}};
\node[gp node left,rotate=-90,font={\fontsize{7.0pt}{8.4pt}\selectfont}] at (8.669,5.721) {\functionId{33}\functionName{elimination_effects}};
\node[gp node left,rotate=-90,font={\fontsize{7.0pt}{8.4pt}\selectfont}] at (8.901,5.721) {\functionId{34}\functionName{gen_reload}};
\node[gp node left,rotate=-90,font={\fontsize{7.0pt}{8.4pt}\selectfont}] at (9.133,5.721) {\functionId{35}\functionName{reload_cse_simplify_set}};
\node[gp node left,rotate=-90,font={\fontsize{7.0pt}{8.4pt}\selectfont}] at (9.365,5.721) {\functionId{36}\functionName{simplify_binary_is2orm1}};
\node[gp node left,rotate=-90,font={\fontsize{7.0pt}{8.4pt}\selectfont}] at (9.597,5.721) {\functionId{37}\functionName{remove_phi_alternative}};
\node[gp node left,rotate=-90,font={\fontsize{7.0pt}{8.4pt}\selectfont}] at (9.829,5.721) {\functionId{38}\functionName{contains_placeholder_p}};
\node[gp node left,rotate=-90,font={\fontsize{7.0pt}{8.4pt}\selectfont}] at (10.061,5.721) {\functionId{39}\functionName{assemble_end_function}};
\node[gp node left,rotate=-90,font={\fontsize{7.0pt}{8.4pt}\selectfont}] at (10.293,5.721) {\functionId{40}\functionName{default_named_section_asm.}};
\node[gp node left,rotate=-90,font={\fontsize{7.0pt}{8.4pt}\selectfont}] at (10.525,5.721) {\functionId{41}\functionName{sample_unpack_12}};
\node[gp node left,rotate=-90,font={\fontsize{7.0pt}{8.4pt}\selectfont}] at (10.757,5.721) {\functionId{42}\functionName{autohelperattpat10}};
\node[gp node left,rotate=-90,font={\fontsize{7.0pt}{8.4pt}\selectfont}] at (10.989,5.721) {\functionId{43}\functionName{autohelperbarrierspat126}};
\node[gp node left,rotate=-90,font={\fontsize{7.0pt}{8.4pt}\selectfont}] at (11.221,5.721) {\functionId{44}\functionName{atari_atari_attack_callba.}};
\node[gp node left,rotate=-90,font={\fontsize{7.0pt}{8.4pt}\selectfont}] at (11.453,5.721) {\functionId{45}\functionName{compute_aa_status}};
\node[gp node left,rotate=-90,font={\fontsize{7.0pt}{8.4pt}\selectfont}] at (11.685,5.721) {\functionId{46}\functionName{dragon_weak}};
\node[gp node left,rotate=-90,font={\fontsize{7.0pt}{8.4pt}\selectfont}] at (11.917,5.721) {\functionId{47}\functionName{get_saved_worms}};
\node[gp node left,rotate=-90,font={\fontsize{7.0pt}{8.4pt}\selectfont}] at (12.149,5.721) {\functionId{48}\functionName{read_eye}};
\node[gp node left,rotate=-90,font={\fontsize{7.0pt}{8.4pt}\selectfont}] at (12.381,5.721) {\functionId{49}\functionName{topological_eye}};
\node[gp node left,rotate=-90,font={\fontsize{7.0pt}{8.4pt}\selectfont}] at (12.613,5.721) {\functionId{50}\functionName{autohelperowl_attackpat19.}};
\node[gp node left,rotate=-90,font={\fontsize{7.0pt}{8.4pt}\selectfont}] at (12.845,5.721) {\functionId{51}\functionName{autohelperowl_attackpat29.}};
\node[gp node left,rotate=-90,font={\fontsize{7.0pt}{8.4pt}\selectfont}] at (13.077,5.721) {\functionId{52}\functionName{autohelperowl_defendpat28.}};
\node[gp node left,rotate=-90,font={\fontsize{7.0pt}{8.4pt}\selectfont}] at (13.309,5.721) {\functionId{53}\functionName{autohelperowl_defendpat38.}};
\node[gp node left,rotate=-90,font={\fontsize{7.0pt}{8.4pt}\selectfont}] at (13.541,5.721) {\functionId{54}\functionName{autohelperpat1114}};
\node[gp node left,rotate=-90,font={\fontsize{7.0pt}{8.4pt}\selectfont}] at (13.773,5.721) {\functionId{55}\functionName{autohelperpat335}};
\node[gp node left,rotate=-90,font={\fontsize{7.0pt}{8.4pt}\selectfont}] at (14.005,5.721) {\functionId{56}\functionName{autohelperpat508}};
\node[gp node left,rotate=-90,font={\fontsize{7.0pt}{8.4pt}\selectfont}] at (14.237,5.721) {\functionId{57}\functionName{autohelperpat83}};
\node[gp node left,rotate=-90,font={\fontsize{7.0pt}{8.4pt}\selectfont}] at (14.469,5.721) {\functionId{58}\functionName{simple_showboard}};
\node[gp node left,rotate=-90,font={\fontsize{7.0pt}{8.4pt}\selectfont}] at (14.701,5.721) {\functionId{59}\functionName{skip_intrabk_SAD}};
\node[gp node left,rotate=-90,font={\fontsize{7.0pt}{8.4pt}\selectfont}] at (14.933,5.721) {\functionId{60}\functionName{free_orig_planes}};
\node[gp node left,rotate=-90,font={\fontsize{7.0pt}{8.4pt}\selectfont}] at (15.165,5.721) {\functionId{61}\functionName{GetSkipCostMB}};
\node[gp node left,rotate=-90,font={\fontsize{7.0pt}{8.4pt}\selectfont}] at (15.397,5.721) {\functionId{62}\functionName{writeSyntaxElement_Level_.}};
\node[gp node left,rotate=-90,font={\fontsize{7.0pt}{8.4pt}\selectfont}] at (15.629,5.721) {\functionId{63}\functionName{GSIAddKeyToIndex}};
\node[gp node left,rotate=-90,font={\fontsize{7.0pt}{8.4pt}\selectfont}] at (15.861,5.721) {\functionId{64}\functionName{EVDBasicFit}};
\node[gp node left,rotate=-90,font={\fontsize{7.0pt}{8.4pt}\selectfont}] at (16.093,5.721) {\functionId{65}\functionName{SampleDirichlet}};
\node[gp node left,rotate=-90,font={\fontsize{7.0pt}{8.4pt}\selectfont}] at (16.325,5.721) {\functionId{66}\functionName{DegenerateSymbolScore}};
\node[gp node left,rotate=-90,font={\fontsize{7.0pt}{8.4pt}\selectfont}] at (16.557,5.721) {\functionId{67}\functionName{Plan7SetCtime}};
\node[gp node left,rotate=-90,font={\fontsize{7.0pt}{8.4pt}\selectfont}] at (16.789,5.721) {\functionId{68}\functionName{MSAToSqinfo}};
\node[gp node left,rotate=-90,font={\fontsize{7.0pt}{8.4pt}\selectfont}] at (17.021,5.721) {\functionId{69}\functionName{null_convert}};
\node[gp node left,rotate=-90,font={\fontsize{7.0pt}{8.4pt}\selectfont}] at (17.253,5.721) {\functionId{70}\functionName{jinit_c_prep_controller}};
\node[gp node left,rotate=-90,font={\fontsize{7.0pt}{8.4pt}\selectfont}] at (17.485,5.721) {\functionId{71}\functionName{glFogf}};
\node[gp node left,rotate=-90,font={\fontsize{7.0pt}{8.4pt}\selectfont}] at (17.717,5.721) {\functionId{72}\functionName{glNormal3d}};
\node[gp node left,rotate=-90,font={\fontsize{7.0pt}{8.4pt}\selectfont}] at (17.949,5.721) {\functionId{73}\functionName{glRasterPos3d}};
\node[gp node left,rotate=-90,font={\fontsize{7.0pt}{8.4pt}\selectfont}] at (18.181,5.721) {\functionId{74}\functionName{glTexCoord2d}};
\node[gp node left,rotate=-90,font={\fontsize{7.0pt}{8.4pt}\selectfont}] at (18.413,5.721) {\functionId{75}\functionName{gl_stippled_bresenham}};
\node[gp node left,rotate=-90,font={\fontsize{7.0pt}{8.4pt}\selectfont}] at (18.646,5.721) {\functionId{76}\functionName{gl_save_Frustum}};
\node[gp node left,rotate=-90,font={\fontsize{7.0pt}{8.4pt}\selectfont}] at (18.878,5.721) {\functionId{77}\functionName{gl_save_LineWidth}};
\node[gp node left,rotate=-90,font={\fontsize{7.0pt}{8.4pt}\selectfont}] at (19.110,5.721) {\functionId{78}\functionName{translate_id}};
\node[gp node left,rotate=-90,font={\fontsize{7.0pt}{8.4pt}\selectfont}] at (19.342,5.721) {\functionId{79}\functionName{gl_Map1f}};
\node[gp node left,rotate=-90,font={\fontsize{7.0pt}{8.4pt}\selectfont}] at (19.574,5.721) {\functionId{80}\functionName{smooth_ci_line}};
\node[gp node left,rotate=-90,font={\fontsize{7.0pt}{8.4pt}\selectfont}] at (19.806,5.721) {\functionId{81}\functionName{free_unified_knots}};
\node[gp node left,rotate=-90,font={\fontsize{7.0pt}{8.4pt}\selectfont}] at (20.038,5.721) {\functionId{82}\functionName{tess_test_polygon}};
\node[gp node left,rotate=-90,font={\fontsize{7.0pt}{8.4pt}\selectfont}] at (20.270,5.721) {\functionId{83}\functionName{auxWireBox}};
\node[gp node left,rotate=-90,font={\fontsize{7.0pt}{8.4pt}\selectfont}] at (20.502,5.721) {\functionId{84}\functionName{gl_ColorPointer}};
\node[gp node left,rotate=-90,font={\fontsize{7.0pt}{8.4pt}\selectfont}] at (20.734,5.721) {\functionId{85}\functionName{r_serial}};
\node[gp node left,rotate=-90,font={\fontsize{7.0pt}{8.4pt}\selectfont}] at (20.966,5.721) {\functionId{86}\functionName{scalar_mult_sub_su3_matri.}};
\node[gp node left,rotate=-90,font={\fontsize{7.0pt}{8.4pt}\selectfont}] at (21.198,5.721) {\functionId{87}\functionName{Decode_MPEG1_Non_Intra_Bl.}};
\node[gp node left,rotate=-90,font={\fontsize{7.0pt}{8.4pt}\selectfont}] at (21.430,5.721) {\functionId{88}\functionName{cpDecodeSecret}};
\node[gp node left,rotate=-90,font={\fontsize{7.0pt}{8.4pt}\selectfont}] at (21.662,5.721) {\functionId{89}\functionName{vlShortLshift}};
\node[gp node left,rotate=-90,font={\fontsize{7.0pt}{8.4pt}\selectfont}] at (21.894,5.721) {\functionId{90}\functionName{encryptfile}};
\node[gp node left,rotate=-90,font={\fontsize{7.0pt}{8.4pt}\selectfont}] at (22.126,5.721) {\functionId{91}\functionName{make_canonical}};
\node[gp node left,rotate=-90,font={\fontsize{7.0pt}{8.4pt}\selectfont}] at (22.358,5.721) {\functionId{92}\functionName{LANG}};
\node[gp node left,rotate=-90,font={\fontsize{7.0pt}{8.4pt}\selectfont}] at (22.590,5.721) {\functionId{93}\functionName{MD5Transform}};
\node[gp node left,rotate=-90,font={\fontsize{7.0pt}{8.4pt}\selectfont}] at (22.822,5.721) {\functionId{94}\functionName{mp_display}};
\node[gp node left,rotate=-90,font={\fontsize{7.0pt}{8.4pt}\selectfont}] at (23.054,5.721) {\functionId{95}\functionName{comp_Jboundaries}};
\node[gp node left,rotate=-90,font={\fontsize{7.0pt}{8.4pt}\selectfont}] at (23.286,5.721) {\functionId{96}\functionName{is_draw}};
\node[gp node left,rotate=-90,font={\fontsize{7.0pt}{8.4pt}\selectfont}] at (23.518,5.721) {\functionId{97}\functionName{push_king}};
\node[gp node left,rotate=-90,font={\fontsize{7.0pt}{8.4pt}\selectfont}] at (23.750,5.721) {\functionId{98}\functionName{stat_retry}};
\node[gp node left,rotate=-90,font={\fontsize{7.0pt}{8.4pt}\selectfont}] at (23.982,5.721) {\functionId{99}\functionName{lextree_subtree_print}};
\node[gp node left,rotate=-90,font={\fontsize{7.0pt}{8.4pt}\selectfont}] at (24.214,5.721) {\functionId{100}\functionName{lm_tg_score}};
\draw[gp path] (1.012,10.131)--(1.012,5.644)--(24.446,5.644)--(24.446,10.131)--cycle;
\gpcolor{rgb color={0.580,0.000,0.827}}
\draw[gp path] (1.012,5.644)--(1.249,5.644)--(1.485,5.644)--(1.722,5.644)--(1.959,5.644)%
  --(2.196,5.644)--(2.432,5.644)--(2.669,5.644)--(2.906,5.644)--(3.142,5.644)--(3.379,5.644)%
  --(3.616,5.644)--(3.852,5.644)--(4.089,5.644)--(4.326,5.644)--(4.563,5.644)--(4.799,5.644)%
  --(5.036,5.644)--(5.273,5.644)--(5.509,5.644)--(5.746,5.644)--(5.983,5.644)--(6.220,5.644)%
  --(6.456,5.644)--(6.693,5.644)--(6.930,5.644)--(7.166,5.644)--(7.403,5.644)--(7.640,5.644)%
  --(7.877,5.644)--(8.113,5.644)--(8.350,5.644)--(8.587,5.644)--(8.823,5.644)--(9.060,5.644)%
  --(9.297,5.644)--(9.533,5.644)--(9.770,5.644)--(10.007,5.644)--(10.244,5.644)--(10.480,5.644)%
  --(10.717,5.644)--(10.954,5.644)--(11.190,5.644)--(11.427,5.644)--(11.664,5.644)--(11.901,5.644)%
  --(12.137,5.644)--(12.374,5.644)--(12.611,5.644)--(12.847,5.644)--(13.084,5.644)--(13.321,5.644)%
  --(13.557,5.644)--(13.794,5.644)--(14.031,5.644)--(14.268,5.644)--(14.504,5.644)--(14.741,5.644)%
  --(14.978,5.644)--(15.214,5.644)--(15.451,5.644)--(15.688,5.644)--(15.925,5.644)--(16.161,5.644)%
  --(16.398,5.644)--(16.635,5.644)--(16.871,5.644)--(17.108,5.644)--(17.345,5.644)--(17.581,5.644)%
  --(17.818,5.644)--(18.055,5.644)--(18.292,5.644)--(18.528,5.644)--(18.765,5.644)--(19.002,5.644)%
  --(19.238,5.644)--(19.475,5.644)--(19.712,5.644)--(19.949,5.644)--(20.185,5.644)--(20.422,5.644)%
  --(20.659,5.644)--(20.895,5.644)--(21.132,5.644)--(21.369,5.644)--(21.606,5.644)--(21.842,5.644)%
  --(22.079,5.644)--(22.316,5.644)--(22.552,5.644)--(22.789,5.644)--(23.026,5.644)--(23.262,5.644)%
  --(23.499,5.644)--(23.736,5.644)--(23.973,5.644)--(24.209,5.644)--(24.446,5.644);
\gpfill{rgb color={0.000,0.000,0.000}} (1.418,5.644)--(1.535,5.644)--(1.535,7.166)--(1.418,7.166)--cycle;
\gpcolor{rgb color={0.000,0.000,0.000}}
\draw[gp path] (1.418,5.644)--(1.418,7.165)--(1.534,7.165)--(1.534,5.644)--cycle;
\gpfill{rgb color={0.000,0.000,0.000}} (1.650,5.644)--(1.767,5.644)--(1.767,7.077)--(1.650,7.077)--cycle;
\draw[gp path] (1.650,5.644)--(1.650,7.076)--(1.766,7.076)--(1.766,5.644)--cycle;
\gpfill{rgb color={0.000,0.000,0.000}} (1.882,5.644)--(1.999,5.644)--(1.999,6.713)--(1.882,6.713)--cycle;
\draw[gp path] (1.882,5.644)--(1.882,6.712)--(1.998,6.712)--(1.998,5.644)--cycle;
\gpfill{rgb color={0.000,0.000,0.000}} (2.114,5.644)--(2.231,5.644)--(2.231,7.390)--(2.114,7.390)--cycle;
\draw[gp path] (2.114,5.644)--(2.114,7.389)--(2.230,7.389)--(2.230,5.644)--cycle;
\gpfill{rgb color={0.000,0.000,0.000}} (2.346,5.644)--(2.463,5.644)--(2.463,7.481)--(2.346,7.481)--cycle;
\draw[gp path] (2.346,5.644)--(2.346,7.480)--(2.462,7.480)--(2.462,5.644)--cycle;
\gpfill{rgb color={0.000,0.000,0.000}} (3.042,5.644)--(3.159,5.644)--(3.159,5.857)--(3.042,5.857)--cycle;
\draw[gp path] (3.042,5.644)--(3.042,5.856)--(3.158,5.856)--(3.158,5.644)--cycle;
\gpfill{rgb color={0.000,0.000,0.000}} (3.506,5.644)--(3.623,5.644)--(3.623,5.933)--(3.506,5.933)--cycle;
\draw[gp path] (3.506,5.644)--(3.506,5.932)--(3.622,5.932)--(3.622,5.644)--cycle;
\gpfill{rgb color={0.000,0.000,0.000}} (3.738,5.644)--(3.855,5.644)--(3.855,6.053)--(3.738,6.053)--cycle;
\draw[gp path] (3.738,5.644)--(3.738,6.052)--(3.854,6.052)--(3.854,5.644)--cycle;
\gpfill{rgb color={0.000,0.000,0.000}} (3.970,5.644)--(4.087,5.644)--(4.087,7.069)--(3.970,7.069)--cycle;
\draw[gp path] (3.970,5.644)--(3.970,7.068)--(4.086,7.068)--(4.086,5.644)--cycle;
\gpfill{rgb color={0.000,0.000,0.000}} (4.434,5.644)--(4.551,5.644)--(4.551,6.647)--(4.434,6.647)--cycle;
\draw[gp path] (4.434,5.644)--(4.434,6.646)--(4.550,6.646)--(4.550,5.644)--cycle;
\gpfill{rgb color={0.000,0.000,0.000}} (5.362,5.644)--(5.479,5.644)--(5.479,5.706)--(5.362,5.706)--cycle;
\draw[gp path] (5.362,5.644)--(5.362,5.705)--(5.478,5.705)--(5.478,5.644)--cycle;
\gpfill{rgb color={0.000,0.000,0.000}} (5.594,5.644)--(5.711,5.644)--(5.711,6.476)--(5.594,6.476)--cycle;
\draw[gp path] (5.594,5.644)--(5.594,6.475)--(5.710,6.475)--(5.710,5.644)--cycle;
\gpfill{rgb color={0.000,0.000,0.000}} (5.826,5.644)--(5.943,5.644)--(5.943,6.357)--(5.826,6.357)--cycle;
\draw[gp path] (5.826,5.644)--(5.826,6.356)--(5.942,6.356)--(5.942,5.644)--cycle;
\gpfill{rgb color={0.000,0.000,0.000}} (6.058,5.644)--(6.175,5.644)--(6.175,5.980)--(6.058,5.980)--cycle;
\draw[gp path] (6.058,5.644)--(6.058,5.979)--(6.174,5.979)--(6.174,5.644)--cycle;
\gpfill{rgb color={0.000,0.000,0.000}} (6.290,5.644)--(6.407,5.644)--(6.407,5.919)--(6.290,5.919)--cycle;
\draw[gp path] (6.290,5.644)--(6.290,5.918)--(6.406,5.918)--(6.406,5.644)--cycle;
\gpfill{rgb color={0.000,0.000,0.000}} (6.522,5.644)--(6.639,5.644)--(6.639,5.739)--(6.522,5.739)--cycle;
\draw[gp path] (6.522,5.644)--(6.522,5.738)--(6.638,5.738)--(6.638,5.644)--cycle;
\gpfill{rgb color={0.000,0.000,0.000}} (6.754,5.644)--(6.872,5.644)--(6.872,5.845)--(6.754,5.845)--cycle;
\draw[gp path] (6.754,5.644)--(6.754,5.844)--(6.871,5.844)--(6.871,5.644)--cycle;
\gpfill{rgb color={0.000,0.000,0.000}} (6.987,5.644)--(7.104,5.644)--(7.104,7.295)--(6.987,7.295)--cycle;
\draw[gp path] (6.987,5.644)--(6.987,7.294)--(7.103,7.294)--(7.103,5.644)--cycle;
\gpfill{rgb color={0.000,0.000,0.000}} (7.683,5.644)--(7.800,5.644)--(7.800,7.454)--(7.683,7.454)--cycle;
\draw[gp path] (7.683,5.644)--(7.683,7.453)--(7.799,7.453)--(7.799,5.644)--cycle;
\gpfill{rgb color={0.000,0.000,0.000}} (7.915,5.644)--(8.032,5.644)--(8.032,7.092)--(7.915,7.092)--cycle;
\draw[gp path] (7.915,5.644)--(7.915,7.091)--(8.031,7.091)--(8.031,5.644)--cycle;
\gpfill{rgb color={0.000,0.000,0.000}} (9.307,5.644)--(9.424,5.644)--(9.424,7.300)--(9.307,7.300)--cycle;
\draw[gp path] (9.307,5.644)--(9.307,7.299)--(9.423,7.299)--(9.423,5.644)--cycle;
\gpfill{rgb color={0.000,0.000,0.000}} (10.235,5.644)--(10.352,5.644)--(10.352,6.731)--(10.235,6.731)--cycle;
\draw[gp path] (10.235,5.644)--(10.235,6.730)--(10.351,6.730)--(10.351,5.644)--cycle;
\gpfill{rgb color={0.000,0.000,0.000}} (10.467,5.644)--(10.584,5.644)--(10.584,5.906)--(10.467,5.906)--cycle;
\draw[gp path] (10.467,5.644)--(10.467,5.905)--(10.583,5.905)--(10.583,5.644)--cycle;
\gpfill{rgb color={0.000,0.000,0.000}} (10.699,5.644)--(10.816,5.644)--(10.816,7.247)--(10.699,7.247)--cycle;
\draw[gp path] (10.699,5.644)--(10.699,7.246)--(10.815,7.246)--(10.815,5.644)--cycle;
\gpfill{rgb color={0.000,0.000,0.000}} (11.627,5.644)--(11.744,5.644)--(11.744,5.980)--(11.627,5.980)--cycle;
\draw[gp path] (11.627,5.644)--(11.627,5.979)--(11.743,5.979)--(11.743,5.644)--cycle;
\gpfill{rgb color={0.000,0.000,0.000}} (12.555,5.644)--(12.672,5.644)--(12.672,6.060)--(12.555,6.060)--cycle;
\draw[gp path] (12.555,5.644)--(12.555,6.059)--(12.671,6.059)--(12.671,5.644)--cycle;
\gpfill{rgb color={0.000,0.000,0.000}} (12.787,5.644)--(12.904,5.644)--(12.904,7.077)--(12.787,7.077)--cycle;
\draw[gp path] (12.787,5.644)--(12.787,7.076)--(12.903,7.076)--(12.903,5.644)--cycle;
\gpfill{rgb color={0.000,0.000,0.000}} (13.019,5.644)--(13.136,5.644)--(13.136,6.927)--(13.019,6.927)--cycle;
\draw[gp path] (13.019,5.644)--(13.019,6.926)--(13.135,6.926)--(13.135,5.644)--cycle;
\gpfill{rgb color={0.000,0.000,0.000}} (13.251,5.644)--(13.368,5.644)--(13.368,6.847)--(13.251,6.847)--cycle;
\draw[gp path] (13.251,5.644)--(13.251,6.846)--(13.367,6.846)--(13.367,5.644)--cycle;
\gpfill{rgb color={0.000,0.000,0.000}} (13.483,5.644)--(13.600,5.644)--(13.600,7.515)--(13.483,7.515)--cycle;
\draw[gp path] (13.483,5.644)--(13.483,7.514)--(13.599,7.514)--(13.599,5.644)--cycle;
\gpfill{rgb color={0.000,0.000,0.000}} (13.715,5.644)--(13.832,5.644)--(13.832,6.076)--(13.715,6.076)--cycle;
\draw[gp path] (13.715,5.644)--(13.715,6.075)--(13.831,6.075)--(13.831,5.644)--cycle;
\gpfill{rgb color={0.000,0.000,0.000}} (13.947,5.644)--(14.064,5.644)--(14.064,6.112)--(13.947,6.112)--cycle;
\draw[gp path] (13.947,5.644)--(13.947,6.111)--(14.063,6.111)--(14.063,5.644)--cycle;
\gpfill{rgb color={0.000,0.000,0.000}} (14.643,5.644)--(14.760,5.644)--(14.760,5.968)--(14.643,5.968)--cycle;
\draw[gp path] (14.643,5.644)--(14.643,5.967)--(14.759,5.967)--(14.759,5.644)--cycle;
\gpfill{rgb color={0.000,0.000,0.000}} (16.035,5.644)--(16.152,5.644)--(16.152,8.024)--(16.035,8.024)--cycle;
\draw[gp path] (16.035,5.644)--(16.035,8.023)--(16.151,8.023)--(16.151,5.644)--cycle;
\gpfill{rgb color={0.000,0.000,0.000}} (16.499,5.644)--(16.616,5.644)--(16.616,6.305)--(16.499,6.305)--cycle;
\draw[gp path] (16.499,5.644)--(16.499,6.304)--(16.615,6.304)--(16.615,5.644)--cycle;
\gpfill{rgb color={0.000,0.000,0.000}} (16.963,5.644)--(17.080,5.644)--(17.080,6.019)--(16.963,6.019)--cycle;
\draw[gp path] (16.963,5.644)--(16.963,6.018)--(17.079,6.018)--(17.079,5.644)--cycle;
\gpfill{rgb color={0.000,0.000,0.000}} (17.427,5.644)--(17.544,5.644)--(17.544,6.767)--(17.427,6.767)--cycle;
\draw[gp path] (17.427,5.644)--(17.427,6.766)--(17.543,6.766)--(17.543,5.644)--cycle;
\gpfill{rgb color={0.000,0.000,0.000}} (17.659,5.644)--(17.776,5.644)--(17.776,5.940)--(17.659,5.940)--cycle;
\draw[gp path] (17.659,5.644)--(17.659,5.939)--(17.775,5.939)--(17.775,5.644)--cycle;
\gpfill{rgb color={0.000,0.000,0.000}} (17.891,5.644)--(18.008,5.644)--(18.008,6.268)--(17.891,6.268)--cycle;
\draw[gp path] (17.891,5.644)--(17.891,6.267)--(18.007,6.267)--(18.007,5.644)--cycle;
\gpfill{rgb color={0.000,0.000,0.000}} (19.748,5.644)--(19.865,5.644)--(19.865,6.927)--(19.748,6.927)--cycle;
\draw[gp path] (19.748,5.644)--(19.748,6.926)--(19.864,6.926)--(19.864,5.644)--cycle;
\gpfill{rgb color={0.000,0.000,0.000}} (20.908,5.644)--(21.025,5.644)--(21.025,5.766)--(20.908,5.766)--cycle;
\draw[gp path] (20.908,5.644)--(20.908,5.765)--(21.024,5.765)--(21.024,5.644)--cycle;
\gpfill{rgb color={0.000,0.000,0.000}} (21.372,5.644)--(21.489,5.644)--(21.489,9.852)--(21.372,9.852)--cycle;
\draw[gp path] (21.372,5.644)--(21.372,9.851)--(21.488,9.851)--(21.488,5.644)--cycle;
\gpfill{rgb color={0.000,0.000,0.000}} (21.604,5.644)--(21.721,5.644)--(21.721,6.580)--(21.604,6.580)--cycle;
\draw[gp path] (21.604,5.644)--(21.604,6.579)--(21.720,6.579)--(21.720,5.644)--cycle;
\gpfill{rgb color={0.000,0.000,0.000}} (22.068,5.644)--(22.185,5.644)--(22.185,6.984)--(22.068,6.984)--cycle;
\draw[gp path] (22.068,5.644)--(22.068,6.983)--(22.184,6.983)--(22.184,5.644)--cycle;
\gpfill{rgb color={0.000,0.000,0.000}} (22.996,5.644)--(23.113,5.644)--(23.113,5.912)--(22.996,5.912)--cycle;
\draw[gp path] (22.996,5.644)--(22.996,5.911)--(23.112,5.911)--(23.112,5.644)--cycle;
\gpfill{rgb color={0.000,0.000,0.000}} (23.228,5.644)--(23.345,5.644)--(23.345,6.041)--(23.228,6.041)--cycle;
\draw[gp path] (23.228,5.644)--(23.228,6.040)--(23.344,6.040)--(23.344,5.644)--cycle;
\gpfill{rgb color={0.800,0.800,0.800}} (1.418,7.165)--(1.535,7.165)--(1.535,7.166)--(1.418,7.166)--cycle;
\gpcolor{rgb color={0.800,0.800,0.800}}
\draw[gp path] (1.418,7.165)--(1.534,7.165)--cycle;
\gpfill{rgb color={0.800,0.800,0.800}} (1.650,7.076)--(1.767,7.076)--(1.767,7.077)--(1.650,7.077)--cycle;
\draw[gp path] (1.650,7.076)--(1.766,7.076)--cycle;
\gpfill{rgb color={0.800,0.800,0.800}} (1.882,6.712)--(1.999,6.712)--(1.999,6.713)--(1.882,6.713)--cycle;
\draw[gp path] (1.882,6.712)--(1.998,6.712)--cycle;
\gpfill{rgb color={0.800,0.800,0.800}} (2.114,7.389)--(2.231,7.389)--(2.231,7.390)--(2.114,7.390)--cycle;
\draw[gp path] (2.114,7.389)--(2.230,7.389)--cycle;
\gpfill{rgb color={0.800,0.800,0.800}} (2.346,7.480)--(2.463,7.480)--(2.463,7.481)--(2.346,7.481)--cycle;
\draw[gp path] (2.346,7.480)--(2.462,7.480)--cycle;
\gpfill{rgb color={0.800,0.800,0.800}} (2.578,5.644)--(2.695,5.644)--(2.695,9.624)--(2.578,9.624)--cycle;
\draw[gp path] (2.578,5.644)--(2.578,9.623)--(2.694,9.623)--(2.694,5.644)--cycle;
\gpfill{rgb color={0.800,0.800,0.800}} (2.810,5.644)--(2.927,5.644)--(2.927,8.001)--(2.810,8.001)--cycle;
\draw[gp path] (2.810,5.644)--(2.810,8.000)--(2.926,8.000)--(2.926,5.644)--cycle;
\gpfill{rgb color={0.800,0.800,0.800}} (3.042,5.856)--(3.159,5.856)--(3.159,5.857)--(3.042,5.857)--cycle;
\draw[gp path] (3.042,5.856)--(3.158,5.856)--cycle;
\gpfill{rgb color={0.800,0.800,0.800}} (3.274,5.644)--(3.391,5.644)--(3.391,7.033)--(3.274,7.033)--cycle;
\draw[gp path] (3.274,5.644)--(3.274,7.032)--(3.390,7.032)--(3.390,5.644)--cycle;
\gpfill{rgb color={0.800,0.800,0.800}} (3.506,5.932)--(3.623,5.932)--(3.623,5.933)--(3.506,5.933)--cycle;
\draw[gp path] (3.506,5.932)--(3.622,5.932)--cycle;
\gpfill{rgb color={0.800,0.800,0.800}} (3.738,6.052)--(3.855,6.052)--(3.855,6.053)--(3.738,6.053)--cycle;
\draw[gp path] (3.738,6.052)--(3.854,6.052)--cycle;
\gpfill{rgb color={0.800,0.800,0.800}} (3.970,7.068)--(4.087,7.068)--(4.087,7.069)--(3.970,7.069)--cycle;
\draw[gp path] (3.970,7.068)--(4.086,7.068)--cycle;
\gpfill{rgb color={0.800,0.800,0.800}} (4.202,5.644)--(4.319,5.644)--(4.319,6.684)--(4.202,6.684)--cycle;
\draw[gp path] (4.202,5.644)--(4.202,6.683)--(4.318,6.683)--(4.318,5.644)--cycle;
\gpfill{rgb color={0.800,0.800,0.800}} (4.434,6.646)--(4.551,6.646)--(4.551,6.647)--(4.434,6.647)--cycle;
\draw[gp path] (4.434,6.646)--(4.550,6.646)--cycle;
\gpfill{rgb color={0.800,0.800,0.800}} (4.898,5.644)--(5.015,5.644)--(5.015,8.794)--(4.898,8.794)--cycle;
\draw[gp path] (4.898,5.644)--(4.898,8.793)--(5.014,8.793)--(5.014,5.644)--cycle;
\gpfill{rgb color={0.800,0.800,0.800}} (5.130,5.644)--(5.247,5.644)--(5.247,6.807)--(5.130,6.807)--cycle;
\draw[gp path] (5.130,5.644)--(5.130,6.806)--(5.246,6.806)--(5.246,5.644)--cycle;
\gpfill{rgb color={0.800,0.800,0.800}} (5.362,5.705)--(5.479,5.705)--(5.479,7.317)--(5.362,7.317)--cycle;
\draw[gp path] (5.362,5.705)--(5.362,7.316)--(5.478,7.316)--(5.478,5.705)--cycle;
\gpfill{rgb color={0.800,0.800,0.800}} (5.594,6.475)--(5.711,6.475)--(5.711,6.476)--(5.594,6.476)--cycle;
\draw[gp path] (5.594,6.475)--(5.710,6.475)--cycle;
\gpfill{rgb color={0.800,0.800,0.800}} (5.826,6.356)--(5.943,6.356)--(5.943,6.357)--(5.826,6.357)--cycle;
\draw[gp path] (5.826,6.356)--(5.942,6.356)--cycle;
\gpfill{rgb color={0.800,0.800,0.800}} (6.058,5.979)--(6.175,5.979)--(6.175,5.980)--(6.058,5.980)--cycle;
\draw[gp path] (6.058,5.979)--(6.174,5.979)--cycle;
\gpfill{rgb color={0.800,0.800,0.800}} (6.290,5.918)--(6.407,5.918)--(6.407,5.919)--(6.290,5.919)--cycle;
\draw[gp path] (6.290,5.918)--(6.406,5.918)--cycle;
\gpfill{rgb color={0.800,0.800,0.800}} (6.522,5.738)--(6.639,5.738)--(6.639,5.739)--(6.522,5.739)--cycle;
\draw[gp path] (6.522,5.738)--(6.638,5.738)--cycle;
\gpfill{rgb color={0.800,0.800,0.800}} (6.754,5.844)--(6.872,5.844)--(6.872,5.845)--(6.754,5.845)--cycle;
\draw[gp path] (6.754,5.844)--(6.871,5.844)--cycle;
\gpfill{rgb color={0.800,0.800,0.800}} (6.987,7.294)--(7.104,7.294)--(7.104,7.295)--(6.987,7.295)--cycle;
\draw[gp path] (6.987,7.294)--(7.103,7.294)--cycle;
\gpfill{rgb color={0.800,0.800,0.800}} (7.219,5.644)--(7.336,5.644)--(7.336,6.559)--(7.219,6.559)--cycle;
\draw[gp path] (7.219,5.644)--(7.219,6.558)--(7.335,6.558)--(7.335,5.644)--cycle;
\gpfill{rgb color={0.800,0.800,0.800}} (7.451,5.644)--(7.568,5.644)--(7.568,7.016)--(7.451,7.016)--cycle;
\draw[gp path] (7.451,5.644)--(7.451,7.015)--(7.567,7.015)--(7.567,5.644)--cycle;
\gpfill{rgb color={0.800,0.800,0.800}} (7.683,7.453)--(7.800,7.453)--(7.800,7.454)--(7.683,7.454)--cycle;
\draw[gp path] (7.683,7.453)--(7.799,7.453)--cycle;
\gpfill{rgb color={0.800,0.800,0.800}} (7.915,7.091)--(8.032,7.091)--(8.032,7.092)--(7.915,7.092)--cycle;
\draw[gp path] (7.915,7.091)--(8.031,7.091)--cycle;
\gpfill{rgb color={0.800,0.800,0.800}} (8.147,5.644)--(8.264,5.644)--(8.264,7.591)--(8.147,7.591)--cycle;
\draw[gp path] (8.147,5.644)--(8.147,7.590)--(8.263,7.590)--(8.263,5.644)--cycle;
\gpfill{rgb color={0.800,0.800,0.800}} (8.379,5.644)--(8.496,5.644)--(8.496,7.285)--(8.379,7.285)--cycle;
\draw[gp path] (8.379,5.644)--(8.379,7.284)--(8.495,7.284)--(8.495,5.644)--cycle;
\gpfill{rgb color={0.800,0.800,0.800}} (8.611,5.644)--(8.728,5.644)--(8.728,7.585)--(8.611,7.585)--cycle;
\draw[gp path] (8.611,5.644)--(8.611,7.584)--(8.727,7.584)--(8.727,5.644)--cycle;
\gpfill{rgb color={0.800,0.800,0.800}} (8.843,5.644)--(8.960,5.644)--(8.960,10.132)--(8.843,10.132)--cycle;
\draw[gp path] (8.843,5.644)--(8.843,10.131)--(8.959,10.131)--(8.959,5.644)--cycle;
\gpfill{rgb color={0.800,0.800,0.800}} (9.075,5.644)--(9.192,5.644)--(9.192,9.232)--(9.075,9.232)--cycle;
\draw[gp path] (9.075,5.644)--(9.075,9.231)--(9.191,9.231)--(9.191,5.644)--cycle;
\gpfill{rgb color={0.800,0.800,0.800}} (9.307,7.299)--(9.424,7.299)--(9.424,8.449)--(9.307,8.449)--cycle;
\draw[gp path] (9.307,7.299)--(9.307,8.448)--(9.423,8.448)--(9.423,7.299)--cycle;
\gpfill{rgb color={0.800,0.800,0.800}} (9.771,5.644)--(9.888,5.644)--(9.888,7.175)--(9.771,7.175)--cycle;
\draw[gp path] (9.771,5.644)--(9.771,7.174)--(9.887,7.174)--(9.887,5.644)--cycle;
\gpfill{rgb color={0.800,0.800,0.800}} (10.003,5.644)--(10.120,5.644)--(10.120,8.234)--(10.003,8.234)--cycle;
\draw[gp path] (10.003,5.644)--(10.003,8.233)--(10.119,8.233)--(10.119,5.644)--cycle;
\gpfill{rgb color={0.800,0.800,0.800}} (10.235,6.730)--(10.352,6.730)--(10.352,6.731)--(10.235,6.731)--cycle;
\draw[gp path] (10.235,6.730)--(10.351,6.730)--cycle;
\gpfill{rgb color={0.800,0.800,0.800}} (10.467,5.905)--(10.584,5.905)--(10.584,5.906)--(10.467,5.906)--cycle;
\draw[gp path] (10.467,5.905)--(10.583,5.905)--cycle;
\gpfill{rgb color={0.800,0.800,0.800}} (10.699,7.246)--(10.816,7.246)--(10.816,7.247)--(10.699,7.247)--cycle;
\draw[gp path] (10.699,7.246)--(10.815,7.246)--cycle;
\gpfill{rgb color={0.800,0.800,0.800}} (10.931,5.644)--(11.048,5.644)--(11.048,10.132)--(10.931,10.132)--cycle;
\draw[gp path] (10.931,5.644)--(10.931,10.131)--(11.047,10.131)--(11.047,5.644)--cycle;
\gpfill{rgb color={0.800,0.800,0.800}} (11.163,5.644)--(11.280,5.644)--(11.280,10.132)--(11.163,10.132)--cycle;
\draw[gp path] (11.163,5.644)--(11.163,10.131)--(11.279,10.131)--(11.279,5.644)--cycle;
\gpfill{rgb color={0.800,0.800,0.800}} (11.395,5.644)--(11.512,5.644)--(11.512,8.685)--(11.395,8.685)--cycle;
\draw[gp path] (11.395,5.644)--(11.395,8.684)--(11.511,8.684)--(11.511,5.644)--cycle;
\gpfill{rgb color={0.800,0.800,0.800}} (11.627,5.979)--(11.744,5.979)--(11.744,5.980)--(11.627,5.980)--cycle;
\draw[gp path] (11.627,5.979)--(11.743,5.979)--cycle;
\gpfill{rgb color={0.800,0.800,0.800}} (11.859,5.644)--(11.976,5.644)--(11.976,9.106)--(11.859,9.106)--cycle;
\draw[gp path] (11.859,5.644)--(11.859,9.105)--(11.975,9.105)--(11.975,5.644)--cycle;
\gpfill{rgb color={0.800,0.800,0.800}} (12.091,5.644)--(12.208,5.644)--(12.208,9.214)--(12.091,9.214)--cycle;
\draw[gp path] (12.091,5.644)--(12.091,9.213)--(12.207,9.213)--(12.207,5.644)--cycle;
\gpfill{rgb color={0.800,0.800,0.800}} (12.323,5.644)--(12.440,5.644)--(12.440,10.132)--(12.323,10.132)--cycle;
\draw[gp path] (12.323,5.644)--(12.323,10.131)--(12.439,10.131)--(12.439,5.644)--cycle;
\gpfill{rgb color={0.800,0.800,0.800}} (12.555,6.059)--(12.672,6.059)--(12.672,6.060)--(12.555,6.060)--cycle;
\draw[gp path] (12.555,6.059)--(12.671,6.059)--cycle;
\gpfill{rgb color={0.800,0.800,0.800}} (12.787,7.076)--(12.904,7.076)--(12.904,7.077)--(12.787,7.077)--cycle;
\draw[gp path] (12.787,7.076)--(12.903,7.076)--cycle;
\gpfill{rgb color={0.800,0.800,0.800}} (13.019,6.926)--(13.136,6.926)--(13.136,6.927)--(13.019,6.927)--cycle;
\draw[gp path] (13.019,6.926)--(13.135,6.926)--cycle;
\gpfill{rgb color={0.800,0.800,0.800}} (13.251,6.846)--(13.368,6.846)--(13.368,6.847)--(13.251,6.847)--cycle;
\draw[gp path] (13.251,6.846)--(13.367,6.846)--cycle;
\gpfill{rgb color={0.800,0.800,0.800}} (13.483,7.514)--(13.600,7.514)--(13.600,7.515)--(13.483,7.515)--cycle;
\draw[gp path] (13.483,7.514)--(13.599,7.514)--cycle;
\gpfill{rgb color={0.800,0.800,0.800}} (13.715,6.075)--(13.832,6.075)--(13.832,6.076)--(13.715,6.076)--cycle;
\draw[gp path] (13.715,6.075)--(13.831,6.075)--cycle;
\gpfill{rgb color={0.800,0.800,0.800}} (13.947,6.111)--(14.064,6.111)--(14.064,6.112)--(13.947,6.112)--cycle;
\draw[gp path] (13.947,6.111)--(14.063,6.111)--cycle;
\gpfill{rgb color={0.800,0.800,0.800}} (14.179,5.644)--(14.296,5.644)--(14.296,10.132)--(14.179,10.132)--cycle;
\draw[gp path] (14.179,5.644)--(14.179,10.131)--(14.295,10.131)--(14.295,5.644)--cycle;
\gpfill{rgb color={0.800,0.800,0.800}} (14.411,5.644)--(14.528,5.644)--(14.528,9.109)--(14.411,9.109)--cycle;
\draw[gp path] (14.411,5.644)--(14.411,9.108)--(14.527,9.108)--(14.527,5.644)--cycle;
\gpfill{rgb color={0.800,0.800,0.800}} (14.643,5.967)--(14.760,5.967)--(14.760,5.968)--(14.643,5.968)--cycle;
\draw[gp path] (14.643,5.967)--(14.759,5.967)--cycle;
\gpfill{rgb color={0.800,0.800,0.800}} (15.107,5.644)--(15.224,5.644)--(15.224,9.632)--(15.107,9.632)--cycle;
\draw[gp path] (15.107,5.644)--(15.107,9.631)--(15.223,9.631)--(15.223,5.644)--cycle;
\gpfill{rgb color={0.800,0.800,0.800}} (15.571,5.644)--(15.688,5.644)--(15.688,7.341)--(15.571,7.341)--cycle;
\draw[gp path] (15.571,5.644)--(15.571,7.340)--(15.687,7.340)--(15.687,5.644)--cycle;
\gpfill{rgb color={0.800,0.800,0.800}} (15.803,5.644)--(15.920,5.644)--(15.920,8.061)--(15.803,8.061)--cycle;
\draw[gp path] (15.803,5.644)--(15.803,8.060)--(15.919,8.060)--(15.919,5.644)--cycle;
\gpfill{rgb color={0.800,0.800,0.800}} (16.035,8.023)--(16.152,8.023)--(16.152,8.024)--(16.035,8.024)--cycle;
\draw[gp path] (16.035,8.023)--(16.151,8.023)--cycle;
\gpfill{rgb color={0.800,0.800,0.800}} (16.267,5.644)--(16.384,5.644)--(16.384,10.132)--(16.267,10.132)--cycle;
\draw[gp path] (16.267,5.644)--(16.267,10.131)--(16.383,10.131)--(16.383,5.644)--cycle;
\gpfill{rgb color={0.800,0.800,0.800}} (16.499,6.304)--(16.616,6.304)--(16.616,6.305)--(16.499,6.305)--cycle;
\draw[gp path] (16.499,6.304)--(16.615,6.304)--cycle;
\gpfill{rgb color={0.800,0.800,0.800}} (16.731,5.644)--(16.848,5.644)--(16.848,7.655)--(16.731,7.655)--cycle;
\draw[gp path] (16.731,5.644)--(16.731,7.654)--(16.847,7.654)--(16.847,5.644)--cycle;
\gpfill{rgb color={0.800,0.800,0.800}} (16.963,6.018)--(17.080,6.018)--(17.080,6.019)--(16.963,6.019)--cycle;
\draw[gp path] (16.963,6.018)--(17.079,6.018)--cycle;
\gpfill{rgb color={0.800,0.800,0.800}} (17.195,5.644)--(17.312,5.644)--(17.312,8.648)--(17.195,8.648)--cycle;
\draw[gp path] (17.195,5.644)--(17.195,8.647)--(17.311,8.647)--(17.311,5.644)--cycle;
\gpfill{rgb color={0.800,0.800,0.800}} (17.427,6.766)--(17.544,6.766)--(17.544,6.767)--(17.427,6.767)--cycle;
\draw[gp path] (17.427,6.766)--(17.543,6.766)--cycle;
\gpfill{rgb color={0.800,0.800,0.800}} (17.659,5.939)--(17.776,5.939)--(17.776,5.940)--(17.659,5.940)--cycle;
\draw[gp path] (17.659,5.939)--(17.775,5.939)--cycle;
\gpfill{rgb color={0.800,0.800,0.800}} (17.891,6.267)--(18.008,6.267)--(18.008,6.268)--(17.891,6.268)--cycle;
\draw[gp path] (17.891,6.267)--(18.007,6.267)--cycle;
\gpfill{rgb color={0.800,0.800,0.800}} (18.355,5.644)--(18.472,5.644)--(18.472,6.915)--(18.355,6.915)--cycle;
\draw[gp path] (18.355,5.644)--(18.355,6.914)--(18.471,6.914)--(18.471,5.644)--cycle;
\gpfill{rgb color={0.800,0.800,0.800}} (18.588,5.644)--(18.705,5.644)--(18.705,10.132)--(18.588,10.132)--cycle;
\draw[gp path] (18.588,5.644)--(18.588,10.131)--(18.704,10.131)--(18.704,5.644)--cycle;
\gpfill{rgb color={0.800,0.800,0.800}} (18.820,5.644)--(18.937,5.644)--(18.937,10.132)--(18.820,10.132)--cycle;
\draw[gp path] (18.820,5.644)--(18.820,10.131)--(18.936,10.131)--(18.936,5.644)--cycle;
\gpfill{rgb color={0.800,0.800,0.800}} (19.284,5.644)--(19.401,5.644)--(19.401,7.276)--(19.284,7.276)--cycle;
\draw[gp path] (19.284,5.644)--(19.284,7.275)--(19.400,7.275)--(19.400,5.644)--cycle;
\gpfill{rgb color={0.800,0.800,0.800}} (19.516,5.644)--(19.633,5.644)--(19.633,7.889)--(19.516,7.889)--cycle;
\draw[gp path] (19.516,5.644)--(19.516,7.888)--(19.632,7.888)--(19.632,5.644)--cycle;
\gpfill{rgb color={0.800,0.800,0.800}} (19.748,6.926)--(19.865,6.926)--(19.865,6.927)--(19.748,6.927)--cycle;
\draw[gp path] (19.748,6.926)--(19.864,6.926)--cycle;
\gpfill{rgb color={0.800,0.800,0.800}} (19.980,5.644)--(20.097,5.644)--(20.097,7.358)--(19.980,7.358)--cycle;
\draw[gp path] (19.980,5.644)--(19.980,7.357)--(20.096,7.357)--(20.096,5.644)--cycle;
\gpfill{rgb color={0.800,0.800,0.800}} (20.676,5.644)--(20.793,5.644)--(20.793,10.132)--(20.676,10.132)--cycle;
\draw[gp path] (20.676,5.644)--(20.676,10.131)--(20.792,10.131)--(20.792,5.644)--cycle;
\gpfill{rgb color={0.800,0.800,0.800}} (20.908,5.765)--(21.025,5.765)--(21.025,5.766)--(20.908,5.766)--cycle;
\draw[gp path] (20.908,5.765)--(21.024,5.765)--cycle;
\gpfill{rgb color={0.800,0.800,0.800}} (21.140,5.644)--(21.257,5.644)--(21.257,7.083)--(21.140,7.083)--cycle;
\draw[gp path] (21.140,5.644)--(21.140,7.082)--(21.256,7.082)--(21.256,5.644)--cycle;
\gpfill{rgb color={0.800,0.800,0.800}} (21.372,9.851)--(21.489,9.851)--(21.489,9.852)--(21.372,9.852)--cycle;
\draw[gp path] (21.372,9.851)--(21.488,9.851)--cycle;
\gpfill{rgb color={0.800,0.800,0.800}} (21.604,6.579)--(21.721,6.579)--(21.721,6.580)--(21.604,6.580)--cycle;
\draw[gp path] (21.604,6.579)--(21.720,6.579)--cycle;
\gpfill{rgb color={0.800,0.800,0.800}} (21.836,5.644)--(21.953,5.644)--(21.953,9.185)--(21.836,9.185)--cycle;
\draw[gp path] (21.836,5.644)--(21.836,9.184)--(21.952,9.184)--(21.952,5.644)--cycle;
\gpfill{rgb color={0.800,0.800,0.800}} (22.068,6.983)--(22.185,6.983)--(22.185,6.984)--(22.068,6.984)--cycle;
\draw[gp path] (22.068,6.983)--(22.184,6.983)--cycle;
\gpfill{rgb color={0.800,0.800,0.800}} (22.300,5.644)--(22.417,5.644)--(22.417,9.855)--(22.300,9.855)--cycle;
\draw[gp path] (22.300,5.644)--(22.300,9.854)--(22.416,9.854)--(22.416,5.644)--cycle;
\gpfill{rgb color={0.800,0.800,0.800}} (22.764,5.644)--(22.881,5.644)--(22.881,7.990)--(22.764,7.990)--cycle;
\draw[gp path] (22.764,5.644)--(22.764,7.989)--(22.880,7.989)--(22.880,5.644)--cycle;
\gpfill{rgb color={0.800,0.800,0.800}} (22.996,5.911)--(23.113,5.911)--(23.113,5.912)--(22.996,5.912)--cycle;
\draw[gp path] (22.996,5.911)--(23.112,5.911)--cycle;
\gpfill{rgb color={0.800,0.800,0.800}} (23.228,6.040)--(23.345,6.040)--(23.345,6.041)--(23.228,6.041)--cycle;
\draw[gp path] (23.228,6.040)--(23.344,6.040)--cycle;
\gpfill{rgb color={0.800,0.800,0.800}} (23.692,5.644)--(23.809,5.644)--(23.809,8.650)--(23.692,8.650)--cycle;
\draw[gp path] (23.692,5.644)--(23.692,8.649)--(23.808,8.649)--(23.808,5.644)--cycle;
\gpfill{rgb color={0.800,0.800,0.800}} (23.924,5.644)--(24.041,5.644)--(24.041,8.771)--(23.924,8.771)--cycle;
\draw[gp path] (23.924,5.644)--(23.924,8.770)--(24.040,8.770)--(24.040,5.644)--cycle;
\gpfill{rgb color={0.800,0.800,0.800}} (24.156,5.644)--(24.273,5.644)--(24.273,7.278)--(24.156,7.278)--cycle;
\draw[gp path] (24.156,5.644)--(24.156,7.277)--(24.272,7.277)--(24.272,5.644)--cycle;
\gpcolor{color=gp lt color border}
\draw[gp path] (1.012,10.131)--(1.012,5.644)--(24.446,5.644)--(24.446,10.131)--cycle;
\node[gp node center] at (12.729,9.682) {\plotLegend{(MIPS)}};
\node[gp node left,font={\fontsize{11.0pt}{13.2pt}\selectfont}] at (1.246,9.817) {\plotLegend{mean improvement: $3.8\%$}};
\node[gp node left,font={\fontsize{11.0pt}{13.2pt}\selectfont}] at (1.246,9.368) {\plotLegend{improved functions: $46\%$}};
\node[gp node left,font={\fontsize{11.0pt}{13.2pt}\selectfont}] at (1.246,8.920) {\plotLegend{mean gap: $10.7\%$}};
\node[gp node left,font={\fontsize{11.0pt}{13.2pt}\selectfont}] at (1.246,8.471) {\plotLegend{optimal functions: $54\%$}};
\gpdefrectangularnode{gp plot 1}{\pgfpoint{1.012cm}{5.644cm}}{\pgfpoint{24.446cm}{10.131cm}}
\end{tikzpicture}

%% file: results/scalability.tex
\begin{tikzpicture}[gnuplot]
\path (0.000,0.000) rectangle (12.500,8.750);
\gpcolor{color=gp lt color border}
\gpsetlinetype{gp lt border}
\gpsetdashtype{gp dt solid}
\gpsetlinewidth{1.00}
\draw[gp path] (1.380,0.985)--(1.560,0.985);
\draw[gp path] (11.947,0.985)--(11.767,0.985);
\node[gp node right,font={\fontsize{12.0pt}{14.4pt}\selectfont}] at (1.380,0.985) {\plotSeconds{0.1}};
\draw[gp path] (1.380,1.542)--(1.470,1.542);
\draw[gp path] (11.947,1.542)--(11.857,1.542);
\draw[gp path] (1.380,1.867)--(1.470,1.867);
\draw[gp path] (11.947,1.867)--(11.857,1.867);
\draw[gp path] (1.380,2.098)--(1.470,2.098);
\draw[gp path] (11.947,2.098)--(11.857,2.098);
\draw[gp path] (1.380,2.277)--(1.470,2.277);
\draw[gp path] (11.947,2.277)--(11.857,2.277);
\draw[gp path] (1.380,2.424)--(1.470,2.424);
\draw[gp path] (11.947,2.424)--(11.857,2.424);
\draw[gp path] (1.380,2.548)--(1.470,2.548);
\draw[gp path] (11.947,2.548)--(11.857,2.548);
\draw[gp path] (1.380,2.655)--(1.470,2.655);
\draw[gp path] (11.947,2.655)--(11.857,2.655);
\draw[gp path] (1.380,2.749)--(1.470,2.749);
\draw[gp path] (11.947,2.749)--(11.857,2.749);
\draw[gp path] (1.380,2.834)--(1.560,2.834);
\draw[gp path] (11.947,2.834)--(11.767,2.834);
\node[gp node right,font={\fontsize{12.0pt}{14.4pt}\selectfont}] at (1.380,2.834) {\plotSeconds{1}};
\draw[gp path] (1.380,3.391)--(1.470,3.391);
\draw[gp path] (11.947,3.391)--(11.857,3.391);
\draw[gp path] (1.380,3.716)--(1.470,3.716);
\draw[gp path] (11.947,3.716)--(11.857,3.716);
\draw[gp path] (1.380,3.947)--(1.470,3.947);
\draw[gp path] (11.947,3.947)--(11.857,3.947);
\draw[gp path] (1.380,4.126)--(1.470,4.126);
\draw[gp path] (11.947,4.126)--(11.857,4.126);
\draw[gp path] (1.380,4.273)--(1.470,4.273);
\draw[gp path] (11.947,4.273)--(11.857,4.273);
\draw[gp path] (1.380,4.397)--(1.470,4.397);
\draw[gp path] (11.947,4.397)--(11.857,4.397);
\draw[gp path] (1.380,4.504)--(1.470,4.504);
\draw[gp path] (11.947,4.504)--(11.857,4.504);
\draw[gp path] (1.380,4.598)--(1.470,4.598);
\draw[gp path] (11.947,4.598)--(11.857,4.598);
\draw[gp path] (1.380,4.683)--(1.560,4.683);
\draw[gp path] (11.947,4.683)--(11.767,4.683);
\node[gp node right,font={\fontsize{12.0pt}{14.4pt}\selectfont}] at (1.380,4.683) {\plotSeconds{10}};
\draw[gp path] (1.380,5.240)--(1.470,5.240);
\draw[gp path] (11.947,5.240)--(11.857,5.240);
\draw[gp path] (1.380,5.565)--(1.470,5.565);
\draw[gp path] (11.947,5.565)--(11.857,5.565);
\draw[gp path] (1.380,5.796)--(1.470,5.796);
\draw[gp path] (11.947,5.796)--(11.857,5.796);
\draw[gp path] (1.380,5.975)--(1.470,5.975);
\draw[gp path] (11.947,5.975)--(11.857,5.975);
\draw[gp path] (1.380,6.122)--(1.470,6.122);
\draw[gp path] (11.947,6.122)--(11.857,6.122);
\draw[gp path] (1.380,6.246)--(1.470,6.246);
\draw[gp path] (11.947,6.246)--(11.857,6.246);
\draw[gp path] (1.380,6.353)--(1.470,6.353);
\draw[gp path] (11.947,6.353)--(11.857,6.353);
\draw[gp path] (1.380,6.447)--(1.470,6.447);
\draw[gp path] (11.947,6.447)--(11.857,6.447);
\draw[gp path] (1.380,6.532)--(1.560,6.532);
\draw[gp path] (11.947,6.532)--(11.767,6.532);
\node[gp node right,font={\fontsize{12.0pt}{14.4pt}\selectfont}] at (1.380,6.532) {\plotSeconds{100}};
\draw[gp path] (1.380,7.089)--(1.470,7.089);
\draw[gp path] (11.947,7.089)--(11.857,7.089);
\draw[gp path] (1.380,7.414)--(1.470,7.414);
\draw[gp path] (11.947,7.414)--(11.857,7.414);
\draw[gp path] (1.380,7.645)--(1.470,7.645);
\draw[gp path] (11.947,7.645)--(11.857,7.645);
\draw[gp path] (1.380,7.824)--(1.470,7.824);
\draw[gp path] (11.947,7.824)--(11.857,7.824);
\draw[gp path] (1.380,7.971)--(1.470,7.971);
\draw[gp path] (11.947,7.971)--(11.857,7.971);
\draw[gp path] (1.380,8.095)--(1.470,8.095);
\draw[gp path] (11.947,8.095)--(11.857,8.095);
\draw[gp path] (1.380,8.202)--(1.470,8.202);
\draw[gp path] (11.947,8.202)--(11.857,8.202);
\draw[gp path] (1.380,8.296)--(1.470,8.296);
\draw[gp path] (11.947,8.296)--(11.857,8.296);
\draw[gp path] (1.380,8.381)--(1.560,8.381);
\draw[gp path] (11.947,8.381)--(11.767,8.381);
\node[gp node right,font={\fontsize{12.0pt}{14.4pt}\selectfont}] at (1.380,8.381) {\plotSeconds{1000}};
\draw[gp path] (1.380,0.985)--(1.380,1.165);
\draw[gp path] (1.380,8.381)--(1.380,8.201);
\node[gp node center,font={\fontsize{12.0pt}{14.4pt}\selectfont}] at (1.380,0.677) {\plotNoInstructions{10}};
\draw[gp path] (2.970,0.985)--(2.970,1.075);
\draw[gp path] (2.970,8.381)--(2.970,8.291);
\draw[gp path] (3.901,0.985)--(3.901,1.075);
\draw[gp path] (3.901,8.381)--(3.901,8.291);
\draw[gp path] (4.561,0.985)--(4.561,1.075);
\draw[gp path] (4.561,8.381)--(4.561,8.291);
\draw[gp path] (5.073,0.985)--(5.073,1.075);
\draw[gp path] (5.073,8.381)--(5.073,8.291);
\draw[gp path] (5.491,0.985)--(5.491,1.075);
\draw[gp path] (5.491,8.381)--(5.491,8.291);
\draw[gp path] (5.845,0.985)--(5.845,1.075);
\draw[gp path] (5.845,8.381)--(5.845,8.291);
\draw[gp path] (6.151,0.985)--(6.151,1.075);
\draw[gp path] (6.151,8.381)--(6.151,8.291);
\draw[gp path] (6.422,0.985)--(6.422,1.075);
\draw[gp path] (6.422,8.381)--(6.422,8.291);
\draw[gp path] (6.664,0.985)--(6.664,1.165);
\draw[gp path] (6.664,8.381)--(6.664,8.201);
\node[gp node center,font={\fontsize{12.0pt}{14.4pt}\selectfont}] at (6.664,0.677) {\plotNoInstructions{100}};
\draw[gp path] (8.254,0.985)--(8.254,1.075);
\draw[gp path] (8.254,8.381)--(8.254,8.291);
\draw[gp path] (9.184,0.985)--(9.184,1.075);
\draw[gp path] (9.184,8.381)--(9.184,8.291);
\draw[gp path] (9.844,0.985)--(9.844,1.075);
\draw[gp path] (9.844,8.381)--(9.844,8.291);
\draw[gp path] (10.357,0.985)--(10.357,1.075);
\draw[gp path] (10.357,8.381)--(10.357,8.291);
\draw[gp path] (10.775,0.985)--(10.775,1.075);
\draw[gp path] (10.775,8.381)--(10.775,8.291);
\draw[gp path] (11.129,0.985)--(11.129,1.075);
\draw[gp path] (11.129,8.381)--(11.129,8.291);
\draw[gp path] (11.435,0.985)--(11.435,1.075);
\draw[gp path] (11.435,8.381)--(11.435,8.291);
\draw[gp path] (11.705,0.985)--(11.705,1.075);
\draw[gp path] (11.705,8.381)--(11.705,8.291);
\draw[gp path] (11.947,0.985)--(11.947,1.165);
\draw[gp path] (11.947,8.381)--(11.947,8.201);
\node[gp node center,font={\fontsize{12.0pt}{14.4pt}\selectfont}] at (11.947,0.677) {\plotNoInstructions{1000}};
\draw[gp path] (1.380,8.381)--(1.380,0.985)--(11.947,0.985)--(11.947,8.381)--cycle;
\node[gp node center] at (6.663,0.215) {\plotAxisLabel{number of input instructions}};
\node[gp node right,font={\fontsize{12.0pt}{14.4pt}\selectfont}] at (10.257,1.390) {\plotShiftRight{\plotLegend{Hexagon}}};
\gpcolor{rgb color={0.000,0.000,0.000}}
\gpsetpointsize{4.00}
\gppoint{gp mark 6}{(3.573,2.324)}
\gppoint{gp mark 6}{(5.812,3.226)}
\gppoint{gp mark 6}{(5.027,3.179)}
\gppoint{gp mark 6}{(4.931,3.276)}
\gppoint{gp mark 6}{(4.780,3.633)}
\gppoint{gp mark 6}{(5.910,3.442)}
\gppoint{gp mark 6}{(4.780,3.144)}
\gppoint{gp mark 6}{(6.208,4.539)}
\gppoint{gp mark 6}{(5.675,3.818)}
\gppoint{gp mark 6}{(6.180,3.595)}
\gppoint{gp mark 6}{(7.230,5.108)}
\gppoint{gp mark 6}{(5.333,4.093)}
\gppoint{gp mark 6}{(4.979,3.529)}
\gppoint{gp mark 6}{(5.675,4.243)}
\gppoint{gp mark 6}{(9.368,6.299)}
\gppoint{gp mark 6}{(8.100,5.685)}
\gppoint{gp mark 6}{(5.207,2.902)}
\gppoint{gp mark 6}{(5.779,5.296)}
\gppoint{gp mark 6}{(5.941,4.748)}
\gppoint{gp mark 6}{(6.422,6.616)}
\gppoint{gp mark 6}{(7.369,6.797)}
\gppoint{gp mark 6}{(7.318,6.763)}
\gppoint{gp mark 6}{(3.743,2.884)}
\gppoint{gp mark 6}{(8.624,6.744)}
\gppoint{gp mark 6}{(8.729,6.802)}
\gppoint{gp mark 6}{(3.743,1.947)}
\gppoint{gp mark 6}{(3.823,1.220)}
\gppoint{gp mark 6}{(6.753,5.175)}
\gppoint{gp mark 6}{(8.847,3.751)}
\gppoint{gp mark 6}{(10.948,3.868)}
\gppoint{gp mark 6}{(10.160,7.897)}
\gppoint{gp mark 6}{(8.277,8.023)}
\gppoint{gp mark 6}{(5.812,4.020)}
\gppoint{gp mark 6}{(4.618,1.546)}
\gppoint{gp mark 6}{(8.063,5.962)}
\gppoint{gp mark 6}{(7.742,5.225)}
\gppoint{gp mark 6}{(5.639,3.572)}
\gppoint{gp mark 6}{(6.396,5.102)}
\gppoint{gp mark 6}{(4.561,3.339)}
\gppoint{gp mark 6}{(5.414,3.506)}
\gppoint{gp mark 6}{(8.219,7.939)}
\gppoint{gp mark 6}{(5.414,3.668)}
\gppoint{gp mark 6}{(5.027,4.616)}
\gppoint{gp mark 6}{(3.823,3.400)}
\gppoint{gp mark 6}{(3.573,2.616)}
\gppoint{gp mark 6}{(4.503,3.211)}
\gppoint{gp mark 6}{(3.573,2.731)}
\gppoint{gp mark 6}{(3.389,2.889)}
\gppoint{gp mark 6}{(5.027,3.758)}
\gppoint{gp mark 6}{(8.196,6.656)}
\gppoint{gp mark 6}{(10.824,7.606)}
\gppoint{gp mark 6}{(5.529,3.063)}
\gppoint{gp mark 6}{(8.524,7.369)}
\gppoint{gp mark 6}{(6.317,2.144)}
\gppoint{gp mark 6}{(5.973,4.744)}
\gppoint{gp mark 6}{(9.138,7.952)}
\gppoint{gp mark 6}{(8.701,6.999)}
\gppoint{gp mark 6}{(7.369,6.151)}
\gppoint{gp mark 6}{(4.120,2.933)}
\gppoint{gp mark 6}{(8.299,6.817)}
\gppoint{gp mark 6}{(4.188,2.439)}
\gppoint{gp mark 6}{(6.775,6.842)}
\gppoint{gp mark 6}{(5.163,3.846)}
\gppoint{gp mark 6}{(3.483,2.824)}
\gppoint{gp mark 6}{(5.779,4.086)}
\gppoint{gp mark 6}{(5.333,3.653)}
\gppoint{gp mark 6}{(8.865,7.153)}
\gppoint{gp mark 6}{(4.255,3.557)}
\gppoint{gp mark 6}{(7.283,6.600)}
\gppoint{gp mark 6}{(6.151,5.097)}
\gppoint{gp mark 6}{(8.985,7.843)}
\gppoint{gp mark 6}{(2.970,1.857)}
\gppoint{gp mark 6}{(10.145,8.217)}
\gppoint{gp mark 6}{(5.779,4.950)}
\gppoint{gp mark 6}{(10.767,8.042)}
\gppoint{gp mark 6}{(10.282,7.817)}
\gppoint{gp mark 6}{(4.882,3.713)}
\gppoint{gp mark 6}{(6.208,5.329)}
\gppoint{gp mark 6}{(6.064,2.414)}
\gppoint{gp mark 6}{(5.639,6.179)}
\gppoint{gp mark 6}{(8.653,7.219)}
\gppoint{gp mark 6}{(3.573,2.211)}
\gppoint{gp mark 6}{(5.812,2.655)}
\gppoint{gp mark 6}{(5.027,2.727)}
\gppoint{gp mark 6}{(4.931,3.018)}
\gppoint{gp mark 6}{(4.780,3.061)}
\gppoint{gp mark 6}{(5.910,3.245)}
\gppoint{gp mark 6}{(11.820,8.018)}
\gppoint{gp mark 6}{(4.780,2.399)}
\gppoint{gp mark 6}{(6.208,2.139)}
\gppoint{gp mark 6}{(5.675,3.274)}
\gppoint{gp mark 6}{(6.180,1.717)}
\gppoint{gp mark 6}{(7.230,4.594)}
\gppoint{gp mark 6}{(5.333,3.131)}
\gppoint{gp mark 6}{(4.979,1.775)}
\gppoint{gp mark 6}{(5.675,3.634)}
\gppoint{gp mark 6}{(9.368,6.438)}
\gppoint{gp mark 6}{(8.100,5.851)}
\gppoint{gp mark 6}{(5.207,2.843)}
\gppoint{gp mark 6}{(5.779,5.252)}
\gppoint{gp mark 6}{(5.941,4.691)}
\gppoint{gp mark 6}{(6.422,6.592)}
\gppoint{gp mark 6}{(7.369,6.770)}
\gppoint{gp mark 6}{(7.318,6.754)}
\gppoint{gp mark 6}{(3.743,2.884)}
\gppoint{gp mark 6}{(8.624,6.697)}
\gppoint{gp mark 6}{(8.729,6.798)}
\gppoint{gp mark 6}{(3.743,1.183)}
\gppoint{gp mark 6}{(3.823,2.616)}
\gppoint{gp mark 6}{(6.753,4.703)}
\gppoint{gp mark 6}{(8.847,6.347)}
\gppoint{gp mark 6}{(10.948,7.428)}
\gppoint{gp mark 6}{(5.812,2.997)}
\gppoint{gp mark 6}{(4.618,3.157)}
\gppoint{gp mark 6}{(8.063,5.555)}
\gppoint{gp mark 6}{(7.742,4.903)}
\gppoint{gp mark 6}{(5.639,2.762)}
\gppoint{gp mark 6}{(6.396,4.672)}
\gppoint{gp mark 6}{(4.561,2.601)}
\gppoint{gp mark 6}{(6.686,4.224)}
\gppoint{gp mark 6}{(8.196,6.321)}
\gppoint{gp mark 6}{(5.414,1.707)}
\gppoint{gp mark 6}{(8.219,6.626)}
\gppoint{gp mark 6}{(5.414,2.984)}
\gppoint{gp mark 6}{(5.027,1.834)}
\gppoint{gp mark 6}{(3.823,3.390)}
\gppoint{gp mark 6}{(3.573,2.125)}
\gppoint{gp mark 6}{(4.503,2.518)}
\gppoint{gp mark 6}{(3.573,1.489)}
\gppoint{gp mark 6}{(3.389,1.317)}
\gppoint{gp mark 6}{(5.027,3.242)}
\gppoint{gp mark 6}{(8.196,5.924)}
\gppoint{gp mark 6}{(5.529,1.831)}
\gppoint{gp mark 6}{(6.317,5.335)}
\gppoint{gp mark 6}{(5.973,4.000)}
\gppoint{gp mark 6}{(4.120,2.494)}
\gppoint{gp mark 6}{(8.299,6.594)}
\gppoint{gp mark 6}{(4.188,1.237)}
\gppoint{gp mark 6}{(6.775,8.116)}
\gppoint{gp mark 6}{(5.163,4.267)}
\gppoint{gp mark 6}{(3.483,2.882)}
\gppoint{gp mark 6}{(5.779,3.487)}
\gppoint{gp mark 6}{(5.333,4.007)}
\gppoint{gp mark 6}{(8.865,2.955)}
\gppoint{gp mark 6}{(4.255,3.489)}
\gppoint{gp mark 6}{(6.151,2.187)}
\gppoint{gp mark 6}{(8.985,7.499)}
\gppoint{gp mark 6}{(2.970,2.267)}
\gppoint{gp mark 6}{(7.120,5.834)}
\gppoint{gp mark 6}{(5.779,4.878)}
\gppoint{gp mark 6}{(10.767,7.018)}
\gppoint{gp mark 6}{(9.951,6.896)}
\gppoint{gp mark 6}{(4.882,3.675)}
\gppoint{gp mark 6}{(6.208,4.718)}
\gppoint{gp mark 6}{(6.064,2.355)}
\gppoint{gp mark 6}{(5.639,6.894)}
\gppoint{gp mark 6}{(9.010,8.249)}
\gppoint{gp mark 6}{(8.653,6.502)}
\gppoint{gp mark 6}{(11.010,1.390)}
\gpcolor{color=gp lt color border}
\node[gp node right,font={\fontsize{12.0pt}{14.4pt}\selectfont}] at (10.257,1.840) {\plotShiftRight{\plotLegend{ARM}}};
\gpcolor{rgb color={0.000,0.000,0.000}}
\gppoint{gp mark 7}{(3.573,1.501)}
\gppoint{gp mark 7}{(6.180,3.966)}
\gppoint{gp mark 7}{(5.027,3.604)}
\gppoint{gp mark 7}{(4.931,3.833)}
\gppoint{gp mark 7}{(4.780,4.780)}
\gppoint{gp mark 7}{(6.003,3.470)}
\gppoint{gp mark 7}{(4.831,4.175)}
\gppoint{gp mark 7}{(6.447,5.339)}
\gppoint{gp mark 7}{(6.034,3.836)}
\gppoint{gp mark 7}{(6.180,3.948)}
\gppoint{gp mark 7}{(5.333,4.001)}
\gppoint{gp mark 7}{(5.453,4.316)}
\gppoint{gp mark 7}{(5.710,3.916)}
\gppoint{gp mark 7}{(9.403,7.503)}
\gppoint{gp mark 7}{(8.483,5.983)}
\gppoint{gp mark 7}{(5.250,3.416)}
\gppoint{gp mark 7}{(5.745,5.986)}
\gppoint{gp mark 7}{(5.941,5.446)}
\gppoint{gp mark 7}{(6.396,7.166)}
\gppoint{gp mark 7}{(7.403,7.230)}
\gppoint{gp mark 7}{(7.266,7.414)}
\gppoint{gp mark 7}{(3.659,2.007)}
\gppoint{gp mark 7}{(8.691,7.327)}
\gppoint{gp mark 7}{(3.743,2.079)}
\gppoint{gp mark 7}{(3.823,1.708)}
\gppoint{gp mark 7}{(7.101,5.788)}
\gppoint{gp mark 7}{(8.441,6.294)}
\gppoint{gp mark 7}{(5.812,4.135)}
\gppoint{gp mark 7}{(4.931,3.526)}
\gppoint{gp mark 7}{(5.414,3.309)}
\gppoint{gp mark 7}{(5.603,4.701)}
\gppoint{gp mark 7}{(4.255,4.236)}
\gppoint{gp mark 7}{(4.049,2.788)}
\gppoint{gp mark 7}{(4.727,3.944)}
\gppoint{gp mark 7}{(3.976,3.806)}
\gppoint{gp mark 7}{(3.659,2.689)}
\gppoint{gp mark 7}{(7.974,7.945)}
\gppoint{gp mark 7}{(5.878,3.415)}
\gppoint{gp mark 7}{(6.944,6.432)}
\gppoint{gp mark 7}{(6.151,5.566)}
\gppoint{gp mark 7}{(3.291,2.831)}
\gppoint{gp mark 7}{(6.151,5.031)}
\gppoint{gp mark 7}{(4.120,1.517)}
\gppoint{gp mark 7}{(5.118,4.851)}
\gppoint{gp mark 7}{(4.319,2.634)}
\gppoint{gp mark 7}{(4.882,4.679)}
\gppoint{gp mark 7}{(5.073,3.232)}
\gppoint{gp mark 7}{(3.483,2.048)}
\gppoint{gp mark 7}{(6.093,5.515)}
\gppoint{gp mark 7}{(5.639,4.728)}
\gppoint{gp mark 7}{(4.255,3.747)}
\gppoint{gp mark 7}{(6.882,5.866)}
\gppoint{gp mark 7}{(6.594,5.986)}
\gppoint{gp mark 7}{(2.970,2.166)}
\gppoint{gp mark 7}{(5.567,4.667)}
\gppoint{gp mark 7}{(6.003,5.406)}
\gppoint{gp mark 7}{(4.882,3.409)}
\gppoint{gp mark 7}{(6.570,7.144)}
\gppoint{gp mark 7}{(5.779,6.419)}
\gppoint{gp mark 7}{(5.973,6.448)}
\gppoint{gp mark 7}{(3.573,2.534)}
\gppoint{gp mark 7}{(6.180,4.085)}
\gppoint{gp mark 7}{(5.027,3.435)}
\gppoint{gp mark 7}{(4.931,3.644)}
\gppoint{gp mark 7}{(4.780,3.991)}
\gppoint{gp mark 7}{(6.003,4.153)}
\gppoint{gp mark 7}{(4.831,3.984)}
\gppoint{gp mark 7}{(6.447,5.711)}
\gppoint{gp mark 7}{(6.034,3.814)}
\gppoint{gp mark 7}{(6.180,3.693)}
\gppoint{gp mark 7}{(7.452,6.239)}
\gppoint{gp mark 7}{(5.333,4.853)}
\gppoint{gp mark 7}{(5.453,4.695)}
\gppoint{gp mark 7}{(5.710,4.160)}
\gppoint{gp mark 7}{(8.483,6.114)}
\gppoint{gp mark 7}{(5.250,3.137)}
\gppoint{gp mark 7}{(5.745,6.001)}
\gppoint{gp mark 7}{(5.941,5.462)}
\gppoint{gp mark 7}{(6.396,7.152)}
\gppoint{gp mark 7}{(7.403,7.081)}
\gppoint{gp mark 7}{(7.266,7.767)}
\gppoint{gp mark 7}{(3.659,2.012)}
\gppoint{gp mark 7}{(3.743,2.394)}
\gppoint{gp mark 7}{(3.823,1.708)}
\gppoint{gp mark 7}{(7.101,6.773)}
\gppoint{gp mark 7}{(4.882,4.238)}
\gppoint{gp mark 7}{(8.441,6.675)}
\gppoint{gp mark 7}{(5.812,4.574)}
\gppoint{gp mark 7}{(6.640,6.204)}
\gppoint{gp mark 7}{(4.931,3.614)}
\gppoint{gp mark 7}{(5.414,2.010)}
\gppoint{gp mark 7}{(5.603,4.751)}
\gppoint{gp mark 7}{(5.292,5.782)}
\gppoint{gp mark 7}{(4.255,3.597)}
\gppoint{gp mark 7}{(4.049,2.840)}
\gppoint{gp mark 7}{(4.727,2.643)}
\gppoint{gp mark 7}{(3.976,3.580)}
\gppoint{gp mark 7}{(3.659,2.730)}
\gppoint{gp mark 7}{(5.333,5.820)}
\gppoint{gp mark 7}{(5.878,3.612)}
\gppoint{gp mark 7}{(6.944,6.841)}
\gppoint{gp mark 7}{(6.151,5.278)}
\gppoint{gp mark 7}{(3.291,3.539)}
\gppoint{gp mark 7}{(6.151,7.369)}
\gppoint{gp mark 7}{(4.120,1.583)}
\gppoint{gp mark 7}{(5.118,6.838)}
\gppoint{gp mark 7}{(4.319,1.624)}
\gppoint{gp mark 7}{(4.882,4.465)}
\gppoint{gp mark 7}{(5.073,3.286)}
\gppoint{gp mark 7}{(3.483,1.612)}
\gppoint{gp mark 7}{(6.093,7.393)}
\gppoint{gp mark 7}{(5.639,4.594)}
\gppoint{gp mark 7}{(4.255,3.445)}
\gppoint{gp mark 7}{(6.882,6.404)}
\gppoint{gp mark 7}{(6.317,6.347)}
\gppoint{gp mark 7}{(6.594,3.865)}
\gppoint{gp mark 7}{(2.970,1.855)}
\gppoint{gp mark 7}{(5.567,5.112)}
\gppoint{gp mark 7}{(6.003,5.289)}
\gppoint{gp mark 7}{(4.882,1.912)}
\gppoint{gp mark 7}{(6.570,6.205)}
\gppoint{gp mark 7}{(6.731,6.458)}
\gppoint{gp mark 7}{(5.779,7.118)}
\gppoint{gp mark 7}{(5.973,6.541)}
\gppoint{gp mark 7}{(11.010,1.840)}
\gpcolor{color=gp lt color border}
\node[gp node right,font={\fontsize{12.0pt}{14.4pt}\selectfont}] at (10.257,2.290) {\plotShiftRight{\plotLegend{MIPS}}};
\gpcolor{rgb color={0.600,0.600,0.600}}
\gppoint{gp mark 7}{(4.120,3.864)}
\gppoint{gp mark 7}{(6.291,5.245)}
\gppoint{gp mark 7}{(5.779,6.513)}
\gppoint{gp mark 7}{(5.207,5.105)}
\gppoint{gp mark 7}{(6.263,5.857)}
\gppoint{gp mark 7}{(5.845,5.917)}
\gppoint{gp mark 7}{(6.686,5.839)}
\gppoint{gp mark 7}{(5.639,6.206)}
\gppoint{gp mark 7}{(4.979,4.582)}
\gppoint{gp mark 7}{(4.831,4.981)}
\gppoint{gp mark 7}{(4.618,4.052)}
\gppoint{gp mark 7}{(8.311,8.048)}
\gppoint{gp mark 7}{(6.775,6.226)}
\gppoint{gp mark 7}{(5.118,6.697)}
\gppoint{gp mark 7}{(5.941,5.377)}
\gppoint{gp mark 7}{(4.049,4.485)}
\gppoint{gp mark 7}{(4.049,3.804)}
\gppoint{gp mark 7}{(3.901,3.636)}
\gppoint{gp mark 7}{(7.785,7.686)}
\gppoint{gp mark 7}{(6.640,6.228)}
\gppoint{gp mark 7}{(4.049,7.161)}
\gppoint{gp mark 7}{(4.979,5.997)}
\gppoint{gp mark 7}{(4.882,5.189)}
\gppoint{gp mark 7}{(4.831,4.401)}
\gppoint{gp mark 7}{(5.250,4.726)}
\gppoint{gp mark 7}{(3.291,2.854)}
\gppoint{gp mark 7}{(6.236,5.510)}
\gppoint{gp mark 7}{(4.882,6.283)}
\gppoint{gp mark 7}{(7.004,6.270)}
\gppoint{gp mark 7}{(6.594,6.195)}
\gppoint{gp mark 7}{(5.639,5.522)}
\gppoint{gp mark 7}{(4.673,4.601)}
\gppoint{gp mark 7}{(6.617,6.474)}
\gppoint{gp mark 7}{(7.248,6.060)}
\gppoint{gp mark 7}{(4.120,3.487)}
\gppoint{gp mark 7}{(6.291,5.207)}
\gppoint{gp mark 7}{(5.779,5.265)}
\gppoint{gp mark 7}{(5.207,5.005)}
\gppoint{gp mark 7}{(5.453,6.951)}
\gppoint{gp mark 7}{(6.263,5.223)}
\gppoint{gp mark 7}{(5.414,6.000)}
\gppoint{gp mark 7}{(6.344,6.701)}
\gppoint{gp mark 7}{(5.845,5.207)}
\gppoint{gp mark 7}{(6.686,5.614)}
\gppoint{gp mark 7}{(5.639,6.516)}
\gppoint{gp mark 7}{(4.979,2.187)}
\gppoint{gp mark 7}{(6.093,5.575)}
\gppoint{gp mark 7}{(6.522,6.728)}
\gppoint{gp mark 7}{(6.731,6.613)}
\gppoint{gp mark 7}{(7.230,7.033)}
\gppoint{gp mark 7}{(8.087,7.312)}
\gppoint{gp mark 7}{(7.895,8.174)}
\gppoint{gp mark 7}{(4.561,5.195)}
\gppoint{gp mark 7}{(4.673,4.664)}
\gppoint{gp mark 7}{(4.831,4.077)}
\gppoint{gp mark 7}{(4.618,2.214)}
\gppoint{gp mark 7}{(6.236,7.109)}
\gppoint{gp mark 7}{(6.775,5.852)}
\gppoint{gp mark 7}{(5.118,5.046)}
\gppoint{gp mark 7}{(5.941,5.170)}
\gppoint{gp mark 7}{(5.603,6.593)}
\gppoint{gp mark 7}{(5.250,7.070)}
\gppoint{gp mark 7}{(4.503,5.944)}
\gppoint{gp mark 7}{(4.049,3.989)}
\gppoint{gp mark 7}{(4.780,5.130)}
\gppoint{gp mark 7}{(4.049,3.731)}
\gppoint{gp mark 7}{(3.901,3.484)}
\gppoint{gp mark 7}{(7.785,7.104)}
\gppoint{gp mark 7}{(6.422,5.393)}
\gppoint{gp mark 7}{(6.640,2.767)}
\gppoint{gp mark 7}{(4.049,5.207)}
\gppoint{gp mark 7}{(4.979,4.294)}
\gppoint{gp mark 7}{(4.882,4.905)}
\gppoint{gp mark 7}{(4.831,4.254)}
\gppoint{gp mark 7}{(4.673,5.863)}
\gppoint{gp mark 7}{(5.250,4.562)}
\gppoint{gp mark 7}{(3.291,2.656)}
\gppoint{gp mark 7}{(6.236,2.639)}
\gppoint{gp mark 7}{(4.882,5.123)}
\gppoint{gp mark 7}{(7.176,6.828)}
\gppoint{gp mark 7}{(7.004,6.008)}
\gppoint{gp mark 7}{(6.594,6.253)}
\gppoint{gp mark 7}{(3.901,4.859)}
\gppoint{gp mark 7}{(5.639,5.459)}
\gppoint{gp mark 7}{(6.546,7.290)}
\gppoint{gp mark 7}{(4.673,5.565)}
\gppoint{gp mark 7}{(6.617,6.299)}
\gppoint{gp mark 7}{(7.248,4.003)}
\gppoint{gp mark 7}{(11.010,2.290)}
\gpcolor{color=gp lt color border}
\draw[gp path] (1.380,8.381)--(1.380,0.985)--(11.947,0.985)--(11.947,8.381)--cycle;
\gpdefrectangularnode{gp plot 1}{\pgfpoint{1.380cm}{0.985cm}}{\pgfpoint{11.947cm}{8.381cm}}
\end{tikzpicture}

%% file: results/combined-scalability.tex
\begin{tikzpicture}[gnuplot]
\path (0.000,0.000) rectangle (12.500,8.750);
\gpcolor{color=gp lt color border}
\gpsetlinetype{gp lt border}
\gpsetdashtype{gp dt solid}
\gpsetlinewidth{1.00}
\draw[gp path] (1.380,0.985)--(1.560,0.985);
\draw[gp path] (11.947,0.985)--(11.767,0.985);
\node[gp node right,font={\fontsize{12.0pt}{14.4pt}\selectfont}] at (1.380,0.985) {\plotSeconds{0.1}};
\draw[gp path] (1.380,1.542)--(1.470,1.542);
\draw[gp path] (11.947,1.542)--(11.857,1.542);
\draw[gp path] (1.380,1.867)--(1.470,1.867);
\draw[gp path] (11.947,1.867)--(11.857,1.867);
\draw[gp path] (1.380,2.098)--(1.470,2.098);
\draw[gp path] (11.947,2.098)--(11.857,2.098);
\draw[gp path] (1.380,2.277)--(1.470,2.277);
\draw[gp path] (11.947,2.277)--(11.857,2.277);
\draw[gp path] (1.380,2.424)--(1.470,2.424);
\draw[gp path] (11.947,2.424)--(11.857,2.424);
\draw[gp path] (1.380,2.548)--(1.470,2.548);
\draw[gp path] (11.947,2.548)--(11.857,2.548);
\draw[gp path] (1.380,2.655)--(1.470,2.655);
\draw[gp path] (11.947,2.655)--(11.857,2.655);
\draw[gp path] (1.380,2.749)--(1.470,2.749);
\draw[gp path] (11.947,2.749)--(11.857,2.749);
\draw[gp path] (1.380,2.834)--(1.560,2.834);
\draw[gp path] (11.947,2.834)--(11.767,2.834);
\node[gp node right,font={\fontsize{12.0pt}{14.4pt}\selectfont}] at (1.380,2.834) {\plotSeconds{1}};
\draw[gp path] (1.380,3.391)--(1.470,3.391);
\draw[gp path] (11.947,3.391)--(11.857,3.391);
\draw[gp path] (1.380,3.716)--(1.470,3.716);
\draw[gp path] (11.947,3.716)--(11.857,3.716);
\draw[gp path] (1.380,3.947)--(1.470,3.947);
\draw[gp path] (11.947,3.947)--(11.857,3.947);
\draw[gp path] (1.380,4.126)--(1.470,4.126);
\draw[gp path] (11.947,4.126)--(11.857,4.126);
\draw[gp path] (1.380,4.273)--(1.470,4.273);
\draw[gp path] (11.947,4.273)--(11.857,4.273);
\draw[gp path] (1.380,4.397)--(1.470,4.397);
\draw[gp path] (11.947,4.397)--(11.857,4.397);
\draw[gp path] (1.380,4.504)--(1.470,4.504);
\draw[gp path] (11.947,4.504)--(11.857,4.504);
\draw[gp path] (1.380,4.598)--(1.470,4.598);
\draw[gp path] (11.947,4.598)--(11.857,4.598);
\draw[gp path] (1.380,4.683)--(1.560,4.683);
\draw[gp path] (11.947,4.683)--(11.767,4.683);
\node[gp node right,font={\fontsize{12.0pt}{14.4pt}\selectfont}] at (1.380,4.683) {\plotSeconds{10}};
\draw[gp path] (1.380,5.240)--(1.470,5.240);
\draw[gp path] (11.947,5.240)--(11.857,5.240);
\draw[gp path] (1.380,5.565)--(1.470,5.565);
\draw[gp path] (11.947,5.565)--(11.857,5.565);
\draw[gp path] (1.380,5.796)--(1.470,5.796);
\draw[gp path] (11.947,5.796)--(11.857,5.796);
\draw[gp path] (1.380,5.975)--(1.470,5.975);
\draw[gp path] (11.947,5.975)--(11.857,5.975);
\draw[gp path] (1.380,6.122)--(1.470,6.122);
\draw[gp path] (11.947,6.122)--(11.857,6.122);
\draw[gp path] (1.380,6.246)--(1.470,6.246);
\draw[gp path] (11.947,6.246)--(11.857,6.246);
\draw[gp path] (1.380,6.353)--(1.470,6.353);
\draw[gp path] (11.947,6.353)--(11.857,6.353);
\draw[gp path] (1.380,6.447)--(1.470,6.447);
\draw[gp path] (11.947,6.447)--(11.857,6.447);
\draw[gp path] (1.380,6.532)--(1.560,6.532);
\draw[gp path] (11.947,6.532)--(11.767,6.532);
\node[gp node right,font={\fontsize{12.0pt}{14.4pt}\selectfont}] at (1.380,6.532) {\plotSeconds{100}};
\draw[gp path] (1.380,7.089)--(1.470,7.089);
\draw[gp path] (11.947,7.089)--(11.857,7.089);
\draw[gp path] (1.380,7.414)--(1.470,7.414);
\draw[gp path] (11.947,7.414)--(11.857,7.414);
\draw[gp path] (1.380,7.645)--(1.470,7.645);
\draw[gp path] (11.947,7.645)--(11.857,7.645);
\draw[gp path] (1.380,7.824)--(1.470,7.824);
\draw[gp path] (11.947,7.824)--(11.857,7.824);
\draw[gp path] (1.380,7.971)--(1.470,7.971);
\draw[gp path] (11.947,7.971)--(11.857,7.971);
\draw[gp path] (1.380,8.095)--(1.470,8.095);
\draw[gp path] (11.947,8.095)--(11.857,8.095);
\draw[gp path] (1.380,8.202)--(1.470,8.202);
\draw[gp path] (11.947,8.202)--(11.857,8.202);
\draw[gp path] (1.380,8.296)--(1.470,8.296);
\draw[gp path] (11.947,8.296)--(11.857,8.296);
\draw[gp path] (1.380,8.381)--(1.560,8.381);
\draw[gp path] (11.947,8.381)--(11.767,8.381);
\node[gp node right,font={\fontsize{12.0pt}{14.4pt}\selectfont}] at (1.380,8.381) {\plotSeconds{1000}};
\draw[gp path] (1.380,0.985)--(1.380,1.165);
\draw[gp path] (1.380,8.381)--(1.380,8.201);
\node[gp node center,font={\fontsize{12.0pt}{14.4pt}\selectfont}] at (1.380,0.677) {\plotNoInstructions{10}};
\draw[gp path] (2.970,0.985)--(2.970,1.075);
\draw[gp path] (2.970,8.381)--(2.970,8.291);
\draw[gp path] (3.901,0.985)--(3.901,1.075);
\draw[gp path] (3.901,8.381)--(3.901,8.291);
\draw[gp path] (4.561,0.985)--(4.561,1.075);
\draw[gp path] (4.561,8.381)--(4.561,8.291);
\draw[gp path] (5.073,0.985)--(5.073,1.075);
\draw[gp path] (5.073,8.381)--(5.073,8.291);
\draw[gp path] (5.491,0.985)--(5.491,1.075);
\draw[gp path] (5.491,8.381)--(5.491,8.291);
\draw[gp path] (5.845,0.985)--(5.845,1.075);
\draw[gp path] (5.845,8.381)--(5.845,8.291);
\draw[gp path] (6.151,0.985)--(6.151,1.075);
\draw[gp path] (6.151,8.381)--(6.151,8.291);
\draw[gp path] (6.422,0.985)--(6.422,1.075);
\draw[gp path] (6.422,8.381)--(6.422,8.291);
\draw[gp path] (6.664,0.985)--(6.664,1.165);
\draw[gp path] (6.664,8.381)--(6.664,8.201);
\node[gp node center,font={\fontsize{12.0pt}{14.4pt}\selectfont}] at (6.664,0.677) {\plotNoInstructions{100}};
\draw[gp path] (8.254,0.985)--(8.254,1.075);
\draw[gp path] (8.254,8.381)--(8.254,8.291);
\draw[gp path] (9.184,0.985)--(9.184,1.075);
\draw[gp path] (9.184,8.381)--(9.184,8.291);
\draw[gp path] (9.844,0.985)--(9.844,1.075);
\draw[gp path] (9.844,8.381)--(9.844,8.291);
\draw[gp path] (10.357,0.985)--(10.357,1.075);
\draw[gp path] (10.357,8.381)--(10.357,8.291);
\draw[gp path] (10.775,0.985)--(10.775,1.075);
\draw[gp path] (10.775,8.381)--(10.775,8.291);
\draw[gp path] (11.129,0.985)--(11.129,1.075);
\draw[gp path] (11.129,8.381)--(11.129,8.291);
\draw[gp path] (11.435,0.985)--(11.435,1.075);
\draw[gp path] (11.435,8.381)--(11.435,8.291);
\draw[gp path] (11.705,0.985)--(11.705,1.075);
\draw[gp path] (11.705,8.381)--(11.705,8.291);
\draw[gp path] (11.947,0.985)--(11.947,1.165);
\draw[gp path] (11.947,8.381)--(11.947,8.201);
\node[gp node center,font={\fontsize{12.0pt}{14.4pt}\selectfont}] at (11.947,0.677) {\plotNoInstructions{1000}};
\draw[gp path] (1.380,8.381)--(1.380,0.985)--(11.947,0.985)--(11.947,8.381)--cycle;
\node[gp node center] at (6.663,0.215) {\plotAxisLabel{number of input instructions}};
\node[gp node right,font={\fontsize{12.0pt}{14.4pt}\selectfont}] at (10.257,1.390) {\plotShiftRight{\plotLegend{without improvements}}};
\gpcolor{rgb color={0.000,0.000,0.000}}
\gpsetpointsize{4.00}
\gppoint{gp mark 7}{(3.573,2.887)}
\gppoint{gp mark 7}{(5.812,3.675)}
\gppoint{gp mark 7}{(5.027,3.882)}
\gppoint{gp mark 7}{(4.931,2.446)}
\gppoint{gp mark 7}{(5.910,5.019)}
\gppoint{gp mark 7}{(4.780,3.239)}
\gppoint{gp mark 7}{(6.208,6.932)}
\gppoint{gp mark 7}{(5.675,4.645)}
\gppoint{gp mark 7}{(6.180,3.689)}
\gppoint{gp mark 7}{(7.230,6.512)}
\gppoint{gp mark 7}{(5.333,4.021)}
\gppoint{gp mark 7}{(4.979,2.302)}
\gppoint{gp mark 7}{(5.675,4.156)}
\gppoint{gp mark 7}{(8.100,5.969)}
\gppoint{gp mark 7}{(5.207,3.930)}
\gppoint{gp mark 7}{(5.941,7.950)}
\gppoint{gp mark 7}{(3.743,3.394)}
\gppoint{gp mark 7}{(3.743,3.019)}
\gppoint{gp mark 7}{(3.823,1.186)}
\gppoint{gp mark 7}{(8.847,3.469)}
\gppoint{gp mark 7}{(10.948,3.838)}
\gppoint{gp mark 7}{(4.618,1.572)}
\gppoint{gp mark 7}{(8.063,6.749)}
\gppoint{gp mark 7}{(5.639,4.939)}
\gppoint{gp mark 7}{(6.396,5.986)}
\gppoint{gp mark 7}{(4.561,3.197)}
\gppoint{gp mark 7}{(8.196,6.632)}
\gppoint{gp mark 7}{(5.414,3.573)}
\gppoint{gp mark 7}{(8.219,6.565)}
\gppoint{gp mark 7}{(5.414,3.545)}
\gppoint{gp mark 7}{(5.027,3.591)}
\gppoint{gp mark 7}{(3.823,5.076)}
\gppoint{gp mark 7}{(3.573,1.697)}
\gppoint{gp mark 7}{(4.503,3.157)}
\gppoint{gp mark 7}{(3.573,1.522)}
\gppoint{gp mark 7}{(3.389,2.946)}
\gppoint{gp mark 7}{(5.027,3.785)}
\gppoint{gp mark 7}{(8.196,7.699)}
\gppoint{gp mark 7}{(5.529,2.190)}
\gppoint{gp mark 7}{(6.317,2.026)}
\gppoint{gp mark 7}{(5.973,4.398)}
\gppoint{gp mark 7}{(4.120,3.239)}
\gppoint{gp mark 7}{(4.188,2.640)}
\gppoint{gp mark 7}{(5.163,6.562)}
\gppoint{gp mark 7}{(3.483,3.472)}
\gppoint{gp mark 7}{(5.779,4.984)}
\gppoint{gp mark 7}{(5.333,2.256)}
\gppoint{gp mark 7}{(8.865,6.574)}
\gppoint{gp mark 7}{(4.255,3.472)}
\gppoint{gp mark 7}{(6.151,4.783)}
\gppoint{gp mark 7}{(2.970,3.637)}
\gppoint{gp mark 7}{(5.779,7.391)}
\gppoint{gp mark 7}{(4.882,6.426)}
\gppoint{gp mark 7}{(6.208,6.234)}
\gppoint{gp mark 7}{(6.064,2.083)}
\gppoint{gp mark 7}{(5.639,5.930)}
\gppoint{gp mark 7}{(3.573,2.879)}
\gppoint{gp mark 7}{(5.812,2.488)}
\gppoint{gp mark 7}{(5.027,3.432)}
\gppoint{gp mark 7}{(4.931,4.362)}
\gppoint{gp mark 7}{(4.780,2.718)}
\gppoint{gp mark 7}{(5.910,3.857)}
\gppoint{gp mark 7}{(4.780,2.222)}
\gppoint{gp mark 7}{(6.208,2.007)}
\gppoint{gp mark 7}{(5.675,4.008)}
\gppoint{gp mark 7}{(6.180,1.624)}
\gppoint{gp mark 7}{(5.333,3.862)}
\gppoint{gp mark 7}{(4.979,1.703)}
\gppoint{gp mark 7}{(5.675,4.171)}
\gppoint{gp mark 7}{(8.100,6.501)}
\gppoint{gp mark 7}{(5.207,3.343)}
\gppoint{gp mark 7}{(5.941,7.065)}
\gppoint{gp mark 7}{(3.743,2.559)}
\gppoint{gp mark 7}{(3.743,2.461)}
\gppoint{gp mark 7}{(3.823,2.693)}
\gppoint{gp mark 7}{(5.812,2.870)}
\gppoint{gp mark 7}{(4.618,3.921)}
\gppoint{gp mark 7}{(7.742,3.143)}
\gppoint{gp mark 7}{(5.639,2.420)}
\gppoint{gp mark 7}{(6.396,5.254)}
\gppoint{gp mark 7}{(4.561,3.390)}
\gppoint{gp mark 7}{(8.196,7.625)}
\gppoint{gp mark 7}{(5.414,1.868)}
\gppoint{gp mark 7}{(5.414,3.539)}
\gppoint{gp mark 7}{(5.027,2.212)}
\gppoint{gp mark 7}{(3.823,6.781)}
\gppoint{gp mark 7}{(3.573,2.950)}
\gppoint{gp mark 7}{(4.503,3.113)}
\gppoint{gp mark 7}{(3.573,1.267)}
\gppoint{gp mark 7}{(3.389,1.172)}
\gppoint{gp mark 7}{(5.027,3.475)}
\gppoint{gp mark 7}{(5.529,1.590)}
\gppoint{gp mark 7}{(6.317,7.477)}
\gppoint{gp mark 7}{(5.973,4.119)}
\gppoint{gp mark 7}{(4.120,3.116)}
\gppoint{gp mark 7}{(4.188,1.169)}
\gppoint{gp mark 7}{(5.163,7.773)}
\gppoint{gp mark 7}{(3.483,3.790)}
\gppoint{gp mark 7}{(5.779,3.042)}
\gppoint{gp mark 7}{(5.333,4.242)}
\gppoint{gp mark 7}{(8.865,2.847)}
\gppoint{gp mark 7}{(4.255,3.490)}
\gppoint{gp mark 7}{(6.151,2.043)}
\gppoint{gp mark 7}{(2.970,5.461)}
\gppoint{gp mark 7}{(10.767,7.837)}
\gppoint{gp mark 7}{(4.882,6.261)}
\gppoint{gp mark 7}{(6.064,2.023)}
\gppoint{gp mark 7}{(3.573,1.424)}
\gppoint{gp mark 7}{(6.180,4.014)}
\gppoint{gp mark 7}{(5.027,3.647)}
\gppoint{gp mark 7}{(4.931,4.060)}
\gppoint{gp mark 7}{(6.003,3.991)}
\gppoint{gp mark 7}{(4.831,4.287)}
\gppoint{gp mark 7}{(6.447,5.328)}
\gppoint{gp mark 7}{(6.034,4.889)}
\gppoint{gp mark 7}{(6.180,3.819)}
\gppoint{gp mark 7}{(7.452,7.058)}
\gppoint{gp mark 7}{(5.333,3.987)}
\gppoint{gp mark 7}{(5.453,4.171)}
\gppoint{gp mark 7}{(5.710,4.326)}
\gppoint{gp mark 7}{(9.403,7.451)}
\gppoint{gp mark 7}{(8.483,5.916)}
\gppoint{gp mark 7}{(5.250,3.660)}
\gppoint{gp mark 7}{(5.941,7.735)}
\gppoint{gp mark 7}{(3.659,4.261)}
\gppoint{gp mark 7}{(3.743,3.369)}
\gppoint{gp mark 7}{(3.823,2.617)}
\gppoint{gp mark 7}{(7.101,6.536)}
\gppoint{gp mark 7}{(8.441,5.998)}
\gppoint{gp mark 7}{(5.812,3.979)}
\gppoint{gp mark 7}{(4.931,3.434)}
\gppoint{gp mark 7}{(5.414,3.606)}
\gppoint{gp mark 7}{(5.603,6.091)}
\gppoint{gp mark 7}{(5.292,4.298)}
\gppoint{gp mark 7}{(4.255,7.148)}
\gppoint{gp mark 7}{(4.049,3.130)}
\gppoint{gp mark 7}{(4.727,4.596)}
\gppoint{gp mark 7}{(3.976,3.074)}
\gppoint{gp mark 7}{(3.659,4.144)}
\gppoint{gp mark 7}{(5.878,7.048)}
\gppoint{gp mark 7}{(6.151,4.431)}
\gppoint{gp mark 7}{(3.291,3.594)}
\gppoint{gp mark 7}{(6.151,6.485)}
\gppoint{gp mark 7}{(4.120,1.359)}
\gppoint{gp mark 7}{(4.319,2.954)}
\gppoint{gp mark 7}{(4.882,4.358)}
\gppoint{gp mark 7}{(5.073,4.906)}
\gppoint{gp mark 7}{(3.483,2.030)}
\gppoint{gp mark 7}{(6.093,7.560)}
\gppoint{gp mark 7}{(5.639,4.266)}
\gppoint{gp mark 7}{(9.603,7.119)}
\gppoint{gp mark 7}{(4.255,3.539)}
\gppoint{gp mark 7}{(6.594,5.076)}
\gppoint{gp mark 7}{(2.970,4.682)}
\gppoint{gp mark 7}{(5.567,5.214)}
\gppoint{gp mark 7}{(4.882,4.070)}
\gppoint{gp mark 7}{(6.731,4.691)}
\gppoint{gp mark 7}{(5.779,6.542)}
\gppoint{gp mark 7}{(5.973,4.983)}
\gppoint{gp mark 7}{(3.573,2.552)}
\gppoint{gp mark 7}{(6.180,4.173)}
\gppoint{gp mark 7}{(5.027,3.629)}
\gppoint{gp mark 7}{(4.931,3.813)}
\gppoint{gp mark 7}{(6.003,4.230)}
\gppoint{gp mark 7}{(4.831,3.862)}
\gppoint{gp mark 7}{(6.447,5.755)}
\gppoint{gp mark 7}{(6.034,4.469)}
\gppoint{gp mark 7}{(6.180,3.908)}
\gppoint{gp mark 7}{(5.333,4.106)}
\gppoint{gp mark 7}{(5.453,4.304)}
\gppoint{gp mark 7}{(5.710,4.407)}
\gppoint{gp mark 7}{(8.483,6.101)}
\gppoint{gp mark 7}{(5.250,3.395)}
\gppoint{gp mark 7}{(3.659,2.522)}
\gppoint{gp mark 7}{(3.743,4.185)}
\gppoint{gp mark 7}{(3.823,2.606)}
\gppoint{gp mark 7}{(7.101,7.503)}
\gppoint{gp mark 7}{(4.882,5.592)}
\gppoint{gp mark 7}{(8.441,6.794)}
\gppoint{gp mark 7}{(5.812,4.365)}
\gppoint{gp mark 7}{(4.931,3.759)}
\gppoint{gp mark 7}{(5.414,1.808)}
\gppoint{gp mark 7}{(5.603,5.733)}
\gppoint{gp mark 7}{(5.292,4.033)}
\gppoint{gp mark 7}{(4.255,6.708)}
\gppoint{gp mark 7}{(4.049,3.096)}
\gppoint{gp mark 7}{(4.727,3.278)}
\gppoint{gp mark 7}{(3.976,4.221)}
\gppoint{gp mark 7}{(3.659,3.161)}
\gppoint{gp mark 7}{(5.333,5.565)}
\gppoint{gp mark 7}{(5.878,2.370)}
\gppoint{gp mark 7}{(3.291,6.773)}
\gppoint{gp mark 7}{(4.120,1.447)}
\gppoint{gp mark 7}{(4.319,1.308)}
\gppoint{gp mark 7}{(4.882,3.746)}
\gppoint{gp mark 7}{(5.073,3.630)}
\gppoint{gp mark 7}{(3.483,1.399)}
\gppoint{gp mark 7}{(5.639,2.160)}
\gppoint{gp mark 7}{(4.255,3.646)}
\gppoint{gp mark 7}{(6.594,3.605)}
\gppoint{gp mark 7}{(2.970,5.332)}
\gppoint{gp mark 7}{(5.567,4.743)}
\gppoint{gp mark 7}{(4.882,1.712)}
\gppoint{gp mark 7}{(6.570,7.359)}
\gppoint{gp mark 7}{(6.731,5.369)}
\gppoint{gp mark 7}{(4.120,3.549)}
\gppoint{gp mark 7}{(6.263,6.728)}
\gppoint{gp mark 7}{(6.686,4.972)}
\gppoint{gp mark 7}{(5.639,5.229)}
\gppoint{gp mark 7}{(4.979,4.531)}
\gppoint{gp mark 7}{(6.263,5.104)}
\gppoint{gp mark 7}{(4.618,5.051)}
\gppoint{gp mark 7}{(5.118,4.954)}
\gppoint{gp mark 7}{(5.941,4.703)}
\gppoint{gp mark 7}{(4.049,4.282)}
\gppoint{gp mark 7}{(4.780,8.033)}
\gppoint{gp mark 7}{(6.422,5.167)}
\gppoint{gp mark 7}{(6.640,7.636)}
\gppoint{gp mark 7}{(6.903,6.244)}
\gppoint{gp mark 7}{(4.979,6.739)}
\gppoint{gp mark 7}{(4.882,7.477)}
\gppoint{gp mark 7}{(4.831,3.743)}
\gppoint{gp mark 7}{(4.673,5.732)}
\gppoint{gp mark 7}{(5.250,3.936)}
\gppoint{gp mark 7}{(3.291,3.200)}
\gppoint{gp mark 7}{(6.236,5.168)}
\gppoint{gp mark 7}{(9.862,7.467)}
\gppoint{gp mark 7}{(4.882,4.182)}
\gppoint{gp mark 7}{(7.004,5.329)}
\gppoint{gp mark 7}{(6.594,5.371)}
\gppoint{gp mark 7}{(5.639,6.800)}
\gppoint{gp mark 7}{(6.617,7.416)}
\gppoint{gp mark 7}{(7.248,5.468)}
\gppoint{gp mark 7}{(6.344,6.237)}
\gppoint{gp mark 7}{(4.120,3.613)}
\gppoint{gp mark 7}{(6.291,7.578)}
\gppoint{gp mark 7}{(5.414,8.052)}
\gppoint{gp mark 7}{(6.686,4.740)}
\gppoint{gp mark 7}{(5.639,5.619)}
\gppoint{gp mark 7}{(4.979,2.123)}
\gppoint{gp mark 7}{(4.831,7.597)}
\gppoint{gp mark 7}{(4.618,4.434)}
\gppoint{gp mark 7}{(6.775,5.654)}
\gppoint{gp mark 7}{(5.118,6.757)}
\gppoint{gp mark 7}{(5.941,4.623)}
\gppoint{gp mark 7}{(5.603,7.220)}
\gppoint{gp mark 7}{(5.250,6.892)}
\gppoint{gp mark 7}{(4.049,4.477)}
\gppoint{gp mark 7}{(4.780,5.430)}
\gppoint{gp mark 7}{(4.049,4.944)}
\gppoint{gp mark 7}{(3.901,4.698)}
\gppoint{gp mark 7}{(6.422,5.514)}
\gppoint{gp mark 7}{(6.640,2.584)}
\gppoint{gp mark 7}{(6.903,6.804)}
\gppoint{gp mark 7}{(4.979,7.783)}
\gppoint{gp mark 7}{(4.882,4.964)}
\gppoint{gp mark 7}{(4.831,3.780)}
\gppoint{gp mark 7}{(4.673,5.927)}
\gppoint{gp mark 7}{(5.250,3.930)}
\gppoint{gp mark 7}{(3.291,3.231)}
\gppoint{gp mark 7}{(6.370,6.822)}
\gppoint{gp mark 7}{(6.236,2.491)}
\gppoint{gp mark 7}{(9.862,7.700)}
\gppoint{gp mark 7}{(4.882,4.177)}
\gppoint{gp mark 7}{(7.004,5.362)}
\gppoint{gp mark 7}{(6.594,5.216)}
\gppoint{gp mark 7}{(5.639,4.982)}
\gppoint{gp mark 7}{(7.248,3.815)}
\gppoint{gp mark 7}{(11.010,1.390)}
\gpcolor{color=gp lt color border}
\node[gp node right,font={\fontsize{12.0pt}{14.4pt}\selectfont}] at (10.257,1.840) {\plotShiftRight{\plotLegend{with improvements}}};
\gpcolor{rgb color={0.000,0.000,0.000}}
\gppoint{gp mark 6}{(3.573,2.324)}
\gppoint{gp mark 6}{(5.812,3.226)}
\gppoint{gp mark 6}{(5.027,3.179)}
\gppoint{gp mark 6}{(4.931,3.276)}
\gppoint{gp mark 6}{(4.780,3.633)}
\gppoint{gp mark 6}{(5.910,3.442)}
\gppoint{gp mark 6}{(4.780,3.144)}
\gppoint{gp mark 6}{(6.208,4.539)}
\gppoint{gp mark 6}{(5.675,3.818)}
\gppoint{gp mark 6}{(6.180,3.595)}
\gppoint{gp mark 6}{(7.230,5.108)}
\gppoint{gp mark 6}{(5.333,4.093)}
\gppoint{gp mark 6}{(4.979,3.529)}
\gppoint{gp mark 6}{(5.675,4.243)}
\gppoint{gp mark 6}{(9.368,6.299)}
\gppoint{gp mark 6}{(8.100,5.685)}
\gppoint{gp mark 6}{(5.207,2.902)}
\gppoint{gp mark 6}{(5.779,5.296)}
\gppoint{gp mark 6}{(5.941,4.748)}
\gppoint{gp mark 6}{(6.422,6.616)}
\gppoint{gp mark 6}{(7.369,6.797)}
\gppoint{gp mark 6}{(7.318,6.763)}
\gppoint{gp mark 6}{(3.743,2.884)}
\gppoint{gp mark 6}{(8.624,6.744)}
\gppoint{gp mark 6}{(8.729,6.802)}
\gppoint{gp mark 6}{(3.743,1.947)}
\gppoint{gp mark 6}{(3.823,1.220)}
\gppoint{gp mark 6}{(6.753,5.175)}
\gppoint{gp mark 6}{(8.847,3.751)}
\gppoint{gp mark 6}{(10.948,3.868)}
\gppoint{gp mark 6}{(10.160,7.897)}
\gppoint{gp mark 6}{(8.277,8.023)}
\gppoint{gp mark 6}{(5.812,4.020)}
\gppoint{gp mark 6}{(4.618,1.546)}
\gppoint{gp mark 6}{(8.063,5.962)}
\gppoint{gp mark 6}{(7.742,5.225)}
\gppoint{gp mark 6}{(5.639,3.572)}
\gppoint{gp mark 6}{(6.396,5.102)}
\gppoint{gp mark 6}{(4.561,3.339)}
\gppoint{gp mark 6}{(5.414,3.506)}
\gppoint{gp mark 6}{(8.219,7.939)}
\gppoint{gp mark 6}{(5.414,3.668)}
\gppoint{gp mark 6}{(5.027,4.616)}
\gppoint{gp mark 6}{(3.823,3.400)}
\gppoint{gp mark 6}{(3.573,2.616)}
\gppoint{gp mark 6}{(4.503,3.211)}
\gppoint{gp mark 6}{(3.573,2.731)}
\gppoint{gp mark 6}{(3.389,2.889)}
\gppoint{gp mark 6}{(5.027,3.758)}
\gppoint{gp mark 6}{(8.196,6.656)}
\gppoint{gp mark 6}{(10.824,7.606)}
\gppoint{gp mark 6}{(5.529,3.063)}
\gppoint{gp mark 6}{(8.524,7.369)}
\gppoint{gp mark 6}{(6.317,2.144)}
\gppoint{gp mark 6}{(5.973,4.744)}
\gppoint{gp mark 6}{(9.138,7.952)}
\gppoint{gp mark 6}{(8.701,6.999)}
\gppoint{gp mark 6}{(7.369,6.151)}
\gppoint{gp mark 6}{(4.120,2.933)}
\gppoint{gp mark 6}{(8.299,6.817)}
\gppoint{gp mark 6}{(4.188,2.439)}
\gppoint{gp mark 6}{(6.775,6.842)}
\gppoint{gp mark 6}{(5.163,3.846)}
\gppoint{gp mark 6}{(3.483,2.824)}
\gppoint{gp mark 6}{(5.779,4.086)}
\gppoint{gp mark 6}{(5.333,3.653)}
\gppoint{gp mark 6}{(8.865,7.153)}
\gppoint{gp mark 6}{(4.255,3.557)}
\gppoint{gp mark 6}{(7.283,6.600)}
\gppoint{gp mark 6}{(6.151,5.097)}
\gppoint{gp mark 6}{(8.985,7.843)}
\gppoint{gp mark 6}{(2.970,1.857)}
\gppoint{gp mark 6}{(10.145,8.217)}
\gppoint{gp mark 6}{(5.779,4.950)}
\gppoint{gp mark 6}{(10.767,8.042)}
\gppoint{gp mark 6}{(10.282,7.817)}
\gppoint{gp mark 6}{(4.882,3.713)}
\gppoint{gp mark 6}{(6.208,5.329)}
\gppoint{gp mark 6}{(6.064,2.414)}
\gppoint{gp mark 6}{(5.639,6.179)}
\gppoint{gp mark 6}{(8.653,7.219)}
\gppoint{gp mark 6}{(3.573,2.211)}
\gppoint{gp mark 6}{(5.812,2.655)}
\gppoint{gp mark 6}{(5.027,2.727)}
\gppoint{gp mark 6}{(4.931,3.018)}
\gppoint{gp mark 6}{(4.780,3.061)}
\gppoint{gp mark 6}{(5.910,3.245)}
\gppoint{gp mark 6}{(11.820,8.018)}
\gppoint{gp mark 6}{(4.780,2.399)}
\gppoint{gp mark 6}{(6.208,2.139)}
\gppoint{gp mark 6}{(5.675,3.274)}
\gppoint{gp mark 6}{(6.180,1.717)}
\gppoint{gp mark 6}{(7.230,4.594)}
\gppoint{gp mark 6}{(5.333,3.131)}
\gppoint{gp mark 6}{(4.979,1.775)}
\gppoint{gp mark 6}{(5.675,3.634)}
\gppoint{gp mark 6}{(9.368,6.438)}
\gppoint{gp mark 6}{(8.100,5.851)}
\gppoint{gp mark 6}{(5.207,2.843)}
\gppoint{gp mark 6}{(5.779,5.252)}
\gppoint{gp mark 6}{(5.941,4.691)}
\gppoint{gp mark 6}{(6.422,6.592)}
\gppoint{gp mark 6}{(7.369,6.770)}
\gppoint{gp mark 6}{(7.318,6.754)}
\gppoint{gp mark 6}{(3.743,2.884)}
\gppoint{gp mark 6}{(8.624,6.697)}
\gppoint{gp mark 6}{(8.729,6.798)}
\gppoint{gp mark 6}{(3.743,1.183)}
\gppoint{gp mark 6}{(3.823,2.616)}
\gppoint{gp mark 6}{(6.753,4.703)}
\gppoint{gp mark 6}{(8.847,6.347)}
\gppoint{gp mark 6}{(10.948,7.428)}
\gppoint{gp mark 6}{(5.812,2.997)}
\gppoint{gp mark 6}{(4.618,3.157)}
\gppoint{gp mark 6}{(8.063,5.555)}
\gppoint{gp mark 6}{(7.742,4.903)}
\gppoint{gp mark 6}{(5.639,2.762)}
\gppoint{gp mark 6}{(6.396,4.672)}
\gppoint{gp mark 6}{(4.561,2.601)}
\gppoint{gp mark 6}{(6.686,4.224)}
\gppoint{gp mark 6}{(8.196,6.321)}
\gppoint{gp mark 6}{(5.414,1.707)}
\gppoint{gp mark 6}{(8.219,6.626)}
\gppoint{gp mark 6}{(5.414,2.984)}
\gppoint{gp mark 6}{(5.027,1.834)}
\gppoint{gp mark 6}{(3.823,3.390)}
\gppoint{gp mark 6}{(3.573,2.125)}
\gppoint{gp mark 6}{(4.503,2.518)}
\gppoint{gp mark 6}{(3.573,1.489)}
\gppoint{gp mark 6}{(3.389,1.317)}
\gppoint{gp mark 6}{(5.027,3.242)}
\gppoint{gp mark 6}{(8.196,5.924)}
\gppoint{gp mark 6}{(5.529,1.831)}
\gppoint{gp mark 6}{(6.317,5.335)}
\gppoint{gp mark 6}{(5.973,4.000)}
\gppoint{gp mark 6}{(4.120,2.494)}
\gppoint{gp mark 6}{(8.299,6.594)}
\gppoint{gp mark 6}{(4.188,1.237)}
\gppoint{gp mark 6}{(6.775,8.116)}
\gppoint{gp mark 6}{(5.163,4.267)}
\gppoint{gp mark 6}{(3.483,2.882)}
\gppoint{gp mark 6}{(5.779,3.487)}
\gppoint{gp mark 6}{(5.333,4.007)}
\gppoint{gp mark 6}{(8.865,2.955)}
\gppoint{gp mark 6}{(4.255,3.489)}
\gppoint{gp mark 6}{(6.151,2.187)}
\gppoint{gp mark 6}{(8.985,7.499)}
\gppoint{gp mark 6}{(2.970,2.267)}
\gppoint{gp mark 6}{(7.120,5.834)}
\gppoint{gp mark 6}{(5.779,4.878)}
\gppoint{gp mark 6}{(10.767,7.018)}
\gppoint{gp mark 6}{(9.951,6.896)}
\gppoint{gp mark 6}{(4.882,3.675)}
\gppoint{gp mark 6}{(6.208,4.718)}
\gppoint{gp mark 6}{(6.064,2.355)}
\gppoint{gp mark 6}{(5.639,6.894)}
\gppoint{gp mark 6}{(9.010,8.249)}
\gppoint{gp mark 6}{(8.653,6.502)}
\gppoint{gp mark 6}{(3.573,1.501)}
\gppoint{gp mark 6}{(6.180,3.966)}
\gppoint{gp mark 6}{(5.027,3.604)}
\gppoint{gp mark 6}{(4.931,3.833)}
\gppoint{gp mark 6}{(4.780,4.780)}
\gppoint{gp mark 6}{(6.003,3.470)}
\gppoint{gp mark 6}{(4.831,4.175)}
\gppoint{gp mark 6}{(6.447,5.339)}
\gppoint{gp mark 6}{(6.034,3.836)}
\gppoint{gp mark 6}{(6.180,3.948)}
\gppoint{gp mark 6}{(5.333,4.001)}
\gppoint{gp mark 6}{(5.453,4.316)}
\gppoint{gp mark 6}{(5.710,3.916)}
\gppoint{gp mark 6}{(9.403,7.503)}
\gppoint{gp mark 6}{(8.483,5.983)}
\gppoint{gp mark 6}{(5.250,3.416)}
\gppoint{gp mark 6}{(5.745,5.986)}
\gppoint{gp mark 6}{(5.941,5.446)}
\gppoint{gp mark 6}{(6.396,7.166)}
\gppoint{gp mark 6}{(7.403,7.230)}
\gppoint{gp mark 6}{(7.266,7.414)}
\gppoint{gp mark 6}{(3.659,2.007)}
\gppoint{gp mark 6}{(8.691,7.327)}
\gppoint{gp mark 6}{(3.743,2.079)}
\gppoint{gp mark 6}{(3.823,1.708)}
\gppoint{gp mark 6}{(7.101,5.788)}
\gppoint{gp mark 6}{(8.441,6.294)}
\gppoint{gp mark 6}{(5.812,4.135)}
\gppoint{gp mark 6}{(4.931,3.526)}
\gppoint{gp mark 6}{(5.414,3.309)}
\gppoint{gp mark 6}{(5.603,4.701)}
\gppoint{gp mark 6}{(4.255,4.236)}
\gppoint{gp mark 6}{(4.049,2.788)}
\gppoint{gp mark 6}{(4.727,3.944)}
\gppoint{gp mark 6}{(3.976,3.806)}
\gppoint{gp mark 6}{(3.659,2.689)}
\gppoint{gp mark 6}{(7.974,7.945)}
\gppoint{gp mark 6}{(5.878,3.415)}
\gppoint{gp mark 6}{(6.944,6.432)}
\gppoint{gp mark 6}{(6.151,5.566)}
\gppoint{gp mark 6}{(3.291,2.831)}
\gppoint{gp mark 6}{(6.151,5.031)}
\gppoint{gp mark 6}{(4.120,1.517)}
\gppoint{gp mark 6}{(5.118,4.851)}
\gppoint{gp mark 6}{(4.319,2.634)}
\gppoint{gp mark 6}{(4.882,4.679)}
\gppoint{gp mark 6}{(5.073,3.232)}
\gppoint{gp mark 6}{(3.483,2.048)}
\gppoint{gp mark 6}{(6.093,5.515)}
\gppoint{gp mark 6}{(5.639,4.728)}
\gppoint{gp mark 6}{(4.255,3.747)}
\gppoint{gp mark 6}{(6.882,5.866)}
\gppoint{gp mark 6}{(6.594,5.986)}
\gppoint{gp mark 6}{(2.970,2.166)}
\gppoint{gp mark 6}{(5.567,4.667)}
\gppoint{gp mark 6}{(6.003,5.406)}
\gppoint{gp mark 6}{(4.882,3.409)}
\gppoint{gp mark 6}{(6.570,7.144)}
\gppoint{gp mark 6}{(5.779,6.419)}
\gppoint{gp mark 6}{(5.973,6.448)}
\gppoint{gp mark 6}{(3.573,2.534)}
\gppoint{gp mark 6}{(6.180,4.085)}
\gppoint{gp mark 6}{(5.027,3.435)}
\gppoint{gp mark 6}{(4.931,3.644)}
\gppoint{gp mark 6}{(4.780,3.991)}
\gppoint{gp mark 6}{(6.003,4.153)}
\gppoint{gp mark 6}{(4.831,3.984)}
\gppoint{gp mark 6}{(6.447,5.711)}
\gppoint{gp mark 6}{(6.034,3.814)}
\gppoint{gp mark 6}{(6.180,3.693)}
\gppoint{gp mark 6}{(7.452,6.239)}
\gppoint{gp mark 6}{(5.333,4.853)}
\gppoint{gp mark 6}{(5.453,4.695)}
\gppoint{gp mark 6}{(5.710,4.160)}
\gppoint{gp mark 6}{(8.483,6.114)}
\gppoint{gp mark 6}{(5.250,3.137)}
\gppoint{gp mark 6}{(5.745,6.001)}
\gppoint{gp mark 6}{(5.941,5.462)}
\gppoint{gp mark 6}{(6.396,7.152)}
\gppoint{gp mark 6}{(7.403,7.081)}
\gppoint{gp mark 6}{(7.266,7.767)}
\gppoint{gp mark 6}{(3.659,2.012)}
\gppoint{gp mark 6}{(3.743,2.394)}
\gppoint{gp mark 6}{(3.823,1.708)}
\gppoint{gp mark 6}{(7.101,6.773)}
\gppoint{gp mark 6}{(4.882,4.238)}
\gppoint{gp mark 6}{(8.441,6.675)}
\gppoint{gp mark 6}{(5.812,4.574)}
\gppoint{gp mark 6}{(6.640,6.204)}
\gppoint{gp mark 6}{(4.931,3.614)}
\gppoint{gp mark 6}{(5.414,2.010)}
\gppoint{gp mark 6}{(5.603,4.751)}
\gppoint{gp mark 6}{(5.292,5.782)}
\gppoint{gp mark 6}{(4.255,3.597)}
\gppoint{gp mark 6}{(4.049,2.840)}
\gppoint{gp mark 6}{(4.727,2.643)}
\gppoint{gp mark 6}{(3.976,3.580)}
\gppoint{gp mark 6}{(3.659,2.730)}
\gppoint{gp mark 6}{(5.333,5.820)}
\gppoint{gp mark 6}{(5.878,3.612)}
\gppoint{gp mark 6}{(6.944,6.841)}
\gppoint{gp mark 6}{(6.151,5.278)}
\gppoint{gp mark 6}{(3.291,3.539)}
\gppoint{gp mark 6}{(6.151,7.369)}
\gppoint{gp mark 6}{(4.120,1.583)}
\gppoint{gp mark 6}{(5.118,6.838)}
\gppoint{gp mark 6}{(4.319,1.624)}
\gppoint{gp mark 6}{(4.882,4.465)}
\gppoint{gp mark 6}{(5.073,3.286)}
\gppoint{gp mark 6}{(3.483,1.612)}
\gppoint{gp mark 6}{(6.093,7.393)}
\gppoint{gp mark 6}{(5.639,4.594)}
\gppoint{gp mark 6}{(4.255,3.445)}
\gppoint{gp mark 6}{(6.882,6.404)}
\gppoint{gp mark 6}{(6.317,6.347)}
\gppoint{gp mark 6}{(6.594,3.865)}
\gppoint{gp mark 6}{(2.970,1.855)}
\gppoint{gp mark 6}{(5.567,5.112)}
\gppoint{gp mark 6}{(6.003,5.289)}
\gppoint{gp mark 6}{(4.882,1.912)}
\gppoint{gp mark 6}{(6.570,6.205)}
\gppoint{gp mark 6}{(6.731,6.458)}
\gppoint{gp mark 6}{(5.779,7.118)}
\gppoint{gp mark 6}{(5.973,6.541)}
\gppoint{gp mark 6}{(4.120,3.864)}
\gppoint{gp mark 6}{(6.291,5.245)}
\gppoint{gp mark 6}{(5.779,6.513)}
\gppoint{gp mark 6}{(5.207,5.105)}
\gppoint{gp mark 6}{(6.263,5.857)}
\gppoint{gp mark 6}{(5.845,5.917)}
\gppoint{gp mark 6}{(6.686,5.839)}
\gppoint{gp mark 6}{(5.639,6.206)}
\gppoint{gp mark 6}{(4.979,4.582)}
\gppoint{gp mark 6}{(4.831,4.981)}
\gppoint{gp mark 6}{(4.618,4.052)}
\gppoint{gp mark 6}{(8.311,8.048)}
\gppoint{gp mark 6}{(6.775,6.226)}
\gppoint{gp mark 6}{(5.118,6.697)}
\gppoint{gp mark 6}{(5.941,5.377)}
\gppoint{gp mark 6}{(4.049,4.485)}
\gppoint{gp mark 6}{(4.049,3.804)}
\gppoint{gp mark 6}{(3.901,3.636)}
\gppoint{gp mark 6}{(7.785,7.686)}
\gppoint{gp mark 6}{(6.640,6.228)}
\gppoint{gp mark 6}{(4.049,7.161)}
\gppoint{gp mark 6}{(4.979,5.997)}
\gppoint{gp mark 6}{(4.882,5.189)}
\gppoint{gp mark 6}{(4.831,4.401)}
\gppoint{gp mark 6}{(5.250,4.726)}
\gppoint{gp mark 6}{(3.291,2.854)}
\gppoint{gp mark 6}{(6.236,5.510)}
\gppoint{gp mark 6}{(4.882,6.283)}
\gppoint{gp mark 6}{(7.004,6.270)}
\gppoint{gp mark 6}{(6.594,6.195)}
\gppoint{gp mark 6}{(5.639,5.522)}
\gppoint{gp mark 6}{(4.673,4.601)}
\gppoint{gp mark 6}{(6.617,6.474)}
\gppoint{gp mark 6}{(7.248,6.060)}
\gppoint{gp mark 6}{(4.120,3.487)}
\gppoint{gp mark 6}{(6.291,5.207)}
\gppoint{gp mark 6}{(5.779,5.265)}
\gppoint{gp mark 6}{(5.207,5.005)}
\gppoint{gp mark 6}{(5.453,6.951)}
\gppoint{gp mark 6}{(6.263,5.223)}
\gppoint{gp mark 6}{(5.414,6.000)}
\gppoint{gp mark 6}{(6.344,6.701)}
\gppoint{gp mark 6}{(5.845,5.207)}
\gppoint{gp mark 6}{(6.686,5.614)}
\gppoint{gp mark 6}{(5.639,6.516)}
\gppoint{gp mark 6}{(4.979,2.187)}
\gppoint{gp mark 6}{(6.093,5.575)}
\gppoint{gp mark 6}{(6.522,6.728)}
\gppoint{gp mark 6}{(6.731,6.613)}
\gppoint{gp mark 6}{(7.230,7.033)}
\gppoint{gp mark 6}{(8.087,7.312)}
\gppoint{gp mark 6}{(7.895,8.174)}
\gppoint{gp mark 6}{(4.561,5.195)}
\gppoint{gp mark 6}{(4.673,4.664)}
\gppoint{gp mark 6}{(4.831,4.077)}
\gppoint{gp mark 6}{(4.618,2.214)}
\gppoint{gp mark 6}{(6.236,7.109)}
\gppoint{gp mark 6}{(6.775,5.852)}
\gppoint{gp mark 6}{(5.118,5.046)}
\gppoint{gp mark 6}{(5.941,5.170)}
\gppoint{gp mark 6}{(5.603,6.593)}
\gppoint{gp mark 6}{(5.250,7.070)}
\gppoint{gp mark 6}{(4.503,5.944)}
\gppoint{gp mark 6}{(4.049,3.989)}
\gppoint{gp mark 6}{(4.780,5.130)}
\gppoint{gp mark 6}{(4.049,3.731)}
\gppoint{gp mark 6}{(3.901,3.484)}
\gppoint{gp mark 6}{(7.785,7.104)}
\gppoint{gp mark 6}{(6.422,5.393)}
\gppoint{gp mark 6}{(6.640,2.767)}
\gppoint{gp mark 6}{(4.049,5.207)}
\gppoint{gp mark 6}{(4.979,4.294)}
\gppoint{gp mark 6}{(4.882,4.905)}
\gppoint{gp mark 6}{(4.831,4.254)}
\gppoint{gp mark 6}{(4.673,5.863)}
\gppoint{gp mark 6}{(5.250,4.562)}
\gppoint{gp mark 6}{(3.291,2.656)}
\gppoint{gp mark 6}{(6.236,2.639)}
\gppoint{gp mark 6}{(4.882,5.123)}
\gppoint{gp mark 6}{(7.176,6.828)}
\gppoint{gp mark 6}{(7.004,6.008)}
\gppoint{gp mark 6}{(6.594,6.253)}
\gppoint{gp mark 6}{(3.901,4.859)}
\gppoint{gp mark 6}{(5.639,5.459)}
\gppoint{gp mark 6}{(6.546,7.290)}
\gppoint{gp mark 6}{(4.673,5.565)}
\gppoint{gp mark 6}{(6.617,6.299)}
\gppoint{gp mark 6}{(7.248,4.003)}
\gppoint{gp mark 6}{(11.010,1.840)}
\gpcolor{color=gp lt color border}
\draw[gp path] (1.380,8.381)--(1.380,0.985)--(11.947,0.985)--(11.947,8.381)--cycle;
\gpdefrectangularnode{gp plot 1}{\pgfpoint{1.380cm}{0.985cm}}{\pgfpoint{11.947cm}{8.381cm}}
\end{tikzpicture}

%% file: results/base-speedup.tex
\begin{tikzpicture}[gnuplot]
\path (0.000,0.000) rectangle (12.500,8.750);
\gpcolor{color=gp lt color border}
\gpsetlinetype{gp lt border}
\gpsetdashtype{gp dt solid}
\gpsetlinewidth{1.00}
\draw[gp path] (1.196,5.460)--(1.376,5.460);
\draw[gp path] (24.446,5.460)--(24.266,5.460);
\node[gp node right,font={\fontsize{8pt}{9.6pt}\selectfont}] at (1.012,5.460) {\plotPercentage{-20}};
\draw[gp path] (1.196,6.102)--(1.376,6.102);
\draw[gp path] (24.446,6.102)--(24.266,6.102);
\node[gp node right,font={\fontsize{8pt}{9.6pt}\selectfont}] at (1.012,6.102) {\plotPercentage{-10}};
\draw[gp path] (1.196,6.744)--(1.376,6.744);
\draw[gp path] (24.446,6.744)--(24.266,6.744);
\node[gp node right,font={\fontsize{8pt}{9.6pt}\selectfont}] at (1.012,6.744) {\plotPercentage{0}};
\draw[gp path] (1.196,7.386)--(1.376,7.386);
\draw[gp path] (24.446,7.386)--(24.266,7.386);
\node[gp node right,font={\fontsize{8pt}{9.6pt}\selectfont}] at (1.012,7.386) {\plotPercentage{10}};
\draw[gp path] (1.196,8.028)--(1.376,8.028);
\draw[gp path] (24.446,8.028)--(24.266,8.028);
\node[gp node right,font={\fontsize{8pt}{9.6pt}\selectfont}] at (1.012,8.028) {\plotPercentage{20}};
\draw[gp path] (1.196,8.671)--(1.376,8.671);
\draw[gp path] (24.446,8.671)--(24.266,8.671);
\node[gp node right,font={\fontsize{8pt}{9.6pt}\selectfont}] at (1.012,8.671) {\plotPercentage{30}};
\draw[gp path] (1.196,9.313)--(1.376,9.313);
\draw[gp path] (24.446,9.313)--(24.266,9.313);
\node[gp node right,font={\fontsize{8pt}{9.6pt}\selectfont}] at (1.012,9.313) {\plotPercentage{40}};
\draw[gp path] (1.196,9.955)--(1.376,9.955);
\draw[gp path] (24.446,9.955)--(24.266,9.955);
\node[gp node right,font={\fontsize{8pt}{9.6pt}\selectfont}] at (1.012,9.955) {\plotPercentage{50}};
\draw[gp path] (1.196,10.597)--(1.376,10.597);
\draw[gp path] (24.446,10.597)--(24.266,10.597);
\node[gp node right,font={\fontsize{8pt}{9.6pt}\selectfont}] at (1.012,10.597) {\plotPercentage{60}};
\draw[gp path] (1.196,11.239)--(1.376,11.239);
\draw[gp path] (24.446,11.239)--(24.266,11.239);
\node[gp node right,font={\fontsize{8pt}{9.6pt}\selectfont}] at (1.012,11.239) {\plotPercentage{70}};
\draw[gp path] (1.196,11.881)--(1.376,11.881);
\draw[gp path] (24.446,11.881)--(24.266,11.881);
\node[gp node right,font={\fontsize{8pt}{9.6pt}\selectfont}] at (1.012,11.881) {\plotPercentage{80}};
\node[gp node left,rotate=-90,font={\fontsize{7pt}{8.4pt}\selectfont}] at (1.436,5.460) {\functionIdActual{101}\functionName{adpcm_coder}};
\node[gp node left,rotate=-90,font={\fontsize{7pt}{8.4pt}\selectfont}] at (1.675,5.460) {\functionIdActual{102}\functionName{main}};
\node[gp node left,rotate=-90,font={\fontsize{7pt}{8.4pt}\selectfont}] at (2.394,5.460) {\functionIdActual{103}\functionName{adpcm_decoder}};
\node[gp node left,rotate=-90,font={\fontsize{7pt}{8.4pt}\selectfont}] at (2.634,5.460) {\functionIdActual{104}\functionName{main}};
\node[gp node left,rotate=-90,font={\fontsize{7pt}{8.4pt}\selectfont}] at (3.114,5.460) {\functionIdActual{105}\functionName{quantize_image}};
\node[gp node left,rotate=-90,font={\fontsize{7pt}{8.4pt}\selectfont}] at (3.353,5.460) {\functionIdActual{106}\functionName{run_length_encode_zeros}};
\node[gp node left,rotate=-90,font={\fontsize{7pt}{8.4pt}\selectfont}] at (3.593,5.460) {\functionIdActual{107}\functionName{encode_stream}};
\node[gp node left,rotate=-90,font={\fontsize{7pt}{8.4pt}\selectfont}] at (3.833,5.460) {\functionIdActual{108}\functionName{ReadMatrixFromPGMStream}};
\node[gp node left,rotate=-90,font={\fontsize{7pt}{8.4pt}\selectfont}] at (4.072,5.460) {\functionIdActual{109}\functionName{main}};
\node[gp node left,rotate=-90,font={\fontsize{7pt}{8.4pt}\selectfont}] at (4.552,5.460) {\functionIdActual{110}\functionName{main}};
\node[gp node left,rotate=-90,font={\fontsize{7pt}{8.4pt}\selectfont}] at (4.791,5.460) {\functionIdActual{111}\functionName{unquantize_image}};
\node[gp node left,rotate=-90,font={\fontsize{7pt}{8.4pt}\selectfont}] at (5.031,5.460) {\functionIdActual{112}\functionName{read_and_huffman_decode}};
\node[gp node left,rotate=-90,font={\fontsize{7pt}{8.4pt}\selectfont}] at (5.271,5.460) {\functionIdActual{113}\functionName{write_pgm_image}};
\node[gp node left,rotate=-90,font={\fontsize{7pt}{8.4pt}\selectfont}] at (5.510,5.460) {\functionIdActual{114}\functionName{internal_int_transpose}};
\node[gp node left,rotate=-90,font={\fontsize{7pt}{8.4pt}\selectfont}] at (5.990,5.460) {\functionIdActual{115}\functionName{quan}};
\node[gp node left,rotate=-90,font={\fontsize{7pt}{8.4pt}\selectfont}] at (6.230,5.460) {\functionIdActual{116}\functionName{fmult}};
\node[gp node left,rotate=-90,font={\fontsize{7pt}{8.4pt}\selectfont}] at (6.469,5.460) {\functionIdActual{117}\functionName{update}};
\node[gp node left,rotate=-90,font={\fontsize{7pt}{8.4pt}\selectfont}] at (6.709,5.460) {\functionIdActual{118}\functionName{g721_encoder}};
\node[gp node left,rotate=-90,font={\fontsize{7pt}{8.4pt}\selectfont}] at (6.949,5.460) {\functionIdActual{119}\functionName{predictor_zero}};
\node[gp node left,rotate=-90,font={\fontsize{7pt}{8.4pt}\selectfont}] at (7.428,5.460) {\functionIdActual{120}\functionName{quan}};
\node[gp node left,rotate=-90,font={\fontsize{7pt}{8.4pt}\selectfont}] at (7.668,5.460) {\functionIdActual{121}\functionName{fmult}};
\node[gp node left,rotate=-90,font={\fontsize{7pt}{8.4pt}\selectfont}] at (7.907,5.460) {\functionIdActual{122}\functionName{update}};
\node[gp node left,rotate=-90,font={\fontsize{7pt}{8.4pt}\selectfont}] at (8.147,5.460) {\functionIdActual{123}\functionName{g721_decoder}};
\node[gp node left,rotate=-90,font={\fontsize{7pt}{8.4pt}\selectfont}] at (8.387,5.460) {\functionIdActual{124}\functionName{predictor_zero}};
\node[gp node left,rotate=-90,font={\fontsize{7pt}{8.4pt}\selectfont}] at (8.866,5.460) {\functionIdActual{125}\functionName{Calculation_of_the_LTP_pa.}};
\node[gp node left,rotate=-90,font={\fontsize{7pt}{8.4pt}\selectfont}] at (9.106,5.460) {\functionIdActual{126}\functionName{Short_term_analysis_filte.}};
\node[gp node left,rotate=-90,font={\fontsize{7pt}{8.4pt}\selectfont}] at (9.345,5.460) {\functionIdActual{127}\functionName{Gsm_Preprocess}};
\node[gp node left,rotate=-90,font={\fontsize{7pt}{8.4pt}\selectfont}] at (9.585,5.460) {\functionIdActual{128}\functionName{Weighting_filter}};
\node[gp node left,rotate=-90,font={\fontsize{7pt}{8.4pt}\selectfont}] at (9.825,5.460) {\functionIdActual{129}\functionName{Autocorrelation}};
\node[gp node left,rotate=-90,font={\fontsize{7pt}{8.4pt}\selectfont}] at (10.304,5.460) {\functionIdActual{130}\functionName{Short_term_synthesis_filt.}};
\node[gp node left,rotate=-90,font={\fontsize{7pt}{8.4pt}\selectfont}] at (10.544,5.460) {\functionIdActual{131}\functionName{Gsm_Long_Term_Synthesis_F.}};
\node[gp node left,rotate=-90,font={\fontsize{7pt}{8.4pt}\selectfont}] at (10.784,5.460) {\functionIdActual{132}\functionName{Postprocessing}};
\node[gp node left,rotate=-90,font={\fontsize{7pt}{8.4pt}\selectfont}] at (11.023,5.460) {\functionIdActual{133}\functionName{APCM_inverse_quantization}};
\node[gp node left,rotate=-90,font={\fontsize{7pt}{8.4pt}\selectfont}] at (11.263,5.460) {\functionIdActual{134}\functionName{gsm_asr}};
\node[gp node left,rotate=-90,font={\fontsize{7pt}{8.4pt}\selectfont}] at (11.742,5.460) {\functionIdActual{135}\functionName{forward_DCT}};
\node[gp node left,rotate=-90,font={\fontsize{7pt}{8.4pt}\selectfont}] at (11.982,5.460) {\functionIdActual{136}\functionName{rgb_ycc_convert}};
\node[gp node left,rotate=-90,font={\fontsize{7pt}{8.4pt}\selectfont}] at (12.222,5.460) {\functionIdActual{137}\functionName{encode_one_block}};
\node[gp node left,rotate=-90,font={\fontsize{7pt}{8.4pt}\selectfont}] at (12.461,5.460) {\functionIdActual{138}\functionName{jpeg_fdct_islow}};
\node[gp node left,rotate=-90,font={\fontsize{7pt}{8.4pt}\selectfont}] at (12.701,5.460) {\functionIdActual{139}\functionName{emit_bits}};
\node[gp node left,rotate=-90,font={\fontsize{7pt}{8.4pt}\selectfont}] at (13.181,5.460) {\functionIdActual{140}\functionName{ycc_rgb_convert}};
\node[gp node left,rotate=-90,font={\fontsize{7pt}{8.4pt}\selectfont}] at (13.420,5.460) {\functionIdActual{141}\functionName{jpeg_idct_islow}};
\node[gp node left,rotate=-90,font={\fontsize{7pt}{8.4pt}\selectfont}] at (13.660,5.460) {\functionIdActual{142}\functionName{h2v2_fancy_upsample}};
\node[gp node left,rotate=-90,font={\fontsize{7pt}{8.4pt}\selectfont}] at (13.900,5.460) {\functionIdActual{143}\functionName{decode_mcu}};
\node[gp node left,rotate=-90,font={\fontsize{7pt}{8.4pt}\selectfont}] at (14.139,5.460) {\functionIdActual{144}\functionName{jpeg_fill_bit_buffer}};
\node[gp node left,rotate=-90,font={\fontsize{7pt}{8.4pt}\selectfont}] at (14.619,5.460) {\functionIdActual{145}\functionName{fdct}};
\node[gp node left,rotate=-90,font={\fontsize{7pt}{8.4pt}\selectfont}] at (14.858,5.460) {\functionIdActual{146}\functionName{fullsearch}};
\node[gp node left,rotate=-90,font={\fontsize{7pt}{8.4pt}\selectfont}] at (15.098,5.460) {\functionIdActual{147}\functionName{dist1}};
\node[gp node left,rotate=-90,font={\fontsize{7pt}{8.4pt}\selectfont}] at (15.338,5.460) {\functionIdActual{148}\functionName{putbits}};
\node[gp node left,rotate=-90,font={\fontsize{7pt}{8.4pt}\selectfont}] at (15.577,5.460) {\functionIdActual{149}\functionName{calcSNR1}};
\node[gp node left,rotate=-90,font={\fontsize{7pt}{8.4pt}\selectfont}] at (16.057,5.460) {\functionIdActual{150}\functionName{conv420to422}};
\node[gp node left,rotate=-90,font={\fontsize{7pt}{8.4pt}\selectfont}] at (16.297,5.460) {\functionIdActual{151}\functionName{form_component_prediction}};
\node[gp node left,rotate=-90,font={\fontsize{7pt}{8.4pt}\selectfont}] at (16.536,5.460) {\functionIdActual{152}\functionName{putbyte}};
\node[gp node left,rotate=-90,font={\fontsize{7pt}{8.4pt}\selectfont}] at (16.776,5.460) {\functionIdActual{153}\functionName{Add_Block}};
\node[gp node left,rotate=-90,font={\fontsize{7pt}{8.4pt}\selectfont}] at (17.016,5.460) {\functionIdActual{154}\functionName{idctcol}};
\node[gp node left,rotate=-90,font={\fontsize{7pt}{8.4pt}\selectfont}] at (17.495,5.460) {\functionIdActual{155}\functionName{gfAddMul}};
\node[gp node left,rotate=-90,font={\fontsize{7pt}{8.4pt}\selectfont}] at (17.735,5.460) {\functionIdActual{156}\functionName{gfMultiply}};
\node[gp node left,rotate=-90,font={\fontsize{7pt}{8.4pt}\selectfont}] at (17.974,5.460) {\functionIdActual{157}\functionName{squareEncrypt}};
\node[gp node left,rotate=-90,font={\fontsize{7pt}{8.4pt}\selectfont}] at (18.214,5.460) {\functionIdActual{158}\functionName{gfInvert}};
\node[gp node left,rotate=-90,font={\fontsize{7pt}{8.4pt}\selectfont}] at (18.454,5.460) {\functionIdActual{159}\functionName{gfSquare}};
\node[gp node left,rotate=-90,font={\fontsize{7pt}{8.4pt}\selectfont}] at (18.933,5.460) {\functionIdActual{160}\functionName{gfAddMul}};
\node[gp node left,rotate=-90,font={\fontsize{7pt}{8.4pt}\selectfont}] at (19.173,5.460) {\functionIdActual{161}\functionName{gfMultiply}};
\node[gp node left,rotate=-90,font={\fontsize{7pt}{8.4pt}\selectfont}] at (19.412,5.460) {\functionIdActual{162}\functionName{squareDecrypt}};
\node[gp node left,rotate=-90,font={\fontsize{7pt}{8.4pt}\selectfont}] at (19.652,5.460) {\functionIdActual{163}\functionName{gfInit}};
\node[gp node left,rotate=-90,font={\fontsize{7pt}{8.4pt}\selectfont}] at (19.892,5.460) {\functionIdActual{164}\functionName{gfInvert}};
\node[gp node left,rotate=-90,font={\fontsize{7pt}{8.4pt}\selectfont}] at (20.371,5.460) {\functionIdActual{165}\functionName{mp_smul}};
\node[gp node left,rotate=-90,font={\fontsize{7pt}{8.4pt}\selectfont}] at (20.611,5.460) {\functionIdActual{166}\functionName{longest_match}};
\node[gp node left,rotate=-90,font={\fontsize{7pt}{8.4pt}\selectfont}] at (20.851,5.460) {\functionIdActual{167}\functionName{fill_window}};
\node[gp node left,rotate=-90,font={\fontsize{7pt}{8.4pt}\selectfont}] at (21.090,5.460) {\functionIdActual{168}\functionName{deflate}};
\node[gp node left,rotate=-90,font={\fontsize{7pt}{8.4pt}\selectfont}] at (21.330,5.460) {\functionIdActual{169}\functionName{mp_compare}};
\node[gp node left,rotate=-90,font={\fontsize{7pt}{8.4pt}\selectfont}] at (21.809,5.460) {\functionIdActual{170}\functionName{mp_smul}};
\node[gp node left,rotate=-90,font={\fontsize{7pt}{8.4pt}\selectfont}] at (22.049,5.460) {\functionIdActual{171}\functionName{mp_compare}};
\node[gp node left,rotate=-90,font={\fontsize{7pt}{8.4pt}\selectfont}] at (22.289,5.460) {\functionIdActual{172}\functionName{ideaCipher}};
\node[gp node left,rotate=-90,font={\fontsize{7pt}{8.4pt}\selectfont}] at (22.528,5.460) {\functionIdActual{173}\functionName{mp_quo_digit}};
\node[gp node left,rotate=-90,font={\fontsize{7pt}{8.4pt}\selectfont}] at (22.768,5.460) {\functionIdActual{174}\functionName{MD5Transform}};
\node[gp node left,rotate=-90,font={\fontsize{7pt}{8.4pt}\selectfont}] at (23.248,5.460) {\functionIdActual{175}\functionName{audspec}};
\node[gp node left,rotate=-90,font={\fontsize{7pt}{8.4pt}\selectfont}] at (23.487,5.460) {\functionIdActual{176}\functionName{det}};
\node[gp node left,rotate=-90,font={\fontsize{7pt}{8.4pt}\selectfont}] at (23.727,5.460) {\functionIdActual{177}\functionName{FORD2}};
\node[gp node left,rotate=-90,font={\fontsize{7pt}{8.4pt}\selectfont}] at (23.967,5.460) {\functionIdActual{178}\functionName{filt}};
\node[gp node left,rotate=-90,font={\fontsize{7pt}{8.4pt}\selectfont}] at (24.206,5.460) {\functionIdActual{179}\functionName{fft_pow}};
\draw[gp path] (1.196,11.881)--(1.196,5.460)--(24.446,5.460)--(24.446,11.881)--cycle;
\gpfill{rgb color={0.000,0.000,0.000}} (1.382,6.744)--(1.491,6.744)--(1.491,6.935)--(1.382,6.935)--cycle;
\gpcolor{rgb color={0.000,0.000,0.000}}
\draw[gp path] (1.382,6.744)--(1.382,6.934)--(1.490,6.934)--(1.490,6.744)--cycle;
\gpfill{rgb color={0.000,0.000,0.000}} (1.621,6.744)--(1.730,6.744)--(1.730,9.423)--(1.621,9.423)--cycle;
\draw[gp path] (1.621,6.744)--(1.621,9.422)--(1.729,9.422)--(1.729,6.744)--cycle;
\gpfill{rgb color={0.000,0.000,0.000}} (2.341,6.744)--(2.449,6.744)--(2.449,7.334)--(2.341,7.334)--cycle;
\draw[gp path] (2.341,6.744)--(2.341,7.333)--(2.448,7.333)--(2.448,6.744)--cycle;
\gpfill{rgb color={0.000,0.000,0.000}} (2.580,6.744)--(2.689,6.744)--(2.689,11.073)--(2.580,11.073)--cycle;
\draw[gp path] (2.580,6.744)--(2.580,11.072)--(2.688,11.072)--(2.688,6.744)--cycle;
\gpfill{rgb color={0.000,0.000,0.000}} (3.060,6.744)--(3.168,6.744)--(3.168,6.745)--(3.060,6.745)--cycle;
\draw[gp path] (3.060,6.744)--(3.167,6.744)--cycle;
\gpfill{rgb color={0.000,0.000,0.000}} (3.299,6.032)--(3.408,6.032)--(3.408,6.745)--(3.299,6.745)--cycle;
\draw[gp path] (3.299,6.744)--(3.299,6.032)--(3.407,6.032)--(3.407,6.744)--cycle;
\gpfill{rgb color={0.000,0.000,0.000}} (3.539,6.744)--(3.648,6.744)--(3.648,6.745)--(3.539,6.745)--cycle;
\draw[gp path] (3.539,6.744)--(3.647,6.744)--cycle;
\gpfill{rgb color={0.000,0.000,0.000}} (3.779,6.744)--(3.888,6.744)--(3.888,7.560)--(3.779,7.560)--cycle;
\draw[gp path] (3.779,6.744)--(3.779,7.559)--(3.887,7.559)--(3.887,6.744)--cycle;
\gpfill{rgb color={0.000,0.000,0.000}} (4.018,6.744)--(4.127,6.744)--(4.127,7.229)--(4.018,7.229)--cycle;
\draw[gp path] (4.018,6.744)--(4.018,7.228)--(4.126,7.228)--(4.126,6.744)--cycle;
\gpfill{rgb color={0.000,0.000,0.000}} (4.498,6.744)--(4.607,6.744)--(4.607,6.745)--(4.498,6.745)--cycle;
\draw[gp path] (4.498,6.744)--(4.606,6.744)--cycle;
\gpfill{rgb color={0.000,0.000,0.000}} (4.737,6.744)--(4.846,6.744)--(4.846,6.745)--(4.737,6.745)--cycle;
\draw[gp path] (4.737,6.744)--(4.845,6.744)--cycle;
\gpfill{rgb color={0.000,0.000,0.000}} (4.977,6.744)--(5.086,6.744)--(5.086,6.910)--(4.977,6.910)--cycle;
\draw[gp path] (4.977,6.744)--(4.977,6.909)--(5.085,6.909)--(5.085,6.744)--cycle;
\gpfill{rgb color={0.000,0.000,0.000}} (5.217,6.744)--(5.326,6.744)--(5.326,7.111)--(5.217,7.111)--cycle;
\draw[gp path] (5.217,6.744)--(5.217,7.110)--(5.325,7.110)--(5.325,6.744)--cycle;
\gpfill{rgb color={0.000,0.000,0.000}} (5.457,6.651)--(5.565,6.651)--(5.565,6.745)--(5.457,6.745)--cycle;
\draw[gp path] (5.457,6.744)--(5.457,6.651)--(5.564,6.651)--(5.564,6.744)--cycle;
\gpfill{rgb color={0.000,0.000,0.000}} (5.936,6.744)--(6.045,6.744)--(6.045,6.745)--(5.936,6.745)--cycle;
\draw[gp path] (5.936,6.744)--(6.044,6.744)--cycle;
\gpfill{rgb color={0.000,0.000,0.000}} (6.176,6.744)--(6.284,6.744)--(6.284,7.357)--(6.176,7.357)--cycle;
\draw[gp path] (6.176,6.744)--(6.176,7.356)--(6.283,7.356)--(6.283,6.744)--cycle;
\gpfill{rgb color={0.000,0.000,0.000}} (6.415,6.744)--(6.524,6.744)--(6.524,6.894)--(6.415,6.894)--cycle;
\draw[gp path] (6.415,6.744)--(6.415,6.893)--(6.523,6.893)--(6.523,6.744)--cycle;
\gpfill{rgb color={0.000,0.000,0.000}} (6.655,6.744)--(6.764,6.744)--(6.764,7.934)--(6.655,7.934)--cycle;
\draw[gp path] (6.655,6.744)--(6.655,7.933)--(6.763,7.933)--(6.763,6.744)--cycle;
\gpfill{rgb color={0.000,0.000,0.000}} (6.895,6.744)--(7.004,6.744)--(7.004,8.350)--(6.895,8.350)--cycle;
\draw[gp path] (6.895,6.744)--(6.895,8.349)--(7.003,8.349)--(7.003,6.744)--cycle;
\gpfill{rgb color={0.000,0.000,0.000}} (7.374,6.744)--(7.483,6.744)--(7.483,6.745)--(7.374,6.745)--cycle;
\draw[gp path] (7.374,6.744)--(7.482,6.744)--cycle;
\gpfill{rgb color={0.000,0.000,0.000}} (7.614,6.744)--(7.723,6.744)--(7.723,7.357)--(7.614,7.357)--cycle;
\draw[gp path] (7.614,6.744)--(7.614,7.356)--(7.722,7.356)--(7.722,6.744)--cycle;
\gpfill{rgb color={0.000,0.000,0.000}} (7.853,6.744)--(7.962,6.744)--(7.962,6.979)--(7.853,6.979)--cycle;
\draw[gp path] (7.853,6.744)--(7.853,6.978)--(7.961,6.978)--(7.961,6.744)--cycle;
\gpfill{rgb color={0.000,0.000,0.000}} (8.093,6.744)--(8.202,6.744)--(8.202,8.420)--(8.093,8.420)--cycle;
\draw[gp path] (8.093,6.744)--(8.093,8.419)--(8.201,8.419)--(8.201,6.744)--cycle;
\gpfill{rgb color={0.000,0.000,0.000}} (8.333,6.744)--(8.442,6.744)--(8.442,8.350)--(8.333,8.350)--cycle;
\draw[gp path] (8.333,6.744)--(8.333,8.349)--(8.441,8.349)--(8.441,6.744)--cycle;
\gpfill{rgb color={0.000,0.000,0.000}} (8.812,6.744)--(8.921,6.744)--(8.921,7.727)--(8.812,7.727)--cycle;
\draw[gp path] (8.812,6.744)--(8.812,7.726)--(8.920,7.726)--(8.920,6.744)--cycle;
\gpfill{rgb color={0.000,0.000,0.000}} (9.052,6.744)--(9.161,6.744)--(9.161,7.830)--(9.052,7.830)--cycle;
\draw[gp path] (9.052,6.744)--(9.052,7.829)--(9.160,7.829)--(9.160,6.744)--cycle;
\gpfill{rgb color={0.000,0.000,0.000}} (9.292,6.689)--(9.400,6.689)--(9.400,6.745)--(9.292,6.745)--cycle;
\draw[gp path] (9.292,6.744)--(9.292,6.689)--(9.399,6.689)--(9.399,6.744)--cycle;
\gpfill{rgb color={0.000,0.000,0.000}} (9.531,6.744)--(9.640,6.744)--(9.640,7.561)--(9.531,7.561)--cycle;
\draw[gp path] (9.531,6.744)--(9.531,7.560)--(9.639,7.560)--(9.639,6.744)--cycle;
\gpfill{rgb color={0.000,0.000,0.000}} (9.771,6.744)--(9.880,6.744)--(9.880,6.745)--(9.771,6.745)--cycle;
\draw[gp path] (9.771,6.744)--(9.879,6.744)--cycle;
\gpfill{rgb color={0.000,0.000,0.000}} (10.250,6.744)--(10.359,6.744)--(10.359,7.411)--(10.250,7.411)--cycle;
\draw[gp path] (10.250,6.744)--(10.250,7.410)--(10.358,7.410)--(10.358,6.744)--cycle;
\gpfill{rgb color={0.000,0.000,0.000}} (10.490,6.744)--(10.599,6.744)--(10.599,7.224)--(10.490,7.224)--cycle;
\draw[gp path] (10.490,6.744)--(10.490,7.223)--(10.598,7.223)--(10.598,6.744)--cycle;
\gpfill{rgb color={0.000,0.000,0.000}} (10.730,6.744)--(10.839,6.744)--(10.839,6.997)--(10.730,6.997)--cycle;
\draw[gp path] (10.730,6.744)--(10.730,6.996)--(10.838,6.996)--(10.838,6.744)--cycle;
\gpfill{rgb color={0.000,0.000,0.000}} (10.969,6.171)--(11.078,6.171)--(11.078,6.745)--(10.969,6.745)--cycle;
\draw[gp path] (10.969,6.744)--(10.969,6.171)--(11.077,6.171)--(11.077,6.744)--cycle;
\gpfill{rgb color={0.000,0.000,0.000}} (11.209,6.744)--(11.318,6.744)--(11.318,6.765)--(11.209,6.765)--cycle;
\draw[gp path] (11.209,6.744)--(11.209,6.764)--(11.317,6.764)--(11.317,6.744)--cycle;
\gpfill{rgb color={0.000,0.000,0.000}} (11.688,6.744)--(11.797,6.744)--(11.797,6.745)--(11.688,6.745)--cycle;
\draw[gp path] (11.688,6.744)--(11.796,6.744)--cycle;
\gpfill{rgb color={0.000,0.000,0.000}} (11.928,6.744)--(12.037,6.744)--(12.037,6.962)--(11.928,6.962)--cycle;
\draw[gp path] (11.928,6.744)--(11.928,6.961)--(12.036,6.961)--(12.036,6.744)--cycle;
\gpfill{rgb color={0.000,0.000,0.000}} (12.168,6.744)--(12.277,6.744)--(12.277,7.643)--(12.168,7.643)--cycle;
\draw[gp path] (12.168,6.744)--(12.168,7.642)--(12.276,7.642)--(12.276,6.744)--cycle;
\gpfill{rgb color={0.000,0.000,0.000}} (12.408,6.744)--(12.516,6.744)--(12.516,7.152)--(12.408,7.152)--cycle;
\draw[gp path] (12.408,6.744)--(12.408,7.151)--(12.515,7.151)--(12.515,6.744)--cycle;
\gpfill{rgb color={0.000,0.000,0.000}} (12.647,6.744)--(12.756,6.744)--(12.756,7.933)--(12.647,7.933)--cycle;
\draw[gp path] (12.647,6.744)--(12.647,7.932)--(12.755,7.932)--(12.755,6.744)--cycle;
\gpfill{rgb color={0.000,0.000,0.000}} (13.127,6.736)--(13.235,6.736)--(13.235,6.745)--(13.127,6.745)--cycle;
\draw[gp path] (13.127,6.744)--(13.127,6.736)--(13.234,6.736)--(13.234,6.744)--cycle;
\gpfill{rgb color={0.000,0.000,0.000}} (13.366,6.744)--(13.475,6.744)--(13.475,7.241)--(13.366,7.241)--cycle;
\draw[gp path] (13.366,6.744)--(13.366,7.240)--(13.474,7.240)--(13.474,6.744)--cycle;
\gpfill{rgb color={0.000,0.000,0.000}} (13.606,6.744)--(13.715,6.744)--(13.715,6.800)--(13.606,6.800)--cycle;
\draw[gp path] (13.606,6.744)--(13.606,6.799)--(13.714,6.799)--(13.714,6.744)--cycle;
\gpfill{rgb color={0.000,0.000,0.000}} (13.846,6.744)--(13.955,6.744)--(13.955,6.745)--(13.846,6.745)--cycle;
\draw[gp path] (13.846,6.744)--(13.954,6.744)--cycle;
\gpfill{rgb color={0.000,0.000,0.000}} (14.085,6.640)--(14.194,6.640)--(14.194,6.745)--(14.085,6.745)--cycle;
\draw[gp path] (14.085,6.744)--(14.085,6.640)--(14.193,6.640)--(14.193,6.744)--cycle;
\gpfill{rgb color={0.000,0.000,0.000}} (14.565,6.744)--(14.674,6.744)--(14.674,6.745)--(14.565,6.745)--cycle;
\draw[gp path] (14.565,6.744)--(14.673,6.744)--cycle;
\gpfill{rgb color={0.000,0.000,0.000}} (14.804,6.744)--(14.913,6.744)--(14.913,6.745)--(14.804,6.745)--cycle;
\draw[gp path] (14.804,6.744)--(14.912,6.744)--cycle;
\gpfill{rgb color={0.000,0.000,0.000}} (15.044,6.744)--(15.153,6.744)--(15.153,6.745)--(15.044,6.745)--cycle;
\draw[gp path] (15.044,6.744)--(15.152,6.744)--cycle;
\gpfill{rgb color={0.000,0.000,0.000}} (15.284,6.744)--(15.393,6.744)--(15.393,6.745)--(15.284,6.745)--cycle;
\draw[gp path] (15.284,6.744)--(15.392,6.744)--cycle;
\gpfill{rgb color={0.000,0.000,0.000}} (15.524,6.744)--(15.632,6.744)--(15.632,7.245)--(15.524,7.245)--cycle;
\draw[gp path] (15.524,6.744)--(15.524,7.244)--(15.631,7.244)--(15.631,6.744)--cycle;
\gpfill{rgb color={0.000,0.000,0.000}} (16.003,6.744)--(16.112,6.744)--(16.112,6.995)--(16.003,6.995)--cycle;
\draw[gp path] (16.003,6.744)--(16.003,6.994)--(16.111,6.994)--(16.111,6.744)--cycle;
\gpfill{rgb color={0.000,0.000,0.000}} (16.243,6.221)--(16.351,6.221)--(16.351,6.745)--(16.243,6.745)--cycle;
\draw[gp path] (16.243,6.744)--(16.243,6.221)--(16.350,6.221)--(16.350,6.744)--cycle;
\gpfill{rgb color={0.000,0.000,0.000}} (16.482,6.744)--(16.591,6.744)--(16.591,6.746)--(16.482,6.746)--cycle;
\draw[gp path] (16.482,6.744)--(16.482,6.745)--(16.590,6.745)--(16.590,6.744)--cycle;
\gpfill{rgb color={0.000,0.000,0.000}} (16.722,6.744)--(16.831,6.744)--(16.831,6.745)--(16.722,6.745)--cycle;
\draw[gp path] (16.722,6.744)--(16.830,6.744)--cycle;
\gpfill{rgb color={0.000,0.000,0.000}} (16.962,6.744)--(17.071,6.744)--(17.071,7.913)--(16.962,7.913)--cycle;
\draw[gp path] (16.962,6.744)--(16.962,7.912)--(17.070,7.912)--(17.070,6.744)--cycle;
\gpfill{rgb color={0.000,0.000,0.000}} (17.441,6.744)--(17.550,6.744)--(17.550,6.936)--(17.441,6.936)--cycle;
\draw[gp path] (17.441,6.744)--(17.441,6.935)--(17.549,6.935)--(17.549,6.744)--cycle;
\gpfill{rgb color={0.000,0.000,0.000}} (17.681,6.542)--(17.790,6.542)--(17.790,6.745)--(17.681,6.745)--cycle;
\draw[gp path] (17.681,6.744)--(17.681,6.542)--(17.789,6.542)--(17.789,6.744)--cycle;
\gpfill{rgb color={0.000,0.000,0.000}} (17.920,6.744)--(18.029,6.744)--(18.029,6.745)--(17.920,6.745)--cycle;
\draw[gp path] (17.920,6.744)--(18.028,6.744)--cycle;
\gpfill{rgb color={0.000,0.000,0.000}} (18.160,6.673)--(18.269,6.673)--(18.269,6.745)--(18.160,6.745)--cycle;
\draw[gp path] (18.160,6.744)--(18.160,6.673)--(18.268,6.673)--(18.268,6.744)--cycle;
\gpfill{rgb color={0.000,0.000,0.000}} (18.400,6.744)--(18.509,6.744)--(18.509,6.839)--(18.400,6.839)--cycle;
\draw[gp path] (18.400,6.744)--(18.400,6.838)--(18.508,6.838)--(18.508,6.744)--cycle;
\gpfill{rgb color={0.000,0.000,0.000}} (18.879,6.744)--(18.988,6.744)--(18.988,6.978)--(18.879,6.978)--cycle;
\draw[gp path] (18.879,6.744)--(18.879,6.977)--(18.987,6.977)--(18.987,6.744)--cycle;
\gpfill{rgb color={0.000,0.000,0.000}} (19.119,6.537)--(19.228,6.537)--(19.228,6.745)--(19.119,6.745)--cycle;
\draw[gp path] (19.119,6.744)--(19.119,6.537)--(19.227,6.537)--(19.227,6.744)--cycle;
\gpfill{rgb color={0.000,0.000,0.000}} (19.359,6.728)--(19.467,6.728)--(19.467,6.745)--(19.359,6.745)--cycle;
\draw[gp path] (19.359,6.744)--(19.359,6.728)--(19.466,6.728)--(19.466,6.744)--cycle;
\gpfill{rgb color={0.000,0.000,0.000}} (19.598,6.744)--(19.707,6.744)--(19.707,7.137)--(19.598,7.137)--cycle;
\draw[gp path] (19.598,6.744)--(19.598,7.136)--(19.706,7.136)--(19.706,6.744)--cycle;
\gpfill{rgb color={0.000,0.000,0.000}} (19.838,6.668)--(19.947,6.668)--(19.947,6.745)--(19.838,6.745)--cycle;
\draw[gp path] (19.838,6.744)--(19.838,6.668)--(19.946,6.668)--(19.946,6.744)--cycle;
\gpfill{rgb color={0.000,0.000,0.000}} (20.317,6.744)--(20.426,6.744)--(20.426,8.256)--(20.317,8.256)--cycle;
\draw[gp path] (20.317,6.744)--(20.317,8.255)--(20.425,8.255)--(20.425,6.744)--cycle;
\gpfill{rgb color={0.000,0.000,0.000}} (20.557,6.744)--(20.666,6.744)--(20.666,7.404)--(20.557,7.404)--cycle;
\draw[gp path] (20.557,6.744)--(20.557,7.403)--(20.665,7.403)--(20.665,6.744)--cycle;
\gpfill{rgb color={0.000,0.000,0.000}} (20.797,6.744)--(20.906,6.744)--(20.906,11.322)--(20.797,11.322)--cycle;
\draw[gp path] (20.797,6.744)--(20.797,11.321)--(20.905,11.321)--(20.905,6.744)--cycle;
\gpfill{rgb color={0.000,0.000,0.000}} (21.036,6.744)--(21.145,6.744)--(21.145,6.745)--(21.036,6.745)--cycle;
\draw[gp path] (21.036,6.744)--(21.144,6.744)--cycle;
\gpfill{rgb color={0.000,0.000,0.000}} (21.276,6.744)--(21.385,6.744)--(21.385,6.745)--(21.276,6.745)--cycle;
\draw[gp path] (21.276,6.744)--(21.384,6.744)--cycle;
\gpfill{rgb color={0.000,0.000,0.000}} (21.755,6.744)--(21.864,6.744)--(21.864,8.256)--(21.755,8.256)--cycle;
\draw[gp path] (21.755,6.744)--(21.755,8.255)--(21.863,8.255)--(21.863,6.744)--cycle;
\gpfill{rgb color={0.000,0.000,0.000}} (21.995,6.744)--(22.104,6.744)--(22.104,6.745)--(21.995,6.745)--cycle;
\draw[gp path] (21.995,6.744)--(22.103,6.744)--cycle;
\gpfill{rgb color={0.000,0.000,0.000}} (22.235,6.744)--(22.344,6.744)--(22.344,6.825)--(22.235,6.825)--cycle;
\draw[gp path] (22.235,6.744)--(22.235,6.824)--(22.343,6.824)--(22.343,6.744)--cycle;
\gpfill{rgb color={0.000,0.000,0.000}} (22.475,6.744)--(22.583,6.744)--(22.583,7.239)--(22.475,7.239)--cycle;
\draw[gp path] (22.475,6.744)--(22.475,7.238)--(22.582,7.238)--(22.582,6.744)--cycle;
\gpfill{rgb color={0.000,0.000,0.000}} (22.714,6.744)--(22.823,6.744)--(22.823,8.560)--(22.714,8.560)--cycle;
\draw[gp path] (22.714,6.744)--(22.714,8.559)--(22.822,8.559)--(22.822,6.744)--cycle;
\gpfill{rgb color={0.000,0.000,0.000}} (23.194,6.744)--(23.302,6.744)--(23.302,7.156)--(23.194,7.156)--cycle;
\draw[gp path] (23.194,6.744)--(23.194,7.155)--(23.301,7.155)--(23.301,6.744)--cycle;
\gpfill{rgb color={0.000,0.000,0.000}} (23.433,6.744)--(23.542,6.744)--(23.542,8.003)--(23.433,8.003)--cycle;
\draw[gp path] (23.433,6.744)--(23.433,8.002)--(23.541,8.002)--(23.541,6.744)--cycle;
\gpfill{rgb color={0.000,0.000,0.000}} (23.673,6.744)--(23.782,6.744)--(23.782,6.745)--(23.673,6.745)--cycle;
\draw[gp path] (23.673,6.744)--(23.781,6.744)--cycle;
\gpfill{rgb color={0.000,0.000,0.000}} (23.913,6.587)--(24.022,6.587)--(24.022,6.745)--(23.913,6.745)--cycle;
\draw[gp path] (23.913,6.744)--(23.913,6.587)--(24.021,6.587)--(24.021,6.744)--cycle;
\gpfill{rgb color={0.000,0.000,0.000}} (24.152,6.744)--(24.261,6.744)--(24.261,6.882)--(24.152,6.882)--cycle;
\draw[gp path] (24.152,6.744)--(24.152,6.881)--(24.260,6.881)--(24.260,6.744)--cycle;
\draw[gp path] (1.196,6.744)--(1.431,6.744)--(1.666,6.744)--(1.901,6.744)--(2.135,6.744)%
  --(2.370,6.744)--(2.605,6.744)--(2.840,6.744)--(3.075,6.744)--(3.310,6.744)--(3.544,6.744)%
  --(3.779,6.744)--(4.014,6.744)--(4.249,6.744)--(4.484,6.744)--(4.719,6.744)--(4.954,6.744)%
  --(5.188,6.744)--(5.423,6.744)--(5.658,6.744)--(5.893,6.744)--(6.128,6.744)--(6.363,6.744)%
  --(6.598,6.744)--(6.832,6.744)--(7.067,6.744)--(7.302,6.744)--(7.537,6.744)--(7.772,6.744)%
  --(8.007,6.744)--(8.241,6.744)--(8.476,6.744)--(8.711,6.744)--(8.946,6.744)--(9.181,6.744)%
  --(9.416,6.744)--(9.651,6.744)--(9.885,6.744)--(10.120,6.744)--(10.355,6.744)--(10.590,6.744)%
  --(10.825,6.744)--(11.060,6.744)--(11.294,6.744)--(11.529,6.744)--(11.764,6.744)--(11.999,6.744)%
  --(12.234,6.744)--(12.469,6.744)--(12.704,6.744)--(12.938,6.744)--(13.173,6.744)--(13.408,6.744)%
  --(13.643,6.744)--(13.878,6.744)--(14.113,6.744)--(14.348,6.744)--(14.582,6.744)--(14.817,6.744)%
  --(15.052,6.744)--(15.287,6.744)--(15.522,6.744)--(15.757,6.744)--(15.991,6.744)--(16.226,6.744)%
  --(16.461,6.744)--(16.696,6.744)--(16.931,6.744)--(17.166,6.744)--(17.401,6.744)--(17.635,6.744)%
  --(17.870,6.744)--(18.105,6.744)--(18.340,6.744)--(18.575,6.744)--(18.810,6.744)--(19.044,6.744)%
  --(19.279,6.744)--(19.514,6.744)--(19.749,6.744)--(19.984,6.744)--(20.219,6.744)--(20.454,6.744)%
  --(20.688,6.744)--(20.923,6.744)--(21.158,6.744)--(21.393,6.744)--(21.628,6.744)--(21.863,6.744)%
  --(22.098,6.744)--(22.332,6.744)--(22.567,6.744)--(22.802,6.744)--(23.037,6.744)--(23.272,6.744)%
  --(23.507,6.744)--(23.741,6.744)--(23.976,6.744)--(24.211,6.744)--(24.446,6.744);
\gpcolor{color=gp lt color border}
\draw[gp path] (1.196,11.881)--(1.196,5.460)--(24.446,5.460)--(24.446,11.881)--cycle;
\node[gp node center] at (23.727,0.138) {\appName{rasta}};
\node[gp node center] at (22.289,0.138) {\appName{pgp-}};
\node[gp node center] at (20.851,0.138) {\appName{pgp+}};
\node[gp node center] at (19.412,0.138) {\appName{pegwit-}};
\node[gp node center] at (17.974,0.138) {\appName{pegwit+}};
\node[gp node center] at (16.536,0.138) {\appName{mpeg2-}};
\node[gp node center] at (15.098,0.138) {\appName{mpeg2+}};
\node[gp node center] at (13.660,0.138) {\appName{jpeg-}};
\node[gp node center] at (12.222,0.138) {\appName{jpeg+}};
\node[gp node center] at (10.784,0.138) {\appName{gsm-}};
\node[gp node center] at (9.345,0.138) {\appName{gsm+}};
\node[gp node center] at (7.907,0.138) {\appName{g721-}};
\node[gp node center] at (6.469,0.138) {\appName{g721+}};
\node[gp node center] at (5.031,0.138) {\appName{epic-}};
\node[gp node center] at (3.593,0.138) {\appName{epic+}};
\node[gp node center] at (2.514,0.138) {\appName{adpcm-}};
\node[gp node center] at (1.556,0.138) {\appName{adpcm+}};
\gpdefrectangularnode{gp plot 1}{\pgfpoint{1.196cm}{5.460cm}}{\pgfpoint{24.446cm}{11.881cm}}
\end{tikzpicture}

%% file: results/multi-speedup.tex
\begin{tikzpicture}[gnuplot]
\path (0.000,0.000) rectangle (12.500,8.750);
\gpcolor{color=gp lt color border}
\gpsetlinetype{gp lt border}
\gpsetdashtype{gp dt solid}
\gpsetlinewidth{1.00}
\draw[gp path] (1.012,1.780)--(1.192,1.780);
\draw[gp path] (11.947,1.780)--(11.767,1.780);
\node[gp node right,font={\fontsize{8pt}{9.6pt}\selectfont}] at (0.828,1.780) {\plotPercentage{-2}};
\draw[gp path] (1.012,2.513)--(1.192,2.513);
\draw[gp path] (11.947,2.513)--(11.767,2.513);
\node[gp node right,font={\fontsize{8pt}{9.6pt}\selectfont}] at (0.828,2.513) {\plotPercentage{0}};
\draw[gp path] (1.012,3.247)--(1.192,3.247);
\draw[gp path] (11.947,3.247)--(11.767,3.247);
\node[gp node right,font={\fontsize{8pt}{9.6pt}\selectfont}] at (0.828,3.247) {\plotPercentage{2}};
\draw[gp path] (1.012,3.980)--(1.192,3.980);
\draw[gp path] (11.947,3.980)--(11.767,3.980);
\node[gp node right,font={\fontsize{8pt}{9.6pt}\selectfont}] at (0.828,3.980) {\plotPercentage{4}};
\draw[gp path] (1.012,4.714)--(1.192,4.714);
\draw[gp path] (11.947,4.714)--(11.767,4.714);
\node[gp node right,font={\fontsize{8pt}{9.6pt}\selectfont}] at (0.828,4.714) {\plotPercentage{6}};
\draw[gp path] (1.012,5.447)--(1.192,5.447);
\draw[gp path] (11.947,5.447)--(11.767,5.447);
\node[gp node right,font={\fontsize{8pt}{9.6pt}\selectfont}] at (0.828,5.447) {\plotPercentage{8}};
\draw[gp path] (1.012,6.181)--(1.192,6.181);
\draw[gp path] (11.947,6.181)--(11.767,6.181);
\node[gp node right,font={\fontsize{8pt}{9.6pt}\selectfont}] at (0.828,6.181) {\plotPercentage{10}};
\draw[gp path] (1.012,6.914)--(1.192,6.914);
\draw[gp path] (11.947,6.914)--(11.767,6.914);
\node[gp node right,font={\fontsize{8pt}{9.6pt}\selectfont}] at (0.828,6.914) {\plotPercentage{12}};
\draw[gp path] (1.012,7.648)--(1.192,7.648);
\draw[gp path] (11.947,7.648)--(11.767,7.648);
\node[gp node right,font={\fontsize{8pt}{9.6pt}\selectfont}] at (0.828,7.648) {\plotPercentage{14}};
\draw[gp path] (1.012,8.381)--(1.192,8.381);
\draw[gp path] (11.947,8.381)--(11.767,8.381);
\node[gp node right,font={\fontsize{8pt}{9.6pt}\selectfont}] at (0.828,8.381) {\plotPercentage{16}};
\node[gp node left,rotate=-90,font={\fontsize{7pt}{8.4pt}\selectfont}] at (1.977,1.780) {\appHistName{adpcm+}};
\node[gp node left,rotate=-90,font={\fontsize{7pt}{8.4pt}\selectfont}] at (2.758,1.780) {\appHistName{adpcm-}};
\node[gp node left,rotate=-90,font={\fontsize{7pt}{8.4pt}\selectfont}] at (3.539,1.780) {\appHistName{g721+}};
\node[gp node left,rotate=-90,font={\fontsize{7pt}{8.4pt}\selectfont}] at (4.320,1.780) {\appHistName{g721-}};
\node[gp node left,rotate=-90,font={\fontsize{7pt}{8.4pt}\selectfont}] at (5.101,1.780) {\appHistName{gsm+}};
\node[gp node left,rotate=-90,font={\fontsize{7pt}{8.4pt}\selectfont}] at (5.882,1.780) {\appHistName{gsm-}};
\node[gp node left,rotate=-90,font={\fontsize{7pt}{8.4pt}\selectfont}] at (6.664,1.780) {\appHistName{jpeg+}};
\node[gp node left,rotate=-90,font={\fontsize{7pt}{8.4pt}\selectfont}] at (7.445,1.780) {\appHistName{jpeg-}};
\node[gp node left,rotate=-90,font={\fontsize{7pt}{8.4pt}\selectfont}] at (8.226,1.780) {\appHistName{mpeg2-}};
\node[gp node left,rotate=-90,font={\fontsize{7pt}{8.4pt}\selectfont}] at (9.007,1.780) {\appHistName{pegwit+}};
\node[gp node left,rotate=-90,font={\fontsize{7pt}{8.4pt}\selectfont}] at (9.788,1.780) {\appHistName{pegwit-}};
\node[gp node left,rotate=-90,font={\fontsize{7pt}{8.4pt}\selectfont}] at (10.569,1.780) {\appHistName{pgp+}};
\node[gp node left,rotate=-90,font={\fontsize{7pt}{8.4pt}\selectfont}] at (11.350,1.780) {\appHistName{pgp-}};
\draw[gp path] (1.012,8.381)--(1.012,1.780)--(11.947,1.780)--(11.947,8.381)--cycle;
\gpfill{rgb color={0.000,0.000,0.000}} (1.806,2.513)--(2.041,2.513)--(2.041,3.576)--(1.806,3.576)--cycle;
\gpcolor{rgb color={0.000,0.000,0.000}}
\draw[gp path] (1.806,2.513)--(1.806,3.575)--(2.040,3.575)--(2.040,2.513)--cycle;
\gpfill{rgb color={0.000,0.000,0.000}} (2.587,2.513)--(2.822,2.513)--(2.822,5.638)--(2.587,5.638)--cycle;
\draw[gp path] (2.587,2.513)--(2.587,5.637)--(2.821,5.637)--(2.821,2.513)--cycle;
\gpfill{rgb color={0.000,0.000,0.000}} (3.368,2.513)--(3.604,2.513)--(3.604,3.366)--(3.368,3.366)--cycle;
\draw[gp path] (3.368,2.513)--(3.368,3.365)--(3.603,3.365)--(3.603,2.513)--cycle;
\gpfill{rgb color={0.000,0.000,0.000}} (4.149,2.513)--(4.385,2.513)--(4.385,3.475)--(4.149,3.475)--cycle;
\draw[gp path] (4.149,2.513)--(4.149,3.474)--(4.384,3.474)--(4.384,2.513)--cycle;
\gpfill{rgb color={0.000,0.000,0.000}} (4.930,2.513)--(5.166,2.513)--(5.166,6.336)--(4.930,6.336)--cycle;
\draw[gp path] (4.930,2.513)--(4.930,6.335)--(5.165,6.335)--(5.165,2.513)--cycle;
\gpfill{rgb color={0.000,0.000,0.000}} (5.711,2.513)--(5.947,2.513)--(5.947,5.433)--(5.711,5.433)--cycle;
\draw[gp path] (5.711,2.513)--(5.711,5.432)--(5.946,5.432)--(5.946,2.513)--cycle;
\gpfill{rgb color={0.000,0.000,0.000}} (6.493,2.513)--(6.728,2.513)--(6.728,4.554)--(6.493,4.554)--cycle;
\draw[gp path] (6.493,2.513)--(6.493,4.553)--(6.727,4.553)--(6.727,2.513)--cycle;
\gpfill{rgb color={0.000,0.000,0.000}} (7.274,2.513)--(7.509,2.513)--(7.509,3.300)--(7.274,3.300)--cycle;
\draw[gp path] (7.274,2.513)--(7.274,3.299)--(7.508,3.299)--(7.508,2.513)--cycle;
\gpfill{rgb color={0.000,0.000,0.000}} (8.055,2.513)--(8.290,2.513)--(8.290,2.897)--(8.055,2.897)--cycle;
\draw[gp path] (8.055,2.513)--(8.055,2.896)--(8.289,2.896)--(8.289,2.513)--cycle;
\gpfill{rgb color={0.000,0.000,0.000}} (8.836,2.456)--(9.071,2.456)--(9.071,2.514)--(8.836,2.514)--cycle;
\draw[gp path] (8.836,2.513)--(8.836,2.456)--(9.070,2.456)--(9.070,2.513)--cycle;
\gpfill{rgb color={0.000,0.000,0.000}} (9.617,2.513)--(9.852,2.513)--(9.852,2.612)--(9.617,2.612)--cycle;
\draw[gp path] (9.617,2.513)--(9.617,2.611)--(9.851,2.611)--(9.851,2.513)--cycle;
\gpfill{rgb color={0.000,0.000,0.000}} (10.398,2.513)--(10.633,2.513)--(10.633,7.835)--(10.398,7.835)--cycle;
\draw[gp path] (10.398,2.513)--(10.398,7.834)--(10.632,7.834)--(10.632,2.513)--cycle;
\gpfill{rgb color={0.000,0.000,0.000}} (11.179,2.513)--(11.414,2.513)--(11.414,8.303)--(11.179,8.303)--cycle;
\draw[gp path] (11.179,2.513)--(11.179,8.302)--(11.413,8.302)--(11.413,2.513)--cycle;
\draw[gp path] (1.012,2.513)--(1.122,2.513)--(1.233,2.513)--(1.343,2.513)--(1.454,2.513)%
  --(1.564,2.513)--(1.675,2.513)--(1.785,2.513)--(1.896,2.513)--(2.006,2.513)--(2.117,2.513)%
  --(2.227,2.513)--(2.337,2.513)--(2.448,2.513)--(2.558,2.513)--(2.669,2.513)--(2.779,2.513)%
  --(2.890,2.513)--(3.000,2.513)--(3.111,2.513)--(3.221,2.513)--(3.332,2.513)--(3.442,2.513)%
  --(3.552,2.513)--(3.663,2.513)--(3.773,2.513)--(3.884,2.513)--(3.994,2.513)--(4.105,2.513)%
  --(4.215,2.513)--(4.326,2.513)--(4.436,2.513)--(4.547,2.513)--(4.657,2.513)--(4.767,2.513)%
  --(4.878,2.513)--(4.988,2.513)--(5.099,2.513)--(5.209,2.513)--(5.320,2.513)--(5.430,2.513)%
  --(5.541,2.513)--(5.651,2.513)--(5.762,2.513)--(5.872,2.513)--(5.982,2.513)--(6.093,2.513)%
  --(6.203,2.513)--(6.314,2.513)--(6.424,2.513)--(6.535,2.513)--(6.645,2.513)--(6.756,2.513)%
  --(6.866,2.513)--(6.977,2.513)--(7.087,2.513)--(7.197,2.513)--(7.308,2.513)--(7.418,2.513)%
  --(7.529,2.513)--(7.639,2.513)--(7.750,2.513)--(7.860,2.513)--(7.971,2.513)--(8.081,2.513)%
  --(8.192,2.513)--(8.302,2.513)--(8.412,2.513)--(8.523,2.513)--(8.633,2.513)--(8.744,2.513)%
  --(8.854,2.513)--(8.965,2.513)--(9.075,2.513)--(9.186,2.513)--(9.296,2.513)--(9.407,2.513)%
  --(9.517,2.513)--(9.627,2.513)--(9.738,2.513)--(9.848,2.513)--(9.959,2.513)--(10.069,2.513)%
  --(10.180,2.513)--(10.290,2.513)--(10.401,2.513)--(10.511,2.513)--(10.622,2.513)--(10.732,2.513)%
  --(10.842,2.513)--(10.953,2.513)--(11.063,2.513)--(11.174,2.513)--(11.284,2.513)--(11.395,2.513)%
  --(11.505,2.513)--(11.616,2.513)--(11.726,2.513)--(11.837,2.513)--(11.947,2.513);
\gpcolor{color=gp lt color border}
\draw[gp path] (1.012,8.381)--(1.012,1.780)--(11.947,1.780)--(11.947,8.381)--cycle;
\gpdefrectangularnode{gp plot 1}{\pgfpoint{1.012cm}{1.780cm}}{\pgfpoint{11.947cm}{8.381cm}}
\end{tikzpicture}

%% file: results/speedup-multi-correlation.tex
\begin{tikzpicture}[gnuplot]
\path (0.000,0.000) rectangle (12.500,8.750);
\gpcolor{color=gp lt color border}
\gpsetlinetype{gp lt border}
\gpsetdashtype{gp dt solid}
\gpsetlinewidth{1.00}
\draw[gp path] (1.504,0.985)--(1.684,0.985);
\draw[gp path] (11.947,0.985)--(11.767,0.985);
\node[gp node right,font={\fontsize{8pt}{9.6pt}\selectfont}] at (1.320,0.985) {\plotPercentage{-20}};
\draw[gp path] (1.504,1.657)--(1.684,1.657);
\draw[gp path] (11.947,1.657)--(11.767,1.657);
\node[gp node right,font={\fontsize{8pt}{9.6pt}\selectfont}] at (1.320,1.657) {\plotPercentage{-10}};
\draw[gp path] (1.504,2.330)--(1.684,2.330);
\draw[gp path] (11.947,2.330)--(11.767,2.330);
\node[gp node right,font={\fontsize{8pt}{9.6pt}\selectfont}] at (1.320,2.330) {\plotPercentage{0}};
\draw[gp path] (1.504,3.002)--(1.684,3.002);
\draw[gp path] (11.947,3.002)--(11.767,3.002);
\node[gp node right,font={\fontsize{8pt}{9.6pt}\selectfont}] at (1.320,3.002) {\plotPercentage{10}};
\draw[gp path] (1.504,3.674)--(1.684,3.674);
\draw[gp path] (11.947,3.674)--(11.767,3.674);
\node[gp node right,font={\fontsize{8pt}{9.6pt}\selectfont}] at (1.320,3.674) {\plotPercentage{20}};
\draw[gp path] (1.504,4.347)--(1.684,4.347);
\draw[gp path] (11.947,4.347)--(11.767,4.347);
\node[gp node right,font={\fontsize{8pt}{9.6pt}\selectfont}] at (1.320,4.347) {\plotPercentage{30}};
\draw[gp path] (1.504,5.019)--(1.684,5.019);
\draw[gp path] (11.947,5.019)--(11.767,5.019);
\node[gp node right,font={\fontsize{8pt}{9.6pt}\selectfont}] at (1.320,5.019) {\plotPercentage{40}};
\draw[gp path] (1.504,5.692)--(1.684,5.692);
\draw[gp path] (11.947,5.692)--(11.767,5.692);
\node[gp node right,font={\fontsize{8pt}{9.6pt}\selectfont}] at (1.320,5.692) {\plotPercentage{50}};
\draw[gp path] (1.504,6.364)--(1.684,6.364);
\draw[gp path] (11.947,6.364)--(11.767,6.364);
\node[gp node right,font={\fontsize{8pt}{9.6pt}\selectfont}] at (1.320,6.364) {\plotPercentage{60}};
\draw[gp path] (1.504,7.036)--(1.684,7.036);
\draw[gp path] (11.947,7.036)--(11.767,7.036);
\node[gp node right,font={\fontsize{8pt}{9.6pt}\selectfont}] at (1.320,7.036) {\plotPercentage{70}};
\draw[gp path] (1.504,7.709)--(1.684,7.709);
\draw[gp path] (11.947,7.709)--(11.767,7.709);
\node[gp node right,font={\fontsize{8pt}{9.6pt}\selectfont}] at (1.320,7.709) {\plotPercentage{80}};
\draw[gp path] (1.504,8.381)--(1.684,8.381);
\draw[gp path] (11.947,8.381)--(11.767,8.381);
\node[gp node right,font={\fontsize{8pt}{9.6pt}\selectfont}] at (1.320,8.381) {\plotPercentage{90}};
\draw[gp path] (1.504,0.985)--(1.504,1.165);
\draw[gp path] (1.504,8.381)--(1.504,8.201);
\node[gp node center,font={\fontsize{8pt}{9.6pt}\selectfont}] at (1.504,0.677) {\plotPercentage{0}};
\draw[gp path] (2.809,0.985)--(2.809,1.165);
\draw[gp path] (2.809,8.381)--(2.809,8.201);
\node[gp node center,font={\fontsize{8pt}{9.6pt}\selectfont}] at (2.809,0.677) {\plotPercentage{10}};
\draw[gp path] (4.115,0.985)--(4.115,1.165);
\draw[gp path] (4.115,8.381)--(4.115,8.201);
\node[gp node center,font={\fontsize{8pt}{9.6pt}\selectfont}] at (4.115,0.677) {\plotPercentage{20}};
\draw[gp path] (5.420,0.985)--(5.420,1.165);
\draw[gp path] (5.420,8.381)--(5.420,8.201);
\node[gp node center,font={\fontsize{8pt}{9.6pt}\selectfont}] at (5.420,0.677) {\plotPercentage{30}};
\draw[gp path] (6.726,0.985)--(6.726,1.165);
\draw[gp path] (6.726,8.381)--(6.726,8.201);
\node[gp node center,font={\fontsize{8pt}{9.6pt}\selectfont}] at (6.726,0.677) {\plotPercentage{40}};
\draw[gp path] (8.031,0.985)--(8.031,1.165);
\draw[gp path] (8.031,8.381)--(8.031,8.201);
\node[gp node center,font={\fontsize{8pt}{9.6pt}\selectfont}] at (8.031,0.677) {\plotPercentage{50}};
\draw[gp path] (9.336,0.985)--(9.336,1.165);
\draw[gp path] (9.336,8.381)--(9.336,8.201);
\node[gp node center,font={\fontsize{8pt}{9.6pt}\selectfont}] at (9.336,0.677) {\plotPercentage{60}};
\draw[gp path] (10.642,0.985)--(10.642,1.165);
\draw[gp path] (10.642,8.381)--(10.642,8.201);
\node[gp node center,font={\fontsize{8pt}{9.6pt}\selectfont}] at (10.642,0.677) {\plotPercentage{70}};
\draw[gp path] (11.947,0.985)--(11.947,1.165);
\draw[gp path] (11.947,8.381)--(11.947,8.201);
\node[gp node center,font={\fontsize{8pt}{9.6pt}\selectfont}] at (11.947,0.677) {\plotPercentage{80}};
\draw[gp path] (1.504,8.381)--(1.504,0.985)--(11.947,0.985)--(11.947,8.381)--cycle;
\node[gp node center,font={\fontsize{12pt}{14.4pt}\selectfont}] at (4.637,6.902) {\plotLegend{(underestimation)}};
\node[gp node center,font={\fontsize{12pt}{14.4pt}\selectfont}] at (7.248,3.204) {\plotLegend{(overestimation)}};
\node[gp node center,rotate=27,font={\fontsize{12pt}{14.4pt}\selectfont}] at (7.926,6.384) {\plotLegend{perfect estimation}};
\node[gp node center,rotate=-270,font={\fontsize{12pt}{14.4pt}\selectfont}] at (0.246,4.683) {\plotAxisLabel{actual speedup}};
\node[gp node center,font={\fontsize{12pt}{14.4pt}\selectfont}] at (6.725,0.215) {\plotAxisLabel{estimated speedup}};
\gpcolor{rgb color={0.000,0.000,0.000}}
\draw[gp path] (1.504,2.330)--(1.606,2.382)--(1.709,2.435)--(1.811,2.488)--(1.913,2.541)%
  --(2.016,2.593)--(2.118,2.646)--(2.221,2.699)--(2.323,2.752)--(2.425,2.804)--(2.528,2.857)%
  --(2.630,2.910)--(2.732,2.962)--(2.835,3.015)--(2.937,3.068)--(3.040,3.121)--(3.142,3.173)%
  --(3.244,3.226)--(3.347,3.279)--(3.449,3.332)--(3.551,3.384)--(3.654,3.437)--(3.756,3.490)%
  --(3.859,3.542)--(3.961,3.595)--(4.063,3.648)--(4.166,3.701)--(4.268,3.753)--(4.370,3.806)%
  --(4.473,3.859)--(4.575,3.912)--(4.678,3.964)--(4.780,4.017)--(4.882,4.070)--(4.985,4.123)%
  --(5.087,4.175)--(5.189,4.228)--(5.292,4.281)--(5.394,4.333)--(5.497,4.386)--(5.599,4.439)%
  --(5.701,4.492)--(5.804,4.544)--(5.906,4.597)--(6.008,4.650)--(6.111,4.703)--(6.213,4.755)%
  --(6.315,4.808)--(6.418,4.861)--(6.520,4.913)--(6.623,4.966)--(6.725,5.019)--(6.827,5.072)%
  --(6.930,5.124)--(7.032,5.177)--(7.134,5.230)--(7.237,5.283)--(7.339,5.335)--(7.442,5.388)%
  --(7.544,5.441)--(7.646,5.493)--(7.749,5.546)--(7.851,5.599)--(7.953,5.652)--(8.056,5.704)%
  --(8.158,5.757)--(8.261,5.810)--(8.363,5.863)--(8.465,5.915)--(8.568,5.968)--(8.670,6.021)%
  --(8.772,6.073)--(8.875,6.126)--(8.977,6.179)--(9.080,6.232)--(9.182,6.284)--(9.284,6.337)%
  --(9.387,6.390)--(9.489,6.443)--(9.591,6.495)--(9.694,6.548)--(9.796,6.601)--(9.899,6.654)%
  --(10.001,6.706)--(10.103,6.759)--(10.206,6.812)--(10.308,6.864)--(10.410,6.917)--(10.513,6.970)%
  --(10.615,7.023)--(10.717,7.075)--(10.820,7.128)--(10.922,7.181)--(11.025,7.234)--(11.127,7.286)%
  --(11.229,7.339)--(11.332,7.392)--(11.434,7.444)--(11.536,7.497)--(11.639,7.550);
\gpcolor{color=gp lt color border}
\node[gp node right,font={\fontsize{12pt}{14.4pt}\selectfont}] at (10.257,1.840) {\plotShiftRight{\plotLegend{original processor}}};
\gpcolor{rgb color={0.000,0.000,0.000}}
\gpsetpointsize{3.60}
\gppoint{gp mark 7}{(1.504,2.330)}
\gppoint{gp mark 7}{(1.504,2.330)}
\gppoint{gp mark 7}{(1.504,2.330)}
\gppoint{gp mark 7}{(1.504,2.330)}
\gppoint{gp mark 7}{(1.504,2.330)}
\gppoint{gp mark 7}{(1.504,2.330)}
\gppoint{gp mark 7}{(1.504,2.330)}
\gppoint{gp mark 7}{(1.504,2.351)}
\gppoint{gp mark 7}{(1.504,2.330)}
\gppoint{gp mark 7}{(1.504,2.330)}
\gppoint{gp mark 7}{(1.504,2.330)}
\gppoint{gp mark 7}{(1.504,2.330)}
\gppoint{gp mark 7}{(1.504,2.330)}
\gppoint{gp mark 7}{(1.504,2.330)}
\gppoint{gp mark 7}{(1.504,2.330)}
\gppoint{gp mark 7}{(1.504,2.330)}
\gppoint{gp mark 7}{(1.504,2.330)}
\gppoint{gp mark 7}{(1.504,2.330)}
\gppoint{gp mark 7}{(1.504,2.330)}
\gppoint{gp mark 7}{(1.514,1.782)}
\gppoint{gp mark 7}{(1.517,2.118)}
\gppoint{gp mark 7}{(1.517,2.113)}
\gppoint{gp mark 7}{(1.532,2.473)}
\gppoint{gp mark 7}{(1.640,3.270)}
\gppoint{gp mark 7}{(1.650,2.232)}
\gppoint{gp mark 7}{(1.650,1.584)}
\gppoint{gp mark 7}{(1.657,2.330)}
\gppoint{gp mark 7}{(1.657,2.313)}
\gppoint{gp mark 7}{(1.767,2.759)}
\gppoint{gp mark 7}{(1.778,2.593)}
\gppoint{gp mark 7}{(1.839,2.530)}
\gppoint{gp mark 7}{(1.839,2.574)}
\gppoint{gp mark 7}{(1.880,2.428)}
\gppoint{gp mark 7}{(1.884,2.575)}
\gppoint{gp mark 7}{(1.884,2.486)}
\gppoint{gp mark 7}{(1.974,2.557)}
\gppoint{gp mark 7}{(1.986,2.740)}
\gppoint{gp mark 7}{(2.137,2.413)}
\gppoint{gp mark 7}{(2.144,3.647)}
\gppoint{gp mark 7}{(2.165,3.573)}
\gppoint{gp mark 7}{(2.223,2.321)}
\gppoint{gp mark 7}{(2.300,3.027)}
\gppoint{gp mark 7}{(2.316,1.730)}
\gppoint{gp mark 7}{(2.335,2.756)}
\gppoint{gp mark 7}{(2.342,2.387)}
\gppoint{gp mark 7}{(2.371,2.255)}
\gppoint{gp mark 7}{(2.371,2.250)}
\gppoint{gp mark 7}{(2.402,2.165)}
\gppoint{gp mark 7}{(2.435,2.272)}
\gppoint{gp mark 7}{(2.445,2.591)}
\gppoint{gp mark 7}{(2.451,2.832)}
\gppoint{gp mark 7}{(2.508,2.847)}
\gppoint{gp mark 7}{(2.515,2.528)}
\gppoint{gp mark 7}{(2.540,3.183)}
\gppoint{gp mark 7}{(2.570,2.713)}
\gppoint{gp mark 7}{(2.636,3.184)}
\gppoint{gp mark 7}{(2.690,2.853)}
\gppoint{gp mark 7}{(2.761,2.220)}
\gppoint{gp mark 7}{(2.806,2.849)}
\gppoint{gp mark 7}{(3.243,2.502)}
\gppoint{gp mark 7}{(3.281,2.946)}
\gppoint{gp mark 7}{(3.419,3.466)}
\gppoint{gp mark 7}{(3.438,2.331)}
\gppoint{gp mark 7}{(3.536,3.020)}
\gppoint{gp mark 7}{(3.640,2.970)}
\gppoint{gp mark 7}{(3.640,2.970)}
\gppoint{gp mark 7}{(3.721,2.837)}
\gppoint{gp mark 7}{(3.800,3.358)}
\gppoint{gp mark 7}{(4.291,3.552)}
\gppoint{gp mark 7}{(4.668,3.912)}
\gppoint{gp mark 7}{(4.668,3.912)}
\gppoint{gp mark 7}{(4.909,4.230)}
\gppoint{gp mark 7}{(5.521,4.011)}
\gppoint{gp mark 7}{(5.521,4.011)}
\gppoint{gp mark 7}{(6.455,4.084)}
\gppoint{gp mark 7}{(6.538,3.575)}
\gppoint{gp mark 7}{(8.031,5.133)}
\gppoint{gp mark 7}{(10.206,6.861)}
\gppoint{gp mark 7}{(11.639,7.123)}
\gppoint{gp mark 7}{(11.010,1.840)}
\gpcolor{color=gp lt color border}
\node[gp node right,font={\fontsize{12pt}{14.4pt}\selectfont}] at (10.257,1.390) {\plotShiftRight{\plotLegend{ideal processor}}};
\gpcolor{rgb color={0.000,0.000,0.000}}
\gppoint{gp mark 6}{(1.504,2.330)}
\gppoint{gp mark 6}{(1.504,2.330)}
\gppoint{gp mark 6}{(1.504,2.330)}
\gppoint{gp mark 6}{(1.504,2.330)}
\gppoint{gp mark 6}{(1.504,2.330)}
\gppoint{gp mark 6}{(1.504,2.330)}
\gppoint{gp mark 6}{(1.504,2.330)}
\gppoint{gp mark 6}{(1.504,2.330)}
\gppoint{gp mark 6}{(1.504,2.330)}
\gppoint{gp mark 6}{(1.504,2.330)}
\gppoint{gp mark 6}{(1.504,2.330)}
\gppoint{gp mark 6}{(1.504,2.330)}
\gppoint{gp mark 6}{(1.504,2.330)}
\gppoint{gp mark 6}{(1.504,2.330)}
\gppoint{gp mark 6}{(1.504,2.330)}
\gppoint{gp mark 6}{(1.504,2.330)}
\gppoint{gp mark 6}{(1.504,2.330)}
\gppoint{gp mark 6}{(1.504,2.330)}
\gppoint{gp mark 6}{(1.504,2.330)}
\gppoint{gp mark 6}{(1.514,2.380)}
\gppoint{gp mark 6}{(1.517,2.338)}
\gppoint{gp mark 6}{(1.517,2.338)}
\gppoint{gp mark 6}{(1.532,2.272)}
\gppoint{gp mark 6}{(1.640,2.360)}
\gppoint{gp mark 6}{(1.650,3.273)}
\gppoint{gp mark 6}{(1.650,2.236)}
\gppoint{gp mark 6}{(1.657,2.408)}
\gppoint{gp mark 6}{(1.657,2.408)}
\gppoint{gp mark 6}{(1.767,2.459)}
\gppoint{gp mark 6}{(1.778,2.471)}
\gppoint{gp mark 6}{(1.839,2.606)}
\gppoint{gp mark 6}{(1.839,2.606)}
\gppoint{gp mark 6}{(1.880,2.700)}
\gppoint{gp mark 6}{(1.884,2.549)}
\gppoint{gp mark 6}{(1.884,2.474)}
\gppoint{gp mark 6}{(2.371,2.565)}
\gppoint{gp mark 6}{(1.974,2.574)}
\gppoint{gp mark 6}{(1.986,2.517)}
\gppoint{gp mark 6}{(2.137,2.555)}
\gppoint{gp mark 6}{(2.144,2.710)}
\gppoint{gp mark 6}{(2.165,3.546)}
\gppoint{gp mark 6}{(2.223,2.634)}
\gppoint{gp mark 6}{(2.300,2.729)}
\gppoint{gp mark 6}{(2.316,2.256)}
\gppoint{gp mark 6}{(2.335,2.781)}
\gppoint{gp mark 6}{(2.342,2.452)}
\gppoint{gp mark 6}{(2.371,2.565)}
\gppoint{gp mark 6}{(2.402,2.628)}
\gppoint{gp mark 6}{(2.435,2.576)}
\gppoint{gp mark 6}{(2.445,2.624)}
\gppoint{gp mark 6}{(2.451,2.893)}
\gppoint{gp mark 6}{(2.508,2.847)}
\gppoint{gp mark 6}{(2.515,2.707)}
\gppoint{gp mark 6}{(2.540,2.330)}
\gppoint{gp mark 6}{(2.570,3.077)}
\gppoint{gp mark 6}{(2.636,2.912)}
\gppoint{gp mark 6}{(2.690,2.847)}
\gppoint{gp mark 6}{(2.761,5.003)}
\gppoint{gp mark 6}{(2.806,3.027)}
\gppoint{gp mark 6}{(3.243,3.230)}
\gppoint{gp mark 6}{(3.281,3.361)}
\gppoint{gp mark 6}{(3.419,3.280)}
\gppoint{gp mark 6}{(3.438,2.331)}
\gppoint{gp mark 6}{(3.536,2.748)}
\gppoint{gp mark 6}{(3.640,3.338)}
\gppoint{gp mark 6}{(3.640,3.339)}
\gppoint{gp mark 6}{(3.721,7.543)}
\gppoint{gp mark 6}{(3.800,3.995)}
\gppoint{gp mark 6}{(4.291,3.701)}
\gppoint{gp mark 6}{(4.668,3.972)}
\gppoint{gp mark 6}{(4.668,3.972)}
\gppoint{gp mark 6}{(4.909,4.084)}
\gppoint{gp mark 6}{(5.521,4.399)}
\gppoint{gp mark 6}{(5.521,4.399)}
\gppoint{gp mark 6}{(6.455,4.683)}
\gppoint{gp mark 6}{(6.538,4.668)}
\gppoint{gp mark 6}{(8.031,5.669)}
\gppoint{gp mark 6}{(10.206,6.755)}
\gppoint{gp mark 6}{(11.639,7.757)}
\gppoint{gp mark 6}{(11.010,1.390)}
\gpcolor{color=gp lt color border}
\draw[gp path] (1.504,8.381)--(1.504,0.985)--(11.947,0.985)--(11.947,8.381)--cycle;
\gpdefrectangularnode{gp plot 1}{\pgfpoint{1.504cm}{0.985cm}}{\pgfpoint{11.947cm}{8.381cm}}
\end{tikzpicture}

%% file: results/functions.tex
\begin{tabular}{llcccccc}
\hline
\textbf{id} & \textbf{name (app)} & \textbf{I} & \textbf{I/B} & \textbf{RP} & \textbf{ILP} & \textbf{CR} & \textbf{CS}\\\hline
1 & \path{handle_noinline_attr..} (\path{gcc}) & 28 & 7 & 0 & 2.2 & 3.6 & 44 \\
2 & \path{control_flow_insn_p} (\path{gcc}) & 78 & 7 & 0 & 2 & 5.1 & 144 \\
3 & \path{insert_insn_on_edge} (\path{gcc}) & 55 & 7 & 0 & 1.5 & 11.1 & 49 \\
4 & \path{update_br_prob_note} (\path{gcc}) & 49 & 5 & 0 & 1.8 & 4.1 & 145 \\
5 & \path{_cpp_init_internal_p..} (\path{gcc}) & 49 & 49 & 0 & 2.1 & 10.4 & 42 \\
6 & \path{lex_macro_node} (\path{gcc}) & 77 & 7 & 0 & 1.8 & 6.5 & 172 \\
7 & \path{cse_basic_block} (\path{gcc}) & 779 & 5 & 2.9 & 1.8 & 4 & 62 \\
8 & \path{rtx_equal_for_cselib_p} (\path{gcc}) & 299 & 5 & 1.6 & 1.6 & 3 & 83 \\
9 & \path{debug_df_chain} (\path{gcc}) & 49 & 12 & 0 & 2.2 & 8.3 & 146 \\
10 & \path{modified_type_die} (\path{gcc}) & 743 & 7 & 0.4 & 1.7 & 5.1 & 85 \\
11 & \path{emit_note} (\path{gcc}) & 87 & 6 & 0 & 1.4 & 2.3 & 85 \\
12 & \path{gen_sequence} (\path{gcc}) & 70 & 6 & 0 & 1.9 & 2.9 & 125 \\
13 & \path{subreg_hard_regno} (\path{gcc}) & 88 & 8 & 0 & 1.6 & 7.7 & 108 \\
14 & \path{split_double} (\path{gcc}) & 142 & 8 & 0 & 1.7 & 5 & 125 \\
15 & \path{add_to_mem_set_list} (\path{gcc}) & 59 & 6 & 0 & 1.6 & 3.4 & 142 \\
16 & \path{find_regno_partial} (\path{gcc}) & 52 & 4 & 0 & 1.4 & 0 & 91 \\
17 & \path{use_return_register} (\path{gcc}) & 72 & 6 & 0 & 1.4 & 5.7 & 95 \\
18 & \path{ix86_expand_move} (\path{gcc}) & 347 & 6 & 0 & 1.6 & 4.9 & 89 \\
19 & \path{legitimate_pic_addre..} (\path{gcc}) & 208 & 5 & 0.1 & 1.6 & 1 & 81 \\
20 & \path{gen_extendsfdf2} (\path{gcc}) & 62 & 15 & 0 & 1.9 & 11.7 & 93 \\
21 & \path{gen_mulsidi3} (\path{gcc}) & 76 & 75 & 0 & 2.9 & 10.7 & 87 \\
22 & \path{gen_peephole2_1255} (\path{gcc}) & 83 & 82 & 0 & 2.5 & 12.4 & 66 \\
23 & \path{gen_peephole2_1271} (\path{gcc}) & 102 & 101 & 0 & 2.3 & 13.1 & 68 \\
24 & \path{gen_peephole2_1277} (\path{gcc}) & 153 & 51 & 0 & 2.4 & 12 & 51 \\
25 & \path{gen_pfnacc} (\path{gcc}) & 145 & 144 & 0 & 2.7 & 11.2 & 55 \\
26 & \path{gen_rotlsi3} (\path{gcc}) & 32 & 31 & 0 & 2.3 & 13 & 120 \\
27 & \path{gen_split_1001} (\path{gcc}) & 264 & 29 & 0 & 2 & 11.5 & 33 \\
28 & \path{gen_split_1028} (\path{gcc}) & 271 & 269 & 0 & 2.5 & 12.7 & 17 \\
29 & \path{gen_sse_nandti3} (\path{gcc}) & 33 & 32 & 0 & 3 & 9.5 & 126 \\
30 & \path{gen_sunge} (\path{gcc}) & 34 & 10 & 0 & 2.1 & 15.2 & 39 \\
31 & \path{insert_loop_mem} (\path{gcc}) & 117 & 6 & 0 & 1.7 & 1.7 & 133 \\
32 & \path{eiremain} (\path{gcc}) & 236 & 9 & 0.5 & 1.9 & 2.7 & 88 \\
33 & \path{elimination_effects} (\path{gcc}) & 522 & 5 & 1.3 & 1.7 & 1.6 & 69 \\
34 & \path{gen_reload} (\path{gcc}) & 503 & 9 & 0.2 & 1.7 & 7.4 & 93 \\
35 & \path{reload_cse_simplify_..} (\path{gcc}) & 222 & 7 & 1.7 & 1.7 & 5.9 & 80 \\
36 & \path{simplify_binary_is2o..} (\path{gcc}) & 62 & 62 & 0.8 & 2.1 & 3.2 & 18 \\
37 & \path{remove_phi_alternative} (\path{gcc}) & 43 & 5 & 0 & 1.7 & 0 & 145 \\
38 & \path{contains_placeholder_p} (\path{gcc}) & 202 & 4 & 0.1 & 1.5 & 4.5 & 93 \\
39 & \path{assemble_end_function} (\path{gcc}) & 179 & 10 & 0 & 1.7 & 8.5 & 116 \\
40 & \path{default_named_sectio..} (\path{gcc}) & 72 & 13 & 0 & 1.9 & 8.4 & 155 \\
41 & \path{sample_unpac..} (\path{ghostscript}) & 98 & 7 & 0 & 2.1 & 0 & 61 \\
42 & \path{autohelperattpat10} (\path{gobmk}) & 46 & 11 & 0.3 & 2 & 6.6 & 175 \\
43 & \path{autohelperbarriers..} (\path{gobmk}) & 111 & 15 & 2 & 2.2 & 4.5 & 54 \\
44 & \path{atari_atari_attack..} (\path{gobmk}) & 515 & 6 & 3.3 & 1.7 & 3.8 & 73 \\
45 & \path{compute_aa_status} (\path{gobmk}) & 206 & 5 & 4.3 & 1.6 & 2.9 & 84 \\
46 & \path{dragon_weak} (\path{gobmk}) & 63 & 9 & 0 & 1.7 & 2.7 & 124 \\
47 & \path{get_saved_worms} (\path{gobmk}) & 135 & 7 & 6.1 & 1.7 & 3.1 & 46 \\
48 & \path{read_eye} (\path{gobmk}) & 170 & 7 & 2.3 & 1.8 & 2.6 & 95 \\
49 & \path{topological_eye} (\path{gobmk}) & 465 & 7 & 7.1 & 1.9 & 2.3 & 54 \\
50 & \path{autohelperowl_atta..} (\path{gobmk}) & 61 & 12 & 0 & 2.2 & 4.9 & 218 \\
\end{tabular}
&
\begin{tabular}{llcccccc}
\hline
\textbf{id} & \textbf{name (app)} & \textbf{I} & \textbf{I/B} & \textbf{RP} & \textbf{ILP} & \textbf{CR} & \textbf{CS}\\\hline
51 & \path{autohelperowl_atta..} (\path{gobmk}) & 53 & 14 & 0 & 2.1 & 3.8 & 110 \\
52 & \path{autohelperowl_defe..} (\path{gobmk}) & 34 & 34 & 0 & 2.4 & 8.9 & 108 \\
53 & \path{autohelperowl_defe..} (\path{gobmk}) & 30 & 8 & 0 & 2.3 & 6.7 & 180 \\
54 & \path{autohelperpat1114} (\path{gobmk}) & 42 & 12 & 0 & 1.9 & 4.8 & 140 \\
55 & \path{autohelperpat335} (\path{gobmk}) & 30 & 30 & 0 & 2.7 & 3.4 & 178 \\
56 & \path{autohelperpat508} (\path{gobmk}) & 27 & 27 & 0 & 2.4 & 3.7 & 142 \\
57 & \path{autohelperpat83} (\path{gobmk}) & 53 & 16 & 0 & 2.4 & 3.8 & 148 \\
58 & \path{simple_showboard} (\path{gobmk}) & 208 & 8 & 2.9 & 1.8 & 5.8 & 68 \\
59 & \path{skip_intrabk_SAD} (\path{h264ref}) & 318 & 11 & 0 & 1.6 & 0 & 37 \\
60 & \path{free_orig_planes} (\path{h264ref}) & 74 & 12 & 0 & 1.6 & 9.7 & 62 \\
61 & \path{GetSkipCostMB} (\path{h264ref}) & 248 & 16 & 11.1 & 1.9 & 2.3 & 31 \\
62 & \path{writeSyntaxEleme..} (\path{h264ref}) & 99 & 8 & 0 & 2 & 0 & 76 \\
63 & \path{GSIAddKeyToIndex} (\path{hmmer}) & 88 & 12 & 0 & 1.7 & 5.8 & 75 \\
64 & \path{EVDBasicFit} (\path{hmmer}) & 229 & 17 & 0.2 & 2 & 5.1 & 42 \\
65 & \path{SampleDirichlet} (\path{hmmer}) & 99 & 10 & 0 & 2.1 & 8.5 & 81 \\
66 & \path{DegenerateSymbolSc..} (\path{hmmer}) & 98 & 11 & 0.1 & 2.1 & 4.7 & 64 \\
67 & \path{Plan7SetCtime} (\path{hmmer}) & 38 & 12 & 0 & 1.6 & 13.6 & 40 \\
68 & \path{MSAToSqinfo} (\path{hmmer}) & 229 & 8 & 0.5 & 1.6 & 5.7 & 60 \\
69 & \path{null_convert} (\path{jpeg}) & 110 & 6 & 0.3 & 2.7 & 0 & 29 \\
70 & \path{jinit_c_prep_contro..} (\path{jpeg}) & 193 & 13 & 3.3 & 2.3 & 3.8 & 49 \\
71 & \path{glFogf} (\path{mesa}) & 38 & 8 & 0 & 1.6 & 7.9 & 151 \\
72 & \path{glNormal3d} (\path{mesa}) & 64 & 59 & 0.5 & 2.4 & 5.5 & 51 \\
73 & \path{glRasterPos3d} (\path{mesa}) & 52 & 10 & 0 & 2 & 7.7 & 146 \\
74 & \path{glTexCoord2d} (\path{mesa}) & 24 & 24 & 0 & 2.4 & 6.7 & 97 \\
75 & \path{gl_stippled_bresenham} (\path{mesa}) & 223 & 11 & 6.5 & 2.2 & 1.3 & 59 \\
76 & \path{gl_save_Frustum} (\path{mesa}) & 109 & 10 & 0.2 & 2.1 & 5.4 & 110 \\
77 & \path{gl_save_LineWidth} (\path{mesa}) & 78 & 7 & 0 & 1.8 & 5.2 & 114 \\
78 & \path{translate_id} (\path{mesa}) & 68 & 5 & 0 & 1.8 & 0.6 & 93 \\
79 & \path{gl_Map1f} (\path{mesa}) & 341 & 6 & 0.1 & 1.9 & 5 & 41 \\
80 & \path{smooth_ci_line} (\path{mesa}) & 186 & 12 & 2.5 & 2.1 & 3.1 & 60 \\
81 & \path{free_unified_knots} (\path{mesa}) & 39 & 4 & 0 & 1.5 & 10.5 & 77 \\
82 & \path{tess_test_polygon} (\path{mesa}) & 641 & 7 & 1.3 & 1.9 & 4.7 & 31 \\
83 & \path{auxWireBox} (\path{mesa}) & 122 & 12 & 0.1 & 2.6 & 8.4 & 26 \\
84 & \path{gl_ColorPointer} (\path{mesa}) & 94 & 6 & 0 & 1.5 & 3.3 & 45 \\
85 & \path{r_serial} (\path{milc}) & 536 & 15 & 4.1 & 2 & 7.3 & 34 \\
86 & \path{scalar_mult_sub_su3..} (\path{milc}) & 156 & 137 & 0 & 2.6 & 4.2 & 14 \\
87 & \path{Decode_MPEG1_Non_I..} (\path{mpeg2}) & 238 & 6 & 3.7 & 1.7 & 3.8 & 64 \\
88 & \path{cpDecodeSecret} (\path{pegwit}) & 23 & 23 & 0 & 2.6 & 13.3 & 31 \\
89 & \path{vlShortLshift} (\path{pegwit}) & 83 & 6 & 0 & 2.3 & 1.3 & 41 \\
90 & \path{encryptfile} (\path{pgp}) & 411 & 13 & 0.8 & 1.9 & 7.4 & 46 \\
91 & \path{make_canonical} (\path{pgp}) & 79 & 18 & 0 & 2 & 11.6 & 75 \\
92 & \path{LANG} (\path{pgp}) & 658 & 10 & 1.4 & 1.8 & 7.7 & 28 \\
93 & \path{MD5Transform} (\path{pgp}) & 551 & 546 & 4.2 & 1.7 & 0 & 5 \\
94 & \path{mp_display} (\path{pgp}) & 255 & 13 & 0.8 & 2 & 9.2 & 55 \\
95 & \path{comp_Jboundaries} (\path{rasta}) & 45 & 9 & 0 & 1.8 & 5.1 & 71 \\
96 & \path{is_draw} (\path{sjeng}) & 92 & 6 & 0 & 1.8 & 0 & 64 \\
97 & \path{push_king} (\path{sjeng}) & 103 & 9 & 0 & 1.4 & 0 & 29 \\
98 & \path{stat_retry} (\path{sphinx3}) & 70 & 9 & 0 & 1.8 & 8.6 & 81 \\
99 & \path{lextree_subtree_..} (\path{sphinx3}) & 146 & 11 & 0 & 2.2 & 7.6 & 63 \\
100 & \path{lm_tg_score} (\path{sphinx3}) & 261 & 7 & 0.2 & 1.7 & 3.5 & 45 \\
\end{tabular}

%% file: results/mediabench-functions.tex
\begin{tabular}{llcccccc}
\hline
\textbf{id} & \textbf{name} & \textbf{I} & \textbf{I/B} & \textbf{RP} & \textbf{ILP} & \textbf{CR} & \textbf{EX}\\\hline
\rowcolor{tblrow} \multicolumn{8}{l}{\path{adpcm+}}\\
\rowcolor{white}
101 & \path{adpcm_coder} & 87 & 8 & 0 & 2.1 & 0 & 96.9 \\
\rowcolor{white}
102 & \path{main} & 83 & 14 & 0 & 1.7 & 10.8 & 0.1 \\
\rowcolor{tblrow} \multicolumn{8}{l}{\path{adpcm-}}\\
\rowcolor{white}
103 & \path{adpcm_decoder} & 66 & 11 & 0 & 1.9 & 0 & 95.2 \\
\rowcolor{white}
104 & \path{main} & 75 & 12 & 0 & 1.6 & 10.7 & 0.1 \\
\rowcolor{tblrow} \multicolumn{8}{l}{\path{epic+}}\\
\rowcolor{white}
105 & \path{quantize_image} & 751 & 40 & 0.5 & 2.3 & 11.9 & 0.2 \\
\rowcolor{white}
106 & \path{run_length_encode_zeros} & 123 & 7 & 0.8 & 1.8 & 0.8 & 0.1 \\
\rowcolor{white}
107 & \path{encode_stream} & 67 & 5 & 0 & 1.7 & 3 & 0.1 \\
\rowcolor{white}
108 & \path{ReadMatrixFromPGMStream} & 191 & 11 & 0 & 1.9 & 8.9 & 0.1 \\
\rowcolor{white}
109 & \path{main} & 775 & 60 & 0 & 2.2 & 11.5 & 0 \\
\rowcolor{tblrow} \multicolumn{8}{l}{\path{epic-}}\\
\rowcolor{white}
110 & \path{main} & 567 & 24 & 0.1 & 1.9 & 10.1 & 1.6 \\
\rowcolor{white}
111 & \path{unquantize_image} & 301 & 9 & 0 & 1.7 & 11.3 & 1.1 \\
\rowcolor{white}
112 & \path{read_and_huffman_decode} & 236 & 8 & 0 & 1.7 & 2.1 & 0.8 \\
\rowcolor{white}
113 & \path{write_pgm_image} & 130 & 13 & 0 & 2.1 & 6.9 & 0.6 \\
\rowcolor{white}
114 & \path{internal_int_transpose} & 203 & 5 & 0 & 2 & 4.4 & 0.5 \\
\rowcolor{tblrow} \multicolumn{8}{l}{\path{g721+}}\\
\rowcolor{white}
115 & \path{quan} & 21 & 4 & 0 & 1.8 & 0 & 46.4 \\
\rowcolor{white}
116 & \path{fmult} & 54 & 18 & 0 & 2.1 & 1.9 & 19.3 \\
\rowcolor{white}
117 & \path{update} & 383 & 7 & 0.5 & 2 & 0.8 & 11.8 \\
\rowcolor{white}
118 & \path{g721_encoder} & 94 & 16 & 0 & 1.9 & 8.5 & 3.2 \\
\rowcolor{white}
119 & \path{predictor_zero} & 61 & 61 & 0 & 2.3 & 9.8 & 2 \\
\rowcolor{tblrow} \multicolumn{8}{l}{\path{g721-}}\\
\rowcolor{white}
120 & \path{quan} & 21 & 4 & 0 & 1.8 & 0 & 44.5 \\
\rowcolor{white}
121 & \path{fmult} & 54 & 18 & 0 & 2.1 & 1.9 & 19.4 \\
\rowcolor{white}
122 & \path{update} & 383 & 7 & 0.5 & 2 & 0.8 & 11.7 \\
\rowcolor{white}
123 & \path{g721_decoder} & 87 & 14 & 0 & 1.8 & 8 & 2.8 \\
\rowcolor{white}
124 & \path{predictor_zero} & 61 & 61 & 0 & 2.3 & 9.8 & 2 \\
\rowcolor{tblrow} \multicolumn{8}{l}{\path{gsm+}}\\
\rowcolor{white}
125 & \path{Calculation_of_the_LTP_pa..} & 504 & 14 & 5.5 & 2 & 1.8 & 40.8 \\
\rowcolor{white}
126 & \path{Short_term_analysis_filte..} & 121 & 20 & 0 & 2.6 & 0 & 25.3 \\
\rowcolor{white}
127 & \path{Gsm_Preprocess} & 107 & 7 & 1.2 & 1.6 & 2.8 & 8.8 \\
\rowcolor{white}
128 & \path{Weighting_filter} & 61 & 20 & 0 & 2.1 & 0 & 3.9 \\
\rowcolor{white}
129 & \path{Autocorrelation} & 675 & 22 & 0.2 & 2.5 & 0.6 & 3.8 \\
\rowcolor{tblrow} \multicolumn{8}{l}{\path{gsm-}}\\
\rowcolor{white}
130 & \path{Short_term_synthesis_filt..} & 109 & 6 & 0 & 1.7 & 0 & 73.7 \\
\rowcolor{white}
131 & \path{Gsm_Long_Term_Synthesis_F..} & 158 & 14 & 0 & 1.9 & 1.3 & 6 \\
\rowcolor{white}
132 & \path{Postprocessing} & 119 & 40 & 0 & 2.2 & 0 & 5.7 \\
\rowcolor{white}
133 & \path{APCM_inverse_quantization} & 92 & 10 & 0 & 1.7 & 7.6 & 2.5 \\
\rowcolor{white}
134 & \path{gsm_asr} & 22 & 3 & 0 & 1.8 & 0 & 1 \\
\rowcolor{tblrow} \multicolumn{8}{l}{\path{jpeg+}}\\
\rowcolor{white}
135 & \path{forward_DCT} & 287 & 9 & 0 & 1.7 & 3.1 & 22.5 \\
\rowcolor{white}
136 & \path{rgb_ycc_convert} & 146 & 15 & 0 & 2.1 & 0 & 21.1 \\
\rowcolor{white}
137 & \path{encode_one_block} & 145 & 7 & 0 & 1.9 & 4.1 & 17.6 \\
\rowcolor{white}
138 & \path{jpeg_fdct_islow} & 135 & 27 & 0 & 2.7 & 0 & 14.5 \\
\rowcolor{white}
139 & \path{emit_bits} & 82 & 7 & 0 & 1.7 & 3.7 & 8.5 \\
\end{tabular}
&
\begin{tabular}{llcccccc}
\hline
\textbf{id} & \textbf{name} & \textbf{I} & \textbf{I/B} & \textbf{RP} & \textbf{ILP} & \textbf{CR} & \textbf{EX}\\\hline
\rowcolor{tblrow} \multicolumn{8}{l}{\path{jpeg-}}\\
\rowcolor{white}
140 & \path{ycc_rgb_convert} & 158 & 14 & 0 & 1.7 & 0 & 29.3 \\
\rowcolor{white}
141 & \path{jpeg_idct_islow} & 235 & 21 & 0.9 & 2.1 & 0 & 28.9 \\
\rowcolor{white}
142 & \path{h2v2_fancy_upsample} & 384 & 16 & 0 & 2.3 & 0 & 12.5 \\
\rowcolor{white}
143 & \path{decode_mcu} & 413 & 8 & 1.4 & 1.8 & 2.4 & 12.1 \\
\rowcolor{white}
144 & \path{jpeg_fill_bit_buffer} & 118 & 6 & 0 & 1.5 & 2.5 & 3.6 \\
\rowcolor{tblrow} \multicolumn{8}{l}{\path{mpeg2+}}\\
\rowcolor{white}
145 & \path{fdct} & 772 & 86 & 1 & 2.5 & 14.2 & 0.8 \\
\rowcolor{white}
146 & \path{fullsearch} & 192 & 8 & 6.2 & 1.8 & 1.6 & 0.4 \\
\rowcolor{white}
147 & \path{dist1} & 337 & 13 & 0 & 3 & 0 & 0.2 \\
\rowcolor{white}
148 & \path{putbits} & 149 & 8 & 0 & 2.2 & 3.4 & 0.2 \\
\rowcolor{white}
149 & \path{calcSNR1} & 552 & 61 & 0.8 & 2.5 & 14.3 & 0.1 \\
\rowcolor{tblrow} \multicolumn{8}{l}{\path{mpeg2-}}\\
\rowcolor{white}
150 & \path{conv420to422} & 260 & 22 & 3.2 & 2.7 & 0 & 20.1 \\
\rowcolor{white}
151 & \path{form_component_prediction} & 313 & 6 & 0 & 2.2 & 0 & 13.8 \\
\rowcolor{white}
152 & \path{putbyte} & 20 & 7 & 0 & 1.5 & 5 & 13.7 \\
\rowcolor{white}
153 & \path{Add_Block} & 288 & 16 & 0 & 1.9 & 0 & 10.2 \\
\rowcolor{white}
154 & \path{idctcol} & 103 & 26 & 0 & 1.7 & 0 & 8.9 \\
\rowcolor{tblrow} \multicolumn{8}{l}{\path{pegwit+}}\\
\rowcolor{white}
155 & \path{gfAddMul} & 179 & 6 & 0 & 1.5 & 0.6 & 42.6 \\
\rowcolor{white}
156 & \path{gfMultiply} & 334 & 6 & 0 & 1.6 & 2.1 & 29.7 \\
\rowcolor{white}
157 & \path{squareEncrypt} & 451 & 451 & 0 & 2.6 & 0 & 6.1 \\
\rowcolor{white}
158 & \path{gfInvert} & 253 & 11 & 0 & 1.8 & 8.3 & 3.2 \\
\rowcolor{white}
159 & \path{gfSquare} & 191 & 5 & 0 & 1.5 & 2.1 & 2.5 \\
\rowcolor{tblrow} \multicolumn{8}{l}{\path{pegwit-}}\\
\rowcolor{white}
160 & \path{gfAddMul} & 179 & 6 & 0 & 1.5 & 0.6 & 41.1 \\
\rowcolor{white}
161 & \path{gfMultiply} & 334 & 6 & 0 & 1.6 & 2.1 & 28.6 \\
\rowcolor{white}
162 & \path{squareDecrypt} & 451 & 451 & 0 & 2.6 & 0 & 11.4 \\
\rowcolor{white}
163 & \path{gfInit} & 143 & 11 & 0 & 2 & 2.1 & 4.2 \\
\rowcolor{white}
164 & \path{gfInvert} & 253 & 11 & 0 & 1.8 & 8.3 & 3.1 \\
\rowcolor{tblrow} \multicolumn{8}{l}{\path{pgp+}}\\
\rowcolor{white}
165 & \path{mp_smul} & 30 & 8 & 0 & 1.7 & 0 & 50.6 \\
\rowcolor{white}
166 & \path{longest_match} & 80 & 5 & 0 & 1.5 & 0 & 13.2 \\
\rowcolor{white}
167 & \path{fill_window} & 255 & 23 & 8 & 2.5 & 0.8 & 4.3 \\
\rowcolor{white}
168 & \path{deflate} & 331 & 11 & 2.2 & 2.2 & 2.4 & 3.7 \\
\rowcolor{white}
169 & \path{mp_compare} & 31 & 6 & 0 & 2 & 0 & 3.5 \\
\rowcolor{tblrow} \multicolumn{8}{l}{\path{pgp-}}\\
\rowcolor{white}
170 & \path{mp_smul} & 30 & 8 & 0 & 1.7 & 0 & 66.2 \\
\rowcolor{white}
171 & \path{mp_compare} & 31 & 6 & 0 & 2 & 0 & 4.6 \\
\rowcolor{white}
172 & \path{ideaCipher} & 204 & 6 & 0 & 1.8 & 0 & 4 \\
\rowcolor{white}
173 & \path{mp_quo_digit} & 30 & 30 & 0 & 1.9 & 0 & 2.8 \\
\rowcolor{white}
174 & \path{MD5Transform} & 484 & 484 & 8 & 1.7 & 0 & 2.5 \\
\rowcolor{tblrow} \multicolumn{8}{l}{\path{rasta}}\\
\rowcolor{white}
175 & \path{audspec} & 330 & 12 & 0.1 & 2.7 & 7.3 & 0.1 \\
\rowcolor{white}
176 & \path{det} & 149 & 25 & 1.2 & 2.7 & 6 & 0.1 \\
\rowcolor{white}
177 & \path{FORD2} & 353 & 7 & 10.8 & 1.8 & 0 & 0.1 \\
\rowcolor{white}
178 & \path{filt} & 410 & 12 & 0.2 & 2.5 & 6.8 & 0.1 \\
\rowcolor{white}
179 & \path{fft_pow} & 443 & 15 & 0 & 2.1 & 9 & 0.1 \\
\end{tabular}